\documentclass[12pt,fleqn,final]{osudissert96-single}
 
\usepackage{amsthm,mathrsfs}
\usepackage{subfigure}
 
\usepackage{mathrsfs}
\usepackage[active]{srcltx}
\usepackage{epsfig}
\usepackage{amssymb}
\usepackage{amsfonts}
\usepackage{latexsym}
\usepackage{graphicx}
\usepackage{amsmath}
\usepackage{pdfsync}
\usepackage{amsfonts}
\usepackage{amsthm}
\usepackage{bm}
\usepackage{epstopdf}
\usepackage{longtable}
\usepackage{float}
\usepackage{cite}

\usepackage[margin=0.8in]{geometry}

\def\em{\it}

\def\la{\langle}
\def\ra{\rangle}

\def\Np{N_{\mathrm{part}}}
\newcommand{\dNdeta}{dN_\mathrm{ch}/d\eta}

\newcommand{\Tdec}{T_\mathrm{dec}}

\newcommand{\Tchem}{T_\mathrm{chem}}

\def\ve{\varepsilon}
\def\vt#1{v_#1\{2\}}
\def\vf#1{v_#1\{4\}}
\def\et#1{\ve_#1\{2\}}
\def\ef#1{\ve_#1\{4\}}

\def\ecc{{\varepsilon}}
\def\epart{{\ecc_{\mathrm{part}}}}
\def\Ecc{\mathcal{E}}

\def\EP{\mathrm{EP}}
\def\Np{N_{\mathrm{part}}}
\def\pT{p_\mathrm{T}}

\def\ve{\varepsilon}
\def\vt#1{v_#1\{2\}}
\def\vf#1{v_#1\{4\}}
\def\et#1{\ve_#1\{2\}}
\def\ef#1{\ve_#1\{4\}}

\def\EP{\mathrm{EP}}
\def\Np{N_{\mathrm{part}}}
\def\pT{p_\mathrm{T}}

\def\ve{\varepsilon}
\def\vt#1{v_{#1}\{2\}}
\def\vf#1{v_{#1}\{4\}}
\def\et#1{\ve_{#1}\{2\}}
\def\ef#1{\ve_{#1}\{4\}}

\def\sec#1{Sec.~\ref{sec:#1}}
\def\chap#1{Chap.~\ref{chap:#1}}

\newcommand{\qtable}[1]{Table~\ref{tab:#1}}
\newcommand{\eq}{{\,=\,}}

\def\fig#1{Fig.~\ref{fig:#1}}
\def\Fig#1{Fig.~\ref{fig:#1}}
\def\equ#1{eq.~(\ref{eq:#1})}
\def\Equ#1{Eq.~(\ref{eq:#1})}

\def\la{\langle}
\def\ra{\rangle}

\def\ecc{{\varepsilon}}
\def\epart{{\ecc_{\mathrm{part}}}}

\def\PsiPP#1{\Phi_{#1}}
\def\PsiEP#1{\Psi_{#1}}
\def\PsiPPP#1{\Phi'_{#1}}

\def\xT{x_\perp}
\def\vxT{\vec{x}_\perp}
\def\pT{p_\perp}
\def\vpT{\vec{p}_\perp}
\def\esp{\eta_s'}
\def\ts{\tilde{\sigma}}
\def\foi{l_{\mr{conv}}}
\def\pit{\pi^{\mu\nu}}
\def\Emax{E_\mr{max}}
\def\Gmax{G_\mr{max}}

\def\mb#1{\mbox{#1}}
\def\mr#1{\mathrm{#1}}
\def\itt#1{{\it#1}}

\def\iEBEURL{https://bitbucket.org/qiu\_24/iebe}

\usepackage{color}
 
\title{Event-by-event Hydrodynamic Simulations for Relativistic Heavy-ion Collisions}
\author{Zhi Qiu}
\authordegrees{B.S., Ph.D.}  
\unit{Physics}
 
\advisorname{Professor Ulrich W. Heinz}
 
\member{Professor Yuri Kovchegov}
 
\member{Professor Eric Braaten}
 
\member{Professor Chris Hammel}
 
\begin{document}
 
 
\maketitle
 
\disscopyright
 
\begin{abstract}
 
\hspace{.2in}
 
 In this thesis, I show my Ph.D. work on event-by-event hydrodynamic simulations for relativistic heavy-ion collision. I show that event-by-event hydrodynamic simulations have become an indispensable tool for studying relativistic heavy-ion collisions and how it can be used to explain many phenomena.
 
 In \chap{1104.0650}, I compare the previously dominating single-shot hydrodynamics with event-by-event hydrodynamic simulations which are now becoming mainstream. The event-by-event simulations are more realistic, but they are also very time consuming; the single-shot simulations are economical, but then the question arises as to whether they can be used as a sufficiently precise replacement for event-by-event simulations. I will compare these two simulation types for two popular initial condition models.
 
 I show that for the event averages of the multiplicity and elliptic and triangular flows, the time consuming event-by-event hydrodynamic simulations can, to a good approximation, be replaced by single-shot ones, when using properly constructed initial conditions. For higher-order flows such as $v_{4,5}$ the single-shot simulations are shown to be incapable of reproducing those from event-by-event simulations.
 
 In \chap{e2e3}, we show that the elliptic and triangular flow data measured by the ALICE collaboration at the LHC prefer a small specific shear viscosity close to $\eta/s=0.08$, when considering the MC-Glauber and MC-KLN  models. In order to allow for a much larger $\eta/s$ value, the initial condition model must feature triangularity values $\sim 50\%$ larger than the ones provided by the MC-Glauber and MC-KLN models.
 
 \chap{correlation} focuses on correlations between event-plane angles. We show that the event-plane angle correlation measurements by the ATLAS collaboration can be explained by hydrodynamic simulations. The same correlation patterns cannot be explained directly from the initial conditions.
 
 In \chap{resonance}, we show that including only $\sim20$ out of $319$ carefully chosen resonances can already yield spectra and flow results within $1\%$ relative error. Such a treatment can be used to shorten the currently lengthy ($\sim$ 3 hours) resonance decay calculations by a factor of ten.
 
 \chap{flow_angle} focuses on various ways of calculating and measuring anisotropic flows, and their comparisons. We point out that the event planes angle $\PsiEP{n}$ fluctuates from event to event, and that the differential event-plane angle $\PsiEP{n}(p_T)-\PsiEP{n}$ also fluctuates from event to event. We show that traces of such angular fluctuations can be measured experimentally, and explain how such measurements allow the study of the fluctuation of the flow orientations in addition to the fluctuation of the flow magnitudes.
 
 In \chap{iSS} we give a thorough description of the general sampling methodology, and show that it can be applied in an efficient way to sample particles whose emission is determined by the Cooper-Frye formula, using the simplification of longitudinal boost invariance.

\end{abstract}

\newpage
\dedication{\it Dedicated to my family}
 
\begin{acknowledgements}
 
\hspace{.2in}
 
First, I want to sincerely thank my advisor, Prof. Ulrich W. Heinz, for all his selfless teaching and support during my Ph.D. period. Not only did I learn a great amount of knowledge from him, but I also acquired the ability to learn, to think, and to question by myself, to talk and to communicate with people in a professional and concise way, to be strict and cautious in research but respectful to people's work at the same time, and last but not least, the attitude that ``one should either not do something or give it his best". His wisdom, sense of humor, strictness to his students but also constant support at the same time, all make an invaluable mix from which I, as his student, benefited a lot. I feel so lucky to have him as my advisor.
  
Next I want to thank all other professors in the Physics Department at OSU, especially those from whom I took courses and those on my Ph.D. committee, for providing such a friendly, active, and supportive environment to all students. I enjoyed very much taking classes from various professors, questioning them or being questioned by them, sharing their visions on the current status and the future of physics. Their energy and friendliness formed another unforgettable experience.

I also want to thank all my friends, especially Chun Shen; our constant discussions and collaborations were a great help to my work.

Finally I want to thank my family. Their consideration for my situation and their unceasing support is what made my work possible.

\end{acknowledgements}
 
\begin{vita}
  \dateitem{Sep.2000-Sep.2004}{Bachelor in Mathematics\\
  University of Science and Technology of China}
  
  \dateitem{Sep.2004-Sep.2009}{Ph.D. in Mathematics\\
  Research and Teaching Assistant\\
  Art and Science College \\
  The Ohio State University}
  
  \dateitem{Sep.2009-Aug.2013}{Research and Teaching Assistant\\
  Art and Science College \\
  The Ohio State University}
 
\begin{publist}
	\pubitem{Zhi Qiu, Ulrich W. Heinz
	{\em Event-by-event shape and flow fluctuations of relativistic heavy-ion collision fireballs}
	Physics Review C84 (2011) 024911}
	
	\pubitem{Zhi Qiu, Chun Shen, Ulrich Heinz
	{\em Hydrodynamic elliptic and triangular flow in Pb-Pb collisions at sqrt(s)=2.76ATeV}
	Physics Letter B707 (2012) 151-155}
	
	\pubitem{Chun Shen, Steffen A. Bass, Tetsufumi Hirano, Pasi Huovinen, Zhi Qiu, Huichao Song, Ulrich Heinz
	{\em The QGP shear viscosity: Elusive goal or just around the corner?}
	Journal of Physics, G38 (2011) 124045}
	
	\pubitem{Zhi Qiu, Ulrich Heinz
	{\em Hydrodynamic event-plane correlations in Pb+Pb collisions at sqrt(s)=2.76ATeV} Physics Letter B717 (2012) 261-265}
	
	\pubitem{Zhi Qiu, Ulrich Heinz
	{\em Event-by-event hydrodynamics for heavy-ion collisions}
	AIP Conference Proceeding, 1441 (2012) 774-776}
	
	\pubitem{J.Scott Moreland, Zhi Qiu, Ulrich W. Heinz
	{\em Imprinting quantum fluctuations on hydrodynamic initial conditions}
	e-Print: arXiv:1210.5508}
	
	\pubitem{Zhi Qiu, Chun Shen, Ulrich W. Heinz
	{\em Resonance decay contributions to higher-order anisotropic flow coefficients} Phys.Rev. C86 (2012) 064906}
	
	\pubitem{Ovidiu Costin, Min Huang, Zhi Qiu
	{\em Ionization in damped time-harmonic fields}
	Journal of Physics. A42 (2009) 325202}
\end{publist}
 
\begin{fieldsstudy}
    \majorfield{Physics}
  \end{fieldsstudy}
\end{vita}

\tableofcontents


\chapter{Introduction}
\label{chap:introduction}

\section{Disclaimer}
\label{sec:disklaimer}

I would like to start with a quote from the book \cite{MathematicaForTheoreticalPhysics}:
 
 ``The aim of classical mechanics and theoretical physics is to provide and develop a self-consistent mathematical structure which runs so closely parallel to the development of physical phenomena that, starting from a minimum number of hypotheses, it may be used to accurately describe and even predict the results of all carefully controlled experiments.''
 
I quote it here to as a reminder that the development of theoretical physics is the process of improving the {\em mathematical model} behind the material world; thus, however successful the current achievement in theoretical physics becomes, whether it can be categorized as ``the truth'' depends on how well it stands the test of time. In this thesis, I use the words ``truth'', ``nature'', etc., to mean the current state-of-art understanding of theoretical physics.

\section{Relativistic heavy-ion collision experiments}
\label{sec:explanation}

``Heavy ions'' are the atomic nuclei of atoms with mass number $A \geq 4$. The relativistic heavy-ion collision program was initiated partially in the hope to study a new type of matter, the quark-gluon plasma, which otherwise can only be found shortly after the big bang or (possibly) in the core of neutron starts. In this section a brief introduction is given to make the thesis self-contained; for a more thorough motivation and background of heavy-ion collision experiments, see for example \cite{Yagi:2005yb}, \cite{Heinz:2004qz}, and \cite{Heinz:2013wva}.

Relativistic heavy-ion collision experiments are mainly performed at the Relativistic Heavy-Ion Collider (RHIC) at Brookhaven National Lab (BNL) near New York and at the Large Hadron Collider (LHC) operated by European Organization for Nuclear Research (CERN) near Geneva. In relativistic heavy-ion collision experiments, one generates beams of bare heavy nuclei (``heavy ions'') and accelerates them to close to the speed of light: RHIC: $v>99\%c$; LHC: $v>99.9\%c$.

The two beams moving around the accelerator ring anti-parallel to each other are brought to collision in the so-called interaction regions. Those interaction regions are instrumented with detectors. Both RHIC and LHC have multiple detectors. Each detector is a complicated engineering piece of art that is a product of, and operated and maintained by a large collaboration of physicists and engineers (ranging from $\sim$ 500 to $\sim$ 3500 members each).

The collision systems that have been run at RHIC include proton-proton (p+p), deuterium-gold (d+Au), copper-copper (Cu+Cu), gold-gold (Au+Au), uranium-uranium (U+U), and most recently copper-gold (Cu+Au) collisions. At the LHC, so far only proton-proton (p+p), lead-lead (Pb+Pb), and proton-lead (p-Pb) have been studied. Among these nuclei, Cu, Au, Pb, and U are heavy nuclei, and others are light ones. The light nuclei collisions such as p+p and mixed-type collision like d+Au are often used to ``calibrate'' the heavy-ion collisions. It is worth mentioning that Cu, Au, and Pb nuclei are almost perfectly spherical while U nuclei have an ellipsoidal shape in their ground states.

The detectors can be used to identify the species of particles, and measure their energy and momentum. What we do not know about the particles are their creation times and their emission locations --- the resolution of the measurement is much coarser than the actual scale of the collision zone; each collision is, from the detector's point of view, point-like and instantaneous.
 
When the two heavy nuclei collide, their nucleons are shattered. Their constituents, the quarks and gluons, are spilled out for a brief amount of time and form a new type of matter: the quark gluon matter. This matter can exist for only very short amount of time ($\sim10^{-23}$ second) before it ``evaporates'': the quarks and gluons quickly recombine into hadrons and fly into the detectors.
 
It is not only heavy-ion collision that can shatter nucleons into quarks and gluons; proton-proton collisions collisions can also break the proton cage to free the quarks and gluons at high enough energy: in all such collisions the quark-gluon matter is created. However one peculiar property of the quark-gluon matter created by heavy-ion collision is that the created matter can actually equilibrate, meaning that the quarks and gluons in the matter can almost reach local equilibrium, which is another way to say that the particle number probability density of finding quarks and gluons with given energy satisfy their corresponding Boltzmann distributions. This thermalized medium is the quark-gluon plasma (QGP).

The information on the out-going particles can be analyzed to form observables, which will be compared with various theoretical results in order to learn about the properties of the QGP.
 
Another concept that needs to be introduced is ``centrality''. A collision where the two nuclei hit head-on is very different from another collision where the two nuclei only graze each other; for this reason, collisions are usually sub-divided into ``centrality classes'', where each class contains collisions under similar conditions. The conventional quantity that can be used theoretically to perform such a division is the impact parameter, which is defined as the closest distance between the tracks of the centers of the two nuclei. However, as mentioned above, in heavy-ion collisions no detailed information including impact parameter can be directly obtained, therefore in measurements another quantity is used to define the ``centrality'' of the collision. Let us explain the idea: it is plausible, at least statistically, that the more central a collision is, the more particles it will produce. Based on this observation, the number of produced particles can be used to indicate how central a collision is. Operationally, all collisions are ordered by the number of charged hadrons they produce, and the rank of a collision is its ``centrality''. For example, the top 5\% of all the events in this ordered list form the 0-5\% centrality class (or centrality bin). The smaller the centrality class, the more central the collisions it contains\footnote{This procedure does not work for p-p collisions, due to multiplicity fluctuations. It only works for nuclear collisions where the monotonic increase of multiplicity with the number of nucleons participating in the collision overwhelms the fluctuations in individual nucleon-nucleon collisions.}.

\section{Relativistic hydrodynamic simulations}
\label{sec:hydrodynamics_simulations}
 
\begin{figure*}[h!]
  \center{\includegraphics[width=\textwidth]{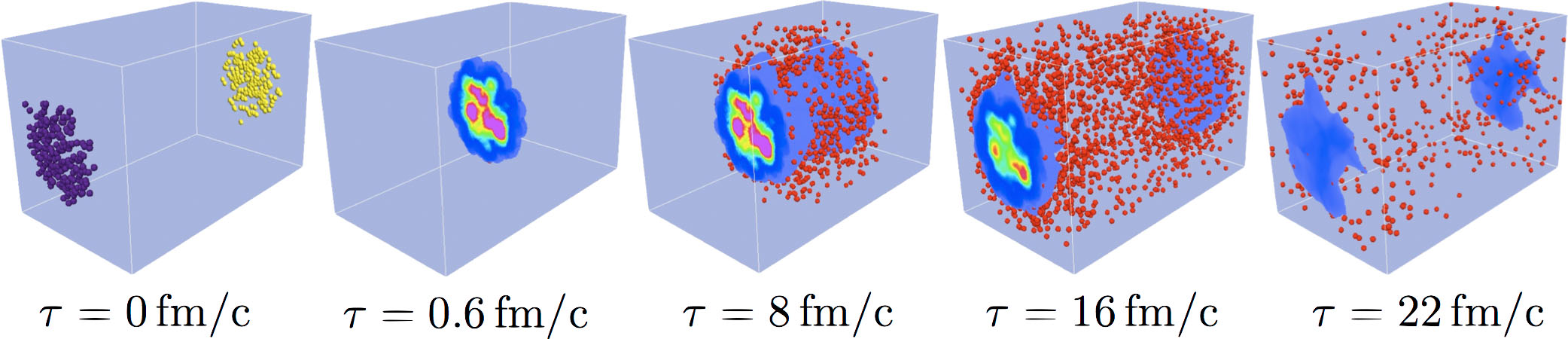}}
  \caption{Snapshots of a simulated event at indicated time. The two groups of colored balls are the colliding nuclei, the color patches are the quark-gluon plasma, and the red balls are the final hadrons emitted from the quark-gluon plasma.}
  \label{fig:hs:1}
\end{figure*}

The evolution of the quark-gluon plasma is simulated using relativistic hydrodynamics. There are several stages involved in this process. For an illustration, a typical simulation is visualized in \fig{hs:1}.

The first stage is to generate the initial state for the quark-gluon plasma, which is usually represented as an energy density profile. In principle, there is a pre-equilibrium stage between the time of collision and the starting time of the quark-gluon plasma, during which the quark matter quickly evolves towards local equilibrium. As a result of this pre-equilibrium evolution, the initial hydrodynamic energy density profile is not the same profile as {\em just} after the collision. Relatively little is known about this pre-equilibrium stage and its study is new and still on-going. However, the pre-equilibrium stage is believed to last only for a very brief of amount of time ($\sim 1$ fm/c) and to contribute at most minor changes to observables I studied. Therefore, in all of my simulations, this stage is assumed to be absent, and energy density profiles produced by models which strictly speaking should only be applied at the beginning of the pre-equilibrium stage are used as the final state of the pre-equilibrium stage --- that is, as the initial state of the hydrodynamic simulation. Because of the pre-equilibrium stage, there could be collective particle motion already before the system equilibrates; in my simulations such initial flow is ignored.

\begin{figure*}[h!]
  \center{\includegraphics[width=0.9\linewidth]{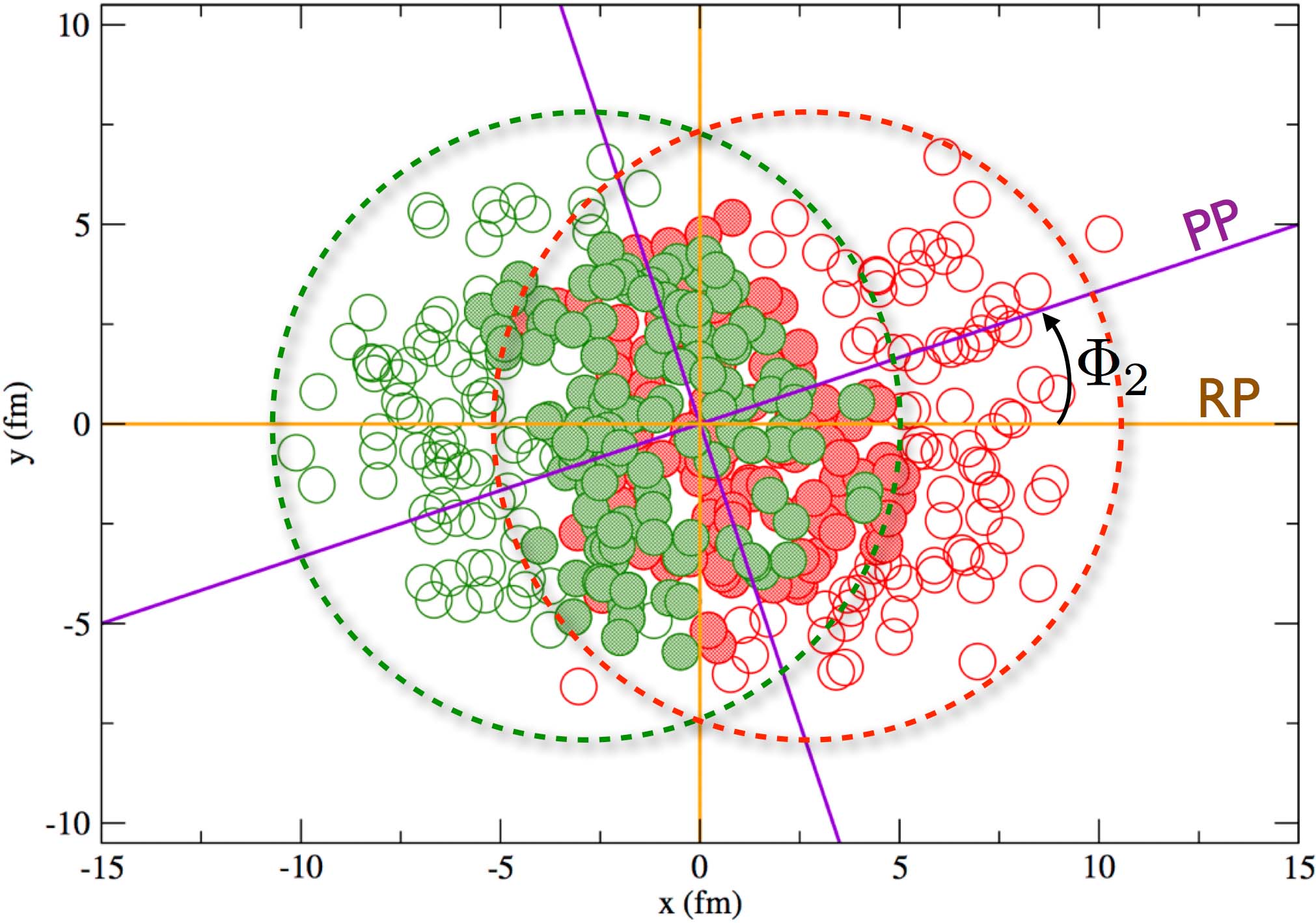}}
  \caption{(From \cite{Qin:2010pf}) Sample nucleon profiles of two colliding nuclei in the transverse plane. The dashed circle indicates the location and the boundary of the two colliding nuclei. The small disks represent nucleons, and solid colored small disks are the wounded ones. Here the frame is the reaction plane frame where the $x$ direction is the direction of the impact parameter, and the $y$ direction is its perpendicular direction in the transverse plane. The plane formed by the $x$ direction and the beam direction is the reaction plane (RP). The direction that maximizes the second order deformation has angle $\PsiPP{2}$ relative to the $x$ direction, and the plane formed by this direction and the beam direction is the (second-order) participant plane (PP).}
  \label{fig:hs:6}
\end{figure*}

The initial condition models used in my simulations are the Glauber model \cite{Miller:2007ri, Hirano:2009ah} and the fKLN model \cite{Hirano:2009ah, KLN,Drescher:2006ca, Hirano:2010jg}. Both models start with generating the three dimensional nucleon position configurations assuming independent draws from the Woods-Saxon density distributions for the colliding nuclei. Next, the density distributions of these nucleons are projected onto the plane perpendicular to the beam direction (the transverse plane). \Fig{hs:6} is an illustration for such a projection. The big dashed circles indicate the boundaries of the two colliding nuclei, and the smaller disks are the nucleons inside each nucleus, identified by their color.

The Glauber model first simulates how the nucleons from one nucleus overlap in the transverse plane with nucleons from the other nucleus. Each overlap of a pair of nucleons gives one binary collision, producing two wounded (participant) nucleons. If a wounded nucleon scatters again with another nucleon from the other nucleus, it remains wounded, wounds the other nucleon (if not already wounded), and contributes one to the binary collision count. In \fig{hs:6}, the solid colored small disks are the wounded nucleons. The Glauber model then assumes that the initially produced entropy density at a given location in the transverse plane is proportional to a linear combination of the wounded nucleon and the binary collision densities (``two-component'' model). Both the proportionality factor and the mixing factor are fitted later by comparing simulated observables to experimental data.

The KLN model uses the nuclear density functions from the nucleon profiles of the two colliding nuclei to determine their local saturation scale $Q_s$, which is then used to determine the unintegrated nuclear gluon density distribution; the unintegrated gluon distributions from the two colliding nuclei can then be used to generated the density of the gluons produced in the collision \cite{KLN,Drescher:2006ca,Hirano:2009ah,Hirano:2010jg}. This produced gluon density is then assumed to be proportional to the initial entropy density, and the proportionality factor, as well as a model parameter $\lambda$ used in determining $Q_s$ from nucleon density functions, are fitted to experimental data.

For both models, the initial entropy density profiles are translated into energy density profiles using the s95p-PCE equation of state (EOS) \cite{Huovinen:2009yb,Shen:2010uy}. Our group uses the superMC code (modified from the rcBk model \cite{Albacete:2013tpa} and the MC-KLN code \cite{Hirano:2009ah,Drescher:2006ca}\footnote{http://www.aiu.ac.jp/\textasciitilde{}ynara/.}) to generate both types of initial conditions. The actual choice of parameters varies between different simulations and will be reported in the corresponding chapters and in Table \ref{tab:1} in the Appendix.

\begin{figure*}[h!]
  \center{\includegraphics[width=\textwidth]{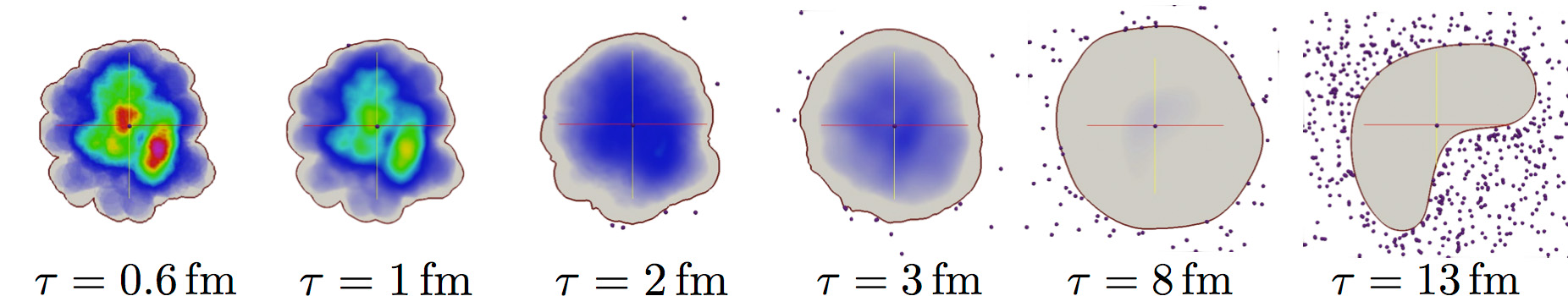}}
  \caption{The evolution of energy density in the transverse plane from a simulated event, at time $\tau$ as indicated. Red indicates high and grey indicates low energy density. The QGP is assumed to convert to hadrons at $e_\mathrm{hadron}=0.508\,\mathrm{GeV}/\mathrm{fm}^3$ (indicated by the thin purple line at the edge of the grey region).} 
  \label{fig:hs:2}
\end{figure*}

The second stage is to evolve the initial energy density profile hydrodynamically by numerically solving the relativistic hydrodynamic evolution differential equations. Our group solves the Israel-Stewart second-order hydrodynamic equations \cite{Heinz:2005bw} under the assumption of longitudinal boost-invariance\footnote{The longitudinal boost-invariant assumption assumes that the system is invariant under a boost in the longitudinal (beam) direction. The experimentally measured rapidity region where this assumption approximately holds is known as the ``plateau''; it covers roughly $\pm 0.5$ units of rapidity at RHIC energy and $\pm 2.5$ units of rapidity at LHC energy, and these regions are the validity regions of our simulations. In particular, our simulations can be applied to the important ``mid-rapidity'' region (around zero rapidity) where most experiments have the best sensitivities.}, using the VISH2+1 code \cite{Song:2007ux}. There are several tunable parameters that describe the properties of the quark-gluon plasma, among which the most famous one is the specific shear viscosity $\eta/s$, the ratio of shear viscosity $\eta$ to entropy density $s$. These parameters are not well-known theoretically and will be adjusted to experimental data. As we will see, different initial conditions require different $\eta/s$ values to describe the same data. This will be discussed later in this thesis. \Fig{hs:2} is an illustration of the evolution of the energy density profile from one simulation.

The VISH2+1 code simulates the evolution of the energy density distribution, and outputs information like flow velocity, energy density, etc. along a constant temperature (isothermal) freeze-out surface, whose functionality depends on the type of simulations:

\begin{enumerate}
\item 
In a purely hydrodynamic simulation, both the quark-gluon plasma and the re-scattering of the emitted hadrons are simulated using hydrodynamics. In such an approach, the freeze-out surface is defined as the surface outside which hadrons cease to interact and reach the detectors by streaming freely.
\item
In a hybrid hydrodynamic simulation, only the quark-gluon plasma is simulated hydrodynamically, and the scattering of the hadrons, after they materialize from the quark-gluon plasma, is simulated using a hadron re-scattering simulator based on a transport approach. In such an approach, the freeze-out surface (or better ``switching surface") is the surface that separates the quark-gluon plasma phase from the hadronic phase.
\end{enumerate}

In both types of simulation, the temperature of the freeze-out surface is a tunable parameter, although for the hybrid simulations people choose it according to the results from lattice QCD calculations \cite{Bazavov:2009zn,Petreczky:2013qj}.

The next stage of the simulation is to generate the momentum distributions of particles from the freeze-out surface. The Cooper-Frye formula (see \sec{iSS:momentum_distribution} for details) is used to calculate the momentum distribution of the emitted particles from the surface.

For purely hydrodynamic simulations, knowing these distributions as continuous functions enables one to calculate many observables. However, in real experiments, each collision only emits a limited number of particles and calculations done using the continuous distribution function do not allow one to study the fluctuations caused by finite statistics. To study finite-statistics fluctuations, one can also simulate a finite number of particles emitted from the freeze-out surface by Monte-Carlo sampling the continuous Cooper-Frye distribution.

For hybrid simulations one must in any case simulate the emission of finite numbers of particles, which is required by the hadron re-scattering simulator.

There is one more subtlety: even for purely hydrodynamic simulations, although the emitted particles are assumed to stop interacting, unstable particles continue to decay into lighter ones before they reach the detectors, and this process changes the momentum distributions of the light particles. This process is called resonance decay and it needs to be additionally computed. For hybrid simulations, the decay of unstable particles is usually included in the hadronic re-scattering simulator and it does not need to be computed separately.

Our group uses the iS code to calculate the continuous distributions of emitted particles and their resonance decays, and the iSS code to simulate the emission of a finite discrete number of particles. The methodology used for sampling in the iSS code will be explained in chapter \chap{iSS}.

For purely hydrodynamic simulations, this is the end of the simulation process. All of my publications are based on purely hydrodynamic simulations, but since part of the work I have contributed is a package for hybrid calculations, I will explain it briefly.

Once the emissions of hadrons has been successfully simulated, they can then be passed to the hadronic re-scattering simulator, from which the final-state momenta of the particles are produced, which can be analyzed to generate simulated observables. Our group uses the UrQMD code \cite{Bass:1998ca} to simulate the hadronic re-scattering. The particle information generated from UrQMD is huge and to efficiently compute simulated observables from it, a code binUtilities has been developed.

\begin{figure*}[h!]
  \includegraphics[width=\linewidth]{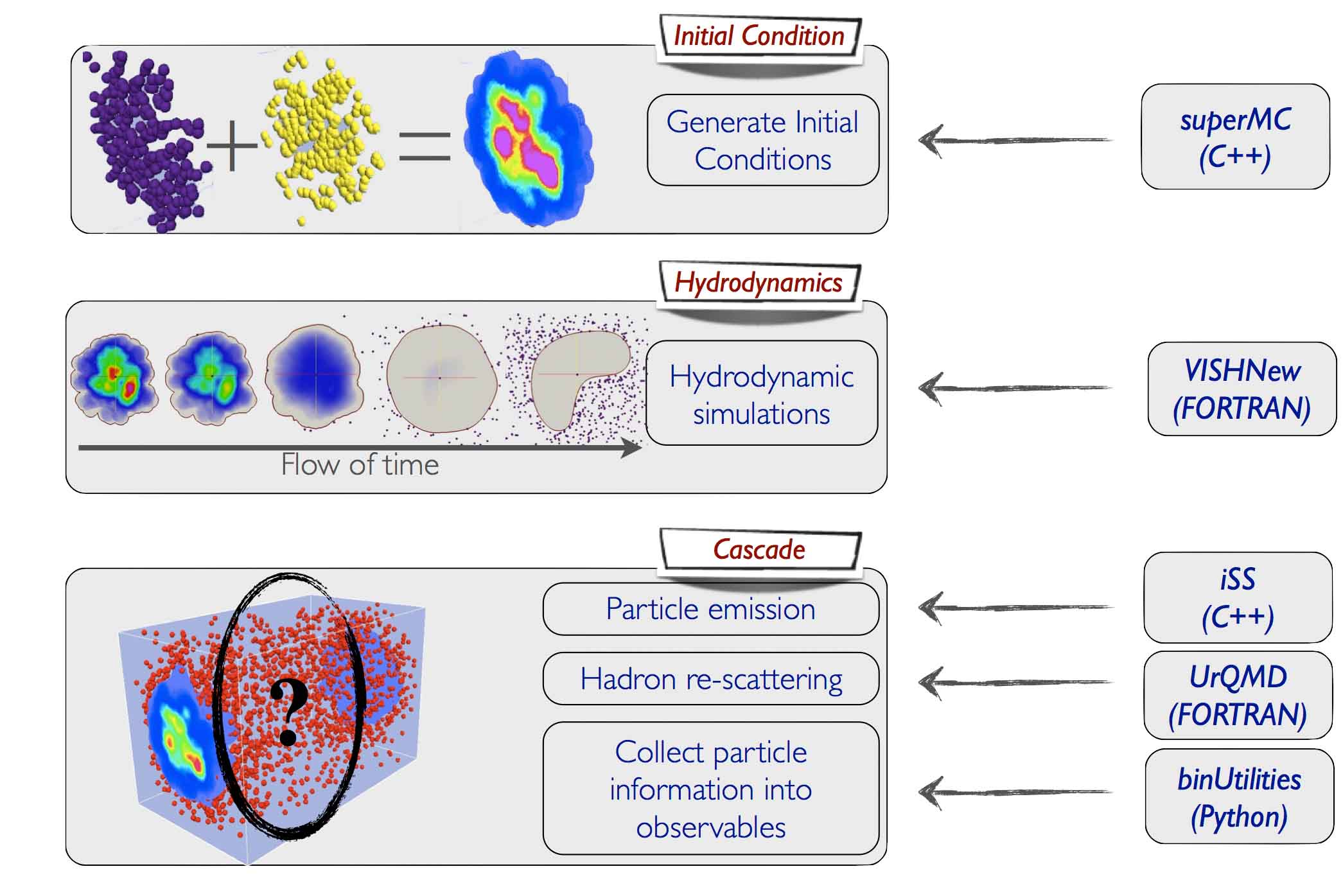}
  \caption{The three-step process for a hydrodynamics simulation (see text for details).}
  \label{fig:hs:3}
\end{figure*}

The stages involved in hydrodynamic simulations, and the corresponding codes used by our group used are schematically summarized in \fig{hs:3}.

\section{Observables}
\label{sec:observables}

\begin{figure*}[h!]
  \includegraphics[width=\linewidth]{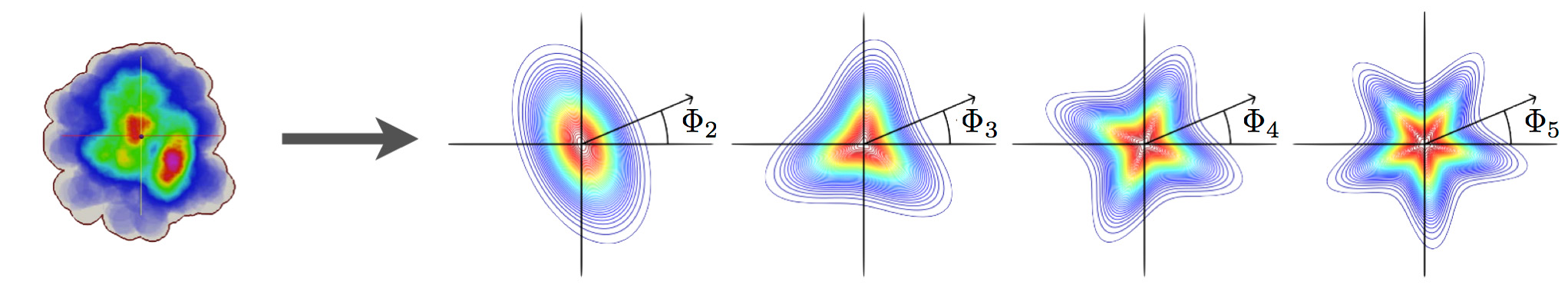}
  \caption{Decomposition of one initial condition into its first 4 harmonic deformations. The $\PsiPP{n}$ are the participant plane angles for each deformation relative to the impact-parameter direction.}
  \label{fig:hs:5}
\end{figure*}

There are many interesting observables that can be studied using hydrodynamic simulations. One type of observables that are particularly important are the anisotropic flows $v_n$, which are the harmonic Fourier coefficients that describe the anisotropy of the particle emission distribution in the transverse direction. Another set of theoretically interesting quantities are the initial eccentricities $\varepsilon_{n}$, which describe in the form of Fourier coefficients the anisotropies of the initial density distributions. \Fig{hs:5} is a schematic illustration of the decomposition of one initial condition into its first 4 harmonics deformations.

It is important to point out, that ``anisotropy'' is a vector, not a scalar, because it carries the information for the orientation of the deformation as well as its magnitude. Because of this, when people report a single scalar as the ``anisotropy coefficient'', they are implicitly projecting the vector to a referencing direction and only report its component along that direction. The choice of the referencing direction is not unique, and we will be explicit on our choice of references when used. For each given harmonic order, there is no ambiguity about the direction of the deformation or the direction that ``maximizes the anisotropy''; these directions can be used to define important planes. The plane spanned by the direction $\Psi_n$ of the anisotropic flow $v_n$ and the beam direction is the $n$-th order {\em event plane} (EP). The plane spanned by the direction $\Phi_n$ of the complex eccentricity vector $\Ecc_n$ (see \equ{1104.0650:15} for formal definition) and the beam direction is the {\em participant plane} (PP). As a quick illustration in \fig{hs:5}, the participant plane angles are marked on the figures for each decomposed deformation. \Fig{hs:6} has an illustration of the 2nd-order participant plane angle. Precise formal definitions will be given in each chapter when needed. 

\begin{figure*}[h!]
  \includegraphics[width=\linewidth]{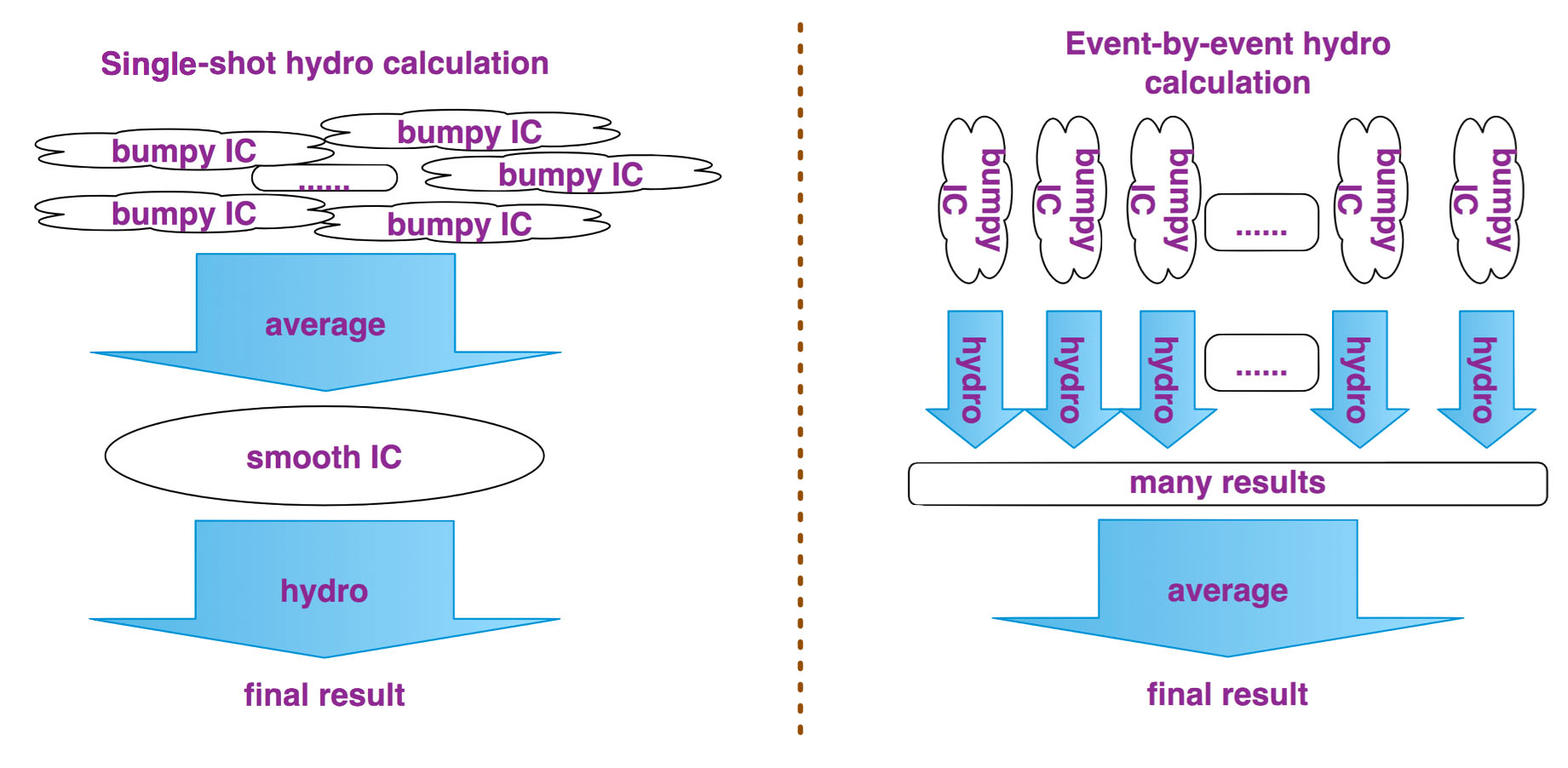}
  \caption{Single-shot hydrodynamic simulations versus event-by-event hydrodynamic simulations.}
  \label{fig:hs:4}
\end{figure*}

Another classification separates hydrodynamic simulations into single-shot simulations and event-by-event simulations. The two procedures and their differences are summarized in \fig{hs:4}.

Event-by-event simulation is de facto the standard simulation type for heavy-ion collisions today, whose goal is to simulate each individual collision, then statistically construct simulated observables using all the simulated events in the same way as the experimentalists do with measured events. In this type of simulations, the initial energy density profile for each simulated collision is Monte-Carlo generated and propagated through hydrodynamic simulations, and the results are analyzed from a collection of particles summed over all simulated events. However such simulations are extremely resource demanding, and they have only been extensively applied in the last few years.

The single-shot simulation was dominantly used half a decade ago, when due to technological limitations people tried to study results from multiple collisions using only a single hydrodynamic simulation. The underlying logic is the following: since event-by-event evolution of many fluctuating initial conditions is expensive but most observables are anyhow measured by summing over many events, we might as well average over the fluctuations in the initial state and evolve only a single, averaged and smooth initial profile, computing the observables from the single final state.

This, unfortunately, ignores the nonlinearities in the hydrodynamic evolution, which leads to characteristic differences between the initial and final fluctuation distributions. I found in my work specific limitations of the single-shot approach and thereby established the need for event-by-event hydrodynamics unambiguously.

It is obvious that the applicability of the single-shot approach depends on the type of observables, since by definition only observables that are direct averages have a chance to be amenable to single-shot hydrodynamics.

There are many ways to generate initial conditions for single-shot calculations. The most common choice of averaging initial conditions is to rotate the initial conditions to align their orientations $\PsiPP{n}$ before averaging. This operation can be done for anisotropies of any given harmonic order $n$, and the resulting averaged profiles are different for different $n$. For example, aligning the second-order harmonic deformation will result in an ellipse-shaped averaged profile while aligning the third-order harmonics will result in a triangular-shaped averaged profile. More details will be given explicitly in related chapters.

The classification of hydrodynamic simulations into single-shot and event-by-event ones is independent of their classification into pure hydrodynamic and hybrid simulations: single-shot pure hydrodynamic simulations dominated half a decade ago, event-by-event pure hydrodynamic simulations are the basis of this thesis, but single-shot hybrid simulations with a hybrid code (e.g. VISHNU\cite{Song:2012mu}) and event-by-event hybrid simulations will be the mainstream in the future.

The event-by-event simulations and the data collection and management processes are tedious to perform manually. To automate these processes, we have collaborated with the QCD group at Duke University to develop the iEBE package\footnote{\iEBEURL.} which allows one to easily set up and perform event-by-event simulations locally or on a cluster; simulated observables can be calculated from the final results using a simple one-line-command interface. The details of this package will be explained in the Appendix \sec{iEBE}.


\chapter{Comparison between event-by-event and single-shot hydrodynamic simulations}
\label{chap:1104.0650}

This chapter focuses on the comparison of event-by-event hydrodynamic simulations and the single-shot ones. The material is largely based on previous work reported in \cite{Qiu:2011iv}, but is supplemented by calculations for nonzero shear viscosities. (The work in \cite{Qiu:2011iv} was based on ideal fluid dynamics.)

\section{Definitions}
\label{sec:1104.0650:2}

In this section, we formally define the harmonic flow 
and eccentricity coefficients and briefly describe the models used in 
computing the initial entropy and energy density profiles.

\subsection{Ellipticity}
\label{sec:1104.0650:2a}

The ``ellipticity''
$\ecc\equiv\ecc_2$ of a given matter distribution in the transverse 
$(x,y)$ plane is defined in terms of its $r^2$-weighted second azimuthal moment 
\cite{Alver:2010gr,Alver:2010dn},
\begin{equation} 
\label{eq:1104.0650:1}
  \Ecc_2 := \ecc_2\, e^{i 2\PsiPP{2}} 
  = -\frac{\int dx\,dy\, r^2 e^{i 2\phi}\,e(x,y)}
{\int dx\,dy\, r^2\,e(x,y)},
\end{equation}
where $x\eq{r}\cos\phi$, $y\eq{r}\sin\phi$, and $\Ecc_2$ is the complex ellipticity. This formula assumes that the 
origin is the center of the distribution $e(x,y)$. In a Monte Carlo approach 
for generating the initial distribution $e(x,y)$ (see Sec.~\ref{sec:1104.0650:2d}), 
this must be ensured by recentering each event before using Eq.~(\ref{eq:1104.0650:1}). 
By default, we characterize in Eq.~(\ref{eq:1104.0650:1}) the matter distribution by 
its energy density $e(x,y)$ \cite{Kolb:2003dz}. In 
Sec.~\ref{sec:1104.0650:3b}, we compare energy- and entropy-weighted ellipticities.

In Eq.~(\ref{eq:1104.0650:1}), $x$ and $y$ are ``reaction plane'' (RP) coordinates:
the reaction plane is the $(x,z)$ plane, with $z$ pointing along the beam 
and $x$ pointing along the direction of the impact parameter $\bm{b}$ 
between the colliding nuclei, while $y$ is perpendicular to the reaction plane.
Because of the minus sign on the right hand side of Eq.~(\ref{eq:1104.0650:1}), the angle 
$\PsiPP{2}$ on the left hand side of Eq.~(\ref{eq:1104.0650:1}) points in the direction of the 
minor axis of the corresponding ellipse. For an elliptically deformed
Gaussian density distribution, this is the direction of the largest 
density gradient and thus of the largest hydrodynamic acceleration and also
of the finally observed elliptic flow. The direction of this minor axis together with the beam direction $z$ defines the participant plane (PP). It is tilted relative to the reaction plane by $\PsiPP{2}$. 
The label ``participant'' is motivated by the fact that the initial energy 
and entropy density distributions of the collision fireball reflect (more 
or less directly, depending on the model for secondary particle creation) 
the transverse distribution of the nucleons participating in the particle 
production process. The ellipticity $\ecc_2$ in Eq.~(\ref{eq:1104.0650:1}) is 
correspondingly called {\em ``participant eccentricity''} and also denoted as
$\epart$.\footnote{Traditionally $\epart$ is defined in terms of the 
  transverse density of wounded nucleons, but since what matters for the 
  subsequent hydrodynamic evolution is not the distribution of wounded 
  nucleons themselves but of the matter generated by the wounded nucleons, 
  we use the name $\epart$ for the ellipticity characterizing the 
  thermalized matter.}  
It can be written as
\begin{eqnarray}
\label{eq:1104.0650:2}
  \epart &\equiv& \ecc_2 = |\ecc_2\, e^{i 2 \PsiPP{2}}| 
\nonumber
\\
  &=& \frac{\sqrt{\{r^2\cos(2\phi)\} + \{r^2\sin(2\phi)\}}}
       {\{r^2\}} 
\\\nonumber
  &=& \frac{\sqrt{\{y^2{-}x^2\}^2 + 4{\{xy\}^2}}}
 {\{y^2{+}x^2\}}.
\end{eqnarray}
Here $\{\dots\}=\int dx\,dy\,(\dots)\,e(x,y)=\int rdr\,d\phi\,(\dots)\,e(r,\phi)$ defines the ``event average''
over the matter distribution $e(x,y)$ in a single collision event
\cite{Alver:2008zza}. Equivalently, the participant eccentricity can be 
written as 
\begin{equation}
\label{eq:1104.0650:3}
 \ecc_\mathrm{part} = \frac{\{y^2{-}x^2\}'}{\{y^2{+}x^2\}'}
\end{equation}
where $\{\dots\}'\eq\int dx\,dy\,(\dots)\,e'(x,y)$ indicates the average
over a rotated event with energy density $e'(x,y)=e\bigl(x
\cos\PsiPP{2}{-}y\sin\PsiPP{2},x\sin\PsiPP{2}{+}y
\cos\PsiPP{2}\bigr)$ whose minor and major axes now align with $x$ 
and $y$. 

The event-average $\{\dots\}$ is to be distinguished from the ``ensemble
average'' $\la\dots\ra\eq\frac{1}{N}\sum_{n \eq 1}^N \{\dots\}_n$ where $N$ 
is the total number of events and $\{\dots\}_n$ is the event-average over 
the energy density $e_n(x,y)$ in event number $n$. The {\em average 
participant eccentricity} is thus defined as
\begin{equation}
\label{eq:1104.0650:4}
  \la\epart\ra = \frac{1}{N}\sum_{n \eq 1}^N (\epart)_n.
\end{equation}
This differs from the {\em mean eccentricity} $\bar{\ecc}_\mathrm{part}$
of the average (recentered and rotated by $\PsiPP{2}$) 
energy density $\bar{e}'(x,y)\eq\frac{1}{N}\sum_{n \eq 1}^N 
e'_n(x,y)$ which can be written in the following equivalent ways: 
\begin{equation}
\label{eq:1104.0650:5}
 \bar{\ecc}_\mathrm{part} = 
 \frac{\sqrt{\la\{y^2{-}x^2\}\ra^2 + 4\la\{xy\}\ra^2}}
  {\la\{y^2{+}x^2\}\ra} = 
 \frac{\la\{y^2{-}x^2\}'\ra}{\la\{y^2{+}x^2\}'\ra}.
\end{equation}
In contrast to (\ref{eq:1104.0650:4}), one here ensemble-averages over numerator 
and denominator separately before forming the ratio. 

In event-by-event simulations, the hydrodynamic forces
generate in each event an elliptic component $v_2$ of the anisotropic flow,
which is causally related to the specific initial ellipticity $\epart$ in
that event. In single-shot simulations, fluctuating initial conditions are averaged into 
a single smooth initial distribution $\bar{e}(x,y)$, which is then evolved 
hydrodynamically and from which the {\em mean elliptic 
flow} $\bar{v}_2$ is extracted, corresponding to the {\em mean eccentricity} 
$\bar{\ecc}_\mathrm{part}$ of that averaged source distribution. Obviously, 
$\bar{v}_2$ is a deterministic consequence of $\bar{\ecc}_\mathrm{part}$
and does not fluctuate at all; it can not be measured experimentally. What 
could be measured experimentally \cite{Ollitrault:2009ie,Jia:2013tja} is the {\em average elliptic flow} 
$\la v_2\ra$ of a large ensemble of collision events. This observable
is conceptually more closely related to $\la\epart\ra$ than to
$\bar{\ecc}_\mathrm{part}$; for an exactly linear hydrodynamic response
$v_2\sim\epart$, one has 
$\la v_2\ra/\la\epart\ra\eq\bar{v}_2/\bar{\ecc}_\mathrm{part}$ 
\cite{Bhalerao:2006tp}. We will explore the differences between 
$\bar{\ecc}_\mathrm{part}$ and $\la\epart\ra$ and discuss consequences
for the theoretically computed $\bar{v}_2$ as opposed to the measured $\la v_2\ra$ in Secs.~\ref{sec:1104.0650:3a} and \ref{sec:1104.0650:5}.

In addition to these ``participant eccentricities'', one can also define
``reaction plane eccentricities''. For a single event, the {\em reaction 
plane eccentricity} $\ecc_\mathrm{RP}$ is defined by
\begin{equation}
\label{eq:1104.0650:6}
 \ecc_\mathrm{RP} = \frac{\{y^2{-}x^2\}}{\{y^2{+}x^2\}}
\end{equation}
in terms of an event average over the (properly centered) energy density
$e(x,y)$. The so-called {\em standard eccentricity} is defined as the
analogous ratio of expectation values taken with a smooth average
energy density $\bar{e}(x,y)\eq\frac{1}{N}\sum_{n \eq 1}^N 
e_n(x,y)$ obtained by superimposing many events {\em without} rotating 
them from the participant to the reaction plane:
\begin{equation}
\label{eq:1104.0650:7}
 \ecc_s \equiv \bar{\ecc}_\mathrm{RP} 
  = \frac{\la\{y^2{-}x^2\}\ra}{\la\{y^2{+}x^2\}\ra}.
\end{equation}
In other words, the standard eccentricity is the {\em mean reaction-plane
eccentricity}. In contrast, the {\em average reaction-plane eccentricity}
is defined by
\begin{equation}
\label{eq:1104.0650:8}
 \la\ecc_\mathrm{RP}\ra 
  = \left\la\frac{\{y^2{-}x^2\}}{\{y^2{+}x^2\}}\right\ra.
\end{equation}

In real experiments, the actual measured quantities are $v_2\{\mathrm{EP}\}$, $v_2\{2\}$,
and $v_2\{4\}$ (defined below) that, even if non-flow contributions could be
completely ignored, are affected by event-by-event $v_2$-fluctuations
and thus differ from $\la v_2\ra$. $\la v_2\ra$ can be reconstructed
from the experimental measurements with some additional assumptions 
\cite{Ollitrault:2009ie}, which on the surface look harmless but should be
further tested, and we will come back to this point in \chap{flow_angle}. Motivated by the hypothesis of linear hydrodynamic response,
$v_2\sim\epart$, these $v_2$ measures motivate the definition of 
corresponding ellipticity measures \cite{Bhalerao:2006tp}, the
2$^\mathrm{nd}$ and 4$^\mathrm{th}$ order cumulants:
\begin{equation} 
\label{eq:1104.0650:9}
  \ecc\{2\} = \sqrt{\la \ecc_{\mathrm{part}}^2 \ra}
\end{equation}
and
\begin{equation} 
\label{eq:1104.0650:10}
  \ecc\{4\} = \left(\la\ecc^2_\mathrm{part}\ra^2 
  - (\la\ecc^4_\mathrm{part}\ra{-}\la\ecc^2_\mathrm{part}\ra^2)\right)^{1/4}.
\end{equation}
Note that the last expression involves the difference of two positive-definite quantities, which itself does not need to be positive definite.
If fluctuations get large, the expression under the fourth root can become
negative, leaving $\ecc\{4\}$ undefined. We will see that this can happen 
in the most central and the most peripheral centrality bins.

It was shown in \cite{Voloshin:2007pc} that in the MC-Glauber model,
the real and imaginary parts of the complex ellipticity defined by 
Eq.~(\ref{eq:1104.0650:1}), with the wounded nucleon density as weight function on 
the right hand side, both have approximately Gaussian fluctuations, with equal
widths $\sigma_\ecc$. If this is the case, the magnitude $\ecc_2$ of
this ellipticity exhibits fluctuations of Bessel-Gaussian
type\footnote{This 
    takes into account that $\ecc_2$ can never fluctuate to negative 
    values.} 
\cite{Voloshin:1994mz}, leading to the identity \cite{Voloshin:2007pc} 
\begin{equation}
\label{eq:1104.0650:11}
  \ecc\{4\} = \la\ecc_\mathrm{RP}\ra.
\end{equation}
For sufficiently large average ellipticities $\la\ecc_2\ra$ (i.e. 
sufficiently large impact parameters), one may hope to be able to ignore
the restriction that $\ecc_2$ can never fluctuate to negative values, 
and correspondingly assume that $\ecc_2$ exhibits Gaussian (instead of
Bessel-Gaussian) fluctuations. In this case one has \cite{Voloshin:2007pc}
\begin{eqnarray}
\label{eq:1104.0650:11a}
\nonumber
  \ecc\{2\}^2 &=& \la\epart\ra^2 +\sigma^2_\ecc,
\\
  \ecc\{4\}^2 &=& \sqrt{(\la\epart\ra^2 - \sigma^2_\ecc)^2-2\sigma_\ecc^4},
\end{eqnarray}
from which it follows that $\la\ecc_2\eq\ecc_\mathrm{part}\ra^4$ is the 
arithmetic mean of $\ecc\{2\}^4$ and $\ecc\{4\}^4$:
\begin{equation}
\label{eq:1104.0650:11b}
  \frac{\ecc\{2\}^4{+}\ecc\{4\}^4}{2\la\ecc_\mathrm{part}\ra^4}=1. 
\end{equation}
We will use Eqs.~(\ref{eq:1104.0650:11}) and (\ref{eq:1104.0650:11b}) (which hold irrespective 
of the fluctuation width $\sigma_\ecc$) in Sec.~\ref{sec:1104.0650:3a}, and their 
analogues for the elliptic flow $v_2$ in Sec.~\ref{sec:1104.0650:5c}, to test the 
assumptions of Bessel-Gaussian and Gaussian fluctuations of the 
event-by-event ellipticity and elliptic-flow fluctuations using the 
Monte Carlo Glauber (MC-Glauber) and Monte Carlo fKLN (MC-KLN) models.

If the hydrodynamic response were indeed linear, $v_2\sim\epart$, and 
non-flow effects could be ignored, the following identities would hold: 
\begin{equation} 
\label{eq:1104.0650:12}
  \frac{\la v_2\ra}{\la\epart\ra} = \frac{\bar{v}_2}{\bar{\ecc}_\mathrm{part}}
  = \frac{v_2\{2\}}{\ecc\{2\}} = \frac{v_2\{4\}}{\ecc\{4\}}.
\end{equation}
Assuming linear
hydrodynamic response, one can compare the theoretically computed
$\la v_2\ra$ with the experimentally measured $v_2\{2\}$ or $v_2\{4\}$
if one normalizes the former by $\la\ecc_\mathrm{part}\ra$ and the latter 
by $\ecc\{2\}$ or $\ecc\{4\}$, respectively, calculated {\em from the same 
initial-state model} \cite{Song:2010mg,Song:2011hk}. In this context, the 
identity $\ecc\{4\}\eq\la\ecc_\mathrm{RP}\ra$ (which holds if the 
ellipticity fluctuations are Gaussian) becomes particularly useful 
because it suggests that the measured $v_2\{4\}$ can be directly
compared with a single-shot hydrodynamic $v_2$ obtained from a smooth 
reaction-plane-averaged initial density of ellipticity 
$\la\ecc_\mathrm{RP}\ra$, without any corrections for flow fluctuations. 
Even better, $v_2\{4\}$ can be shown to be completely free of two-particle 
non-flow contributions \cite{Bhalerao:2006tp,Voloshin:2007pc}. These
arguments have been used in \cite{Hirano:2010jg} and provide a strong 
motivation for us to test the underlying assumptions (Gaussian ellipticity 
fluctuations and linear hydrodynamic elliptic flow response) in the 
present work.

We close this subsection by recalling the expression for the participant 
plane angle of a given event (see e.g. \cite{Alver:2008zza})
\begin{equation}
\label{eq:1104.0650:13}
  \PsiPP{2}= \frac{1}{2}
  \tan^{-1}\left(\frac{2\{xy\}}{\{y^2{-}x^2\}}\right)
\end{equation}
and for its transverse area
\begin{equation}
\label{eq:1104.0650:14}
   S = \pi \sqrt{\{x^2\}'\{y^2\}'}.
\end{equation}
Both expressions assume that the events are properly centered at the origin.

\subsection{Higher order eccentricity coefficients}
\label{sec:1104.0650:2b}

The definition (\ref{eq:1104.0650:1}) can be generalized to higher ($n\geq 3$) harmonic
eccentricity coefficients \cite{Alver:2010gr,Alver:2010dn}:
\begin{equation} 
\label{eq:1104.0650:15}
 \Ecc_n := \ecc_n\, e^{i n\PsiPP{n}} 
  = -\frac{\int dx\,dy\, r^2 e^{i n\phi}\,e(x,y)}
{\int dx\,dy\, r^2\,e(x,y)}
\end{equation}
where we call $\Ecc_n$ the $n$-th order complex eccentricity.
Alternatively one can use $r^n$ instead of $r^2$ as radial weight 
on the right hand side \cite{Qin:2010pf}:
\begin{equation} 
\label{eq:1104.0650:16}
 \Ecc'_n := \ecc'_n\, e^{i n\PsiPPP{n}} 
  = -\frac{\int dx\,dy\, r^n e^{i n\phi}\,e(x,y)}
{\int dx\,dy\, r^n\,e(x,y)}.
\end{equation}
Still another variant uses the entropy density $s(x,y)$ instead of
the energy density $e(x,y)$ as weight function:
\begin{eqnarray}
\label{eq:1104.0650:17}
 \Ecc_n(s) := \ecc_n(s)\, e^{i n\PsiPP{n}(s)} 
  &=& -\frac{\int dx\,dy\, r^2 e^{i n\phi}\,s(x,y)}
  {\int dx\,dy\, r^2\,s(x,y)},
\\
\label{eq:1104.0650:18}
 \Ecc'_n(s) := \ecc'_n(s)\, e^{i n\PsiPPP{n}(s)} 
  &=& -\frac{\int dx\,dy\, r^n e^{i n\phi}\,s(x,y)}
  {\int dx\,dy\, r^n\,s(x,y)}.
\end{eqnarray}
We note that the $r^2$-weighted eccentricity coefficients $\ecc_n$ fall 
off faster with increasing harmonic order $n$ than the $r^n$-weighted 
eccentricities $\ecc'_n$ (see section \ref{sec:1104.0650:appa}). Also, as in \Equ{1104.0650:1}, the minus 
sign in Eqs.~(\ref{eq:1104.0650:15})-(\ref{eq:1104.0650:18}) guarantees that, for a Gaussian density 
distribution that has only $n^\mathrm{th}$-order eccentricity $\ecc_n$, 
the angle $\PsiPP{n}$ points in the direction of the steepest 
density gradient, and thus in the direction of the corresponding 
hydrodynamically generated $n^\mathrm{th}$-order harmonic flow $v_n$ 
(see next subsection). It can be written as $-1\eq{e}^{-in(\pi/n)}$ and 
amounts to a rotation of $\PsiPP{n}$ by $\pi/n$. For example, 
if the profile is square-shaped, $\PsiPP{4}$ points to the 
sides instead of its corners.

As stated before, we will use the energy density as the default weight 
function; in cases of possible ambiguity, we will use the notations 
$\ecc_n(e)$, $\ecc_n(s)$ etc. to distinguish between energy- and entropy-density-weighted eccentricity coefficients and angles. Eccentricities 
$\ecc$ without harmonic index $n$ denote ellipicities (i.e. in the 
absence of $n$, $n \eq 2$ is implied).

The coefficients $\ecc_n$ and angles $\PsiPP{n}$ define the 
eccentricies and angles of the matter distribution in the participant plane. 
We note that the participant plane angles $\PsiPP{n}$ associated 
with eccentricity coefficients of different harmonic order $n$ do not, in 
general, agree (see Sec.~\ref{sec:1104.0650:4a}). We will not study higher harmonic 
generalizations of the reaction-plane ellipticity (\ref{eq:1104.0650:6}).

\subsection{Harmonic flow coefficients}
\label{sec:1104.0650:2c}

We characterize the final observed momentum distribution $dN/(dy\,p_T
dp_T\,d\phi_p)$ by ``harmonic flow coefficients'' constructed in analogy
to Eq.~(\ref{eq:1104.0650:15}), but without the extra minus sign:
\begin{eqnarray} 
\label{eq:1104.0650:19}
  &&\!\!\!\!\!\!
  V_n(y,p_T) := v_n(y,p_T)\, e^{i n \PsiEP{n}(y,p_T)} = 
  \frac{\int d\phi_p\, e^{i n\phi_p}\,\frac{dN}{dy\,p_Tdp_T\,d\phi_p}}
       {\frac{dN}{dy\,p_Tdp_T}},\quad
\\
\label{eq:1104.0650:20}
  &&\!\!\!\!\!\!
  V_n(y) := v_n(y)\, e^{i n \PsiEP{n}(y)} = 
  \frac{\int p_Tdp_T\,d\phi_p\, e^{i n\phi_p}\,\frac{dN}{dy\,p_Tdp_T\,d\phi_p}}
       {\frac{dN}{dy}}.\quad
\end{eqnarray}
where $V_n$ are the complex flows.

In boost-invariant hydrodynamics, they are rapidity-independent, so 
we drop the argument $y$ and keep in mind that we should only compare
with midrapidity data at $y \eq 0$ where the assumption of boost-invariant 
longitudinal expansion is most justified. The spectra $\frac{dN}{dy\,p_T
dp_T\,d\phi_p}$ are computed from the hydrodynamic output with the
Cooper-Frye prescription \cite{Cooper:1974mv} along an isothermal 
kinetic decoupling surface whose temperature is chosen in this chapter as $T_\mathrm{dec} = 140$\,MeV.
Equation~(\ref{eq:1104.0650:19}) defines the $p_T$-differential harmonic flow 
$v_n(p_T)$ and flow angle $\PsiEP{n}(p_T)$, whereas 
Eq.~(\ref{eq:1104.0650:20}) gives their $p_T$-integrated values $v_n$ and 
$\PsiEP{n}$. The orientation of the $n$th-order harmonics of the final momentum distribution 
defines the $n$th-order ``event plane''. Again, different 
harmonic flows are usually associated with differently-oriented event 
planes. The first three harmonic flow coefficients are the directed flow 
($v_1$), elliptic flow ($v_2$), and triangular flow ($v_3$).

\subsection{Initial-state models}
\label{sec:1104.0650:2d}

We use Monte Carlo versions \cite{Hirano:2009ah,Hirano:2010jg} of the 
Glauber \cite{Miller:2007ri} and {\tt fKLN} \cite{Drescher:2006ca} models
to generate fluctuating initial conditions for the entropy density in 
$200\,A$\,GeV Au+Au collisions. For the MC-Glauber model we assume a 
two-component (soft+hard) model with a small hard fraction 
($\delta \eq 0.14$ \cite{Hirano:2009ah}); we also use a Woods-Saxon profile 
for the distribution of nucleon centers whose radius and surface-thickness parameters have been corrected for the finite nucleon size 
\cite{Hirano:2009ah}. The resulting entropy density profile is normalized 
to the final charged-hadron multiplicity density $dN_\mathrm{ch}/dy$ in 
central collisions; after this normalization, the centrality dependence 
of the initial entropy production is fixed by the model (MC-Glauber or 
MC-KLN). To convert the initial entropy density to energy density, we 
use the equation of state (EOS) s95p-PCE which matches Lattice QCD data 
at high temperatures to a chemically frozen hadron resonance gas at low 
temperatures \cite{Huovinen:2009yb,Shen:2010uy}, using 
$T_\mathrm{chem} \eq 165$\,MeV as chemical freeze-out temperature.

In the following we compute harmonic eccentricity and flow coefficients
as functions of impact parameter $b$ and collision centrality.
The centrality classes are defined in terms of percentages of the total
inelastic cross section, calculated from the distribution of the number of 
wounded nucleons $dN_\mathrm{event}/dN_\mathrm{part}$ in the optical Glauber 
model (i.e.\ without accounting for fluctuations in $N_\mathrm{part}$
at given impact parameter). Each centrality class is thus characterized by 
a range of impact parameters $b_\mathrm{min}<b<b_\mathrm{max}$ and an 
average value $\bar{b}$, together with a mean number of wounded nucleons 
$\bar{N}_\mathrm{part}$. They are listed in Table~\ref{T1} 
\cite{Hirano:2009ah}.  

\subsection{Averaging procedures for the initial profiles}
\label{sec:1104.0650:2e}

In this section, we will compare results obtained from event-by-event 
hydrodynamic simulations with traditional single-shot hydrodynamic simulations,
where one first averages over many fluctuating initial profiles to obtain 
a smooth average profile, then evolves this smooth profile 
hydrodynamically. The question addressed in this comparison is to what 
extent the average harmonic-flow coefficients from event-by-event 
hydrodynamics can (or cannot) be faithfully represented by the harmonic-flow 
coefficients extracted (at much lower numerical expense) from the 
hydrodynamic evolution of an ``average event".

Taking the initial density profiles from the Monte Carlo generator and 
superimposing them directly without additional manipulations (such that the impact parameter directions for each collision are aligned) generates a 
``reaction-plane averaged'' profile with ellipticity 
$\bar{\ecc}_\mathrm{RP}$ (Eq.~(\ref{eq:1104.0650:7})). After recentering 
each event to the origin of the $x$-$y$-plane, we can compute event
by event the reaction and participant plane ellipticities (Eqs.~(\ref{eq:1104.0650:6}) 
and (\ref{eq:1104.0650:2},\ref{eq:1104.0650:3})) and evaluate their ensemble averages (\ref{eq:1104.0650:8}) 
and (\ref{eq:1104.0650:4}), respectively. To generate a smooth average profile with
ellipticity $\bar{\ecc}_\mathrm{part}$ (Eq.~\ref{eq:1104.0650:5}), we rotate each
recentered event by the angle $\PsiPP{2}(e)$ ($\PsiPP{2}(s)$)
if we want to determine the eccentricity of the average energy (entropy)
density. For the calculation of entropy-weighted average eccentricities, 
we perform any ensemble average first and convert the result to energy 
density later; in this case all events are rotated by their 
$\PsiPP{2}(s)$ angles. For energy-weighted ensemble averages, we 
convert $s$ to $e$ in each event 
first, rotate by $\PsiPP{2}(e)$, and perform the ensemble average 
last. Other sequences or mixtures of these steps are technically possible 
but physically not meaningful. Note that the processes of computing the 
energy density from the entropy density via the EOS and of averaging the 
event profiles do not commute: The energy density obtained via the EOS 
from the ensemble-averaged entropy density profile is not the same as the 
ensemble-averaged energy density where the EOS is used in each event to 
convert $s$ to $e$. 

\vspace*{-3mm}
\section{Eccentricities}
\label{sec:1104.0650:3}
\vspace*{-3mm}

\subsection{Centrality dependence of different ellipticities}
\label{sec:1104.0650:3a}
\vspace*{-3mm}

%
\begin{figure*}[h!]
\begin{center}
 \includegraphics[width=0.45\linewidth]{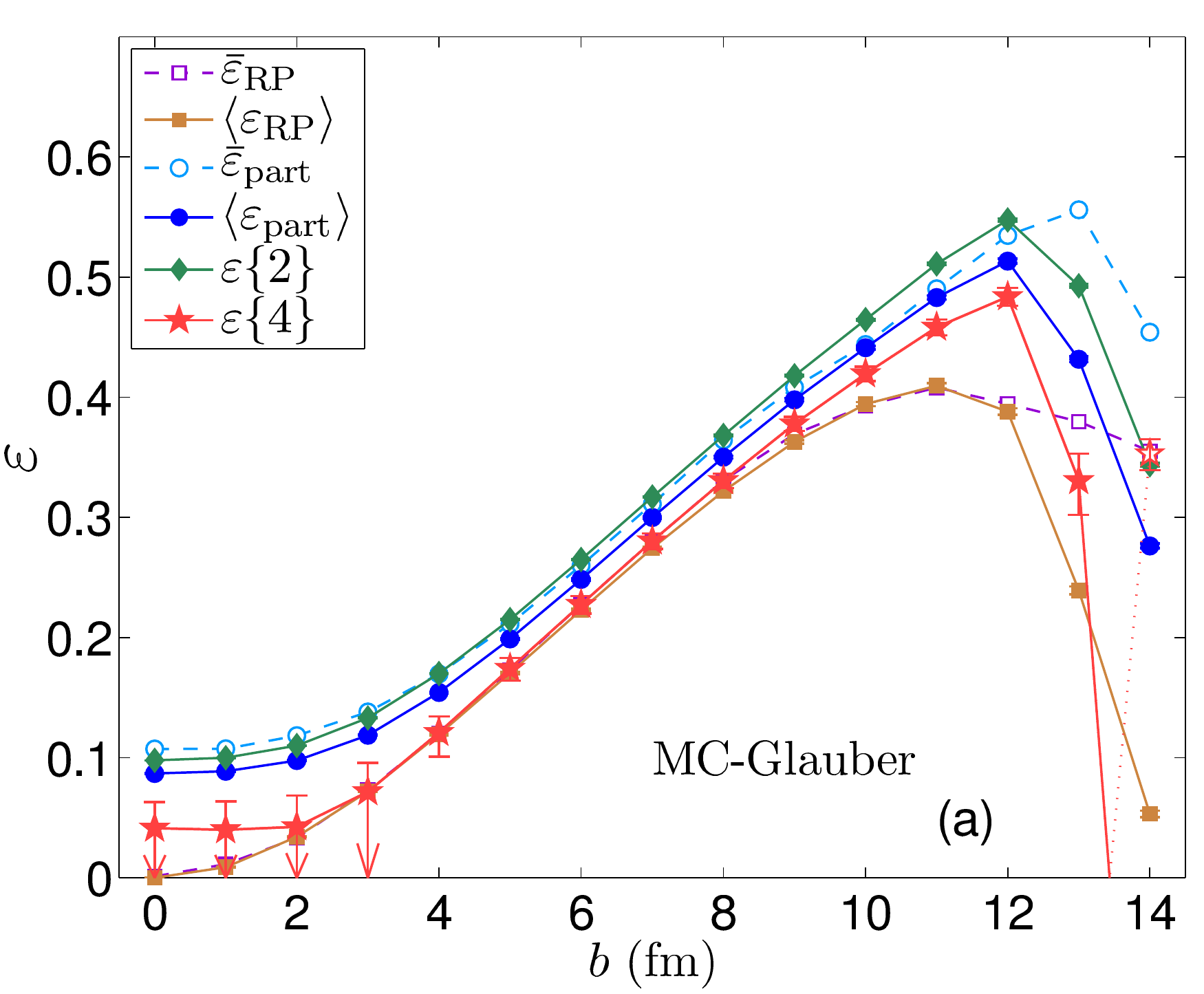}
 \includegraphics[width=0.45\linewidth]{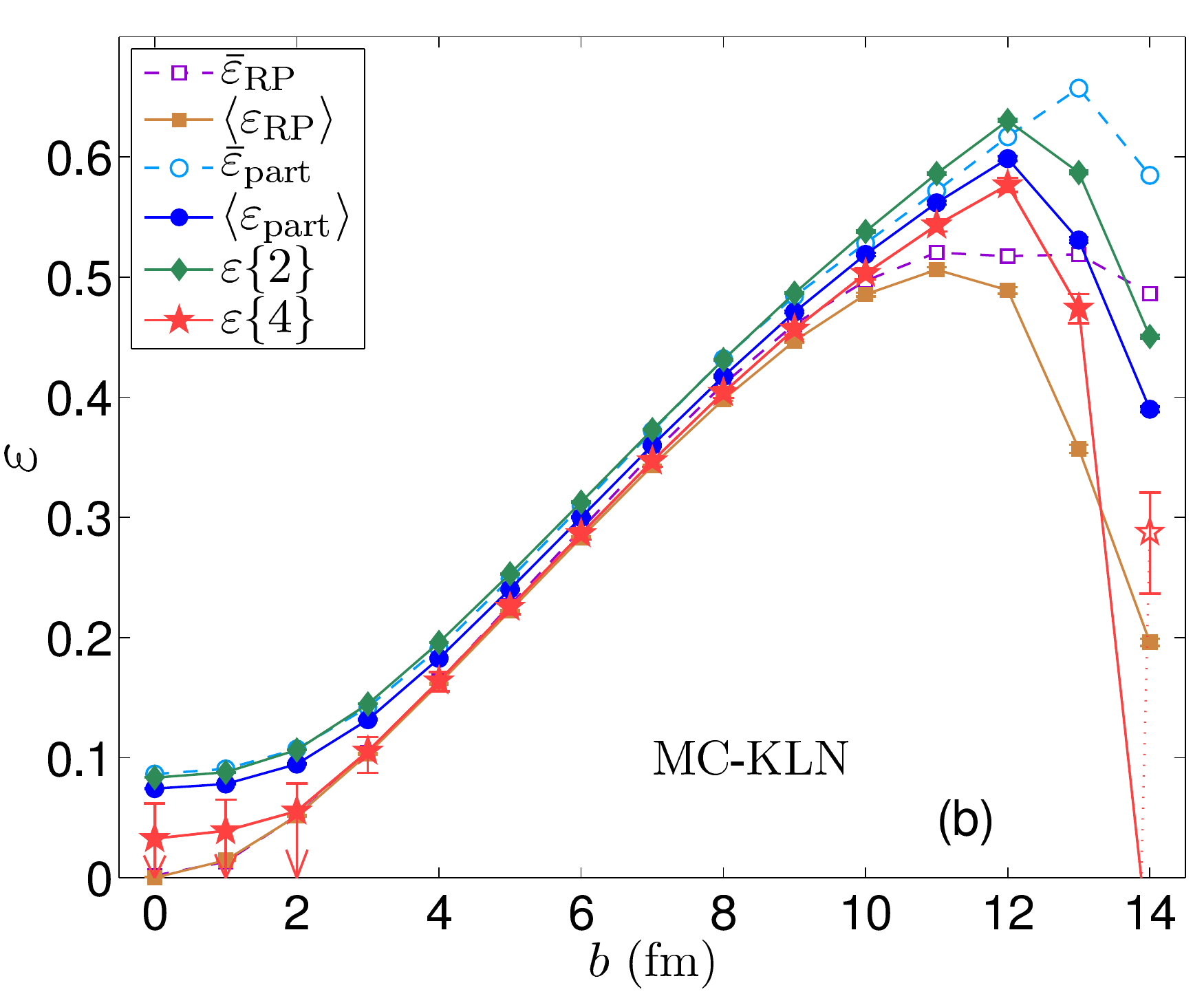}\\
 \includegraphics[width=0.45\linewidth]{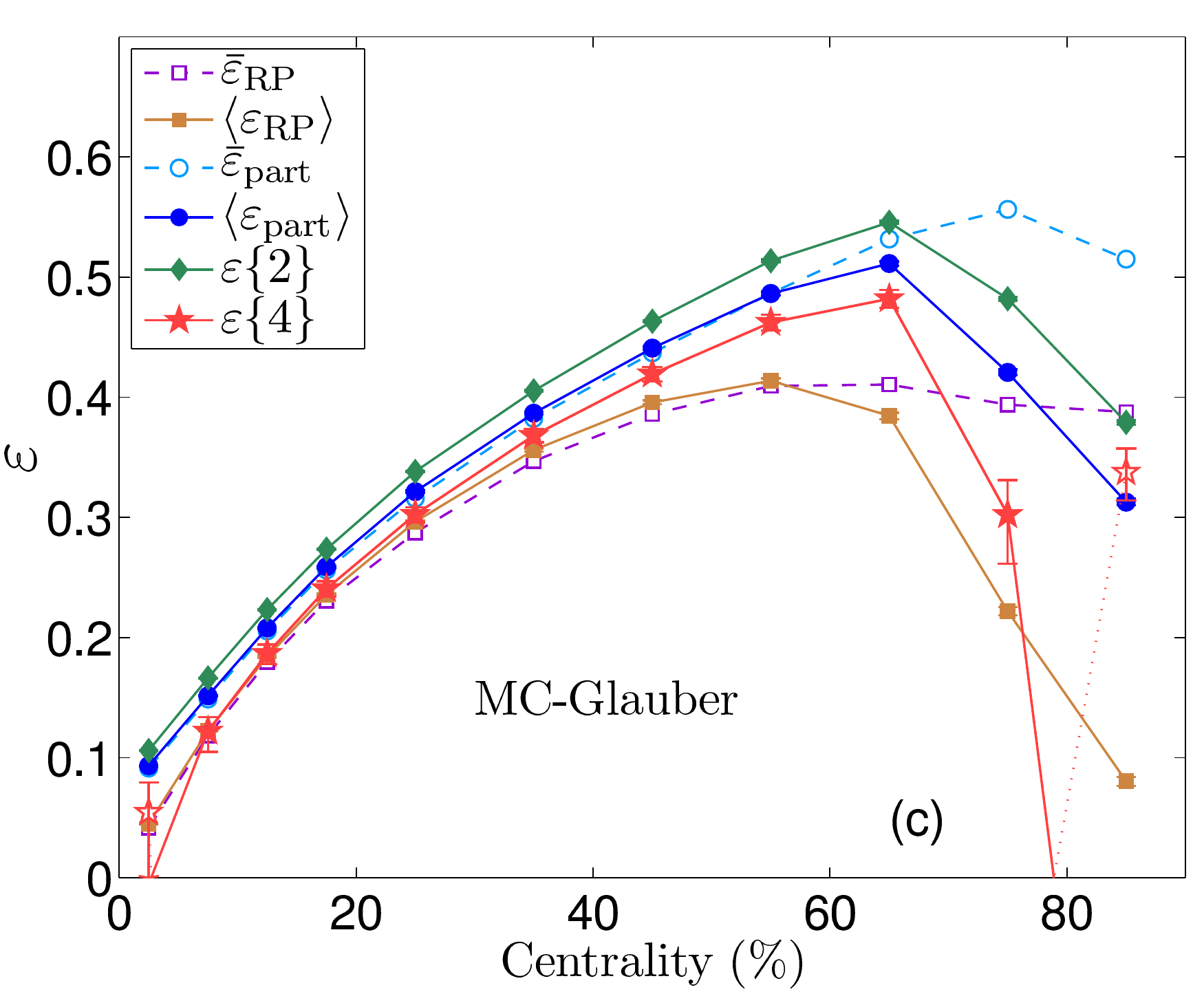}
 \includegraphics[width=0.45\linewidth]{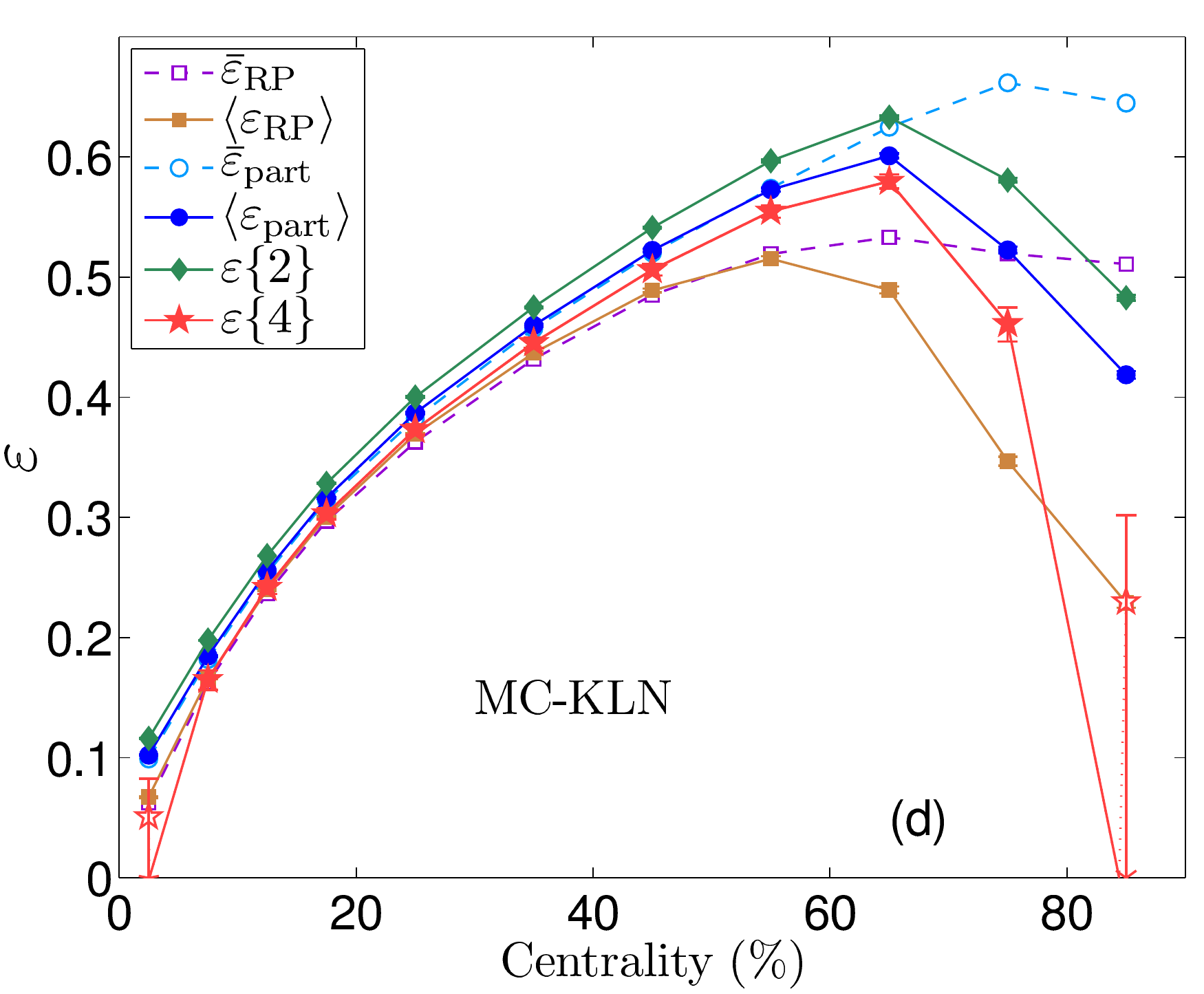}
\caption{Different ellipticities as a function of
   impact parameter (top row) or collision centrality (bottom
   row), for the MC-Glauber (panels (a) and (c)) and the MC-KLN
   model (panels (b) and (d)). Panels (a,b) show $e$-weighted, panels
   (c,d) show $s$-weighted ellipticities. (See Fig.~\ref{1104.0650:F4} 
   below for a direct comparison between $e$- and $s$-weighted eccentricities.)
   Open stars indicate negative values for $\ecc\{4\}^4$. Taken from \cite{Qiu:2011iv}.
   \label{1104.0650:F1}
}
\end{center}
\end{figure*}
%

Fig.~\ref{1104.0650:F1} shows a comparison between the different ellipticities 
defined in Sec.~\ref{sec:1104.0650:2a} as functions of the impact parameter $b$
in panels (a) and (b) and as functions of collision centrality (as defined
in Sec~\ref{sec:1104.0650:2d}) in panels (c) and (d). For panels (a) and (b), we 
generated 10,000 initial profiles for each impact parameter (except for 
$b \eq 0,\,1,$ and 2\,fm for which we generated 30,000 events each), and the smooth initial conditions are averaged from them; for 
panels (c) and (d) we generated and averaged over 10,000 profiles for each centrality 
bin. Within the centrality bins, the impact parameters were sampled between 
$b_\mathrm{min}$ and $b_\mathrm{max}$ with $b\,db$ weight. Compared to 
panels (a) and (b), this leads to additional ellipticity fluctuations 
related to the fluctuating impact parameter, whereas in Fig.~\ref{1104.0650:F1}a,b, 
only $N_\mathrm{part}$ fluctuations at fixed $b$ contribute.

As discussed in Sec.~\ref{sec:1104.0650:2a}, Eq.~(\ref{eq:1104.0650:10}), $\ecc\{4\}^4$ can 
become negative when fluctuations grow large. Whenever this happens, we 
replace $\ecc\{4\}$ by $\sqrt[4]{\left|\ecc\{4\}^4\right|}$ and indicate 
this by an open star in Fig.~\ref{1104.0650:F1} (connected by dotted lines to other
points in the graph). One sees that $\ecc\{4\}^4$ has a tendency
to turn negative in the most peripheral collisions. In very central
collisions, $\ecc\{4\}^4$ becomes very small, with central values that
can have either sign depending on whether we keep the impact parameter
fixed (Figs.~\ref{1104.0650:F1}a,b) or average over events with different impact 
parameters in a given centrality bin (see the $0{-}5\%$ centrality values
in Figs.~\ref{1104.0650:F1}c,d). Statistical errors are large, however, and within 
errors $\ecc\{4\}^4$ is compatible with zero for impact parameters 
$b<3$\,fm, i.e. in the most central ($0{-}5\%$ centrality) collisions.
We indicate this by open-ended error bars for 
$\sqrt[4]{\left|\ecc\{4\}^4\right|}$, pointing from its upper limit all 
the way to zero. 

%
\begin{figure}[b!]
 \includegraphics[width=0.9\linewidth]{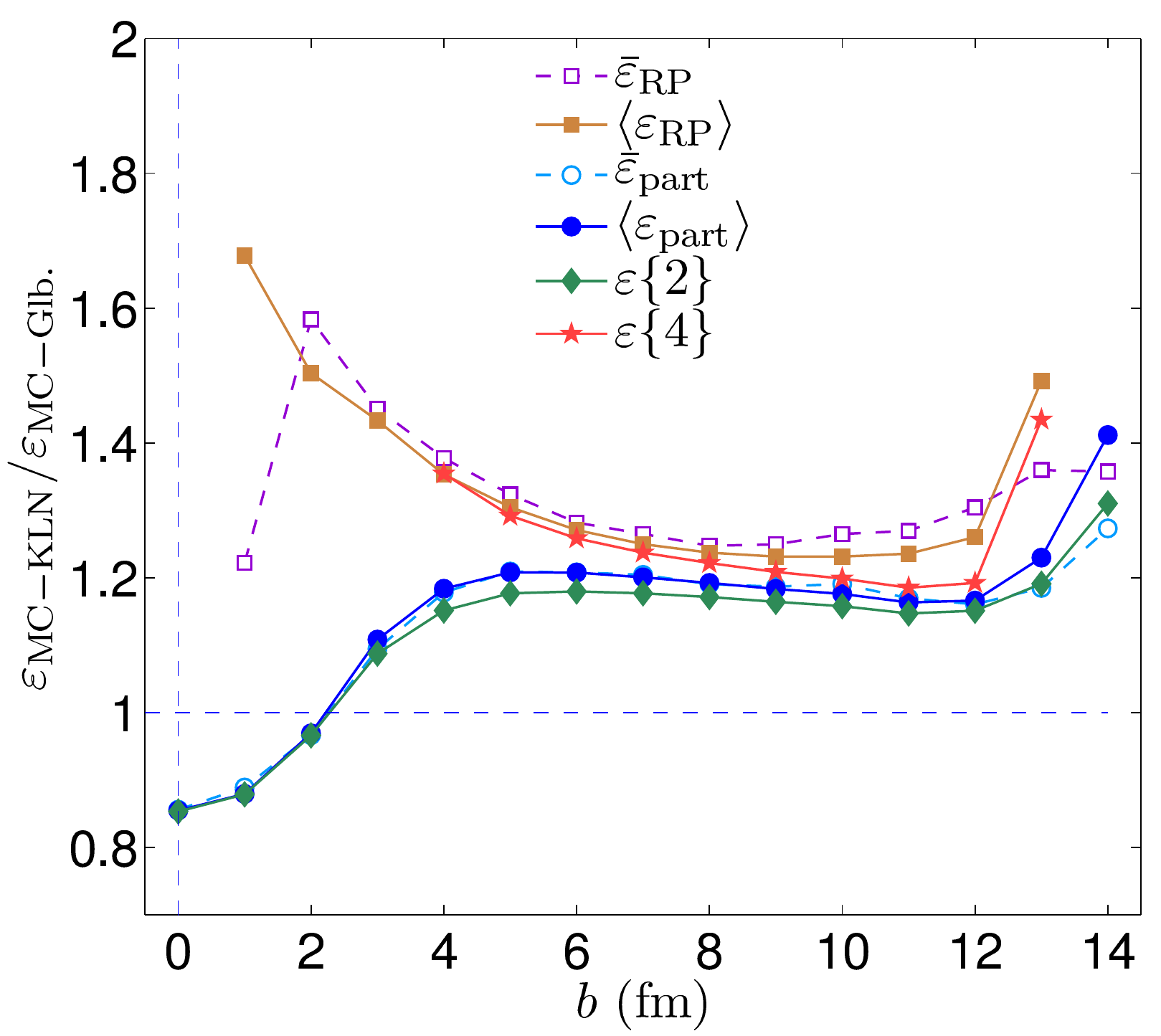}
\caption{Impact-parameter dependence of the ratio of 
   ellipticities obtained from the MC-KLN and MC-Glauber models 
   as shown in Figs.~\ref{1104.0650:F1}a,b. Taken from \cite{Qiu:2011iv}.
   \label{1104.0650:F2}
}
\end{figure}
%

Comparing panels (a,c) for the MC-Glauber model with panels (b,d) for
the MC-KLN model we see great similarities in shape, but systematic
differences in magnitude of the ellipticities. The ratio of the MC-KLN
and MC-Glauber ellipticities is shown in Fig.~\ref{1104.0650:F2}. Except for the 
most central and most peripheral collisions, the MC-KLN ellipticities
exceed the MC-Glauber ones by an approximately constant factor of around
1.2. Please note the difference in the ratios for the reaction plane and 
participant eccentricities at small $b$. (The point for 
$\bar{\ecc}_\mathrm{RP}$ at $b \eq 1$\,fm is obtained from a ratio of very 
small numbers and probably not statistically robust -- we had only 30,000 
events to determine the ensemble-averaged density profile.) For the 
$\ecc\{4\}$ ratio we dropped all points where the statistical error 
for $\ecc\{4\}^4$ extended into the region of negative values for either
the MC-Glauber or MC-KLN model.

Figure~\ref{1104.0650:F1} shows that, for central and mid-peripheral collisions,
the ensemble-averaged participant and reaction plane eccentricities 
$\la\ecc_\mathrm{part}\ra$ and $\la\ecc_\mathrm{RP}\ra$ agree very well
with the mean eccentricities $\bar{\ecc}_\mathrm{part}$ and 
$\bar{\ecc}_\mathrm{RP}$ of the corresponding ensemble-averaged profiles.
For strongly peripheral collisions ($b\gtrsim10$\,fm), however, the average
of the ratio (Eqs.~(\ref{eq:1104.0650:3},\ref{eq:1104.0650:4},\ref{eq:1104.0650:8})) differs strongly from 
the ratio of averages (Eqs.~(\ref{eq:1104.0650:5},\ref{eq:1104.0650:7})), indicating strong 
event-by-event fluctuations. We note that in very peripheral collisions, the 
average event ellipticity drops quickly with increasing $b$ while the
ellipticity of the average profile remains large; single-shot hydrodynamic
calculations based on a smooth average initial profile thus overestimate
the effective initial source ellipticity and produce more elliptic flow 
than expected from event-by-event hydrodynamic evolution of individual
peripheral events. Still, as first emphasized in \cite{Song:2011hk}, the 
calculated $v_2$ from single-shot hydrodynamics decreases steeply at large 
collision centralities \cite{Hirano:2010jg,Song:2011qa,%
Shen:2011eg}, due to the decreasing fireball lifetime, which contrasts 
with the initially reported experimentally observed behaviour 
\cite{:2008ed,Aamodt:2010pa}, but agrees qualitatively with a 
reanalysis \cite{ALICE:2011ab} where non-flow effects have been
largely eliminated and/or corrected for. We do point out that our 
Monte-Carlo simulations do not include fluctuations in the amount of 
entropy generated per nucleon-nucleon collision \cite{Qin:2010pf} which have important effects on the ellipticities in very peripheral 
collisions. 

\begin{figure}[h!]
 \includegraphics[width=0.9\linewidth]{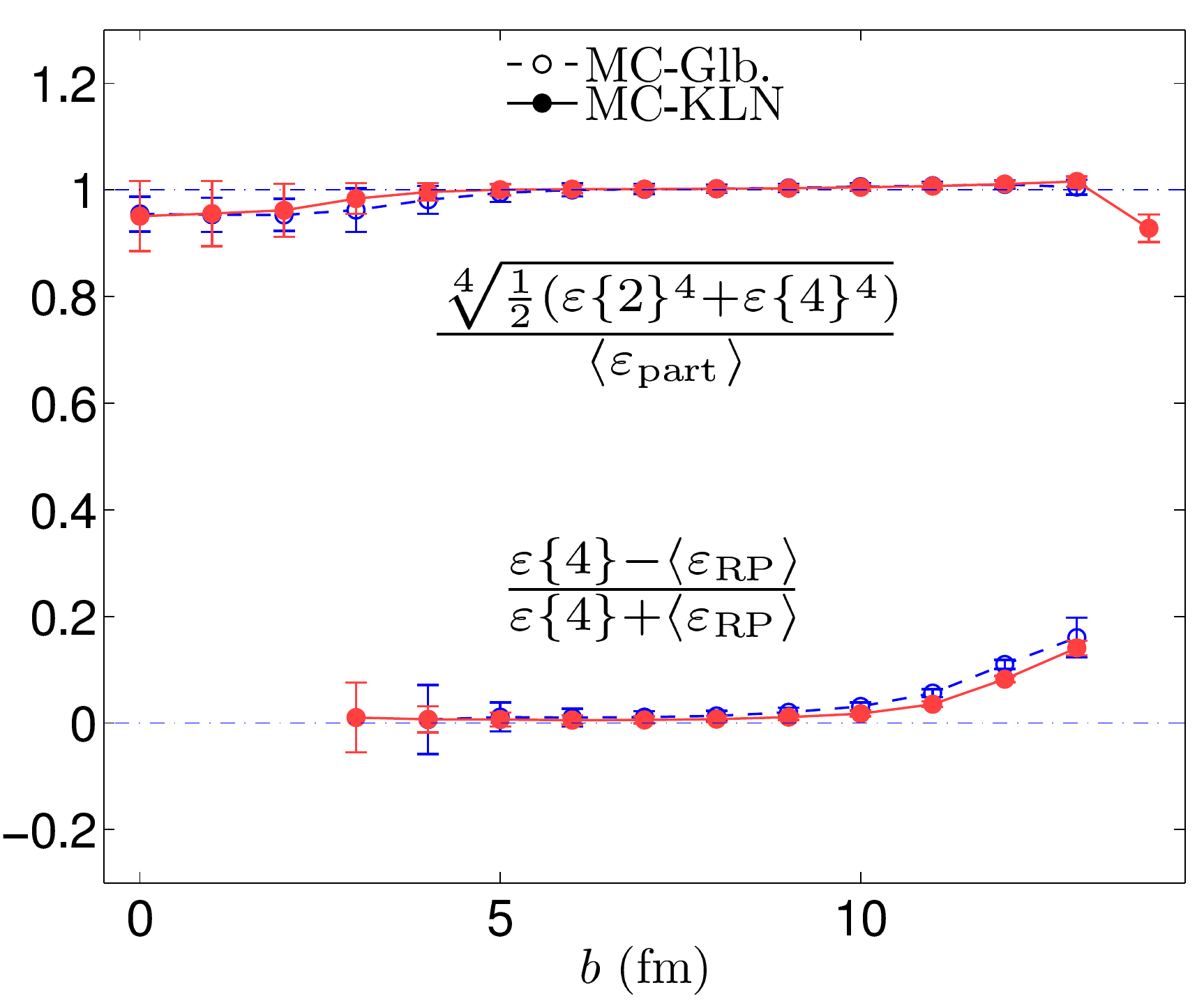}
 \caption{Checks of the assumption of Gaussian
    and Bessel-Gaussian fluctuations for $\epart$ (see text for 
    discussion). Taken from \cite{Qiu:2011iv}.
 \label{1104.0650:F3}
 }
\end{figure}

Comparing the curves for $\la\ecc_\mathrm{part}\ra$, $\ecc\{2\}$, and
$\ecc\{4\}$ in Fig.~\ref{1104.0650:F1}, we see that (as is manifest in the Gaussian 
model analysis in Eq.~(\ref{eq:1104.0650:11a})) $\ecc\{2\}$ receives a positive and 
$\ecc\{4\}$ receives a negative contribution from event-by-event 
ellipticity fluctuations. In Fig.~\ref{1104.0650:F3} we check, as a function of 
impact parameter, the validity of the identities (\ref{eq:1104.0650:11}) and 
%
%
(\ref{eq:1104.0650:11b}), which follow from Bessel-Gaussian and Gaussian $\epart$ 
distributions, respectively. We see that both hold with good accuracy in 
the mid-centrality range ($b{\,\lesssim\,}10$\,fm for Eq.~(\ref{eq:1104.0650:11}),
$5{\,\lesssim\,}b{\,\lesssim\,}11$\,fm for Eq.~(\ref{eq:1104.0650:11b})) but break down in 
the most peripheral collisions. Both the Gaussian and Bessel-Gaussian 
hypotheses work slightly better for the MC-KLN than for the MC-Glauber 
model. Consistent with the analysis in Ref.~\cite{Voloshin:2007pc}, the 
Gaussian fluctuation hypothesis for $\epart$ breaks down at small impact 
parameters, whereas (as theoretically expected \cite{Voloshin:2007pc}) the 
Bessel-Gaussian hypothesis appears to continue to hold, although we are 
unable to make this statement with statistical confidence. (For the ratio 
$(\ecc\{4\}{-}\la\ecc_\mathrm{RP}\ra)/(\ecc\{4\}{+}\la\ecc_\mathrm{RP}\ra)$,
we again dropped all points for which the error band for $\ecc\{4\}^4$
reaches into negative territory.)

The assumption of Gaussian fluctuations of the real and imaginary parts 
of the complex ellipticity (\ref{eq:1104.0650:1}) is often used to argue that the 
average reaction-plane ellipticity $\la\ecc_\mathrm{RP}\ra$ can serve as 
a proxy for $\ecc\{4\}$ (see Eq.~(\ref{eq:1104.0650:11})), and that therefore 
reaction-plane averaged initial density profiles can be used in 
single-shot hydrodynamics (which ignores event-by-event fluctuations) 
to simulate the experimentally measured $v_2\{4\}$ values. Fig.~\ref{1104.0650:F1} 
and the bottom curves in Fig.~\ref{1104.0650:F3} show that $v_2\{4\}$ values obtained 
from single-shot hydrodynamic simulations with reaction-plane averaged 
initial conditions \cite{Hirano:2010jg,Song:2011qa} should not be trusted 
quantitatively for centralities $>40\%$.

To summarize this subsection, all the simplifying assumptions that allow one
to focus attention on the three quantities $\la\ecc_\mathrm{part}\ra$,
$\ecc\{2\}$, and $\ecc\{4\}$ only (by substituting $\la\ecc_\mathrm{part}\ra$
for $\bar{\ecc}_\mathrm{part}$ and $\ecc\{4\}$ for $\bar{\ecc}_\mathrm{RP}$
or $\la\ecc_\mathrm{RP}\ra$) hold well for central to mid-central collisions
(${\,\leq\,}40\%$ centrality) but break down for peripheral collisions.
For ${\,>\,}40\%$ centrality there exists no substitute for event-by-event 
hydrodynamics if one aims for quantitative precision in the comparison with 
experimental elliptic flow data, since the latter are strongly affected by non-Gaussian 
event-by-event fluctuations at those centralities. 

\subsection{Ellipticities with different weight functions}
\label{sec:1104.0650:3b}

%
\begin{figure*}[h!]
 \center{
 	\includegraphics[width=0.7\linewidth]{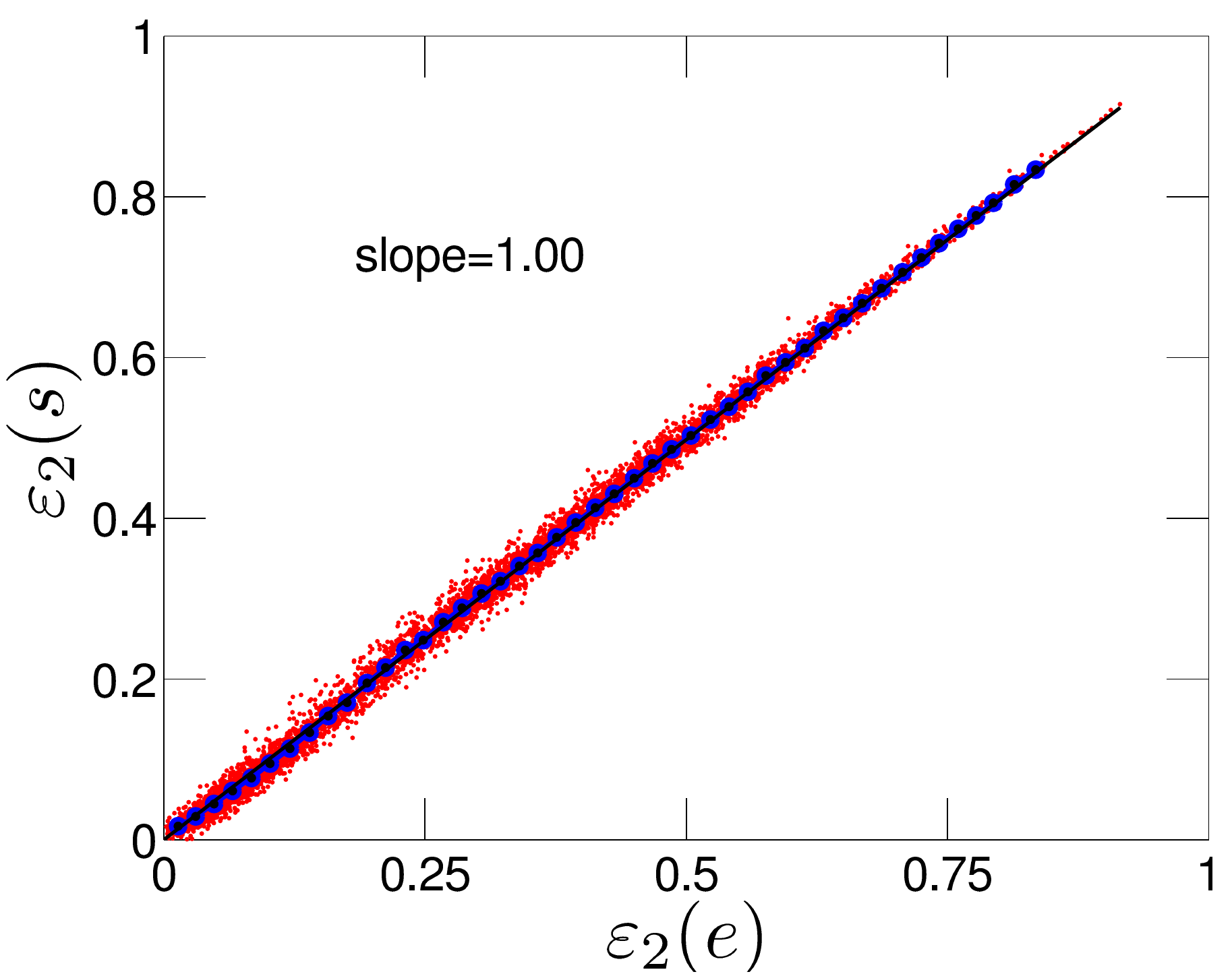}
 }
 \caption{$\ecc_\mathrm{part}(e)$ vs. $\ecc_\mathrm{part}(s)$ for 6000 initial profiles from the MC-KLN model (1000 each for $b \eq 0$\,fm, $0{-}5\%$, $15{-}20\%$, $20{-}30\%$, $30{-}40\%$, and $50{-}60\%$ centrality). Taken from \cite{Qiu:2011iv}.
 \label{1104.0650:F4}
 }
\end{figure*}
%

Figure~\ref{1104.0650:F4} shows a comparison between the energy- and entropy-weighted
ellipticities of the initial profiles generated with the MC-KLN model on 
an event-by-event basis. The scatter plot is based on 6000 events, 1000 
each for $b=0$ and for the following finite-width centrality bins: 
$0{-}5\%$, $15{-}20\%$, $20{-}30\%$, $30{-}40\%$, and $50{-}60\%$.
This is not a realistic mix in the experimental sense, but permits us to 
explore the full range from very small to very large event ellipticities.
The blue dots in Fig.~\ref{1104.0650:F4} represent bin averages, and the solid black
line is a linear fit through the origin. The fitted slope is 1.00, the 
scatter plot is seen to be tightly clustered around this fitted line, and 
only at small ellipticities $\ecc_2{\,<\,}20\%$ the $e$-weighted values are 
seen to be slightly larger on average than their $s$-weighted counterparts
(see also Fig.~\ref{1104.0650:F5}a below).  

\subsection{Higher order harmonics}
\label{sec:1104.0650:3c}

%
\begin{figure*}[h!]
 \includegraphics[width=0.49\linewidth]{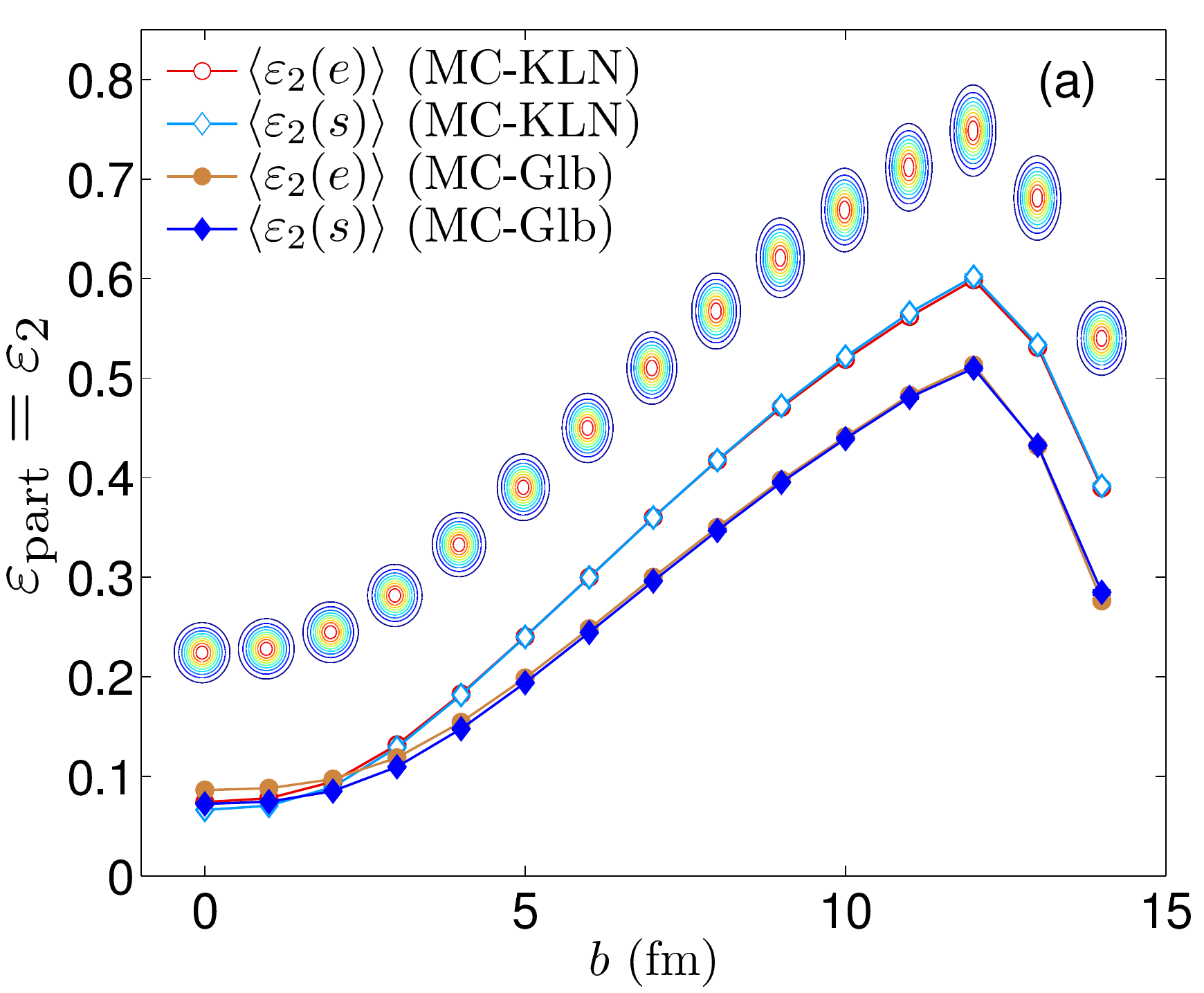}
 \includegraphics[width=0.50\linewidth]{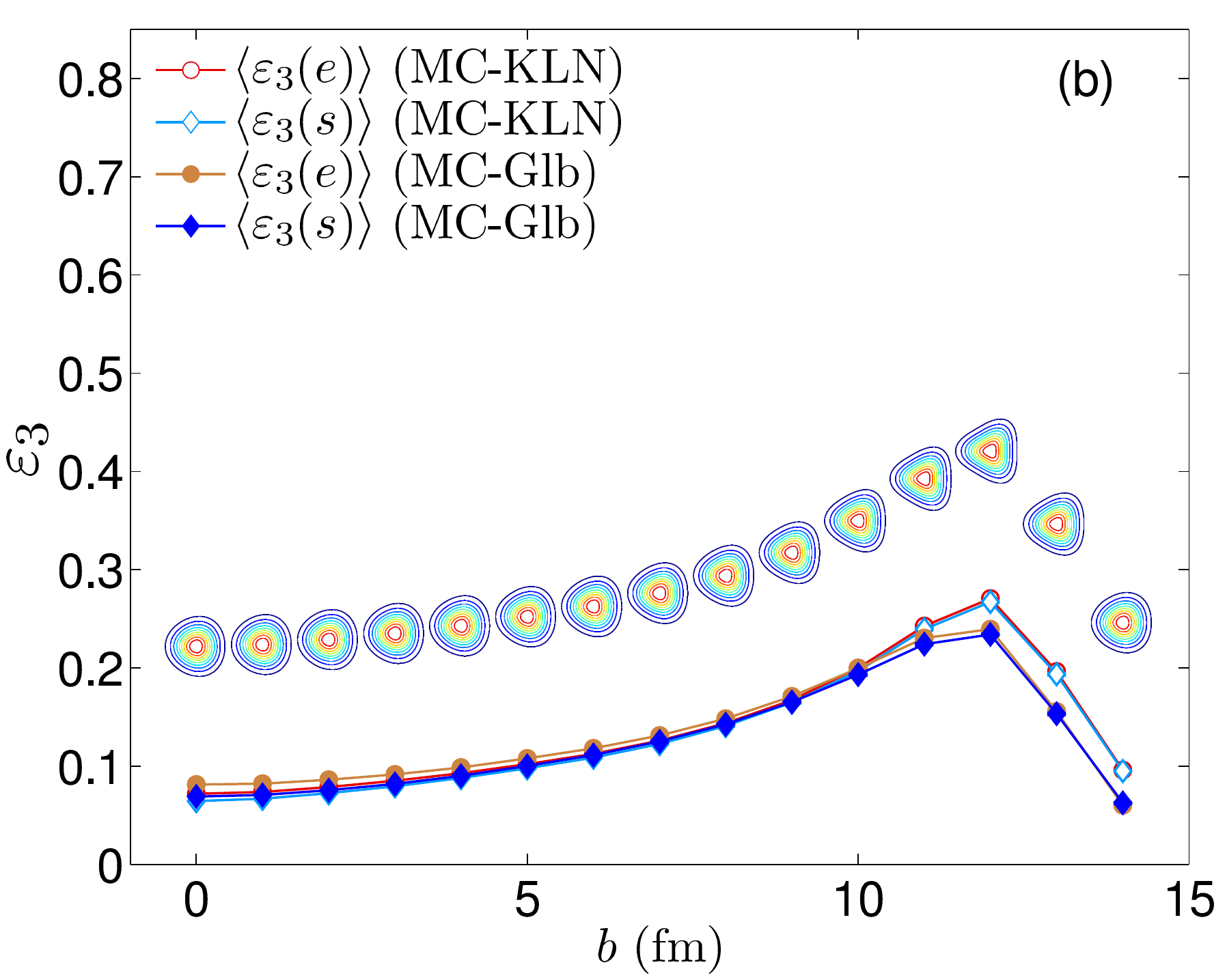}\\
 \includegraphics[width=0.49\linewidth]{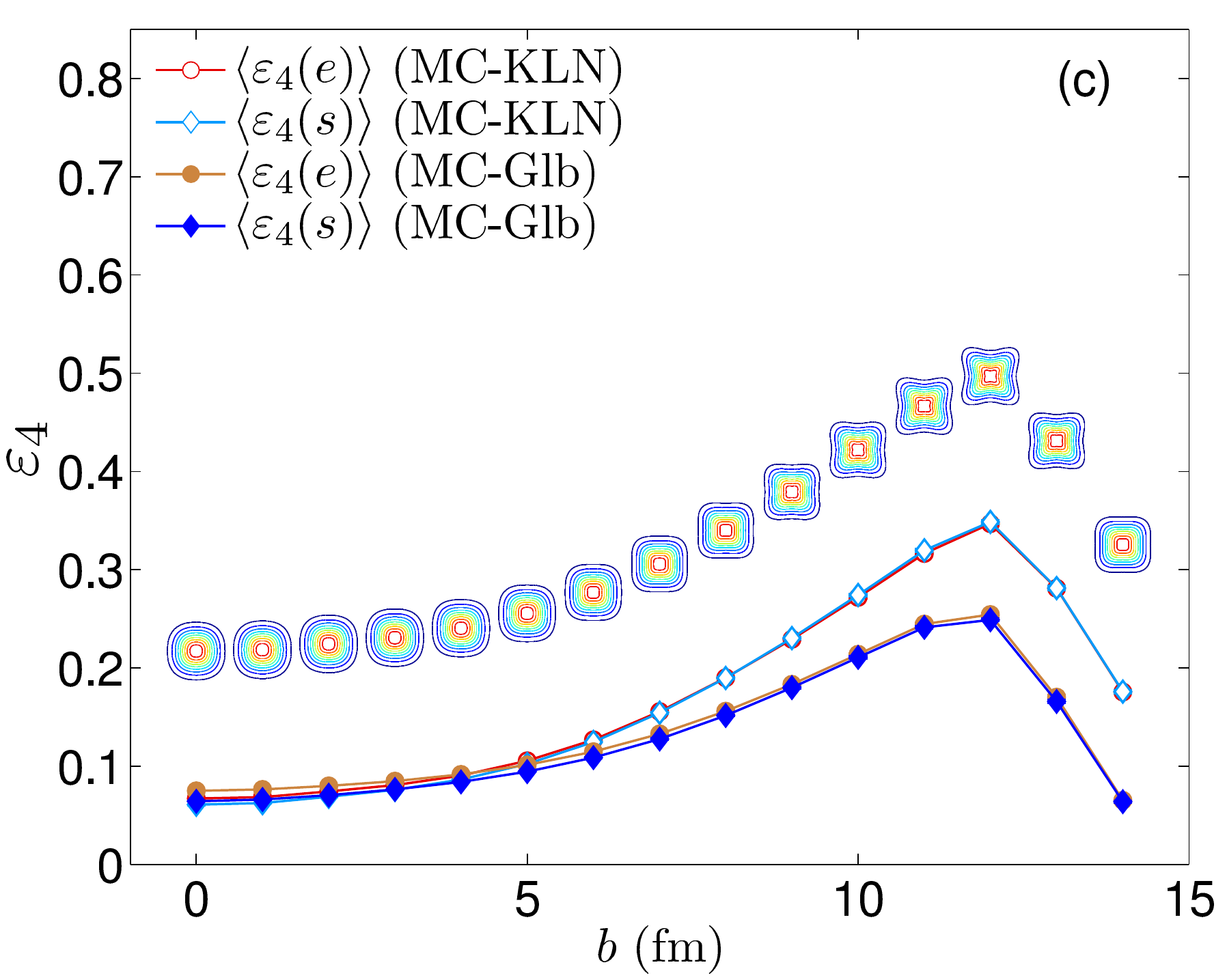}
 \includegraphics[width=0.49\linewidth]{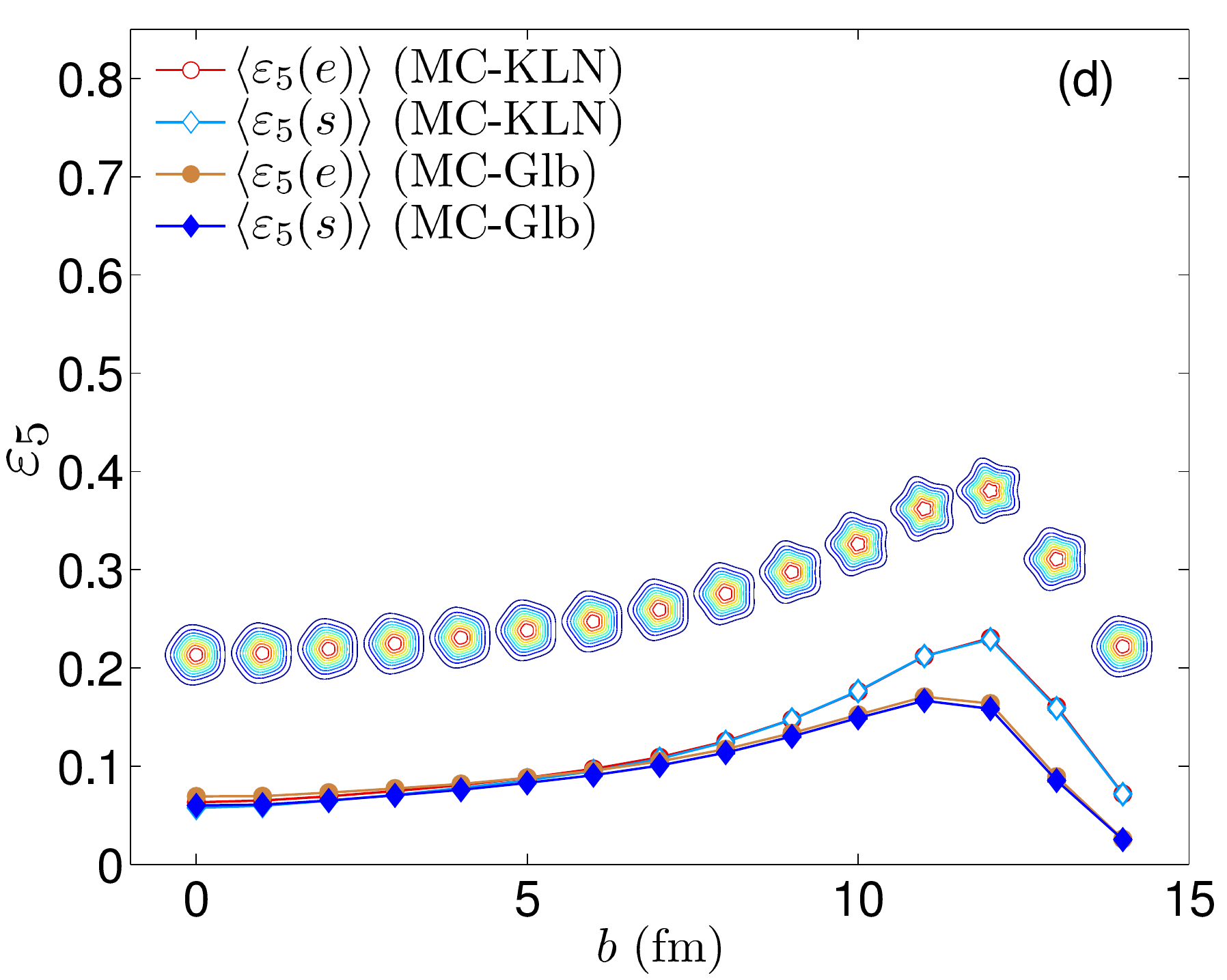}
 \caption{Harmonic eccentricity coefficients $\ecc_2\eq\ecc_\mathrm{part}$ (a), $\ecc_3$ (b), $\ecc_4$ (c) and $\ecc_4$ (d) as functions of impact parameter, calculated from the MC-Glauber (filled symbols, solid lines) and MC-KLN models (open symbols, dashed lines), using the energy density (circles) or entropy density (triangles) as weight function. The contour plots illustrate deformed Gaussian profiles $e(r,\phi)\eq{e}_0\,\exp\left[-\frac{r^2}{2\rho^2} \bigl(1{+}\ecc_n\cos(n\phi)\bigr)\right]$, with eccentricity $\ecc_n(e)$ taken from the MC-KLN model at the corrsponding impact parameter. Taken from \cite{Qiu:2011iv}.
 \label{1104.0650:F5}
 }
\end{figure*}
%

In Figs.~\ref{1104.0650:F5}a-d we compare the centrality dependences of the 
ensemble-averaged second to fifth harmonic eccentricity coefficients 
(energy- and entropy-weighted) from the MC-Glauber and MC-KLN models. 
The contour plots give a visual impression of the degree of deformation 
corresponding to the (larger) MC-KLN eccentricities, assuming (for 
illustration) the absence of any other eccentricity coefficients than 
the one shown in the particular panel.

First, one observes very little difference between the eccentricities of 
the entropy and energy density profiles, except for very central collisions 
($b{\,\lesssim\,}5$\,fm for the MC-Glauber, $b{\,\lesssim\,}3$\,fm for the MC-KLN 
model) where the energy-weighted eccentricities lie systematically somewhat 
above the entropy-weighted ones (for all orders $n$ studied here). The 
difference between $s$- and $e$-weighted eccentricities at small $b$ is 
bigger in the MC-Glauber than in the MC-KLN model.

Next, one notes the significantly larger ellipticities and quadrangularities
of the MC-KLN distributions compared to those from the MC-Glauber model
for all but the most central collisions. These are driven by geometry, i.e.
by the almond-shaped deformation of the nuclear overlap zone in non-central
collisions, which in the KLN-model is more eccentric than in the Glauber
model. The third and fifth order harmonics, which are entirely due to
fluctuations (and whose associated angles $\PsiPP{n}$ are therefore
completely uncorrelated to the reaction plane -- see Ref.~\cite{Qin:2010pf} 
and discussion below), show remarkably similar eccentricity values in the 
two initialization models, except for the most peripheral events. 
Comparing the viscous suppression of elliptic and triangular flow thus allow to distinguish experimentally between the MC-Glauber and MC-KLN models (see \chap{e2e3}).
 
Third, in central collisions all four eccentricity coefficients are
roughly of the same size. In peripheral collisions, the fluctuation-dominated
eccentricity coefficients ($\ecc_3$ and $\ecc_5$) are generically smaller 
than the geometry-dominated ones ($\ecc_2$, but also to some extent 
$\ecc_4$).\footnote{We checked that the centrality dependences
  of the ratios $\ecc_n/\ecc_2$ agree qualitatively, but not
  quantitatively, with Fig.~3 in Ref.~\cite{Lacey:2010hw}. We suspect
  that the differences, which are larger for the MC-Glauber than the
  MC-KLN model, are due to somewhat different Woods-Saxon and (in the 
  MC-Glauber case) fluctuation-size parameters used in 
  Ref.~\cite{Lacey:2010hw}.} 
This is less obvious when one defines the higher order eccentricities
with $r^n$ instead of $r^2$ weight \cite{Qin:2010pf}, which tends to
increase the values of the higher harmonics in peripheral collisions.

Even with ``only'' an $r^2$ weight, $\ecc_4$ and $\ecc_5$ are seen to 
become large enough around $b\sim10-13$\,fm that, if collective acceleration
happens predominantly in the directions of steepest descent of the 
density profile, one has to expect cross-currents in the developing 
anisotropic flow patterns. These can lead to destructive interference
and a correspondingly reduced efficiency of converting $n^\mathrm{th}$-order
eccentricities $\ecc_n$ into $n^\mathrm{th}$-order harmonic flows $v_n$
\cite{Alver:2010dn}. In realistic situations this issue is exacerbated
by the simultaneous presence of {\em several} large eccentricity components
$\ecc_n$, which is expected to lead to a strongly nondiagonal and
probably nonlinear response matrix relating $v_n$ to $\ecc_n$ 
\cite{Qin:2010pf}. This will be discussed in Sec.~\ref{sec:1104.0650:4}, and more in \sec{misc:best_estimator}. 
  
\subsection{Eccentricity correlations}
\label{sec:1104.0650:3d}

%
\begin{figure*}
 \includegraphics[width=0.32\linewidth]{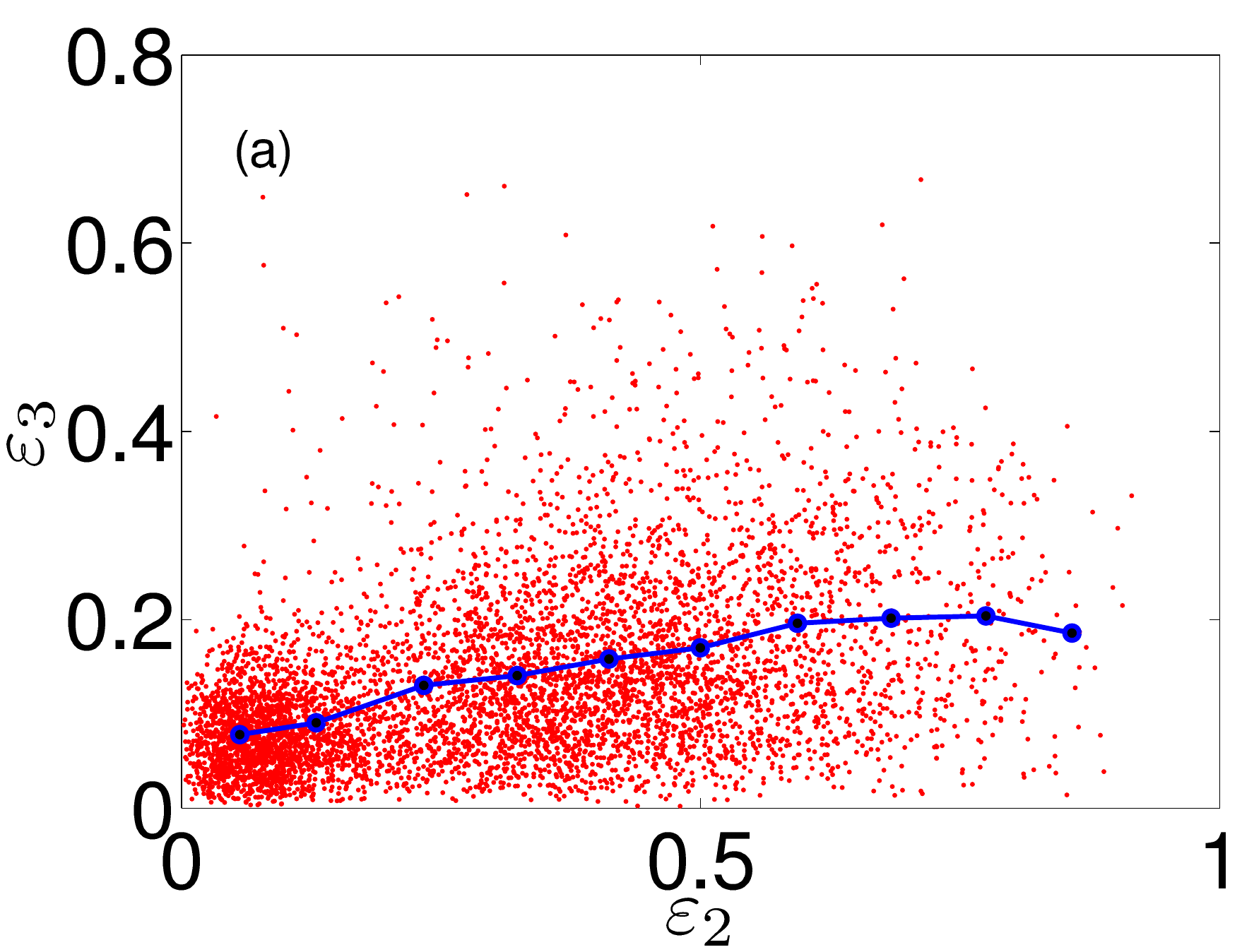}
 \includegraphics[width=0.32\linewidth]{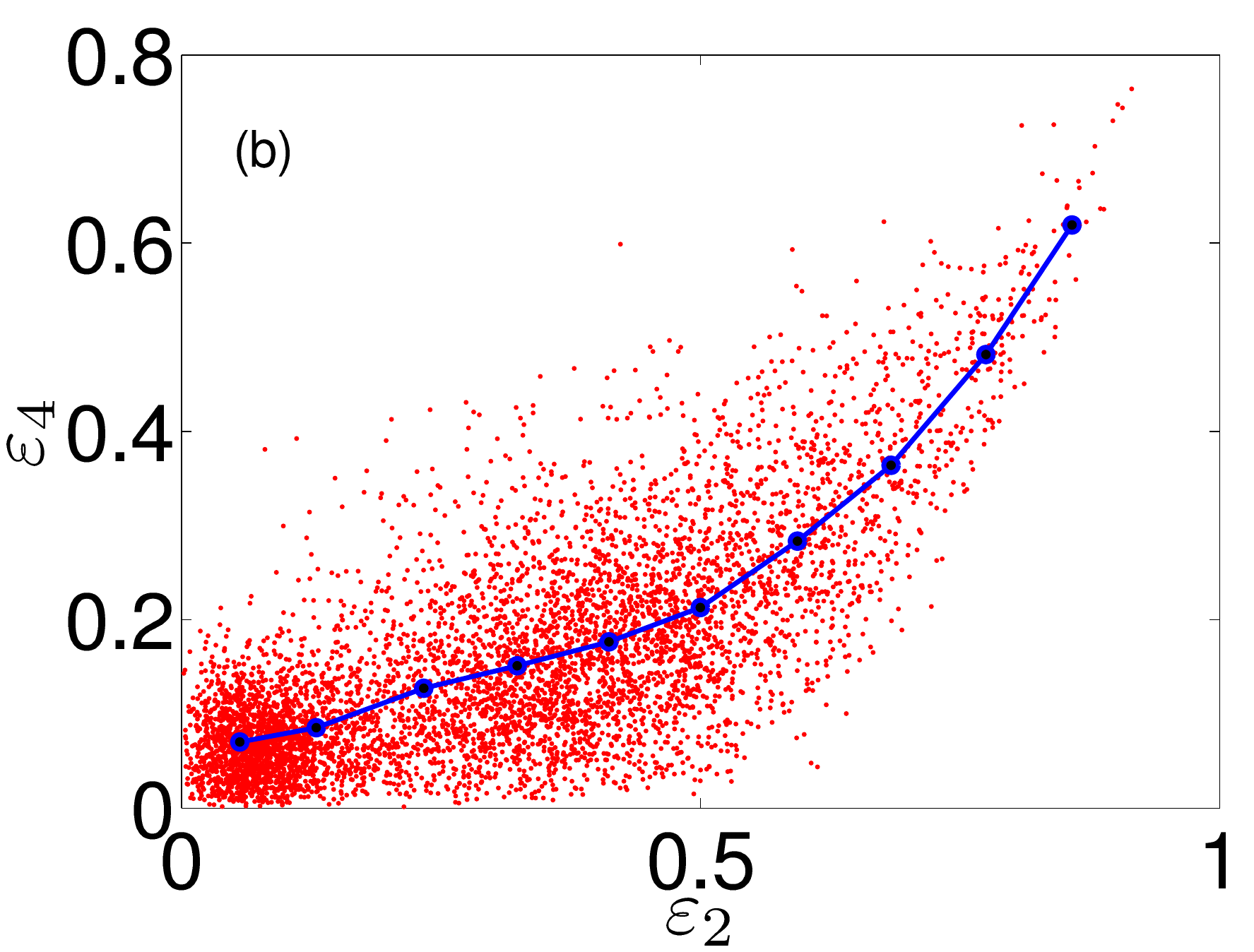}
 \includegraphics[width=0.32\linewidth]{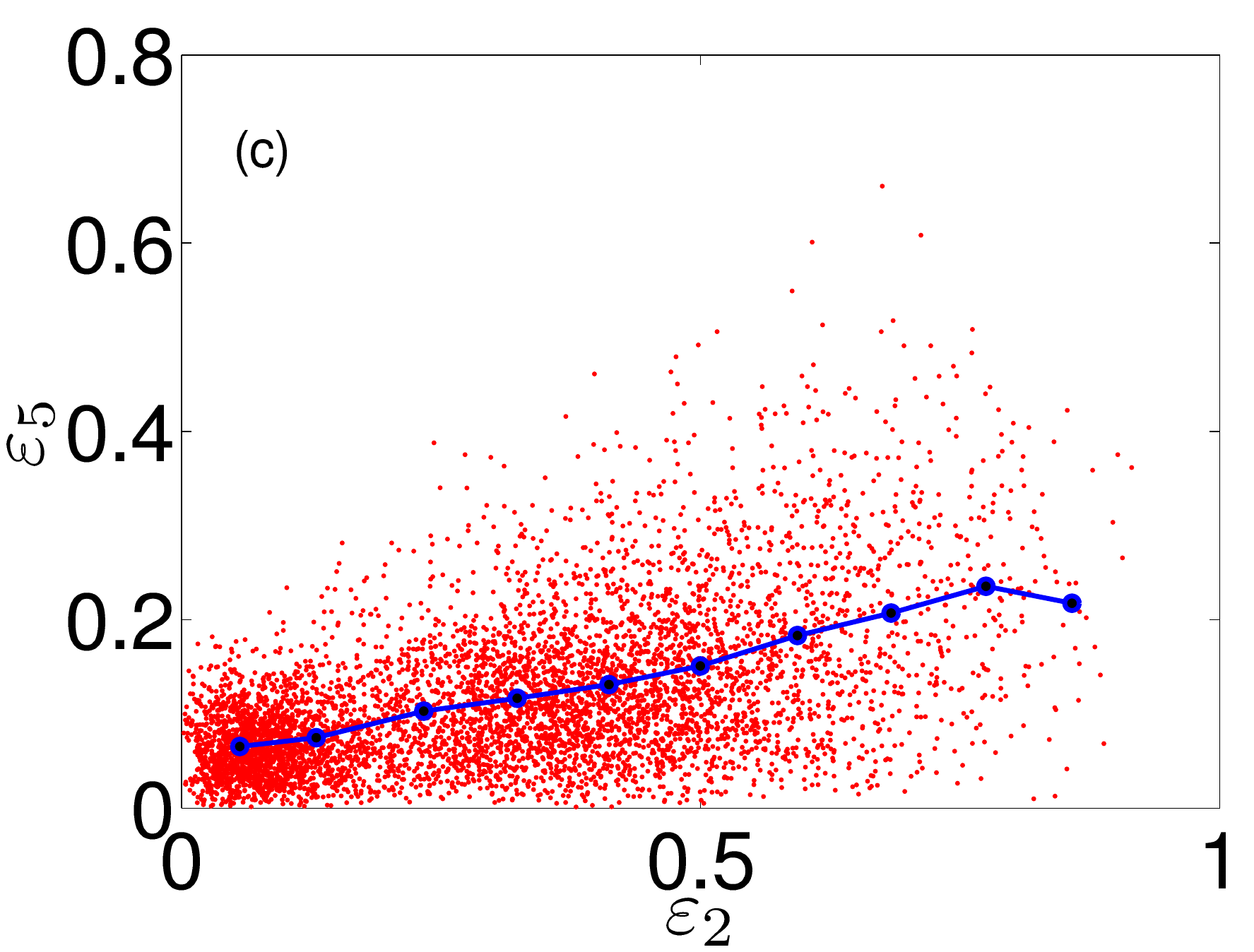}
 \caption{Scatter plots illustrating the event-by-event
   correlation of $\ecc_3$ (a), $\ecc_4$ (b), and $\ecc_5$ (c) with the 
   fireball ellipticity $\ecc_2\eq\ecc_\mathrm{part}$, for the same set
   of 6000 event profiles analyzed in Fig.~\ref{1104.0650:F4}. The blue dots
   connected by a line indicate $\ecc_2$-bin averages to guide the eye. Taken from \cite{Qiu:2011iv}.
 \label{1104.0650:F6}
 }
\end{figure*}
%

It is reasonable to ask whether and how the different harmonic 
eccentricity coefficients $\ecc_n$ are correlated with each other.
Figure~\ref{1104.0650:F6} shows scatter plots of the correlations between 
$\ecc_{3,4,5}$ and the ellipticity $\ecc_2$, which, for large $\ecc_2$
values, is dominated by geometric overlap effects. We note that,
according to the definition (\ref{eq:1104.0650:15}), all eccentricity coefficients
are positive definite, $\ecc_n{\,\geq\,}0$. Keeping this in mind, 
Figs.~\ref{1104.0650:F6}a,c show that $\ecc_3$ and $\ecc_5$ are uncorrelated
with the fireball ellipticity; the slight growth of $\la\ecc_{3,5}\ra$
with increasing $\ecc_2$ is related to the growth of the variances of
their distributions in more peripheral collisions. 

In contrast, the quadrangularity $\ecc_4$ shows a clear positive 
correlation with the ellipticity, see Fig.~\ref{1104.0650:F6}b. It is of geometrical 
origin: it reflects the football or almond shape of the overlap zone in 
non-central collisions which is a little sharper than a pure $\cos(2\phi)$ 
deformation. This is corroborated by the behavior of angle $\PsiPP{4}$ shown 
in Fig.~\ref{1104.0650:F7}a below, which, on average, points $45^\circ$ relative to 
$\PsiPP{2}$ (which again points in $x$-direction). This means 
that the quadrangular component of the initial fireball definition is 
oriented like a diamond, with its corners on the $x$ and $y$ axes. 
Superimposing it on a pure $\cos(2\phi)$ deformation leads to a somewhat 
sharper shape of the density distribution.

\section{Event-by-event hydrodynamics and flow fluctuations}
\label{sec:1104.0650:4}

In this section we analyze the results from event-by-event
hydrodynamic evolution of the fluctuating initial profiles studied in
the previous section. We focus on the anisotropic flow coefficients $v_n$,
their relationship to the initial eccentricity coefficients $\ecc_n$,
and the correlation between the $n^\mathrm{th}$-order flow angles 
$\PsiEP{n}$ and the corresponding $n^\mathrm{th}$-order 
participant-plane angles $\PsiPP{n}$ associated with $\ecc_n$.

\subsection{Correlations between participant plane, event plane, and
  reaction plane}
\label{sec:1104.0650:4a}

One of the key characteristics of fluid dynamics is its ability
to transform initial geometric deformation into a deformation of
the final momentum distribution, via collective flow. This happens
through spatially anisotropic hydrodynamic forces (i.e. pressure
gradients) which cause anisotropic acceleration of the fluid. As a 
result, correlations between participant and event planes are expected:
The angle $\PsiPP{n}$ points in the direction of the largest
pressure gradient associated with the $n^\mathrm{th}$ harmonic component
of the spatial deformation of the initial density distribution, while
$\PsiEP{n}$ points into the direction where the $n^\mathrm{th}$ 
harmonic component of the final collective flow is largest. Without
interference between harmonics of different order, we would thus expect
$\PsiPP{n}$ and $\PsiEP{n}$ to point, on average
and up to event-by-event fluctuations, in the same direction.

%
\begin{figure*}[h!]
 \includegraphics[width=0.32\linewidth]{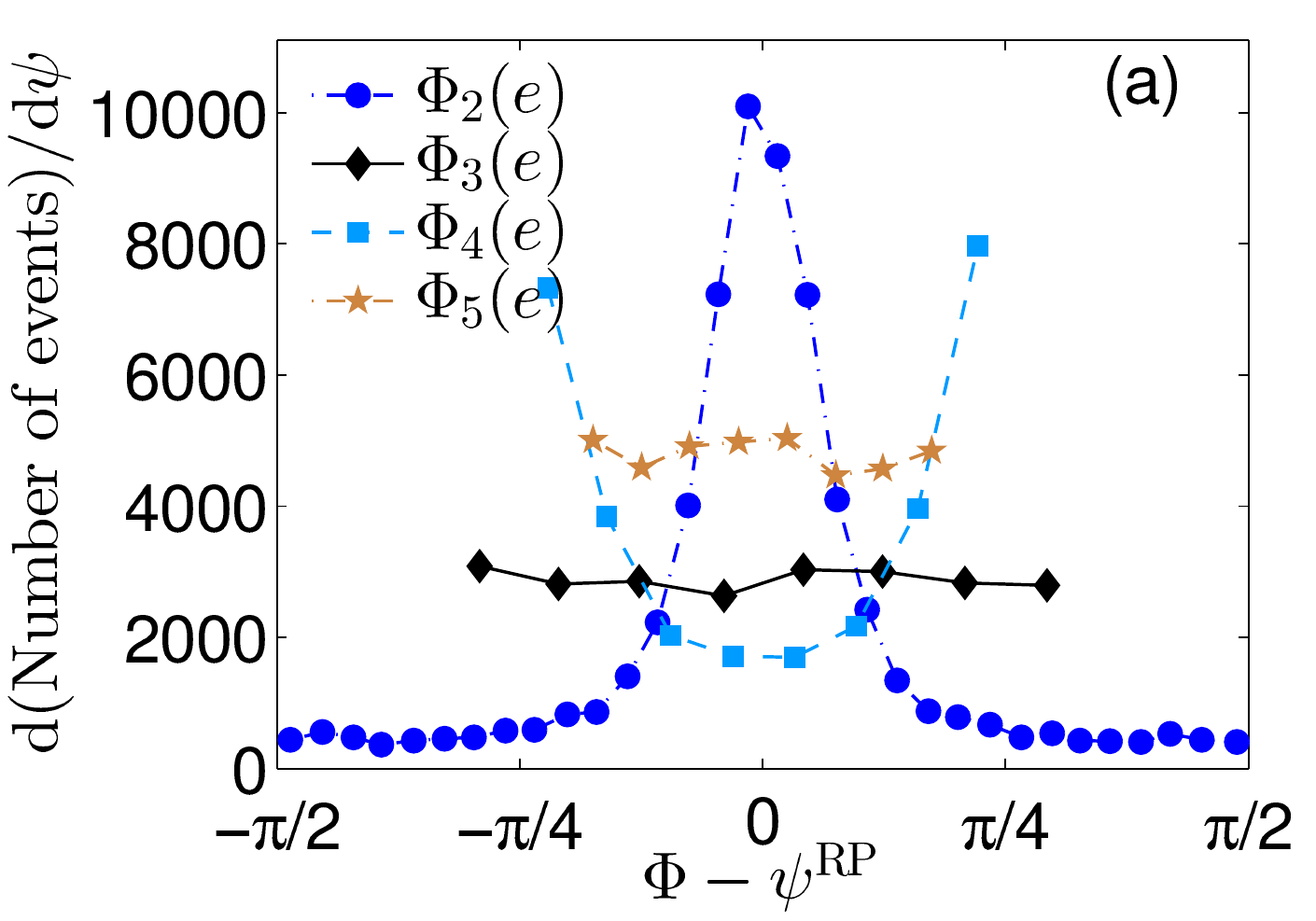}
 \includegraphics[width=0.32\linewidth]{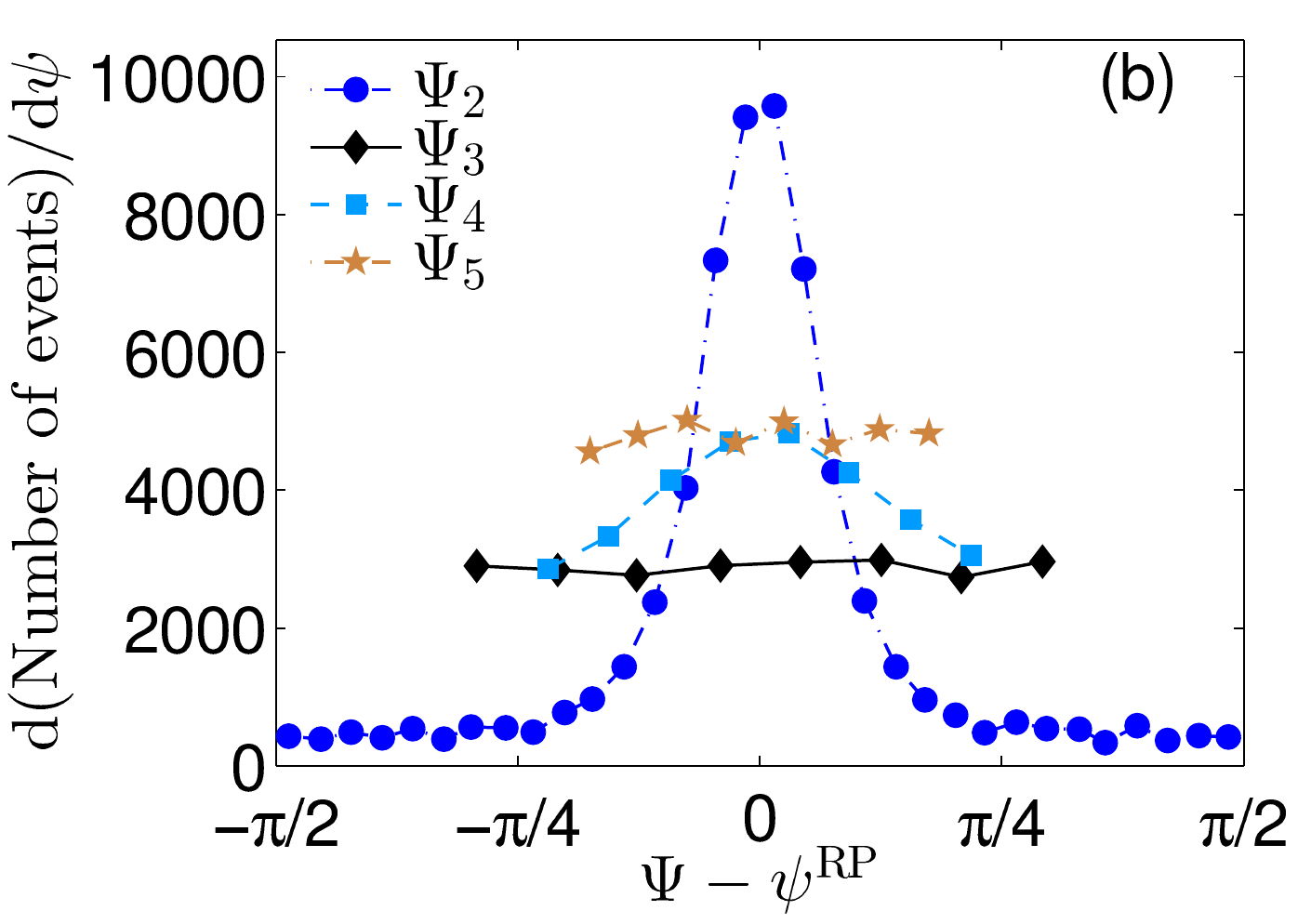}
 \includegraphics[width=0.33\linewidth]{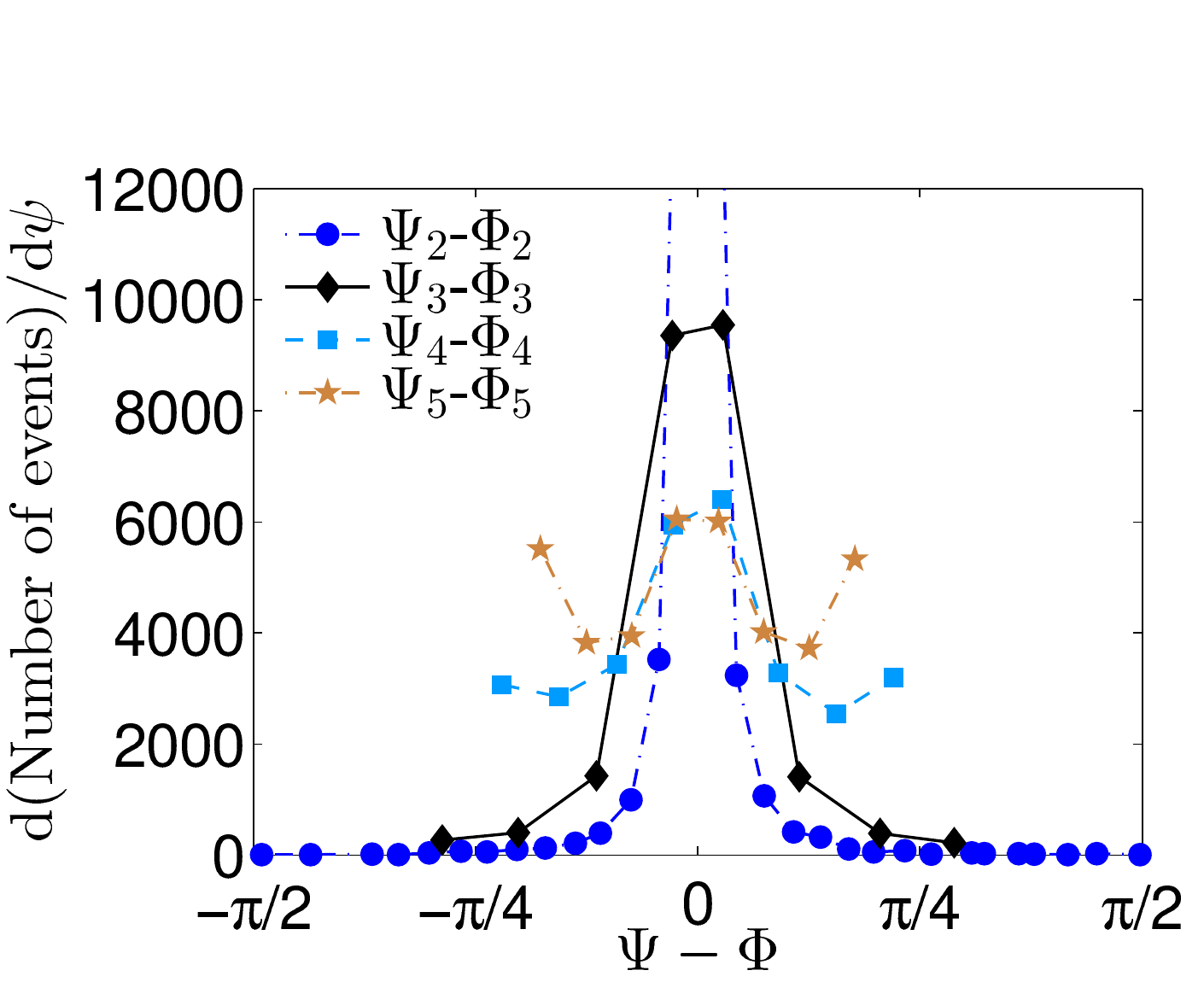} 
 \caption{Event-by-event correlation of the participant 
    plane (a) and event plane (b) angles with the reaction plane 
    ($\psi^\mathrm{RP}$), as well as the correlation between participant and event plane 
    angles (c) for different harmonic eccentricity and flow coefficients.
    The same 6000 events as in Fig.~\ref{1104.0650:F4} were analyzed after hydrodynamic evolution with $\eta/s=0$ (ideal fluid). Taken from \cite{Qiu:2011iv}.
 \label{1104.0650:F7}
 }
\end{figure*}
%

In Figs.~\ref{1104.0650:F7}a,b we show the distribution of participant- and event-plane 
angles associated with the $n^\mathrm{th}$-order eccentricities and 
harmonic flows, relative to the $x$-$z$ reaction plane. The analysis
uses the same 6000 events as before, and evolves them with ideal fluid dynamics ($\eta/s=0$). In panel (a) we see that 
$\PsiPP{3,5}$ are completely uncorrelated with the reaction 
plane \cite{Qin:2010pf}, as expected from the fact that the corresponding 
eccentricities are entirely fluctuation-driven, without contribution from 
the collision geometry. Panel (b) shows that the same holds true
for $\PsiEP{3,5}$, which is (at least superficially) consistent
with the expectation that $v_3$ is mostly or entirely driven by $\ecc_3$,
and $v_5$ by $\ecc_5$. We will revisit this below. $\PsiPP{2}$
and $\PsiEP{2}$ are strongly correlated with the reaction plane $\psi^\mathrm{RP}$,
at least for this mixed-centrality set of events. This is expected since,
for non-central collisions, $\ecc_2$ is mostly controlled by the almond-shaped
overlap geometry, and $v_2$ is mostly a collective flow response to this
geometric deformation; event-by-event fluctuations contribute to $\ecc_2$
(and thus $v_2$), but in general do not dominate them.

The behavior of $\PsiPP{4}$ in Fig.~\ref{1104.0650:F7}a is interesting 
because it is on average strongly ``anti-correlated'' with the reaction 
plane, in the sense that it points (on average) at $45^\circ$ relative to 
the $x$-axis. The geometric reason for this has already been discussed above 
in subsection \ref{sec:1104.0650:3d}. On the other hand, Fig.~\ref{1104.0650:F7}b shows that
the angle $\PsiEP{4}$ points on average {\em into} the reaction
plane. This correlation of  $\PsiEP{4}$ with the reaction plane 
is somewhat weaker than the anti-correlation of $\PsiPP{4}$ with
that plane seen in panel (a). Still, it suggests that quadrangular flow 
$v_4$ does not, on average, develop predominantly in the direction of the 
steepest pressure gradient associated with $\ecc_4$, but in the direction 
of steepest $\ecc_2$-induced pressure gradient.

Figure~\ref{1104.0650:F7}c, however, in which we analyze directly the correlation 
between the event and participant plane angles, paints a more subtle 
picture. It shows, surprisingly, a correlation peak at zero relative
angle between $\PsiEP{4}$ and $\PsiPP{4}$, whereas the
above discussion should have led us to expect a correlation peak at $45^\circ$.
The resolution of this paradox is presented in the next subsection:  
the relative importance of geometric and fluctuation-induced contributions
to $\ecc_n$, $v_n$, and their associated angles changes with collision 
centrality, with geometry playing a relatively larger role in peripheral 
collisions. One should therefore look at the angle correlations as a
function of collision centrality. One finds that the correlation function
peaks in Figs.~\ref{1104.0650:F7}a,b for the $4^\mathrm{th}$-order angles relative 
to the reaction plane are almost entirely due to geometric effects in 
peripheral collisions, while in central collisions both $\PsiPP{4}$ 
and $\PsiEP{4}$ are fluctuation-dominated and thus essentially
uncorrelated with the reaction plane. On the other hand, precisely because
in central collisions geometric effects such as geometrically driven elliptic
flow do not dominate the hydrodynamic response to the fluctuation-driven 
higher-order eccentricities, $\PsiEP{4}$ and $\PsiPP{4}$ 
remain relatively strongly correlated in near-central collisions. This 
is the reason for the peak at $0^\circ$ for $n \eq 4$ in Fig.~\ref{1104.0650:F7}c. 
(A hint of the ``anti-correlation'' at $45^\circ$ is still visible in 
Fig.~\ref{1104.0650:F7}c, and it would be stronger if we had not (for unrelated 
reasons) strongly oversampled central collisions in our mixed-centrality 
sample.)

We close this discussion with the following additional observations
about Fig.~\ref{1104.0650:F7}c: (i) The second-order participant and event planes
are much more strongly correlated with each other than either one of them
is with the reaction plane. This shows that even in very central collisions, 
where the source ellipticity is mostly fluctuation-driven and its angle 
therefore only weakly correlated with the reaction plane, elliptic flow 
develops event-by-event in the direction of the short axis of the 
ellipsoid. (ii) Even though the angles associated with $\ecc_3$ and $v_3$
are uncorrelated with the reaction plane (Figs.~\ref{1104.0650:F7}a,b), they
are strongly correlated with each other. This indicates that $v_3$
is mostly driven by $\ecc_3$, especially in the more central collisions, 
with relatively little interference from other harmonics. (iii) The  
$5^\mathrm{th}$-order event and participant plane angles show correlation 
peaks both at $0$ and $\pi/5$. As we will see in the following subsection, 
the former results from central and the latter from peripheral collisions.
The peak at $\pi/5$ indicates significant cross-feeding between modes 
with $n \eq 2,\,3$, and 5.

\subsection{Centrality dependence of event and participant plane 
  correlations}
\label{sec:1104.0650:4b}

%
\begin{figure*}
 \includegraphics[width=0.45\linewidth]{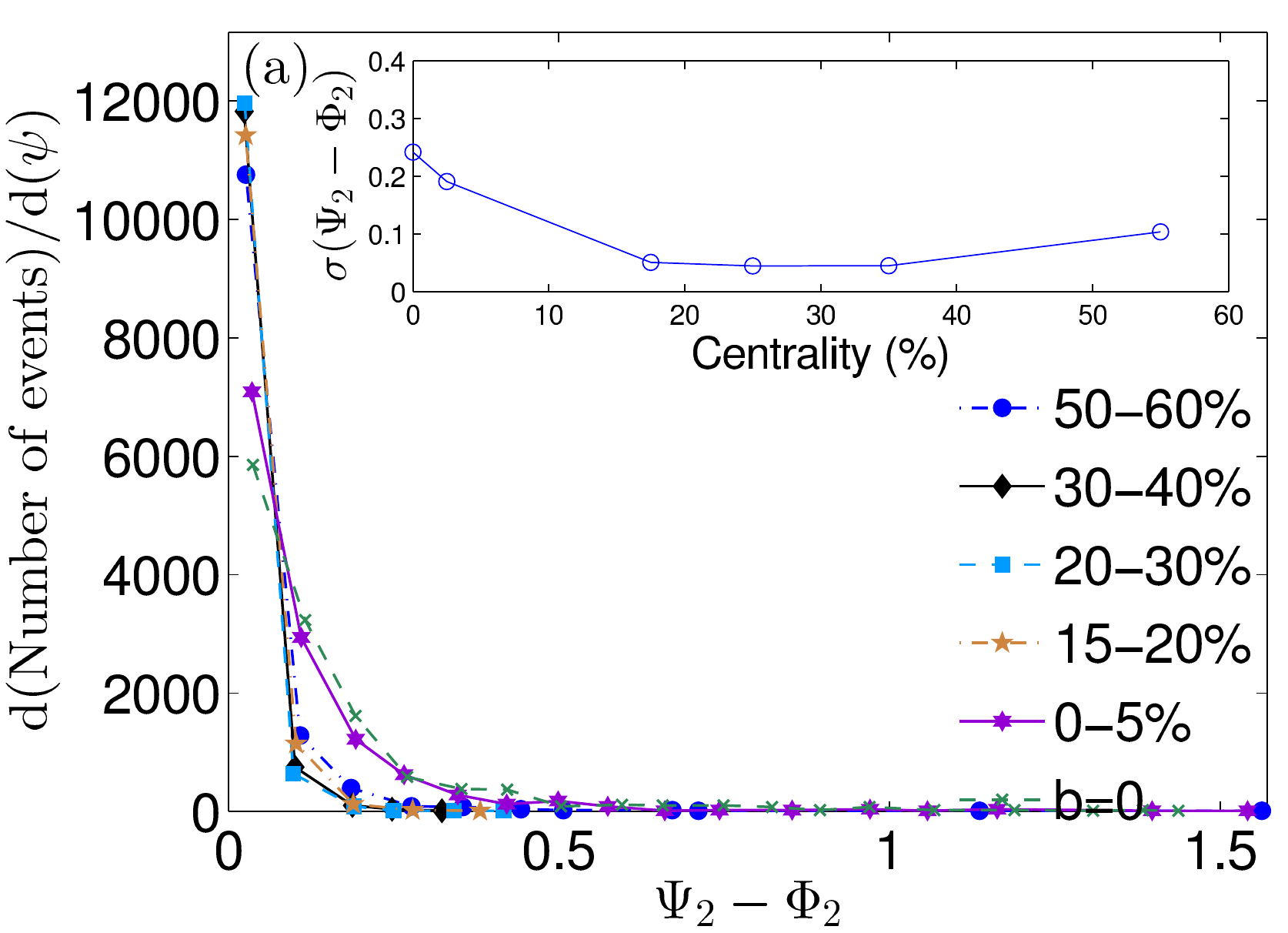}
 \includegraphics[width=0.44\linewidth]{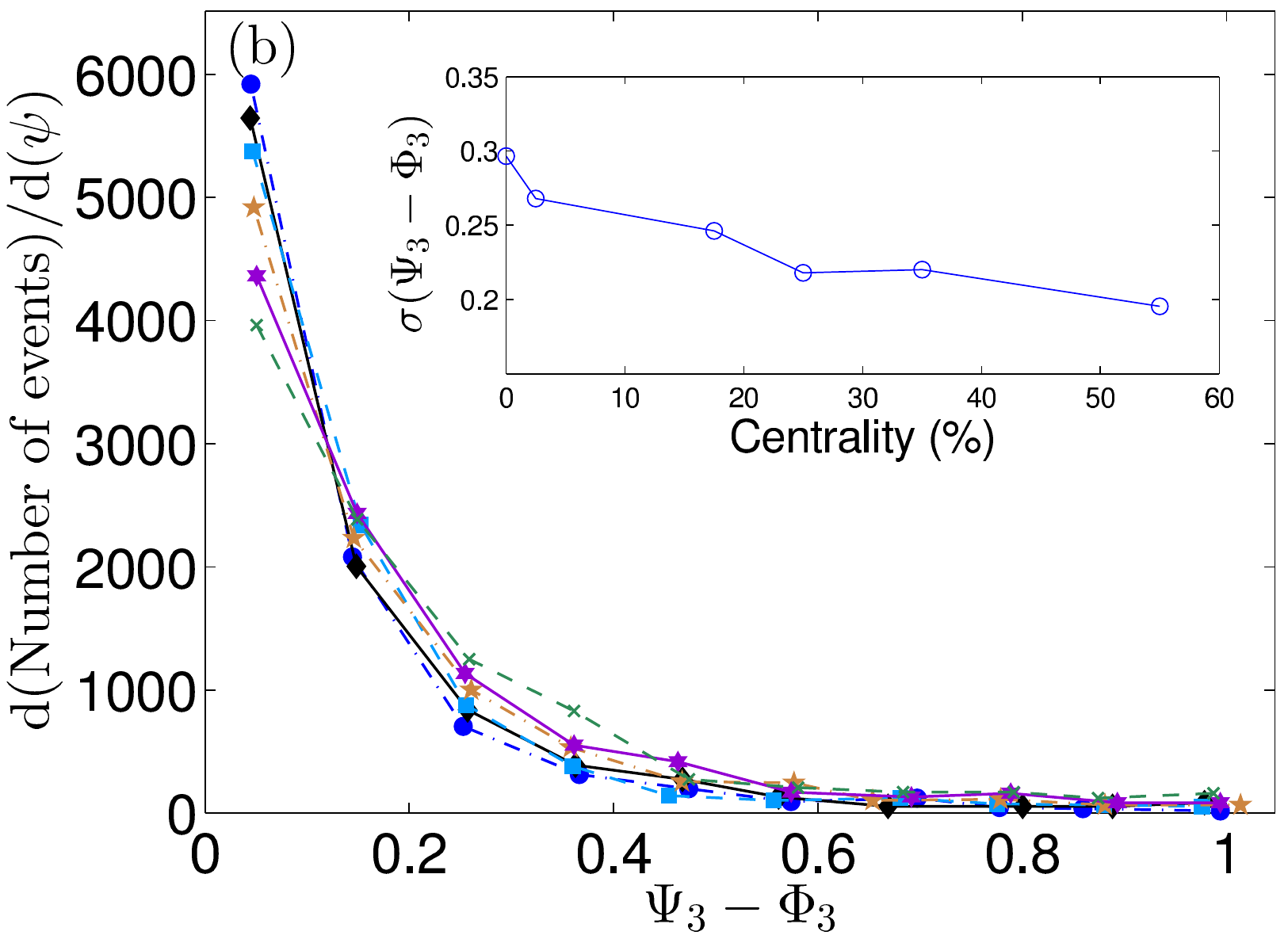}\\
 \includegraphics[width=0.45\linewidth]{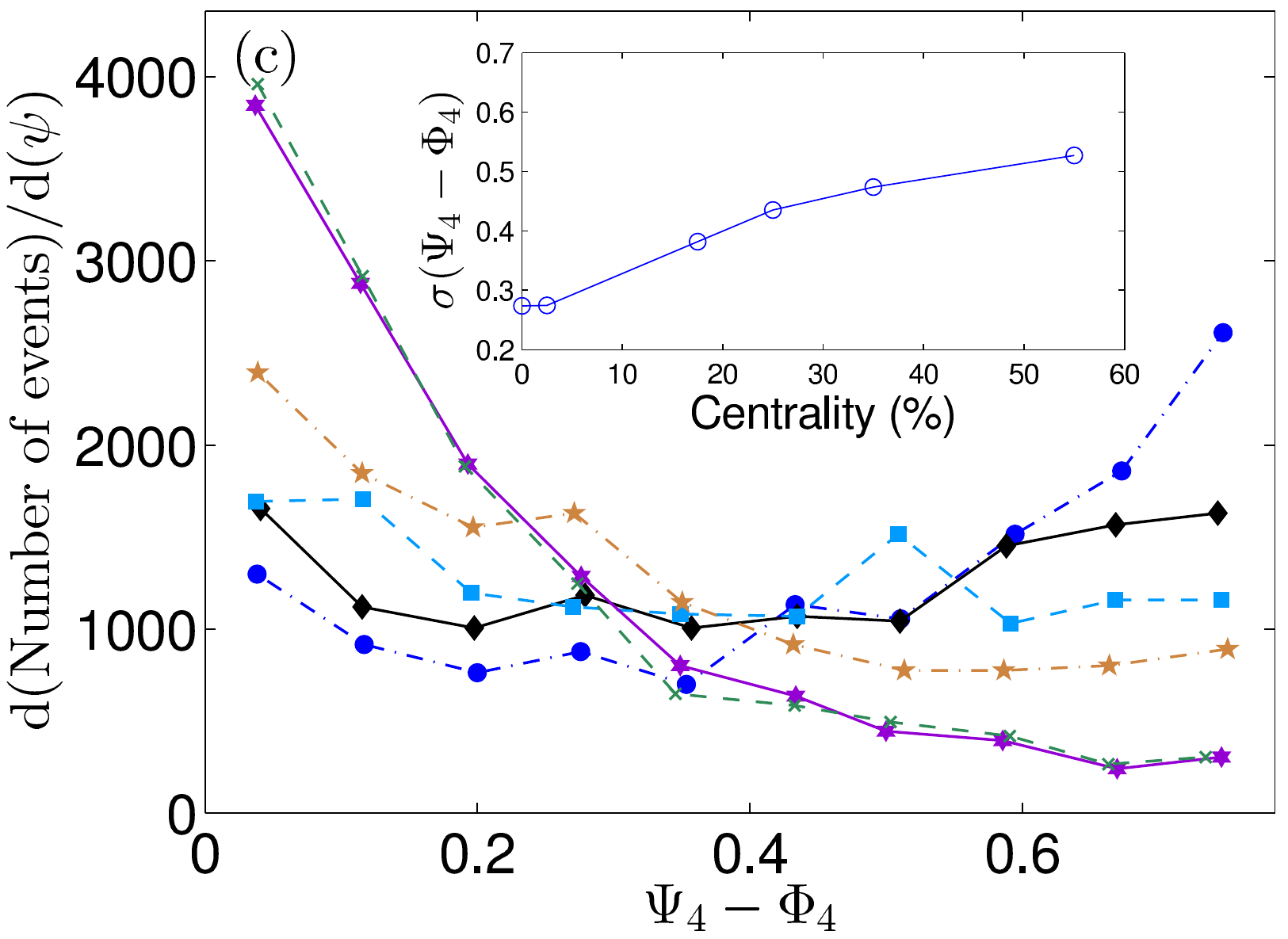}
 \includegraphics[width=0.45\linewidth]{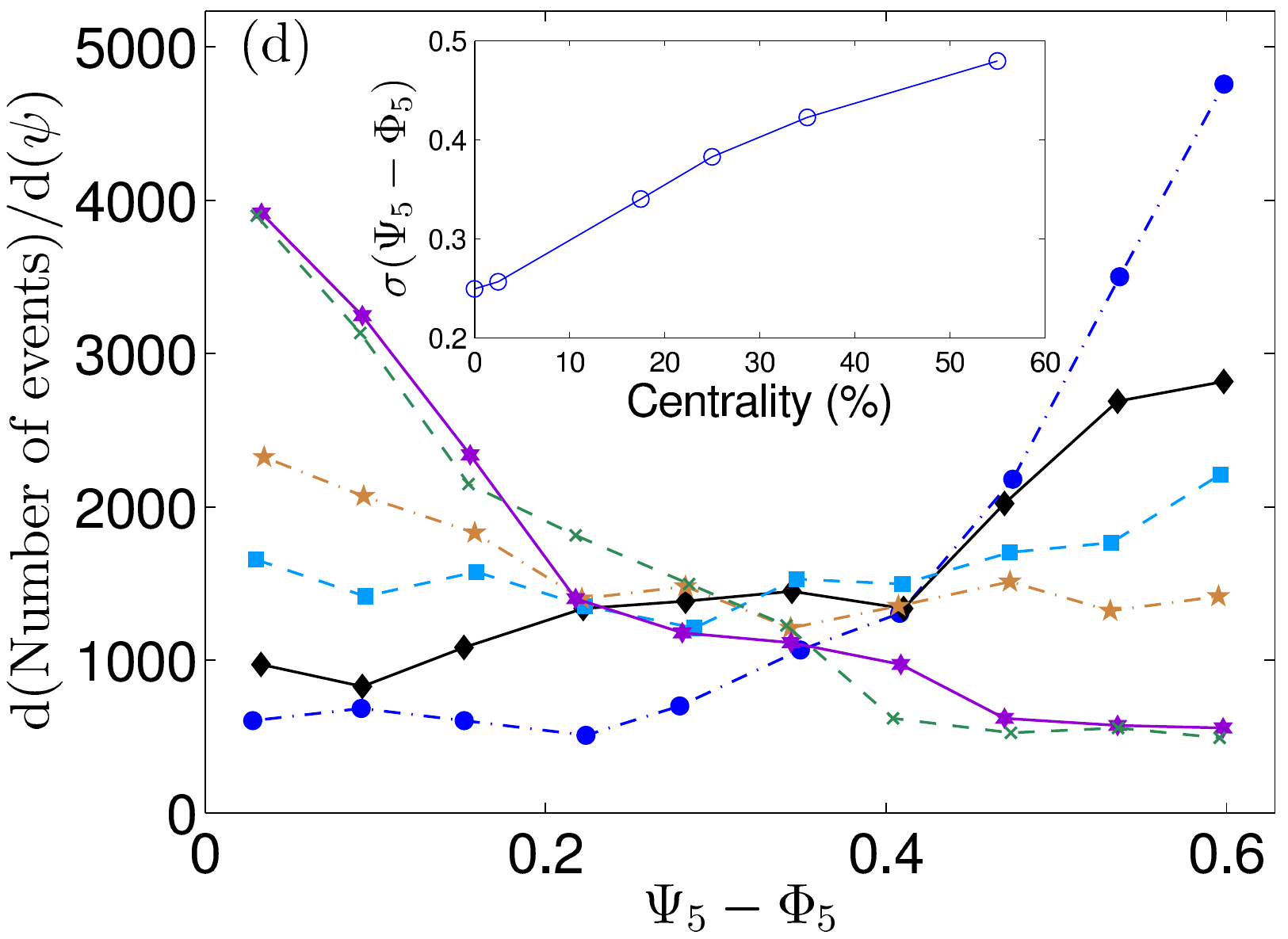}
 \caption{ Event-by-event correlation between the absolute 
   values of the event plane ($\PsiEP{n}$) and participant plane ($\PsiPP{n}$) angles for
   the harmonics of order $2-5$ (panels (a)-(d)), for events in different 
   centrality classes as indicated in the legend. (The same set of MC-KLN 
   events as in Fig.~\ref{1104.0650:F4} was used.) The insets show the centrality dependence of the widths of these correlations around zero. All fireballs are evolved with $\eta/s=0$ (ideal fluid hydrodynamics). Taken from \cite{Qiu:2011iv}.
   \label{1104.0650:F8}
 }
\end{figure*}
%

Figure~\ref{1104.0650:F8} looks at the correlation between the $n^\mathrm{th}$-order 
EP and PP angles at different collision centralities. This generalizes a 
similar analysis for $n \eq 2$ in Ref.~\cite{Holopainen:2010gz} to higher 
harmonics. Plotted are the distributions of the absolute value of the 
difference between the two angles in the main graph and the root mean square of this 
distribution (i.e. the width around zero of the correlation) in the inset, 
as a function of collision centrality. Panel (a) shows that the 
second-order participant and event planes are strongly correlated at all 
collision centralities. This demonstrates that elliptic flow is 
generated almost exclusively by the source ellipticity. The variance of 
the correlation is $\sim0.05$\,rad in the mid-central range (15-40\% 
centrality) and increases in very central and very peripheral collisions 
due to growing ellipticity fluctuations.

A similar correlation exists for the $3^\mathrm{rd}$-order participant
and event planes, at all collision centralities, but with a larger
variance of order $0.2-0.3$\,rad (depending on centrality). The relatively
strong correlation suggests that $\ecc_3$ is the dominant driver for $v_3$ 
\cite{Alver:2010gr}.

For the $4^\mathrm{th}$- and $5^\mathrm{th}$-order participant and event
planes the situation is complicated, as seen in panels (c) and (d). The
planes are correlated with each other (i.e. the distributions peak
at zero difference angle) in central collisions, become essentially 
uncorrelated in mid-central collisions and anti-correlated (i.e.
peaked at a difference angle of $\pi/n$, $n \eq 4,5$) in peripheral collisions.
The anti-correlation in peripheral collisions indicates strong mode-mixing,
driven by the large ellipticity $\ecc_2$ and strong elliptic flow $v_2$
at large impact parameters, which generates $v_4$ and $v_5$ contributions
by coupling to lower harmonics, as described in the previous subsection.
For $v_4$ in particular, a strong $\cos(2\phi)$ component in the collective
flow velocity generates a $v_4$ of the final momentum distribution, without
any need for nonzero $\ecc_4$. At large impact parameters, $\ecc_2$-induced
quadrupolar flow from the initial elliptic deformation of the overlap 
region thus dominates over any contribution from initial quadrangular 
deformation. However, there are additional mode-coupling effects arising from the nonlinear hydrodynamic evolution. As we will see in \chap{correlation}, the flow angles $\PsiEP{n}$ reflected in the measured momentum distribution closely correlate with the angles of the hydrodynamic flow anisotropies, so there is a nonlinear contribution to the 4th harmonic of the collective-flow velocity. In near-central collisions, on the other hand, where all 
$\ecc_n$ stem mostly from shape fluctuations, $v_{4,5}$ are dominantly 
driven by $\ecc_{4,5}$.

%
\begin{figure*}[h!]
\begin{center}
 \includegraphics[width=0.45\linewidth]{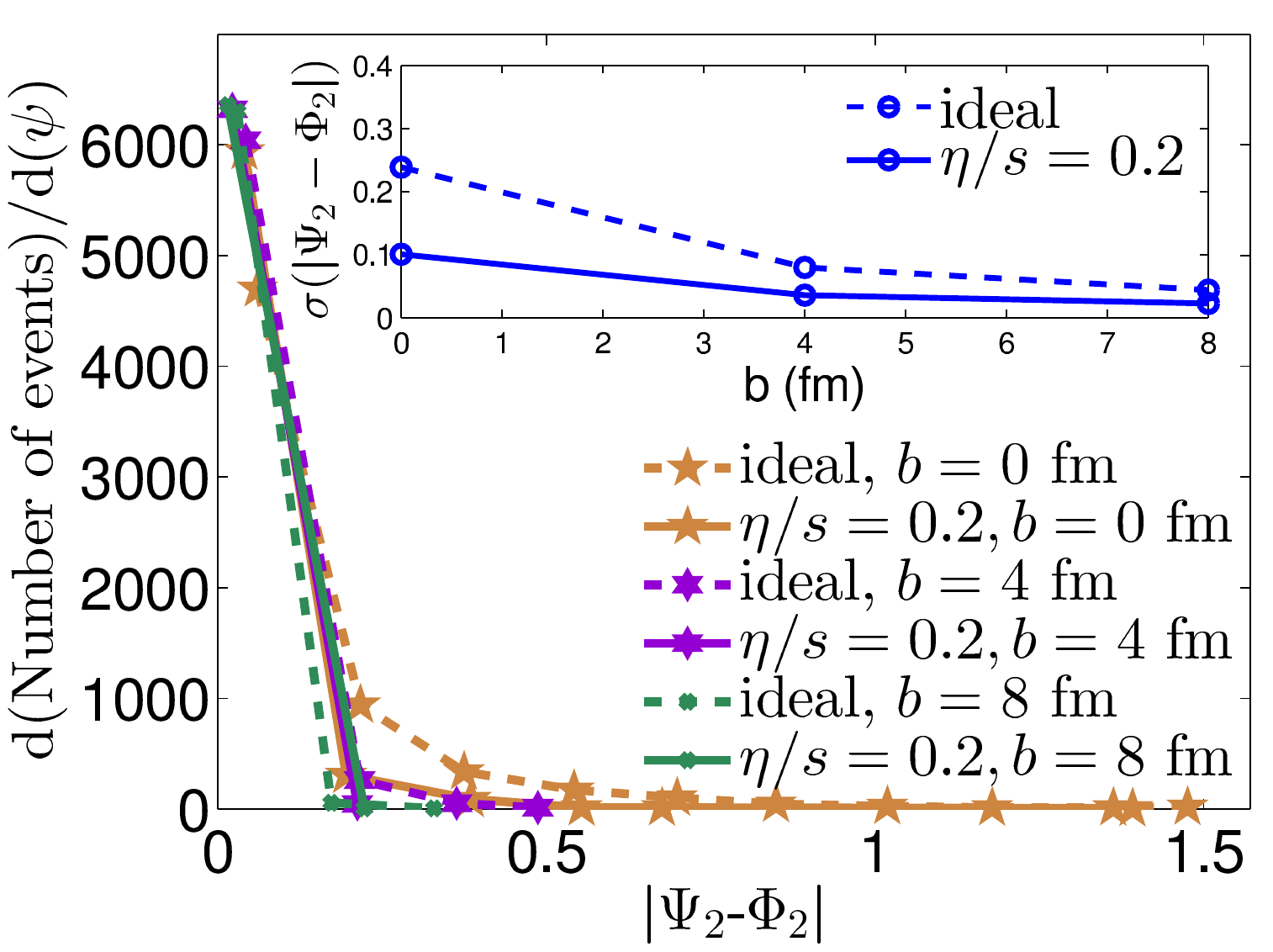}
 \includegraphics[width=0.45\linewidth]{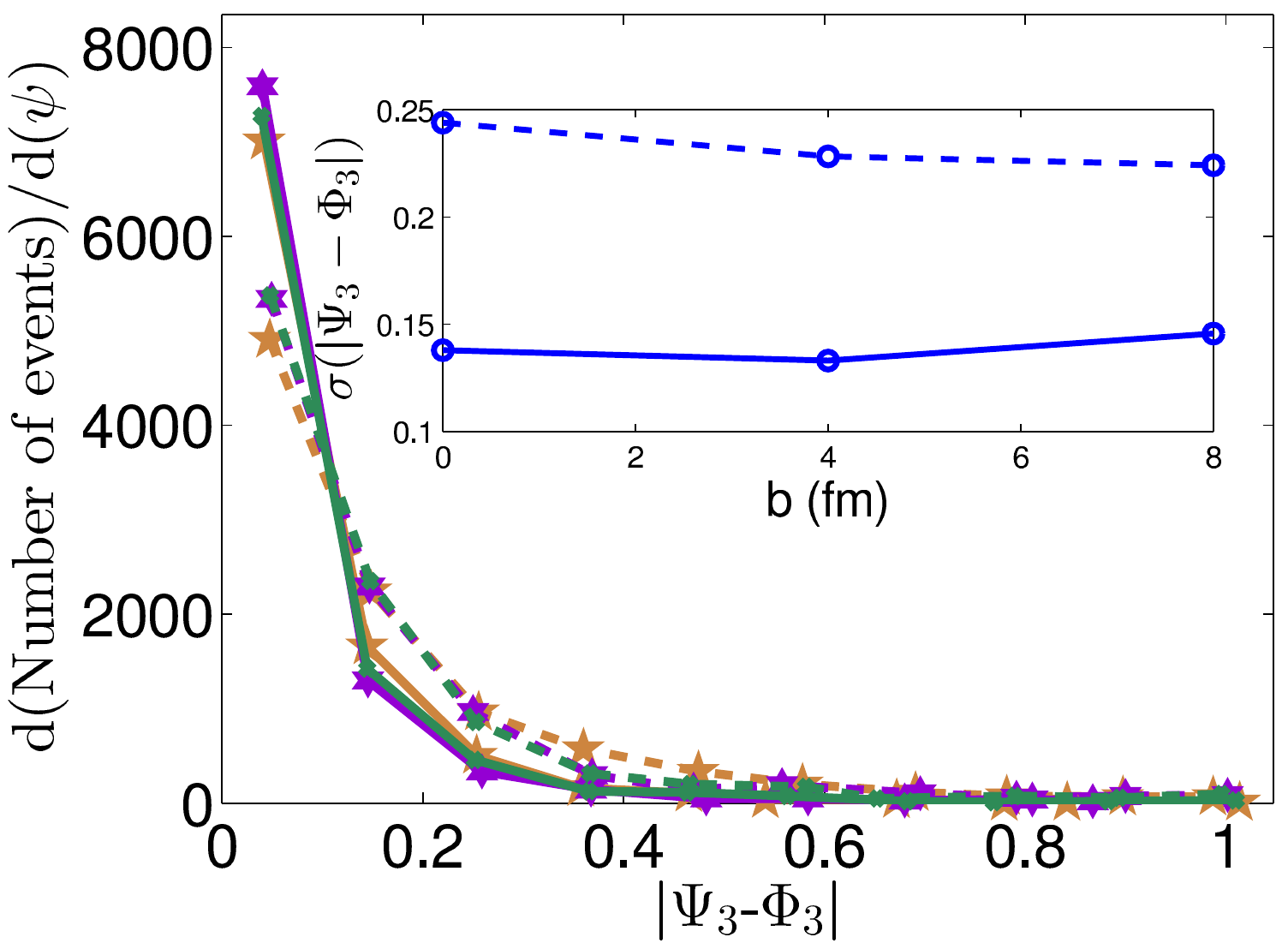}\\
 \includegraphics[width=0.45\linewidth]{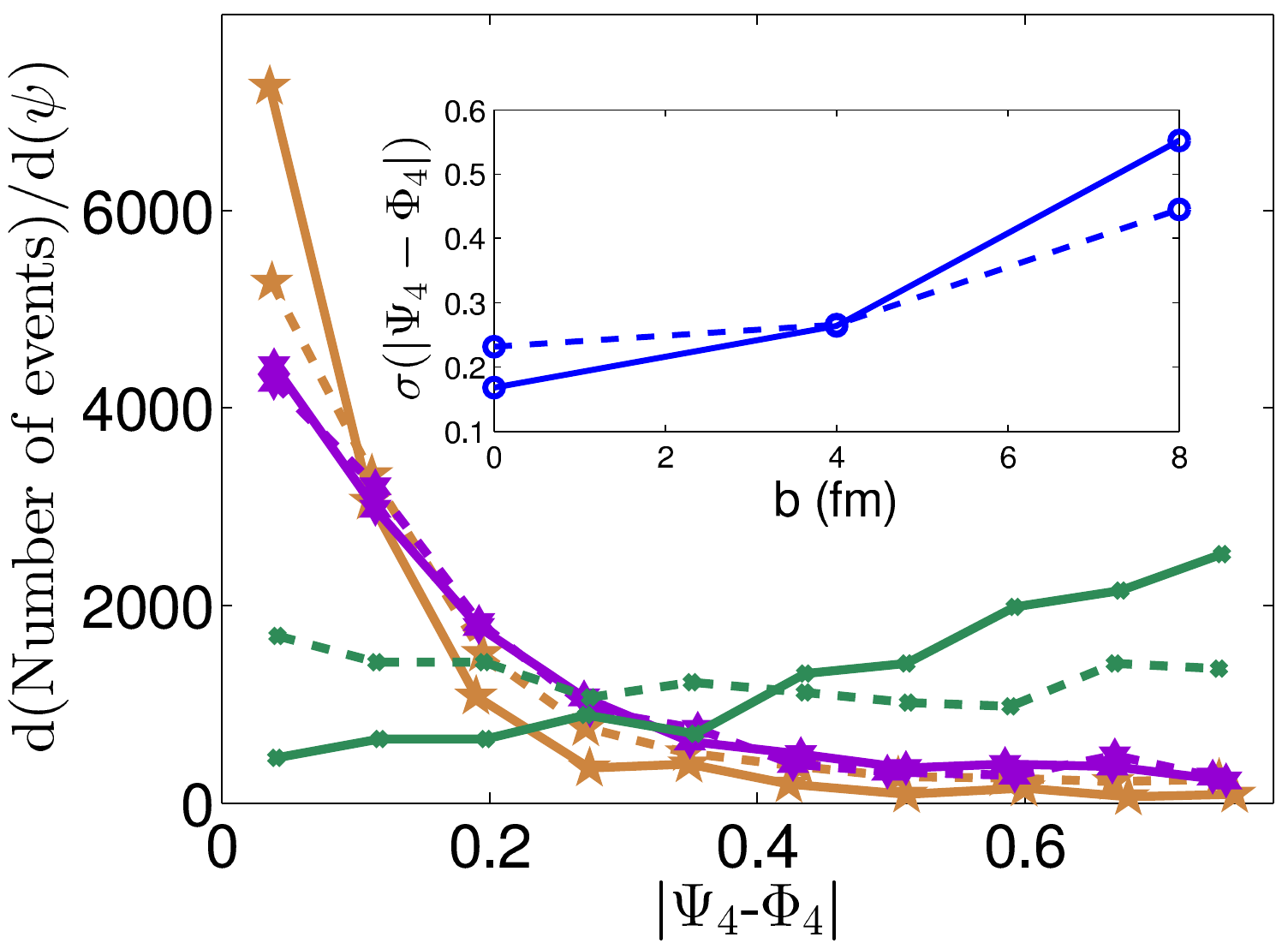}
 \includegraphics[width=0.45\linewidth]{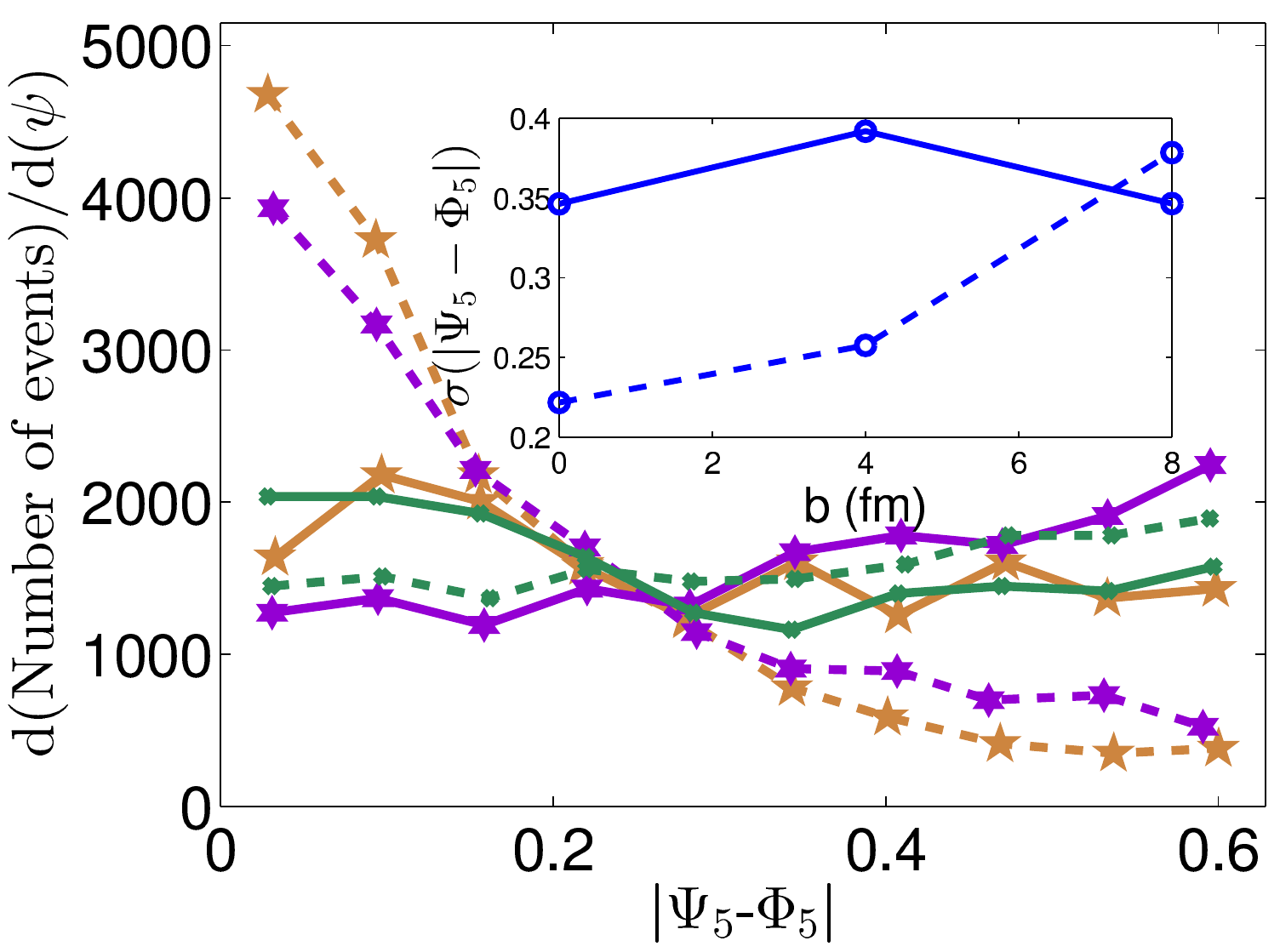}
 \caption{Event-by-event correlation between the absolute 
    values of the event plane ($\PsiEP{n}$) and participant plane ($\PsiPP{n}$) angles for
    the harmonics of order $2-5$. 1000 events are simulated at each impact parameter, using both ideal hydrodynamics and viscous ones with $\eta/s=0.2$, and grouped in eccentricity bins of sufficient width to have reasonable statistics in each bin.
 \label{1104.0650:F8.1}
 }
\end{center}
\end{figure*}
%

To study viscous effects we show Fig.~\ref{1104.0650:F8.1}, which is similar to Fig.~\ref{1104.0650:F8}, but using simulated data from both ideal and viscous ($\eta/s=0.2$) hydrodynamic simulations, starting from identical initial conditions. Each impact parameter group has 1000 simulated events.

It can be seen that nonzero viscosity manifests itself in two distinct ways: (1) For flows that are not strongly affected by mode couplings, for example $v_{2,3}$ at any $b$, and $v_4$ at $b=0$, the effect of shear viscosity is to reduce the fluctuations in $|\PsiEP{n}-\PsiPP{n}|$. Viscosity suppresses the randomness in the flow profile caused by the existence of hotspots in the initial density distribution, thus helping to build $v_n$ up from $\varepsilon_n$ and to align $\PsiEP{n}$ with $\PsiPP{n}$. (2) Viscosity also increases the strength of mode-coupling between different harmonics. In those cases where $v_n$ receives large contributions from $\varepsilon_m$, $m\neq n$ (e.g. $v_4$ at large $b$, $v_5$ in general), the fluctuations in the angle difference $|\PsiEP{n}-\PsiPP{n}|$ receive multiple contributions, with each contributor possibly contributing positively or negatively, resulting in hard-to-predict net effects. 

\vspace*{-3mm}
\subsection{Harmonic flows and their corresponding initial 
eccentricities: nonlinear hydrodynamic response}
\label{sec:1104.0650:4c}
\vspace*{-3mm}

\begin{figure*}[h!]
\begin{center}
 \includegraphics[width=0.45\linewidth]{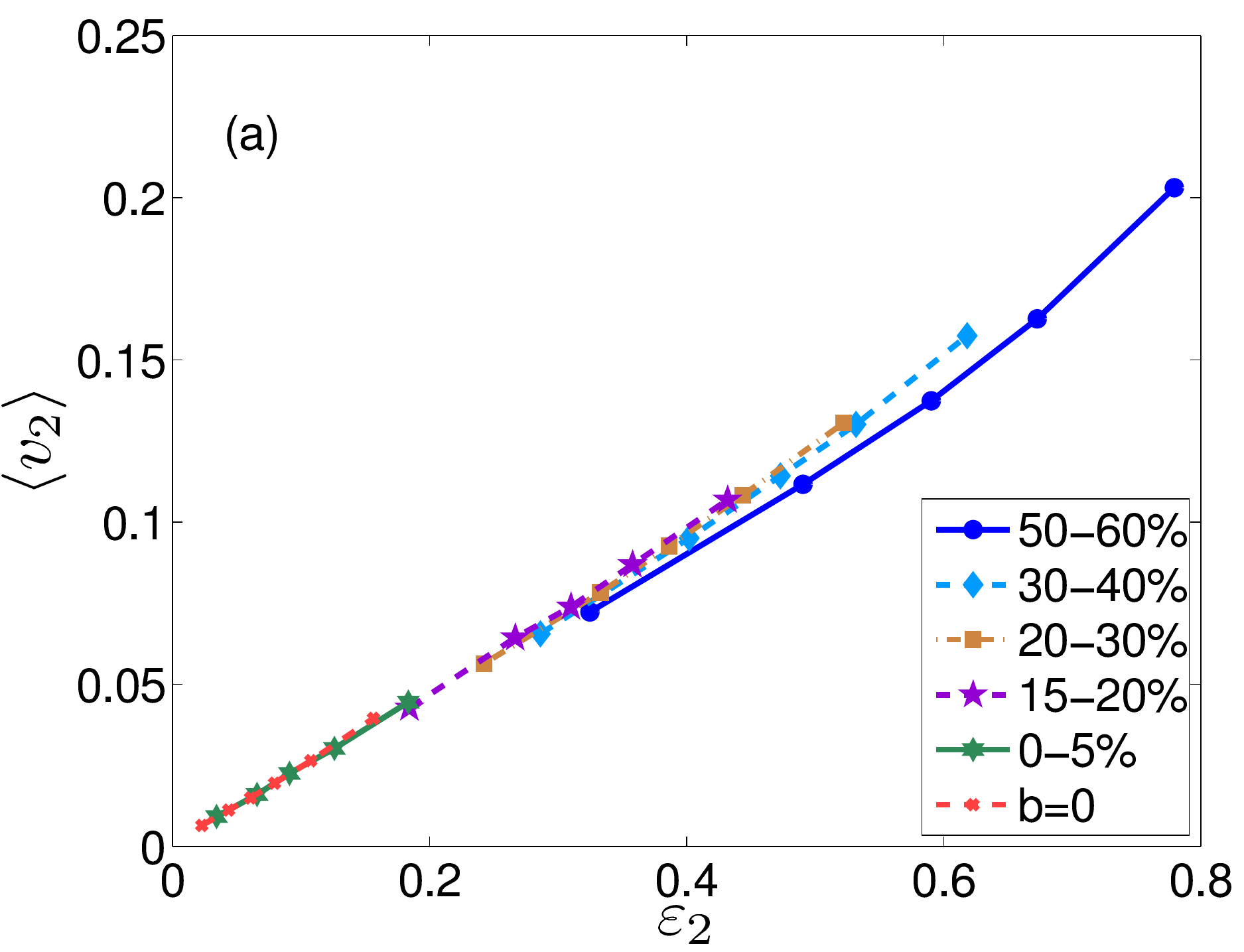}
 \includegraphics[width=0.445\linewidth]{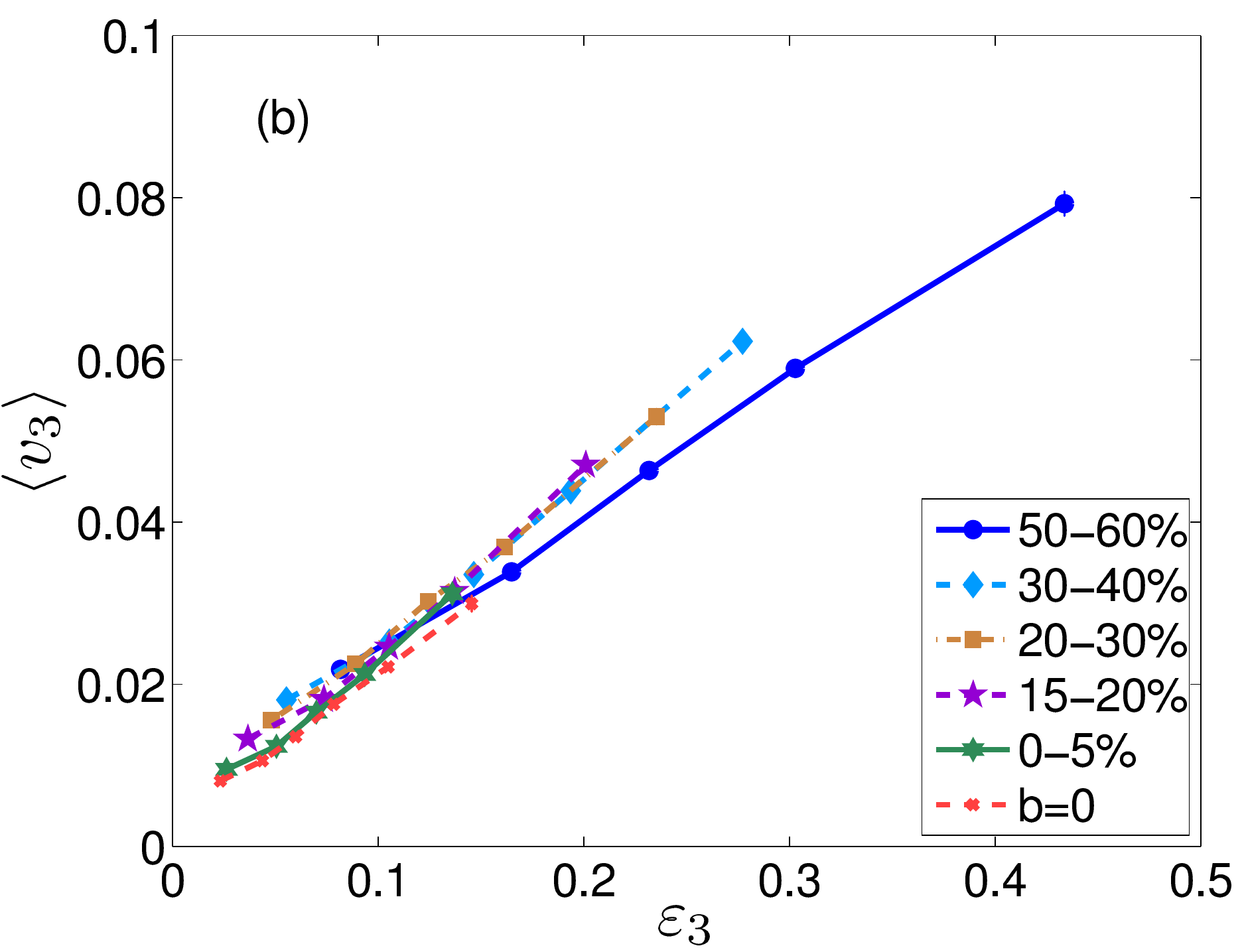}\\
 \includegraphics[width=0.45\linewidth]{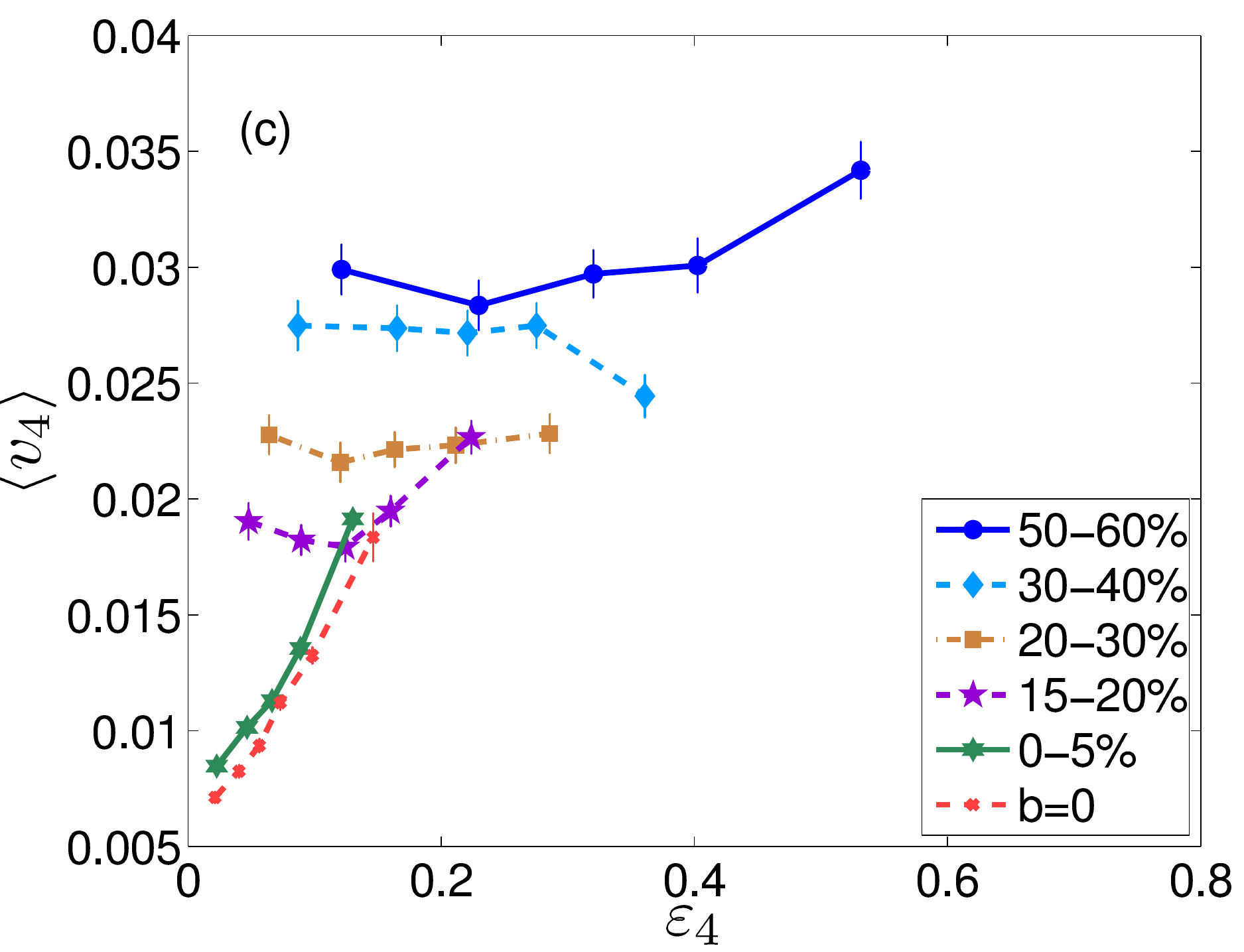}
 \includegraphics[width=0.45\linewidth]{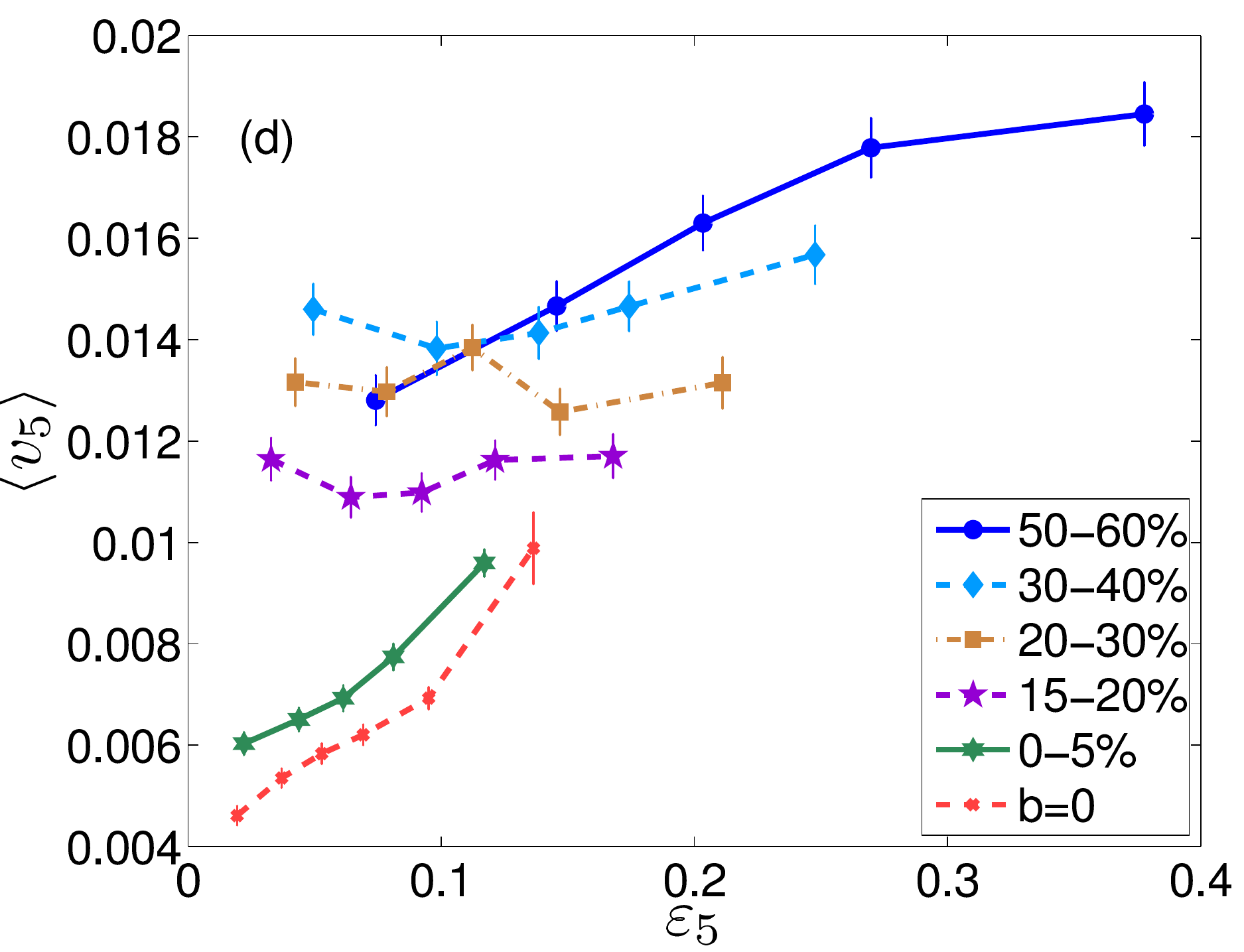}
 \caption{$\la v_n\ra(\ecc_n)$ for $n \eq 2,3,4,5$ 
   (panels (a)-(d)), using ideal fluid dynamics. As in Fig.~\ref{1104.0650:F4}, each centrality 
   class contains 1000 MC-KLN events, grouped in eccentricity bins 
   of sufficient width to have reasonable statistics in each bin. Taken from \cite{Qiu:2011iv}.
 \label{1104.0650:F9}
 }
\end{center}
\end{figure*}

It is often assumed that the harmonic
flows $v_n$ respond linearly to the eccentricities $\ecc_n$, at least as
long as the latter are small. This assumption receives support from
hydrodynamic simulations \cite{Alver:2010dn} as long as one probes 
deformed initial profiles with only a single non-vanishing harmonic 
eccentricity coefficient. In Fig.~\ref{1104.0650:F9}, we investigate the validity of 
this assumption with fluctuating MC-KLN events which feature nonzero 
$\ecc_n$ values for all $n$. 

%
%
  
Figure~\ref{1104.0650:F9}a generally provides support for the assumption of a
linear dependence of the elliptic flow $v_2$ on initial ellipticity 
$\ecc_2$, with two important caveats: 
\begin{itemize}
\item[(i)]
At small and large ellipticities,
$v_2$ deviates upward from a best-fit line through the origin, indicating
additional contributors to the elliptic flow. Indeed, for zero ellipticity
$\ecc_2 \eq 0$ we find a nonzero average $\la v_2\ra$. These are events 
with typically large nonzero values for eccentricities of higher harmonic
order, which generate elliptic flow through mode-mixing (e.g. between $\ecc_3$
and $\ecc_5$). We see that this happens at all centralities, even for
$b \eq 0$, due to event-by-event fluctuations of the eccentricity coefficients.
\item[(ii)]
The slope of the curve $\la v_2\ra(\ecc_2)$ decreases in very peripheral
collisions, indicating destructive interference via mode-mixing from other
harmonics in the hydrodynamic evolution of the small and highly fluctuating
fireballs created at large impact parameters.
\end{itemize} 

The $\ecc_3$-dependence of triangular flow $\la v_3\ra$, shown in 
Fig.~\ref{1104.0650:F9}b, shows a qualitatively similar story, but the deviations
from linear response are stronger, with significant nonzero triangular
flow in events with zero initial triangularity, especially for larger
impact parameters. 

For $\la v_4\ra$ and $\la v_5\ra$, shown in Figs.~\ref{1104.0650:F9}c and \ref{1104.0650:F9}d, 
mode-mixing effects are very strong, and a linear response of $v_n$ to
$\ecc_n$ ($n=4,5$) can no longer be claimed. This is quite different from
the results in \cite{Alver:2010dn}, where $v_4$ was studied for a source
that had only $\ecc_4$ deformation: in this case $v_4(\ecc_4)$ was found to
be approximately linear for small $\ecc_4$, with a downward bend at larger
$\ecc_4$ values due to negative interference from cross-currents for
sources with large quadrangularities. (This approximately linear dependence
survived in the $p_T$-integrated $v_4$, even though it was noticed in a 
related study \cite{Luzum:2010ae} that, for mid-central collisions, the 
differential quadrangular flow $v_4(p_T)$ at high $p_T$ appears to be 
mostly determined by the elliptic deformation of the hydrodynamic flow 
profile generated by $\ecc_2$.) Our study 
shows that it is unlikely that the anisotropic flow resulting from highly 
inhomogeneous initial profiles with nonzero eccentricity coefficients of 
all harmonic orders can be obtained by some sort of linear superposition 
of flows generated from sources with only a single nonzero harmonic 
eccentricity coefficient, as suggested in \cite{Teaney:2010vd}. (We will discuss  this topic in more detail in \chap{correlation}.) The 
hydrodynamic response $\{v_n\}$ to a set of initial eccentricity 
coefficients $\{\ecc_n\}$ is not only nondiagonal, but also (via 
mode-mixing) nonlinear, and there is no suitable single-shot substitute 
for event-by-event hydrodynamic evolution of fluctuating initial 
conditions.

%
\begin{figure*}[h!]
\begin{center}
 \includegraphics[width=0.45\linewidth]{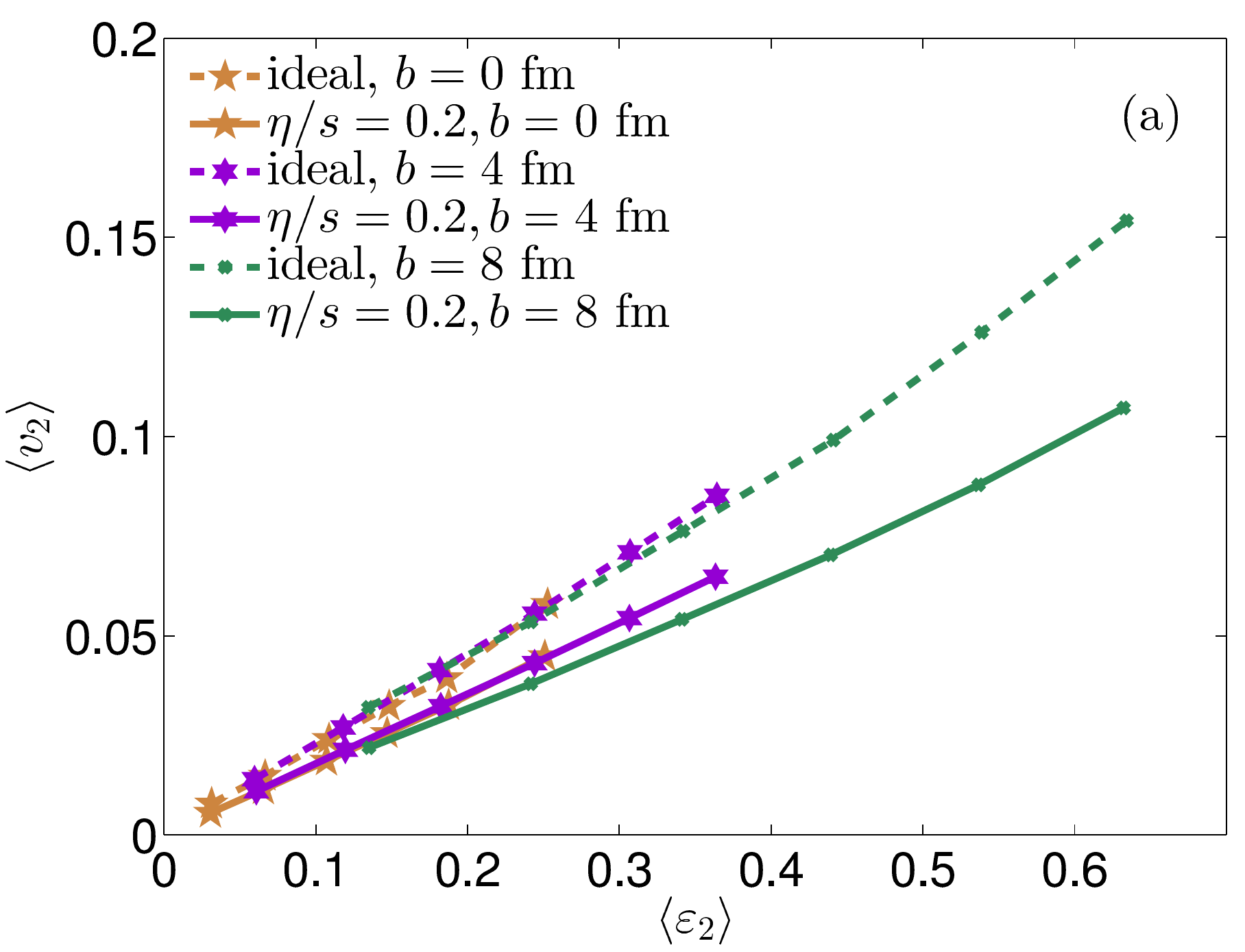}
 \includegraphics[width=0.45\linewidth]{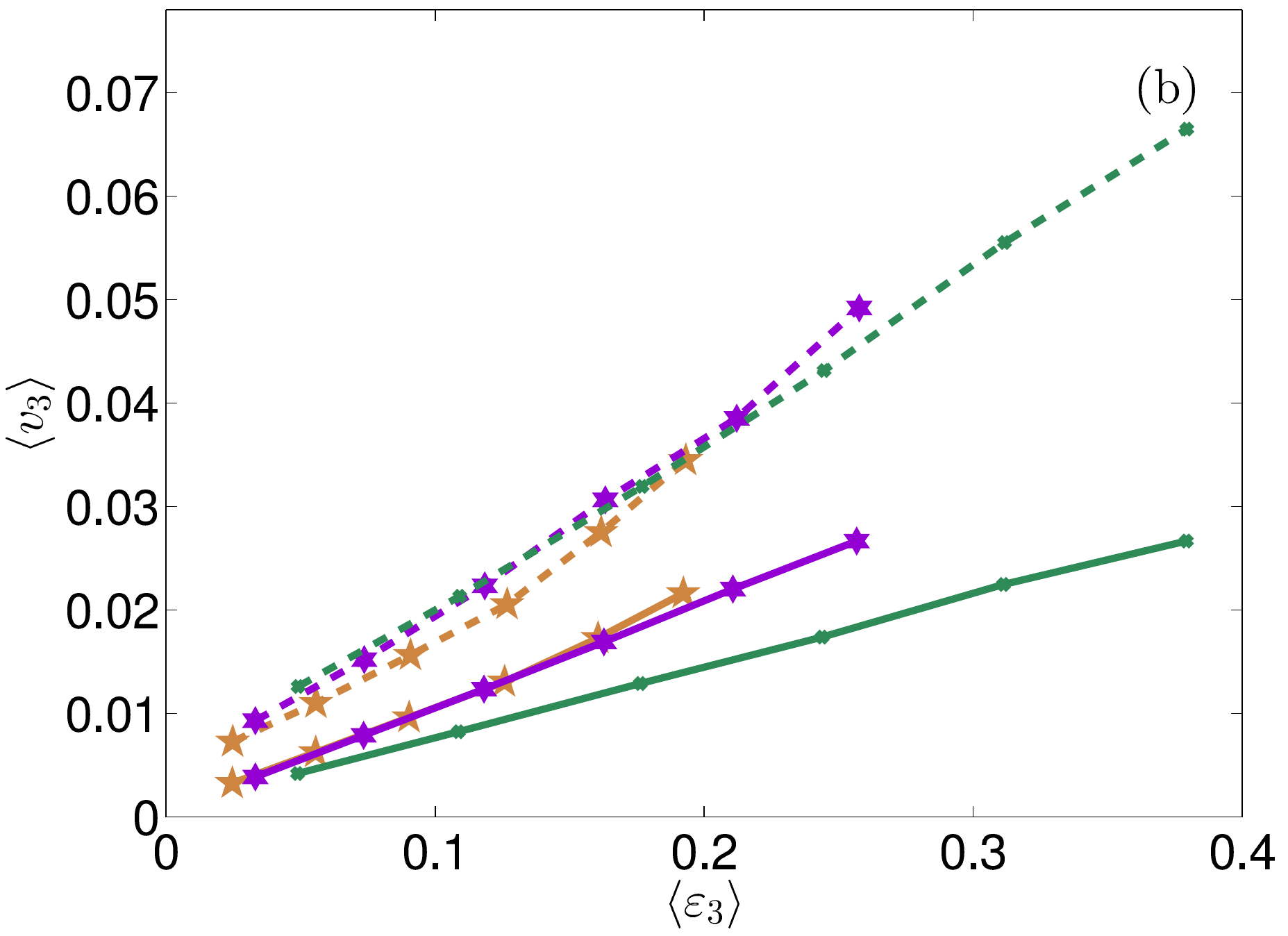}\\
 \includegraphics[width=0.45\linewidth]{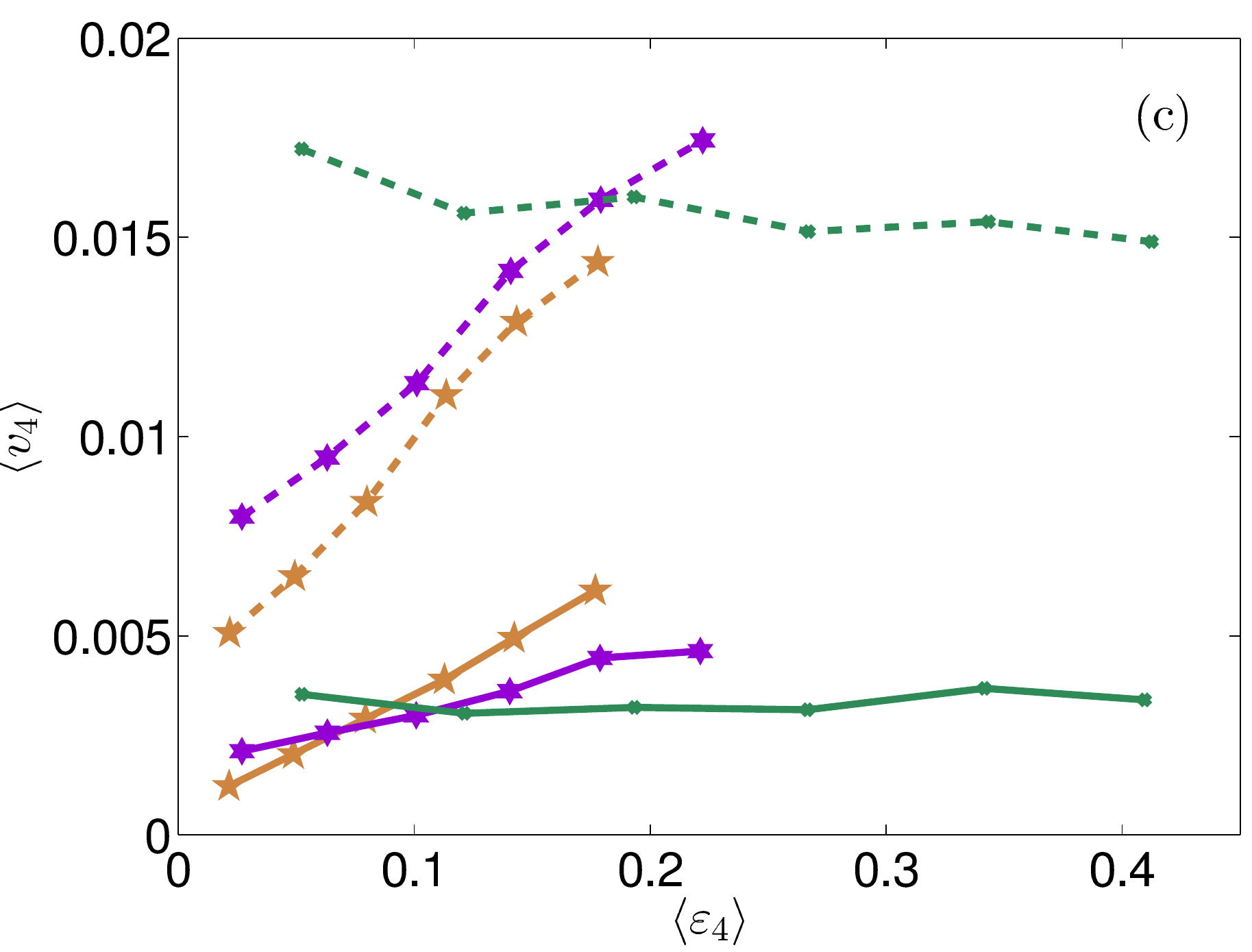}
 \includegraphics[width=0.45\linewidth]{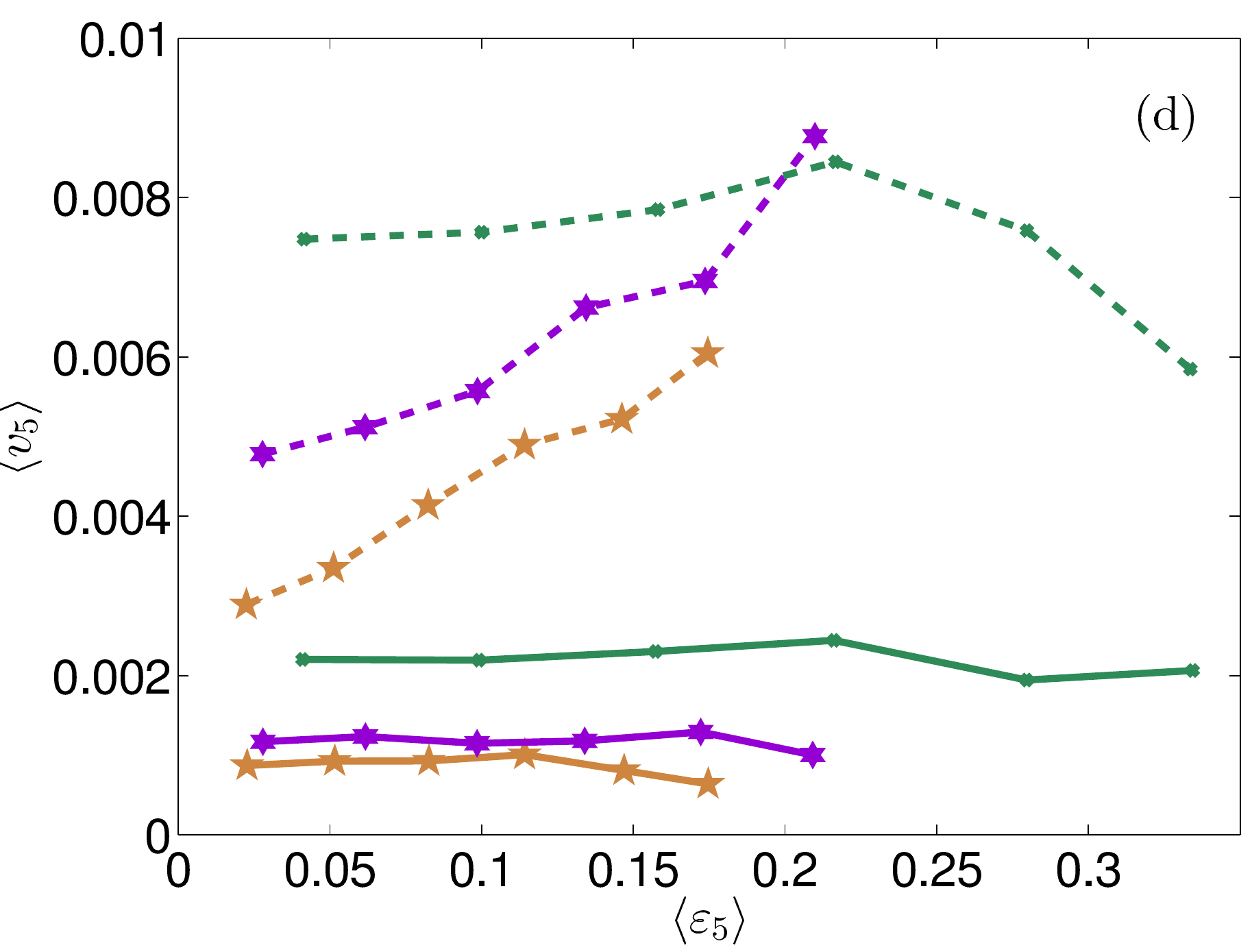}
 \caption{$\la v_n\ra(\la \ecc_n \ra)$ for $n \eq 2,3,4,5$ 
   (panels (a)-(d)). 1000 initial conditions are used in simulations for each impact parameter, using both ideal hydrodynamics and viscous runs with $\eta/s=0.2$. Results are grouped in eccentricity bins of sufficient width to have reasonable statistics in each bin.
 \label{1104.0650:F9.1}
 }
\end{center}
\end{figure*}
%

The effect of viscosity can be studied using Fig.~\ref{1104.0650:F9.1}, which contains curves corresponding to both ideal and viscous ($\eta/s=0.2$) hydrodynamic evolutions. In contrast to Fig.~\ref{1104.0650:F9}, the events are grouped directly by impact parameters. Again 1000 events are simulated for each impact parameter.

First it is clearly seen that the existence of viscosity reduces the response of $v_n$ from $\varepsilon_n$, and this effect increases with increasing $n$. For elliptic and triangular flow, viscosity suppresses the conversion coefficient $v_{2,3}/\varepsilon_{2,3}$ more strongly for larger impact parameters, indicating larger viscous effects in more peripheral collisions. For $v_{4,5}$ viscosity wipes out the monotonicity of its dependence on $\varepsilon_{4,5}$ in central and semi-central collisions, meaning that it receives larger contributions from other sources than $\varepsilon_{4,5}$ --- this is another way to state that viscosity increases nonlinear mode-coupling effects.

We note that in both ideal and viscous cases, nonlinear mode-mixing effects appear to be 
minimal for the elliptic and triangular flow (Figs.~\ref{1104.0650:F9}a,b, Figs.~\ref{1104.0650:F9.1}a,b). $v_2$
and $v_3$ remain therefore the best candidates for an extraction of the 
fluid's viscosity, by studying (with quantitative precision) the fluid's 
efficiency in converting initial spatial deformations into final momentum 
anisotropies and anisotropic flows. We will further elaborate on this 
theme in the next section.

\section{Single-shot versus event-by-event hydrodynamics}
\label{sec:1104.0650:5}

We now discuss the effects of event-by-event initial-state fluctuations 
on the finally observed pion and proton $p_T$-spectra and anisotropic flow, 
comparing traditional single-shot hydrodynamic evolution of an 
appropriately constructed smooth average initial profile with
event-by-event evolution of fluctuating initial conditions (with an 
ensemble average taken at the end). Since the calculation of resonance 
decay feeddown corrections is computationally expensive but not expected 
to cause qualitative changes, we here concentrate on directly emitted 
(``thermal'') pions and protons. For the graphs shown in this section, 
we generated for each impact parameter 1000 fluctuating events from the MC-KLN model and
propagated them either event-by-event or via a single-shot hydrodynamic simulation run down to a decoupling temperature of 140\,MeV.

\subsection{Transverse momentum spectra}
\label{sec:1104.0650:5a}

%
\begin{figure*}[h!]
\begin{center}
 \includegraphics[width=0.45\linewidth]{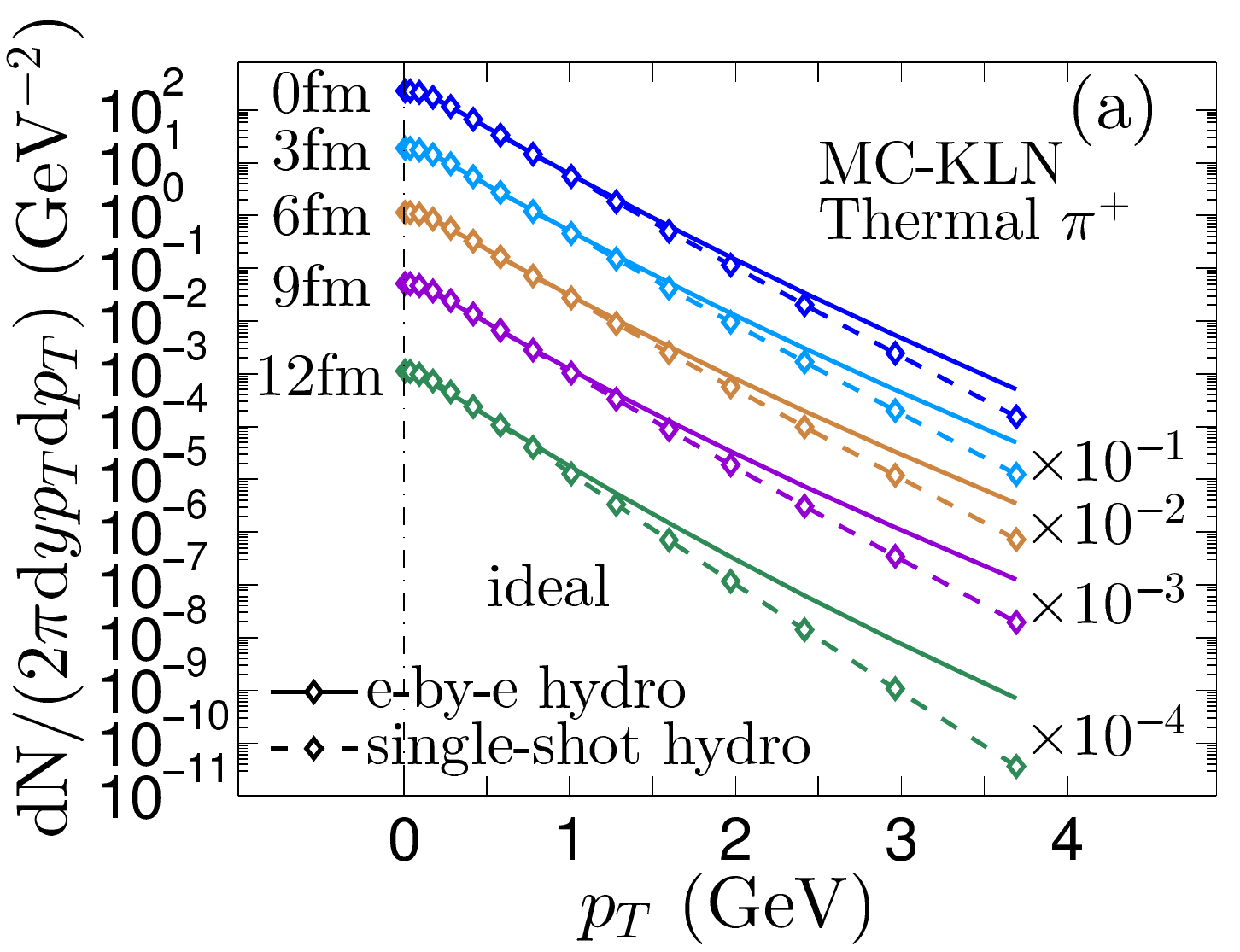}
 \includegraphics[width=0.45\linewidth]{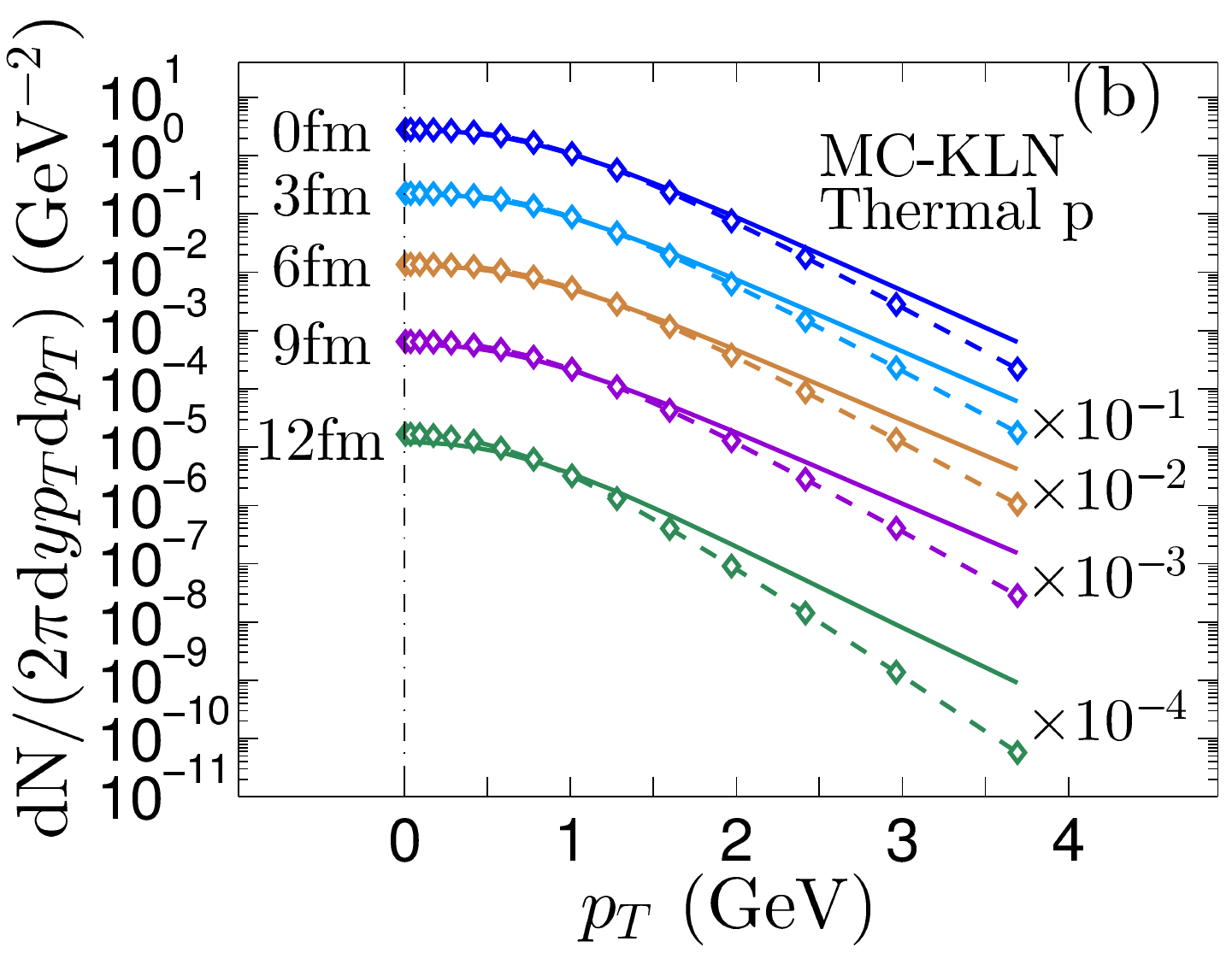}
 \caption{Transverse momentum spectra for directly
 emitted $\pi^+$ (left) and protons (right) from event-by-event (solid lines)
 and single-shot (dashed lines) ideal hydrodynamic simulations with MC-KLN initial conditions, for $200\,A$\,GeV Au+Au 
 collisions at six selected impact parameters. 
 \label{1104.0650:F10}
 }
\end{center}
\end{figure*}
%

In \cite{Chatterjee:2011dw}, Chatterjee 
{\it et al.\ } showed that thermal photon spectra from exploding heavy-ion
collision fireballs with fluctuating initial conditions which were 
hydrodynamically evolved event-by-event are significantly harder than
those obtained from single-shot hydrodynamic evolution of the corresponding 
ensemble-averaged smoother initial profiles. The 
authors of \cite{Chatterjee:2011dw} attributed this effect to the existence 
of ``hot spots'' in the fluctuating initial conditions that radiate photons 
at a higher-than-average temperature. Figure~\ref{1104.0650:F10} shows that the same
hardening effect occurs in the pion and proton spectra even though these strongly
interacting hadrons are emitted only at freeze-out, with the same decoupling 
temperature assumed in both types of evolution.\footnote{A similar 
  effect was also seen in \cite{Holopainen:2010gz}, whose authors further 
  pointed out that the strength of this ``hardening effect'' depends on 
  the fluctuation size parameter in the initial conditions (i.e. the area 
  over which the entropy produced in a nucleon-nucleon collision is 
  distributed).}
This proves that the effect 
is due to stronger radial flow in the event-by-event evolved fluctuating 
fireballs, driven by the stronger than average pressure gradients associated 
with the ``hot spots'' (i.e. over-dense regions) in the initial profile. The
importance of initial-state fluctuation effects on the final $p_T$-spectra
becomes stronger in peripheral collisions, where the initial fireballs are 
smaller and ``hot spots'' have a relatively larger influence. If stronger
radial flow is the explanation of the fluctuation-driven hardening of the
pion and proton spectra observed in Fig.~\ref{1104.0650:F10}, it is probably also an important 
contributor to the hardening of the photon spectra noted in 
Ref.~\cite{Chatterjee:2011dw}, at least for low $p_T$ (i.e. in
the hydrodynamic regime).

%
\begin{figure*}[t]
\begin{center}
 \includegraphics[width=0.45\linewidth]{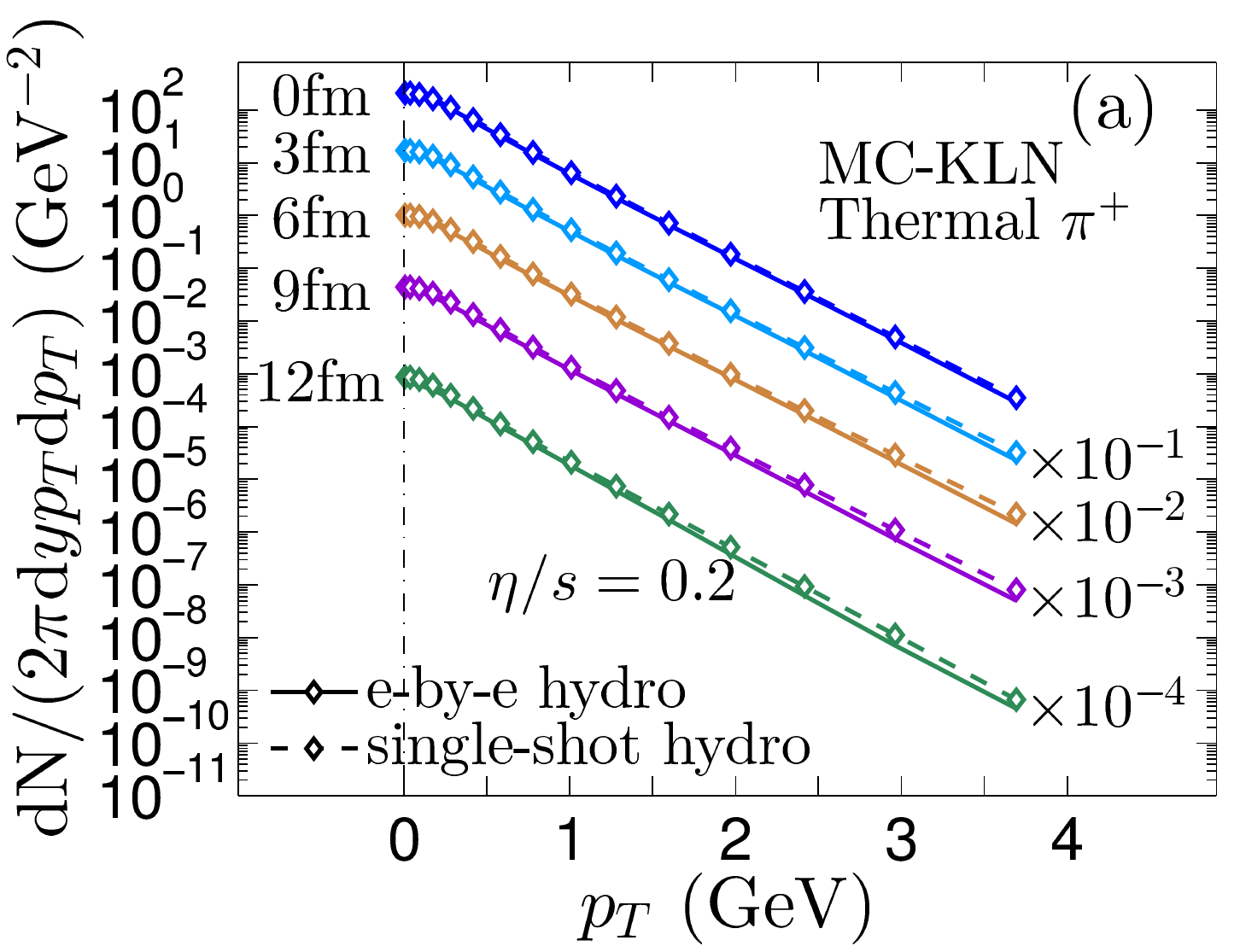}
 \includegraphics[width=0.45\linewidth]{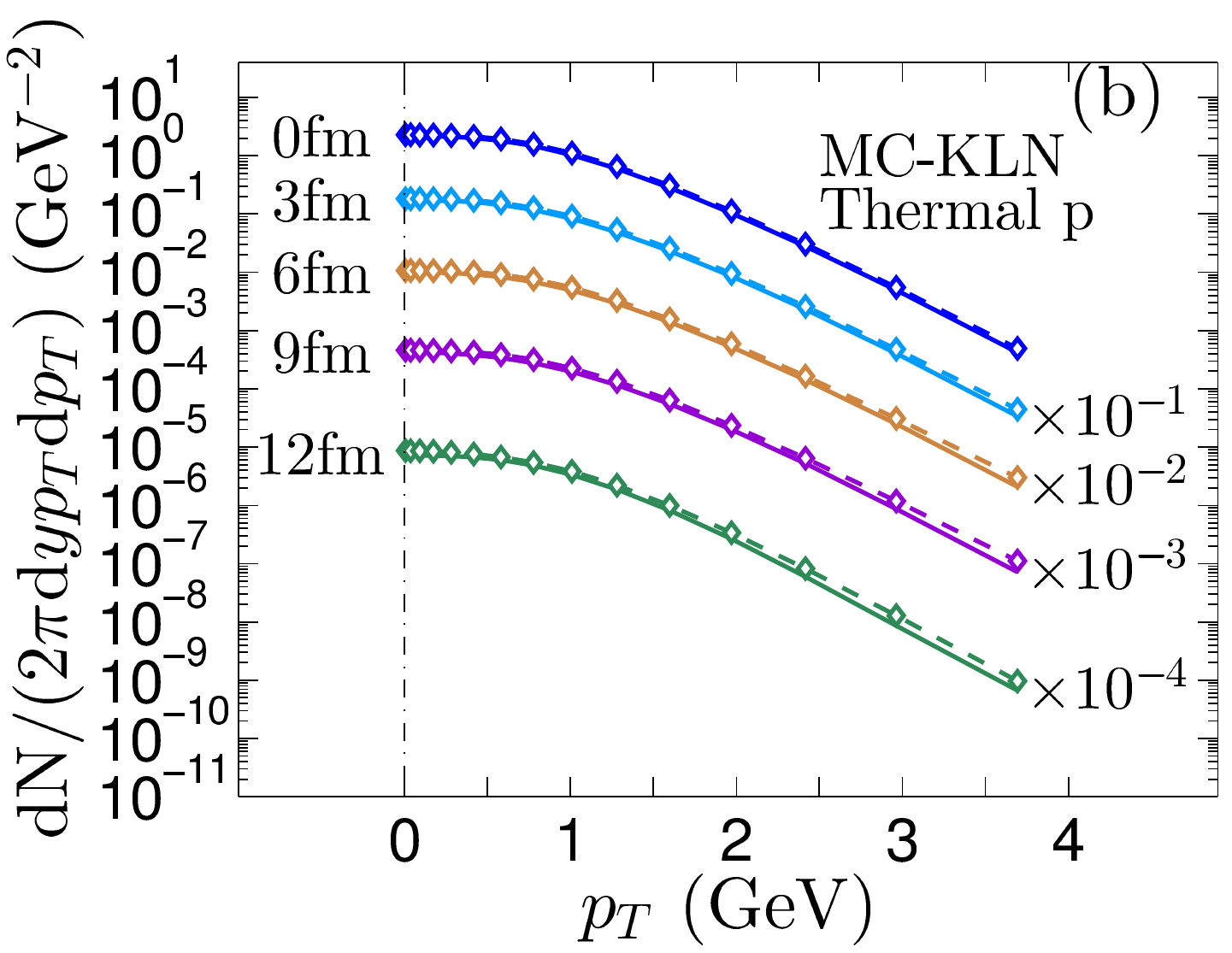} 
 \caption{Same as Fig.~\ref{1104.0650:F10}, but for viscous hydrodynamics with $\eta/s=0.2$. 
 \label{1104.0650:F10.1}
 }
\end{center}
\end{figure*}
%

Nonzero viscosity softens the hardening effect, as shown in Fig.~\ref{1104.0650:F10.1}. It is clearly seen that the difference in $p_T$-spectra between the two types of simulations is reduced in the viscous case. We attribute this reduction to the fact that the nonzero viscosity tends to quickly wash out density inhomogeneities from hot spots by dissipation, thereby reducing the difference between fluctuating and smooth initial conditions in the later stages of the expansion.

\subsection{Elliptic and triangular flow}
\label{sec:1104.0650:5b}

%
\begin{figure*}[h!]
 \includegraphics[width=0.45\linewidth]{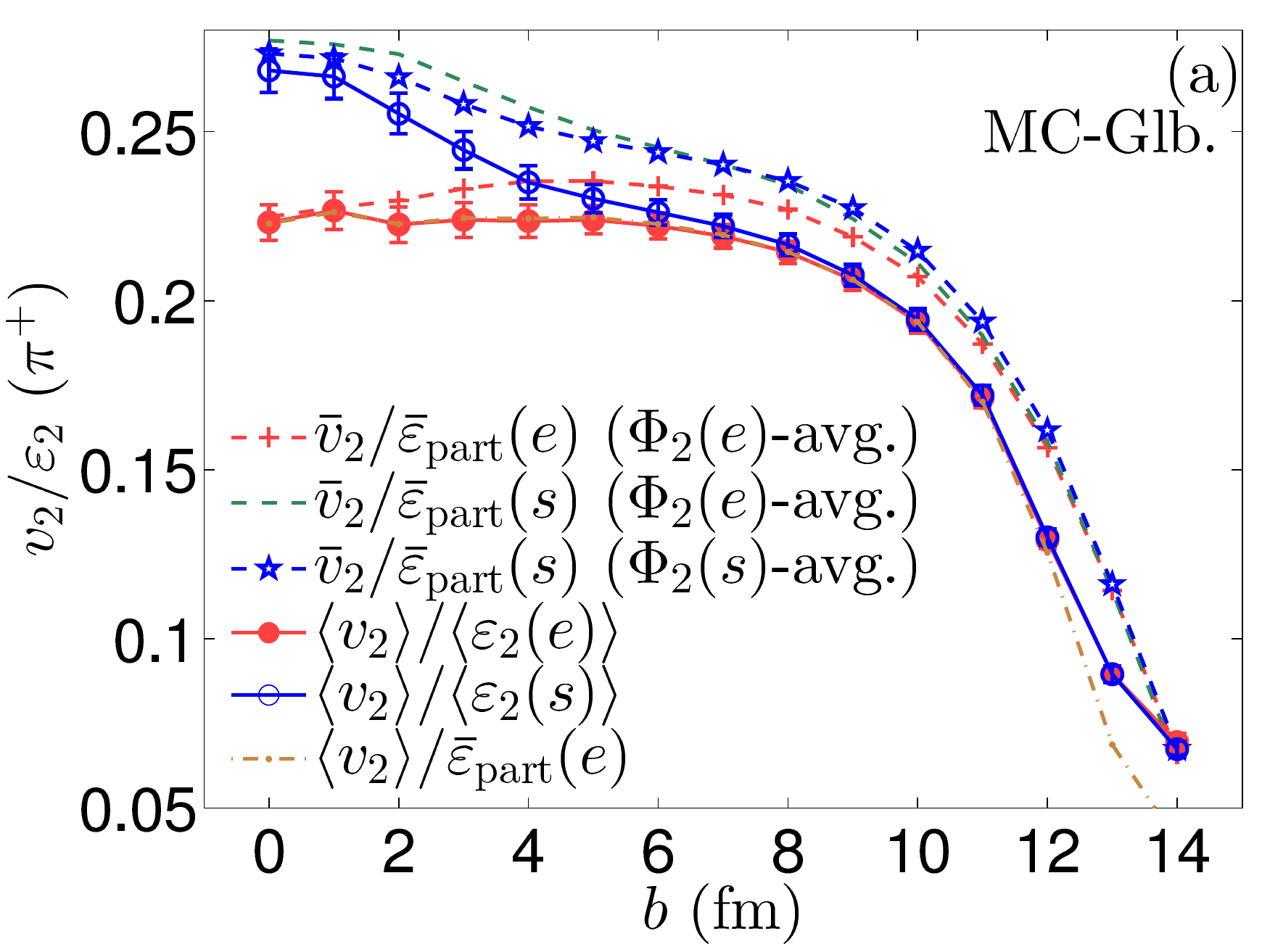}
 \includegraphics[width=0.45\linewidth]{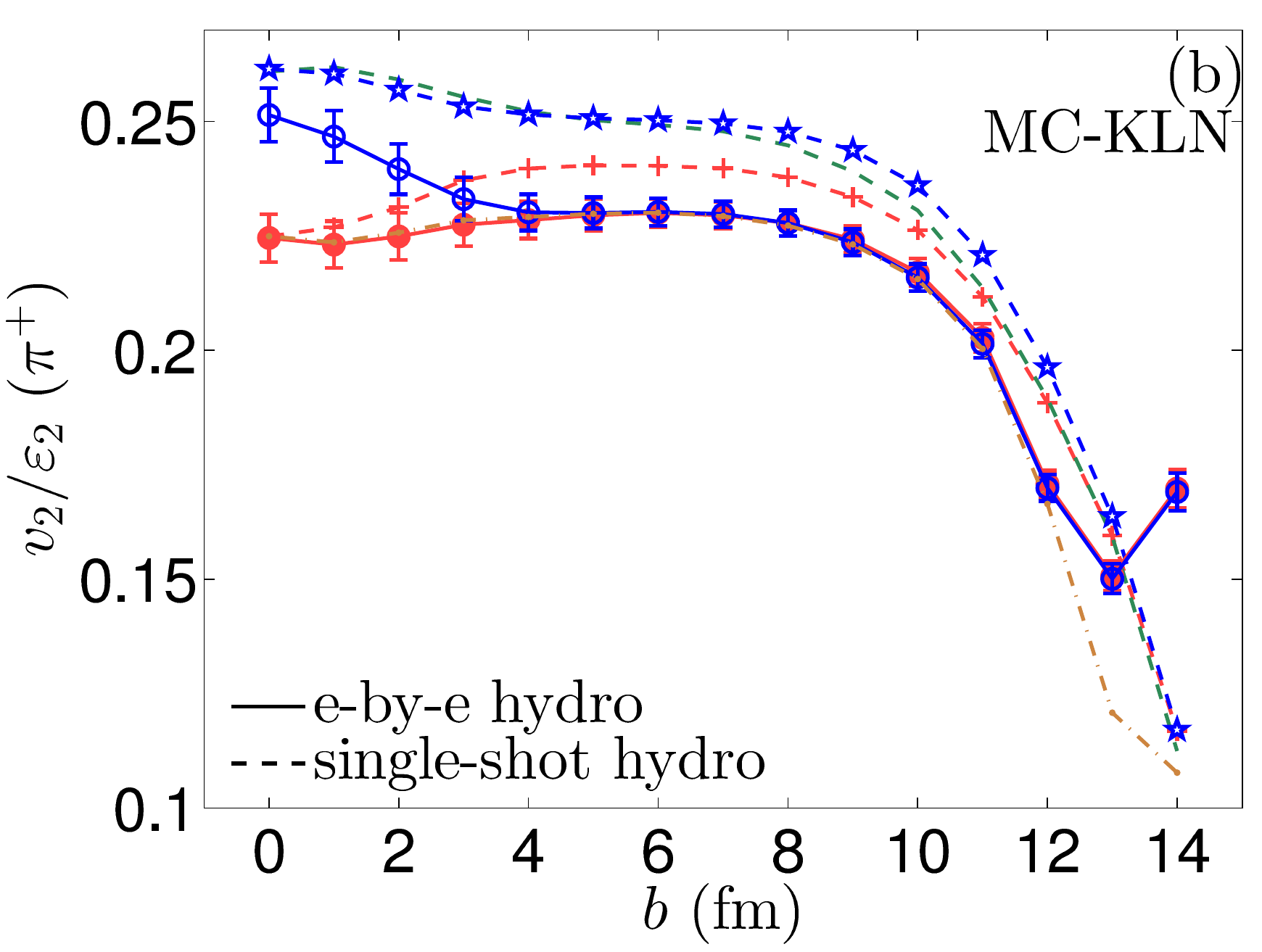}\\
 \includegraphics[width=0.45\linewidth]{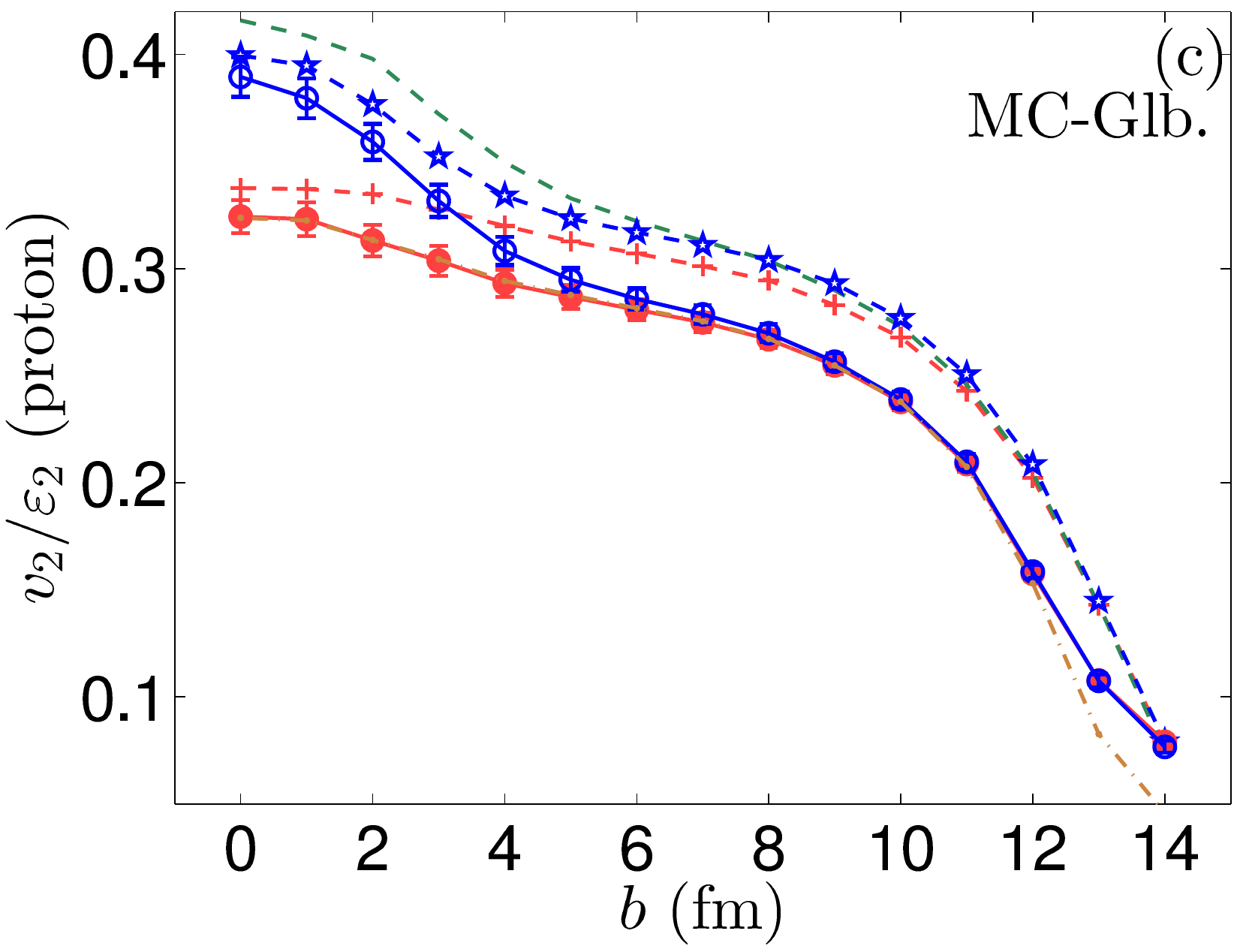}
 \includegraphics[width=0.45\linewidth]{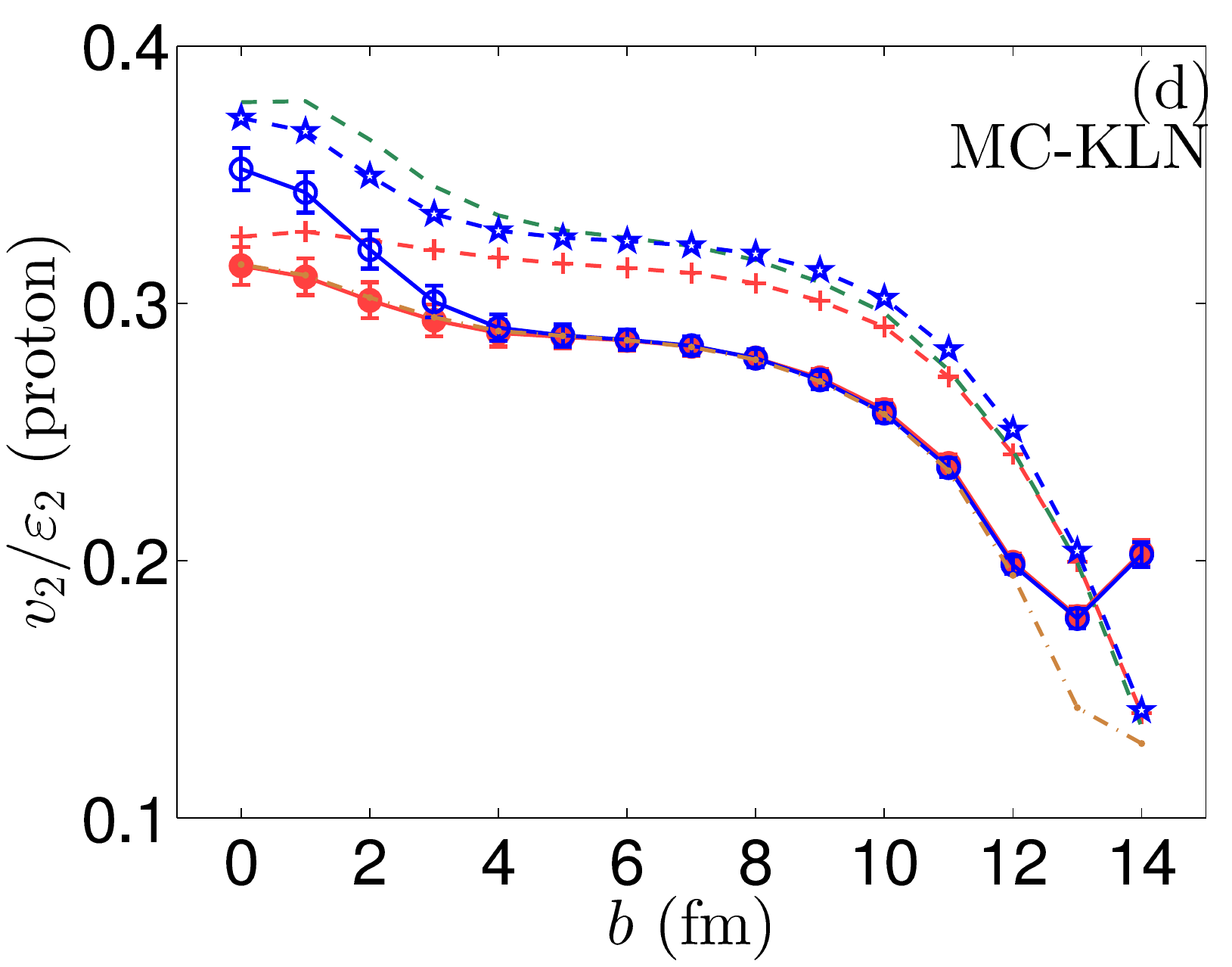}
 \caption{Eccentricity-scaled elliptic flow $v_2/\ecc_2$ as 
  function of impact parameter, for pions (panels (a,b)) and protons 
  (panels (c,d)), from the ideal fluid dynamic evolution of initial
  MC-Glauber (a,c) and MC-KLN (b,d) density profiles. Solid (dashed)
  lines correspond to event-by-event (single-shot) hydrodynamics.
  See text for discussion. Taken from \cite{Qiu:2011iv}.
 \label{1104.0650:F11}
 }
\end{figure*}
%

%
\begin{figure*}[h!]
 \includegraphics[width=0.45\linewidth]{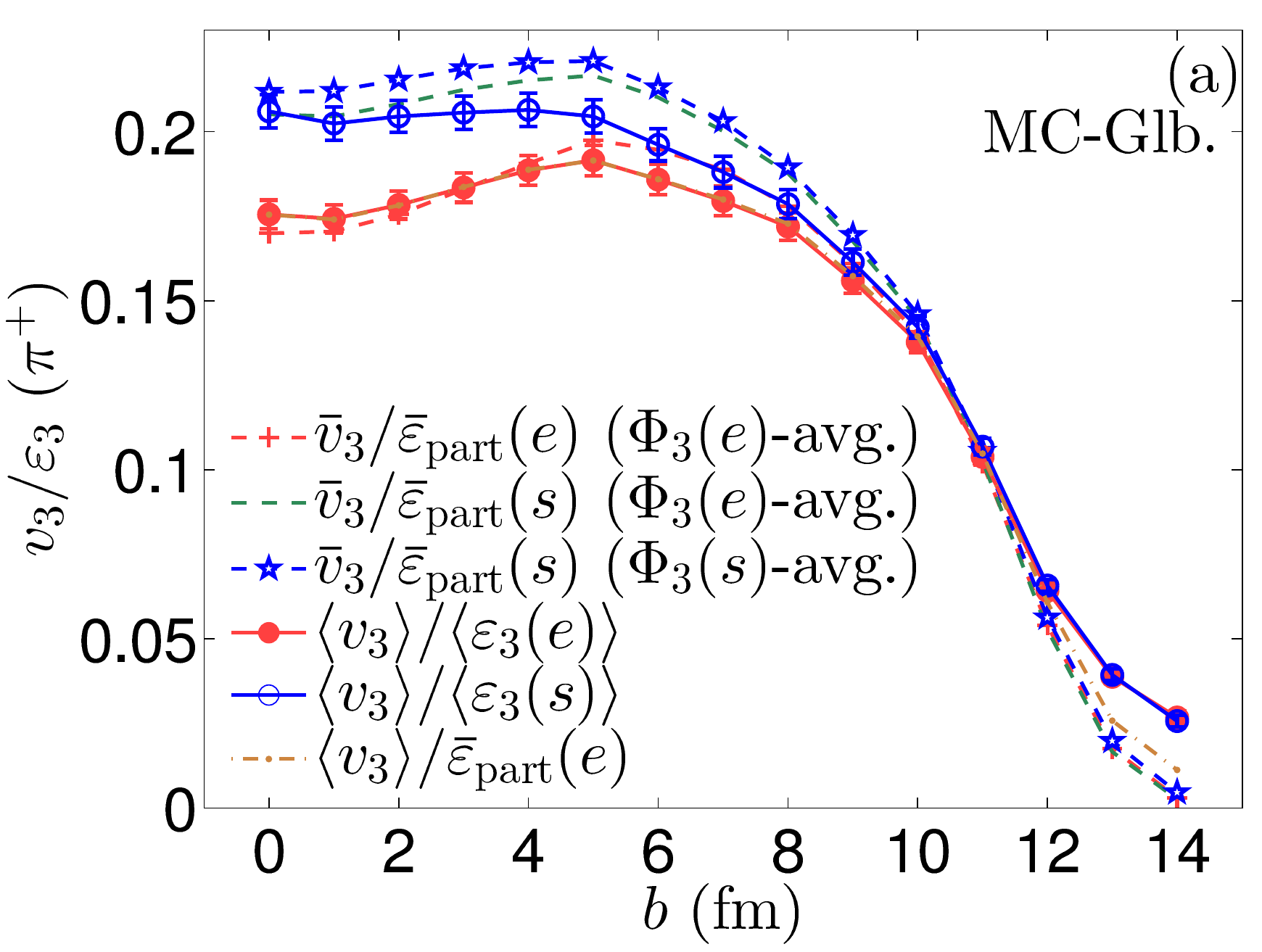}
 \includegraphics[width=0.45\linewidth]{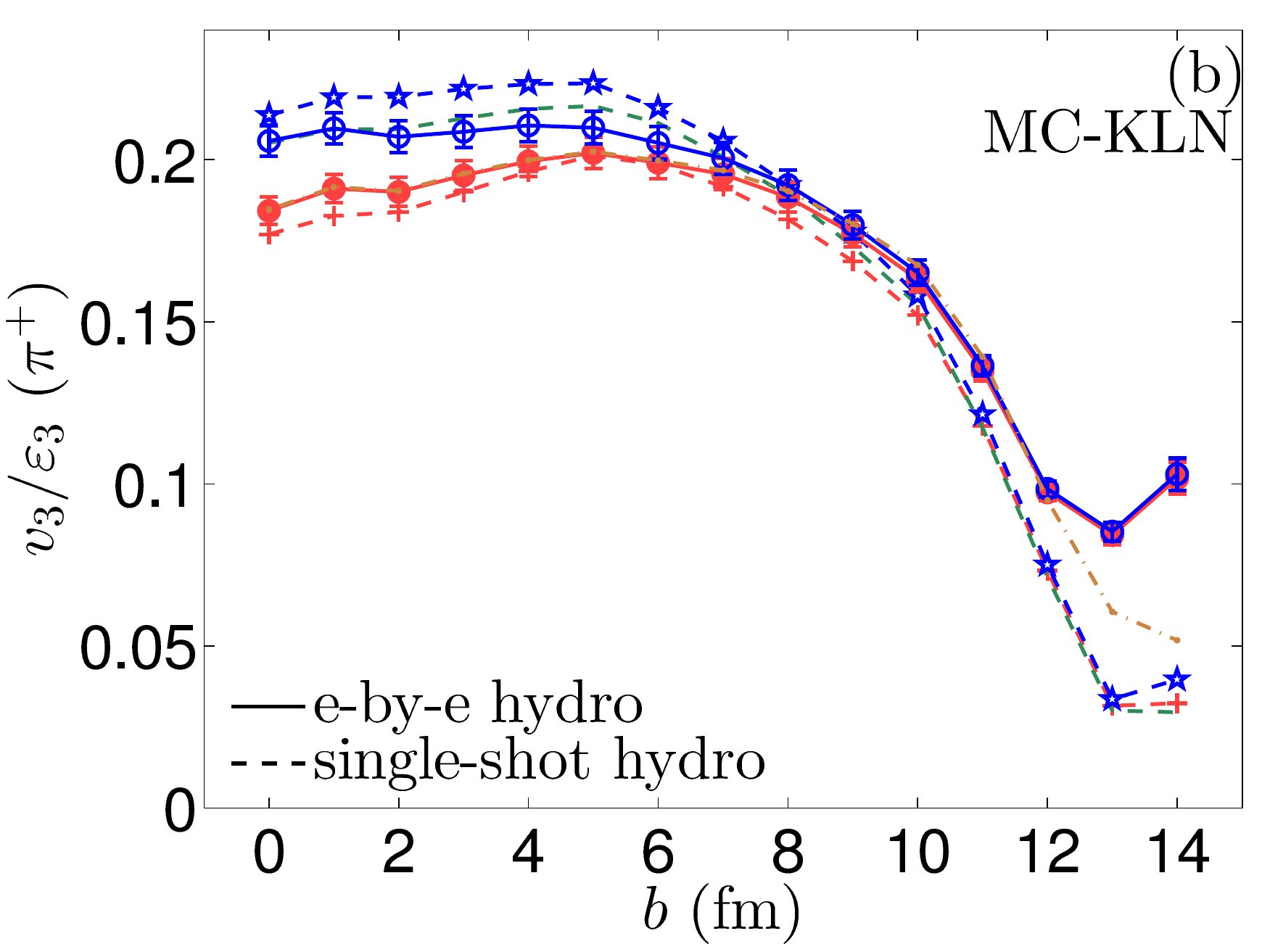}\\
 \includegraphics[width=0.45\linewidth]{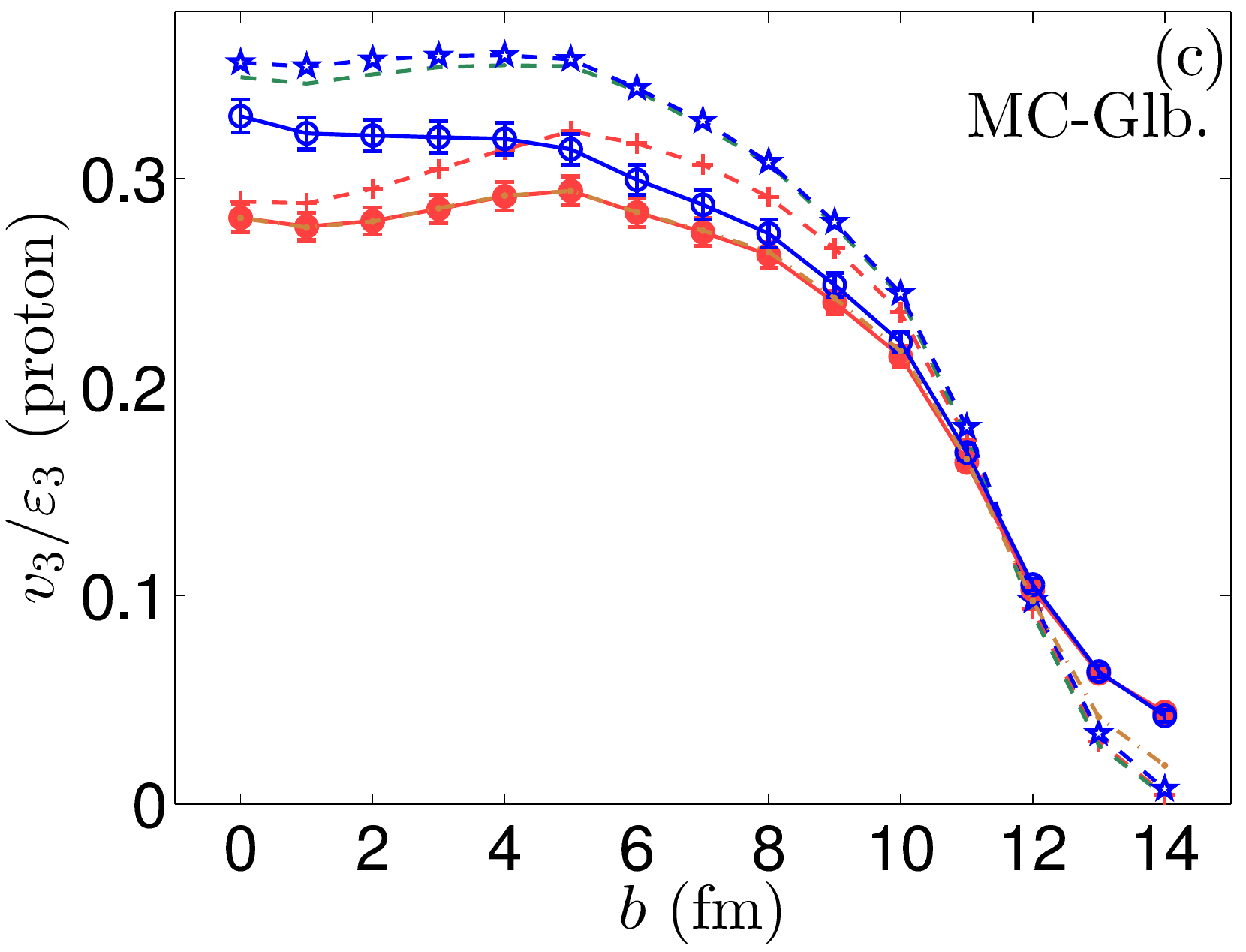}
 \includegraphics[width=0.45\linewidth]{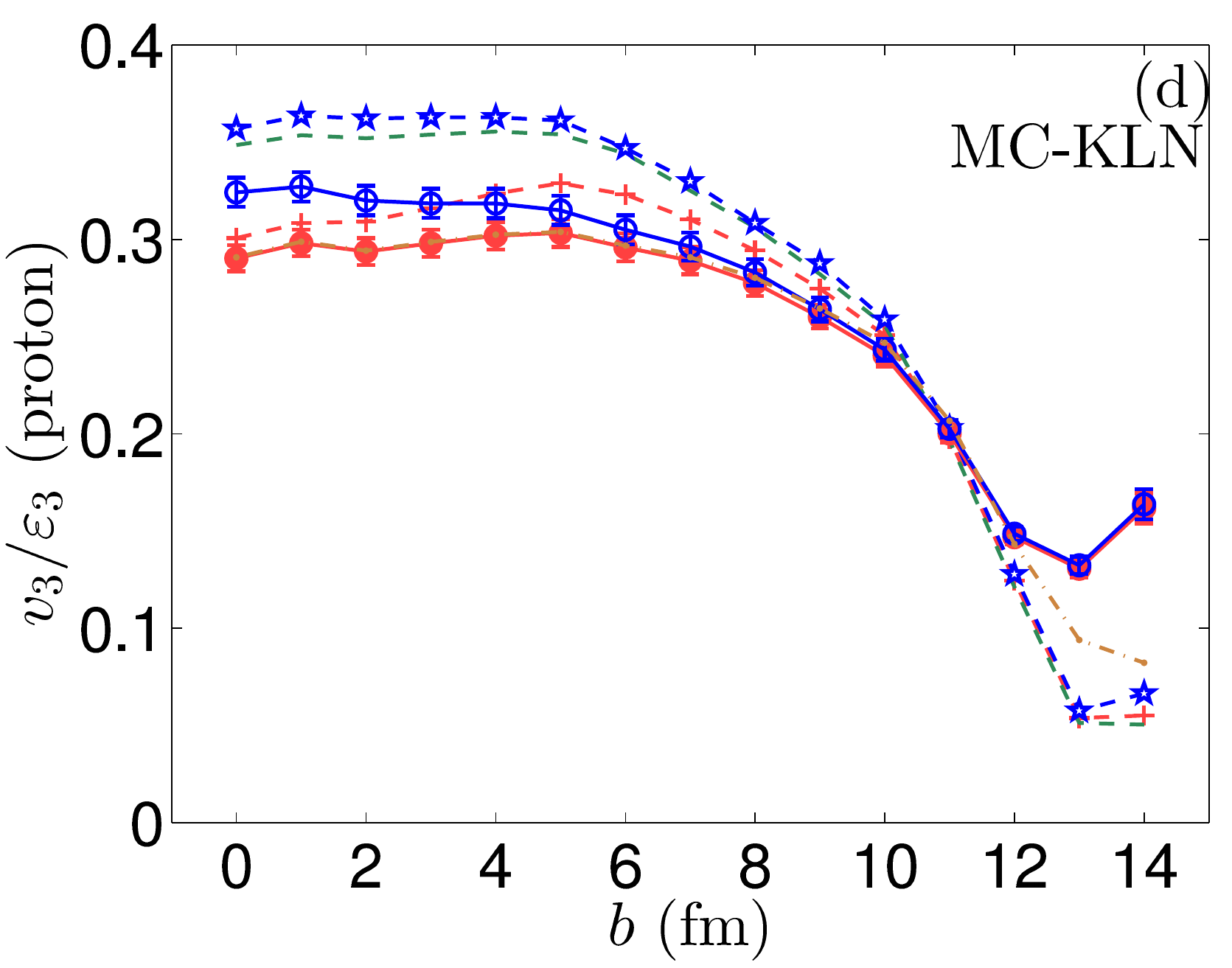} 
 \caption{Same as Fig.~\ref{1104.0650:F11}, but for the
 eccentricity-scaled triangular flow $v_3/\ecc_3$. Taken from \cite{Qiu:2011iv}.
 \label{1104.0650:F12}
 }
\end{figure*}
%

In Figures~\ref{1104.0650:F11} and \ref{1104.0650:F12}, we compare the eccentricity-scaled 
elliptic and triangular flows, $v_2/\ecc_2$ and $v_3/\ecc_3$, for pions 
and protons as a function of impact parameter, from single-shot (dashed
lines) and event-by-event hydrodynamics (solid lines). These ratios 
represent the efficiency of the fluid for converting initial spatial 
deformations into final-state momentum anisotropies. This conversion
efficiency is affected (to be more precise, reduced) by shear viscosity, so these 
ratios form the basis of many analyses that aim to extract this 
transport coefficient from experimental heavy-ion data. 

For event-by-event hydrodynamics, we show two curves, using either the 
entropy- (blue open circles) or the energy-weighted (red 
solid circles) average eccentricities to normalize the average final 
flow $\la v_n\ra$. For the ellipticity (Fig.~\ref{1104.0650:F11}), this choice is seen 
to make a difference only in rather central collisions ($b{\,<\,}4$\,fm), 
but for the triangularity, the differences are significant out to average 
impact parameters probed in minimum-bias samples, $b{\,\lesssim\,}8$\,fm. As 
stated earlier, we prefer the energy-weighted eccentricities 
(solid circles) as deformation measures because energy density and 
pressure are closely related through the EOS, and it is the pressure 
gradients (and their anisotropies) that drive the collective flow (and its anisotropies).

For the single-shot hydrodynamic simulations, a question arises as to
how exactly one should construct the ensemble-averaged smooth initial 
profile which is then evolved hydrodynamically. We have explored three 
reasonable procedures (variations of which have been used in the 
literature) and show them as dashed lines in Figs.~\ref{1104.0650:F11} and \ref{1104.0650:F12}. 
For the {\em lines labeled by stars}, we rotate the entropy density for 
each fluctuating event by the corresponding entropy-weighted 
participant-plane angle $\PsiPP{n}(s)$ ($n \eq 2,3$, see 
Eq.~(\ref{eq:1104.0650:17})),\footnote{Note that for computation of $\bar{v}_3$, 
  we rotate the events by a different angle before averaging than
  for $\bar{v}_2$, i.e. $\bar{v}_3$ and $\bar{v}_2$ are obtained from
  two different single-shot hydrodynamic runs, starting from different
  averaged initial-energy-density profiles.}
then average the rotated entropy profiles, compute the eccentricity 
$\bar{\ecc}_n(s)$ of the resulting average entropy density profile and 
convert it to energy density using the EOS for input into the 
hydrodynamic code. For the {\em lines labeled by crosses}, we rotate the 
energy density for each fluctuating event (obtained from the EOS) by the 
corresponding energy-weighted participant-plane angle $\PsiPP{n}(e)$
(see Eqs.~(\ref{eq:1104.0650:1},\ref{eq:1104.0650:2})), compute the averaged rotated energy 
density profile and its eccentricity $\bar{\ecc}_n(e)$, and use it 
directly as hydrodynamic input. For the {\em dashed lines without symbols}, 
finally, the averaged initial energy density (and therefore the final 
$\bar{v}_n$) are exactly the same as for the lines with crosses, but the 
final $\bar{v}_n$ is scaled by the entropy-weighted (rather than 
energy-weighted) eccentricity of the averaged initial profile, where the 
entropy density is obtained from the smooth averaged energy density via 
the EOS. 

The differences between the different dashed lines illustrate the 
uncertainties associated with the choice of averaging procedure for the 
initial state. Keeping in mind that a 20\% reduction in $v_2/\ecc_2$ 
corresponds (very roughly) to an increase of $\eta/s$ by an additive term $1/4\pi$ 
\cite{Song:2010mg}, one sees that these differences are not negligible 
if one aims for quantitative precision in the extraction of the specific 
shear viscosity. Comparing the three dashed lines, we see that it doesn't 
make much difference whether we use the $s$-weighted or $e$-weighted
participant-plane angles to rotate the events before superimposing them
(the dashed lines without symbols and with stars are all very close to 
each other), but that in the more central collisions we obtain significantly
different values for the conversion efficiencies $\bar{v}_n/\bar{\ecc}_n$ 
if we normalize by $e$- or $s$-weighted mean eccentricities. Even though 
they look similar in Fig.~\ref{1104.0650:F5}a, at small impact parameters 
$\bar{\ecc}_\mathrm{part}(e)$ and $\la\ecc_2(e)\ra$ are larger than 
$\bar{\ecc}_\mathrm{part}(s)$ and $\la\ecc_2(s)\ra$, respectively, and 
this is the main reason why the red and blue lines in Fig.~\ref{1104.0650:F11} 
diverge at small $b$, for both event-by-event (solid lines) and 
single-shot hydrodynamics (dashed lines). 

An apples-to-apples comparison between event-by-event and single-shot 
hydrodynamics (and between theory and experimental data) therefore must 
ensure that the same (or at least conceptually compatible) eccentricities 
are used to normalize the anisotropic flow coefficients that are to be 
compared. In Figs.~\ref{1104.0650:F11},\,\ref{1104.0650:F12}, we should therefore compare
blue solid with blue dashed or red solid with red dashed lines, but
not curves of different colors. 

\begin{figure*}[h!]
 \includegraphics[width=0.48\linewidth]{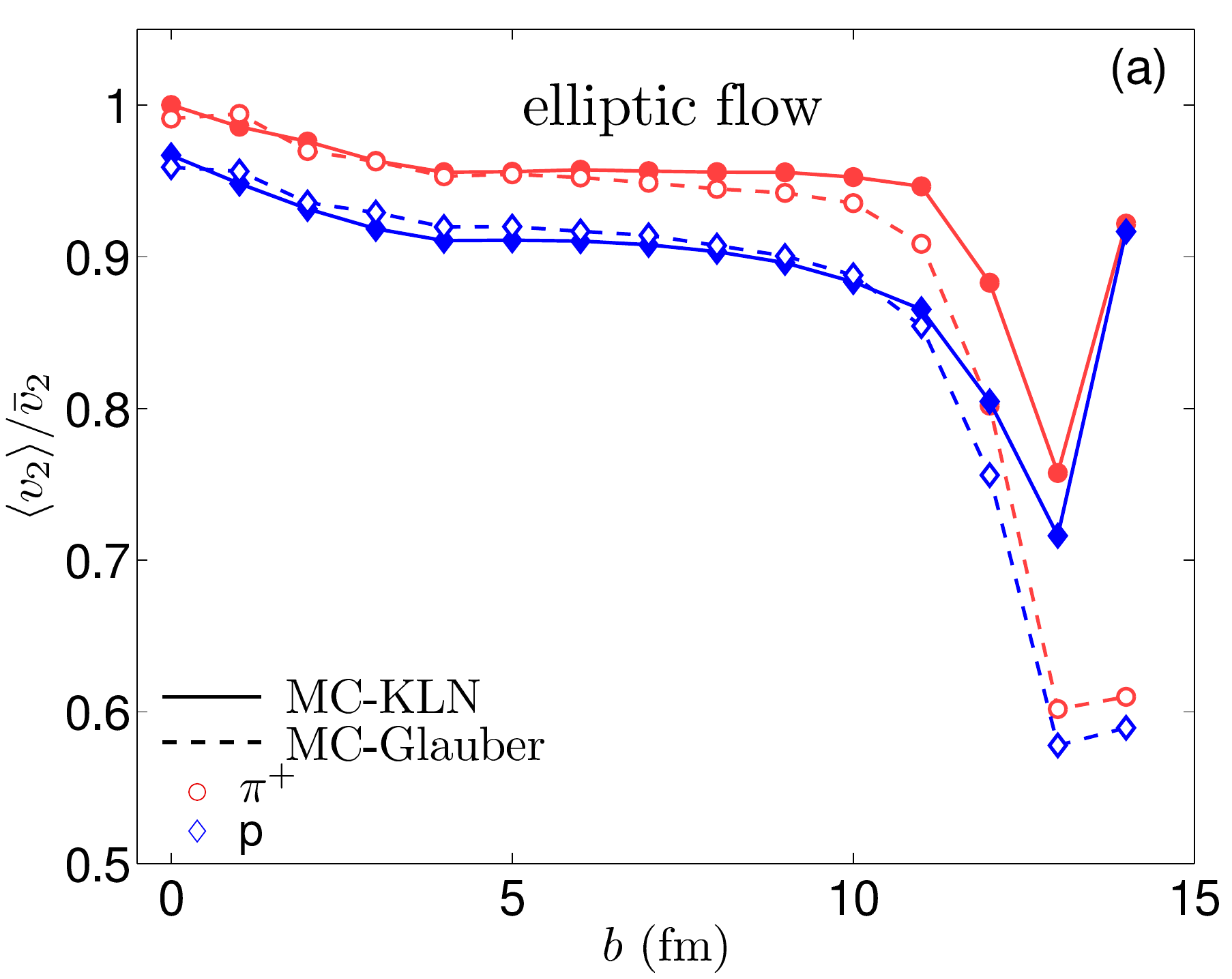}
 \includegraphics[width=0.48\linewidth]{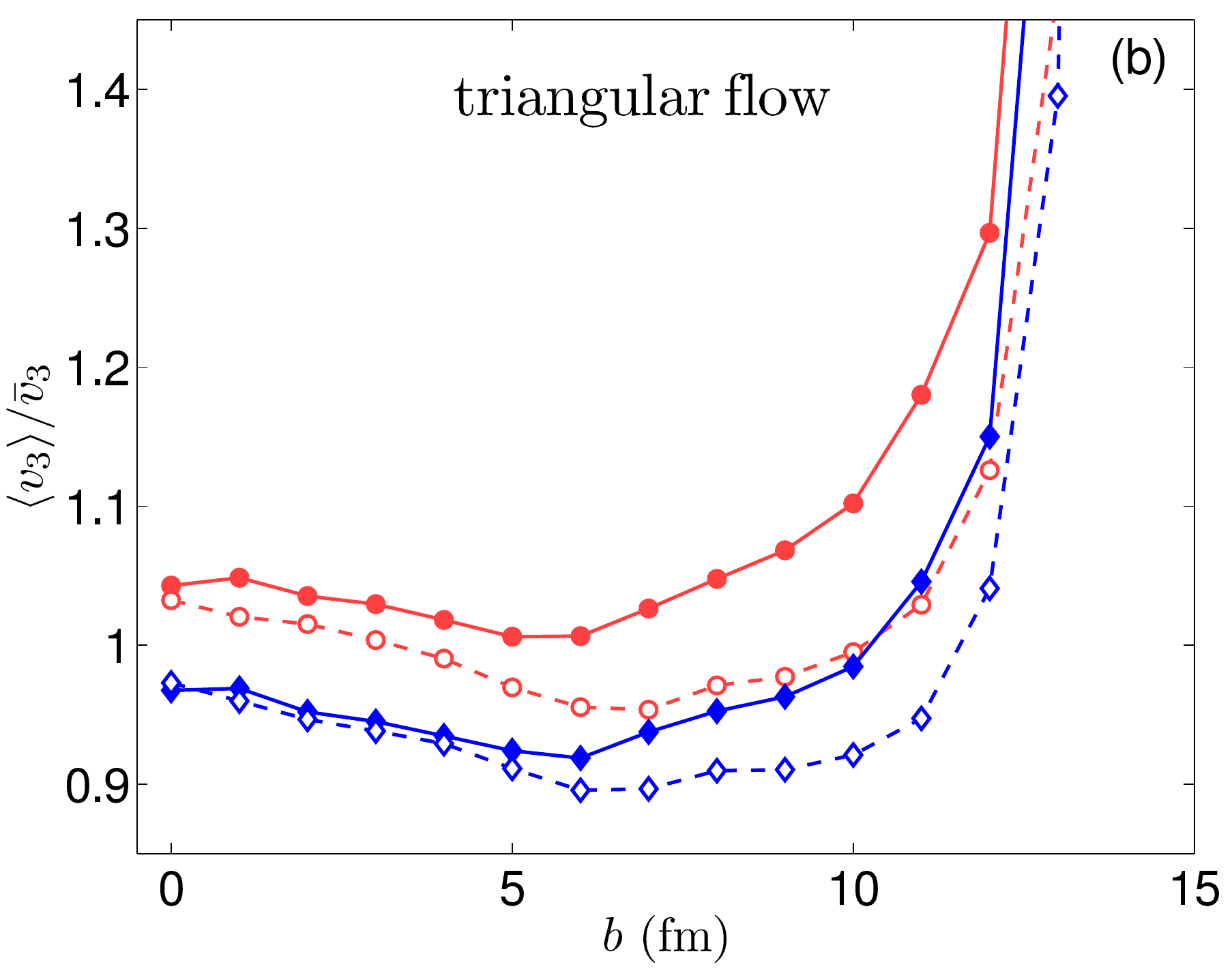}
 \caption{Ratio of the average flow coefficient
  $\la v_n\ra$ from event-by-event ideal hydrodynamics and the corresponding
  mean $\bar{v}_n$ from single-shot ideal hydrodynamics, as a function of
  impact parameter in $200\,A$\,GeV Au+Au collisions, for $n \eq 2$ (a)
  and $n \eq 3$ (b). Shown are the ratios for directly emitted pions 
  (circles) and protons (diamonds) from fluctuating events using the 
  MC-Glauber (dashed lines) and MC-KLN models (solid lines). Average 
  events for computing $\bar{v}_n$ using single-shot hydrodyamics
  were obtained by rotating the energy density of each event by 
  $\PsiPP{n}(e)$ before superimposing them. Taken from \cite{Qiu:2011iv}.
 \label{1104.0650:F13}
 }
\end{figure*}

Even this is not good enough if one wants to accurately assess the relative 
space-to-momentum anisotropy conversion efficiency in single-shot and 
event-by-event hydrodynamics: in the single-shot hydro curves, we use 
$\bar{\ecc}_\mathrm{part}$ to normalize the final elliptic flow, whereas 
the event-by-event hydro results were normalized with 
$\la\ecc_2\ra{\,\equiv\,}\la\ecc_\mathrm{part}\ra$. While each of these 
eccentricity measures makes perfect sense in its own context, they differ 
at large impact parameters, $\bar{\ecc}_\mathrm{part}$ being larger (see 
Figs.~\ref{1104.0650:F1}a,b). To avoid this problem, we have added in Figs.~\ref{1104.0650:F11} 
and \ref{1104.0650:F12} an additional ``mixed ratio'' (dash-dotted purple line), which 
normalizes the ensemble-averaged anisotropic flow $\la v_n\ra$ ($n \eq 2,3$) 
from event-by-event hydrodynamics (used in the ratio $\la v_n\ra/\la\ecc_n\ra$ 
denoted by solid lines with solid red circles) by the mean $e$-weighted
eccentricity $\bar{\ecc}_n$ from single-shot hydrodynamics (used in the 
ratio $\bar{v}_n/\bar{\ecc}_n$ denoted by dashed lines with crosses). This
dot-dashed purple line agrees almost perfectly with the solid red line
with circles over most of the impact parameter range, except for peripheral
collisions with $b{\,\gtrsim\,}10$\,fm where $\bar{\ecc}_n$ and $\la\ecc_n\ra$
begin to diverge. The red dashed lines with crosses and purple dash-dotted 
lines show the anisotropic flows from single-shot and event-by-event 
hydrodynamics normalized by the {\em same} eccentricity measure 
characterizing the fluctuating event sample. Their comparison allows 
an unambiguous assessment of the different efficiencies of single-shot 
and event-by-event hydrodynamics in converting initial eccentricities 
to final momentum anisotropies. Their ratio is shown in Fig.~\ref{1104.0650:F13}.

%
%

From Fig.~\,\ref{1104.0650:F13}a, one concludes that, for ideal hydrodynamics,
event-by-event fluctuations on average reduce the efficiency of the 
fluid in converting initial source ellipticity into elliptic flow.
Over most of the centrality range, this reduction is about 4\% for pions
and about twice as large for protons and it is similar for MC-KLN
and MC-Glauber initial profiles. In very central collisions, the ratio
of conversion efficiencies for event-by-event vs.\ single-shot hydrodynamics
is closer to 1, but it degrades strongly in very peripheral collisions
where event-by-event evolution generates on average $30{-}40\%$ less
elliptic flow than single-shot hydrodynamics. The generic tendency of the 
event-by-event hydrodynamic evolution of fluctuating initial profiles 
to generate less elliptic flow than expected from hydrodynamic 
evolution of the corresponding smooth average profile has been observed 
before \cite{Andrade:2006yh,Schenke:2010rr}; our systematic study in 
Fig.~\ref{1104.0650:F13}a quantifies this effect over the full range of collision 
centralities.

The situation with triangular flow, shown in Fig.~\ref{1104.0650:F13}b, is quite 
different: event-by-event propagation of initial-state flutuations can
lead to an increase or decrease of the triangular flow compared to
single-shot hydrodynamics, depending on particle mass (pions or protons), 
the nature of the fluctuations (MC-Glauber or MC-KLN), and collision 
centrality. Contrary to elliptic flow, in peripheral collisions 
event-by-event evolution leads to significantly {\em larger} average 
triangular flow than single-shot hydrodynamics.

%
\begin{figure*}[h!]
 \includegraphics[width=0.45\linewidth]{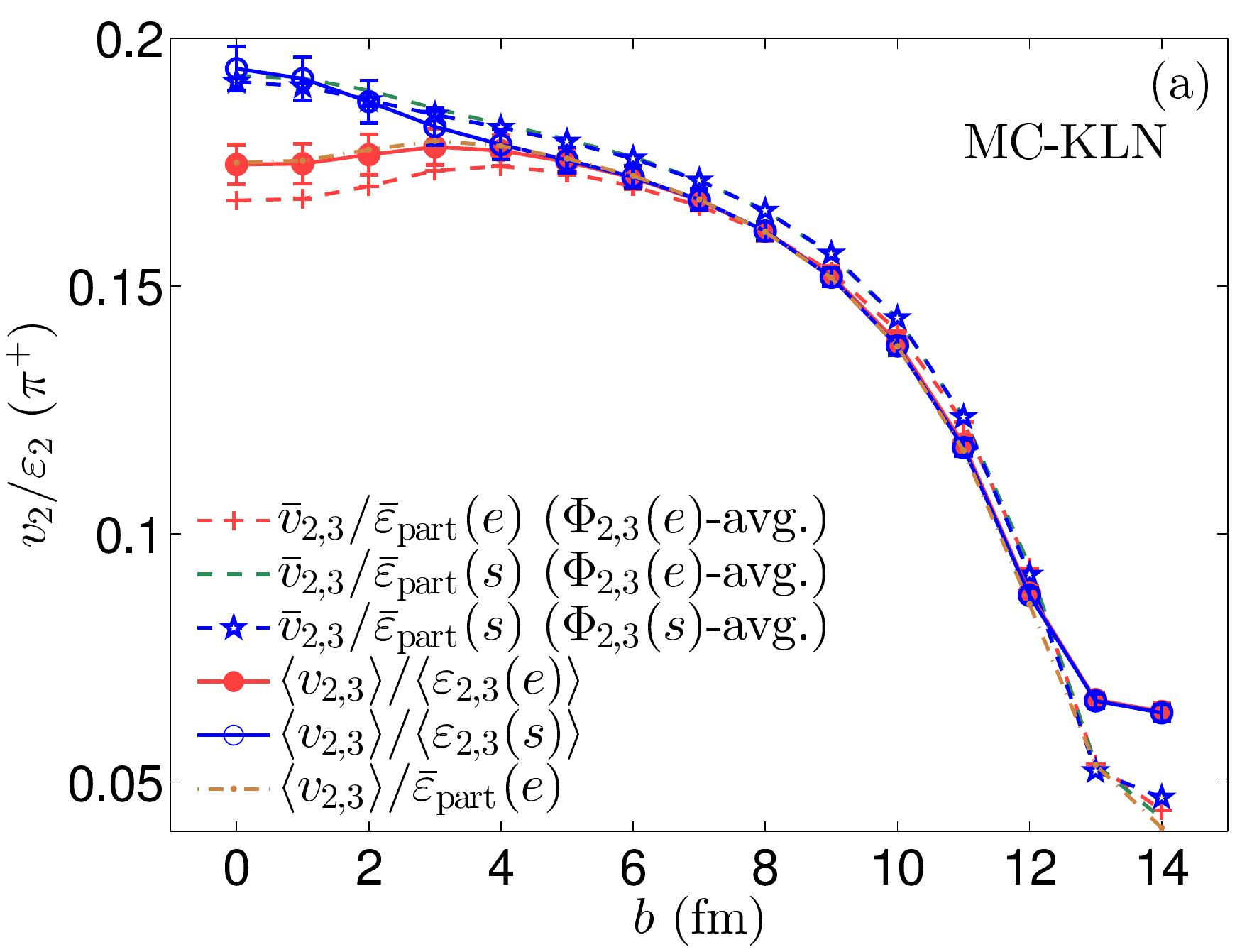}
 \includegraphics[width=0.45\linewidth]{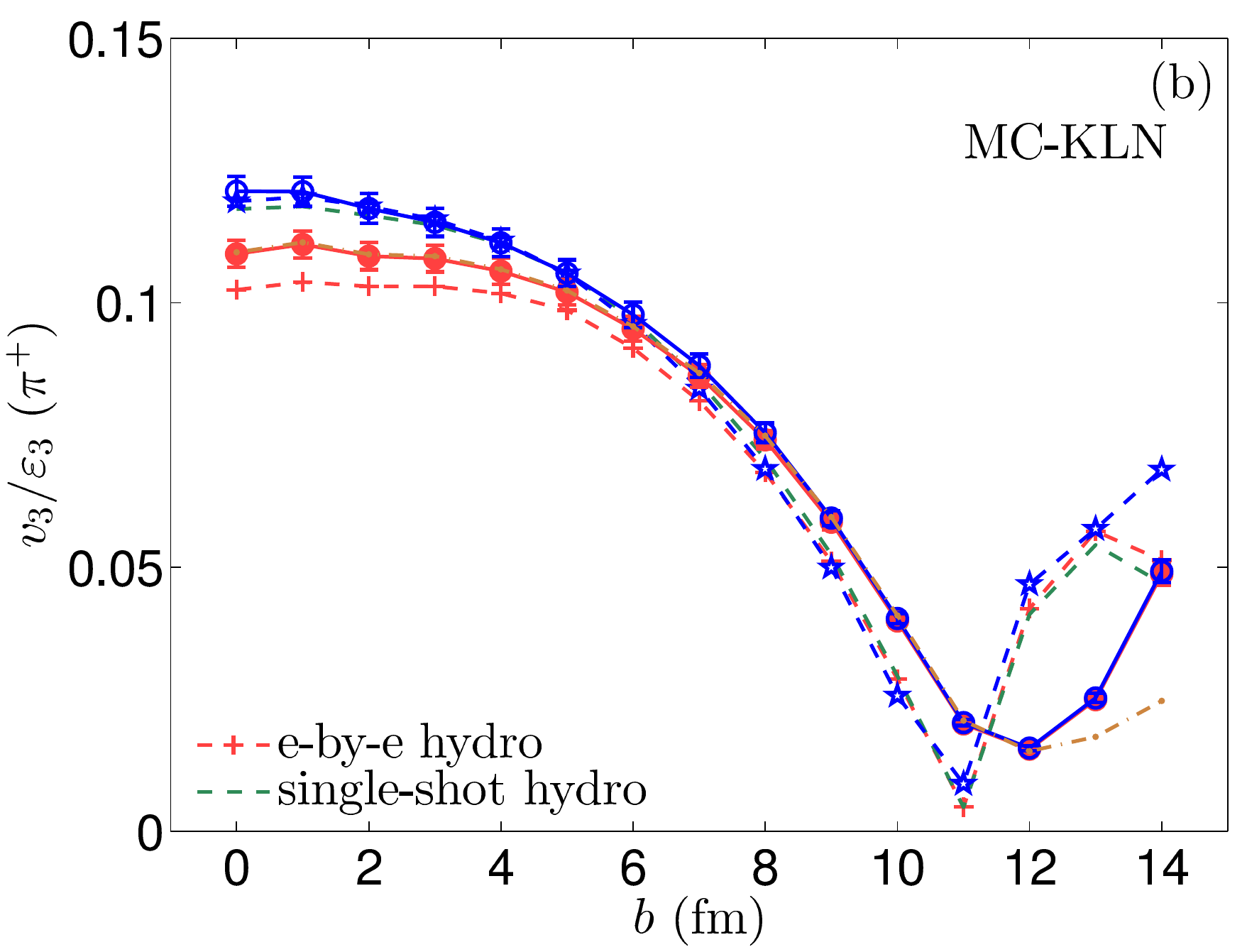}\\
 \includegraphics[width=0.45\linewidth]{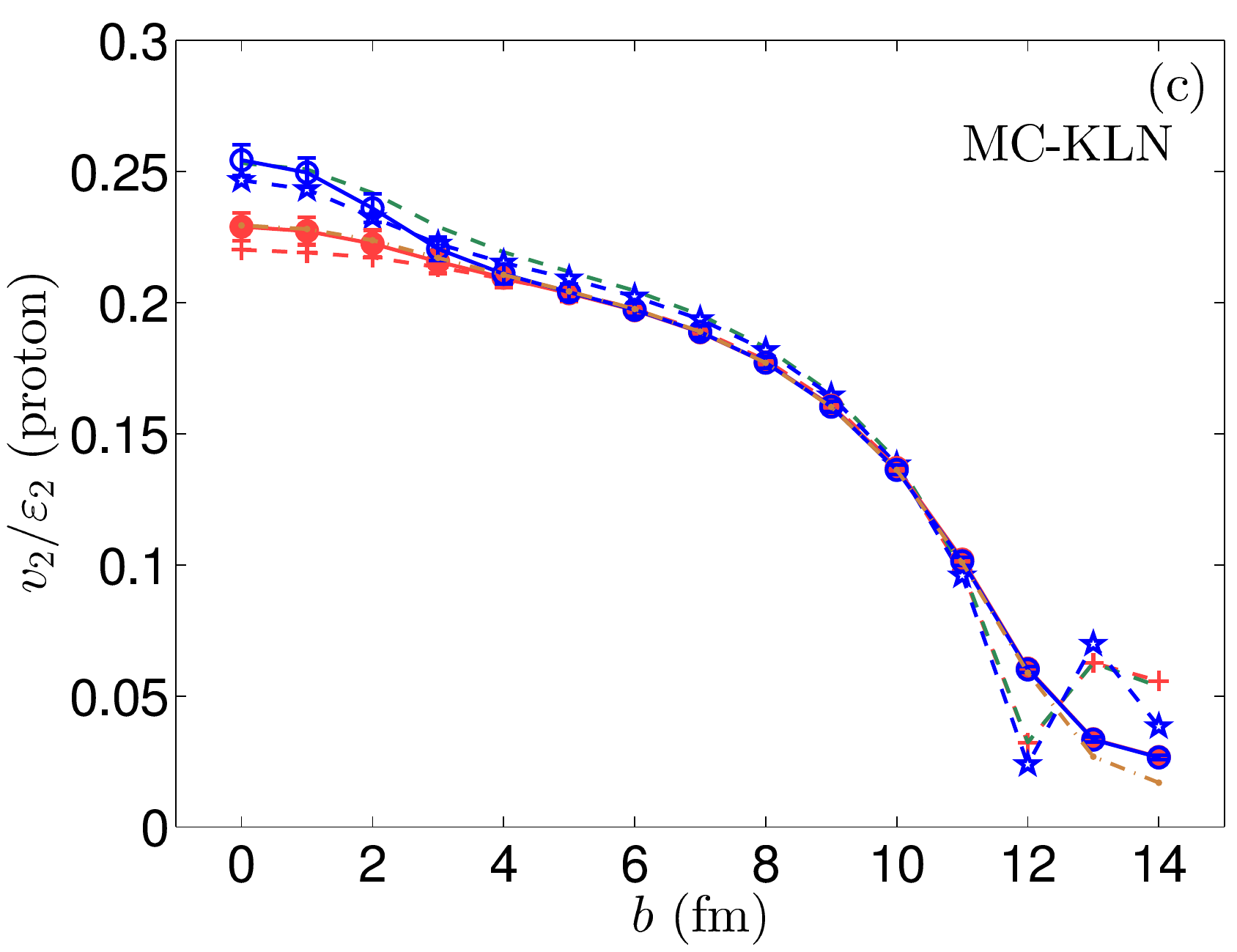}
 \includegraphics[width=0.45\linewidth]{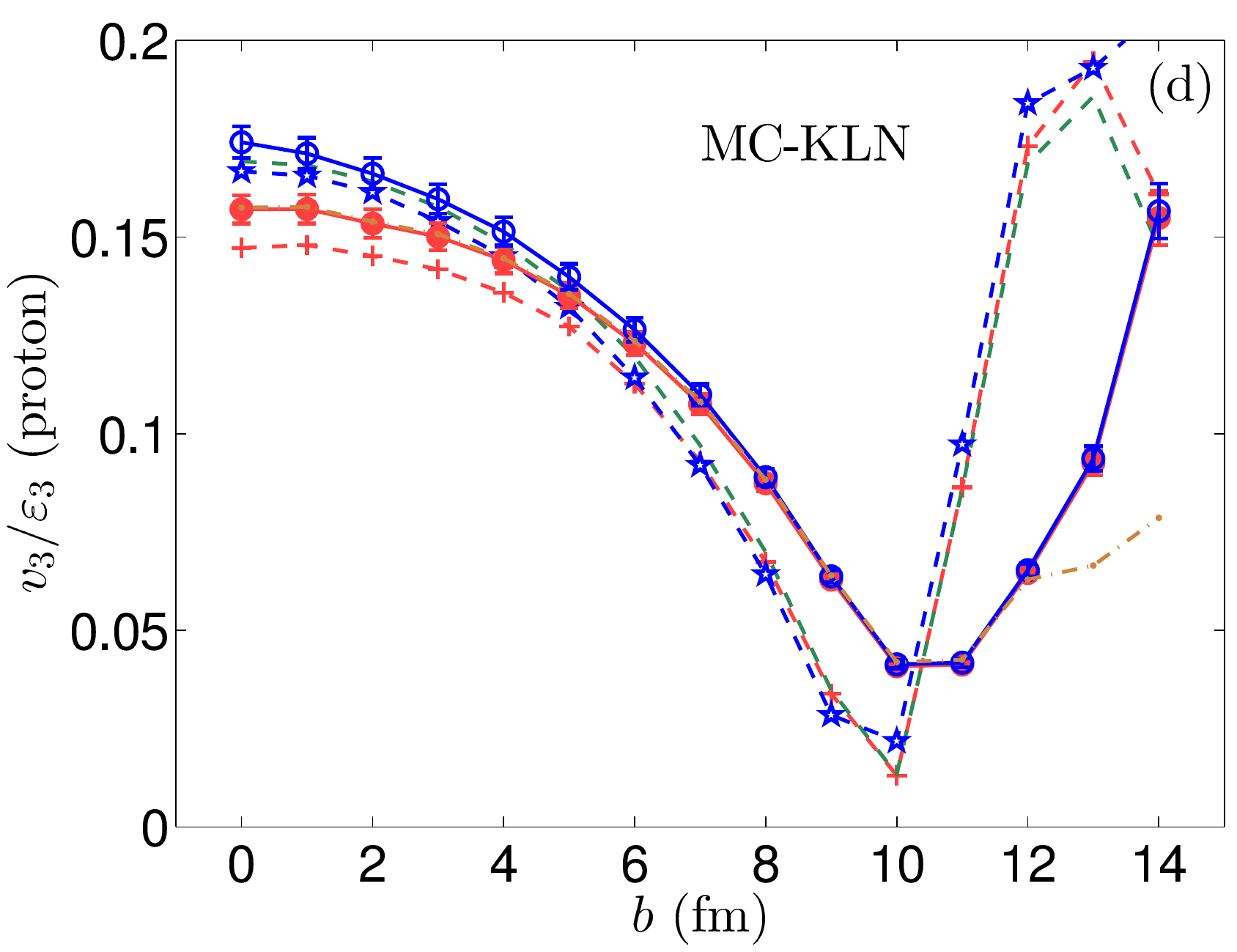}\\
 \caption{Eccentricity-scaled elliptic flow $v_2/\ecc_2$ (left) and $v_3/\ecc_3$ (right) as 
  functions of impact parameter, for pions (panels (a,b)) and protons 
  (panels (c,d)), from viscous ($\eta/s=0.2$) fluid dynamic evolution of initial
  MC-KLN density profiles. Solid (dashed)
  lines correspond to event-by-event (single-shot) hydrodynamics.
  See text for discussion, and Fig.~\ref{1104.0650:F11}, \ref{1104.0650:F12} for a comparison with ideal fluid dynamics.
  \label{1104.0650:F12.1}
 }
\end{figure*}
%

%
\begin{figure*}[h!]
 \includegraphics[width=0.45\linewidth]{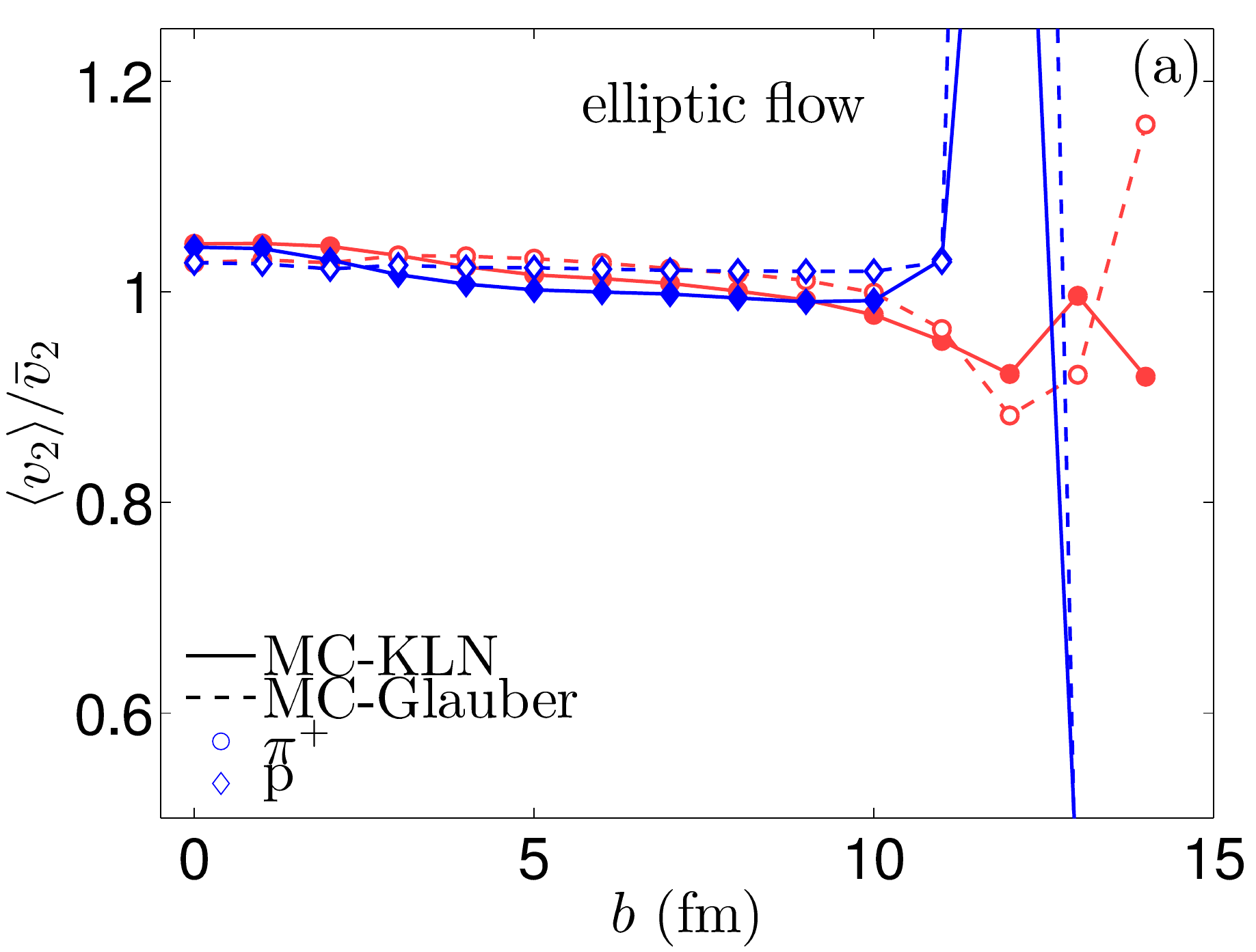}
 \includegraphics[width=0.45\linewidth]{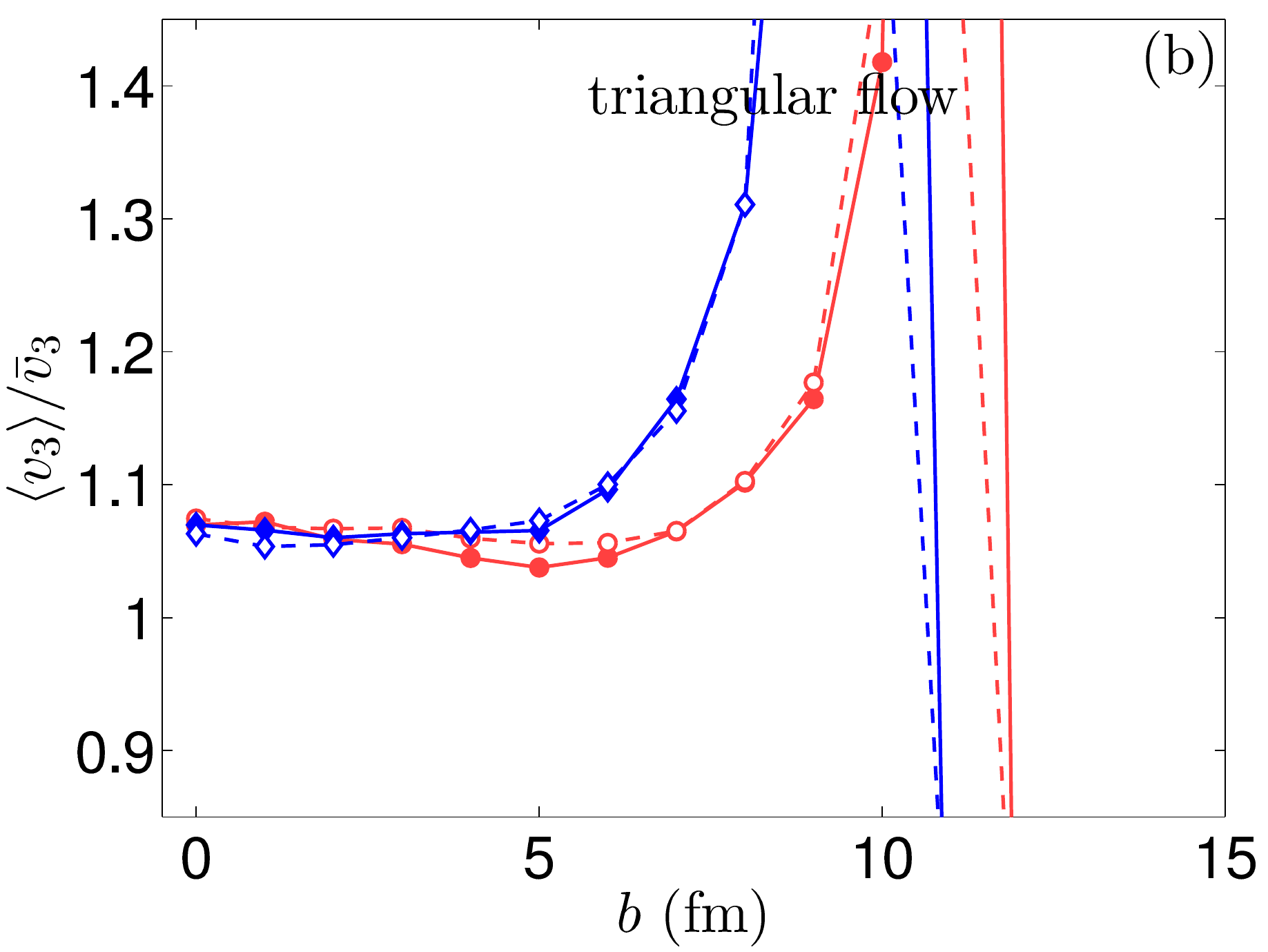}
 \caption{
  Same as Fig.~\ref{1104.0650:F13}, but for viscous hydrodynamic evolution with $\eta/s=0.2$.
  \label{1104.0650:F13.1}
 }
\end{figure*}
%

The analogous plots to Fig.~\ref{1104.0650:F11} and Fig.~\ref{1104.0650:F12} using viscous ($\eta/s=0.2$) hydrodynamic simulations with MC-KLN initial conditions are shown in Fig.~\ref{1104.0650:F12.1}. Comparing the ideal and viscous results, we see that viscosity greatly reduces the gap between event-by-event and single-shot simulations, both for $\pi^+$ and protons, for both elliptic and triangular flow. This is consistent with viscosity damping inhomogeneities by dissipation, making the density distributions smoother, at least at the late times.

To make the statement more quantitative, we show the corresponding ratio plot in Fig.~\ref{1104.0650:F13.1}. Comparing to Fig.~\ref{1104.0650:F13}, it is seen from panel (a) that, except for extremely peripheral collisions ($b\geq 10$ fm), the $\sim 10\%$ gap between the elliptic flows calculated from the two types of simulations is reduced to $\sim 3\%$. For the triangular flow, the story is somewhat different: although the curves from event-by-event and single-shot simulations appear to be closer to each other in Fig.~\ref{1104.0650:F12.1}b,d than in Fig.~\ref{1104.0650:F12}b,d, the relative difference between the two types of simulations is actually slightly larger, rising from $\pm 5\%$ to $\sim 8\%$.

If one aims for quantitative calculations with accuracy better than $5\%$ or for a study of higher-order flows, event-by-event hydrodynamic evolution is an essential and indispensable ingredient; however if one has larger accuracy tolerance, one might try to extract $\eta/s$ from $v_2$ and $v_3$ using single-shot simulations. Unfortunately, it turns out that $v_2$ and $v_3$ are not enough to constrain $\eta/s$ very well unless one has complete control over the initial fluctuation spectrum which, as already shown in this chapter, is not true for the MC-KLN and MC-Glauber models. This will be discussed in more detail in the following chapters.

\vspace*{-3mm}
\subsection{Elliptic flow fluctuations}
\label{sec:1104.0650:5c}
\vspace*{-3mm}

%
\begin{figure*}
 \center{
 \includegraphics[width=0.45\linewidth]{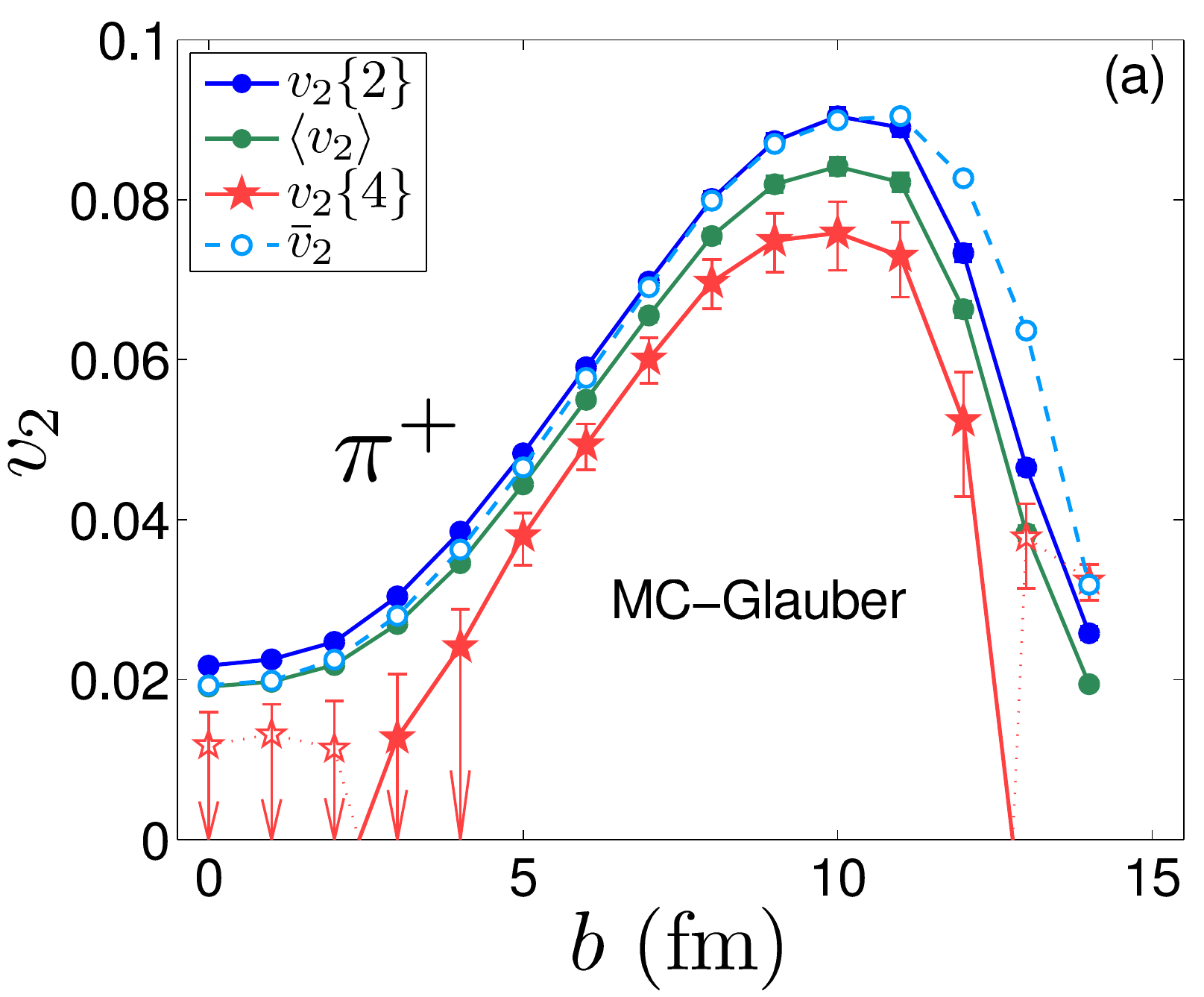}
 \includegraphics[width=0.45\linewidth]{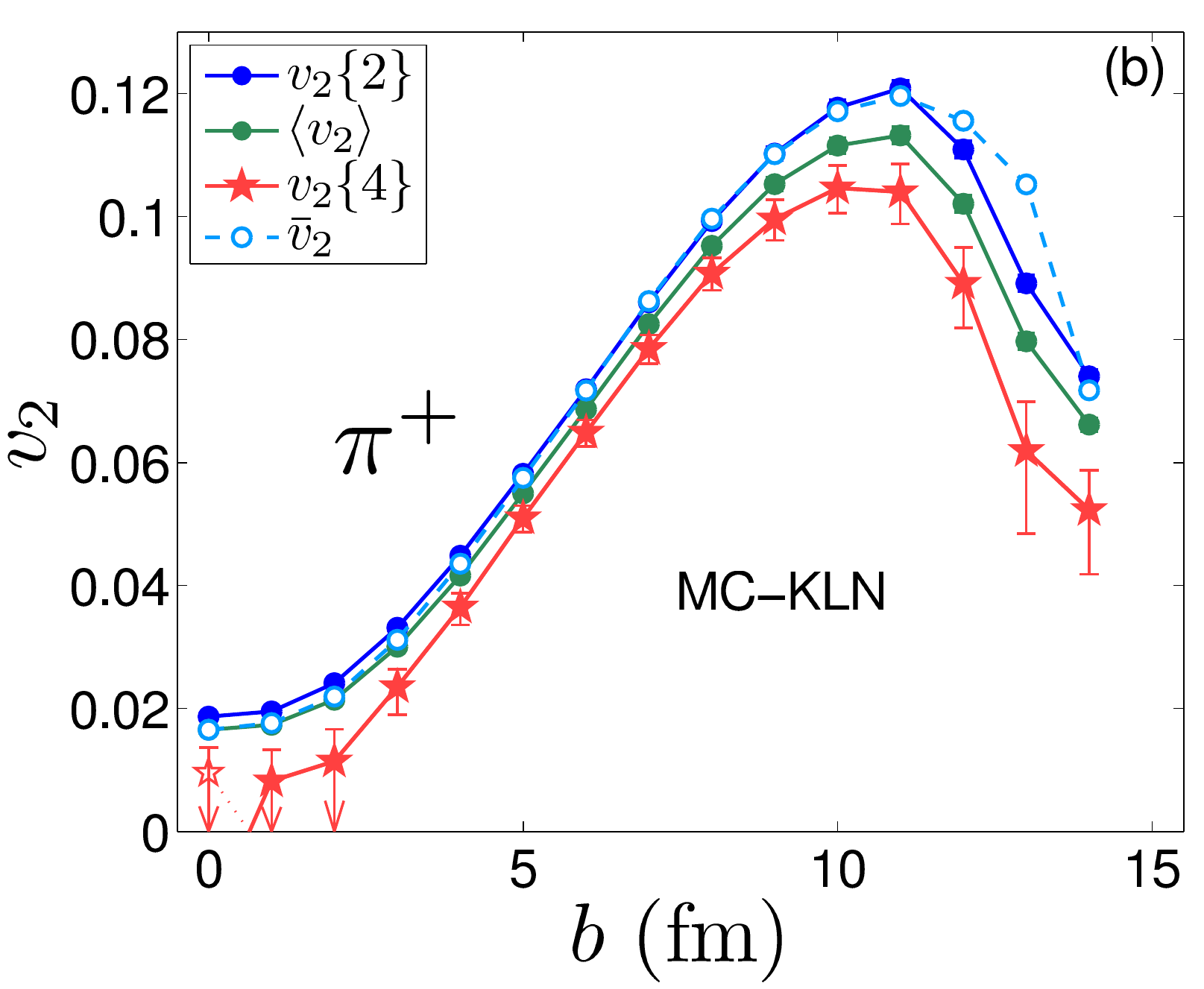}\\
 \includegraphics[width=0.45\linewidth]{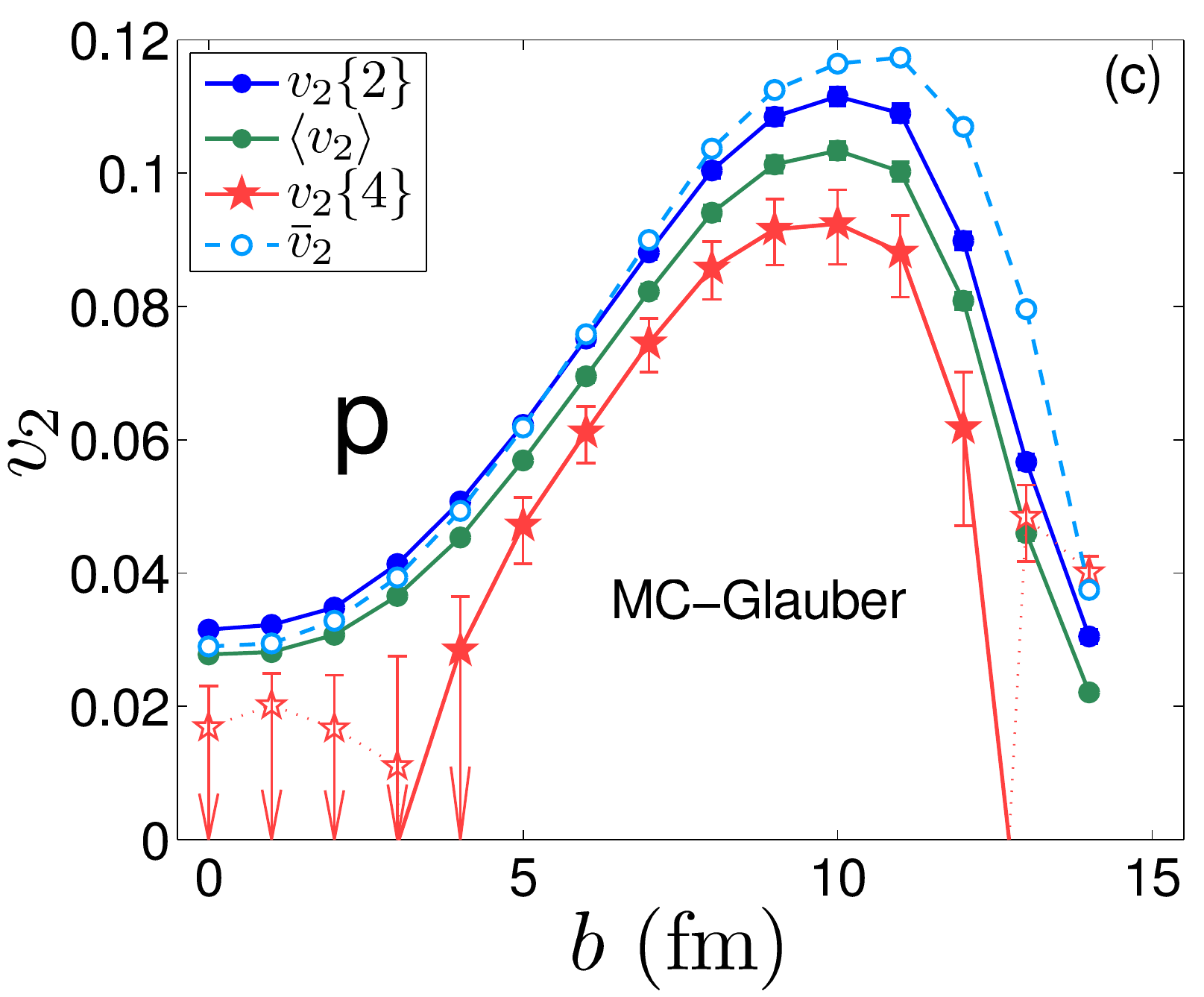}
 \includegraphics[width=0.45\linewidth]{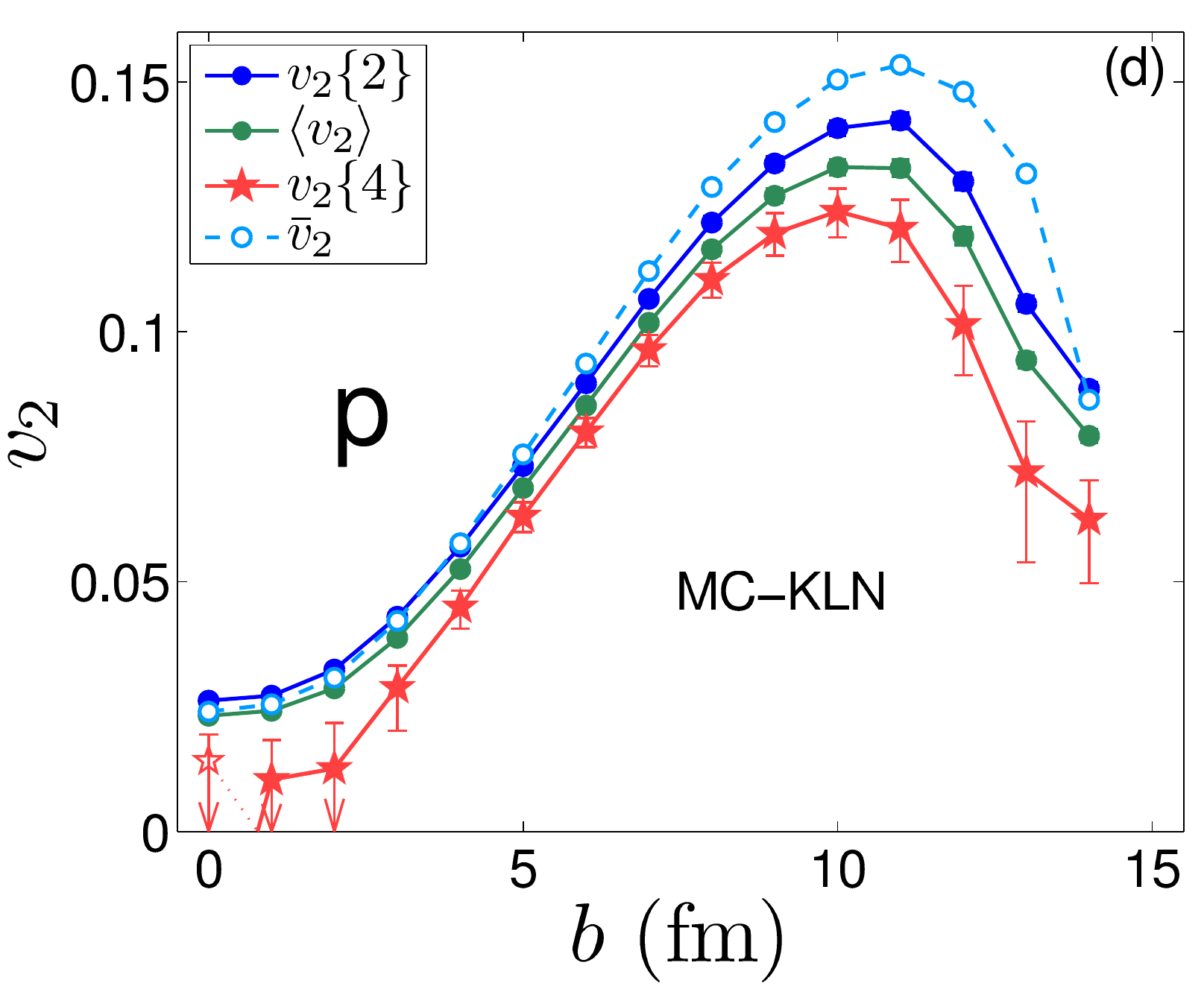}
 }
 \caption{Different measures for the final elliptic
    flow $v_2$ (similar to Fig.~\ref{1104.0650:F1}) for directly emitted pions 
    (a,b) and protons (c,d) as functions of impact parameter from 
    event-by-event ideal fluid dynamics for $T_\mathrm{dec}=140$MeV, using MC-Glauber (a,c) and 
    MC-KLN (b,d) initial conditions for $200\,A$\,GeV Au+Au collisions. Taken from \cite{Qiu:2011iv}.
 \label{1104.0650:F14}
 }
\end{figure*}
%

Similar to what is shown in Figs.~\ref{1104.0650:F1}a,b for the initial source 
ellipticities, Fig.~\ref{1104.0650:F14} shows the elliptic flow measures $\la v_2\ra$, 
$v_2\{2\}$, and $v_2\{4\}$ from event-by-event hydrodynamics, together with
$\bar{v}_2$ from single-shot hydrodynamic evolution of the corresponding
averaged initial profile, for pions and protons, using MC-Glauber and
MC-KLN initializations, respectively. $v_2\{2\}$ and $v_2\{4\}$ are defined
in analogy to Eqs.~(\ref{eq:1104.0650:9},\ref{eq:1104.0650:10}) by
\begin{eqnarray}
 \label{eq:1104.0650:23}
 v_2\{2\}^2 &=& \la v_2^2 \ra,
\\
 \label{eq:1104.0650:24}
 v_2\{4\}^4 &=& 2\la v_2^2\ra^2 - \la v_2^4\ra.
\end{eqnarray}
Here $v_2$ is calculated event by event via Eq.~(\ref{eq:1104.0650:20}) from the 
Cooper-Frye spectrum at freeze-out (with zero statistical uncertainties
since it is determined with mathematical precision by the event-by-event 
hydrodynamic output). 

As in Fig.~\ref{1104.0650:F1}, open stars show the central values for 
$\sqrt[4]{|v_2\{4\}^4|}$ whenever $v_2\{4\}^4$ turns negative, and 
open-ended error bars indicate that the error band for $v_2\{4\}^4$ 
ranges from positive to negative values. Similar to the ellipticities
shown in Fig.~\ref{1104.0650:F1}, the latter happens at small impact parameters,
but for the MC-Glauber model the $b$-range over which this happens
for $v_2\{4\}$ (for both pions and protons) is somewhat larger than 
for $\ecc\{4\}$. Still, $v_2\{4\}$ is compatible with zero over this
entire range, and we do not find statistically significant negative 
values for $v_2\{4\}$ at small impact parameters. At large $b{\,>\,}12$\,fm,
$v_2\{4\}^4$ turns negative for both pions and protons when we use 
MC-Glauber initial conditions, whereas it remains positive for MC-KLN 
initial profiles.   

By comparing $\bar{v}_2$ (open circles in Fig.~\ref{1104.0650:F14}) with 
$\la v_2\ra$ (solid green circles), one sees that in mid-central to 
peripheral collisions the $v_2$-suppression from event-by-event 
hydrodynamic evolution is of the same order as or (especially for 
protons) even larger than the difference between $v_2\{2\}$ and 
$\la v_2\ra$ (solid blue vs.\ solid green circles) that arises from 
event-by-event flow fluctuations. As a result, $v_2\{2\}$ from 
event-by-event hydrodynamics lies in peripheral collisions even 
below $\bar{v}_2$ from single-shot hydrodynamics, in spite of its 
fluctuation-induced enhancement.

%
\begin{figure*}
 \center{
 \includegraphics[width=0.45\linewidth]{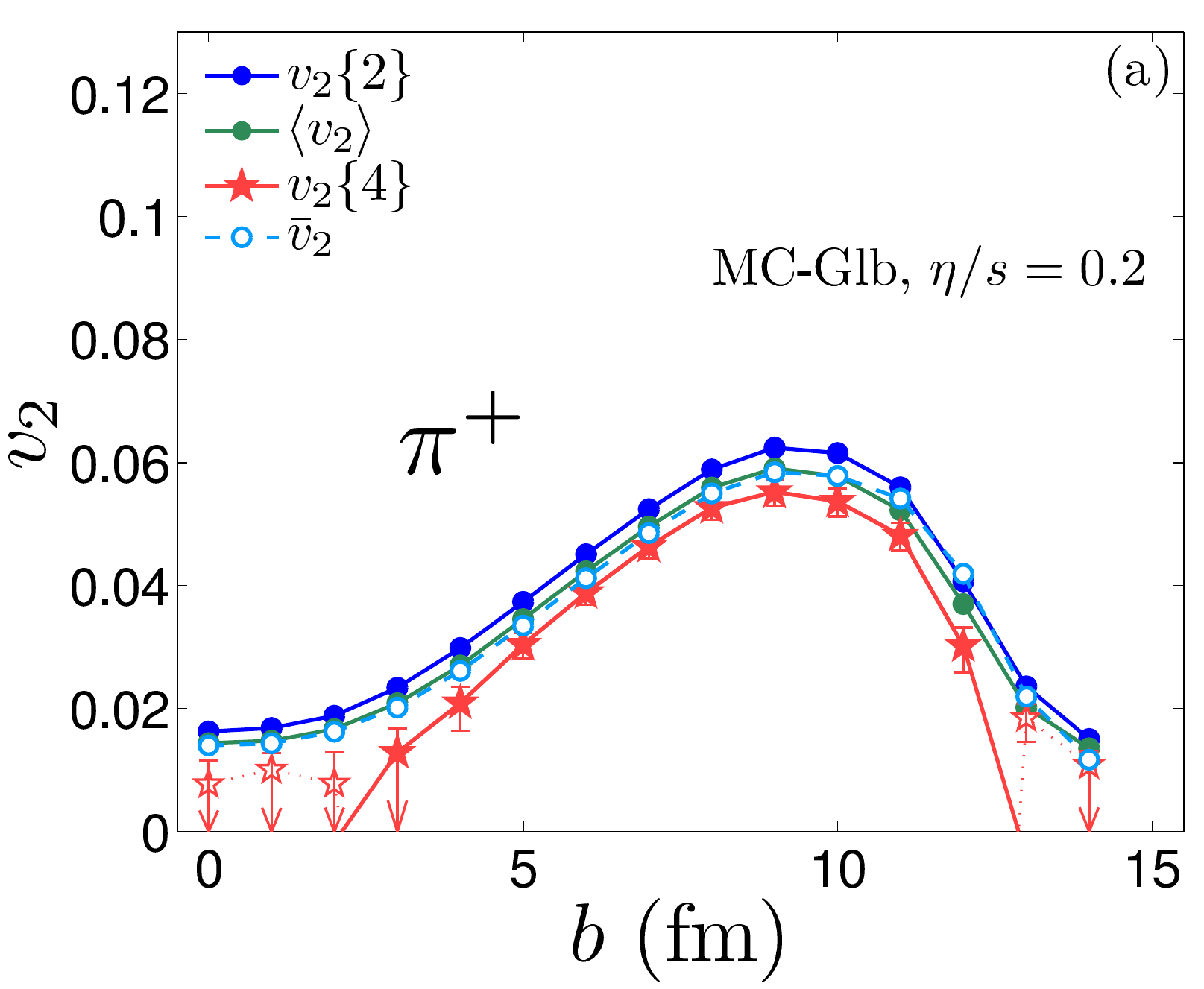}
 \includegraphics[width=0.45\linewidth]{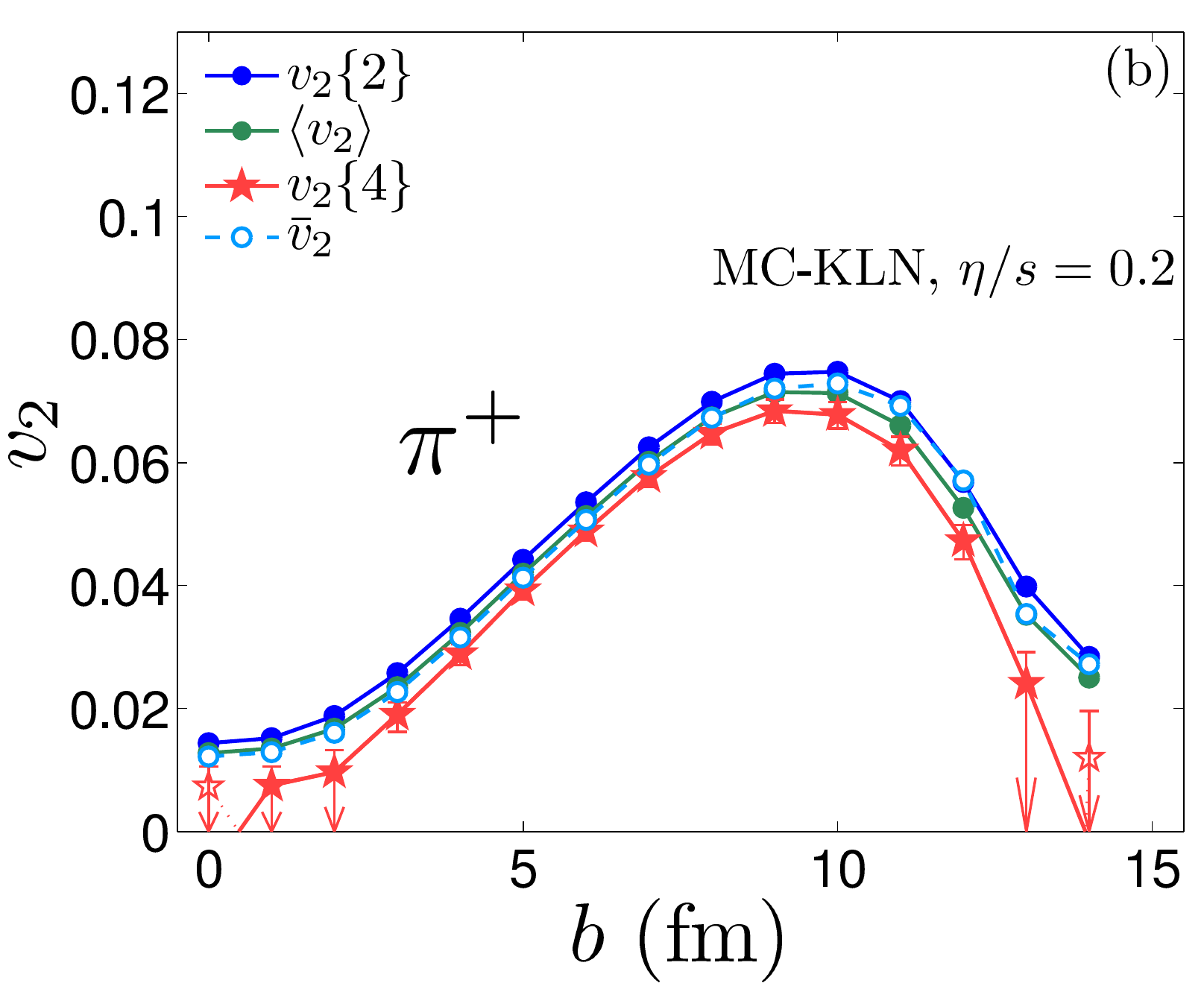}
 \includegraphics[width=0.45\linewidth]{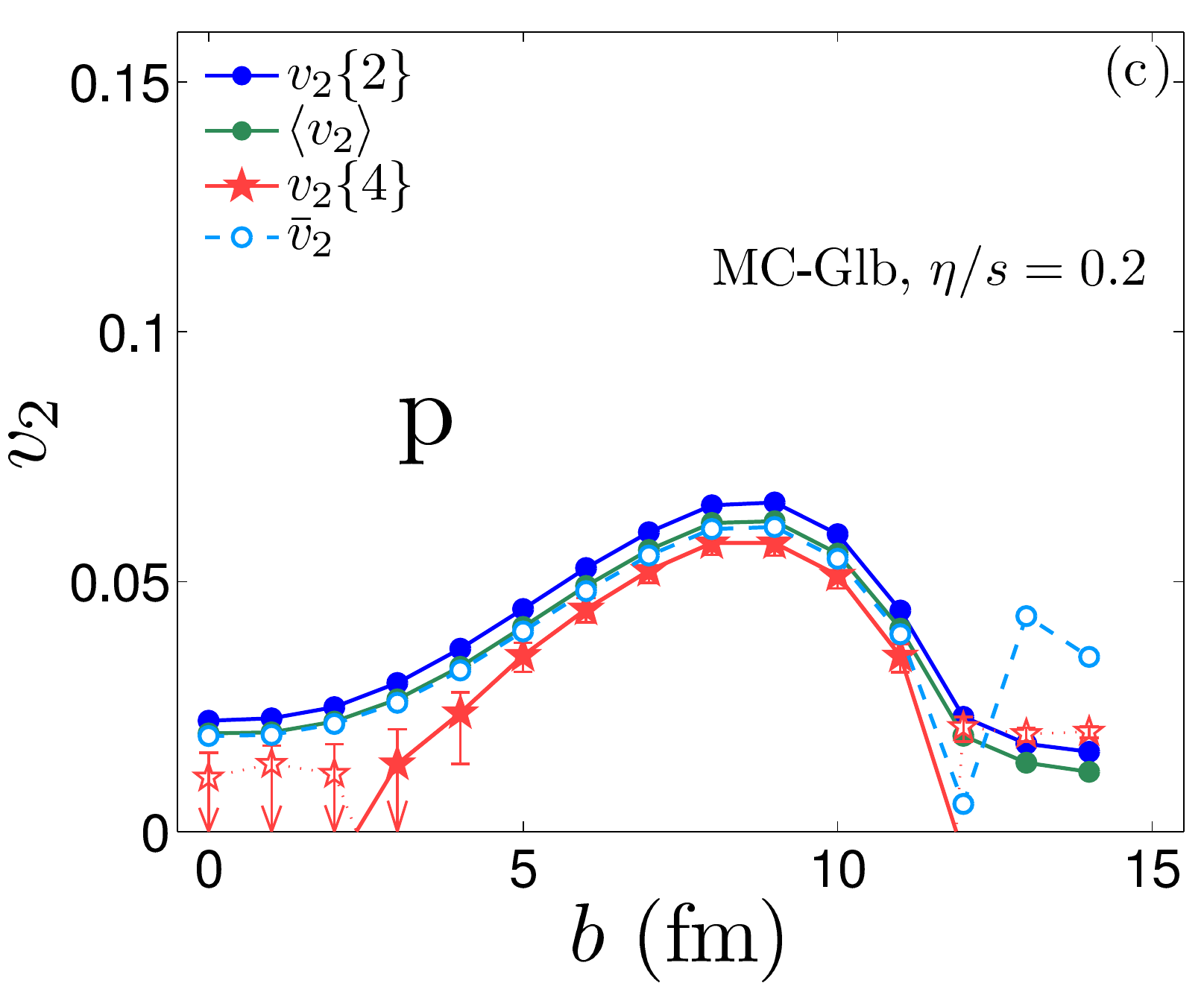}
 \includegraphics[width=0.45\linewidth]{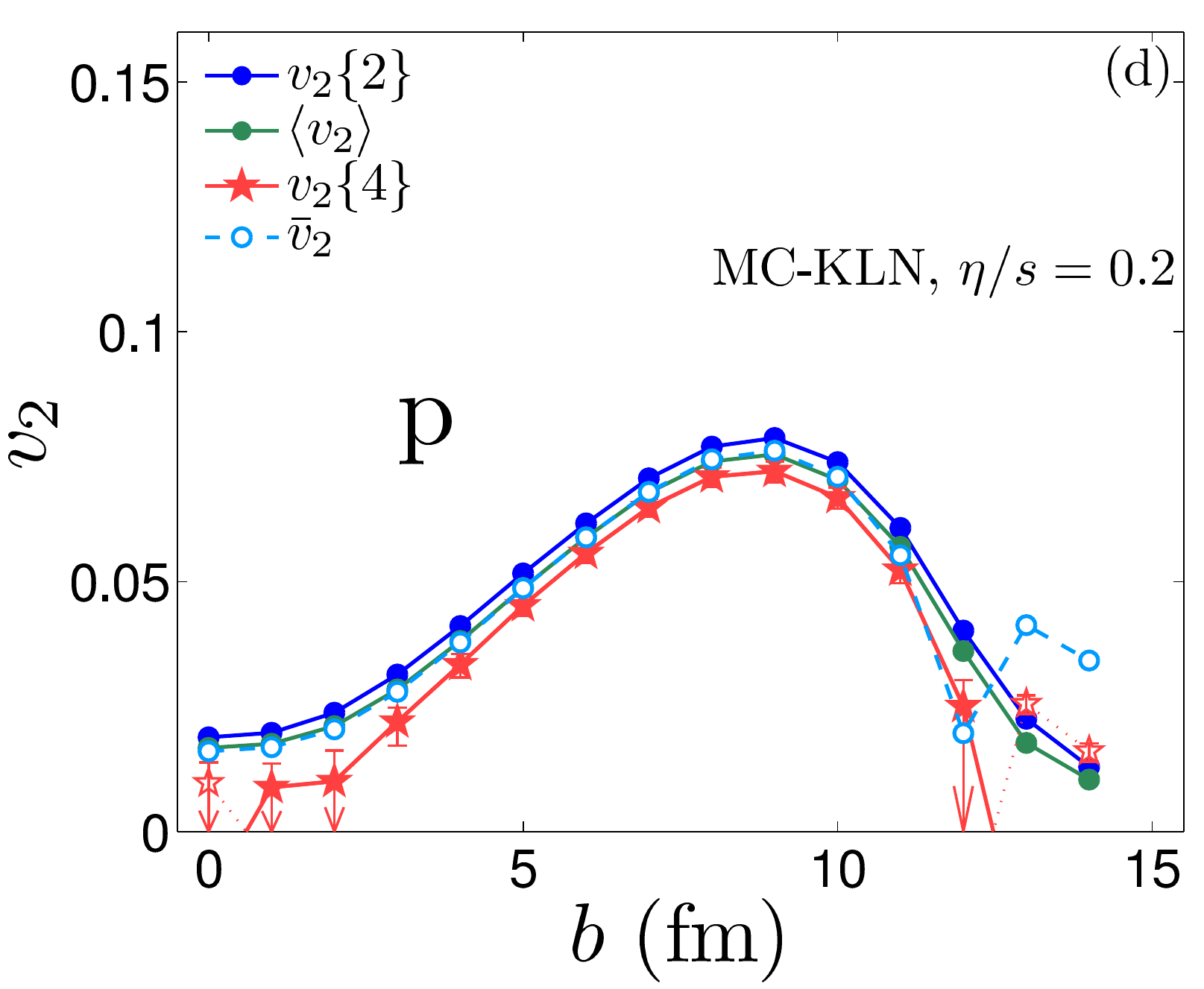}
 }
 \caption{Similar to Fig.~\ref{1104.0650:F14} but with $\eta/s=0.2$.
 \label{1104.0650:F14.1}
 }
\end{figure*}
%

Corresponding results from viscous hydrodynamic simulations with $\eta/s=0.2$ are shown in Fig.~\ref{1104.0650:F14.1}. We see that viscosity suppresses the elliptic flow, consistent with what we concluded in previous sections. We also see that the event-by-event fluctuation of $v_2$ is also suppressed, giving much smaller error bars compared to the ideal case as shown in Fig.~\ref{1104.0650:F14}. Again we attribute this effect to the fact that viscosity dissipates hot-spots and thus to some extent reduces the event-by-event fluctuations caused by the randomness of the initial nucleons positions.

%
\begin{figure}[t]
 \center{\includegraphics[width=0.7\linewidth]{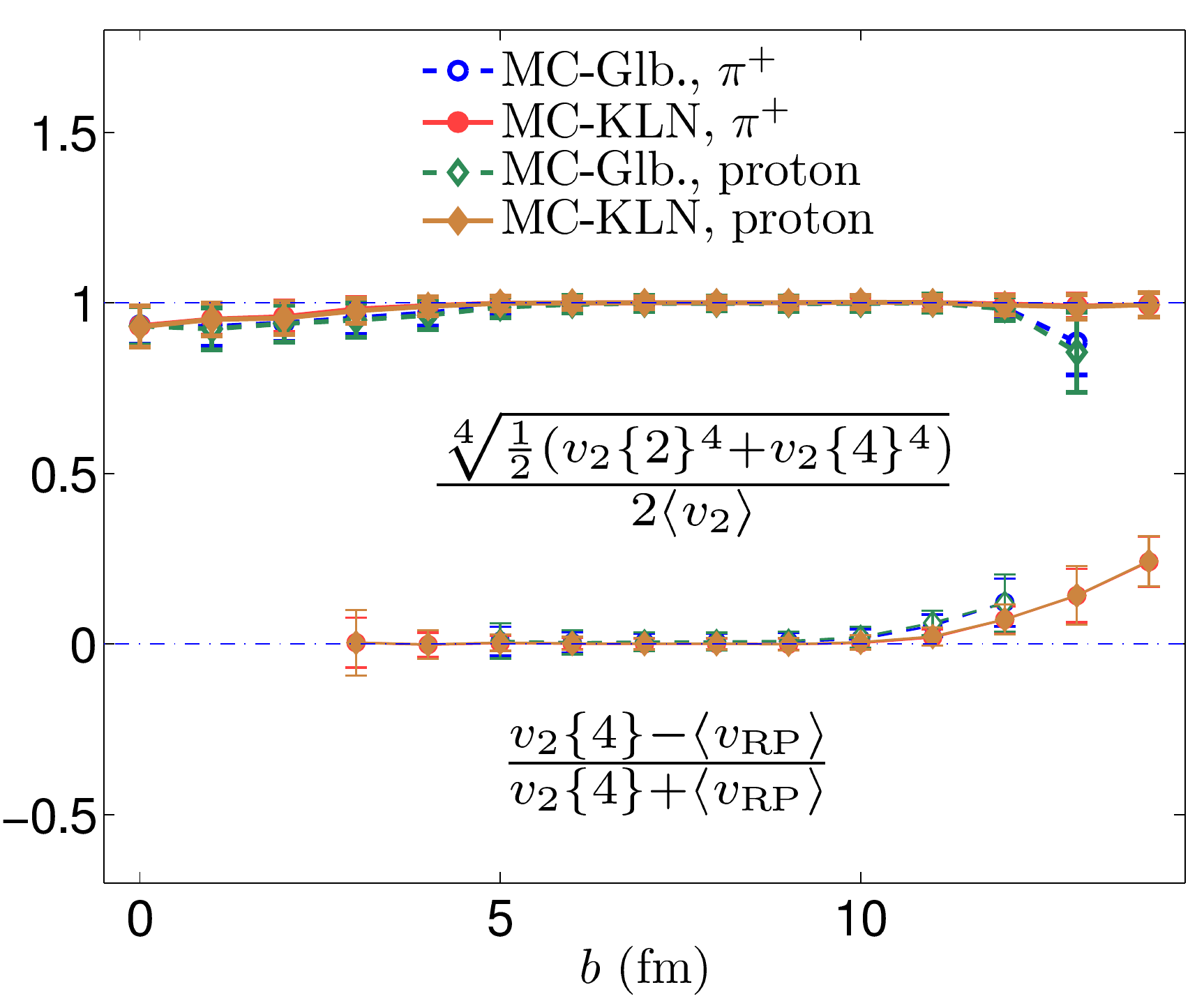}}
 \caption{Similar to Fig.~\ref{1104.0650:F3}, but for the
    elliptic flow $v_2$ of pions and protons using ideal fluid dynamics. See text for discussion. Taken from \cite{Qiu:2011iv}.
 \label{1104.0650:F15}
 }
\end{figure}
%

Similar to Eqs.~(\ref{eq:1104.0650:11})-(\ref{eq:1104.0650:11b}) we can test whether the 
$v_2$ fluctuations from event to event have Gaussian or Bessel-Gaussian
distributions. This is done in Fig.~\ref{1104.0650:F15}. The upper set of curves
(thick lines) tests the $v_2$-analogue of relation (\ref{eq:1104.0650:11b}), whereas
the lower set (thin lines) tests the validity of Eq.~(\ref{eq:1104.0650:11}). (In 
the lower set of curves, we dropped all $b$-values for which the error 
band for $v_2\{4\}^4$ extends to negative values.) Just as we saw for 
the initial ellipticities in Fig.~\ref{1104.0650:F3}, both the Gaussian and
Bessel-Gaussian hypotheses for $v_2$-fluctuations are seen to hold quite 
well in mid-central ($4{\,\lesssim\,}b{\,\lesssim\,}10$\,fm) collisions. The 
Bessel-Gaussian hypothesis breaks down in peripheral collisions
($b{\,>\,}10$\,fm). We expect that the hypothesis holds (as expected \cite{Voloshin:2007pc})
in central collisions although it is not clearly seen from the plot which is generated using only 1000 events per $b$-value. The assumption of Gaussian $v_2$-fluctuations
breaks down in central collisions ($b{\,<\,}5$\,fm), as expected.
For the MC-Glauber model, it also breaks down in very peripheral collisions,
whereas for MC-KLN initial conditions the final elliptic flow exhibits
a nice Gaussian distribution all the way to the largest impact parameters.

We checked that viscosity leads only to small changes in the ratios shown in Fig.~\ref{1104.0650:F15}, and we therefore refrain from plotting them for the viscous simulations.

Overall, a comparison of Figs.~\ref{1104.0650:F15} and \ref{1104.0650:F3} (as well as of
Figs.~\ref{1104.0650:F14} and \ref{1104.0650:F1}) shows that the statistical properties of 
$v_2$ fluctuations are qualitatively similar but quantitatively different 
from those of the initial ellipticity fluctuations. This is consistent 
with the fact that the main driver for elliptic flow is the initial
ellipticity, but that eccentricity coefficients of higher harmonic orders 
affect the evolution of $v_2$ weakly but measurably through nonlinear 
mode-coupling effects.

\vspace*{-3mm}
\section{Chapter summary}
\label{sec:1104.0650:6}
\vspace*{-2mm}
We summarize a few key results from this chapter:

\noindent
-- The average and mean ellipticities $\la \ecc_2\ra$ and $\bar{\ecc}_2$ 
agree with excellent accuracy over a wide range of impact parameters,
but diverge in very peripheral collisions ($\geq 60\%$ centrality) where
$\bar{\ecc}_2{\,>\,}\la \ecc_2\ra$ (both for participant-plane and 
reaction-plane averaged profiles). 

\noindent
-- The average energy and entropy density weighted eccentricities agree 
with excellent accuracy over a wide range of impact parameters, except 
for central collisions ($b{\,\lesssim\,}4$\,fm) where 
$\la\ecc_n(e)\ra{\,>\,}\la\ecc_n(s)\ra$.

\noindent
-- Whether the fluctuating entropy density distributions for individual 
events are first converted to energy density and then rotated by 
$\PsiPP{n}(e)$ and averaged, or first rotated by 
$\PsiPP{n}(s)$ and averaged and then converted to energy density
has very little influence on the shape of the resulting smooth average 
initial energy density profile for single-shot hydrodynamics. We prefer the conversion to energy density as 
the first step, since in event-by-event hydrodynamics the energy density
gradients of each event generate (through the EOS) the pressure gradients 
that drive the evolution of collective flow.

\noindent
-- The shortcut of using reaction-plane averaging to generate a smooth
profile for single-shot hydrodynamics with ellipticity approximately equal 
to $\ecc\{4\}$ of the ensemble, in the hope of generating with a single 
hydrodynamic run an elliptic flow $\bar{v}_2$ that can be directly 
compared with $v_2\{4\}$ measurements, works only in the $0{-}40\%$ 
centrality range. For peripheral collisions, this method cannot be trusted.   

\noindent
-- The assumption of Bessel-Gaussian fluctuations for initial source 
ellipticity and final elliptic flow work well for $b{\,\lesssim\,}10$\,fm 
but breaks down in more peripheral collisions. For more peripheral 
collisions, the hypothesis that $\ecc_2$ and $v_2$ are Gaussian distributed
works better than the Bessel-Gaussian assumption, but it breaks down
for $b{\,<\,}5$\,fm. For MC-Glauber initial conditions, directly emitted 
pions and protons feature negative values of $v_2\{4\}^4$ in very 
peripheral collisions. The fluctuations of initial source ellipticities 
and final elliptic flow values have qualitatively similar but
quantitatively different statistical properties.

\noindent
-- Except for rather central collisions, the eccentricities $\la\ecc_2\ra$, 
$\la\ecc_4\ra$ and $\la\ecc_5\ra$ from the MC-KLN model are all significantly
larger than those from the MC-Glauber model. In contrast, $\la\ecc_3\ra$
is numerically very similar for the two models over most of the impact 
parameter range. The viscous suppression of triangular flow $v_3$ thus 
allows for a determination of the QGP shear viscosity $(\eta/s)_\mathrm{QGP}$ 
that is free from the large model uncertainties that arise from the different 
MC-Glauber and MC-KLN ellipticities when using $v_2$ for such an extraction; or alternatively, an extraction of sheer viscosity using both $v_2$ and $v_3$ simultaneously can be used to study the correctness of initial condition models, see \chap{e2e3}.

\noindent
-- The second- and fourth-order eccentricities $\ecc_2$ and $\ecc_4$
are strongly correlated by collision geometry, and $v_4$ receives strong
contributions even from a purely elliptical deformation of the final flow 
velocity distribution. These complications make $v_4$ a poor candidate for
systematic studies of viscous effects on the evolution of collective
flow. Similar comments apply to $v_5$ since it couples via mode-coupling
to triangularity from fluctuations and to ellipticity from collision 
geometry. This mixture of contributions from conceptually different 
origins complicates a systematic analysis. In general, flow coefficients
$v_n$ of high harmonic order ($n{\,>\,}3$) show poor correlation with the
eccentricity coefficients $\ecc_n$ of the same harmonic order, except
for very central collisions where all eccentricities are driven by
fluctuations alone (and not by overlap geometry).   
 
\noindent
-- In spite of nonlinear mode-coupling effects, the basic response of
elliptic flow $v_2$ to ellipticity $\ecc_2$, and of triangular flow $v_3$
to triangularity $\ecc_3$, is approximately linear. These two observables 
thus remain prime candidates for systematic studies of viscous effects 
on collective hydrodynamic flow.

\noindent
-- Event-by-event hydrodynamics generates harder $p_T$-spectra for
the emitted hadrons than single-shot hydrodynamic evolution of the
corresponding averaged initial profile. This is due to additional
radial flow generated by large pressure gradients arising from ``hot
spots'' in the initial fluctuating density distribution. The hardening
effect is particularly strong in peripheral collisions which produce
small fireballs that fluctuate strongly; it is reduced by shear viscosity.

\noindent
-- Event-by-event hydrodynamic evolution of fluctuating initial conditions
leads to smaller average elliptic flow than obtained by evolving the
corresponding averaged initial condition in a single shot. This 
suppression depends somewhat on collision centrality, and for ideal fluids 
it is generically of order 4-5\% for pions and 8-10\% for protons. The 
effect is sufficiently large to possibly lead to a significant over-estimate of 
the fluid's specific shear viscosity if one extracts it from elliptic 
flow measurements by comparing with single-shot hydrodynamic simulations. 
The discrepancy between event-by-event and 
single-shot hydrodynamics decreases, however, in viscous fluid dynamics. More related studies are reported in \chap{e2e3}.


\chapter{Using simultaneously measured elliptic and triangular flow to resolve initial condition ambiguities}
\label{chap:e2e3}

This chapter focuses on a simultaneous comparison of both elliptic and triangular flow from viscous fluid dynamics with measurements in Pb+Pb collisions at the LHC. Using initial density distributions from the MC-Glauber and MC-KLN models, we show that the data favor a small specific shear viscosity $(\eta/s)_\mathrm{QGP}{\,\sim\,}1/(4\pi)$ for the quark-gluon plasma. Using this viscosity value, the relative magnitude of the elliptic and triangular flow is well described with MC-Glauber initial conditions while the MC-KLN initial conditions require twice as large viscosity to reproduce the elliptic flow and then under-predict triangular flow by about 30\%. We show that compatibility of the experimental data with larger values for the specific shear viscosity of the QGP would require initial-state models whose density distributions fluctuate more strongly, yielding significantly larger triangular deformations on average than those obtained from both the MC-Glauber and MC-KLN models.
The material in this chapter is based on \cite{Qiu:2011hf} and \cite{Qiu:2011fi}.

\section{Introduction} \label{sec:1110.3033:introduction}
Much attention has been given to the extraction of the specific shear viscosity ($\eta/s$) of the quark-gluon plasma from elliptic flow data in relativistic heavy-ion collisions \cite{Teaney:2003kp,Lacey:2006pn,Romatschke:2007mq,Luzum:2008cw,Luzum:2009sb,Song:2010mg,Aamodt:2010pa,Luzum:2010ag,Lacey:2010ej,Bozek:2010wt,Hirano:2010jg,Schenke:2010rr,Schenke:2011tv,Song:2011qa,Shen:2010uy}.

A major road block in this effort is insufficient knowledge of the initial shape of the thermalized fireball created in these collisions, whose initial ellipticity is uncertain by about 20\% \cite{Hirano:2005xf,Drescher:2006pi,Hirano:2009ah,Heinz:2009cv,Qiu:2011iv} (Fig.~\ref{1104.0650:F5}). As shown in \chap{1104.0650}, this induces an ${\cal O}(100\%)$ uncertainty in the value of $(\eta/s)_\mathrm{QGP}$ extracted from elliptic flow \cite{Luzum:2008cw,Song:2010mg}. After the discovery of triangular flow in heavy ion collisions at RHIC \cite{Alver:2010gr,Adare:2011tg,Sorensen:2011fb} and LHC energies \cite{ALICE:2011vk,CMS:2011era,Steinberg:2011dj}, followed by the confirmation of its collective hydrodynamic nature \cite{Alver:2010gr,Alver:2010dn,Petersen:2010cw,Qin:2010pf,Luzum:2010sp,Xu:2011fe,Luzum:2011mm} and the realization that shear viscosity suppresses higher-order harmonic flow  coefficients more strongly than elliptic flow \cite{Alver:2010dn,Schenke:2010rr,Schenke:2011tv,Chaudhuri:2011qm,Schenke:2011bn}, it was suggested \cite{Lacey:2010hw,Adare:2011tg,ALICE:2011vk,Shen:2011zc,Qiu:2011fi} that a combined analysis of the elliptic and triangular flow coefficients $v_2$ and $v_3$ could yield a more precise value for the QGP shear viscosity and thereby reduce or eliminate the model uncertainty in the initial deformation of the QGP fireball and its event-by-event fluctuations. This chapter focuses on such an analysis, using Pb+Pb collision data collected by the ALICE collaboration at the LHC \cite{Aamodt:2010jd}.

\section{Justification for using single-shot simulations} \label{sec:1110.3033:addon}

Event-by-event viscous hydrodynamic simulations with full inclusion of unstable 
resonance decays are at present numerically too costly for systematic flow studies
over a range of viscosities, collision energies, centralities, and collision systems.
To study the possibility of replacing event-by-event simulations by less costly single-shot ones, we draw on our intuition based on previous hydrodynamic simulations for Au+Au collisions at $\sqrt{s}=200\,A$\,GeV (\chap{1104.0650}). The initial conditions under comparison are from the MC-Glauber and the MC-KLN models (see \sec{hydrodynamics_simulations}).


We have shown in \chap{1104.0650} that, for ideal hydrodynamic simulations, the eccentricity-scaled elliptic and triangular flows $v_{2,3}/\ve_{2,3}$ calculated from single-shot and event-by-event hydrodynamics show $5-10\%$ differences (Figs.~\ref{1104.0650:F11},~\ref{1104.0650:F12}), for both light (thermal pions) and heavy (protons) particles; the differences for viscous hydrodynamic simulations with $\eta/s=0.2$ are even smaller (Fig.~\ref{1104.0650:F12.1}). This suggests that for viscous hydrodynamic simulations with sufficiently large viscosity
$\eta/s$, single-shot hydrodynamics can substitute well for event-by-event evolution 
for the purpose of calculating $v_2/\ve_2$ and $v_3/\ve_3$. For ideal hydrodynamic simulations, there is a $\sim 7\%$ inaccuracy caused by this substitution, however this is already much smaller than the $\sim 20\%$ uncertainly in ellipticity induced by the ambiguity between the MC-Glauber and MC-KLN models. Therefore a (relatively) quick study using single-shot simulations with both elliptic and triangular flows becomes a meaningful proposition for gaining new information beyond the previous studies based on $v_2/\ve_2$ alone.

\begin{figure*}[h!]
  \center{
  	\includegraphics[width=0.45\textwidth]{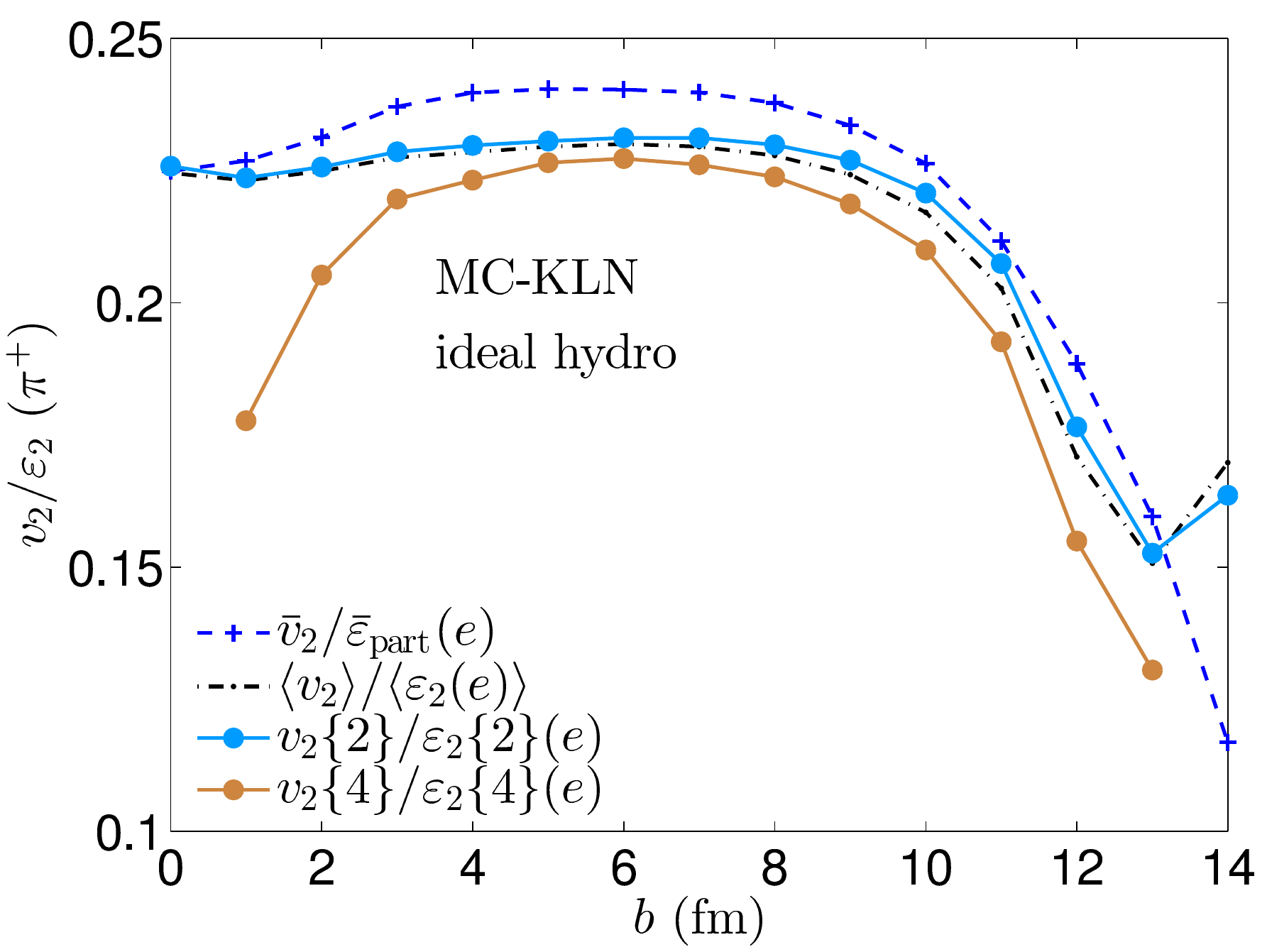}
  	\includegraphics[width=0.45\textwidth]{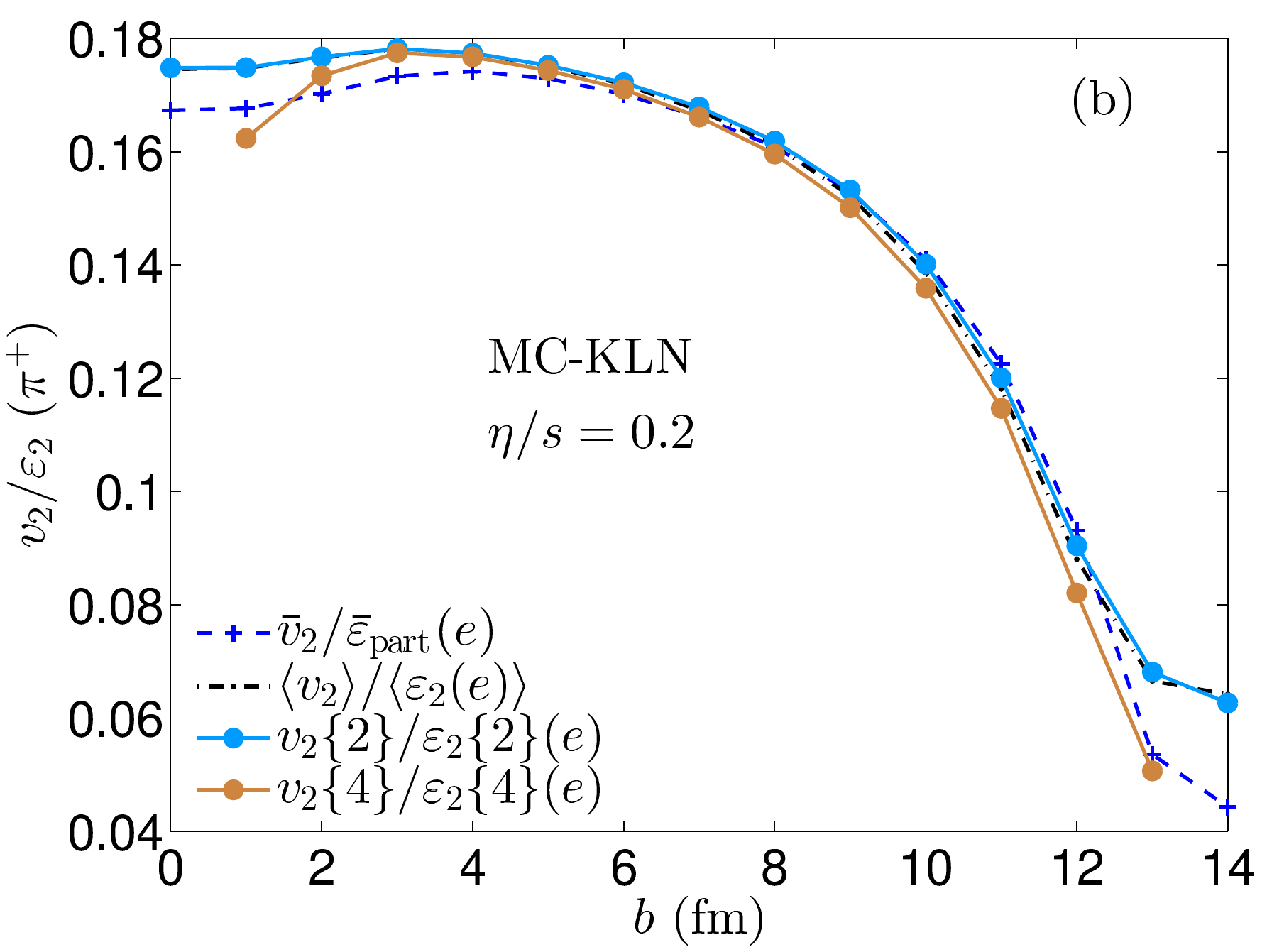}\\
  	\includegraphics[width=0.45\textwidth]{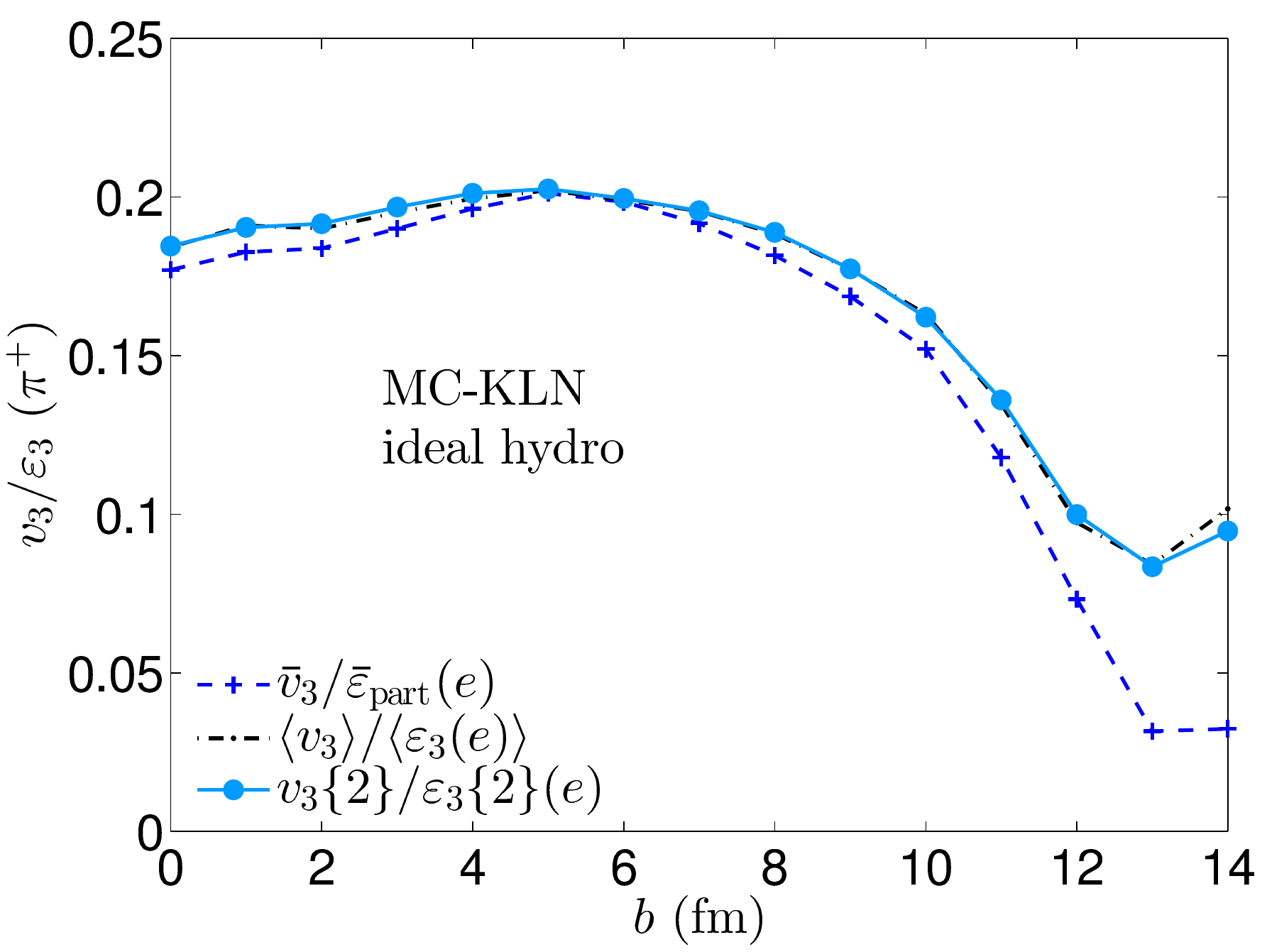}
  	\includegraphics[width=0.45\textwidth]{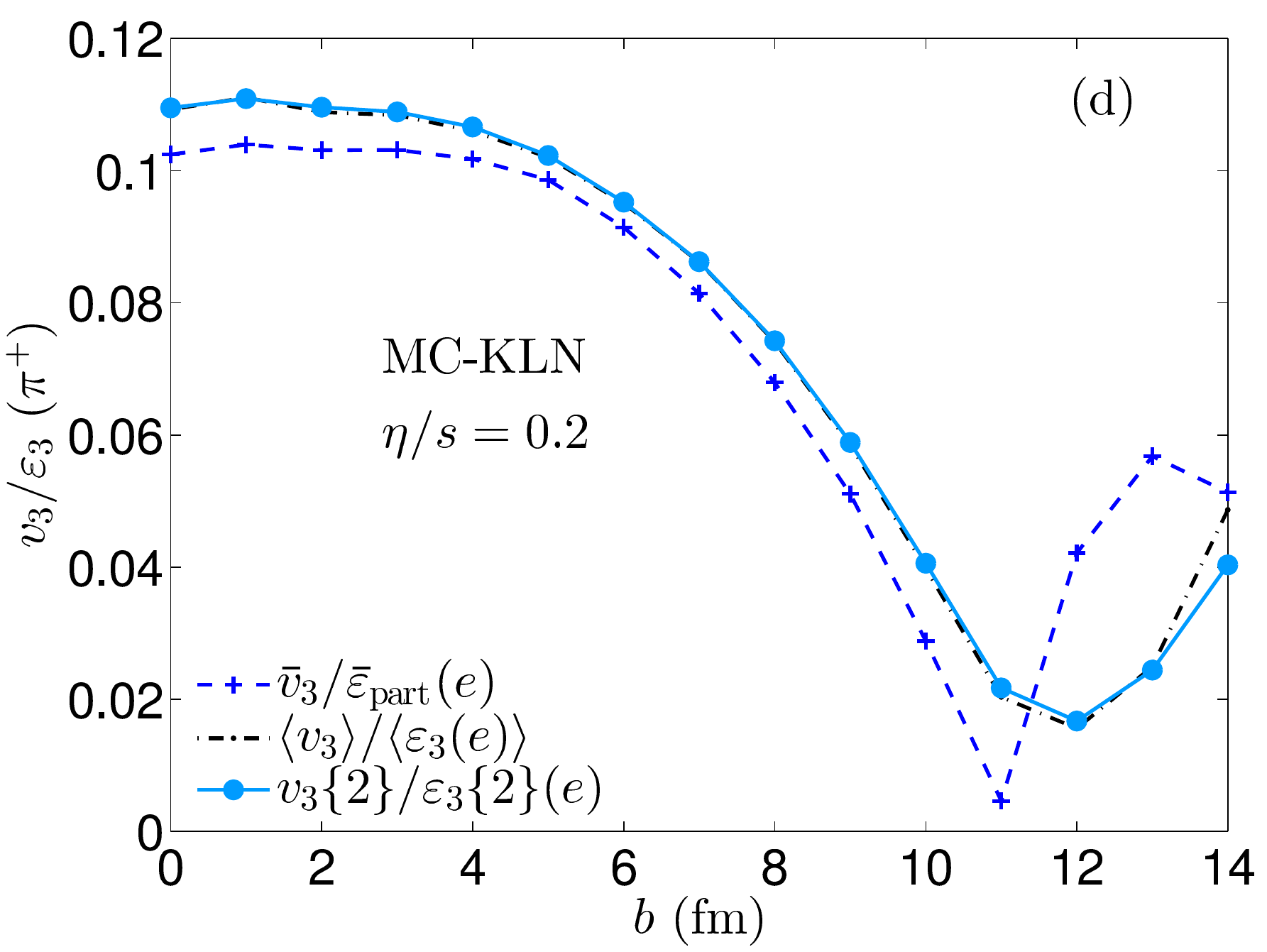}
  }
  \caption{$\la v_n \ra / \ve_n$,  $\vt{n} / \et{n}$,  and $\vf{n} / \ef{n}$ for $n=2$ (ab) and $n=3$ (cd), using thermal pions from ideal hydrodynamics (ac) and viscous ($\eta/s=0.2$) hydrodynamics (bd), from the MC-KLN model calculations for Au+Au collisions at the RHIC energy. The blue dash curves represent results from single-shot calculations and the rest are from event-by-event calculations.
  \label{fig:ebe2:15}}
\end{figure*}

However there is a subtlety: the experimental flow data are mostly determined with two-particle and four-particle correlations methods which give $\vt{2,3}$ and $\vf{2,3}$ instead of $\la v_2 \ra$. Following \cite{Bhalerao:2006tp} we try to normalize the flow by the correspondingly calculated eccentricity, that is, $\la v_n \ra$ by $\la \ve_n \ra$, $\vt{n}$ by $\et{n}$, $\vf{n}$ by $\ef{n}$, and the results are shown in \fig{ebe2:15}.

\fig{ebe2:15}(ab) shows that, for the elliptic flow, the process of normalizing the flow by the correspondingly calculated eccentricity gives very similar scaled flow in the ideal fluid case (\fig{ebe2:15}(a)), and almost completely equal scaled flow in the viscous case (\fig{ebe2:15}(b)). There are $5\%\sim10\%$ differences between $\bar{v}_2/\bar{\ve}_2$ and $\vt2/\et2$ or $\vf2/\ef2$ when small $\eta/s$ is used (\fig{ebe2:15}(a)); with $\eta/s=0.2$, this difference has almost vanished (\fig{ebe2:15}(b)).

The scaled triangular flows are shown in \fig{ebe2:15}(cd). It is clear that $\vt3/\et3$ is very close to $\la v_3 \ra/\la \ve_3 \ra$ and $\bar{v}_3/\bar{\ve}_3$, which are almost identical in both the ideal (\fig{ebe2:15}(c)) and viscous (\fig{ebe2:15}(d)) cases. For triangular flow, we do not show $\vf3/\ef3$ since the values for both numerator and denominator are small and plagued by large statistical errors.

We take the results of this study as justification for applying a similar single-shot approach to the LHC data in the remaining of this chapter.

\section{Setup for the simulations}

We use the (2+1)-dimensional viscous hydrodynamic simulation code VISH2+1, with longitudinal boost-invariance, describing numerically the transverse evolution of the heavy-ion collision fireball near midrapidity. As in past work \cite{Luzum:2008cw,Hirano:2009ah,Luzum:2009sb,Heinz:2009cv,Qiu:2011iv,Song:2010mg}, 
we use the MC-Glauber and the MC-KLN models.

\begin{figure*}[h!]
  \center{
  	\includegraphics[width=0.45\linewidth]{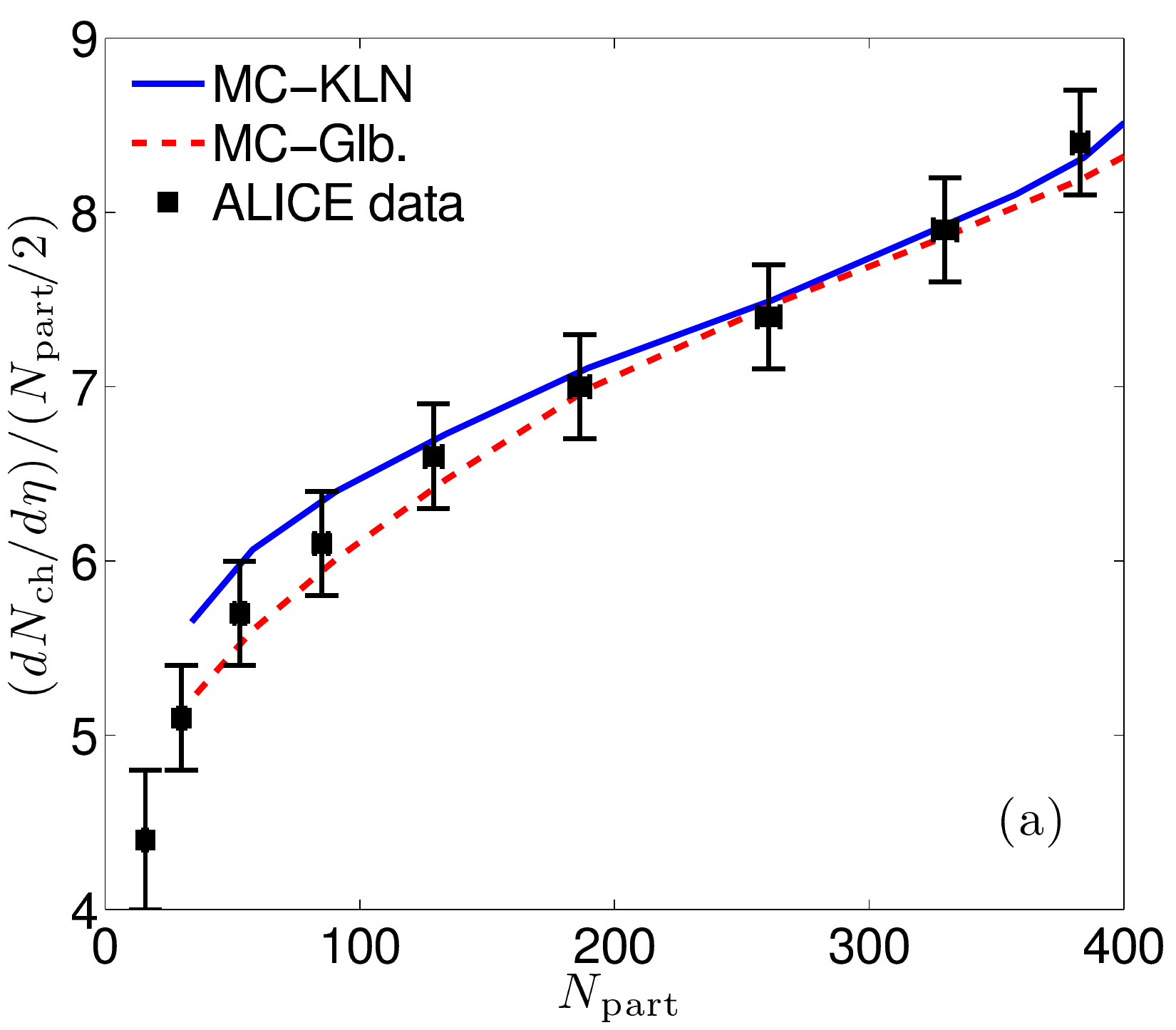}
  	\includegraphics[width=0.45\linewidth]{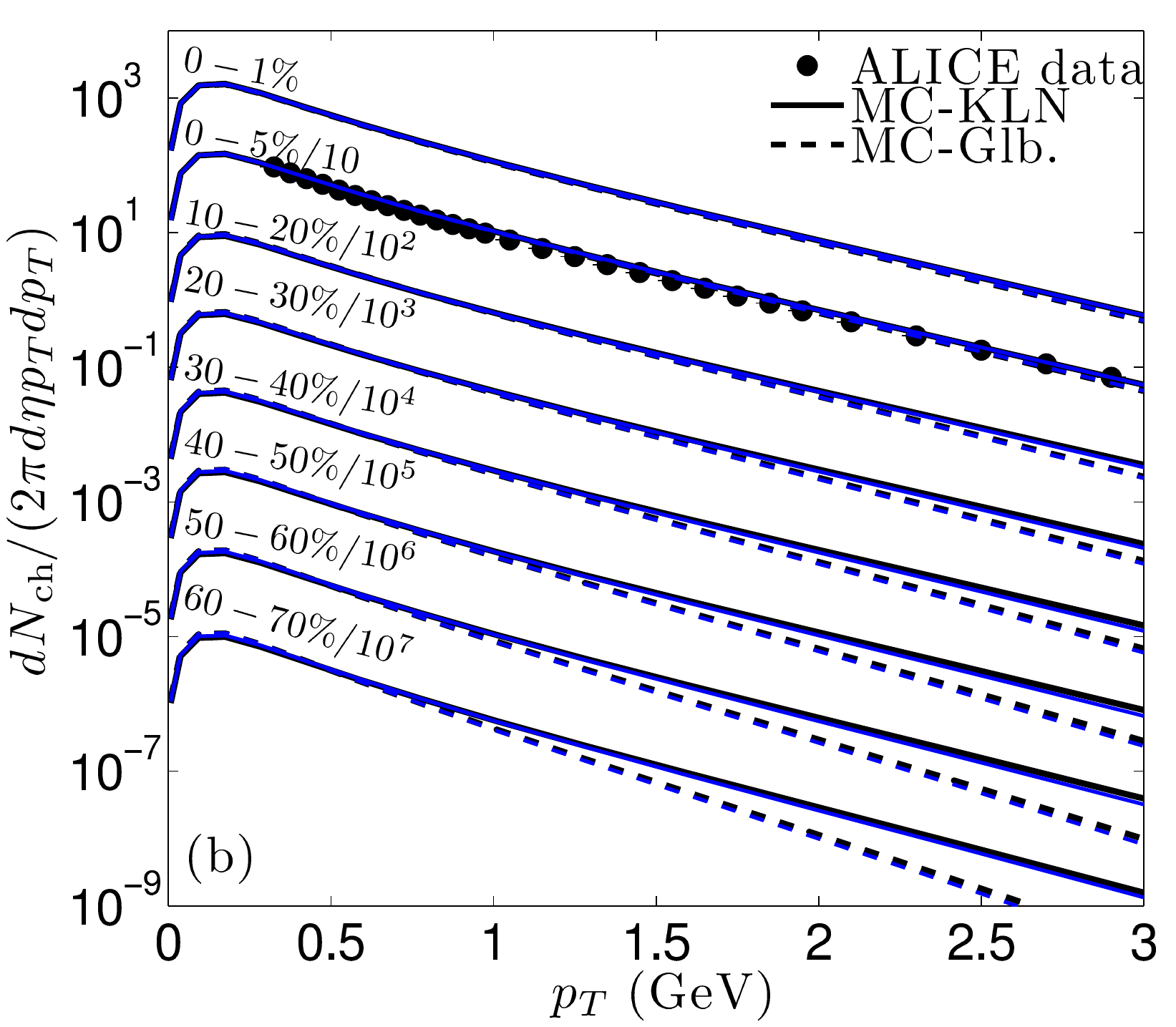}
  }
  \caption{(a) (Color online) Centrality dependence of charged particle multiplicity density 
    as a function of $\Np$ from the MC-Glauber (dashed) and MC-KLN (solid) models, 
    compared with ALICE measurements \cite{Aamodt:2010cz} for 2.76\,$A$\,TeV Pb-Pb 
    collisions. 
    (b) Charged particle $p_T$-spectra from the MC-Glauber and MC-KLN models for 
    different centralities. The most central ($0{-}5\%$) results are compared with 
    ALICE data \cite{Aamodt:2010jd}. From \cite{Qiu:2011hf}.
  \label{fig:1110.3033:1}}
\end{figure*}

The MC-KLN calculations were done using a Monte-Carlo sample of initial state profiles 
with identical properties as those used in \cite{Shen:2011eg} for the calculation
of transverse momentum spectra and elliptic flow in 2.76\,$A$\,TeV Pb-Pb collisions at the LHC. To compute the nuclear thickness function $T_A(\bm{r}_\perp;b)$, we use an inelastic nucleon-nucleon cross section $\sigma_{NN}=62$ mb at LHC energies. For the $x$ dependence of the gluon structure function in the MC-KLN model, we
used the power $\lambda\eq0.28$ \cite{Hirano:2009ah}; the normalization factor for the
initial entropy density was fixed by hand to reproduce the measured charged hadron
multiplicity density $\dNdeta$ for the 5\% most central collisions \cite{Aamodt:2010pb};
the measured dependence of $\dNdeta$ on collision centrality \cite{Aamodt:2010cz}
is then automatically reproduced reasonably well by the model \cite{Shen:2011eg} (see 
Fig.~\ref{fig:1110.3033:1}(a)). MC-KLN runs were done with $\eta/s\eq0.2$ which, for this type of 
initial conditions, was shown to yield a good overall description of the measured 
transverse momentum spectra and elliptic flow in 200\,$A$\,GeV Au-Au collisions 
at RHIC \cite{Shen:2011eg} and gave an impressively accurate prediction for the 
unidentified and identified charged-hadron spectra and elliptic flows in 2.76\,$A$\,TeV 
Pb-Pb collisions at the LHC \cite{Shen:2011eg,Heinz:2011kt}. 

For the MC-Glauber runs, we generated a new set of initial configurations that differ
from those used for 200\,$A$\,GeV Au-Au collisions in \cite{Song:2010mg} by 
the wounded-nucleon-to-binary-collision ratio. Taking the initial entropy density
\begin{equation} \label{eq:Glauber}
s(\bm{r}_\perp;b)\eq\kappa\left(\frac{1{-}x}{2}n_{_\mathrm{WN}}(\bm{r}_\perp;b) + x  
n_{_\mathrm{BC}}(\bm{r}_\perp;b)\right),
\end{equation}
we determine $\kappa$ and $x$ by a two-parameter
fit to the ALICE data \cite{Aamodt:2010cz} shown in Fig.~\ref{fig:1110.3033:1}(a). Due to viscous
entropy production during the hydrodynamic evolution, which itself depends on collision 
centrality, the fitted value for $x$ depends on the assumed shear viscosity. For MC-Glauber 
initial conditions, we took $\eta/s\eq0.08$ since this value was shown in 
\cite{Schenke:2011tv, Schenke:2011bn,ALICE:2011vk} to provide a reasonable 
description of the charged hadron $v_2(p_T)$ and $v_3(p_T)$ data measured by the 
ALICE experiment; this results in $x\eq0.118$ for Pb-Pb collisions at the LHC. Both the 
MC-Glauber and MC-KLN initial conditions are hydrodynamically evolved with equation 
of state (EOS) s95p-PCE \cite{Shen:2010uy}, which matches numerical results from lattice 
QCD at high temperatures to a hadron resonance gas at low temperatures 
\cite{Huovinen:2009yb} and implements chemical freeze-out at $\Tchem\eq165$\,MeV. 
The hydrodynamic output  is converted to final hadron distributions along an isothermal 
decoupling surface of  temperature $\Tdec\eq120$\,MeV, using the Cooper-Frye prescription.

In \chap{1104.0650} \cite{Qiu:2011iv}, we showed that, due to similar fluctuation mechanisms, the
MC-KLN and MC-Glauber models generate similar third-order eccentricities 
$\ve_3$, whereas the ellipticity $\ve_2$, which is mostly controlled by collision geometry,
is about 20\% larger in the MC-KLN model. Event-by-event ideal
and viscous hydrodynamic simulations with both realistically-fluctuating \cite{Qiu:2011iv} (Recall Fig.~\ref{1104.0650:F11},~\ref{1104.0650:F12},~\ref{1104.0650:F12.1}) and doubly-deformed Gaussian initial
conditions \cite{Qiu:2011fi} (with simultaneously nonzero $\ve_2$ and $\ve_3$ 
eccentricities) have shown that the hydrodynamic conversion efficiencies for translating 
initial spatial eccentricities into final flow anisotropies  \cite{Qin:2010pf,Teaney:2010vd,Shuryak:2009cy}, although different for $v_2/ \ve_2$ and $v_3/\ve_3$,
are very similar in the MC-KLN and MC-Glauber models. The similarities in $\ve_3$ and
differences in $\ve_2$ between these models should thus straightforwardly reflect 
themselves in analogous differences in $v_2$ and $v_3$ \cite{Shen:2011zc,Qiu:2011fi}, 
allowing for an experimental distinction between the models.

For all the reasons discussed in \sec{1110.3033:addon}, we use the single-shot approach. For a meaningful comparison with experiment, we include the full cascade of resonance decays in the final state. We compare with recent data from Pb-Pb collisions at the LHC, and we use the properly normalized ratios $v_n/\ve_n$ for our comparison. The results are shown in next few sections. Our approach here differs from that in \cite{Alver:2010dn} by replacing 
the singly-deformed Gaussian parametrization of the initial density used there by
the ensemble average of realistically-fluctuating, non-Gaussian initial profiles and 
from \cite{Schenke:2010rr,Schenke:2011tv,Schenke:2011bn}
by employing a more realistic EOS that accounts for the important effects of chemical
non-equilibrium hadronic evolution on the elliptic flow $v_2$ \cite{Hirano:2002ds}.
In \cite{ALICE:2011vk} it was shown that, with the approach used in \cite{Alver:2010dn},
MC-KLN initial conditions with $\eta/s\eq0.16$ cannot describe the $p_T$-integrated
$v_3$ measured in 2.76\,$A$\,TeV Pb-Pb collisions, whereas the MC-Glauber based
event-by-event calculations (with $\eta/s\eq0.08$) of Schenke {\it et al.}
\cite{Schenke:2011tv} appear to describe $v_3(p_T)$ at selected centralities reasonably 
well.


\section{Transverse momentum spectra} \label{sec:1110.3033:tranverse}
Fig.~\ref{fig:1110.3033:1}(b) shows the charged hadron $p_T$-spectra for 2.76\,$A$\,TeV Pb-Pb 
collisions at different centralities, for both MC-Glauber ($\eta/s\eq0.08$) and MC-KLN 
($\eta/s\eq0.2$) initial conditions. For the most central ($0{-}5\%$) collisions, the spectra
from both models agree well with published ALICE data. In more peripheral collisions,
the MC-KLN spectra are harder than those from MC-Glauber initial conditions. This is
a consequence of larger radial flow caused by larger transverse viscous pressure
gradients in the MC-KLN case, where the fluid is taken to have 2.5 times larger shear
viscosity than for the MC-Glauber simulations in order to obtain the same elliptic flow
\cite{Luzum:2008cw,Song:2010mg}. In peripheral collisions, these viscous effects
are stronger than in more central collisions where the fireball is
larger \cite{Song:2008si}. As shown in \cite{Holopainen:2010gz,Qiu:2011iv} and \chap{1104.0650}, event-by-event evolution of fluctuating initial conditions generates, for small values 
of $\eta/s$, flatter hadron spectra than single-shot hydrodynamics, especially in 
peripheral collisions, due to stronger radial flow driven by hot spots in the fluctuating 
initial states. Proper event-by-event evolution of the latter is therefore expected 
to reduce the difference between the MC-Glauber and MC-KLN curves, as explained in \sec{1110.3033:introduction}, since this effect is relatively strong for $\eta/s\eq0.08$ (MC-Glauber) \cite{Qiu:2011iv} but almost absent for $\eta/s\eq0.2$ (MC-KLN).

\begin{figure*}[h!]
  \center{\includegraphics[width=0.9\linewidth]{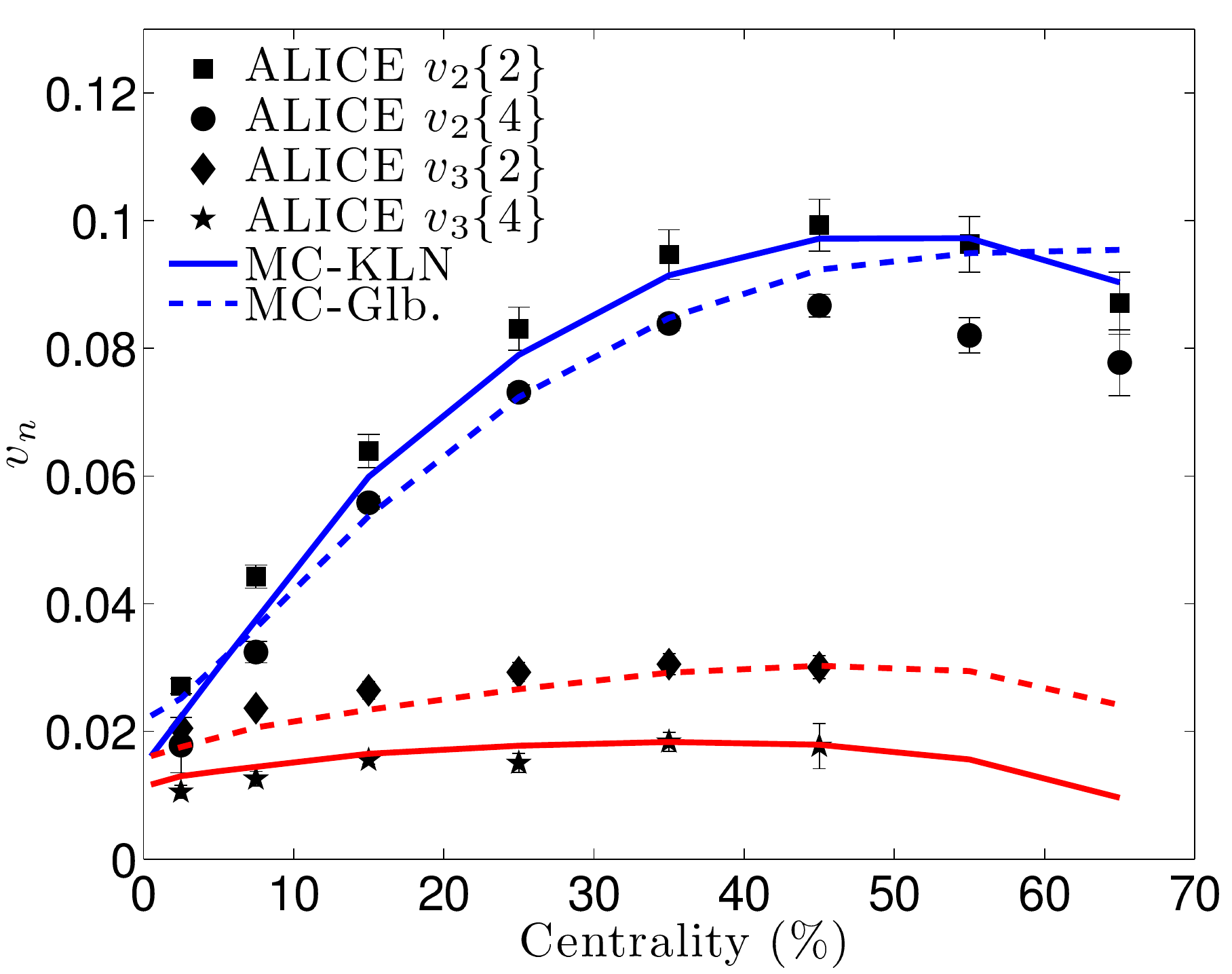}}
  \caption{Charged hadron $\bar{v}_2$ and $\bar{v}_3$ vs.\ centrality, compared with ALICE $\vt{2}$, $\vf{2}$, 
         $\vt{3}$, and $\vf{3}$ data for 2.76\,$A$\,TeV Pb+Pb \cite{ALICE:2011vk}. From \cite{Qiu:2011hf}.
  \label{fig:1110.3033:2}}
\end{figure*}


\section{$p_T$-integrated elliptic and triangular flow} \label{sec:1110.3033:pTIntegrated}

In Fig.~\ref{fig:1110.3033:2}, we compare our $p_T$-integrated $\bar{v}_2$ and $\bar{v}_3$ as functions 
of centrality with ALICE $\vt{2}$, $\vf{2}$, $\vt{3}$, and $\vf{3}$ data, extracted from 
2- and 4-particle correlations \cite{ALICE:2011vk} (also see \chap{1104.0650}). For both models, $\bar{v}_{2,3}$ from the averaged smooth initial conditions lie between the experimental $\vt{{2,3}}$ and $\vf{{2,3}}$
values. This is consistent with the theoretical expectation \cite{Miller:2003kd,Voloshin:2007pc}
that $\vt{n}$ ($\vf{n}$) is shifted up (down) relative to the average flow by event-by-event flow fluctuations and was also found elsewhere \cite{Song:2010mg,Schenke:2011tv,Luzum:2010ag}. Upon closer inspection, however, and recalling that ideal single-shot hydrodynamics with smooth initial condition was shown in \cite{Qiu:2011iv} and in \chap{1104.0650} to generate $\bar{v}_2$ 
similar to $\vt{2}$ from the corresponding event-by-event evolution, it seems that the 
MC-KLN is favored since it produces $\bar{v}_2$ results closer to the $\vt{2}$ data. Unfortunately,
a similar argument using $\bar{v}_3$ can be held against the MC-KLN model. To eliminate the 
interpretation difficulties associated with a comparison of average flows from single-shot 
evolution of averaged initial conditions with data affected irreducibly by naturally existing 
event-by-event fluctuations, we proceed to a comparison of eccentricity-scaled flow 
coefficients.

Assuming linear response of $v_{2,3}$ to their respective eccentricities $\ve_{2,3}$
(which was found to hold in \chap{1104.0650} and in \cite{Qiu:2011iv} with reasonable accuracy for $v_2$ and $v_3$, but not for
higher order anisotropic flows) we follow \cite{Bhalerao:2006tp}
and scale the flow $\bar{v}_{2,3}$ from single-shot hydrodynamics by the eccentricity
$\bar{\ve}_{2,3}$ of the ensemble-averaged smooth initial energy density,
while scaling the experimental  $\vt{{2,3}}$ and $\vf{{2,3}}$ data by the corresponding 
fluctuating eccentricity measures $\et{{2,3}}$ and $\ef{{2,3}}$, respectively, calculated 
from the corresponding models. As shown in \sec{1110.3033:addon}, this procedure is justified 
for $\vt{{2,3}}$ and $\vf{2}$; we do not perform comparisons using $\vf{3}/\ef{3}$ because of the lack of statistical confidence in our theoretical calculations.

\begin{figure*}[h!]
  \center{
  	\includegraphics[width=0.45\linewidth]{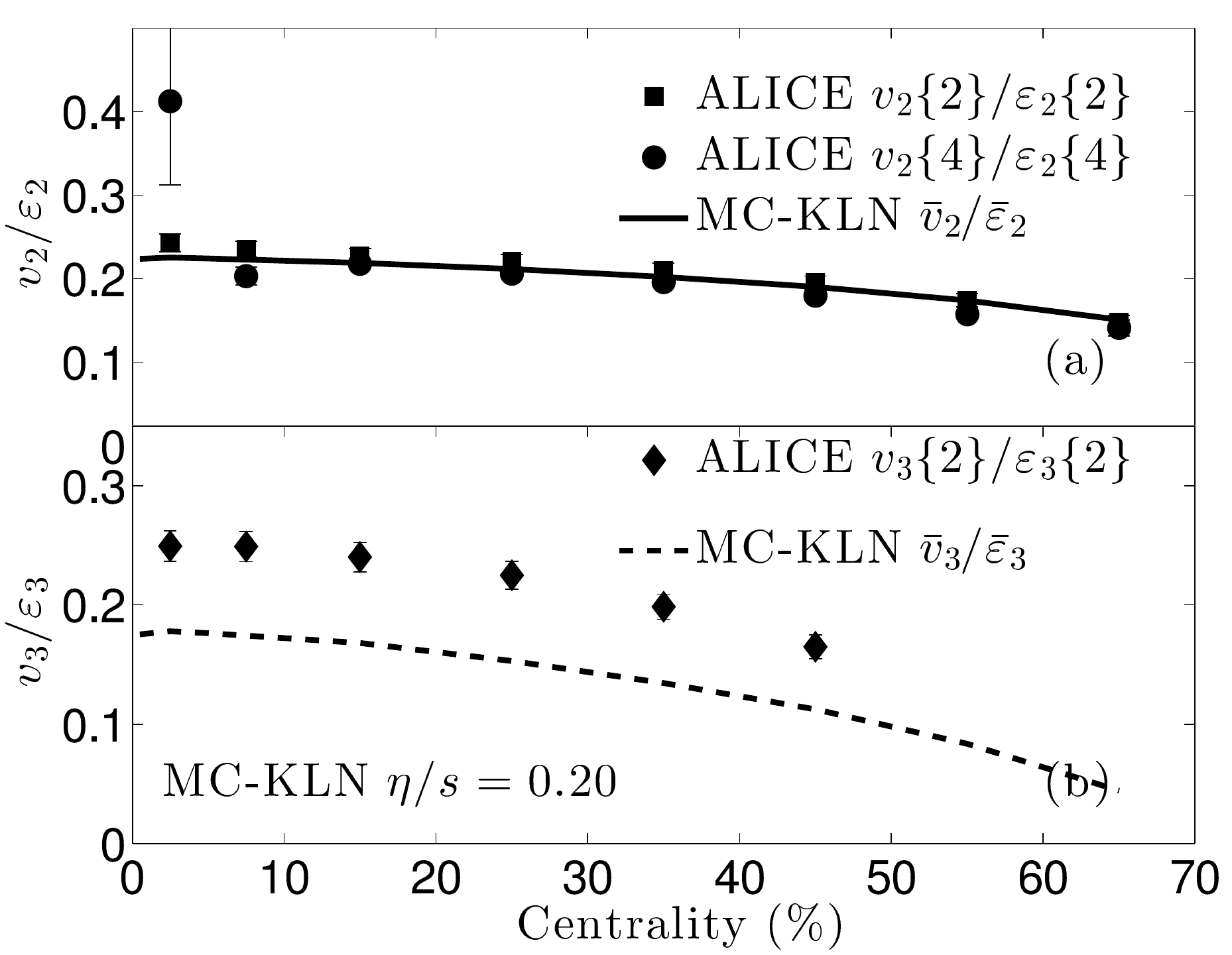}
  	\includegraphics[width=0.45\linewidth]{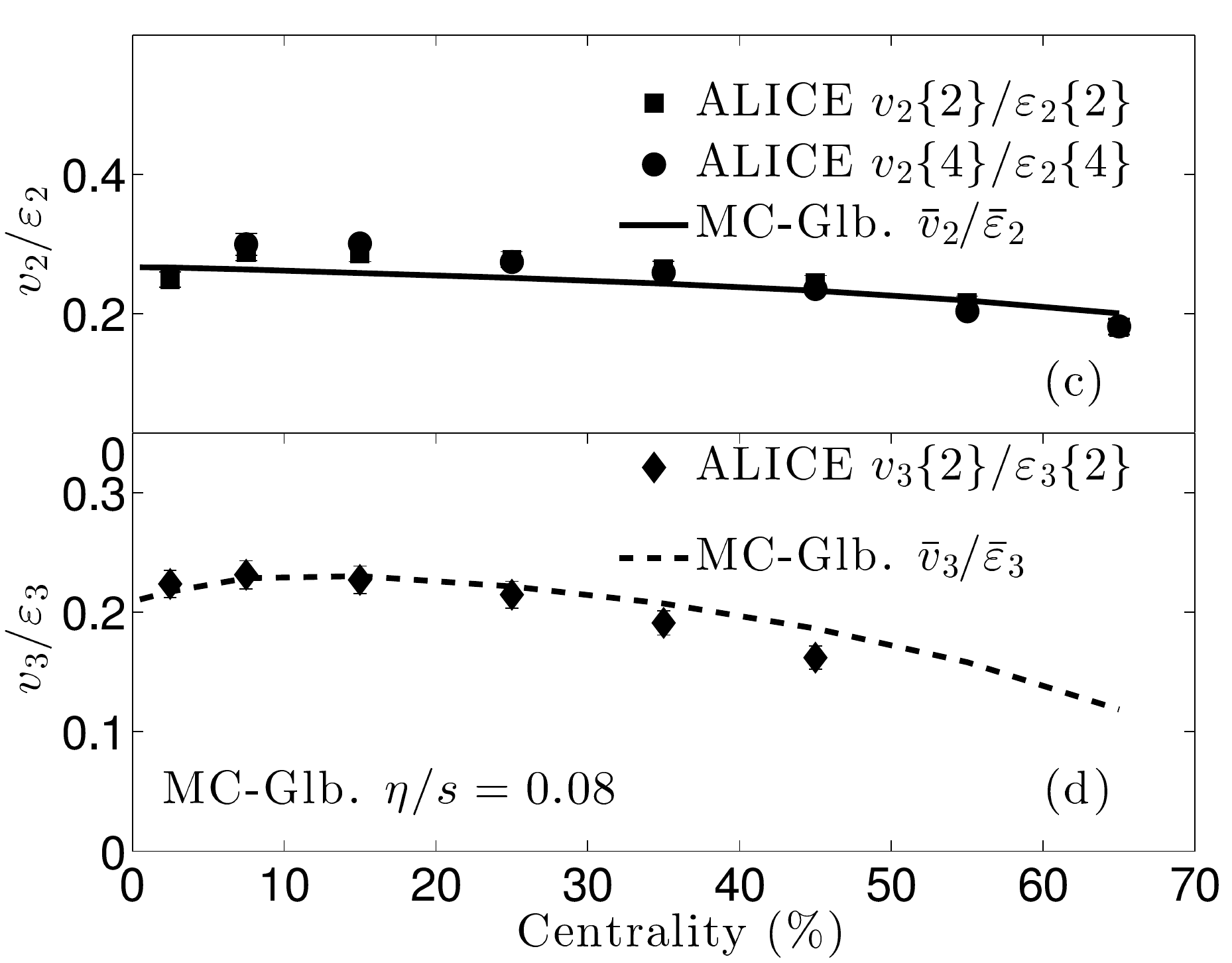}  	
  }
  \caption{Eccentricity-scaled, $p_T$-integrated $\bar{v}_{2,3}$ for the hydrodynamically evolved 
    MC-KLN (a,b) and MC-Glauber (c,d) models, compared with ALICE $v_{2,3}$ data 
    for 2.76\,$A$\,TeV Pb-Pb collisions \cite{ALICE:2011vk} scaled by their corresponding
    eccentricities (see text). From \cite{Qiu:2011hf}.
    \label{fig:1110.3033:3}}
\end{figure*}

The eccentricity-scaled elliptic and triangular flow coefficients for the MC-KLN 
and MC-Glauber models are shown in Figs.~\ref{fig:1110.3033:3}(a,b) and Fig.~\ref{fig:1110.3033:3}(c,d),
respectively, and compared with the corresponding data from ALICE \cite{ALICE:2011vk}. The first thing
to note is the impressively accurate agreement between the experimentally
measured $\vt{2}/\et{2}$ and $\vf{2}/\ef{2}$, showing that for elliptic flow the idea
of scaling ``each flow with its own eccentricity" \cite{Bhalerao:2006tp} works very well. Secondly,
both $\vt{2}/\et{2}$ and $\vf{2}/\ef{2}$ measured by ALICE agree well with the viscous
hydrodynamic calculations for both the MC-Glauber and MC-KLN models, confirming 
that for each model the correct value of $\eta/s$ has been used as far as elliptic flow
is concerned.

The bottom panels in Fig.~\ref{fig:1110.3033:3} show the triangular flow $v_3$. Clearly, with the
viscosities needed to reproduce $v_2$, the MC-KLN model badly disagrees with
the experimental data. The measured triangular flow is too big to accommodate
a specific shear viscosity as large as 0.2. Within the present approach, the only 
possibility to avoid this conclusion is that somehow the MC-Glauber and MC-KLN 
models both under-predict the initial third-order eccentricity $\ve_3$ by about 50\%.
With MC-Glauber initial conditions and $\eta/s\eq0.08$, on the other hand, the 
ALICE data agree well with viscous hydrodynamics, even if the measured centrality 
dependence of $\vt{3}/\et{3}$ is slightly steeper than the calculated one. 
 
Summarizing Fig.~\ref{fig:1110.3033:3}, the only possibility to have a large $\eta/s\sim 0.2$ for the QGP that is compatible with large ALICE $v_3$ data is to require an initial-condition model that produces much larger triangularities than the MC-Glauber and the MC-KLN models \cite{Schenke:2012wb}.


\section{$p_T$-differential elliptic and triangular flow} \label{sec:1110.3033:introduction}

\begin{figure*}
  \center{
  \includegraphics[width=0.45\linewidth]{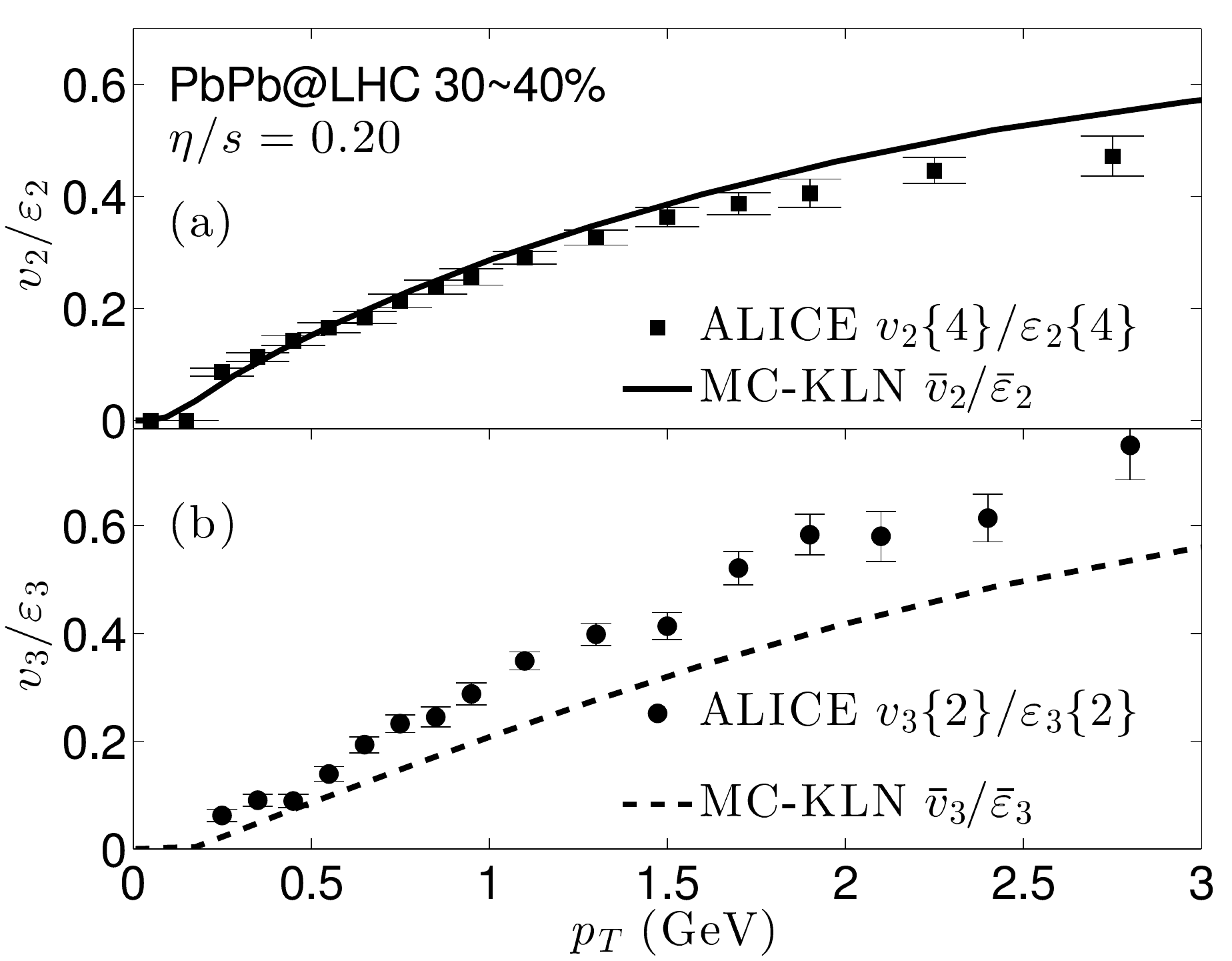}
  \includegraphics[width=0.46\linewidth]{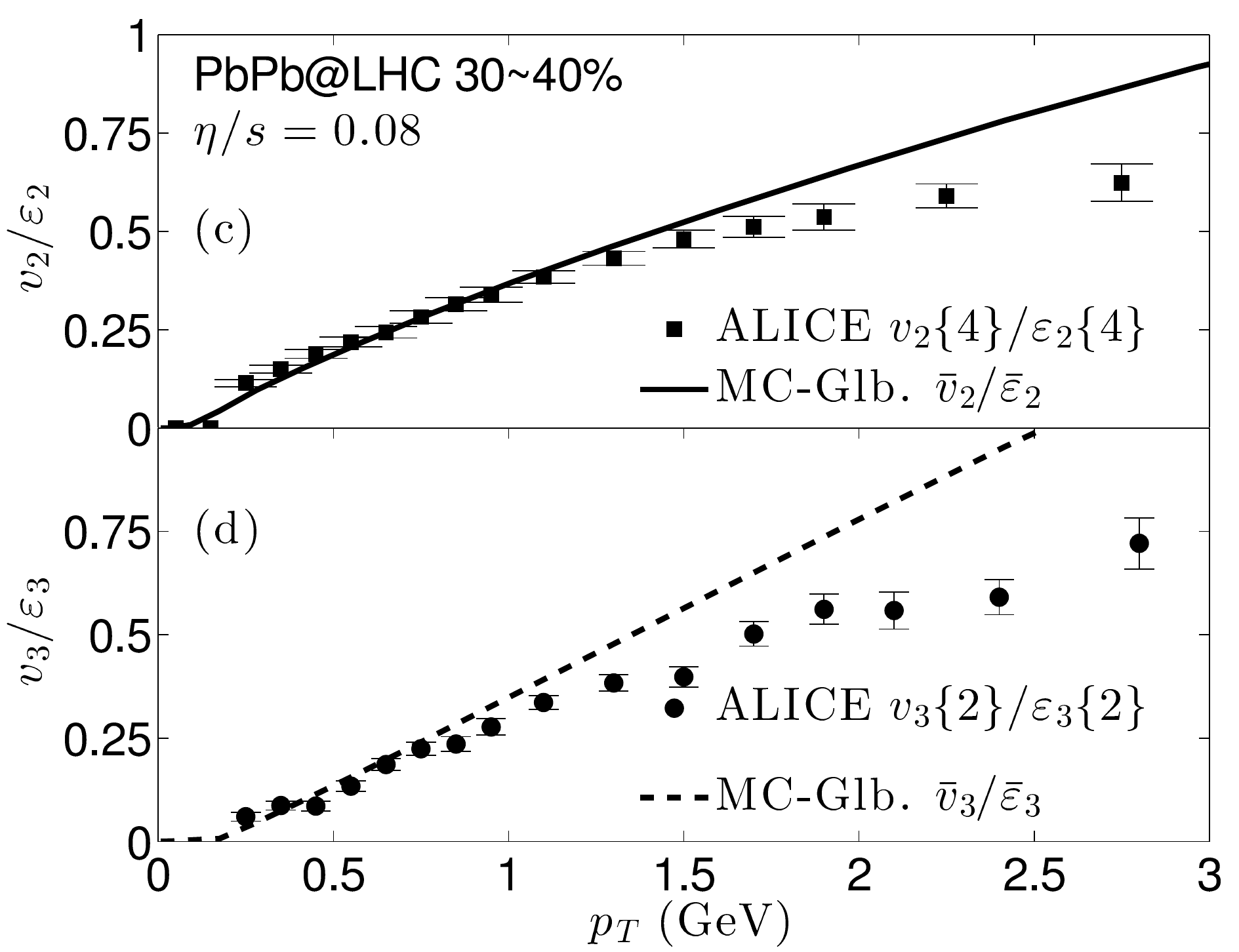}  
  }
  \caption{Eccentricity-scaled, $p_T$-differential elliptic and triangular flow for 
     2.76\,$A$\,TeV Pb-Pb collisions from viscous hydrodynamics with MC-KLN (a,b) 
     and MC-Glauber (c,d) initial conditions. The ALICE data \cite{ALICE:2011vk} are 
     scaled according to their corresponding eccentricities, see text. From \cite{Qiu:2011hf}.
  \label{fig:1110.3033:4}}
\end{figure*}

As a cross-check, we compare our calculations also to the $p_T$-differential anisotropic 
flow data at one collision centrality ($30{-}40\%$) \cite{ALICE:2011vk}. The corresponding comparison between data and theory is shown in Fig.~\ref{fig:1110.3033:4};
as in Fig.~\ref{fig:1110.3033:3}, we compare the eccentricity-scaled flows, plotting $\bar{v}_{2,3}/\bar{\ve}_{2,3}$
for the models and $\vf{2}/\ef{2}$ ($\vt{3}/\et{3}$) for the elliptic (triangular) flow data.
As seen in the upper panels, both initial-state models describe the measured
elliptic flow well up to $p_T{\,\sim\,}1{-}1.5$\,GeV/$c$; at larger $p_T$, they over-predict
$v_2(p_T)$ for charged particles -- a problem noticed before \cite{Song:2011qa,Shen:2011eg}
and possibly related to an imperfect model description of the measured final chemical
composition \cite{Heinz:2011kt}. The disagreement at larger $p_T$ is worse for
MC-Glauber initial conditions; this is likely related to our earlier observation in 
Fig.~\ref{fig:1110.3033:1}(b) that our the MC-Glauber $p_T$-spectra are steeper than the MC-KLN ones
in peripheral collisions -- an artifact of our single-shot approach and possibly remedied
by a proper event-by-event hydrodynamical simulation.

Fig.~\ref{fig:1110.3033:4}(b) shows again the disagreement between theory and
experiment for triangular flow when we use MC-KLN initial conditions: the model
strongly under-predicts the data at all $p_T$, i.e. it gives the wrong slope for
$v_3(p_T)$. With MC-Glauber initial conditions and correspondingly lower
shear viscosity $\eta/s\eq0.08$ (Fig.~\ref{fig:1110.3033:4}(d)), the measured $v_3(p_T)$ is 
well described up to $p_T{\,\sim\,}1$\,GeV/$c$ but over-predicted at larger $p_T$.
Again, the latter can be at least partially attributed to the fact that MC-Glauber 
$p_T$-spectrum from our single-shot hydrodynamic approach is too steep at this 
collision centrality, which can be corrected by performing the
hydrodynamic evolution properly event by event.  
 

\section{Chapter summary} \label{sec:1110.3033:chapter}
Using a single-shot viscous hydrodynamic approach 
without any hadronic after-burner but properly implementing hadronic chemical 
freeze-out at $\Tchem{\,\approx\,}165$\,MeV and including a full set of resonance 
decays, we have shown that a combined analysis of the ALICE data for elliptic 
and triangular flow from 2.76\,$A$\,TeV Pb-Pb collisions leads to a strong preference
for initial conditions from the Monte-Carlo Glauber model, combined with a low
value for the QGP shear viscosity $\eta/s{\,\simeq\,}0.08$, and disfavors the considerably
larger viscosities of $\eta/s{\,\sim\,}0.2$ that are required to reproduce the measured
elliptic flow when assuming the more eccentric Monte-Carlo KLN initial profiles.

The analysis presented in this chapter was restricted to only these two initial-state models, and only to elliptic and triangular flows. While this analysis, published in \cite{Qiu:2011hf}, was the first combined analysis of two different flow harmonics, it was later superseded by analyses that included several higher order harmonics \cite{Gale:2012in, Heinz:2013th, Wei:2012, Heinz:2013wva}. From the later work one must conclude that neither the MC-KLN model with $\eta/s=0.2$ nor the MC-Glauber model with $\eta/s=0.08$ can simultaneously explain all flow harmonics. An additional ingredient is missing from both of these models in order to produce the correct initial density fluctuation spectrum: In addition to the fluctuating nucleon positions, fluctuations of the quark and gluon fields inside the nucleons must be accounted for. Models that do this were recently developed in \cite{Schenke:2012wb, Schenke:2012hg, Gale:2012in}.


\chapter{Hydrodynamic event-plane correlations in Pb+Pb collisions}
\label{chap:correlation}

In this chapter, we show that correlations between the flow angles associated with higher harmonics measured by the ATLAS collaboration have hydrodynamic origin. The correlation strength is found to be sensitive to both the initial conditions and the shear viscosity of the expanding fireball medium. The material in this chapter is based on \cite{Qiu:2012uy}.

\section{Introduction}
In this chapter, we follow the conventions to define the eccentricity coefficients $\ve_n$ with associated participant plane angles $\Phi_n$ \cite{Alver:2010gr,Qin:2010pf,Teaney:2010vd}(\chap{introduction}) as:
\begin{eqnarray}
\label{eq:1208.1200:1}
  &&\Ecc_1 := \ve_1\,e^{i \Phi_1}
  = -\frac{\int r\,dr\,d\phi\, r^3 e^{i \phi}\,e(r,\phi)}{\int r\,dr\,d\phi\, r^3\,e(r,\phi)},
\\ 
  &&\Ecc_n := \ve_n\,e^{i n\Phi_n}
  = -\frac{\int r\,dr\,d\phi\, r^n e^{i n\phi}\,e(r,\phi)}{\int r\,dr\,d\phi\, r^n\,e(r,\phi)},
  \quad (n>1)
\nonumber
\end{eqnarray}
where $e(r,\phi)$ is the initial energy density distribution in the plane transverse to the beam direction at the collision point $z{=}0$. The final momentum distributions of the emitted charged hadrons are characterized, as in previous chapters, by their anisotropic flow coefficients $v_n$ and their associated flow (event plane) angles $\Psi_n$ \cite{Qin:2010pf,Voloshin:2008dg,Qiu:2011iv}(\chap{introduction}):
\begin{equation} 
\label{eq:1208.1200:2}
  V_n := v_n\,e^{i n \Psi_n} =\frac
  {\int p_T\,dp_T\,d\phi_p\, e^{i n\phi_p}\,\frac{dN_\mathrm{ch}}{d\eta\,p_Tdp_T\,d\phi_p}}
  {\int p_Tdp_T\,d\phi_p\,\frac{dN_\mathrm{ch}}{d\eta\,p_Tdp_T\,d\phi_p}}.
\end{equation}

As pointed out in \cite{Mishra:2007tw,Mocsy:2011xx}, a complete understanding of the entire spectrum of harmonic flow coefficients $v_n$ is expected to yield strong constraints on the initial conditions and dynamical evolution of heavy-ion collisions, in particular the transport coefficients of the fireball medium. The authors of \cite{Teaney:2010vd,Staig:2010pn,Nagle:2010zk,Bhalerao:2011yg,Qin:2011uw,Jia:2012ma} added that correlations between the event plane angles $\Psi_n$ of different harmonic order can yield valuable additional insights into the initial conditions. Such correlations were measured with good precision by the ATLAS Collaboration in Pb+Pb collisions at the LHC \cite{Jia:2012sa}. In this chapter, we demonstrate that some of the measured final-state event-plane correlations have a qualitatively different centrality dependence from the corresponding initial-state participant-plane correlations, and that this characteristic change between initial and final state is correctly reproduced by hydrodynamic evolution. This provides additional strong support for the validity of the hydrodynamic paradigm in relativistic heavy-ion collisions. Furthermore, we show that the measured event-plane correlations are not only sensitive to the initial conditions, but also to the shear viscosity of the hydrodynamic medium, thus providing an independent constraint for this key transport coefficient.

\section{Methodology}
We evolve fluctuating initial energy density profiles for Pb+Pb collisions at $\sqrt{s}=2.76$ ATeV using event-by-event viscous hydrodynamics. To explore the sensitivity to model uncertainties in the initial state, we have evolved events from two sets of initial conditions obtained from the Monte-Carlo Glauber and the Monte-Carlo KLN models \cite{Drescher:2006pi,Hirano:2009ah}\footnote{As mentioned in \chap{e2e3}, it was realized later that neither the MC-Glauber model nor the MC-KLN model can give a complete description to all the flow data, but the transition to the emerging new initial-condition models had not happened when this thesis was written; the only initial-condition models that are available to us are the MC-Glauber and the MC-KLN models.}. We divided each set into centrality classes according to the number $N_\mathrm{part}$ of wounded nucleons (see tables \ref{tab:cen:1}-\ref{tab:cen:6}); for each centrality class, we evolved 11,000 events for each of the two initial condition models. Model
parameters were tuned to reproduce the $p_T$ spectra and elliptic flows of unidentified charged particles and identified hadrons, as reported in \cite{Qiu:2011hf,Shen:2011eg}(\chap{e2e3}). As explained in \chap{e2e3}, this results in a choice of specific shear viscosity $\eta/s{=}0.08$ for MC-Glauber initial conditions and the larger value $\eta/s{=}0.2$ for MC-KLN initial conditions. Both the QGP phase and the hadronic phase are evolved hydrodynamically; particle momentum distributions are calculated with the Cooper-Frye prescription, taking into account strong decays of all hadron resonances with masses up to $2.25$\,GeV. \footnote{We found, however, that the event-plane correlations discussed below are almost identical for all particle species, so including resonance decays is not essential for this work.} From the resulting charged hadron distribution we calculate for each event the flow angles $\Psi_n$ according to
\begin{equation} 
\label{eq:1208.1200:3}
  v_n\,e^{i n \Psi_n} =
  \frac{\int_{0.5{<}|\eta|{<}2.5} d\eta \int_{p_\mathrm{min}} p_Tdp_T\,d\phi_p\, e^{i n\phi_p}\,\frac{dN_\mathrm{ch}}{d\eta\,p_Tdp_T\,d\phi_p}}
  {\int_{0.5{<}|\eta|{<}2.5} d\eta \int_{p_\mathrm{min}} p_Tdp_T\,d\phi_p\,\frac{dN_\mathrm{ch}}{d\eta\,p_Tdp_T\,d\phi_p}},
\end{equation}
employing the same pseudorapidity range $0.5 <\ |\eta| < 2.5$ and lower $p_T$ cutoff 
$p_T{\,>\,}p_\mathrm{min}{\,=\,}0.5$\,GeV as used in the experimental analysis \cite{Jia:2012sa}.\footnote{The ATLAS results were obtained with two independent methods:
   (a) using a calorimetric measurement of transverse energy $E_T$ over rapidity range 
   $0.5{\,<\,}|\eta|{\,<\,}4.8$, and (b) using charged particle tracks with $p_T{\,>\,}0.5$\,GeV
   and $0.5{\,<\,}|\eta|{\,<\,}2.5$. The data from method (a) have better precision but 
   are fully compatible with those from method (b), within error bars. Since we cannot 
   simulate the calorimetric response of ATLAS theoretically, we compute the event-plane
   correlations according to method (b), but compare them in the figures to the more
   precise data obtained from method (a).}\footnote{It was however realized only recently that the flow analysis performed by the ATLAS collaboration suffers from inaccuracies caused by the event-plane method, and an analysis with the scalar-product method yields much better agreement between theory and experimental data \cite{Bhalerao:2013ina}.}
From these event plane angles we compute for each event $\cos(k_1 \Psi_{n_1}{+}\dots{+}k_m\Psi_{n_m})$
for the two-plane ($m{=}2$) and three-plane ($m{=}3$) correlations listed in Tables 1 and 2
of Ref.~\cite{Jia:2012sa} and shown in the figures below, and then average this quantity over all events in the given centrality class. We compare these event-plane correlations with the corresponding correlations between the initial-state participant plane angles, $\Phi_n$, calculated from the initial energy density profile of each propagated event according to Eq.~(\ref{eq:1208.1200:1}) and then averaged over events in a similar way. 

\section{Results}
%

\begin{figure*}
  \center{\includegraphics[width=0.9\linewidth]{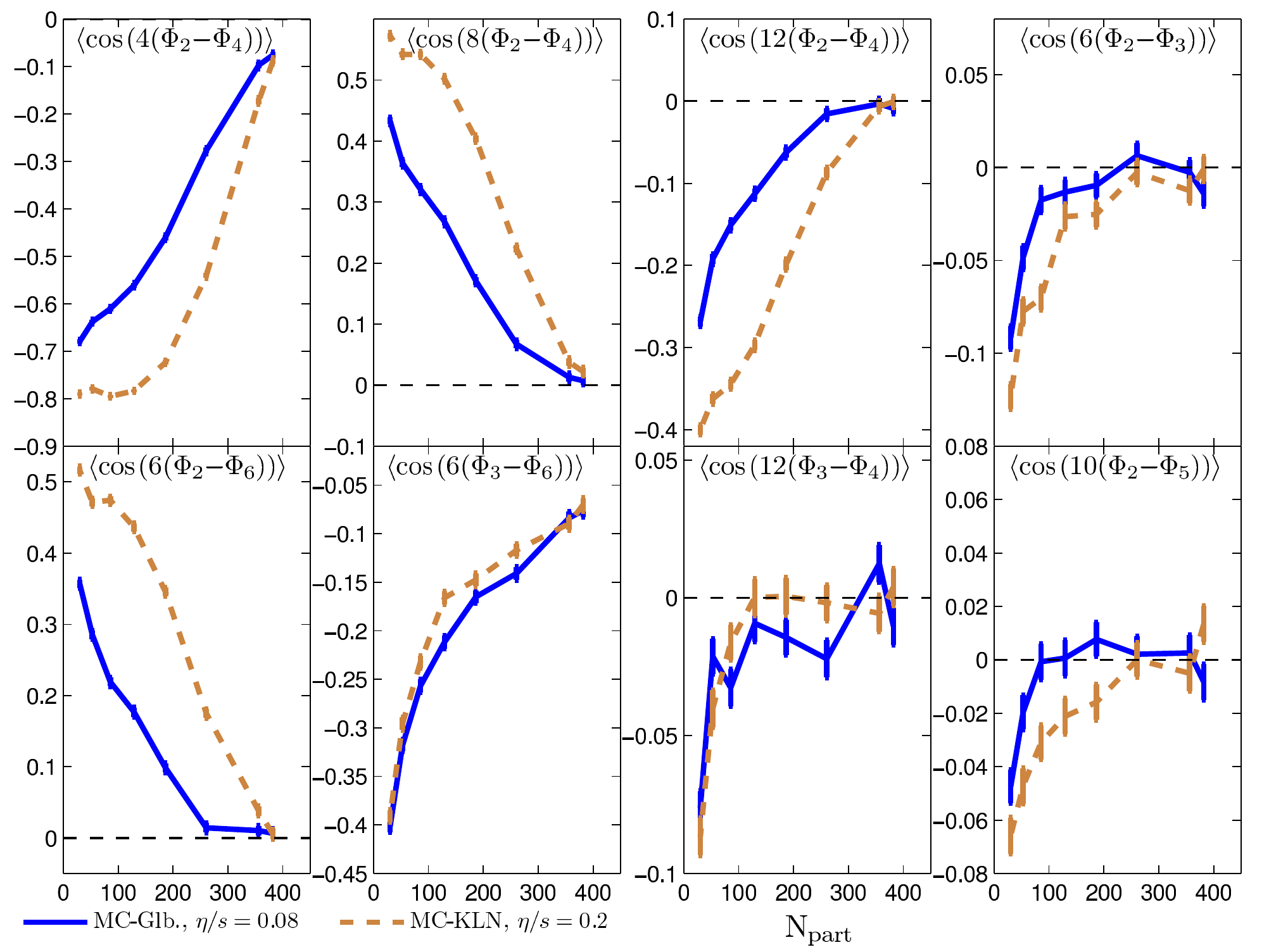}}
  \caption{Two-plane correlations $\langle\cos(jk(\Phi_n{-}\Phi_m))\rangle$, 
     where $j$ is an integer and $k$ is the least common multiple (LCM) of $n$ and $m$ 
     \cite{Bhalerao:2011yg,Jia:2012ma}, between pairs of participant-plane angles $\Phi_{n,m}$ 
     for the harmonics $(n,m)$ and multipliers $j$ studied in Ref.~\cite{Jia:2012sa}. Solid (dashed)
     lines show results for initial density profiles obtained from the MC-Glauber (MC-KLN) model. From \cite{Qiu:2012uy}.
    \label{1208.1200:F1}}
\end{figure*}
%
\begin{figure*}
  \center{\includegraphics[width=0.9\linewidth]{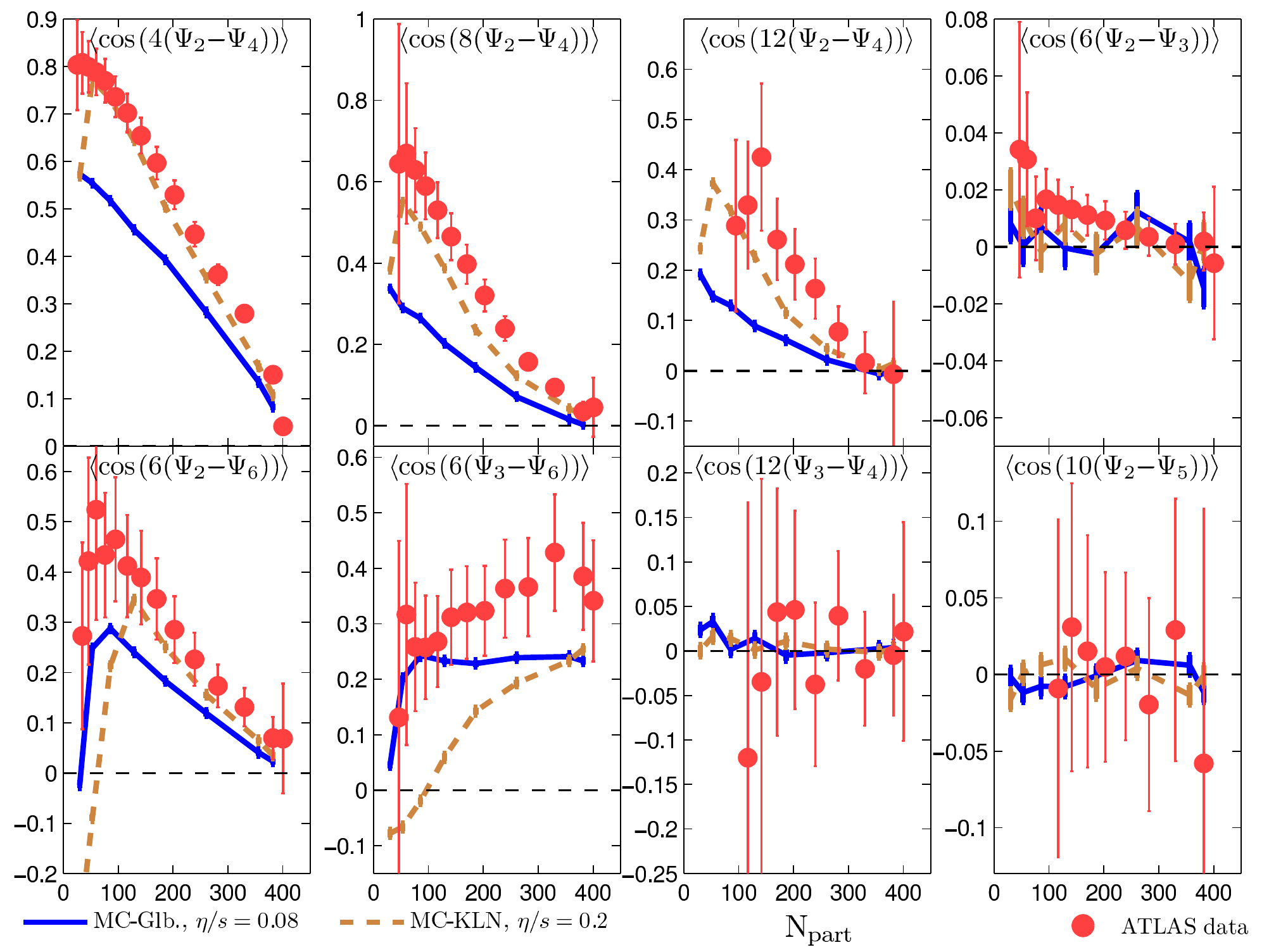}}
  \caption{Similar to Fig.~\ref{1208.1200:F1}, but for the corresponding final-state 
  event-plane angles 
  $\Psi_{n,m}$. Filled circles show the experimental values measured by ATLAS
  \cite{Jia:2012sa}. The MC-Glauber (solid) and MC-KLN (dashed) initial profiles used in 
  Fig.~\ref{1208.1200:F1} were propagated individually using viscous hydrodynamics with $\eta/s{=}0.08$
  and 0.2, respectively. From \cite{Qiu:2012uy}.
  \label{1208.1200:F2}}
\end{figure*}

Figures \ref{1208.1200:F1} and \ref{1208.1200:F2} show the initial and final state two-plane correlations, for the eight
different combinations of angles and weight factors explored by the ATLAS experiment
\cite{Jia:2012sa}. Each correlation function is plotted against collision centrality, with
peripheral collisions (small $\Np$ values) on the left and central collisions (large $\Np$) on the right. Fig.~\ref{1208.1200:F1} shows that several of these correlations are quite sensitive to the model
used to generate the initial energy density profiles (MC-Glauber vs. MC-KLN). These model differences in the initial state manifest themselves in corresponding model differences between the final-state event-plane correlations shown in Fig.~\ref{1208.1200:F2}, but they are additionally modified by the different shear viscosities $\eta/s$ (0.08 and 0.2, respectively) used to evolve the initial conditions from the two models. This is most clearly seen in the ``3-6 correlation", where the two models give almost identical initial-state participant-plane correlations $\langle\cos(6(\Phi_3{-}\Phi_6))\rangle$ (second lower panel from the left in Fig.~\ref{1208.1200:F1}) whereas the corresponding final-state event-plane correlators $\langle\cos(6(\Psi_3{-}\Psi_6))\rangle$ exhibit significant model differences. This demonstrates the sensitivity of these event-plane correlations to the specific shear viscosity of the expanding fireball medium. 

It is worth emphasizing that several of these two-plane correlators exhibit dramatically different centrality dependences for the initial-state participant-plane and the final-state event-plane angles (see, for example, the upper left, two upper right and second lower left panels in Figs.~\ref{1208.1200:F1} and \ref{1208.1200:F2}). The difference is largest in peripheral collisions (small $\Np$). This effect is caused by a dynamical rotation of the event-plane angles during the hydrodynamic evolution, driven by large elliptic flow in non-central collisions which leads to mode coupling between the angles $\Phi_n$ and $\Phi_{n\pm2k}$  (where $k$ is an integer and the largest coupling coefficient should correspond to $k{=}1$).\footnote{This is different from the 
   mode coupling at freeze-out \cite{Borghini:2005kd} caused by an elliptic (quadrupole)
   deformation of the collective flow velocity appearing in the exponent of 
   the Boltzmann factor in the Cooper-Frye expression for the final-particle momentum 
   distribution that couples $v_n$ with $v_{n\pm2k}$. In contrast, in the presence of 
   strong elliptic flow, the nonlinear hydrodynamic evolution {\it before freeze-out} leads to 
   mode coupling between the modes $n$ and $n{\pm}2k$ ($k$ integer) for the entire 
   complex flow vector on the left-hand side of Eq.~(\ref{eq:1208.1200:2}). We will come back to this point later in this chapter.}

\begin{figure*}
  \center{\includegraphics[width=0.9\linewidth]{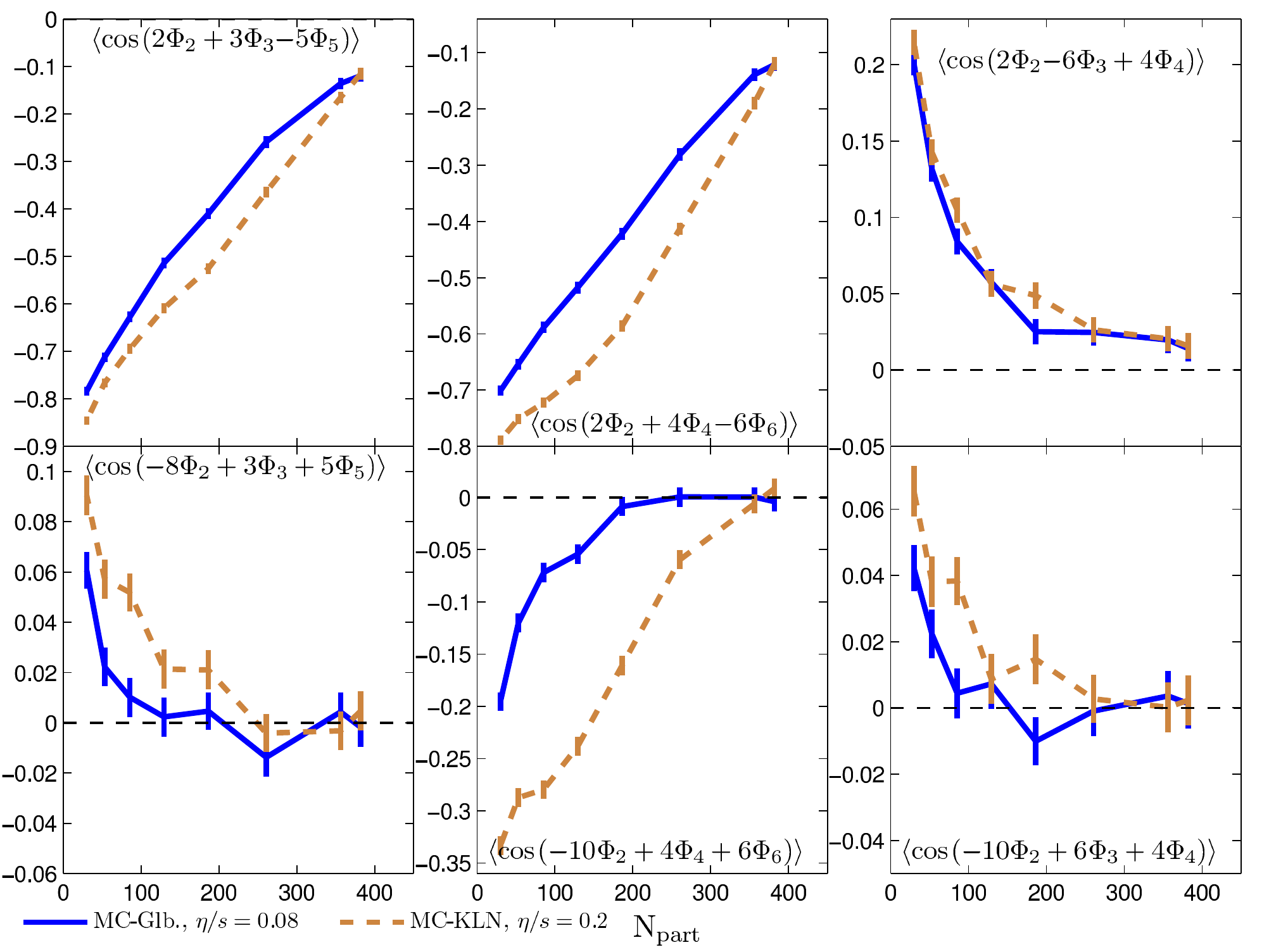}}
  \caption{Similar to Fig.~\ref{1208.1200:F1}, but for selected \cite{Jia:2012sa} 
  three-plane correlators 
  of the form $\langle\cos(c_l l\Phi_l{+}c_n n\Phi_n{+}c_m m\Phi_m)\rangle$, where the 
  $c_i$ are integers satisfying $c_l l{+}c_n n{+}c_m m{=}0$ \cite{Bhalerao:2011yg}. From \cite{Qiu:2012uy}.
  \label{1208.1200:F3}}
\end{figure*}
%
\begin{figure*}
  \center{\includegraphics[width=0.9\linewidth]{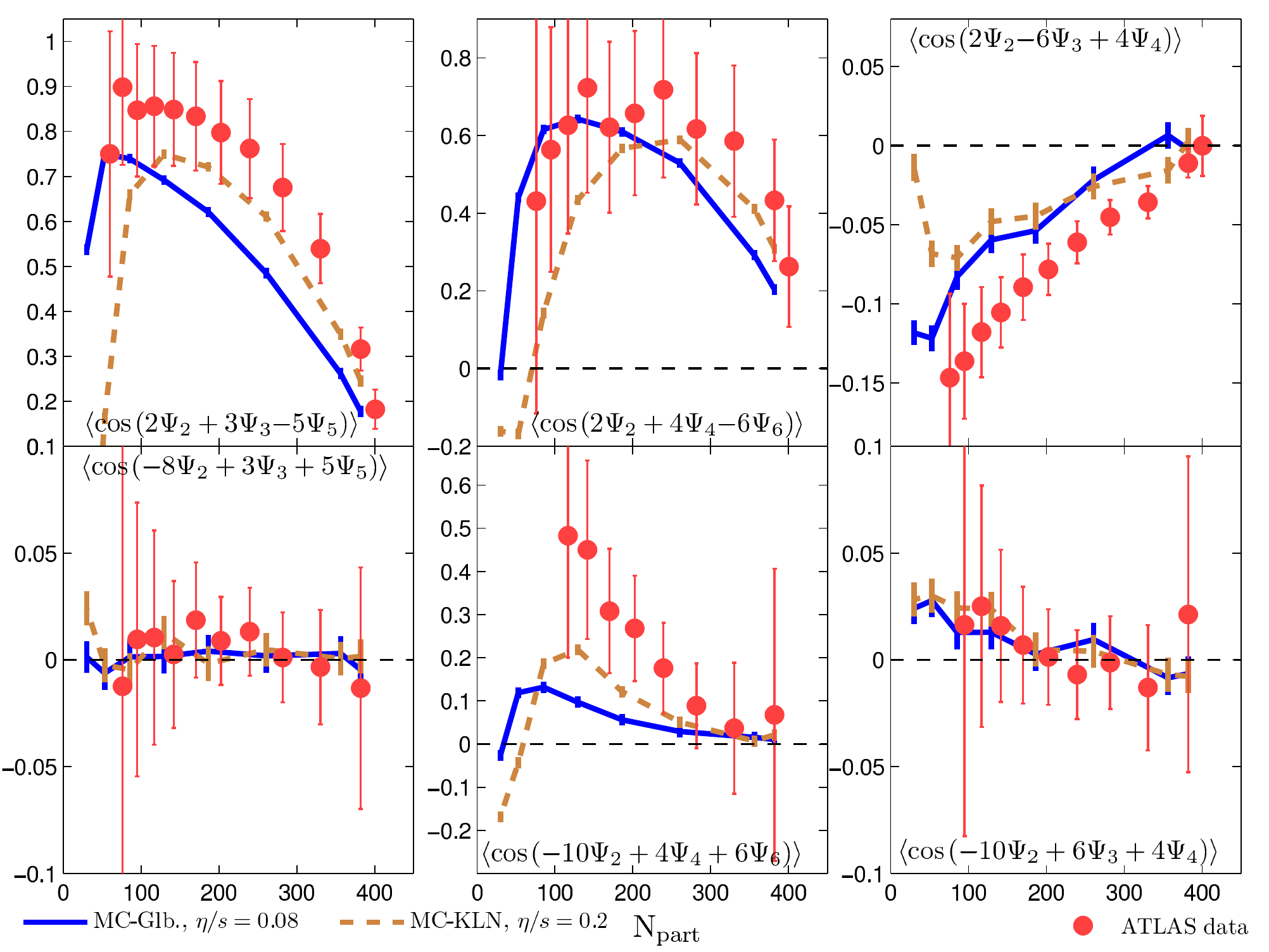}}
  \caption{The final-state event-plane correlators corresponding to the 
  initial-state correlators 
  between three participant planes of different harmonic order shown in Fig.~\ref{1208.1200:F3}. Solid 
  and dashed lines show results from viscous hydrodynamics with MC-Glauber and MC-KLN 
  initial conditions, evolved with $\eta/s{=}0.08$ and 0.2, respectively; filled circles show
  ATLAS data \cite{Jia:2012sa}. From \cite{Qiu:2012uy}.
  \label{1208.1200:F4}}
\end{figure*}

Figures~\ref{1208.1200:F3} and \ref{1208.1200:F4} show a number of three-plane correlations studied by the ATLAS experiment \cite{Jia:2012sa}, with the initial-state participant-plane correlators plotted in Fig.~\ref{1208.1200:F3} and the corresponding final-state event-plane correlators in Fig.~\ref{1208.1200:F4}, together with the experimental data. Again, we observe characteristic sign changes between several of the initial-state correlations and their corresponding final-state correlators. Even if neither of the two initial-state models (MC-Glauber and MC-KLN) reproduces the experimental data exactly, we find it impressive that the hydrodynamic model reproduces all the qualitative features of the centrality dependences of the 14 different measured event-plane correlation functions correctly: where the data show strong (weak) correlations, the same is true for the theoretical results, and where the data show correlations that increase (decrease) from peripheral to central collisions, the same holds for the theoretical predictions, without any parameter tuning. This provides very strong support for the hydrodynamic model description of the fireball evolution from a new set of observables that is quite independent of all previously studied observables ($p_T$-spectra, anisotropic flow coefficients $v_n$, and HBT radii). 

We note that the nonlinear mode coupling first discovered in \cite{Qiu:2011iv} (see \chap{1104.0650}) and the event-plane rotations driven by this nonlinear effect, are key to the qualitative agreement between theory and data in Figs.~\ref{1208.1200:F2} and \ref{1208.1200:F4}. It was recently shown that the measured correlation can also be reproduced with the AMPT model \cite{Bhalerao:2013ina}, a microscopic model that also features strong collective flow and an approach to approximate local thermal equilibrium \cite{Lin:2004en}. We doubt that a similar agreement can be obtained with dynamical models that do not rely on a large degree of local thermalization in the expanding fireball, or from an approach based on linear \cite{Teaney:2010vd,Staig:2010pn,Staig:2011wj} hydrodynamic response to the initial-state density fluctuations. Inclusion of first-order nonlinear terms in the hydrodynamic response \cite{Teaney:2012ke} appears to yield event-plane correlations with qualitatively similar features as shown here \cite{Teaney}, but quantitative success likely requires a numerical approach that fully accounts for the intrinsic nonlinearity of viscous hydrodynamics.

A closer look at Figs.~\ref{1208.1200:F1} and \ref{1208.1200:F3} shows that the MC-KLN model tends to produce stronger correlations between the initial-state participant-plane angles $\Phi_n$ than the MC-Glauber model. We observe that hydrodynamic evolution translates the stronger initial-state participant-angle correlations into stronger final-state event-plane correlations, even though the signs of some of the correlators featuring the strongest correlation strengths flip between initial and final state. This is especially true for the two-plane correlations shown in Fig.~\ref{1208.1200:F1}, while the three-plane correlators exhibit some exceptions to this ``rule" in the most peripheral collisions. The experimental data appear to prefer the stronger angle correlations in the initial profiles from the MC-KLN model, even though this model gives an elliptic-to-triangular flow ratio $v_2/v_3$ that is much larger than measured \cite{Qiu:2011hf}, caused by a larger $\ve_2/\ve_3$ ratio than in the MC-Glauber model \cite{Qiu:2011iv} (see \chap{1104.0650}). These observations show that a combined analysis of both the anisotropic flow coefficients $v_n$ and their associated flow angles $\Psi_n$ (and the correlations among them) promises to yield powerful constraints on initial state models for the fireball energy density profiles created in heavy-ion collisions.

Finally, to demonstrate that the mode-coupling effect responsible for the observed event-plane angle correlations is dynamically generated during the hydrodynamic evolution rather than at the freeze-out stage, we show angle correlations similar to the ones shown above, but calculated using the flow velocity. We define the anisotropies of the transverse fluid velocity along the freeze-out surface as:
\begin{equation} \label{eq:1208.1200:11}
   \bar{V}_{n} e^{i n \bar{\Psi}_n} =  \frac {  \int_\Sigma | \, d^3\sigma| v_\perp^n e^{i n \phi(v_\perp)}  }  {  \int_\Sigma |d^3\sigma| \, v_\perp^n  }
\end{equation}


\begin{figure*}[h!]
  \center{
        \includegraphics[width=0.9\textwidth, height=0.6\textwidth]{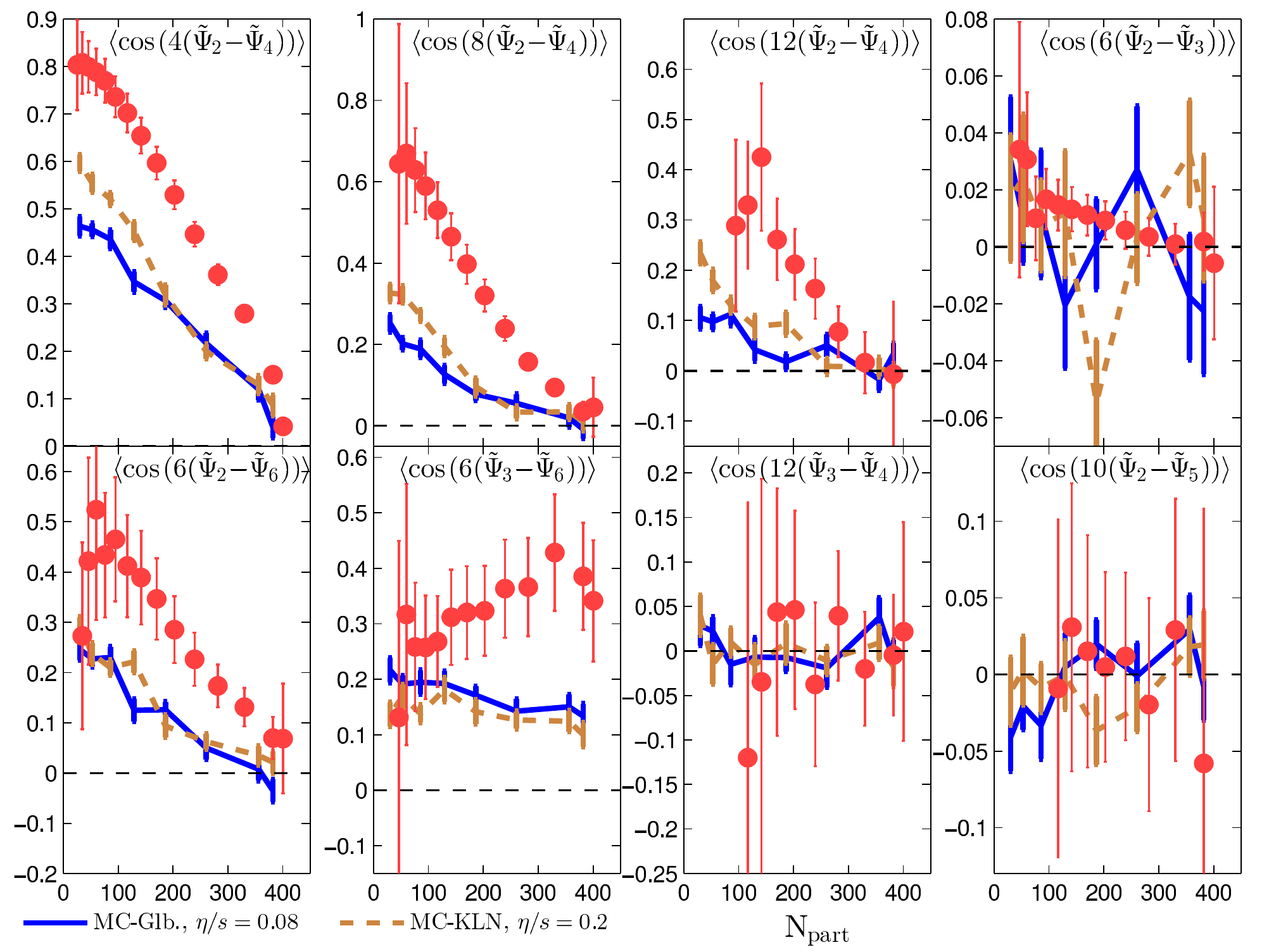}
        \includegraphics[width=0.9\textwidth, height=0.6\textwidth]{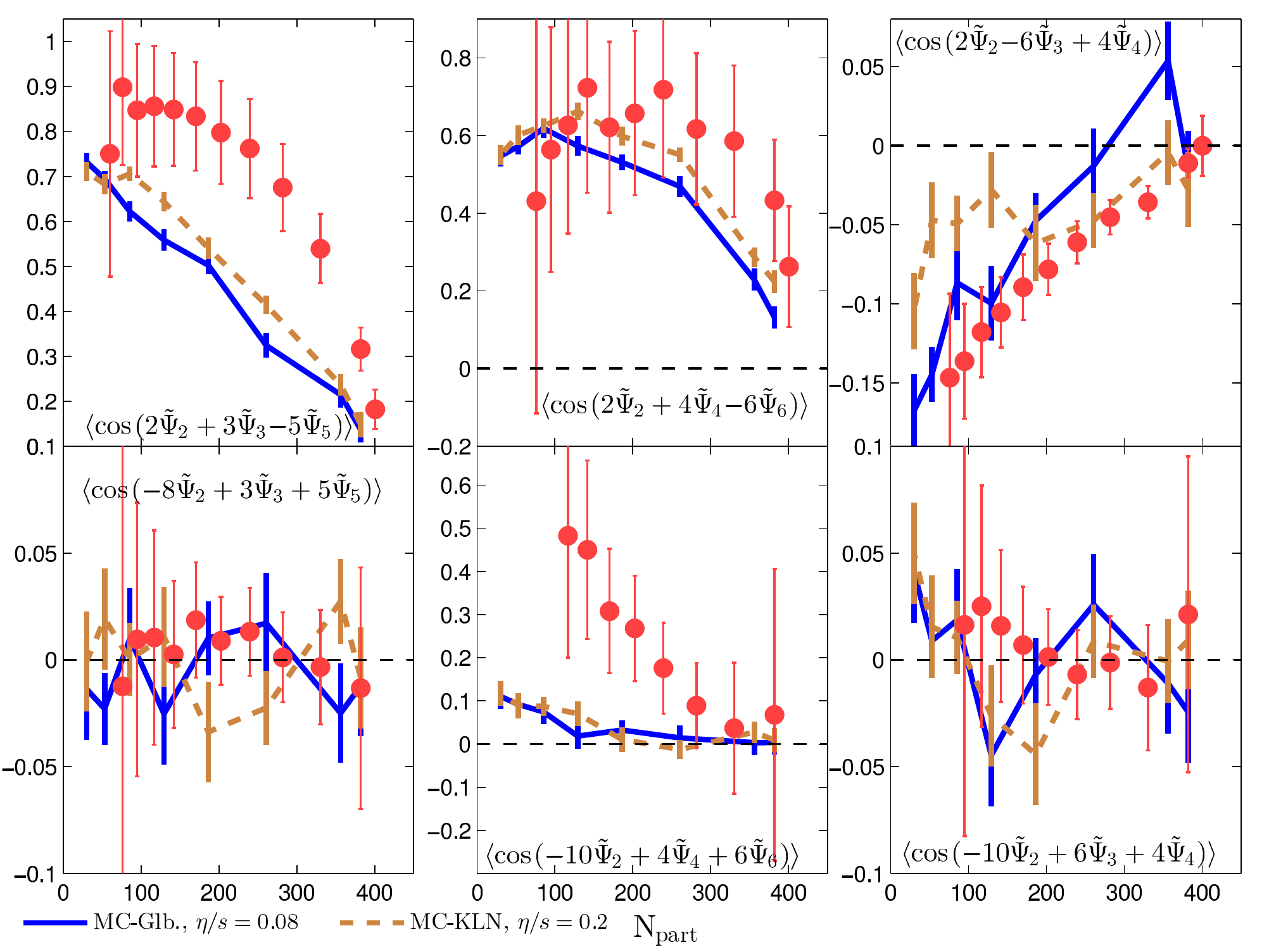}
        }
  \caption{Correlations similar to Fig.\ref{1208.1200:F2} (top) and Fig.\ref{1208.1200:F4} (bottom), but with anisotropy angles calculated from fluid velocity, see text for details.}
  \label{fig:1208.1200:41}
\end{figure*}

\noindent In \Fig{1208.1200:41} we show the correlations between the so-defined flow velocity angles $\bar{\Psi}_n$\footnote{We checked that almost identical results are obtained if the power $v_\perp^n$ is replaced by $v_\perp^2$ in the integrand of the numerator and the denominator in \equ{1208.1200:11}.}. It is seen that the correlations shown in Fig.\ref{fig:1208.1200:41} between the anisotropy angles calculated using the fluid velocity are impressively similar to those shown in Figs.~\ref{1208.1200:F1} and \ref{1208.1200:F2}. They still show differences which should be attributed to either the ``freeze-out mode-coupling" effect or a viscous correction to the equilibrium hadron emission function, or to the resolution effects discussed in \cite{Bhalerao:2013ina}. Still the qualitative similarity between the correlation patterns calculated from the fluid velocity and from anisotropic flows from hydrodynamical simulations is an interesting finding in the following sense: the fluid velocity, which is a result of {\em only} the dynamical process, cares little about the mechanism for particle emissions and thus is not affected by final-stage effects (for example, resonance decay, or the choice of forms for the viscous correction terms to the distribution function, etc.\ ). Therefore the fact that the correlations calculated from the fluid velocity are similar to those from the anisotropic flow data provides strong evidence that the correlations among anisotropic flow angles are generated during the dynamical evolution rather than at the freeze-out stage.

Even though more detailed studies will be necessary to fully explore the event-plane correlations discussed in this chapter, the calculations presented here suggest that very likely neither the MC-Glauber nor the MC-KLN initial conditions will ultimately provide a quantitatively satisfactory description of the experimental data from the ATLAS Collaboration \cite{Jia:2012sa}. While this statement is backed up by the study in \chap{e2e3}, however, a recent study shows that the quantitative disagreement between theory results and experiments is very likely to be an artifact of the event-plane method used in the experimental flow analysis \cite{Bhalerao:2013ina}. We have not performed the same analysis using our own simulated results, but no matter what the outcome might be, the question of how to turn the multitude of anisotropic flow observables (magnitudes and angles) that are already measured and will be measured into a focused search for the correct initial-state model is an interesting and welcome new challenge for the theory community.


\chapter{Resonance decay contributions to higher-order anisotropic flow coefficients}
\label{chap:resonance}

The computation of resonance decays at the end of a hydrodynamic calculation in order to obtain the complete spectra of stable hadrons is numerically very costly and therefore often omitted. In this chapter, we show that, in hydrodynamic simulations for relativistic heavy-ion collisions, strong resonance decay calculations can be performed with fewer species of particle resonances than usually implemented while preserving good accuracy in single-particle spectra and flow anisotropies. Such partial resonance calculations boost computational efficiency by a factor of 10, which is essential for large scale event-by-event simulations. The material in this chapter is based on \cite{Qiu:2012tm}.

\section{Chapter introduction}
\label{sec:1210.7010:1}

In event-by-event hydrodynamic simulations, due to the limited number of final-state particles in each event, observables can be measured with good statistical precision only as ensemble averages, which requires the simulation of a large number of events. Due to algorithmic progress over the last few years, the hydrodynamic evolution part is no longer the bottleneck in such event-by-event studies; at least for (2+1)-dimensional simulations (which assume longitudinal boost-invariance), the largest 
fraction of the computer time is spent converting the hydrodynamic output into final particle distributions, either on a ``switching surface" between a macroscopic hydrodynamic description of the QGP fluid and a microscopic kinetic evolution of the dilute late hadronic-rescattering stage \cite{Hirano:2005xf,Song:2010aq} or on a ``kinetic decoupling" surface marking the transition from a strongly coupled fluid directly to a noninteracting gas of free-streaming hadrons. The high numerical cost of this ``hydro-to-particle conversion" process results from the large number of unstable hadron resonances that need to be included and whose post-freeze-out decays (mostly due to strong-interaction processes, although for some comparisons with experimental data that have not been corrected \cite{Adams:2003xp,Adler:2003cb} for weak-decay feed-down, weak and electromagnetic decays must also be considered) modify the finally observed particle distributions.

The hydro-to-hadron conversion algorithm is based on the Cooper-Frye formula \cite{Cooper:1974mv}, which expresses the final hadron momentum distribution as an integral of the local equilibrium (for ideal fluid dynamics) or slightly off-equilibrium (in viscous fluids) distribution function for the particle species in question over the conversion surface. Contributions to the spectra of experimentally measured {\em stable} particles from the strong decays of unstable resonances are then calculated from the single-particle spectra for the resonances \cite{Sollfrank:1990qz}. This requires the calculation of the directly emitted (``thermal'') particle momentum distributions for all $\sim300$ hadron species with mass typically up to 2 GeV
via Cooper-Frye integrals, followed by the evaluation of the phase-space integrals \cite{Sollfrank:1990qz} for all contributing decay channels. On a typical personal computer with a single CPU core in year 2012 this calculation takes about 2-3 hours, compared to 10-15 minutes for the preceding hydrodynamic evolution.

The $\sim2$GeV cutoff in resonance mass is dictated by requiring convergence of the relative particle yields of the measured hadronic final state after all unstable resonances have been allowed to decay. (The pion yields are especially sensitive to resonance feeddown.) Experimental evidence points to chemical decoupling at a temperature of $T_\mathrm{chem}\approx 165$\,MeV, i.e.\ close to the (pseudo)critical temperature for the quark-hadron phase transition \cite{BraunMunzinger:2003zd}; at this temperature, only resonances with masses above 2\,GeV are sufficiently strongly Boltzmann-suppressed that their decay contributions to stable particle yields can be safely ignored.

Here we show that for an accurate determination of the pion and proton anisotropic flow coefficients $v_n$, a much smaller number of resonances needs to be taken into account than for the hadron yields, and that even the shape of the azimuthally averaged pion and proton transverse momentum spectra can be reliably determined by accounting for only a small subset of the $\sim300$ resonance species mentioned above. These are the observables needed for an extraction of the QGP shear viscosity from heavy-ion collision experiments \cite{Song:2010mg}. In this chapter, we show that by rearranging the resonance decay table in the order of decreasing importance for the calculation of $p_T$-spectra and $v_n$ coefficients instead of increasing mass, good convergence for these observables can be achieved with a significantly reduced set of only about 20-30 resonances. This speeds up the computation by a factor of 10 -- a significant gain in efficiency for the iterative determination of the QGP shear viscosity.

The analysis presented here uses final states generated with the same $(2{+1})$-dimensional boost-invariant viscous hydrodynamic code {\tt VISH2{+}1} for 200\,$A$\,GeV Au+Au collisions at the Relativistic Heavy-Ion Collider (RHIC) and for 2.76\,$A$\,TeV Pb+Pb collisions at the Large Hadron Collider (LHC) at various collision centralities, with previously determined \cite{Song:2010mg,Shen:2011eg,Qiu:2011hf} hydrodynamic input parameters. We find very similar results at both collision energies and therefore show here only plots for LHC collisions. Since the decay contributions from different resonances to the mentioned observables depend only on their decay channels and transverse momentum distributions, we expect little sensitivity to the assumption of longitudinal boost-invariance implicit in our approach and expect our reordered resonance decay tables to perform equally well for both $(2{+}1)$-d and $(3{+}1)$-d hydrodynamic simulations, and for a wide range of input parameters, such as QGP viscosity, 
thermalization time, initial entropy and energy density, etc.

\section{Resonance ordering}
\label{sec:1210.7010:2}

The momentum distributions of directly emitted (``thermal'') resonances of species $i$ are computed from the Cooper-Frye formula \cite{Cooper:1974mv}:
\begin{equation} \label{eq:1210.7010:1}
  E \frac{dN_i}{d^3p} = \frac{dN}{dy\,p_T\,dp_T\,d\phi_p} =
  \frac{g_i}{(2 \pi)^3} \int_\Sigma \,p^\mu d^3\sigma_\mu (f_{i0} + \delta f_i).
\end{equation}
Here $\Sigma$ is the hydro-to-hadron conversion hypersurface, $d^3\sigma_\mu$ is its surface normal vector, $f_{i0}=1/[e^{\beta (p \cdot u - \mu_i)} \mp 1]$ is the Bose or Fermi thermal equilibrium distribution function, and $\delta f_i$ accounts for viscous corrections (driven by the viscous pressure tensor $\pi_{\mu\nu}(x)$ on the conversion surface) of the local phase-space distribution along $\Sigma$. We assume the quadratic form \cite{deltaf, Baier:2006um}:
\begin{equation}
\delta f =\frac{1}{2} f_0 (1\pm f_0) \frac{p^\mu p^\nu}{T^2}\frac{ \pi_{\mu\nu}}{(e+p)}.
\end{equation}

Resonance decays increase the total yields of the stable hadrons and change their momentum distributions. For kinematic reasons, most of the light decay daughters have low transverse momenta, thus modifying the shape of light stable hadrons (pions, kaons) particle spectra mostly in the region $p_T<1.5$\,GeV \cite{Sollfrank:1990qz}. We denote the total decay contribution to the momentum distribution of stable hadron species $i$ by $\delta\!\left(dN_i/(dy d^2p_T)\right)$, and the total spectrum (obtained by adding this to the thermally emitted spectrum $dN_i^\mathrm{th}/(dy d^2p_T)$) by $dN_i^\mathrm{tot}/(dy d^2p_T)$. (We here include only strong and electromagnetic decays.) The $\bm{p}_T$-integrated total yield $\delta(dN_i/dy)$ of decay products of species $i$ is denoted by $\delta N_i$, with $N_i^\mathrm{tot}{\,=\,}N_i^\mathrm{th}{+}\delta N_i{\,=\,}N_i^\mathrm{th}{+}\sum_j \tilde{b}_{j\to i} N_j^\mathrm{th}$, where the sum is over resonances $j$ and $\tilde{b}_{j\to i}$ is the effective branching ratio (see Eq.~(\ref{eq:1210.7010:4}) below) for the decay $j{\,\to\,}i$.

The contribution to $\delta N_i$ from a particular resonance $j$ is not only influenced by its mass (through the Boltzmann suppression factor $\sim e^{-E_j/T}$), but also by its spin degeneracy factor $g_j$ and its branching ratio $\tilde{b}_{j\to i}$ into the decay channel that feeds stable particle species $i$. For each stable hadron species $i$, it is therefore a different set of resonances that makes the most important contributions. Our goal is to order the resonances in decreasing order of importance for each stable particle species $i$. We here assume that the conversion surface has constant temperature $T_\mathrm{conv}$. The different hadron resonances have $T_\mathrm{conv}$-dependent nonequilibrium fugacities $\lambda_j$ that ensure constant stable particle ratios equal to their chemical equilibrium values at $T_\mathrm{chem}$ and $\mu_B{\,=\,}0$, independent of the hydro-to-hadron conversion temperature $T_\mathrm{conv}$. While the actual fractions contributed by each resonance to the stable particle 
yields depend on $T_\mathrm{conv}$, the ordering of these fractions is largely $T_\mathrm{conv}$-independent.

We start from the resonance table in the {\tt AZHYDRO} package,\footnote{{\tt AZHYDRO} is available at {\tt http://www.physics.ohio-state.edu/\~{}froderma/}.} which includes  $319$ species of hadrons (counting different isospin states such as $\pi^+$, $\pi^0$, $\pi^-$ as separate species) with rest masses up to $2.25$ GeV. After fixing the value of $T_\mathrm{conv}$ we look up the non-equilibrium fugacity $\lambda_j$ for each of these 319 species
from the EOS s95p-PCE tables constructed in Ref.~\cite{Huovinen:2009yb}. For each stable particle species $i$, we then generate an ordered list of resonances $j$ that can decay directly into $i$.
Note that, in this ordering, we account not only for direct decay contributions, but also for multi-step decay cascades, where $j$ first decays into an unstable resonance $k$ which further decays (directly or through more intermediate steps) into the stable species $i$.

Table \ref{tab:1210.7010:1} shows the beginning of this contribution table for positively charged pions, for a conversion temperature $T_\mathrm{conv}{\,=\,}120$\,MeV. The ``total contribution'' percentages $c_{j\to i}$ in the third column are computed as
\begin{eqnarray}
  c_{j\to i}&=&\frac{N_i^{(j)}}{\sum_{j'}N_i^{(j')}}
                  =  \frac{\tilde{b}_{j{\to}i}N_j^\mathrm{th}}
                            {\sum_{j'} \tilde{b}_{j'\to i} N_{j'}^\mathrm{th}}, 
\label{eq:1210.7010:2}
\\
 N_{j}^\mathrm{th} &=&
 g_j m_j^2 \sum_{k=1}^\infty \frac{(\pm)^{k+1}}{k} \lambda_j^k K_2\left(k\frac{m_j}{T}\right),
 \label{eq:1210.7010:3}
\end{eqnarray}
where the effective branching ratios $\tilde{b}_{j{\to}i}$ in Eq.~(\ref{eq:1210.7010:2}) account for multi-step decay cascades as follows:
\begin{eqnarray}
\label{eq:1210.7010:4}
     \tilde{b}_{j{\to}i} =  b_{j\to i} + \sum_{k_1} b_{j\to k_1} b_{k_1\to i} 
     \nonumber
   + \sum_{k_1,k_2} b_{j\to k_1} b_{k_1\to k_2} b_{k_2\to i} + \dots\, .\\
\end{eqnarray}
%

\begin{table}
\begin{center}
\begin{tabular}{|c|c|c|}
	\hline
	name & mass (GeV) & total contribution (\%) \\ \hline
	$\omega$ & 0.78260 & 15.398  \\ \hline
	$\rho^0$ & 0.77580 & 11.179  \\ \hline
	$\rho^+$ & 0.77580 & 11.098  \\ \hline
	...... & ...... & ......
\end{tabular}
\end{center}
\caption{Example of the $\pi^+$ contribution table for $T_\mathrm{conv}{\,=\,}120$\,MeV.}
\label{tab:1210.7010:1}
\end{table}
%
The sum over $k$ in (\ref{eq:1210.7010:3}) takes care of quantum statistical effects, with the upper (lower) sign for bosons (fermions). For all hadrons except pions, accurate results can be obtained by keeping 
only the first term $k{\,=\,}1$, i.e. by ignoring quantum statistical effects. Even for pions, a few $k$-terms suffice for good precision (in our calculations we truncate the series in (\ref{eq:1210.7010:3}) at $k{\,=\,}10$). The complete ordered resonance decay contribution tables for $\pi^+$, $K^+$, $p$, $\Lambda$, $\Sigma^+$ and $\Xi^-$ are given in the Appendix (\sec{1210.7010:6}). Horizontal lines in the tables indicate where the cumulative resonance decay contributions $c_i^\mathrm{cut}=\sum_{j{=}1}^{j_\mathrm{cut}} c_{j\to i}$ exceed certain threshold percentages (as indicated) of the total resonance decay contribution to species $i$.

In the following section, we show the stable hadron $p_T$-spectra and their anisotropic flow coefficients as functions of these cumulative decay contribution percentages $c_i^\mathrm{cut}$ in order to assess how many resonances from these ordered decay tables should be included for an accurate computation of these observables.

\section{Results and discussion}
\label{sec:1210.7010:3}

Using the ordered tables described in Sec.~\ref{sec:1210.7010:2} and truncating the sum over resonance decay contributions at $j_\mathrm{cut}$ values corresponding to various different cumulative resonance decay contribution thresholds $c_i^\mathrm{cut}$, we performed calculations for $\pi^+$, $K^+$, and $p$. We tested individual bumpy as well as (ellipticity-aligned and ensemble-averaged) smooth initial conditions at both RHIC and LHC energies for a variety of collision centralities. Since the results were found all to be qualitatively similar, we show only a small selection, focussing on pions and protons from one bumpy Pb-Pb event from the $0{-}10\%$ centrality class and from the smooth averaged initial condition corresponding to the $60{-}70\%$ centrality class, both at LHC energy ($\sqrt{s}{\,=\,}2.76\,A$\,GeV).

%
\begin{figure*}[h!]
  \center{
	  \includegraphics[width=0.45\textwidth,height=0.35\textwidth]{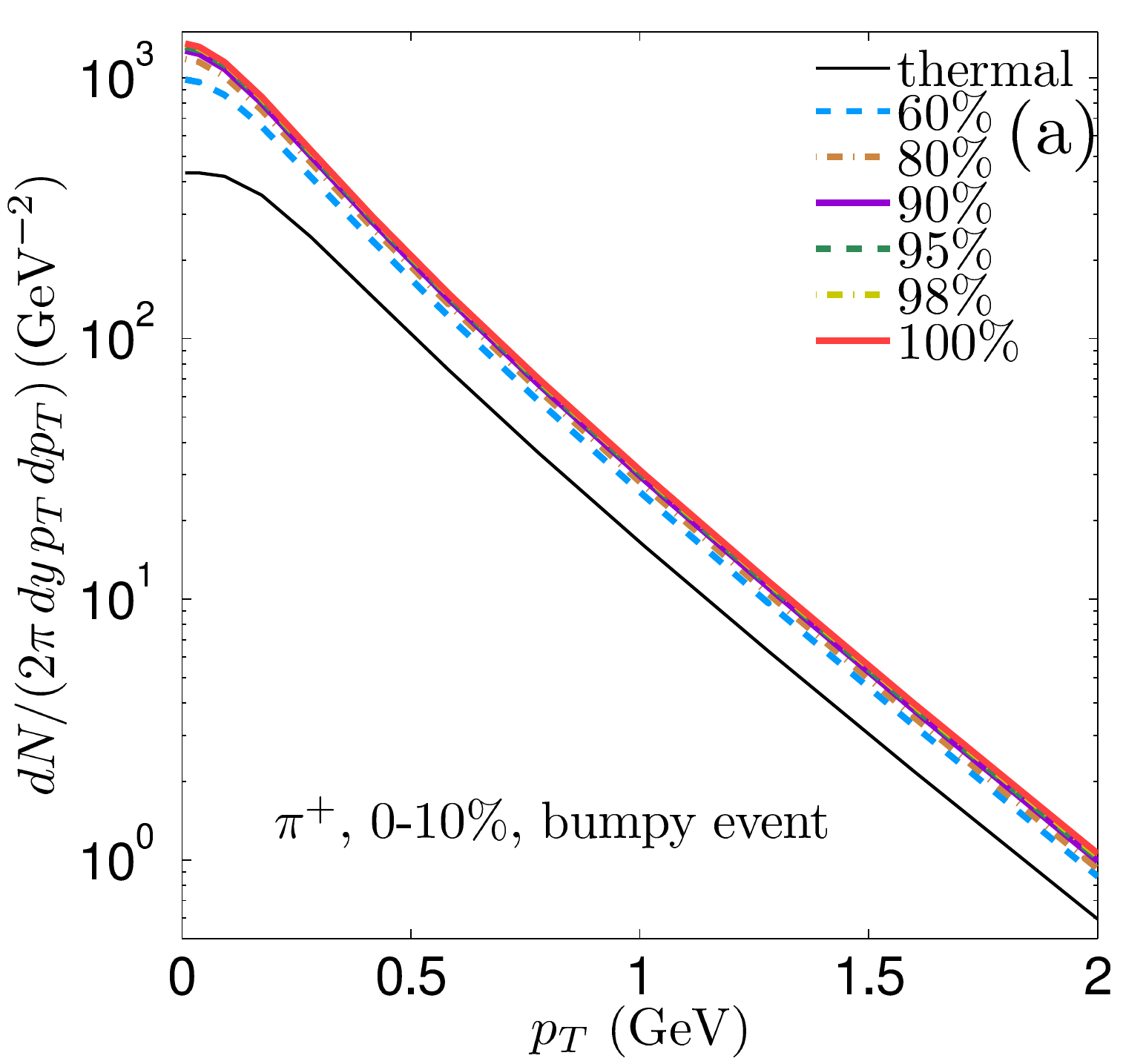}
	  \includegraphics[width=0.45\textwidth,height=0.35\textwidth]{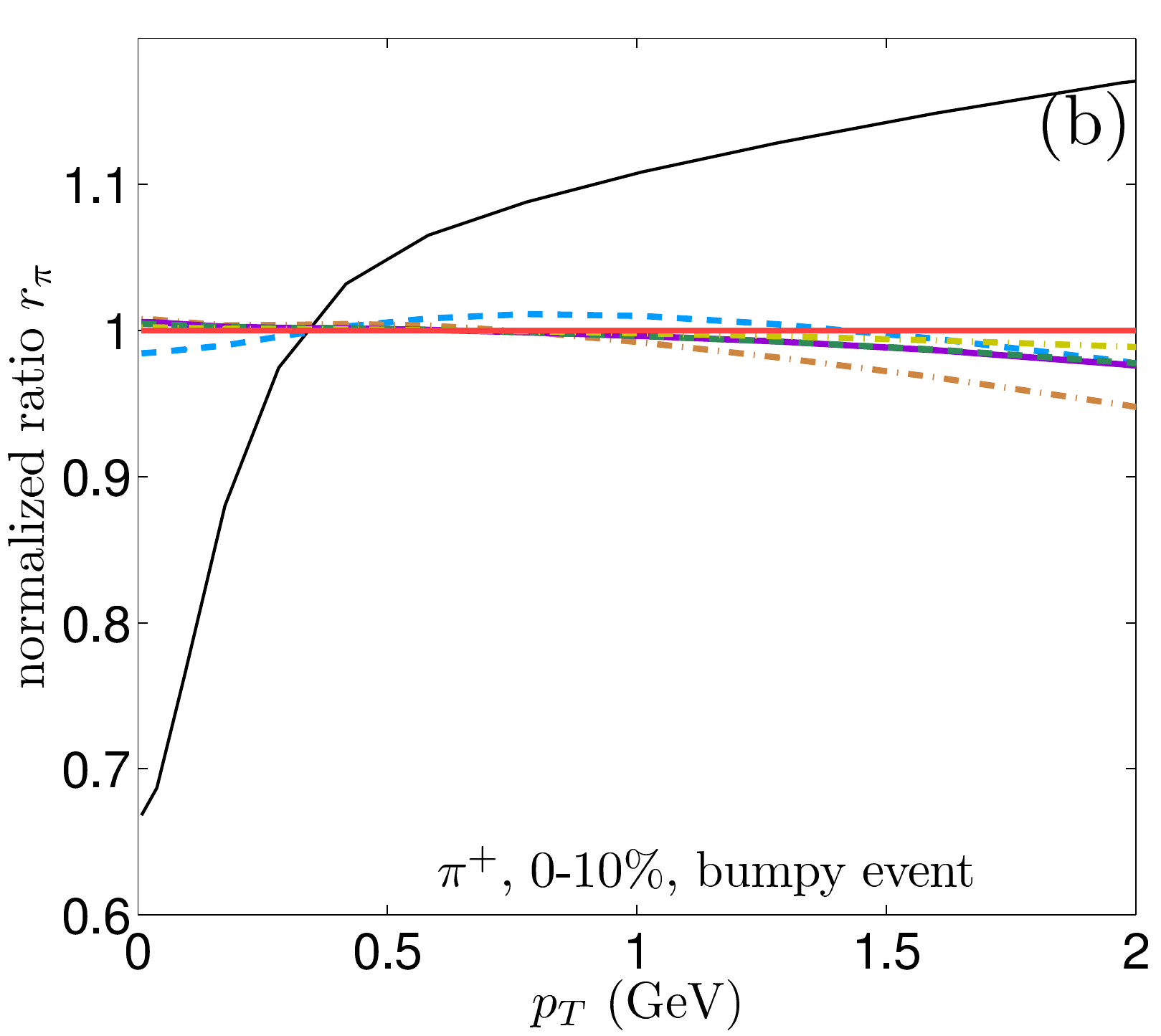}\\
	  \includegraphics[width=0.45\textwidth,height=0.35\textwidth]{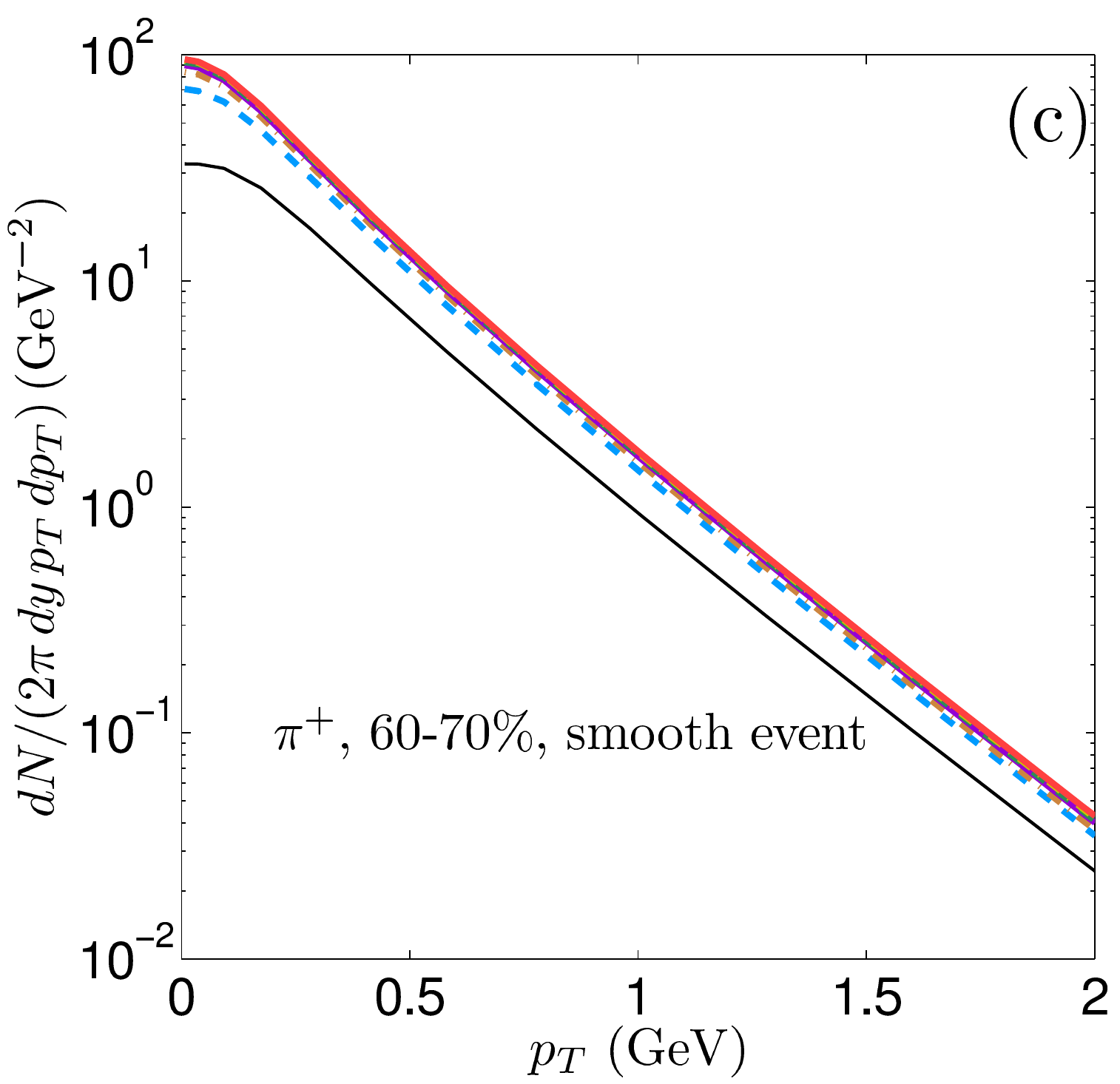}
	  \includegraphics[width=0.45\textwidth,height=0.35\textwidth]{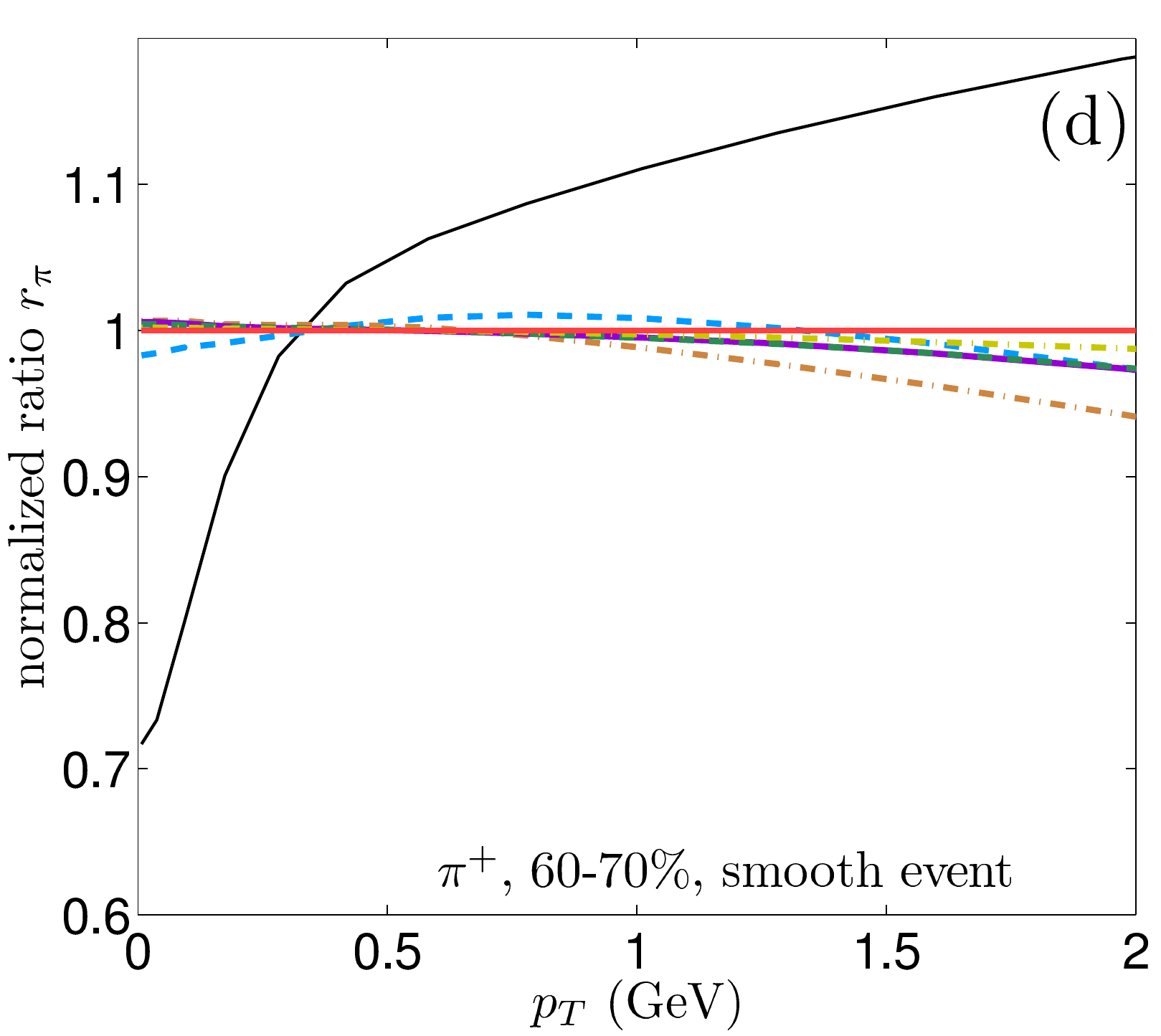}
  }
  \caption{Transverse momentum spectra of $\pi^+$, for a bumpy central 
  (0-10\% centrality, top panels) and a smooth peripheral (60-70\% centrality, bottom panels)
  Pb-Pb collision at LHC energies. 
  Panels (a) and (c) present the absolutely normalized spectra, while panels (b) and (d) show 
  the normalized ratio $r_\pi(p_T)$ defined in Eq.~(\ref{ratio}). Different lines correspond 
  to different cumulative resonance decay contributions between 0\% (``thermal'') and 100\%.
  See text for discussion. From \cite{Qiu:2012tm}.
\label{fig:1210.7010:1}}
\end{figure*}
%

Figure \ref{fig:1210.7010:1} shows the pion $p_T$-spectra for the bumpy central collision in the upper panels and the smooth peripheral event in the lower panels. The left panels show the usual semilogarithmic plots of the absolutely normalized $p_T$-distribution. As is well-known, the directly emitted (``thermal'') pions constitute only about 50-60\% of all observed pions, the rest coming from resonance decays. The ``thermal" spectrum also has the wrong shape: resonance-decay pions predominantly contribute to the low-$p_T$ part of the spectrum, making it steeper. However, this shape difference between the truncated and full resonance-decay spectrum disappears almost completely already when including only the 9 strongest decay channels, accounting for just 60\% of the total pion yield from resonance decays. This is shown in the right panels of Fig.~\ref{fig:1210.7010:1} where we plot the ratio 
\begin{equation}
\label{ratio}
r_i(p_T) = \frac{
                 \frac{dN_i^\mathrm{th}}{dyp_Tdp_T} + 
                 \frac{1}{\sum_{j{=}1}^{j_\mathrm{cut}}c_{j{\to}i}} \sum_{j{=}1}^{j_\mathrm{cut}}
                 \frac{dN_i^{(j)}}{dyp_Tdp_T}}
                 {\frac{dN_i^\mathrm{th}}{dyp_Tdp_T} + 
                 \sum_{j{=}1}^{j_\mathrm{max}}
                 \frac{dN_i^{(j)}}{dyp_Tdp_T}}
\end{equation}
for $i{\,=\,}\pi$ as a function of $p_T$. ($N_i^{(j)}$ is the contribution to particle species $i$ from decays of particle species $j$ (see Eq.~(\ref{eq:1210.7010:2})), and $j_\mathrm{max}$ is the index of the last resonance in the ordered resonance decay table from Sec.~\ref{sec:1210.7010:2}.) The numerator includes only resonance decays up to $j_\mathrm{cut}$, but we renormalize those decay contributions by the cumulative decay contribution $c_i^\mathrm{cut}$ corresponding to the same $j_\mathrm{cut}$ value. ($c_i^\mathrm{cut}$ is easily calculated from Eqs.~(\ref{eq:1210.7010:2},\ref{eq:1210.7010:3}) and directly obtained by summing the entries in the third column of the resonance decay table.) This renormalization corrects for the missing yield from the truncation of the decay table. The remaining effect (after missing yield renormalization) of the truncation on the {\em shape} of the $p_T$-spectrum is seen in panels (b) and (d) of Fig.~\ref{fig:1210.7010:1}: Whereas without any resonance decays the ratio $r_\pi(p_T)$ changes by almost a factor 2 between $p_T{\,=\,}0$ and 2\,GeV, this variation is reduced to less than 5\% already for $c_\pi^\mathrm{cut}{\,=\,}60\%$ for both bumpy and smooth initial conditions in both central and peripheral collisions.

%
\begin{figure*}[h!]
  \center{
	  \includegraphics[width=0.45\textwidth,height=0.35\textwidth]{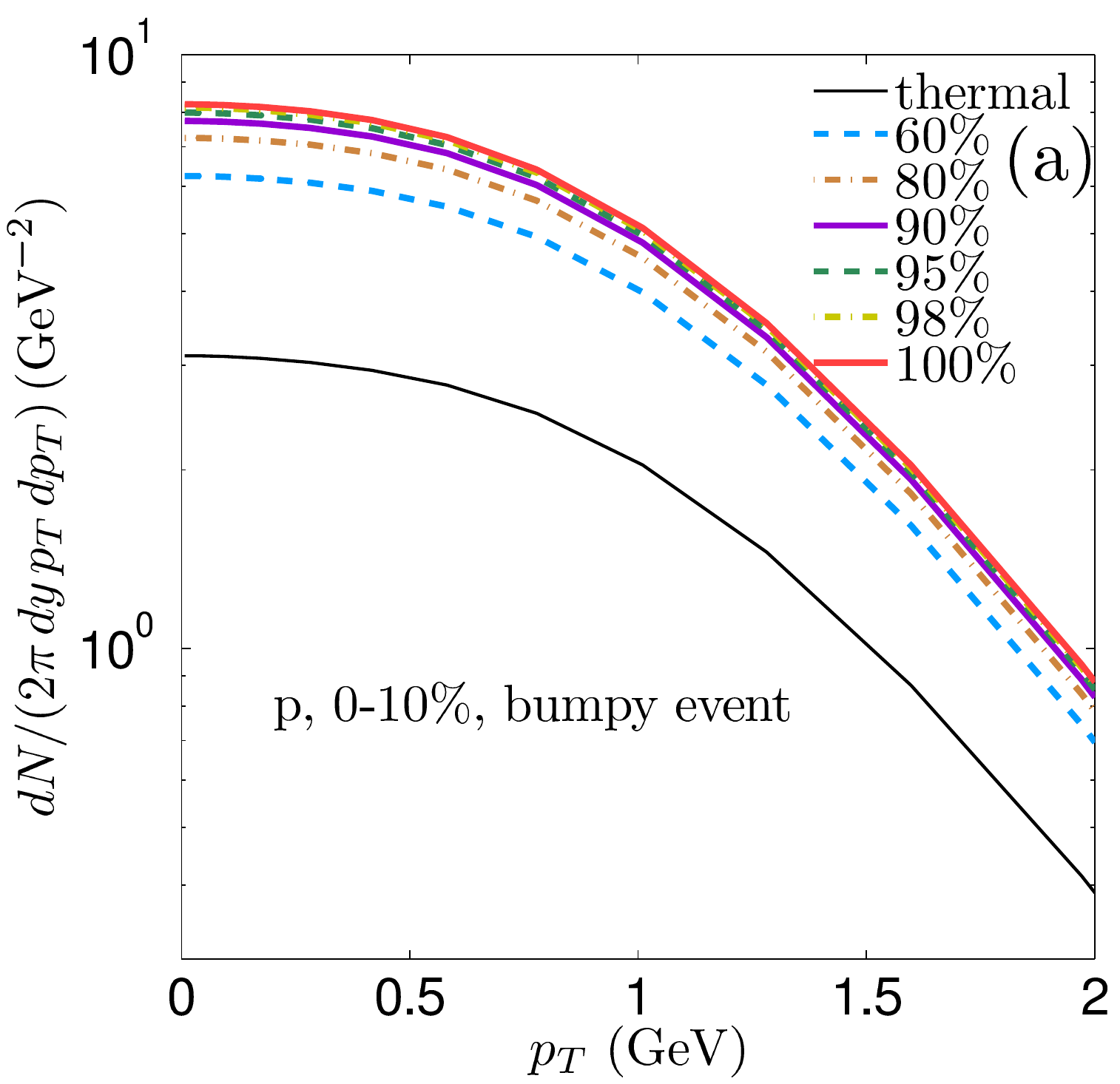}
	  \includegraphics[width=0.45\textwidth,height=0.35\textwidth]{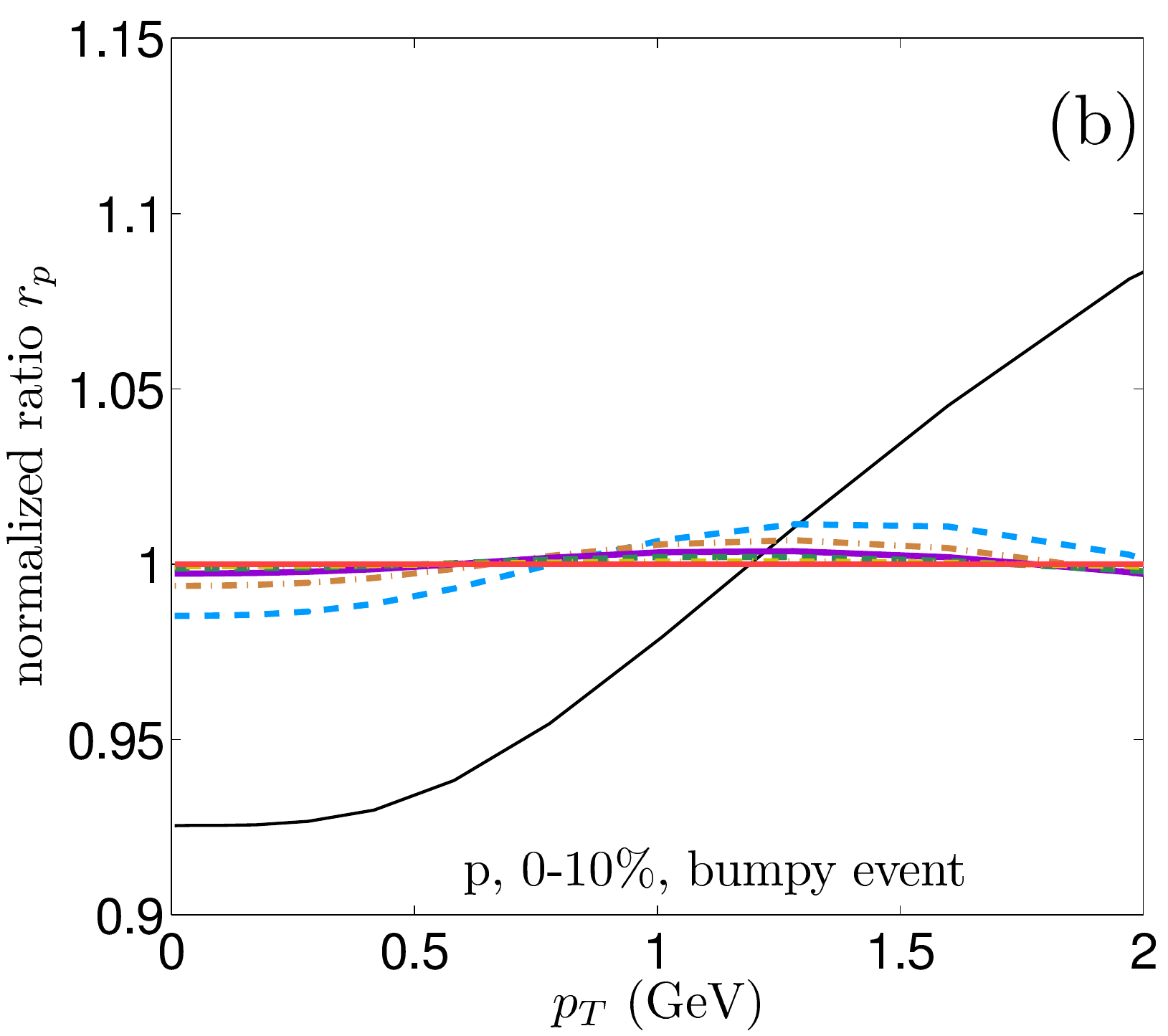}\\
	  \includegraphics[width=0.45\textwidth,height=0.35\textwidth]{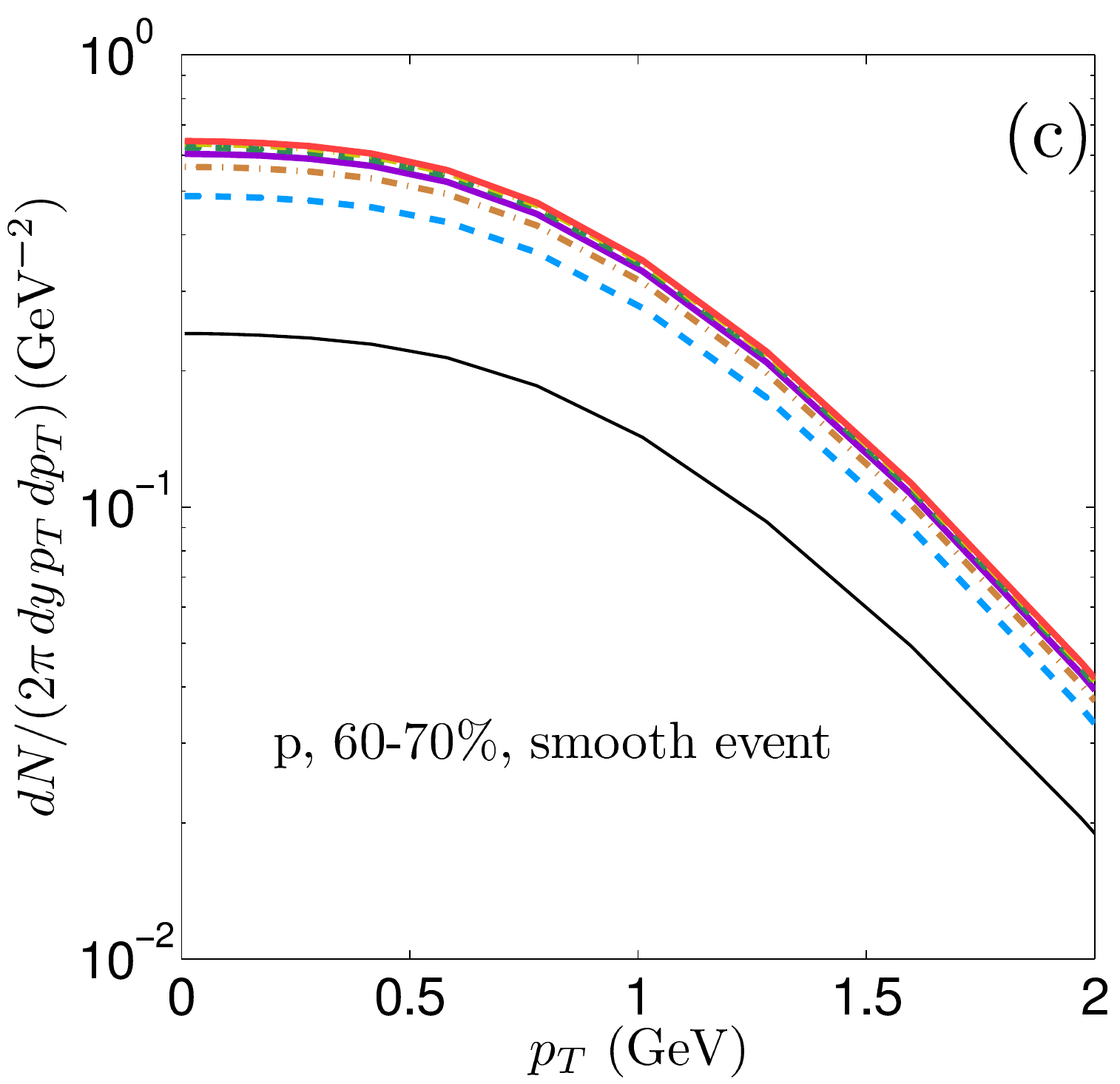}
	  \includegraphics[width=0.45\textwidth,height=0.35\textwidth]{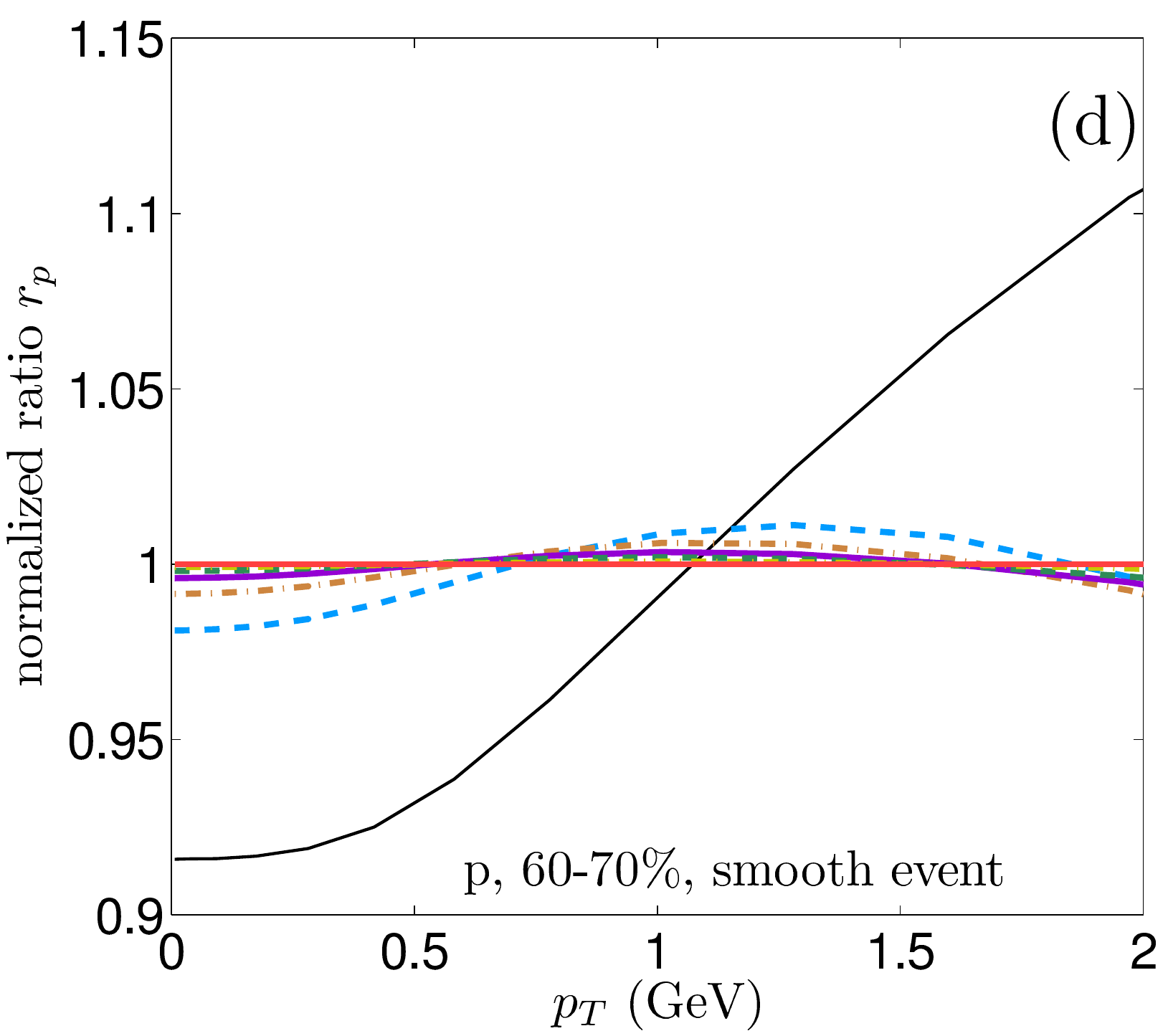}
  }
  \caption{Same as Fig.\ref{fig:1210.7010:1}, but for protons. From \cite{Qiu:2012tm}.
  \label{fig:1210.7010:2}}
\end{figure*}
%

In Fig.~\ref{fig:1210.7010:2}, we show in the same way the proton spectra. Again the shape of the spectra can be accurately reproduced by taking into account a small fraction of all decay contributions (note the expanded vertical scale in Figs.~\ref{fig:1210.7010:2}b,d): after renormalization to account for the missing yield, just the 4 strongest of 75 decay channels (three charge states of the $\Delta(1232)$ resonance and one charge state of $\Delta(1600)$), corresponding to 60\% of the total resonance decay yield for protons, reproduce the full proton spectrum with ${<\,}5\%$ error between $p_T{\,=\,}0$ and 2\,GeV.

We conclude that, by accounting for the missing yield through appropriate renormalization, the correctly normalized total pion and proton spectra can be obtained, with shape errors ${<\,}5\%$, by including only the strongest decay channels accounting for the leading 60\% of the total resonance decay yields. A quick look at the tables in the Appendix shows that this will reduce the number of resonance decays (and thus computer time) by at least a factor 10.

\begin{figure*}[h!]
  \center{
	  \includegraphics[width=0.45\textwidth, height=0.35\textwidth]{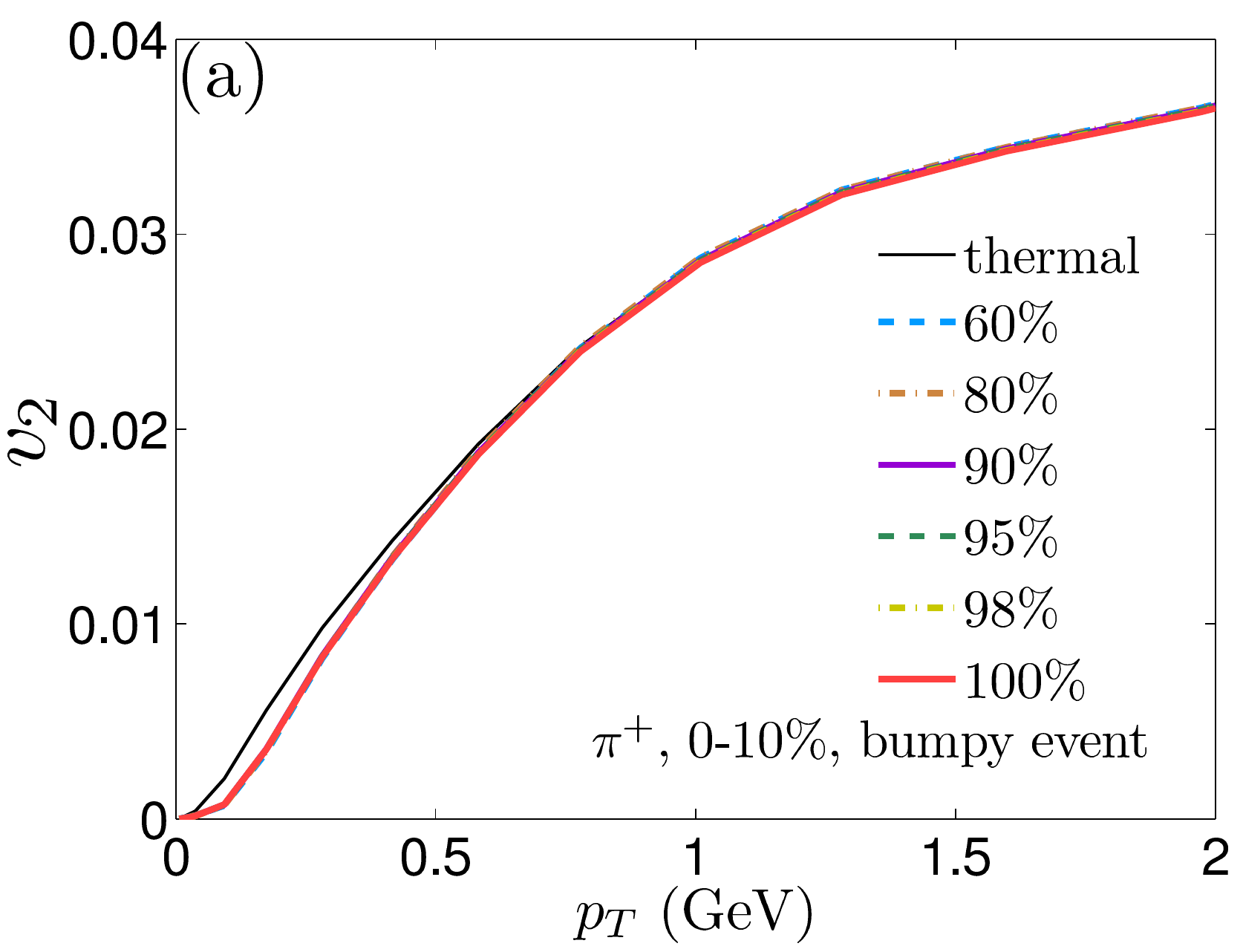}
	  \includegraphics[width=0.45\textwidth, height=0.35\textwidth]{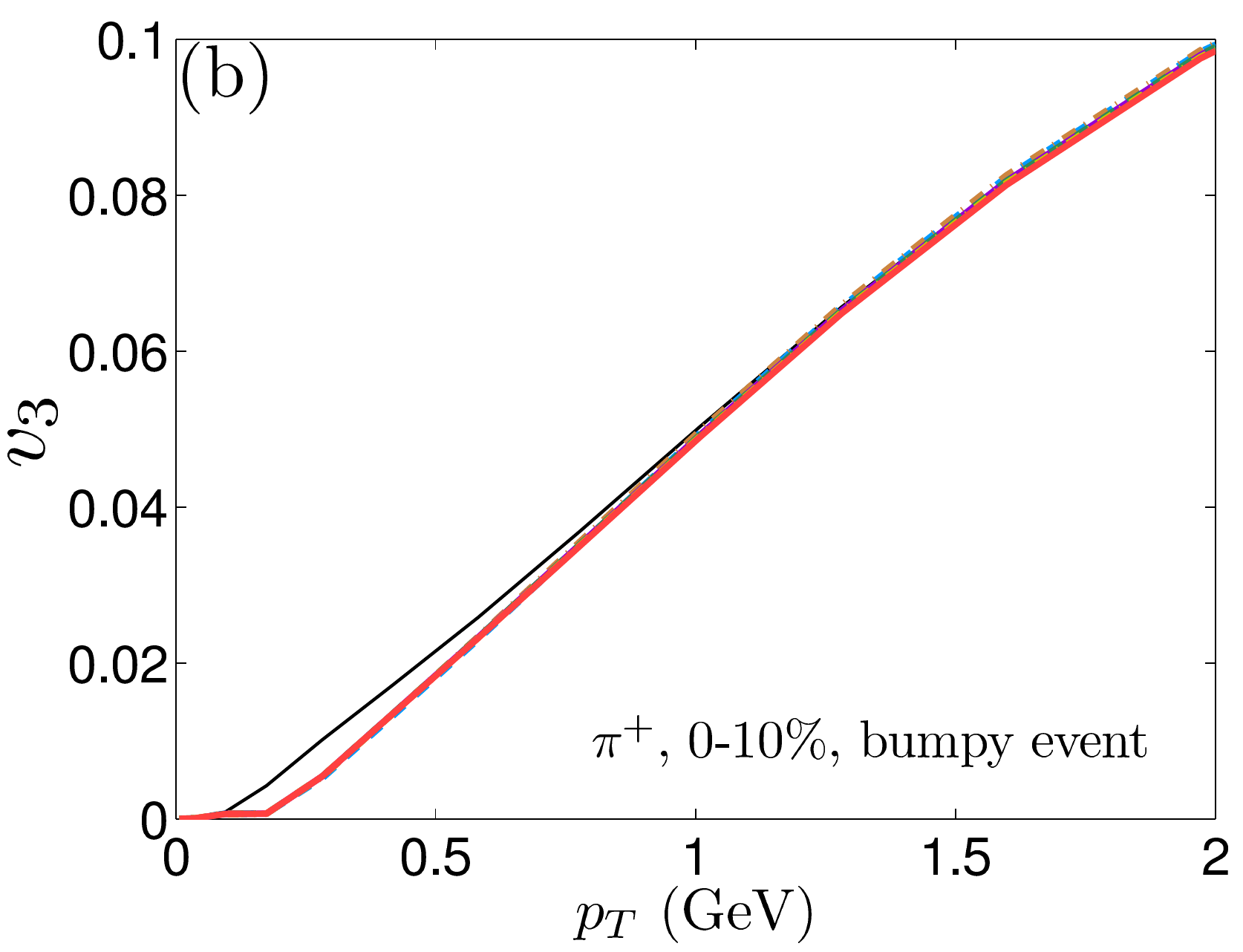}\\
	  \includegraphics[width=0.45\textwidth, height=0.35\textwidth]{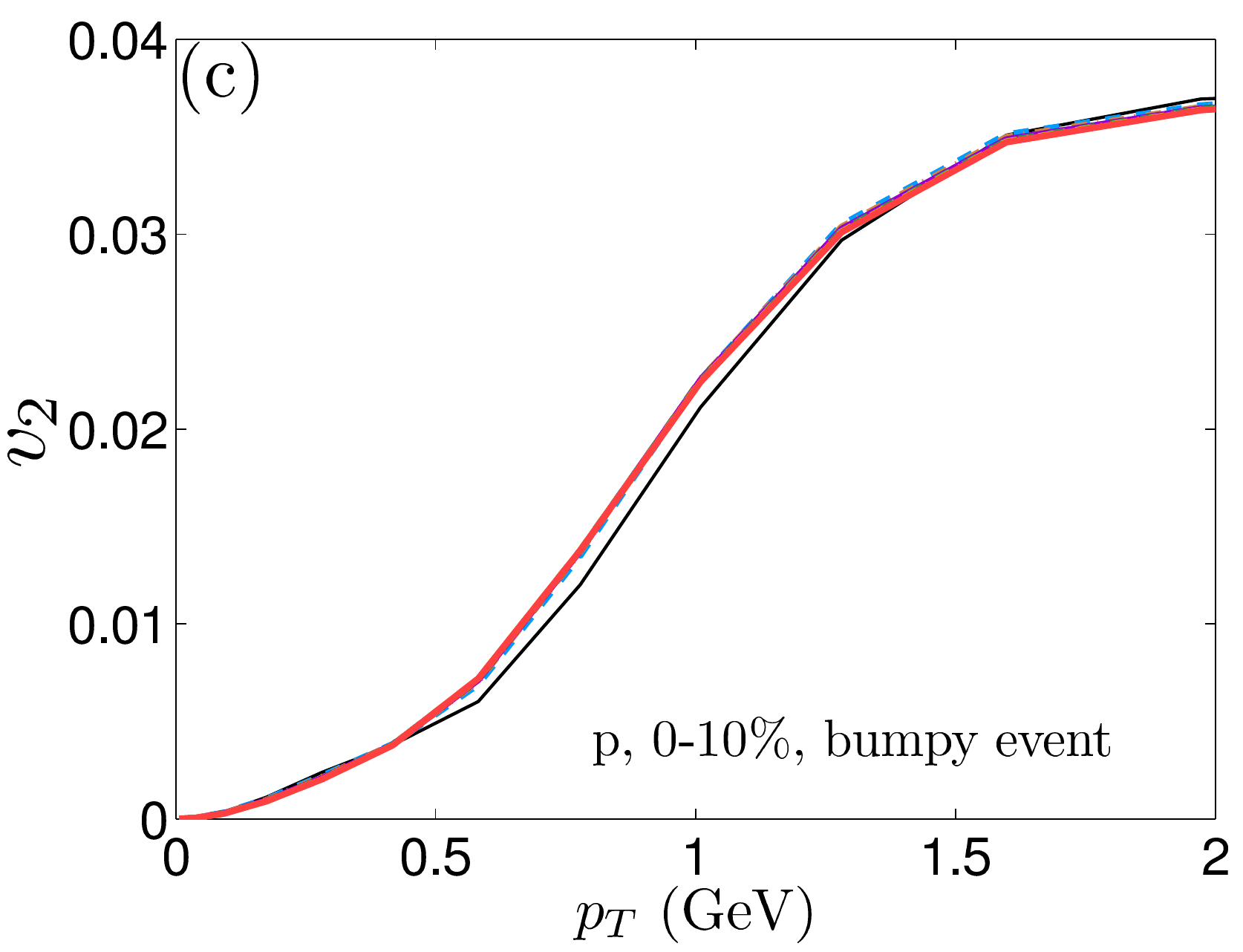}
	  \includegraphics[width=0.45\textwidth, height=0.35\textwidth]{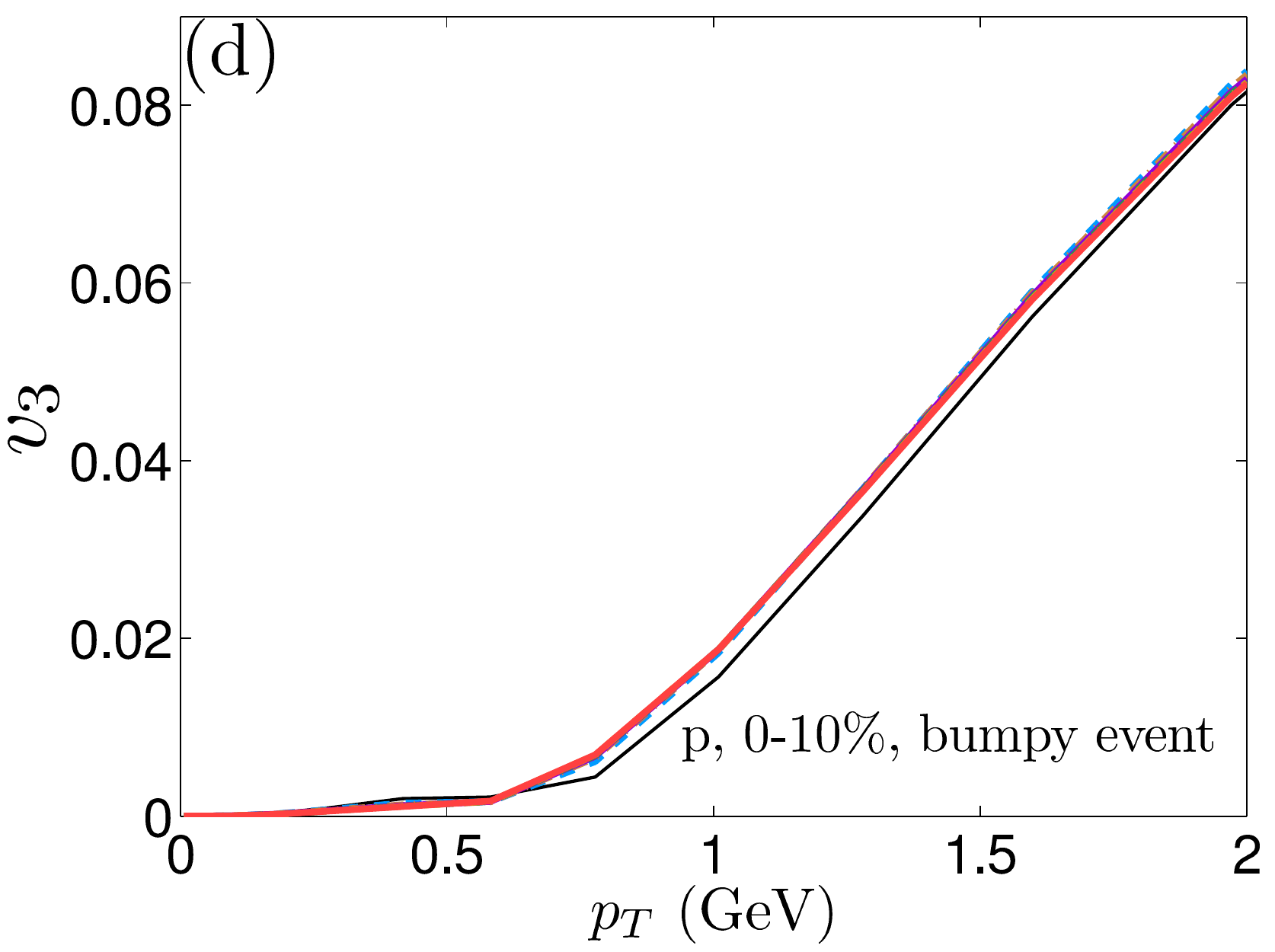}
  }
  \caption{The differential elliptic ($v_2$, left panel) and triangular flow ($v_3$,
   right panel), for $\pi^+$ (upper panels) and $p$ (lower panels), for one bumpy Pb-Pb 
   event from the $0{-}10\%$ centrality class at LHC energy. As in Fig.~\ref{fig:1210.7010:1}, lines of
   different styles and colors correspond to different cumulative resonance decay fractions. From \cite{Qiu:2012tm}.
  \label{fig:1210.7010:3}}
\end{figure*}

\begin{figure*}[h!]
  \center{
	  \includegraphics[width=0.45\textwidth, height=0.35\textwidth]{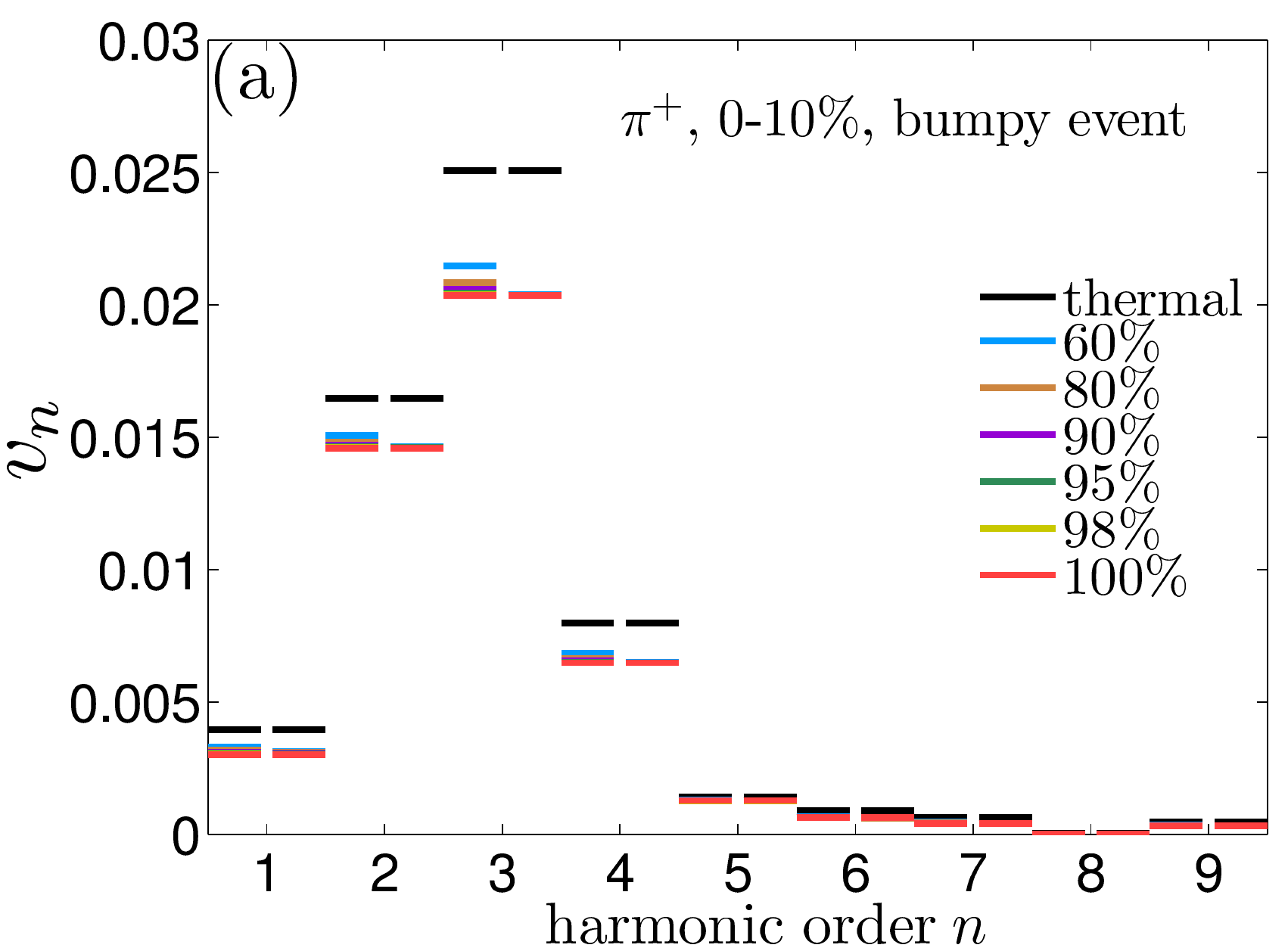}
	  \includegraphics[width=0.45\textwidth, height=0.35\textwidth]{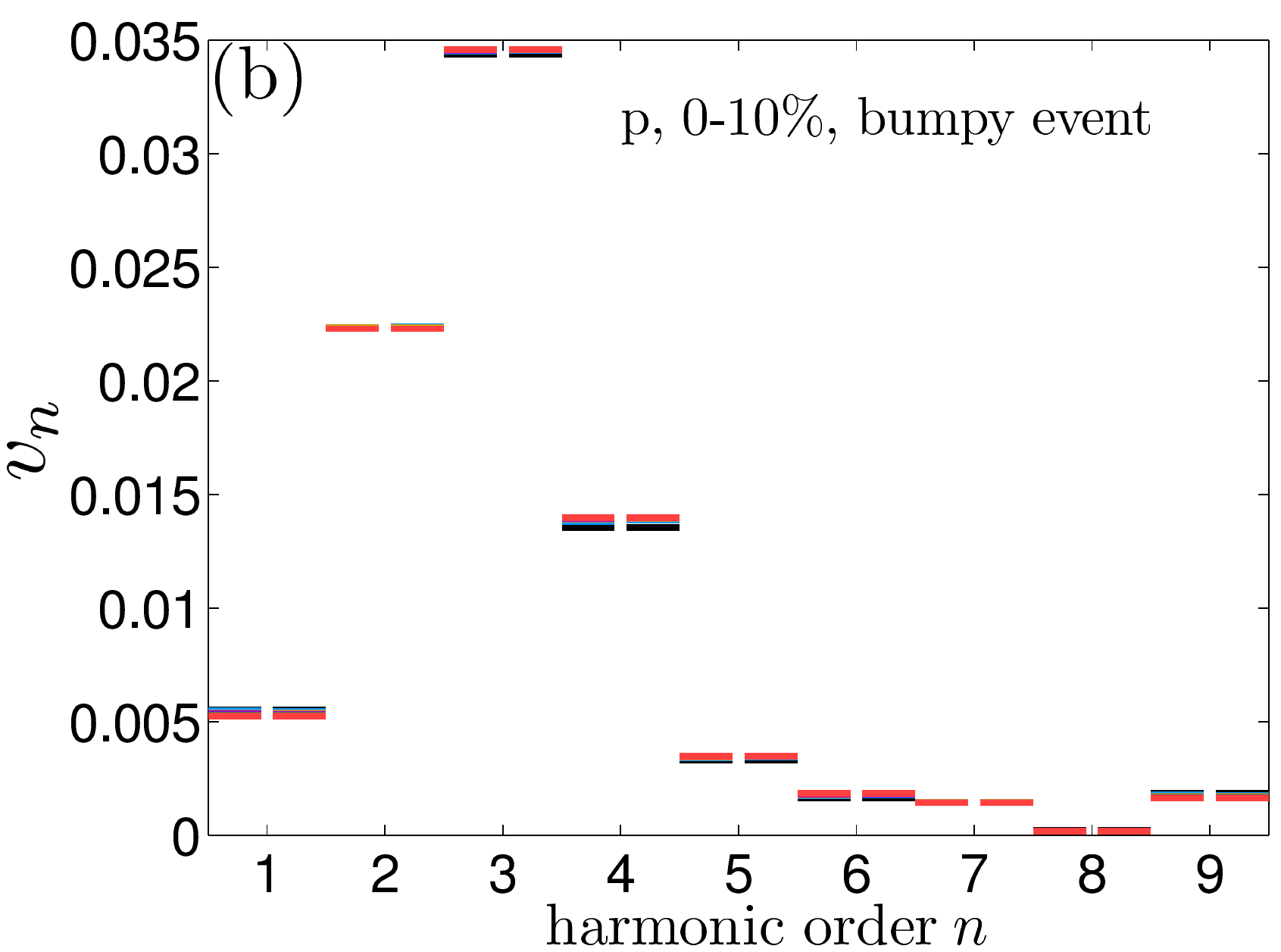}\\
	  \includegraphics[width=0.45\textwidth, height=0.35\textwidth]{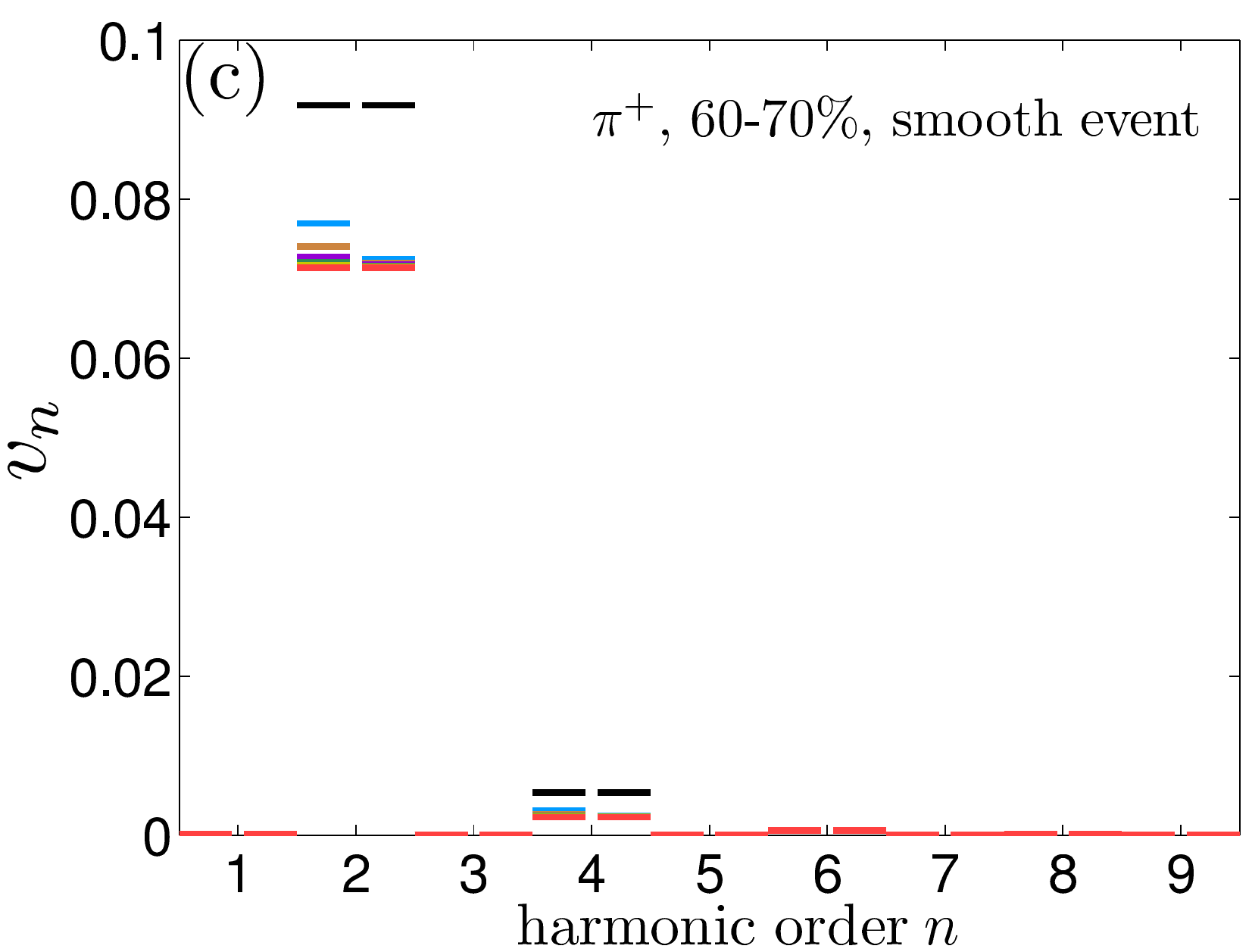}
	  \includegraphics[width=0.45\textwidth, height=0.35\textwidth]{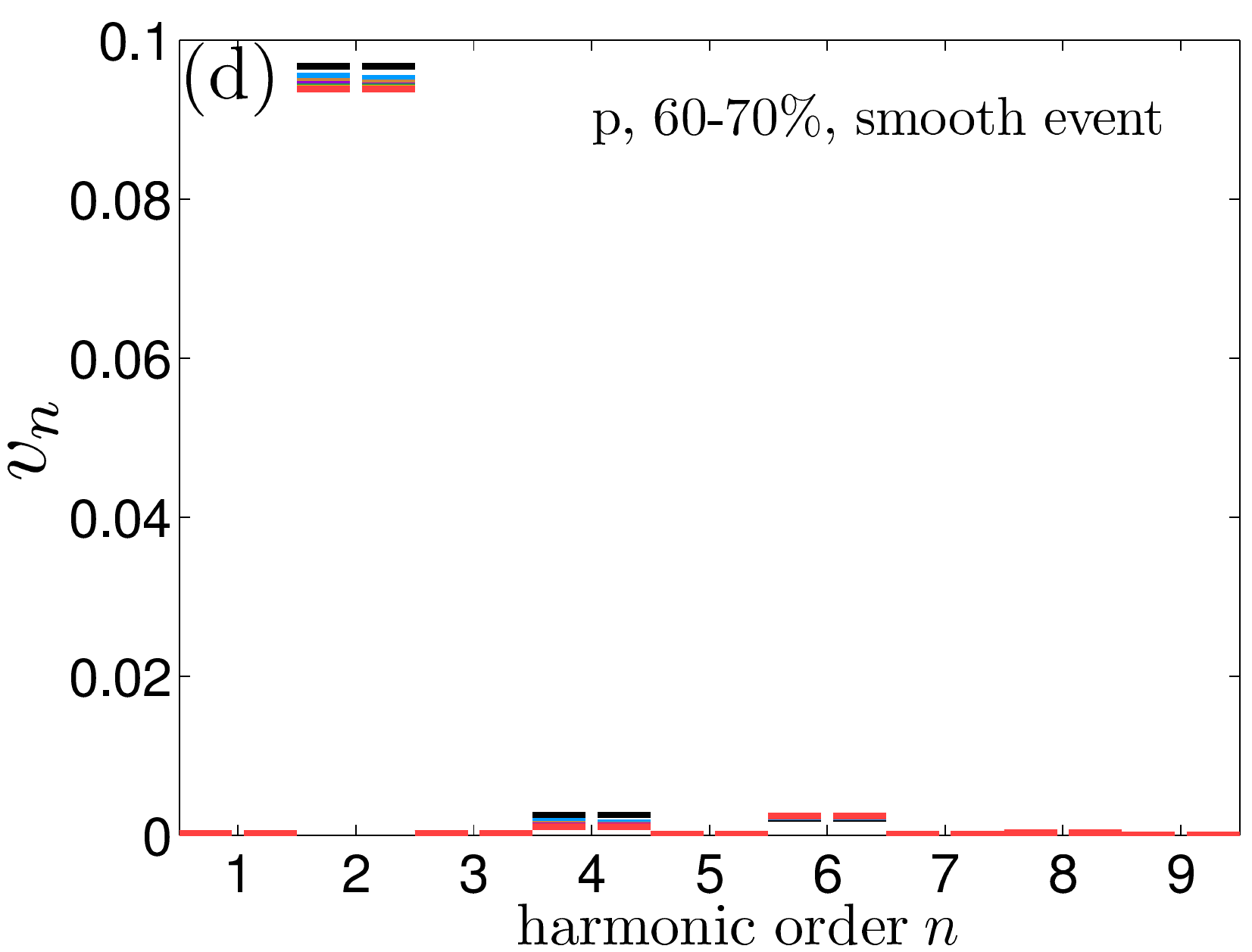}
  }
  \caption{The $p_T$-integrated anisotropic flow coefficients $v_n$, 
  $n{\,=\,}1,\dots,9$, for $\pi^+$ (a,c) and $p$ (b,d), for a bumpy central (0-10\%) collision 
  event (a,b) and a smooth averaged peripheral (60-70\%) collision event (c,d). Line colors 
  and styles as in Figs.~\ref{fig:1210.7010:1}-\ref{fig:1210.7010:3}. For a discussion of the two sets of lines for 
  each harmonic order $n$ see text. From \cite{Qiu:2012tm}.
  \label{fig:1210.7010:4}}
\end{figure*}

We now proceed to a discussion of the differential and $p_T$-integrated anisotropic flow coefficients $v_n$ defined by \equ{1104.0650:19} and \equ{1104.0650:20}.
%
%
In the formula, the spectrum $dN/(dy p_T dp_T d\phi_p)$ includes all contributions from the ordered resonance decay table for the considered stable species up to a certain threshold $j_\mathrm{cut}$, with the truncated resonance decay contribution renormalized for the missing yield by a factor $1/c_i^\mathrm{cut}$ as shown in the numerator of Eq.~(\ref{ratio}). In Figs.~\ref{fig:1210.7010:3} and \ref{fig:1210.7010:4}, we specify the cumulative decay contribution percentage $c_i^\mathrm{cut}$ to indicate the truncation level corresponding to each curve.

%
%

Figure~\ref{fig:1210.7010:3} shows the differential elliptic and triangular flows for pions and protons, for one single bumpy central (0-10\% centrality) event. We see that once again excellent agreement with the full resonance decay calculation is already obtained when including only the small subset of resonances that account for the top 60\% of the resonance decay yields. We checked that this result is generic, i.e. it does not depend on the selected event (although the elliptic and triangular flows do).

%
%

For the $p_T$-integrated harmonic flow coefficients $v_n$, we show in Fig.~\ref{fig:1210.7010:4} results for all harmonic orders from $n{\,=\,}1$ to 9, again for pions and protons and for a bumpy central as well as a smooth peripheral event. For the smooth averaged initial condition, the odd harmonics vanish by symmetry. For fluctuating initial conditions, the $v_n$ values shown here and their relative size depend on the randomly selected event. All plots shown in this chapter are based on one and the same bumpy central collision event. 

For each harmonic order $n$, Fig.~\ref{fig:1210.7010:4} shows two sets of results. The left set corresponds to results obtained by using the truncated resonance decay spectra shown in Figs.~\ref{fig:1210.7010:1}a,c and \ref{fig:1210.7010:2}a,c, without missing yield renormalization. The right set
uses the renormalized truncated decay spectra as defined in the numerator of Eq.~(\ref{ratio}). One observes a much faster convergence towards the full result in the right sets than in the left sets. The reason is that, by renormalizing the truncated resonance decay contributions for the missing yield, the correct mixing ratio between direct thermal and indirect decay contributions is ensured and the shape of the total $p_T$-spectrum is approximated much more accurately than without renormalization (see Figs.~\ref{fig:1210.7010:1}b,d and \ref{fig:1210.7010:2}b,d). Figure~\ref{fig:1210.7010:4} demonstrates that, when using the renormalized truncated decay spectra, accounting for just the top 60\% decay contributions (i.e. including only the 9 strongest decay channels contributing to pions and the 4 strongest decay channels contributing to the proton spectra) reproduces the full results for the harmonic flow coefficients $v_n$ with excellent precision: the lines corresponding to different $c_i^\mathrm{cut}$ values ${\geq}60\%$ are almost indistinguishable.   

Future precision extractions of the QGP viscosity may require highly precise $v_n$ values. For such a purpose one can adjust $j_\mathrm{cut}$ to include a larger fraction of all resonance decays if needed.

For a given precision, the required minimal $j_\mathrm{cut}$ truncation indices and cumulative resonance decay fractions $c_\mathrm{K}^\mathrm{cut}$ for kaons lie between those for pions and protons. The $c_i^\mathrm{cut}$ for $i{\,=\,}\pi$,\,K,\,$p$ are almost identical at RHIC and LHC energies, i.e.\ only weakly sensitive to radial flow.

\section{Chapter conclusions}
\label{sec:1210.7010:4}

In this chapter, we showed that for a sufficiently accurate determination of the differential anisotropic flow coefficients $v_n(p_T)$, only those resonances need to be included that generate the top 60\% of the largest decay contributions to the stable particle yields. For the single particle spectra, correct normalization of the total yield requires a renormalization of the truncated resonance decay yield as given in the numerator of Eq.~(\ref{ratio}). With this renormalization, good convergence of the slope of the pion spectra and of the $p_T$-integrated anisotropic flow coefficients $v_n$ requires inclusion of only the 9 strongest contributing channels for pions and only the 4 strongest channels for protons, accounting in both cases for just 60\% of the total decay yield. This reduces the number of resonance decay channels to be evaluated by a factor ${>}10$, without loss of precision, leading to a similar reduction of the total computing time for the final stable hadron distributions.   

In hybrid model calculations \cite{Song:2010aq}, the late hadronic stage is described microscopically by a Boltzmann cascade that propagates a reduced set of resonances until final kinetic decoupling. In this case, the spectra of all unstable resonances that are explicitly included in the Boltzmann cascade must be generated on the conversion surface. This is still only a small subset of all resonances included in the resonance decay tables. The optimal ordering of the resonance decay tables for the purpose of generating input for the late-stage Boltzmann cascade and the corresponding optimized truncation fractions $c_i^\mathrm{cut}$ are left as an open question.


\chapter{Fluctuating flow angles and anisotropic flow measurements}
\label{chap:flow_angle}

Event-by-event fluctuations in the initial density distributions of the fireballs created in relativistic heavy-ion collisions lead to event-by-event fluctuations of the final anisotropic flow angles, and density inhomogeneities in the initial state cause these flow angles to vary with the transverse momentum of the emitted particles. In this chapter, we show that these effects lead to characteristically different transverse momentum dependencies for anisotropic flow coefficients extracted from different experimental methods. These differences can be used to experimentally constrain flow angle fluctuations in the final state of heavy-ion collisions which, in turn, are sensitive to the initial-state density fluctuations and the shear viscosity of the expanding fireball medium.

The material in this chapter is based on \cite{Heinz:2013bua}; although for completeness, the whole material is included, only part of the content was contributed by the author.

\section{Chapter introduction}
\label{sec:1302.3535:1}

Due to quantum fluctuations of the positions of the nucleons inside the colliding nuclei and of the positions of the colored quark and gluon constituents inside each nucleon, the density of the fireball matter created in collisions between ultra-relativistic heavy ions is highly inhomogeneous in the transverse plane, even for collisions with identical impact parameters. It has been shown in this thesis that the initial density inhomogeneities lead to highly anisotropic pressure gradients, causing an anisotropic collective expansion of the fireball, whose harmonic flow coefficients $v_n$ and associated flow angles $\Psi_n$ (both defined below) fluctuate from collision to collision \cite{Alver:2008zza}.

While $v_n$ fluctuations and the effect of their variance on different methods for measuring $v_n$ have been studied extensively over the last few years, flow angle fluctuations and correlations have only recently found attention \cite{Qiu:2011iv,Teaney:2010vd,Gardim:2011xv,Teaney:2012ke,Jia:2012ma,Jia:2012sa,Qiu:2012uy,Ollitrault:2012cm,Gardim:2012im}. Gardim {\it et al.} \cite{Gardim:2012im} pointed out that, since the fluctuating flow angles $\Psi_n$ depend on transverse momentum $p_T$ and rapidity $y$, the usually assumed (and experimentally observed \cite{Alver:2010rt,Aamodt:2011by,Chatrchyan:2012wg,ATLAS:2012at}) factorization of the azimuthal oscillation amplitudes of the two-particle angular correlations into a product of single-particle flow coefficients is slightly broken even if these correlations are entirely due to collective flow. We show here that the $p_T$-dependence and fluctuating nature of the flow angles $\Psi_n$ also affects the $p_T$-dependence of the experimentally measured differential flow coefficients $v_n(p_T)$, and that it does so in different ways for different experimental methods of determining $v_n(p_T)$. 

For each collision event, the momentum distribution of finally emitted particles can be characterized by a set of harmonic flow coefficients $v_n$ and flow angles $\Psi_n$ through the complex quantities (see also \chap{1104.0650})
\begin{eqnarray} 
  \label{1302.3535:eq1}
  V_n &=& v_n e^{in\Psi_n} 
  := \frac{\int \pT d\pT d\phi\, e^{i n \phi} \,\frac{dN}{dy \pT d\pT d\phi}}
                  {\int \pT d\pT d\phi\, \frac{dN}{dy \pT d\pT d\phi}}
  \nonumber\\
  &\equiv&\{e^{in\phi}\},
  \\
\label{1302.3535:eq2}
   V_n(\pT) &=& v_n(\pT) e^{in\Psi_n(\pT)} 
   := \frac{\int d\phi \, e^{i n \phi}\, \frac{dN}{dy \pT d\pT d\phi}} 
                   {\int d\phi \, \frac{dN}{dy \pT d\pT d\phi}}
   \nonumber\\
   &\equiv&\{e^{in\phi}\}_{\pT}.
\end{eqnarray}

\begin{figure*}[h!]
  \begin{center}
    \includegraphics[width=0.335\linewidth]{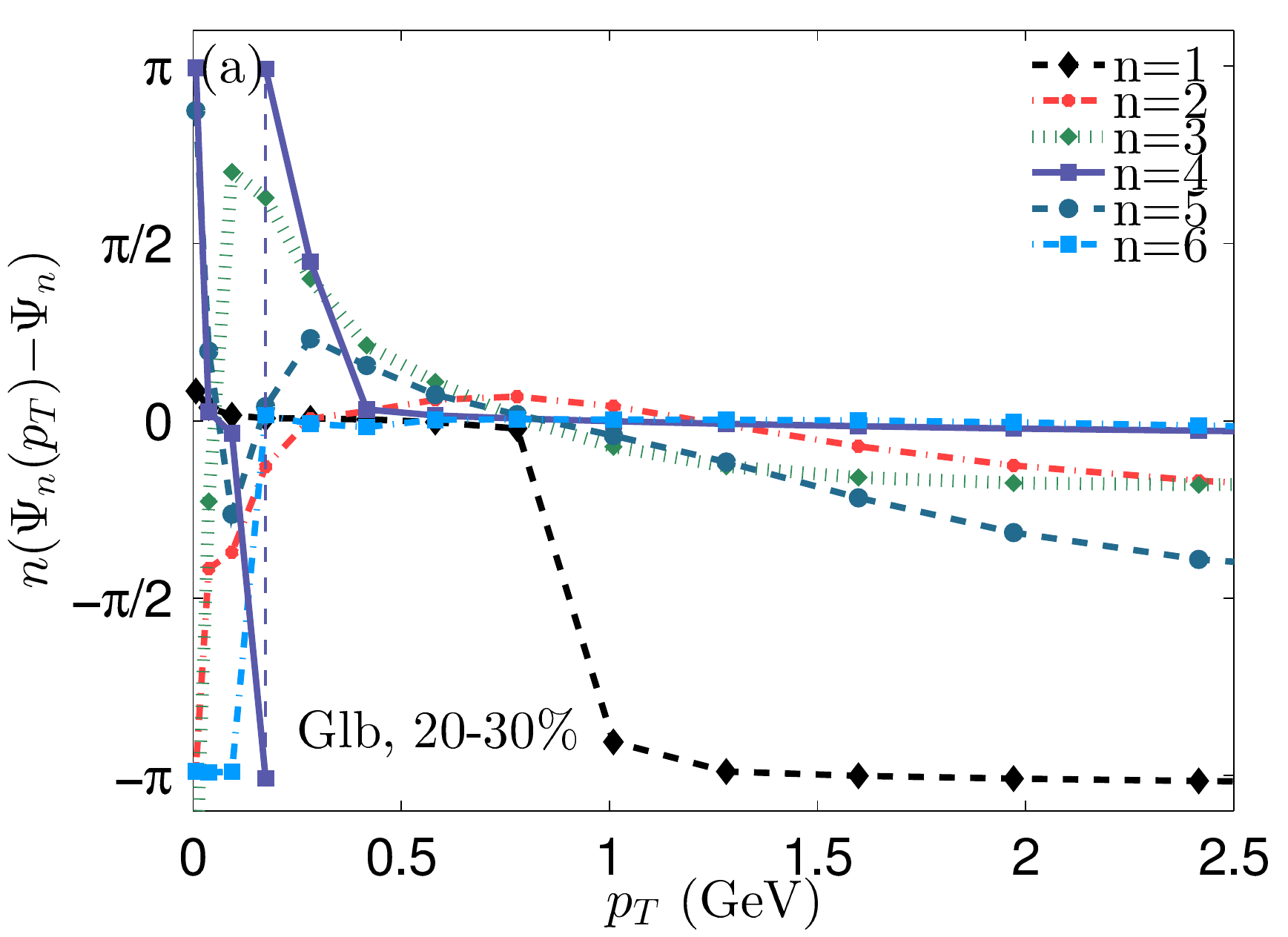}
    \includegraphics[width=0.32\linewidth]{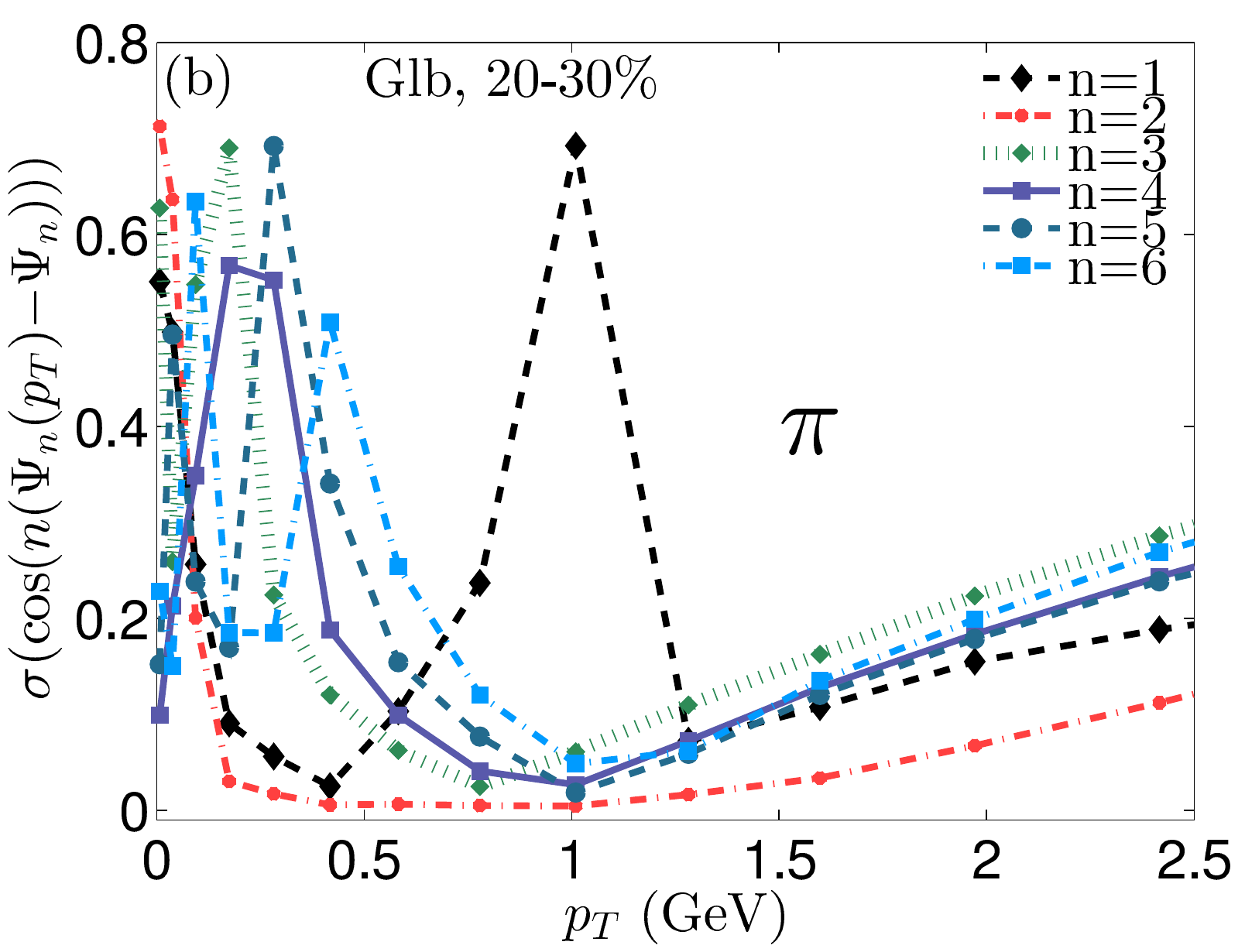}
    \includegraphics[width=0.32\linewidth]{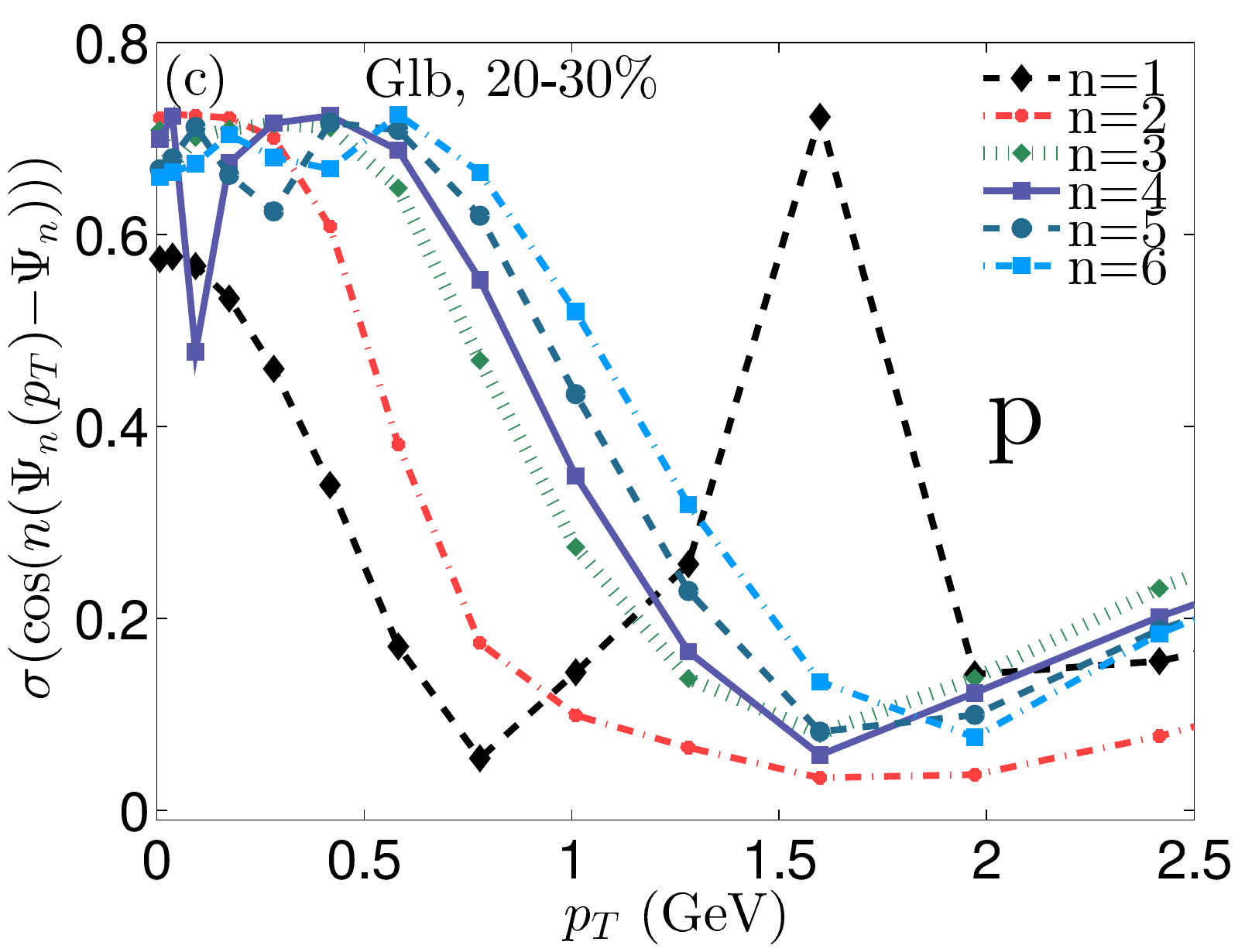}
   \end{center}
  \caption{(a) $\pT$ dependence of the flow angles $\Psi_n(\pT)$ relative to 
    their average value $\Psi_n$ ($n{\,=\,}1,\dots,6$), for pions from a single but
    typical Pb+Pb collision event at LHC energies in the $20{-}30\%$ centrality class, computed 
    with the viscous hydrodynamic code {\tt VISH2{+}1} using an initial density profile from the 
    MC-Glauber model and $\eta/s{\,=\,}0.08$. Note the flip of the dipolar ($n{\,=\,}1$)  
    flow angle from 0 to $-\pi$ (happening around $\pT{\,\sim\,}0.9$\,GeV in this particular 
    event); this reflects the constraint from total transverse momentum conservation 
    \cite{Retinskaya:2012ky}. (b,c) The variance $\sigma$ of the cosine of the flow angle 
    fluctuations (which is free of ambiguities of the angles by multiples of $2\pi/n$) as a function
    of $\pT$, for pions (b) and protons (c) from 11,000 hydrodynamic events in the same 
    centrality class from which the event in (a) was taken. The variance is minimal around 
    $\pT{\,=\,}1$\,GeV for pions and around $\pT{\,=\,}1.5{-}2$\,GeV for protons (except for the 
    dipolar flow angle which flips by $\pi$ exactly in that momentum region, see panel (a)). It is
    smallest for the elliptic flow angle fluctuation $\Psi_2(\pT){-}\Psi_2$, and it becomes large for 
    all flow harmonics at low $\pT$. From \cite{Heinz:2013bua}.
    }
  \label{1302.3535/F1}
\end{figure*}

\noindent Here $\phi$ is the azimuthal angle of the particle's transverse momentum $p_\mathrm{T}$ and the curly brackets denote the average over particles from a single collision.\footnote{The average can include all charged particles or only particles of a specific identified species; we will not clutter our notation to account for these different possibilities.}
Eq.\,(\ref{1302.3535:eq1}) defines the flow coefficients and associated flow angles for the entire event, whereas Eq.\,(\ref{1302.3535:eq2}) is the analogous definition for the subset of particles in the event with a given magnitude of the transverse momentum $\pT$.  We suppress the dependence of both types of flow coefficients on the rapidity $y$. As mentioned before, $v_n$ are the integrated anisotropic flows and $v_n(\pT)$ are the differential flows. By definition, both $v_n$ and $v_n(\pT)$ are positive definite. Hydrodynamic simulations show that in general the flow angles $\Psi_n$ depend on $\pT$ and that, as a function of $\pT$, $\Psi_n(\pT)$ wanders around the ``average angle'' $\Psi_n$ that characterizes the integrated flow $v_n$ of the entire event (see Fig.\,\ref{1302.3535/F1} below
and also Fig.\,2 in Ref.~\cite{Ollitrault:2012cm}). Some theoretical and experimental definitions of $v_2$ have yielded values that turn negative over certain $\pT$ ranges; we will see that this is due to defining the flows of each event relative to a fixed azimuthal angle (for example, relative to the direction of the impact parameter of the collision in theoretical calculations, or relative to the integrated elliptic flow angle $\Psi_2$ in experiment), and that the same thing can happen for higher-order harmonic flow coefficients when defining them relative to a fixed (i.e. $\pT$-independent) flow angle $\Psi_n$. The subject of this chapter is to elucidate the origins of such differences between different anisotropic flow measures and, in particular, the manifestation of event-by-event fluctuations of the $\pT$-dependent flow coefficients $v_n(\pT)$ and flow angles $\Psi_n(\pT)$ in different experimental flow measures. 

\section{Differential flows from the event-plane method and from two-particle correlations}
\label{sec:1302.3535:2}

The key experimental difficulty is that, due to the finite number of particles emitted in each collision, the left hand sides of Eqs.~(\ref{1302.3535:eq1},\ref{1302.3535:eq2}) cannot be determined accurately for a single event. The $V_n$ are characterized by probability distributions that depend on the studied class of events (system size, collision energy and centrality) from which each collision takes a sample. Experimental flow measurements rely on a number of different methods that amount to taking different moments of that probability distribution by averaging over large numbers of events. Understanding the nature of these moments and reconstructing them from theoretical event-by-event dynamical simulations are essential steps in a meaningful comparison between theory and experiment.

Our main interest lies in the event-by-event fluctuations in the initial state of the collision fireball. These are primarily caused by the finite number of nucleons (or effective collision centers) in the colliding nuclei and unrelated to detector capabilities. In addition, there are fluctuations related to the finite number of particles produced (or detected) in the event which depend on collision energy and (in part) on detector capabilities. They reflect the fact that in practice the final state of the fireball evolution, which in principle (with the appropriate dynamical evolution model) can be {\em predicted} from the initial state with perfect precision, cannot be {\em measured} with perfect precision, due to finite sampling statistics. In this chapter, we are not interested in the fluctuations arising from finite sampling statistics; instead we focus on the hydrodynamical consequences of unavoidable event-by-event fluctuations in the initial state over which we have no control since they are rooted in the internal structure of the colliding nuclei, and which we therefore have to live with in any case even after we correct the measurements for finite final-state multiplicity effects.   

The most extensively used experimental methods for measuring anisotropic flows are the event-plane and two-particle correlation methods \cite{Voloshin:2008dg}. We begin with a discussion of the latter. Two-particle azimuthal correlations receive contributions from the anisotropic collective flow as well as from non-flow correlations; the latter can be minimized by appropriate experimental cuts and corrected for \cite{Voloshin:2008dg,Ollitrault:2009ie}. Again, we are not interested in non-flow correlations and will here simply ignore their existence, assuming that they have been corrected for in the experimental analysis.

Two-particle correlation measures of anisotropic flow are based on correlators of the type
\begin{equation}
\label{1302.3535:eq2a}
  \langle\{e^{in(\phi_1{-}\phi_2)}\}\rangle 
\end{equation}
where $\phi_1$ and $\phi_2$ are the azimuthal angles around the beam direction of two particles with transverse momenta $p_\mathrm{T1}$ and $p_\mathrm{T2}$, and $\langle\dots\rangle$ denotes the average over $N_\mathrm{ev}{\,\gg\,}1$ events from a set of given characteristics (e.g. of collisions in a certain centrality bin),
\begin{equation}
\label{1302.3535:eq4}
  \langle O \rangle = \langle\{{\cal O}\}\rangle := 
  \frac{1}{N_\mathrm{ev}}\sum_{i=1}^{N_\mathrm{ev}} \{{\cal O}\}_i,
\end{equation}      
whereas $\{\dots\}_i$ is the average of the observable ${\cal O}$ over all (or a specified subset of all) particle pairs in the event $i$:
\begin{equation}
\label{1302.3535:eq6}
 \{e^{in(\phi_1{-}\phi_2)}\}_i=\frac{1}{N^{(i)}_\mathrm{pairs}}\sum_{\mathrm{pairs}\in i}
   e^{in(\phi_1{-}\phi_2)}\ .
\end{equation} 
Different chosen subsets for the event-wise average $\{\dots\}_i$ define different correlation measures for the anisotropic flow coefficients, as we will explain below. In this chapter, we will always correlate pairs of particles of the same kind (e.g. protons with protons or charged hadrons with charged hadrons, but not protons with charged hadrons), unless specifically stated otherwise. We will also assume that they have the same rapidity $y$; the generalization to particles with different rapidities is straightforward, following the procedure discussed below when we go from particles with the same to particles with different $\pT$.

The magnitudes $v_n(\pT)$ of the anisotropic flow coefficients defined in Eq.~(\ref{1302.3535:eq2}) fluctuate from event to event according to some probability distribution $P(v_n(\pT))$. Let us denote the rms mean of this distribution by $v_n[2](\pT){\,:=\,}\sqrt{\langle v_n^2(\pT)\rangle}$, and similarly the rms mean for the integrated flow $v_n$ by $v_n[2]{\,:=\,}\sqrt{\langle v_n^2\rangle}$. These rms means can be obtained from two-particle correlators of the type (\ref{1302.3535:eq2a}) as follows:
\begin{eqnarray}
\label{1302.3535:eq5}
  &&\!\!\!\!\!\!\!\!
  v_n^2[2](\pT) = \langle\{e^{in(\phi_1{-}\phi_2)}\}_{\pT}\rangle 
                        = \langle\{e^{in\phi_1}\}_{\pT} \{e^{-in\phi_2}\}_{\pT}\rangle ,
  \nonumber\\
  &&\!\!\!\!\!\!\!\!
  v_n^2[2] = \langle\{e^{in(\phi_1{-}\phi_2)}\}\rangle 
                = \langle\{e^{in\phi_1}\} \{e^{-in\phi_2}\}\rangle.
\end{eqnarray}
Note that for the differential flow in the first line of equation (\ref{1302.3535:eq5}), both particles are taken from the same $\pT$ bin, and that the event-wise pair averages $\{e^{in(\phi_1{-}\phi_2)}\}$ factorize in each event due to our assumptions (absence of non-flow two-particle correlations, independent hydrodynamic emission of particles 1 and 2). Due to $1\leftrightarrow2$ symmetry under particle exchange, the exponential can be replaced by the cosine, and we get
\begin{eqnarray}
\label{1302.3535:eq7}
  &&\!\!\!\!\!\!
  v_n^2[2](\pT) =
  \left\langle\frac
  {\int d\Delta\phi \,  \cos(n\Delta\phi)\left. \frac{dN_\mathrm{pairs}}
                                  {dy_1 dy_2 p_\mathrm{T1} dp_\mathrm{T1} p_\mathrm{T2} 
                                  dp_\mathrm{T2} d\Delta\phi}\right|_{p_\mathrm{T1}{=}p_\mathrm{T2}} 
          } 
  {\int d\Delta\phi \left.\frac{dN_\mathrm{pairs}}
                                  {dy_1 dy_2 p_\mathrm{T1} dp_\mathrm{T1} p_\mathrm{T2}
                                  dp_\mathrm{T2} d\Delta\phi}\right|_{p_\mathrm{T1}{=}p_\mathrm{T2}}
          }\right\rangle,
  \nonumber\\
  &&\!\!\!\!\!\!
  v_n^2[2] = \left\langle\frac
  {\int d\Delta\phi \,\cos(n\Delta\phi) \,\frac{dN_\mathrm{pairs}}
                                  {dy_1 dy_2 d\Delta\phi} 
          } 
  {\int d\Delta\phi \frac{dN_\mathrm{pairs}}
                                  {dy_1 dy_2 d\Delta\phi} 
          }\right\rangle,
\end{eqnarray}
where $\Delta\phi{\,=\,}\phi_1{-}\phi_2$ and the pair distribution has already been integrated over the average angle $\tilde\phi{\,\equiv\,}(\phi_1{+}\phi_2)/2$. 

Note that in Eqs.~(\ref{1302.3535:eq7}), the single-event averages are normalized by the number of pairs in the event, before averaging over events. This is important: since the pair multiplicity fluctuates from event to event and within a multiplicity bin, and multiplicity anti-correlates with impact parameter with which the magnitudes of some of the anisotropic flow coefficients are geometrically correlated, this event-wise normalization avoids biasing the measured flow coefficients towards their values in events with larger than average multiplicity. 

Our definition of the integrated flow $v_n[2]$ agrees with the standard definition for the ``two-particle cumulant'' flow $v_n\{2\}$ \cite{Borghini:2000sa,Borghini:2001vi,Bilandzic:2010jr}, but the same is not true for the differential flow $v_n[2](\pT)$ which differs from $v_n\{2\}(\pT)$. The experimental definition of $v_n\{2\}(\pT)$ is \cite{Borghini:2000sa,Borghini:2001vi,Bilandzic:2010jr}
\begin{eqnarray}
\label{1302.3535:eq9a}
  &&v_n\{2\}(\pT) := \langle\{e^{in\phi_1}\}_{p_\mathrm{T1}} \{e^{-in\phi_2}\}\rangle/v_n\{2\}
 \nonumber\\
  && = \Bigl\langle v_n(\pT) v_n \cos[n(\Psi_n(\pT){-}\Psi_n)]\Bigr\rangle/v_n[2]\ .
\end{eqnarray}
Here only the first of the two particles within an event is taken from the desired $\pT$ bin and particle species; it is correlated with {\em all} other particles detected in the event, with obvious statistical advantages compared with $v_n[2](\pT)$ which requires both particles to be of the same kind and from the same $\pT$ bin. The normalization factor is the total rms flow of all charged hadrons. The last expression shows that $v_n\{2\}(\pT)$ reduces to $v_n[2](\pT){\,=\,}\sqrt{\langle v_n^2(\pT)\rangle}$ if and only if the flow angle $\Psi_n$ does not depend on $\pT$, the event-by-event fluctuations of $v_n(\pT)$ affect only its normalization but not the shape of its $\pT$ dependence, and the $v_n$ fluctuations of the particle species of interest are proportional to those of all hadrons. All of these assumptions are violated in hydrodynamic simulations of bumpy expanding fireballs. The difference between $v_n\{2\}(\pT)$ and $v_n[2](\pT)$, is thus sensitive to event-by-event fluctuations of the $\pT$-dependent difference $\Psi_n(\pT){-}\Psi_n$ between the flow angle of particles with momentum $\pT$ and the average event flow angle,\footnote{More precisely, $v_n\{2\}(\pT)$ depends on the difference between $\Psi_n(\pT)$ 
   {\em of the particles of interest} and the average flow angle $\Psi_n$ {\em of all detected 
   particles}. We checked numerically that the average hydrodynamic flow angles $\Psi_n$ 
   for identified pions and protons agree with great precision with the average flow angles for
   all particles in the event: computing the ensemble average of 
   $\langle\cos[n(\Psi_n^\pi{-}\Psi_n^p)]\rangle$ for all harmonics $n$ and all collision 
   centralities, we found deviations of less than $1{-}2\%$ in all cases except for some of the 
   high-order harmonics with $n{\,>\,}6$, whose calculation is plagued by numerical errors at
   low $\pT{\,\lesssim\,}0.2$\,GeV arising from the finite grid spacing of our square numerical
   grid used in solving the hydrodynamic equations.} 
in addition to the (largely independent) fluctuations in the magnitudes of $v_n$ and $v_n(\pT)$.

Another approach to isolating effects arising from the $\pT$-dependence of the flow angles is a comparison of the $\pT$-dependent rms flow $v_n[2](\pT)$ with the so-called event-plane flow\footnote{One can replace the cosine function in this definition by the 
    exponential, omitting taking the real part in the second line, since the flow-angle fluctuations
    are symmetrically distributed such that the imaginary part vanishes after taking the event 
    average (this has been verified numerically).}$^,$\footnote{Note that we define the 
    $n^\mathrm{th}$-order event-plane flow relative to the $n^\mathrm{th}$-order flow 
    plane $\Psi_n$, and not relative to the elliptic flow plane $\Psi_2$ as sometimes done.}
\begin{eqnarray}
\label{1302.3535:eq10}
  &&v_n\{\EP\}(\pT) :=
  \left\la \frac{\int d\phi \, \cos[n(\phi{-}\Psi_n)]\,\frac{dN}{dy \pT d\pT d\phi}}
                    {\int d\phi \, \frac{dN}{dy \pT d\pT d\phi}} \right\ra
  \nonumber\\
  &&= \Re\Bigl\langle \{e^{in\phi}\}_{\pT}e^{-in\Psi_n}\Bigr\rangle 
  = \Re\Bigr\langle v_n(\pT) e^{in(\Psi_n(\pT){-}\Psi_n)}\Bigr\rangle
  \nonumber\\
  &&= \Bigl\langle v_n(\pT) \cos[n(\Psi_n(\pT){-}\Psi_n)]\Bigr\rangle .             
\end{eqnarray}
The equality in the second line arises from Eq.~(\ref{1302.3535:eq2}). Here for each event the ``average flow angle'' $\Psi_n$ is first obtained by computing the $\bm{Q}_n$ vector \cite{Poskanzer:1998yz}
\begin{equation}
\label{1302.3535:eq11}
  \bm{Q}_n = Q_n e^{in\Psi_n} := \frac{1}{N} \sum_{k=1}^N \omega_k\,e^{in\phi_k}
\end{equation}
(where $N$ is the number of detected particles in the event) and determining its phase. In 
principle, different choices for the weights $\omega_k$ can be considered \cite{Voloshin:2008dg}, but for consistency with Eq.~(\ref{1302.3535:eq10}) one must choose $\omega_k{\,=\,}1$. The ``average angle'' $\Psi_n$ for the event extracted from $\bm{Q}_n$ in general depends on the types of particles included in the sum in Eq.~(\ref{1302.3535:eq11}). As noted in footnote 2, however, the average flow angle for particles emitted from a hydrodynamic source is (within numerical precision) the same for all particle species, and the precision of extracting $\Psi_n$ in experiments can thus be maximized by including all detected particles in the $\bm{Q}_n$ vector (\ref{1302.3535:eq11}).\footnote{Since in this paper we ignore finite particle 
    statistical fluctuations in the final state, we know $\Psi_n$ with infinite precision for each
    particle species, and we will simply use these particle-specific values in our numerical 
    results below.}

The last line in Eq.~(\ref{1302.3535:eq10}) makes it clear that the differential event-plane flows $v_n\{\EP\}(\pT)$ are sensitive to the event-by-event fluctuations of the $\pT$-dependent flow angles $\Psi_n(\pT)$ around the ``average flow'' angle $\Psi_n$. Just like the finite-number statistical fluctuations\footnote{Due
    to the finite number of particles detected in each event, the accuracy of determining 
    $\Psi_n$ is limited by finite number statistics, and an accurate experimental estimation 
    of the event-plane flow $v_n\{\EP\}$ requires an ``event-plane resolution correction'' 
    \cite{Voloshin:2008dg}. As shown in \cite{Alver:2008zza,Ollitrault:2009ie} (see also 
    the discussion in \cite{Gardim:2012im}), which moment of the underlying $v_n$ distribution 
    is actually measured by the total event-plane flow $v_n\{\EP\}$ depends on this 
    event-plane resolution: for perfect resolution, $v_n\{\EP\}$ approaches the average flow 
    $\langle v_n\rangle$, whereas in the case of poor resolution, it is closer to the rms flow 
    $v_n[2]{\,=\,}v_n\{2\}$ \cite{Alver:2008zza,Ollitrault:2009ie}. The mathematical analysis 
    in \cite{Ollitrault:2009ie} applies only to the integrated flow which allows one to ignore the 
    $\pT$-dependence of $v_n$ fluctuations as well as initial-state-related, $\pT$-dependent 
    fluctuations of the flow angles that are not caused by finite multiplicity in the final state. 
    In view of the latter, event-plane resolution effects on differential flow measurements 
    and their correction require a new analysis.}
of the flow angle reconstructed from $\bm{Q}_n$ around the ``true'' flow angle of the event, these fluctuations smear out the azimuthal oscillations of the transverse momentum spectra and thus reduce the oscillation amplitudes $v_n\{\EP\}(\pT)$. In contrast to the former, they arise from fluctuations in the initial state and thus cannot be eliminated by improving or accounting for the resolution of the measurement of the final state. They carry valuable physical information about the initial state and the dynamics of its evolution into the final state.

We can remove the sensitivity of the measured quantity to the $\pT$-dependent fluctuations of the flow angle by first computing {\em for each event} the magnitude $v_n(\pT)$ of 
$\{e^{in\phi}\}_{\pT}{\,=\,}V_n(\pT)$, before summing over events:
\begin{eqnarray}
\label{1302.3535:eq12}
 && \langle v_n(\pT) \rangle =
       \left\langle\left\vert\{e^{in\phi}\}_{\pT}e^{-in\Psi_n}\right\vert\right\rangle 
       = \left\langle\left\vert\{e^{in\phi}\}_{\pT}\right\vert\right\rangle 
 \nonumber\\
 &&=\left\langle \sqrt{\{\cos(n\phi)\}^2_{\pT} +
                                \{\sin(n\phi)\}^2_{\pT}} \right\rangle                               
                               .\quad
\end{eqnarray}

Since the quantity inside the event average does not depend on the average flow angle $\Psi_n$, this observable is not subject to an event-plane resolution correction. However, due to finite multiplicity in the final state, the right hand side will still in general be positive and non-zero experimentally even if there is no underlying anisotropic flow in the event. Again, how to properly account for such finite sampling statistical effects requires additional analysis.  

By comparing $\langle v_n(\pT) \rangle$ (\ref{1302.3535:eq12}) with $v_n[2](\pT)$ (\ref{1302.3535:eq5}, \ref{1302.3535:eq7}), $v_n\{2\}(\pT)$ (\ref{1302.3535:eq9a}), and $v_n\{\EP\}(\pT)$ (\ref{1302.3535:eq10}), we can experimentally assess and separate the relative importance of event-by-event fluctuations in the magnitudes and directions of the anisotropic flows as functions of $\pT$.

Let us now proceed to two-particle correlations between particles of different (but specified) momenta. Since in the first line of Eq.~(\ref{1302.3535:eq7}), both particles are taken from the same bin in $\pT$, the flow angle $\Psi_n(\pT)$ drops out from the expression. This is not true for azimuthal correlations between two particles with different $\pT$ \cite{Gardim:2012im}. In this case, one finds \cite{Gardim:2012im,Luzum:2011mm} 
\begin{eqnarray}
\label{1302.3535:eq8}
  &&\tilde{V}_{n\Delta}(p_\mathrm{T1},p_\mathrm{T2}) 
  := \left\langle\{e^{in(\phi_1{-}\phi_2)}\}_{p_\mathrm{T1}p_\mathrm{T2}}\right\rangle
  \nonumber\\
  &&= \left\langle\{e^{in\phi_1}\}_{p_\mathrm{T1}}\{e^{-in\phi_2}\}_{p_\mathrm{T2}}\right\rangle
  = \left\langle V_n(p_\mathrm{T1}) V^*_n(p_\mathrm{T2})\right\rangle
  \nonumber\\
  &&=\left\langle v_n(p_\mathrm{T1})v_n(p_\mathrm{T2}) e^{in(\Psi_n(p_\mathrm{T1}){-}
  \Psi_n(p_\mathrm{T2}))}\right\rangle
  \nonumber\\
  &&=\Bigl\langle v_n(p_\mathrm{T1})v_n(p_\mathrm{T2}) \cos[n(\Psi_n(p_\mathrm{T1}){-}
  \Psi_n(p_\mathrm{T2}))]\Bigr\rangle.\quad\quad
\end{eqnarray}
Due to parity symmetry, $\tilde{V}_{n\Delta}(p_\mathrm{T1},p_\mathrm{T2})$ is real: while the quantity inside the event average $\langle\dots\rangle$ is in general complex for each individual event, its imaginary part averages to zero when summed over many events.

To properly account for multiplicity fluctuations, in Eq.\,(\ref{1302.3535:eq8}) the averages $\{\dots\}_{p_\mathrm{Ti}}$ within an event are once again normalized by the total number of particles included in the average, similar to Eq.~(\ref{1302.3535:eq7}). For this reason, $\tilde{V}_{n\Delta}(p_\mathrm{T1},p_\mathrm{T2}){\,=\,}\left\langle\{\cos(n\Delta\phi)\}_{p_\mathrm{T1},p_\mathrm{T2}}\right\rangle$ defined in Eq.~(\ref{1302.3535:eq8}) is not identical with the experimental quantity $V_{n\Delta}(p_\mathrm{T1},p_\mathrm{T2})$, which is obtained from a Fourier decomposition with respect to the difference angle $\Delta\phi$ of the two-particle distribution obtained by summing over many events, without normalizing the contribution from each event by the corresponding event multiplicity \cite{Alver:2010rt,Aamodt:2011by,Chatrchyan:2012wg,Luzum:2011mm,ATLAS:2012at}. For a meaningful comparison between theory and experiment, one should either normalize on the experimental side the contribution from each event to the two-particle distribution by the number of pairs in the event, or weight the theoretical prediction for $v_n(p_\mathrm{T1})v_n(p_\mathrm{T2}) e^{in(\Psi_n(p_\mathrm{T1}){-}\Psi_n(p_\mathrm{T2}))}$ for each event $i$ with a factor $N^{(i)}_{\mathrm{pairs}}/\langle N_\mathrm{pairs}\rangle$ before summing over events. We prefer the first option, since it avoids the geometric bias arising from the correlation between collision geometry and particle multiplicity.

Equation~(\ref{1302.3535:eq8}) makes it obvious that the two-particle correlation coefficient $\tilde{V}_{n\Delta}(p_\mathrm{T1},p_\mathrm{T2})$ does not factorize into a product of single-particle anisotropic flow coefficients \cite{Gardim:2012im}. There are two contributions to this breaking of factorization: $\pT$-dependent event-by-event fluctuations of the magnitude of the flow coefficient $v_n$, and $\pT$-dependence of the flow angles $\Psi_n$ \cite{Gardim:2012im} (which also fluctuate from event to event). It is possible to define a non-factorizing correlator that is only affected by the fluctuations of $v_n(\pT)$ but is insensitive to the flow angles:
\begin{eqnarray}
\label{1302.3535:eq9}
  &&\Bigl\langle  v_n(p_\mathrm{T1})v_n(p_\mathrm{T2})\Bigr\rangle 
  =
  \left\langle \sqrt{\{\cos(n\Delta\phi)\}^2_{p_\mathrm{T1},p_\mathrm{T2}} +
                            \{\sin(n\Delta\phi)\}^2_{p_\mathrm{T1},p_\mathrm{T2}}}\right\rangle.\qquad
\end{eqnarray}
It is obtained experimentally by first obtaining the magnitude of $\{e^{in\phi_1}\}_{p_\mathrm{T1}} \{e^{-in\phi_2}\}_{p_\mathrm{T2}}$ for each event, normalizing it to the number of pairs used for its computation, and then adding the results for many events. Its sensitivity to finite-number statistical effects should be similar to Eq.~(\ref{1302.3535:eq12}) and needs to be explored. By comparing the quantity $\tilde{V}_{n\Delta}(p_\mathrm{T1},p_\mathrm{T2})$ from Eq.~(\ref{1302.3535:eq8}) with $\langle  v_n(p_\mathrm{T1})v_n(p_\mathrm{T2})\rangle$ from Eq.~(\ref{1302.3535:eq9}) one can assess the importance of the $\pT$-dependence and event-by-event fluctuations of the flow angles $\Psi_n$ (which affect the former but not the latter).

\section{The effect of flow fluctuations on differential $v_n$ measures}
\label{sec:1302.3535:3}

In this section we compare the differential flows $v_n(\pT)$ extracted from the 22,000 viscous hydrodynamic simulations per centrality bin of 2.76\,$A$\,TeV Pb+Pb collisions at the LHC (11,000 each with MC-Glauber and MC-KLN initial density profiles) that were generated \cite{Qiu:2012uy}. We use the Cooper-Frye prescription to compute from the hydrodynamic output on the freeze-out surface the single-particle distributions $dN/(dy\pT d\pT d\phi)$ as continuous functions of $\pT$ and $\phi$ (i.e. we do not sample the distribution to generate a finite number of particles per event, but pretend that the spectrum is sampled infinitely finely -- this avoids the need to correct for effects arising from finite-number statistics, such as imperfect event-plane resolution). All resonance decays are included in the final stable-hadron spectra. The details of the hydrodynamic simulations, initial conditions and freeze-out parameters are not important for the qualitative study presented here, but they can be found in Refs.~\cite{Qiu:2012uy,Qiu:2011hf,Shen:2011eg}. Here we only note that MC-Glauber (MC-KLN) initial conditions were hydrodynamically evolved with specific shear viscosity $\eta/s{\,=\,}0.08$ (0.2). 

%
\begin{figure*}[h!]
  \begin{center}
    \includegraphics[width=0.9\linewidth]{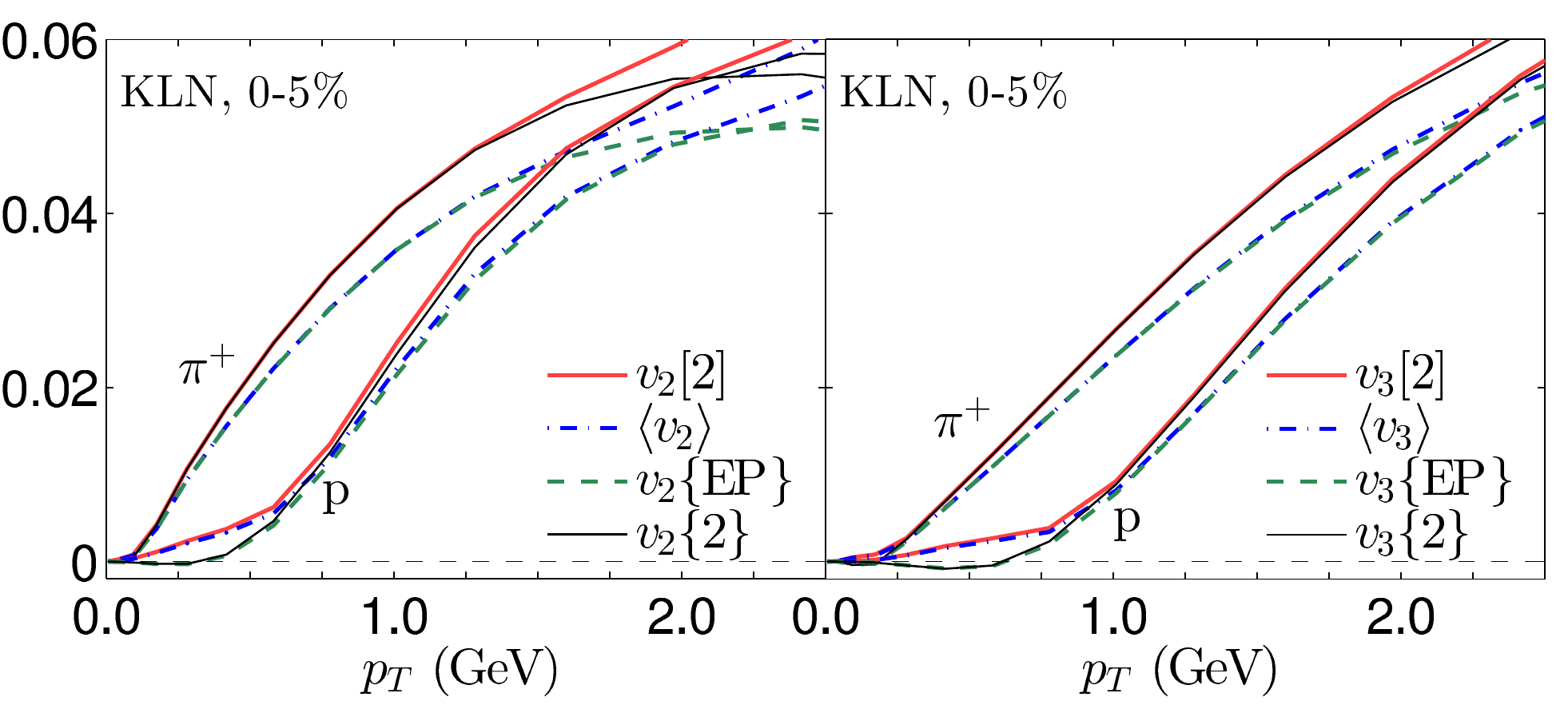}
  \end{center}
  \caption{Comparison between the different definitions of the differential flows 
    $v_n[2](\pT)$ (\ref{1302.3535:eq5},\ref{1302.3535:eq7}), $v_n\{2\}(\pT)$ (\ref{1302.3535:eq9a}), 
    $v_n\{\EP\}(\pT)$ (\ref{1302.3535:eq10}), and $\langle v_n(\pT) \rangle$ (\ref{1302.3535:eq12}),  for pions 
    and protons from central ($0{-}5\%$ centrality) Pb+Pb collisions at 
    $\sqrt{s}{\,=\,}2.76\,A$\,TeV, computed with the viscous hydrodynamic code 
    {\tt VISH2{+}1}. See text for discussion. From \cite{Heinz:2013bua}.
  \label{1302.3535/F2}}
\end{figure*}
%
%
\begin{figure*}[h!]
  \begin{center}
    \includegraphics[width=0.9\linewidth]{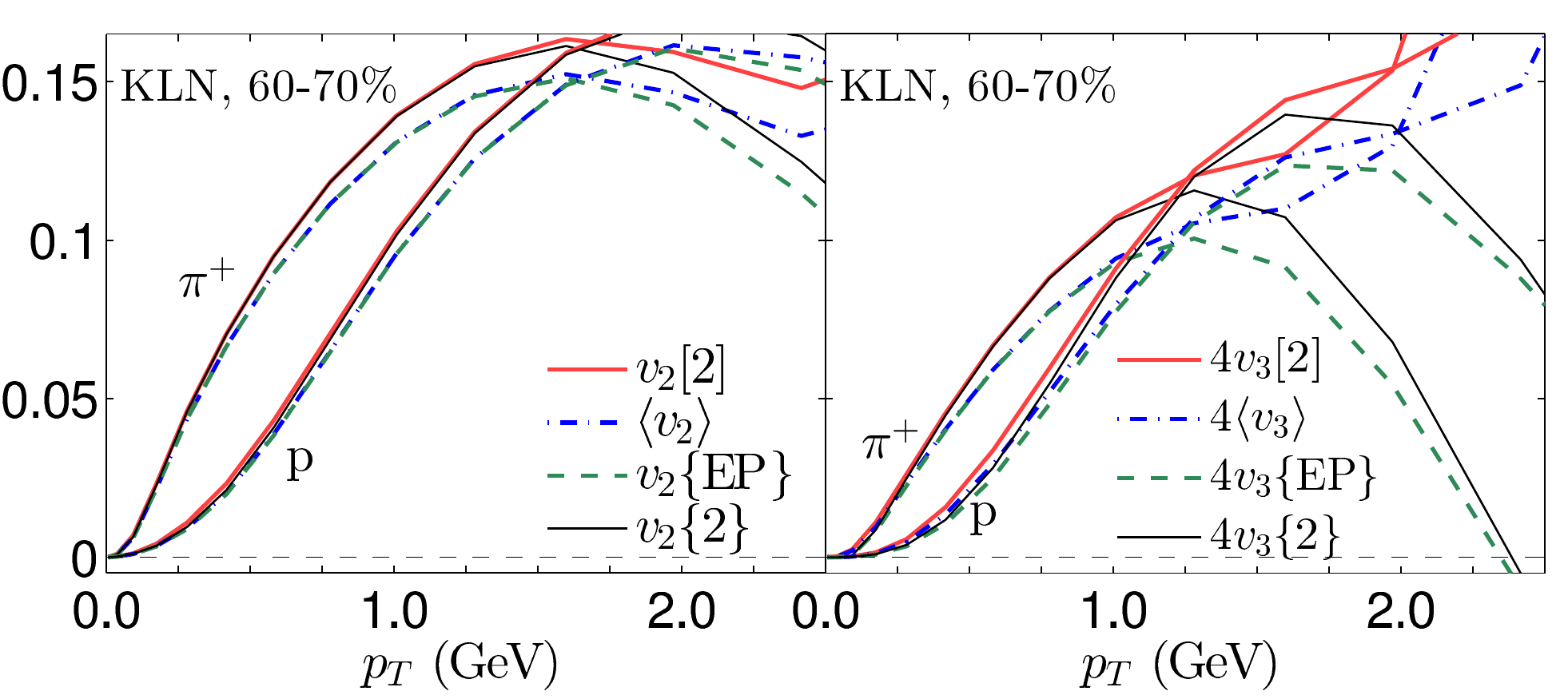}
  \end{center}
  \caption{Same as Fig.~\ref{1302.3535/F2}, but for peripheral Pb+Pb collisions
      at $60{-}70\%$ centrality. From \cite{Heinz:2013bua}.
  \label{1302.3535/F3}}
\end{figure*}
%

We present results for pions and protons, representing light and heavy particle species. Qualitatively, although not quantitatively, the same generic features are observed with MC-KLN and MC-Glauber model initial density profiles, and we show examples of both. Figures ~\ref{1302.3535/F2} and \ref{1302.3535/F3} show elliptic and triangular flows in their left and right panels, for central (Fig.~\ref{1302.3535/F2}) and peripheral (Fig.~\ref{1302.3535/F3}) Pb+Pb collisions. The curves for $v_{2,3}[2]$ and $\langle v_{2,3}\rangle$, which are not affected by flow angle fluctuations (Eqs.~(\ref{1302.3535:eq5}) and (\ref{1302.3535:eq12})), should be compared with those for $v_{2,3}\{2\}$ and $v_{2,3}\{\EP\}$, which {\em are} affected by the $\pT$-dependence of the flow angles $\Psi_n$ and their event-by-event fluctuations (Eqs.~(\ref{1302.3535:eq9a}) and (\ref{1302.3535:eq10})): for protons with transverse momenta below about 1\,GeV, flow angle fluctuations are seen to cause a significant suppression of the latter (in some cases even leading to negative elliptic flow values).\footnote{Note that the factor $\cos[n(\Psi_n(\pT){-}\Psi_n)]$
   in Eqs.~(\ref{1302.3535:eq9a}) and (\ref{1302.3535:eq10}) is maximal if $\Psi_n(\pT)$ is always aligned with 
   $\Psi_n$. The suppression of, say, $v_n\{\EP\}(\pT)$ relative to $\langle v_n \rangle(\pT)$
   does therefore not indicate a definite momentum tilt of the emitting source at a given 
   $\pT$ relative to the average $\Psi_n$, but simply reflects a nonzero difference 
   $\Psi_n(\pT){-}\Psi_n$ that fluctuates from event to event, suppressing the value of  
   $\cos[n(\Psi_n(\pT){-}\Psi_n)]$ for either sign of the difference.}
For the much lighter pions, flow angle fluctuation effects are almost invisible at low $\pT$. For protons they gradually disappear, too, as one goes from central (Fig.~\ref{1302.3535/F2}) to peripheral (Fig.~\ref{1302.3535/F3}) collisions.\footnote{The curves shown in Figs.~\ref{1302.3535/F2},\,\ref{1302.3535/F3} include the
   decay products from unstable hadronic resonances. We have observed that for protons 
   the flow angle fluctuation induced difference at low $\pT$ between ($v_{2,3}[2]$, 
   $\langle v_{2,3}\rangle$) on the one hand and ($v_{2,3}\{2\}$, $v_{2,3}\{\EP\}$) on the other 
   hand doubles if only directly emitted (``thermal'') particles are included in the analysis. 
   Resonance decays thus dilute the sensitivity of the proposed observables to flow angle 
   fluctuations by about 50\%.}

Event-by-event fluctuations of the {\em magnitudes} of $v_{2,3}$ are accessible by comparing $\langle v_{2,3}\rangle$ with $v_{2,3}[2]{\,=\,}\langle v^2_{2,3}\rangle^{1/2}$. When plotting the ratios $v_{2,3}[2](\pT)/\langle v_{2,3}\rangle(\pT)$ for central ($0{-}5\%$ centrality) collisions, where anisotropic flows are caused exclusively by fluctuations, with negligible geometric bias from a non-zero average deformation of the nuclear overlap region, we found for both pions and protons a constant (i.e. $\pT$-independent) value of $2/\sqrt{\pi}{\,\approx\,}1.13$. This is expected \cite{Voloshin:2007pc,Luzum:2012da}: If the flow angle $\Psi_n$ is randomly distributed relative to the reaction plane, the components of $V_n(\pT)$ along and perpendicular to the reaction plane are approximately Gaussian distributed around zero, and the magnitude $v_n(\pT)$ of the complex flow coefficient is Bessel-Gaussian distributed with $\sqrt{\langle v_n^2(\pT)\rangle}{\,=\,}\frac{2}{\sqrt{\pi}}\langle v_n(\pT)\rangle$ (see Eqs.\,(4) and (5) in Ref.~\cite{Voloshin:2007pc}). A similar $\pT$-independent ratio is not observed at larger impact parameters: even for triangular flow, which continues to be fluctuation-dominated also at non-zero impact parameters, we observe deviations of the ratio $v_3[2](\pT)/\langle v_3\rangle(\pT)$ from $2/\sqrt{\pi}$ at both low and high $\pT$; for elliptic flow these deviations are larger and significant at all $\pT$.  

Interestingly, for central collisions we found approximately the same constant value $2/\sqrt{\pi}$ for the ratio $v_{2,3}\{2\}(\pT)/v_{2,3}\{\EP\}(\pT)$ (except near the $\pT$ values where either the numerator or denominator passes through zero). Looking at the definitions (\ref{1302.3535:eq9a}) and (\ref{1302.3535:eq10}), this suggests an approximate factorization of the $\pT$-dependent flow angle fluctuations (which enter through the factor $\cos[n(\Psi_n(\pT){-}\Psi_n)]$ that cancels between numerator and denominator if it fluctuates independently) from the fluctuations of the magnitude $v_n(\pT)$, as well as an approximate $\pT$-independence of the $v_{2,3}$ fluctuations.

To follow up on these observations and gain deeper insight into the relative importance of flow angle fluctuations in different $\pT$ ranges, let us look at Figs.~\ref{1302.3535/F2},\,\ref{1302.3535/F3} and note that the frequently measured quantity $v_{2,3}\{2\}(\pT)$ behaves like the event-plane flow $v_{2,3}\{\EP\}(\pT)$ at low $\pT$ and like the differential rms flow $v_{2,3}[2](\pT)$ at intermediate $\pT$. This suggests that it is dominated by flow angle fluctuations at low $\pT$ and by fluctuations of the magnitude of $v_{2,3}(\pT)$ at higher $\pT$. In central collisions, the proton $v_{2,3}\{2\}(\pT)$ even turns negative at low $\pT$, whereas $v_{2,3}[2](\pT)$ is by definition always positive. A related observation is that the proton event-plane flow $v^p_{2,3}\{\EP\}(\pT)$ in Fig.~\ref{1302.3535/F2} approximately agrees with $v^p_{2,3}\{2\}(\pT)$ at low $\pT$ (where flow angle fluctuations seem to have strong effects) but with the mean flow $\langle v^p_{2,3}(\pT)\rangle$ at higher $\pT$ (where flow angle fluctuation effects are weak). This is reminiscent of the behavior of the {\em $\pT$-integrated} event-plane flow, which approaches the mean flow for good event-plane resolution. Flow angle fluctuations appear to have similar effects on flow measures as a decrease in flow angle resolution. The difference is that the former is a physical effect due to initial-state fluctuations, whereas the latter is a finite-sampling statistical effect in the final state and affected by detector performance.

\begin{figure*}[h!]
  \begin{center}
    \includegraphics[width=0.9\linewidth]{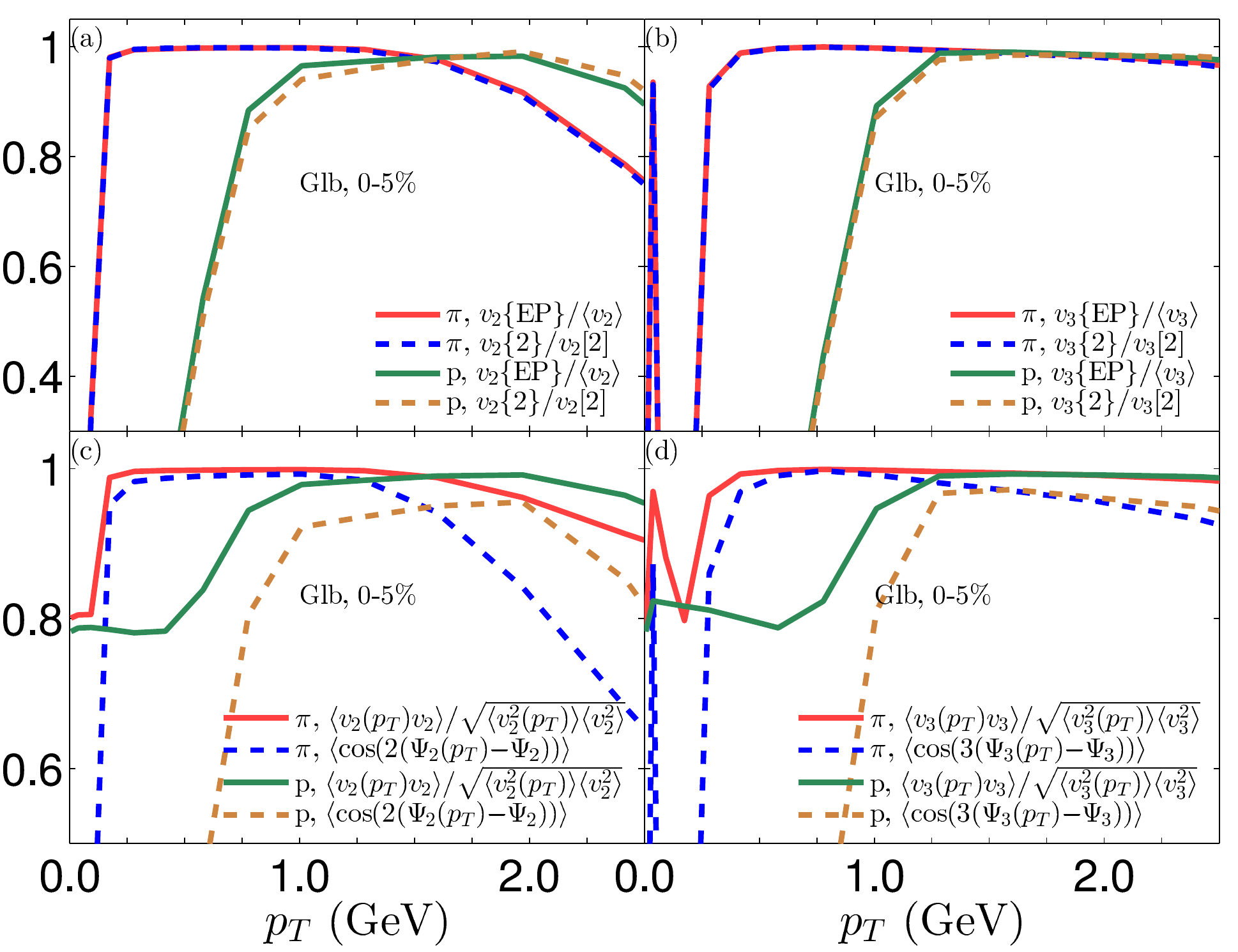}
  \end{center}
  \caption{(a,b) Ratios of differently defined elliptic (a) and triangular (b) flow 
     coefficients for pions and protons as functions of $\pT$. (c,d) $\pT$-dependence of the
     separate fluctuations of the magnitudes $v_n$ and the angles $\Psi_n$ for pions and 
     protons as discussed in the text. All curves from viscous hydrodynamics with 
     $\eta/s{\,=\,}0.08$ for central 2.76\,$A$\,TeV Pb+Pb collisions with MC-Glauber initial
     conditions. Results for MC-KLN initial conditions evolved with $\eta/s{\,=\,}0.2$ look very 
     similar. From \cite{Heinz:2013bua}.
     }
  \label{1302.3535/F4}
\end{figure*}

To make these qualitative observations quantitative, we plot in the upper two panels of Fig.~\ref{1302.3535/F4} the ratios $v_{2,3}\{2\}/v_{2,3}[2]$ and $v_{2,3}\{\EP\}/\langle v_{2,3}\rangle$ as functions of $\pT$, for both pions and protons. (We focus here on the results from Fig.~\ref{1302.3535/F2} for central collisions, where all anisotropic flows are fluctuation-dominated.) In each case the numerator is sensitive to the flow angle fluctuations while the denominator is not. However, numerator and denominator are also differently affected by fluctuations in the magnitudes of $v_n$. Both ratios are seen to behave very similarly, staying close to 1 at intermediate $p_T$ but dropping steeply at low $\pT$ and more moderately at high $\pT$. The steep drop at low transverse momenta sets in at $\pT{\,\sim\,}1$\,GeV for protons, but at much smaller $\pT{\,<\,}0.25$\,GeV for pions. We do not have a full understanding of this mass dependence, beyond the qualitative observation that the minimum of the variance of the flow angle fluctuations shown in Figs.~\ref{1302.3535/F1}b,c is shifted to higher $\pT$ for protons compared to pions, and that quite generally strong radial flow shifts all flow anisotropies to higher $\pT$ values for heavier particles.

The lower two panels of Fig.~\ref{1302.3535/F4} demonstrate that the behavior of the ratios shown in the two upper panels is strongly dominated by flow angle fluctuations. The dashed lines in Figs.~\ref{1302.3535/F4}c,d show the flow angle fluctuations $\langle\cos[n(\Psi_n(\pT){-}\Psi_n)]\rangle$ in isolation. Their $\pT$ dependence alone is almost sufficient to completely explain the shape of the curves
in panels (a) and (b). The solid lines in Figs.~\ref{1302.3535/F4}c,d show that at intermediate $\pT$ fluctuations in the magnitudes of the $\pT$-dependent flow $v_n(\pT)$ and the $\pT$-integrated $v_n$ tend to be correlated with each other ($v_n(\pT)\propto v_n$) while they appear to fluctuate more independently at low and high $\pT$. At high $\pT$ this decorrelation contributes to the suppression of the ratios shown in panels (a,b). At low $\pT$, the decorrelation of the $\pT$-dependent flow {\em magnitude} fluctuations $v_n(\pT)$ from the $\pT$-integrated flow $v_n$ does not become effective until after the ratios have already been suppressed by flow {\em angle} fluctuations, and its effect is therefore subdominant.

In summary, we see for central collisions that at low $\pT$ the differences between $v_n\{2\}(\pT)$ and $v_n[2](\pT)$, as well as between $v_n\{\EP\}(\pT)$ and $\langle v_n(\pT)\rangle$, are dominated by flow angle fluctuations, whereas at high $\pT$ fluctuations of both the flow angles and flow magnitudes must be considered to explain their differences. At intermediate $\pT$, flow angle fluctuations appear to be unimportant, $v_n(\pT)$ fluctuates in sync with the $\pT$-integrated $v_n$, and the differences between $v_n\{2\}(\pT)$ and $v_n[2](\pT)$, as well as between $v_n\{\EP\}(\pT)$ and $\langle v_n(\pT)\rangle$, vanish.

\begin{figure*}[h!]
  \begin{center}
    \includegraphics[width=0.9\linewidth]{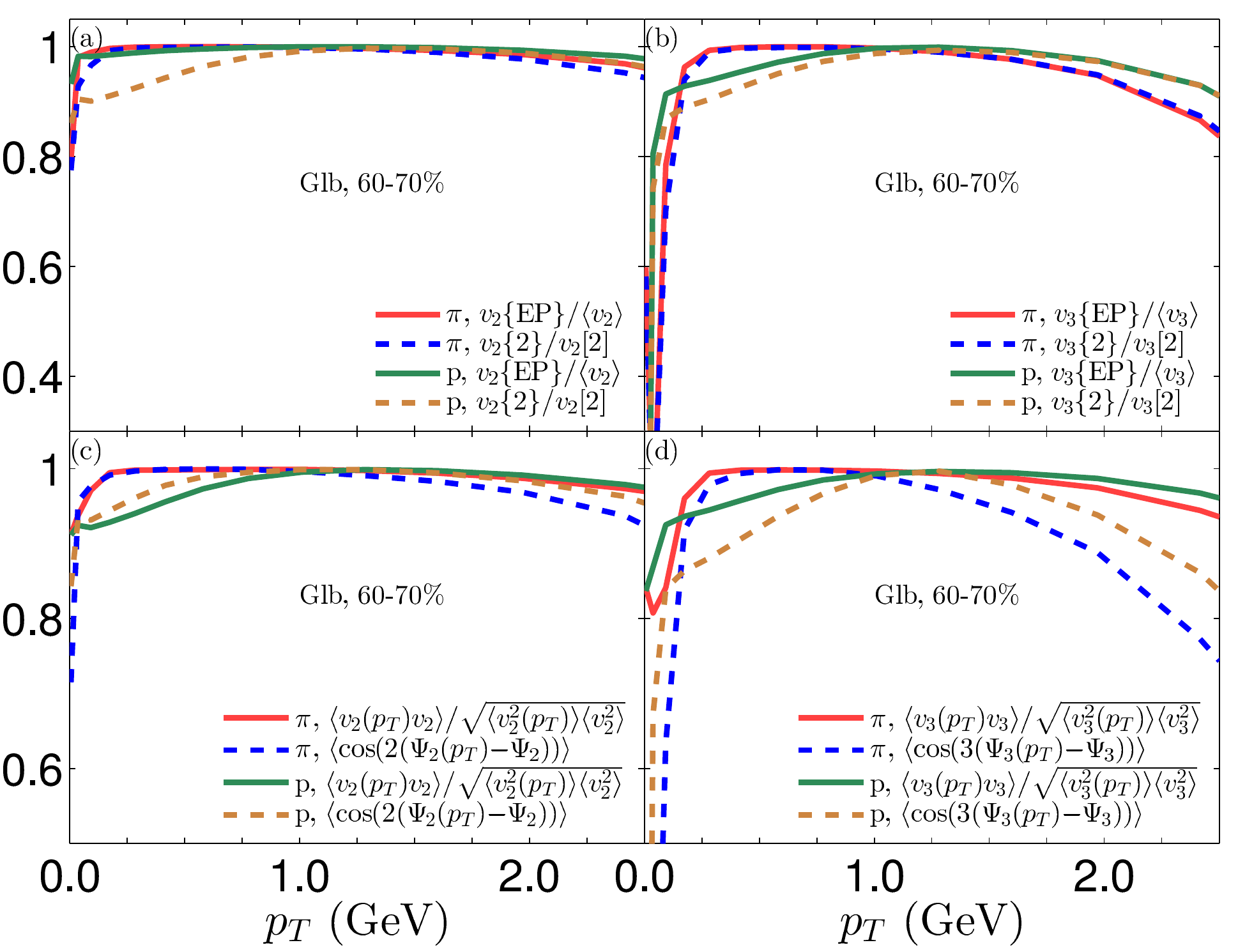}
  \end{center}
  \caption{Same as Fig.~\ref{1302.3535/F4}, but for peripheral collisions at 
     $60{-}70\%$ centrality. From \cite{Heinz:2013bua}.
  \label{1302.3535/F5}}
\end{figure*}

Figure~\ref{1302.3535/F5} shows the same ratios as Fig.~\ref{1302.3535/F4} for peripheral Pb+Pb collisions, again using MC-Glauber initial conditions with $\eta/s{\,=\,}0.08$.\footnote{The main difference 
    with results from MC-KLN initial conditions with $\eta/s{\,=\,}0.2$ (not shown) is that the 
    latter exhibit stronger suppression effects from the flow fluctuation factor 
    $\cos[n(\Psi_n(\pT){-}\Psi_n)]$ in the high-$\pT$ region $\pT{\,\gtrsim\,}1$\,GeV (see 
    also Fig.~\ref{1302.3535/F3}).} 
Compared to central collisions (shown in Fig.~\ref{1302.3535/F4}), the flow angle fluctuation effects at low $\pT$ are much weaker and appear to be shifted to lower transverse momenta, for both pions and protons. At high $\pT{\,\gtrsim\,}1$\,GeV, Figs.~\ref{1302.3535/F5}c,d show that effects from fluctuations of the flow angles (dashed lines) dominate over those from fluctuations of the flow magnitudes (solid lines).

\begin{figure*}[h!]
  \begin{center}
    \includegraphics[width=\linewidth]{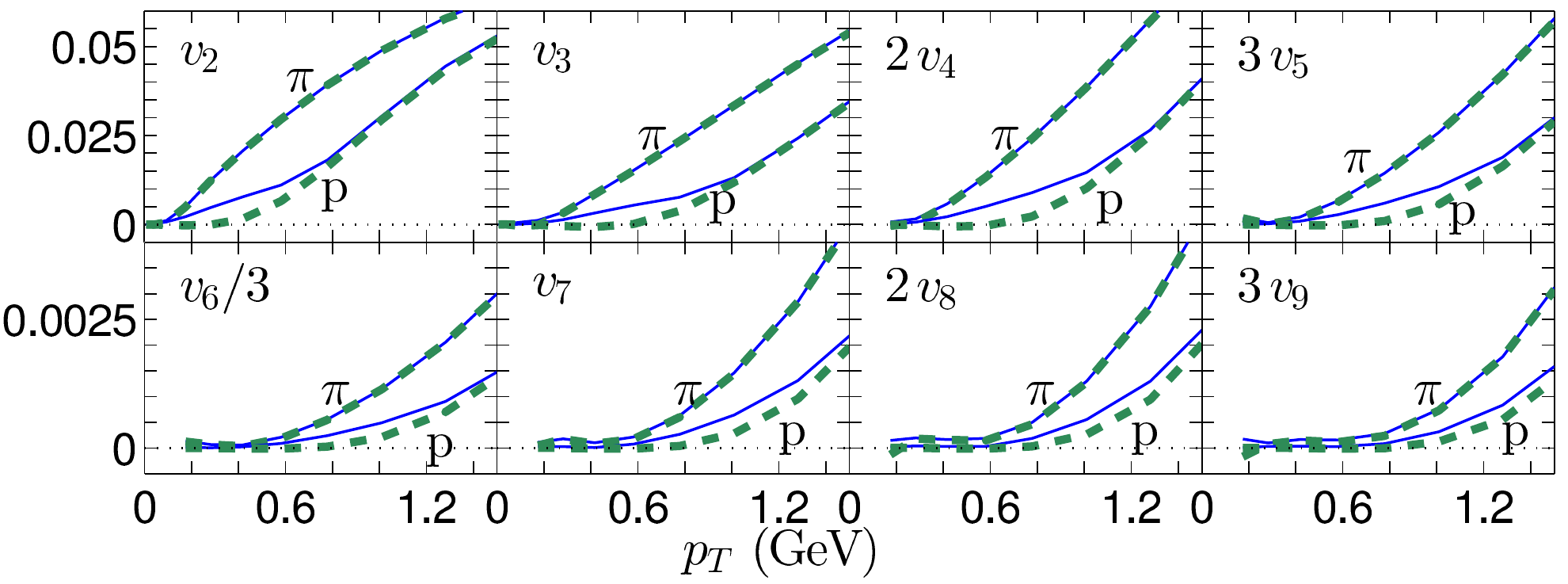}
  \end{center}
  \caption{Similar to Figs.~\ref{1302.3535/F2},\ref{1302.3535/F3}, but with MC-Glauber initial conditions
    for Pb+Pb collisions in the $0{-}10\%$ centrality bin. For clarity only $v_n\{\EP\}(\pT)$ 
    (Eq.~(\ref{1302.3535:eq10}), thick dashed) and $\langle v_n(\pT) \rangle$ (Eq.~(\ref{1302.3535:eq12}), thin solid)
    are shown, but for all harmonics from $n{\,=\,}2$ to $n{\,=\,}9$ (scaled by appropriate factors 
    for best visibility). This set of plots focusses on the low-$\pT$ region $\pT{\,<\,}1.5$\,GeV, 
    where the effects from flow angle fluctuations are strongest. See text for discussion. From \cite{Heinz:2013bua}.
  \label{1302.3535/F6}}
\end{figure*}

Finally, in Figure~\ref{1302.3535/F6} we explore (for near-central collisions) how the flow angle fluctuation effects, which push the event-plane flow $v_n\{\EP\}$ at low-$\pT$ below the value of the average flow $\langle v_n\rangle$, evolve as the harmonic order $n$ increases. (For $n{\,\geq\,}4$, we do not show results below $\pT{\,=\,}0.2$\,GeV for technical reasons explained in footnote 2.) For pions, flow angle fluctuations are invisible in the shown $\pT$ region for all flow harmonics; for protons, they are clearly visible for all harmonic flows. The relative magnitude of their effect on the difference $\langle v_n\rangle(\pT){-}v_n\{\EP\}(\pT)$ at any fixed $\pT$ decreases as $n$ increases, but the difference remains nonzero over a larger $\pT$ range for the higher harmonics. 

\section{Non-factorization of flow-induced two-particle correlations}
\label{sec:1302.3535:4}

%
\begin{figure*}[h!]
  \center{\includegraphics[width=0.9\linewidth]{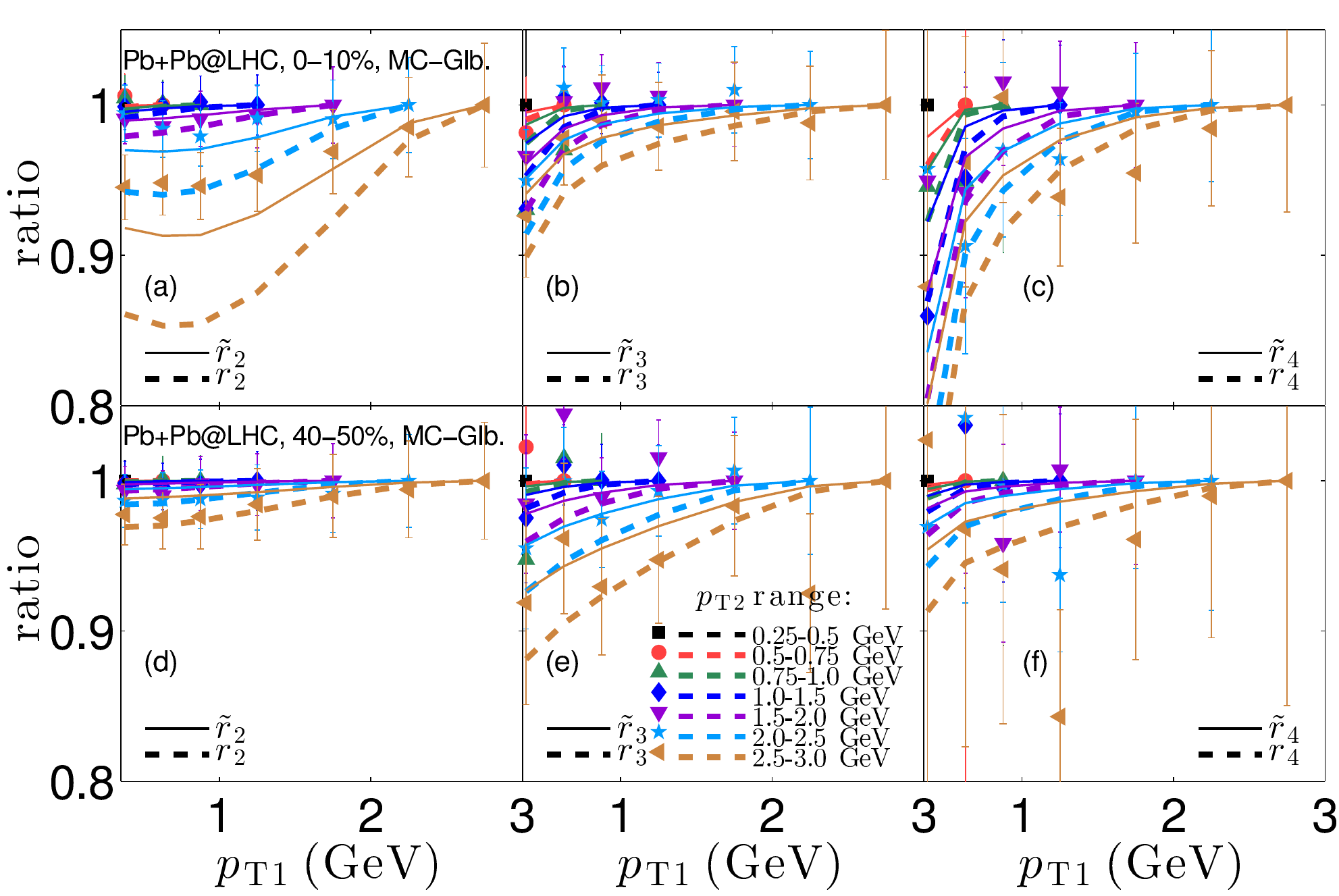}}
  \caption{The ratios $r_{2,3,4}(p_\mathrm{T1},p_\mathrm{T2})$ (thick dashed 
      lines) and $\tilde{r}_{2,3,4}(p_\mathrm{T1},p_\mathrm{T2})$ (thin solid lines), defined in
      Eqs.~(\ref{1302.3535:eq15}) and (\ref{1302.3535:eq16}), as functions of $p_\mathrm{T1}$ for different 
      $p_\mathrm{T2}$ ranges, as indicated. Filled symbols represent ALICE data for
      Pb-Pb collisions at $\sqrt{s}{\,=\,}2.76\,A$\,TeV \cite{Aamodt:2011by}. The lines are the
      corresponding viscous hydrodynamic calculations with MC-Glauber initial conditions, 
      using specific shear viscosity $\eta/s{\,=\,}0.08$. Panels (a,b,c) show $r_{2,3,4}$ and 
      $\tilde{r}_{2,3,4}$ for central ($0{-}10\%$) collisions, panels (d,e,f) show the same ratios 
      for peripheral ($40{-}50\%$) collisions. From \cite{Heinz:2013bua}.
\label{1302.3535/F7}}
\end{figure*}
%

\begin{figure*}[h!]
  \center{\includegraphics[width=0.9\linewidth]{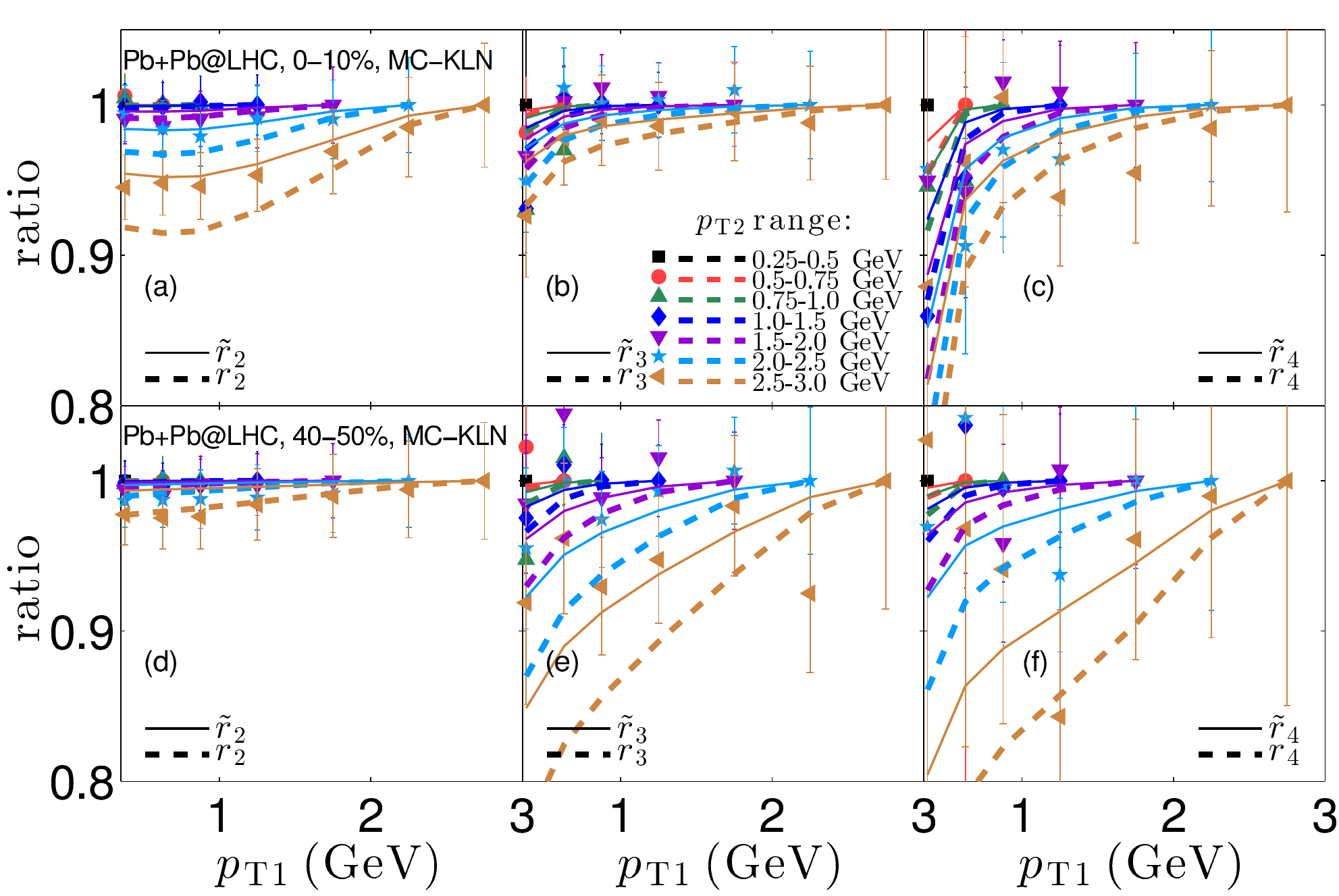}}
  \caption{Same as Fig.~\ref{1302.3535/F6}, but for MC-KLN initial conditions that have 
     been evolved hydrodynamically with $\eta/s{\,=\,}0.2$ (a shear viscosity value that is 2.5
     times larger than the one used in Fig.~\ref{1302.3535/F6}). From \cite{Heinz:2013bua}.
\label{1302.3535/F8}}
\end{figure*}

The breaking of factorization of flow-induced two-particle correlations by flow fluctuations was first emphasized by Gardim {\it et al.} \cite{Gardim:2012im}. Their study was based on simulations using ideal fluid dynamics, which are here repeated with viscous fluid dynamics. A comparison of Figs.~\ref{1302.3535/F7},\,\ref{1302.3535/F8} below with the plots shown in Ref.~\cite{Gardim:2012im} shows that viscous effects reduce the amount by which event-by-event fluctuations break factorization. We here explore the relative role played in this context by fluctuations in the magnitudes and angles of the flows.

To this end we define the following two ratios, both symmetric in $p_\mathrm{T1}$ and $p_\mathrm{T2}$:
\begin{eqnarray}
\label{1302.3535:eq15}
  r_n(p_\mathrm{T1},p_\mathrm{T2}) := 
  && \frac{\tilde{V}_{n\Delta}(p_\mathrm{T1},p_\mathrm{T2})}
         {\sqrt{\tilde{V}_{n\Delta}(p_\mathrm{T1},p_\mathrm{T1})
                   \tilde{V}_{n\Delta}(p_\mathrm{T2},p_\mathrm{T2})}}
 \nonumber\\  
  = && \frac{\la V_n(p_\mathrm{T1}) V_n^*(p_\mathrm{T2}) \ra} 
                {\sqrt{\la |V_n(p_\mathrm{T1})|^2\ra \la |V_n(p_\mathrm{T2})|^2 \ra}}
 \\\nonumber 
 = && \frac{\langle v_n(p_\mathrm{T1})v_n(p_\mathrm{T2}) 
                 \cos[n(\Psi_n(p_\mathrm{T1}){-}\Psi_n(p_\mathrm{T2}))]\rangle}
                {\sqrt{\la v_n^2(p_\mathrm{T1})\ra\la v_n^2(p_\mathrm{T2})\ra}};
 \end{eqnarray}
 
 \begin{equation}
 \\\label{1302.3535:eq16}
 \tilde{r}_n(p_\mathrm{T1},p_\mathrm{T2}) := 
        \frac{\langle v_n(p_\mathrm{T1})v_n(p_\mathrm{T2}) 
                 \cos[n(\Psi_n(p_\mathrm{T1}){-}\Psi_n(p_\mathrm{T2}))]\rangle}
                {\la v_n(p_\mathrm{T1}) v_n(p_\mathrm{T2})\ra}. 
\end{equation}
The ratio $r_n$, first introduced and studied with ideal fluid dynamics in \cite{Gardim:2012im}, is sensitive to fluctuations of both the magnitudes $v_n(\pT)$ and angles $\Psi_n(\pT)$ of the complex anisotropic flow coefficients $V_n(\pT)$ defined in Eq.~(\ref{1302.3535:eq2}). The second ratio $\tilde{r}_n$, on the other hand, differs from unity only on account of flow angle fluctuations. By comparing the two ratios with each other and with experimental data, we can isolate the role played by flow angle fluctuations in the breaking of factorization of the event-averaged two-particle cross section. In the absence of non-flow correlations both ratios are always ${\leq\,}1$.  

Figures~\ref{1302.3535/F7} show these ratios for all charged hadrons as functions of $p_\mathrm{T1}{\,\leq\,}p_\mathrm{T2}$ for fixed ranges of $p_\mathrm{T2}$, indicated by different colors.\footnote{The $p_\mathrm{T2}$ ranges are adjusted to the experimental  
    data, and the ratios were computed by first averaging the numerator and denominator 
    over the given $p_\mathrm{T2}$ range.}
Figs.~\ref{1302.3535/F7}a,b,c focus on central, Figs.~\ref{1302.3535/F7}d,e,f on peripheral collisions; in both cases, we used MC-Glauber initial conditions and evolved them with {\tt VISH2{+}1} using $\eta/s{\,=\,}0.08$ for the specific shear viscosity. In central collisions, the hydrodynamic simulations appear to overpredict the factorization breaking effects, while in peripheral collisions, theory and data agree somewhat better. More precise experimental data would be desirable. The comparison of $r_n$ (dashed lines) with $\tilde{r}_n$ shows that a significant fraction (${\,\sim\,}50\%$ or more) of the effects that cause the breaking of factorization arises from flow angle fluctuations. This seems to hold at all the transverse momenta shown in the figures. A comparison of the top and bottom rows of panels in Fig.~\ref{1302.3535/F7} shows that factorization-breaking effects are stronger for harmonics that are fluctuation dominated ({\it i.e.} all harmonics in central collisions and the odd harmonics (especially $v_3$) in peripheral collisions) and appear to weaken for $v_2$ and $v_4$ in peripheral collisions, where both the magnitudes $v_{2,4}$ and the flow angles
$\Psi_{2,4}$ are mostly controlled by collision geometry.

%
%

To explore the effects of shear viscosity of the expanding fluid on the breaking of factorization, we show in Figure~\ref{1302.3535/F8} the same data as in Fig.~\ref{1302.3535/F7}, but compared with hydrodynamic calculations that use MC-KLN initial conditions evolved with $\eta/s{\,=\,}0.2$ (a 2.5 times larger viscosity than used in Fig.~\ref{1302.3535/F7}). Obviously, the MC-KLN model produces a different initial fluctuation spectrum than the MC-Glauber model, so not all of the differences between Figs.~\ref{1302.3535/F7} and \ref{1302.3535/F8} can be attributed to the larger viscosity. However, in conjunction with the ideal fluid results reported in \cite{Gardim:2012im}, the comparison of these two figures strengthens the conclusion that increased shear viscosity tends to weaken the fluctuation effects that cause the event-averaged two-particle cross section to no longer factorize.  

\section{Chapter summary}
\label{sec:1302.3535:5}

All experimental precision measures of anisotropic flow in relativistic heavy-ion collisions are based on observables that average over many collision events. It has been known for a while that both the magnitudes $v_n$ and flow angles $\Psi_n$ of the complex anisotropic flow coefficients $V_n$ fluctuate from event to event, but only very recently it became clear that not only the $v_n$, but also their associated angles $\Psi_n$ depend on $\pT$, and that the difference $\Psi_n(\pT){-}\Psi_n$ between the $\pT$-dependent and $\pT$-averaged flow angles also fluctuates from event to event. Here in this chapter we pointed out that these flow angle fluctuations leave measurable traces in experimental observables from which the ensemble-averaged $\pT$-dependent anisotropic flows are extracted. We have introduced several new flow measures and shown how their comparison with each other and with flow measures that are already in wide use allows to separately assess the importance of event-by-event fluctuations of the magnitudes and angles of $V_n{\,=\,}v_n e^{in\Psi_n}$ on  experimentally determined flow coefficients. 

Viscous hydrodynamic simulations show that flow angle fluctuations affect the $\pT$-dependent flow coefficients of heavy hadrons (such as protons) more visibly than those of light hadrons (pions). In near-central collisions, where anisotropic flow is dominated by initial density fluctuations rather than overlap geometry, the effects from flow angle fluctuations appear to be strongest for particles with transverse momenta $\pT{\,\lesssim\,}m$. A precise measurement and comparison of $\la v_n(\pT)\rangle$ (Eq. (\ref{1302.3535:eq12})), $v_n\{\EP\}(\pT)$ (Eq.~(\ref{1302.3535:eq10})), $v_n[2](\pT)$ (Eqs.~(\ref{1302.3535:eq5},\ref{1302.3535:eq7})), and $v_n\{2\}(\pT)$ (Eq.~(\ref{1302.3535:eq9a})) for identified pions, kaons and protons with transverse momenta $\pT{\,<\,}2$\,GeV should be performed to confirm the hydrodynamically predicted effects from flow angle fluctuations. The theoretical interpretation of these measurements requires a reanalysis of finite-sampling statistical effects on the {\em $\pT$-dependent differential} flows, stemming from the finite multiplicity of particles of interest in a single event, which we did not consider here. The proposed comparison holds the promise of yielding valuable experimental information to help constrain the distribution of initial density fluctuations in relativistic heavy ion collisions and may prove crucial for a precision determination of the QGP shear viscosity. 

We also showed that flow angle fluctuations are responsible for more than half of the hydrodynamically predicted factorization breaking effects studied in Ref.~\cite{Gardim:2012im} and in Sec.~\ref{sec:1302.3535:4} above, and that these effects are directly sensitive to the shear viscosity of the expanding fluid, decreasing with increasing viscosity. By combining the study of various types of differential anisotropic flow measures with an investigation of the flow-induced breaking of the factorization of two-particle observables into products of single-particle observables, one can hope to independently constrain the fluid's transport coefficients and the initial-state fluctuation spectrum.

\chapter{Sampling particles from the Cooper-Frye distribution}
\label{chap:iSS}

This chapter explains the methodology of sampling particles from the emission function calculated with the Cooper-Frye formula along the freeze-out surface. I will focus on the correctness and efficiency of the sampling algorithm. The material product of this methodology are the iS and the iSS code packages, which are among the core packages used in our group's simulations. For completeness, both the required mathematical background and some technical details are given in this chapter.

The main structure of the chapter is outlined as follows: \sec{iSS:random_variables_and_sampling} provides the background on random variables and general sampling methodologies. The calculation of particle emission is briefly explained in \sec{iSS:momentum_distribution}. The actual sampling algorithm for particle momenta and space-time position is given in \sec{iSS:sampling_emission_function}. Finally, a number of code checks are performed in \sec{iSS:code_verification}.

The material is based on a document for the iSS code that is not yet published. The code was developed together with Chun Shen, who shares equal credit for this effort.

\section{iS and iSS overview} \label{sec:iSS:overview}
The name ``iS'' stands for ``iSpectra''; iS is a fast Cooper-Frye particle momentum distribution calculator along the conversion surface. Its output is a continuous function, evaluated at discrete momenta provided by the users, for the invariant momentum distributions of the desired hadron species. The code ``iSS'', whose name stands for ``iSpectraSampler'', goes one step further by generating individual particles, using the calculated particle momentum distributions as the relative emission probability. iSS is an ``event generator" which generates a complete collision event of emitted hadrons, similar to the events created in the experiment. Both codes are written keeping the following factors in mind:
\begin{itemize}
 \item {\bf Readability and extendability.} The most important goal is to create a cleanly written framework that calculates particle momentum distributions and performs sampling, whose components and output can be used easily for further physics analyses and tests of new physical ideas. To achieve this, the entire program is divided into modules according to their functionalities, the structures and the algorithms are documented with comments, and long but informative names are chosen for variables and function names.
 \item {\bf Efficiency.} Both the iS and iSS codes are written aiming for intensive hybrid event-by-event calculations where every CPU cycle counts. To achieve the necessary degree of efficiency, much effort is put into optimizing the algorithms at different levels of the calculations.
 \item {\bf Easy maintainability and re-usability.} The framework is divided into different carefully chosen functionality modules, for better interoperability and to maximize re-usability.
\end{itemize}


Although the codes are meant to serve as a basic framework upon which additional physical analysis modules can be added, some such modules are already implemented, as will be explained in the following sections. An external parameter file and several tables allow for easy adjustment of parameters and choosing among multiple choices between available modules.


\section{Random variables and sampling methods}
\label{sec:iSS:random_variables_and_sampling}

\subsection{Random variable, PDF and CDF}
\label{subsec:iSS:random_variables}

A random variable $X$ is a variable that takes different values within a set $S$ every time it is sampled. If $S$ is a discrete set, then $X$ can be characterized by specifying the probability of it taking each element in $S$; if $S$ is a continuous set, then $X$ is characterized by the probability density function (PDF) $f(X)$, defined such that $\int_{X\in A} f(X) dX$ gives the probability of finding $X$ in a subset $A$.

The cumulative distribution function (CDF) $F(X)$ is formally defined to be the function that gives the probability of finding $X' \leq X$:
\begin{equation}
 F(X) = \int^X f(X') dX'
\end{equation}
The ``inverse CDF" is the inverse function of $F(X)$.

By definition $f(X)$ is non-negative and it integrates to 1, and $F(X)$ is non-decreasing and $F(X\to-\infty)=0$, $F(X\to\infty)=1$.

In this chapter, a variable is ``$n$-dimensional" if the dimension of the set $S$ is $n$. The PDF and CDF definitions are also valid for multi-dimensional random variables, where for the CDF the notion of $X\leq x$ should be understood component by component.

\subsection{Sampling 1d random variables according to the inverse CDF}
\label{subsec:iSS:inverse_CDF}

The inverse CDF sampling is best understood if we partition the range $[0,1]$ of the CDF into $n$ intervals with equal lengths $R_i$, and their pre-image in the domain space are denoted as $D_i:=F^{-1}(R_i)$. By definition of CDF, the probability that $F(X)$ lies inside any $R_i$ is the same, therefore the probability for $X$ to lie in any $D_i$ is the same. By throwing another uniform random variable in $[0,1]$, we can sample an index $i$ which means that the sampled $X\in D_i$. By taking $n\to \infty$, we see that the process becomes: uniformly randomly sample a value $T\in[0,1]$, then find the sampled $X=F^{-1}(T)$.

If the range of the CDF is not properly normalized to $[0,1]$ but $[A,B]$, the process given above can be extended trivially by shift and scaling.

\subsection{Direct sampling of 1d random variables according to their PDF: special accept-reject method}
\label{subsec:iSS:direct_PDF}

Another intuitive sampling method is to sample the PDF directly using the accept-reject method. It starts with sampling a uniformly randomly chosen value for $X$, then use another continuous random variable $T$ uniform in $[0,1]$ to decide whether to accept or reject the sample: if the sampled trial variable value is greater than $f(X)$, then it is rejected; otherwise it is accepted. By doing so it is ensured that the probability of keeping $X$ is proportional to $f(X)$, which is what we want.

One immediate improvement on efficiency is that instead of comparing the test variable to $f(X)$, we can also compare it to $k\,f(X)$, where $k$ is a factor such that $k\,f(X) \leq 1$ for all $X$. A larger $k$ values means less rejections and larger efficiency, and max efficiency is attained by choosing $k=1/\max(f(X))$.

Like the inverse CDF method, if the PDF is not properly normalized, a scaling in $T$ can be used to generalize the method.

Note that the accept-reject method for sampling PDF does not require the PDF to be properly normalized; the only requirement is that the probabilities used to make accept-reject selections need to be less than $1$.

\subsection{Sampling a 1d random variable using envelope distribution: general accept-reject method}
\label{subsec:iSS:envelope_distribution}

A random variable $Y$ on $S$ with PDF $g(y)$ is considered as an envelope distribution to $X$ on $S$ with PDF $f(x)$ if there exists a constant $k$ so that $k\,g(s)\geq f(s)$, $\forall s\in S$.

Given an envelope distribution, the random variable $X$ can be sampled using a two-step accept-reject procedure. First samples of $Y$ according to the distribution $g(y)$ are generated using any appropriate method, then each sample $Y=s$ is accepted as a sample of $X$ with probability $p(s)=f(s)/(k\,g(s))$.

The reasoning is simply based on a re-expressing the probability of an event as a product of the probability of another event and a conditional probability: the probability of generating a sample $X=s$ is $f(s)$, and it equals the probability $g(s)$ of first generating a sample $Y=s$, multiplied by the conditional probability of getting $X=s$ knowing $Y=s$, which is proportional to $f(s)/\left(k\,g(s)\right)$.

As a special case, when $g$ is a constant, sampling using the envelope distribution reduces to the case of direct sampling of PDF method explained in Sec.~\ref{subsec:iSS:direct_PDF}.

The efficiency of using an envelope distribution function is analyzed in Sec.~\ref{subsec:iSS:sampling_efficiency}.

Often an envelope distribution can be constructed as a sum of step functions, see Sec.~\ref{subsec:iSS:automatic_envelope}.

\subsection{Sampling 1d random variable using grouping}
\label{subsec:iSS:grouping}

Another less intuitive method to sample $X$ is to use the grouping method, which divides $S$ into subsets and first samples the subset the next sample belongs to, then samples the element inside that subset. To illustrate it consider a simple case where $S=\{0,1,2,3\}$ and $f(0)=f(1)=f(2)=f(3)=1/4$.

The way to divide $S$ into groups is not unique but suppose $S$ is divided into $S_0=\{0,1\}$ and $S_1=\{2,3\}$, then the probability of finding $X\in S_0$ is $f(0)+f(1)=1/2$ and the probability of finding $X\in S_1$ is $f(2)+f(3)=1/2$.

The subsets are first sampled, which is equivalent to sampling a random variable $Y$ on the indices $\{0,1\}$ of the subsets with PDF $g(0)=g(1)=1/2$. Next elements in that subset $S_Y$ are sampled according to the PDF $f(x)$ but restricted to $S_Y$.

In general, after $S$ is partitioned into subsets $S_i$, $i=0,...,n$, first subsets are sampled using the induced ``group-index-variable" $Y$ on $\{0,...,n\}$ with PDF $g(i)=\int_{X\in S_i} f(x) dx$, then $X$ is sampled with restricted PDF $f_{\mathrm{restricted}}(x)=f(x)\,\Theta(X\in S_i)$.

The grouping method can be used recursively, where $S_i$ are sub-divided into finer sub-subsets.

How $S$ should be divided into subsets, and the efficiency of the grouping method are explained in Sec.~\ref{subsec:iSS:sampling_efficiency}.

\subsection{Sampling efficiency}
\label{subsec:iSS:sampling_efficiency}

So far, four different sampling methods have been explained:
\begin{enumerate}
 \item sampling using inverse CDF (Sec.~\ref{subsec:iSS:inverse_CDF})
 \item direct PDF sampling (Sec.~\ref{subsec:iSS:direct_PDF})
 \item sampling with envelope distribution (Sec.~\ref{subsec:iSS:envelope_distribution})
 \item sampling using grouping (Sec.~\ref{subsec:iSS:grouping})
\end{enumerate}
Among them, (1) and (2) are direct sampling methods while (3) and (4) are composite in the sense that they involve sampling a second random variable, using any desired direct sampling method. For sampling with the envelope distribution method, the way to sample the envelope distribution is unspecified; and for sampling using grouping, both the way to sample groups and the way to sample elements within the group are left open.

In this section, sampling efficiencies are compared for methods (1)-(4). Here the ``sampling efficiencies'' are measured in terms of the number of times $f(x)$ is evaluated, while all other calculations are assumed to pose no costs. I will abbreviate {\em the number of evaluations of $f(x)$} in this section as ``NOE". When sampling a large number of samples, another good measurement of efficiency is the average number of times $f(x)$ needs to be calculated in order to produce $1$ accepted sample, which is the reciprocal of the ``acceptance rate".

In almost all cases, $f(x)$ either has an analytic expression or can be pre-tabulated. In cases where $F(x)$ can be calculated analytically or pre-tabulated, the time spent on evaluating $F(x)$ is similar to that of evaluating $f(x)$. In such cases, the sampling using inverse CDF method is no doubt the most efficient one, since each time $F(x)$ is evaluated, it gives one sample with $100\%$ acceptance. However in many cases, $F(x)$ cannot be evaluated analytically or pre-tabulated\footnote{One common situation is that $f(x)=f(x;\alpha)$ depends on another continuous parameter and $F(x;\alpha)$ cannot be calculated analytically, thus there are ``infinitely many'' $F(x)$ that need to be tabulated.}, and its calculation requires numerical integration of $f(x)$. It is for these cases that the efficiencies of different methods matters, and the following comparisons are all done under such conditions.

For demonstration, it is assumed that $f(x)$ is discrete and $S$ is a finite set with $N$ elements. Assume also that a total number of $M$ samples is desired. In such a case, the evaluation of $F(x)$ for all $x\in S$ requires NOE=$N$. 

Two extreme scenarios for $X$ are used: scenario (A) assumes that the distribution function $f(x)=1/N$ is uniform, and scenario (B) assumes that the distribution function $f(x)$ is delta-like: $f$ is $1$ for one special element and $0$ for others. A realistic distribution can have any shape in between, and the sampling efficiency of it is expected to also lie in between; a flat distribution should have sampling efficiency close to scenario (A) and a sharply peaked one close to scenario (B).

It can be shown that if an event has success probability $1/N$, then the average waiting time to have one success in a series of repeated independent sampling is $N-1$, that is, on average the $N$-th sample is the successful one.

Starting with scenario (A) where $f(x)$ is uniform, the direct PDF sampling method (2) in the most optimized case where $f_m=\max\{f\}=1/N$ (see Sec.~\ref{subsec:iSS:direct_PDF}) gives full acceptance: every sample of $X$ is accepted because the acceptance probability is $1$. Thus NOE is $M$. The sampling using inverse CDF method (1), as explained requires $N$ NOE for constructing the inverse CDF plus $M$ NOE for actual sampling.

For scenario (B) where $f(x)$ is delta-like, the direct PDF sampling method (2) in the most optimized case requires $M N$ NOE since each sampling has $1/N$ success rate and there are $M$ desired samples. The inverse CDF sampling method (1) requires again $N$ NOE for constructing $F(x)$ plus $M$ NOE for sampling.

The results are summarized in the Table \ref{tab:PDF_vs_insCDF}.

\begin{table}[h]
\begin{center}
\begin{tabular}{c|c|c}
  \hline \hline
  NOE & flat distribution & delta-like distribution \\
  \hline
  direct PDF & $M$ & $M N$\\
  \hline
  inverse CDF & $M+N$ & $M+N$ \\
  \hline \hline
\end{tabular}
\end{center}
\label{tab:PDF_vs_insCDF}
\caption{Summary for waiting times (NOE per $M$ successful sampling) between direct PDF sampling and inverse CDF sampling methods.}
\end{table}

Therefore for a flat distribution, the direct PDF sampling method, when used properly, is most efficient. For distributions with peaks, the sampling using inverse CDF method is always superior than the direct PDF sampling method, except for $M=1$. In general for $M \gg N$ or $M\sim N$, using inverse CDF sampling will practically almost always grant excellent efficiency; only for $M \ll N$ should the direct PDF sampling method be considered.

From our study of scenario (B), it is also clear that an accurate estimate for the maximum of the PDF is crucial for the direct PDF sampling method: an overestimate of the maximum by a factor of $2$ would decrease the success rate by a factor of $2$, thus doubling the sampling time.

As a quick summary, for steep distributions, the sampling using the inverse CDF method is superior; for flat and general unknown distributions, to generate a small number of samples the direct PDF sampling is faster, while for large samples, the inverse CDF method is faster.

The envelope distribution function method (3) first samples according to the envelope distribution, after which it uses the accept-reject method to sample the target distribution (see Sec.~\ref{subsec:iSS:envelope_distribution}). The envelope distribution can be sampled by direct PDF sampling or inverse CDF sampling. However if direct PDF sampling is used for the envelope function, then sampling with the envelope distribution method has no advantage over the direct PDF sampling method since the over-all probability for accepting a sample is still the same as the one in the direct PDF sampling. The sampling using the envelope method is only more efficient when the envelope variable can be sampled using the inverse CDF method and its CDF can be calculated analytically or pre-tabulated, or at least its evaluation takes a much shorter time.

To see why using envelopes improves the sampling efficiency, consider a simple example of scenario (B): $S=\{0,1,2,3\}$ and $X$ has PDF $f(0)=1$, $f(1)=f(2)=f(3)=0$, which is delta-like. The direct PDF sampling method requires on average $N=4$ NOE to acquire one sample. Consider using an envelope distribution $Y$ with PDF $g(0)=g(1)=1/2$, $g(3)=g(4)=0$. By assumption, the sampling of $Y$ takes only $1$ NOE, which yields either $Y=0$ or $Y=1$. Next consider using an accept-reject method to sample $X$, but since the possible sample spaces has been restricted to $\{0,1\}$ from sampling $Y$, the sampling of $X$ using direct PDF sampling has a much larger acceptance rate, and the average NOE is $2$. The over-all required NOE is $1(\mb{for } Y)+2(\mb{for } X)=3$, which is smaller than the $4$ NOE required for direct PDF sampling.

The efficiency of sampling using the envelope method depends on the choice of envelope, and the closer the shape of $g$ is to $f$, the better the efficiency. In the extreme case that the envelope becomes the original distribution, $g=f$, the acceptance becomes $100\%$ and the sampling using the envelope method degenerates into the sampling using the inverse CDF method (assuming the envelope distribution is sampled using the inverse CDF method for efficiency).

As a summary, the efficiency of sampling using the envelope method is greater than the direct PDF sampling method but worse than the inverse CDF sampling method. Only when the inverse CDF cannot be analytically calculated or pre-tabulated and when the NOE spent on constructing CDF is significant should the sampling using envelopes method be tried. The envelope distribution should be so constructed that: (1) its shape resembles the desired distribution\footnote{However there is one constraint: the envelope function, when multiplied by a constant, should be larger than the original distribution (see \ref{subsec:iSS:envelope_distribution}).}; (2) its CDF can be pre-tabulated, calculated analytically, or computed numerically with only low expenses.

Often an envelope function can be numerically constructed as a sum of step functions, which only requires evaluation of $f$ at a few locations, and a general-purpose example is given in Sec.~\ref{subsec:iSS:automatic_envelope}.

The last method to analyze is the grouping method (4). Similar to the envelope method, its efficiency varies depending on the choice of groups.

As a case study, consider the set $S$ is divided into $l_1$ subsets: $S_{i_1}$, $i_1=1,..,l_1$, and each subset is further divided into $l_2$ sub-subsets: $S_{i_1,i_2}$, $i_2=1,...,l_2$, and so on. Assume ``on the ground level'' each sub-sub-...-subset contains only one element, that is, $S_{i_1,i_2,...}$ is an element. Assume that at each level, the probability of $X$ being in any subset is the same: $p(X\in S_{i_1,i_2,...,i_k}) = l^{-1}_1 l^{-1}_2 ... l^{-1}_k$. Assume for now that we use direct PDF sampling method to sample groups at each level. The efficiency for sampling using such a grouping is discussed in the next paragraph.

For sampling subsets at the 1st level $S_{i_1}$, the success rate is $1/l_1$ and the average NOE is $l_1$. Next, knowing which 1st level subset $X$ belongs to, to sample the 2nd level subsets $S_{i_1,i_2}$ requires similarly an average NOE $l_2$. Continuing this it is clear that the total NOE required in the end is $l_1+l_2+...$. Note that $l_j$'s are subject to the constraint $l_1 l_2 ... =N$, and recall that the direct PDF sampling method requires $N$ NOE, it is then clear that the sampling using grouping method is much more efficient.

The best efficiency is attained when $l_1=l_2=...=2$. In such extreme case, the required NOE is naively estimated as $2+2+... \sim 2\log_2(N)$ but in fact it is just $1$. The reason is that $l_j=2$ is different from $l_j>2$ because even a ``failure" in the sampling can also tell us which subset $X$ belongs to. For example, if $X$ does not belong to $S_0$, then it must belong to $S_1$. Therefore both ``success" and ``failure" can be used to inquire the subset $X$ belongs to and in the end only $1$ NOE is required\footnote{Again, all other calculations, like throwing $\log_2(N)$ random dices, are considered to have negligible time here.}.

In fact, when all $l_j$'s are $2$, the grouping sampling method becomes the sampling using inverse CDF method. To see this, first notice that we can re-order the set $S$ so that $S_0=\{x, F(x)<1/2\}$ and $S_0=\{x, F(x)>1/2\}$\footnote{The boundary of the interval is irrelevant only for continuous distribution but it is ignored here for simplicity.} etc., which in general gives that:
\begin{equation} \label{eq:iSS:S_binary}
  S_{i_1,i_2,...,i_k} = \left\{x: \frac{i_1}{2^1} + \frac{i_2}{2^2} + ... + \frac{i_k}{2^k} <F(x)< \frac{i_1}{2^1} + \frac{i_2}{2^2} + ... + \frac{i_k+1}{2^k} \right\}.
\end{equation}
Then it is clear that what the grouping method does is just a binary search for the element $S_{i_1,i_2,...}$ with a given sequence $(i_1 i_2 ...)$. The sequence becomes a real number in $[0,1]$ when its elements are interpreted as digits in binary code: $(i_1 i_2 ...)_2 = i_1 2^{-1} + i_2 2^{-2} + ...$. In this way, the grouping method can be interpreted as: For a uniformly chosen real number $t_0 \in [0,1]$, first express it in binary code, then perform a binary search for the element $x$ that, according to \equ{iSS:S_binary}, satisfies $F(x) = t_0$. Therefore the grouping sampling method is just a particular numerical implementation (binary search for inverse) of the inverse CDF sampling method.

We just showed that in this extreme case, the grouping sampling and inverse CDF methods become identical. For general cases, such as when not all $l_j$'s are $2$'s, or when some of the subsets are sampled using inverse CDF methods, its efficiency lies between the direct PDF sampling and the inverse CDF methods.

In reality, the grouping method is much slower than the sampling using inverse CDF method, since even choosing a random number between $0$ and $1$ for $\log_2(N)$ number of times costs a significant number of computations. The grouping method, like the sampling with envelope method, should only be tried when the inverse CDF cannot be analytically calculated or pre-tabulated. If part of the sampling (usually the sampling of the groups) uses the inverse CDF sampling method, then for a grouping scheme, the larger the portion of the entire sampling that is done with the inverse CDF sampling method, the more efficient the whole process becomes.

As a summary, the efficiency of the grouping sampling method is greater than that of the direct PDF sampling but less than that of the inverse CDF sampling. It should only be attempted if the inverse CDF cannot be calculated analytically or pre-tabulated, or is too time consuming when constructed numerically. When each level of the group is sampled using the direct PDF sampling method, it achieves better efficiency when more levels of groups are used, with fewer groups at each level. Also, the more groups are sampled using the inverse CDF method, the more efficient the method.

%
%
%
%

\subsection{Automatically generated stair function envelope distribution}
\label{subsec:iSS:automatic_envelope}

Many distributions are piecewise monotonic. Within one such monotonic interval $[a,b]$, an envelope function can be constructed using constant function. Taking the proportional constant to be $k=1$ (see Sec.~\ref{subsec:iSS:envelope_distribution}), due to the monotonicity of the PDF, the envelope can be taken as a constant function $g=c$ where $c=\max\{f(a),f(b)\}$. Given that the real axis is divided into a series of intervals on each of which $f$ is monotonic, a stair-shaped envelope function can be generated by using a constant envelop function on each monotonicity interval. Such a construction only involves evaluating $f(x)$ at those boundary points that separate the intervals, and the inverse CDF of a stair function is easily calculable; therefore when the number of monotonicity intervals is small, constructing and sampling with such a stair-shaped envelope function becomes efficient.

For example, assuming one wants to sample a Gaussian distribution  $f(x)=e^{-x^2}$ on $[0,100]$. When using direct PDF sampling, $95\%$ of the time the suggested sample for $X$ lies in the interval $[5,100]$ and its acceptance is below $e^{-25} \ll 1\%$ and it almost always is rejected. However, if a two-stair envelope distribution $Y$ with PDF $g=1$, $x\in[0,5]$, and $g=e^{-25}$, $x\in[5,100]$ is used, then the acceptance rate for $X$ will be greatly increased because sampling $Y$ will only give a tiny chance to suggest a sample above $5$ for $X$. Furthermore, the construction of such an envelope including its inverse CDF only requires $2$ evaluations of $f(x)$, at $x=0$ and $x=5$, which is negligible compared to the NOE required in direct PDF sampling, which is typically $> 10,000$ (inverse acceptance rate).

In many cases, the encountered distribution function has a unique maximum located at its ``mode" and is monotonically decreasing on both sides when $x$ goes away from the mode. Examples include the Gaussian distribution function, the negative binomial distribution (NBD), etc. For such distribution functions, one way to automatically generate an envelope is to start with its mode, and construct stair-shaped envelope functions whose width for each step is, for example, its standard deviation, and which reaches out for a few (e.g. 6) multiples of it. Such a stair-shaped envelope function can be constructed without knowing the analytic form of the inverse CDF and it only requires $\sim 10$ more NOE. For rapidly dropping distributions like Gaussian and NBD, the time constructing the envelope is negligible compared to the time spent on typical direct PDF sampling.

For piecewise monotonic distributions like the Gaussian and NBD distributions, if they do not have {\em parameter dependence}, their inverse CDF should be calculated in advance and pre-tabulated to provide maximum sampling efficiency; if however they do have parameter dependence and tabulating their inverse CDF becomes impossible, the automatically constructed stair-shaped envelope can greatly boost performance compared to a brute-force PDF sampling method.

\subsection{Generating genuine continuous samples}
\label{subsec:iSS:discrete_vs_continuous}

For a variable $X$ defined on a continuous set $S$, it can be sampled using the direct PDF sampling method (or other composite methods) with continuum PDF, which will produce samples that can take any continuous values in $S$\footnote{This is also true if the inverse CDF has an analytical expression, which is really rare.}. However, when $X$ is sampled using numerically constructed discretized inverse CDF, the generated samples are on a discrete set upon which the CDF is constructed. When the discrete set is dense, such a discrete sampling hardly even causes any practical problems. If the discrete set is not so dense and genuine continuous samples are preferred, the continuum can still be approximately restored by an additional random process: instead of using a sample in the discrete set, a ``perturbed'' sample, which is generated by assigning an additional random shift whose range is half of the discretization spacing, can be used. Such a process generates samples close to the continuous sampling method while maintaining the efficiency of the inverse CDF sampling method (the cost for the additional random process is by our assumption ignored, see Sec.~\ref{subsec:iSS:sampling_efficiency}).

\subsection{Sampling multi-dimensional random variables}
\label{subsec:iSS:sampling_multidimensional}

For multi-dimensional random variables, the direct PDF sampling method, the sampling with envelope method, and the grouping sampling method can be generalized naturally, but it is not clear how the most efficient sampling method, the inverse CDF sampling method, can be generalized to multi-dimensional case for a continuous set $S$, while the solution for a discrete $S$ is simple. Since almost all functions are discretized on a computer, in this section the discussion is restricted to the case that $S$ is discrete.

When $S$ is discrete\footnote{The following argument actually only requires a weaker assumption that $S$ can be made into an ordered set.}, it can be ``flattened'' into a 1d set by re-ordering its elements in any desired fashion. In this way, any random variable $X$ on $S$ can be treated as a 1d random variable and all the sampling methods and their sampling efficiency analyses explained in previous sections directly apply.

For example, supposing $S=\{(x,y)\}$ resembles the $[0,1]\times[0,1]$ area, after discretizing along both directions the interval $[0,1]$ into $10$ points, $S=\{(x_i,y_j)\}$, $x_i,y_j=0.1,...,1$. The set $S$ can be ordered, for example, as:
\begin{equation} \label{eq:iSS:S_100}
  S=\left\{(0.1,0.1),...,(1,0.1),(0.2,0.1),...,(0.2,1),...,...,...,(1,1)\right\}
\end{equation}
Therefore any ``2d" random variable $X$ on $S$ can be treated as a 1d random variable on the re-ordered set. Higher-dimensional random variables can be ``flattened" similarly.

As explained (Sec.~\ref{subsec:iSS:sampling_efficiency}), sampling using the inverse CDF is the most efficient method: if it can be calculated analytically or pre-tabulated, or if a large number of samples are desired so the time spent on constructing the CDF numerically is negligible, the sampling using inverse CDF method should always be used. Otherwise the sampling with envelope or grouping methods should be tried before using the direct PDF sampling method.

When applying the grouping sampling method to multi-dimensional variables, it is very natural to divide $S$ into subsets according to one of the dimensions. For example, the set $S$ in \equ{iSS:S_100} can be divided into $10$ groups $S_i=\{(0.1\times i, 0.1), (0.1\times i, 0.2),...\}$. By doing so, the sampling using grouping method essentially samples $X=(X_1,X_2)$ by first sampling $X_1$ then sampling $X_2$.  

For multi-dimensional random variables, there is yet one more practical constraint: even if the CDF of the variable can be pre-calculated, it cannot be tabulated due to memory shortage. In such situations, one practical solution is to tabulate not the CDF for the multi-dimensional variable, but only for the CDF corresponding to higher level groups created in grouping sampling. To give an example, suppose that discretization into $100$ points is too much so that the inverse CDF cannot be stored for $X=(X_1,X_2)$ on $S$ given in \equ{iSS:S_100}, but the inverse CDF for the variable $X_1$, which controls the sampling of groups $S_i$, requires only $10$ points as it has a much larger chance to fit into the memory.

\subsection{Possible issues}
\label{subsec:iSS:possible_issues}

\subsubsection{Deviations due to limited number of samples}
It may seem trivial that, with a limited number of samples, the statistically calculated quantities, for example the mean and standard deviation, have statistical errors compared to the exact values calculated from the distribution function. However during numerical calculations, some situations may still seem to give ``counterintuitive'' results when using exotic distributions. For example, if $X$ takes the value $0$ with $1-10^{-1000}$ probability and $10^{1000}$ with $10^{-1000}$ probability, it has mean $1$ but in almost any actual sampling process the mean is $0$. Therefore when sampling distributions with fast-dropping tails, the mean calculated from samples is expect to deviate from the exact value, especially for a highly skewed distribution.

\subsubsection{Discretizing continuous distribution}
When continuous distributions are discretized, the weight on each discretized point should be the integral of the PDF (probability) around that point, instead of being the PDF itself (probability density) at that point. The difference only matters if the discretization is on an irregular lattice. For example, assuming that $X$ is uniform on $[0,3]$ and one discretization is $[0,2]$ represented by the middle point $x=1$ and $[2,3]$ represented by middle point $x=2.5$, then the weight at $x=1$ should be twice that at $x=2.5$, in order to generate twice as many samples in $[0,2]$ than $[2,3]$. This requires using the integral of the PDF on $[0,2]$ and $[2,3]$ as the sampling weights instead of the values of the PDF at $x=1$ and $x=2.5$ as the sampling weights; that is, weights are probabilities, not probability densities.

One scenario in which irregular lattice discretization is encountered is during the sampling on a lattice specified by Gauss quadrature. Assuming that a function is numerically costly to evaluate but it needs to be integrated, then the numerical integral can be efficiently calculated by summing its values on only a few lattice points given by Gauss quadrature, during which process values of it on the lattice are stored to avoid repeated calculation. If such a discretization of this function is used later as a PDF in sampling, then samples should be generated not using the values of the function on the lattice, but using the integral of it on each interval, that is, with the specified Gauss weight.

\section{Sampling the particle momentum distribution}
\label{sec:iSS:momentum_distribution}

The iS and iSS codes read the hydrodynamic variables along the conversion surface generated from the hydrodynamical simulation, then calculate the particle momentum distribution and observables, and also sample particles accordingly. Both codes calculate the particle momentum distribution using the Cooper-Frye formula for the emission function which, together with the details for calculating related observables and sampling, will be explained in this section.

\subsection{Emission function and Cooper-Frye formula}
\label{subsec:iSS:Cooper_Frye}

The emission function is the function that determines how many particles on average will be emitted from a given location on the conversion surface. In general, it depends on $(x^{\mu},p^{\nu})$, where the $4$-vector $x=(\tau,\vxT,\eta_s)$ has the following components: the proper time $\tau$; the transverse location $\vxT$, and the space-time rapidity $\eta_s$ on the conversion surface. The $4$-vector $p=(E,\vpT,y)$ has the following components: the energy $E$, the transverse momentum $\vpT$, and the particle rapidity $y$. All the out-going particles are on-shell, and the 4-vector $x^{\mu}$ is constrained to be on the conversion surface, so there are only $6$ free components on which the emission function depends.

The conversion surface is conventionally denoted as $\Sigma$; it is a 3-dimensional set of points in 4-dimensional space-time. An infinitesimal piece of $\Sigma$ is characterized by its surface normal vector $d^3\sigma_{\mu}$, whose norm gives the 3-volume of the element and whose components give the direction.

The Cooper-Frye formula has been widely used to calculate the emission function, whose differential form is the following:
\begin{equation} \label{eq:iSS:Cooper_Frye}
  dN(\tau, \vxT, \eta_s, \vpT, y) = g p^{\mu} d^3\sigma_{\mu} \frac{d^3p}{E (2\pi)^3}  f(x,p).
\end{equation}
Here $g$ is the spin degeneracy, and $f$ is the distribution function taking the form\footnote{All through this chapter, the upper sign is for fermions and lower sign is for bosons.}:
\begin{equation} \label{eq:iSS:distribution_function}
  f(x,p) = f_0 + \delta f = \frac{1}{e^{(p\cdot u-\mu)/T}\pm 1} + \frac{\chi(p^2) p^{\mu} p^{\nu} \pi_{\mu\nu}}{2 T^2 (e+p)} f_0 (1\mp f_0).
\end{equation}
$e$, $p$, $T$, $u^{\mu}$ are the energy density, pressure, temperature, and 4-flow velocity of the fluid, respectively; and $\mu$ is the chemical potential of the particle; the function $\chi$ is a function of only $p^2$; $\chi=1$ corresponds to the quadratic ansatz for $\delta f$.

In numerical calculations, it is natural to separate the rapidity dependence from the surface element as $d^3\sigma_{\mu} = \tau d\eta_s d^2\ts_{\mu}$. It is also convenient to use the variable $\esp=y-\eta_s$ and to write $d^2\pT=\pT\,d\pT d\phi$. With these choices, the emission function \equ{iSS:Cooper_Frye} assumes the form:
\begin{equation} \label{eq:iSS:Cooper_Frye_code_style}
  \frac{dN}{dy} (\tau, \vxT, \esp, \vpT, y) = g \left(\tau d\esp d^2\ts_{\mu}(\xT)\right) \frac{\pT d\pT d\phi}{ (2\pi)^3} p^{\mu} f(x,p).
\end{equation}
The emission function drops rapidly with increasing $|\esp|$ or $\pT$, and this influences the sampling efficiency, as will be explained in Sec.~\ref{sec:iSS:sampling_emission_function}.

The differential emission function given by \equ{iSS:Cooper_Frye} is not necessarily always positive since, depending on the shape of $\Sigma$, $p^{\mu}d^3\sigma_{\mu}$ can become negative; for sampling with probability methods, we need, however, a positive-definite emission probability density. This issue and its possible treatments are discussed in Sec.~\ref{sec:iSS:sampling_emission_function}.

\subsection{Spectra and flow calculations}
\label{subsec:iSS:spectra_and_flow_calculations}

The azimuthally averaged spectrum is calculated from $dN/(dy\,\pT d\pT d\phi)(\vpT)$ as:
\begin{equation} \label{eq:iSS:spectra}
  \frac{dN}{2\pi dy\,\pT d\pT} = \frac{1}{2\pi} \int_0^{2\pi} d\phi \frac{dN}{dy\,\pT d\pT d\phi}(\vpT).
\end{equation}
and the differential anisotropic flows $v_n$'s are the Fourier coefficients with respect to the azimuthal angles of the same distribution:
\begin{equation} \label{eq:iSS:anisotropic_flows}
  v_n e^{i n \Psi_n} = \frac{ \int \pT d\pT d\phi \,dN/(dy\,\pT d\pT d\phi) \,e^{i n \phi} }{ \int \pT d\pT d\phi \,dN/(dy\,\pT d\pT d\phi) }.
\end{equation}
Therefore the calculation of spectra and flow requires the calculation of the distribution function $dN/(dy\,\pT d\pT d\phi)$, which from \equ{iSS:Cooper_Frye_code_style} is calculated as:
\begin{equation} \label{eq:iSS:dN_for_spectra_and_flows}
  \frac{dN}{dy\,\pT d\pT d\phi} = \frac{g}{(2\pi)^3} \tau \int_{\Sigma} d\esp d^2\ts_{\mu}  p^{\mu} f(x,p).
\end{equation}

In numerical calculations, the distribution function $dN/(dy\,\pT d\pT d\phi)$ is represented as an array whose elements are calculated using \equ{iSS:dN_for_spectra_and_flows}. Usually to achieve high numerical accuracy, a dense conversion surface discretization is used, and the numerical integrals in \equ{iSS:dN_for_spectra_and_flows}, which involve numerical sums over the conversion surface, become very time consuming. Therefore a quadrature rule that requires a smaller size of the $dN/(dy\,\pT d\pT d\phi)$ array while not losing accuracy means less computing time for the integral over the conversion surface, resulting in better efficiency. Gauss quadrature, which requires much fewer sampling points, is thus used in our codes for both the $\pT$ and $\phi$ integrals.

The integral over $\esp$ in \equ{iSS:dN_for_spectra_and_flows} can also be done using Gauss quadrature. Under the boost-invariant assumption, the integrand is an even function and the integral further reduces to one along only the positive axis, which is then calculated using $15$-point Gauss quadrature.

Further optimization for performing the numerical integral in \equ{iSS:dN_for_spectra_and_flows} involves adjusting the order of the integration loops, using local variables, pre-tabulating mathematical functions, etc. The resulting code iS is $\sim 7$ times faster compared to the previous used AZSpectra code.

Note that a negative differential emission function is physically allowed in all the integrals in this section, since only $dN/dyd^2\pT$ (i.e.\ a momentum distribution integrated over $\Sigma$) and $dN(\vxT, \eta, \tau)$ (i.e.\ the spatial emission density integrated over momentum) have to be always positive definite. However, for the sampling algorithms to work, the integrands need to be replaced by their absolute values, or the negative contributions must be set to zero by hand. While this gives physically incorrect results, it is necessary for testing the sampling algorithm. This is explained in detail in Sec.~\ref{sec:iSS:sampling_emission_function}.

\section{Sampling the emission function}
\label{sec:iSS:sampling_emission_function}

We now discuss how to sample the emission function to generate particles whose space-time and momentum distributions are given by the differential emission function (\ref{eq:iSS:Cooper_Frye_code_style}). Required information are the total emitted number of particles, their species, the emission location, and their momenta. The space-time information of the particles comes from the location on the conversion surface from where they are emitted. Their momentum information is sampled according to the Cooper-Frye formula (\equ{iSS:Cooper_Frye} or (\ref{eq:iSS:Cooper_Frye_code_style})). 

In numerical calculations, the conversion surface is generated from the hydrodynamic calculation, and it is discretized into surface cells which are labeled by the ``conversion surface cell index'', denoted as $\foi$ in this section. The space-time information $\vxT$, $\tau$, and $\eta_s$ on the conversion surface cell are functions of this index. For a boost-invariant system, the conversion surface is invariant under a shift in the $\eta_s$ direction, and only the conversion surface at $\eta_s=0$ is stored. Therefore only the $\vxT$ and $\tau$ information are expressed through the conversion surface cell index, and the $\esp=\eta_s-y$ dependence is kept explicitly. Finally, for boost-invariant systems, the $\esp$-integrated emission function does not depend on particle rapidity; the rapidity $y$ dependence of the particle can always be sampled additionally in the end with a uniform distribution in a given rapidity range, which together with $\esp$ determines $\eta_s$. Therefore the quantities that the differential emission function $dN$ depends on in practice are: the conversion surface cell index $\foi$, the transverse momentum $\pT$ and $\phi$, and the relative space-time rapidity $\esp$.

Of these $4$ quantities, the conversion surface cell index is already discretized, and the other $3$ can be discretized or kept as continuous quantities. As explained in Sec.~\ref{subsec:iSS:discrete_vs_continuous}, continuous samples can still be approximately generated from samples on a discrete set, so the issue of whether the generated samples assume continuous or discrete values will be largely ignored in this section, and the main focus is on the efficiency of the sampling algorithm.

\subsection{The purely numerical approach}
\label{subsec:iSS:numerical_approach}

The straightforward approach is where all required quantities are calculated numerically.

For a given species of particle, the average total yield $N$ is calculated by numerically integrating the differential emission function over all its dependences. By storing all the partial sums encountered in the numerical integration, the inverse CDF for the flattened variable (see Sec.~\ref{subsec:iSS:sampling_multidimensional}) can be generated at the same time. As explained in Sec.~\ref{subsec:iSS:sampling_efficiency}, the inverse CDF sampling method is the most efficient one, and it should be used for the generation of particle samples.

However in practice, the inverse CDF for the full differential $dN$ can hardly fit into the current generation computer memory. In a typical calculation, we have $\sim30,000$ surface cells, $720$ discretized $\vpT$ points, and $40$ discretized $\esp$ points, giving an array of size $864$MB, already dangerously large. In the most central Pb-Pb collisions at $2.76$AGeV energy at the LHC, the number of conversion surface cells exceeds $100,000$, and the memory demand reaches $3$GB. My personal design philosophy is to leave flexible memory spaces to allow for unexpected extreme cases, for possible future modifications, etc. Furthermore, it is soon realized that input/output to files is more costly than the sampling time difference between the inverse CDF sampling method and the one that is currently used, as explained below.

From Sec.~\ref{subsec:iSS:sampling_efficiency}, it is clear that, if the inverse CDF sampling method cannot be applied, then the other two composite sampling methods should be tried. For a multi-dimensional random variable $dN$, it is most natural to use the grouping sampling method (see Sec.~\ref{subsec:iSS:sampling_multidimensional}), with the sampling of groups being efficiently handled by the inverse CDF sampling method.

As explained, the variable $dN$ effectively depends on a tuple of $4$ quantities: the conversion surface cell index $\foi$, the transverse momentum magnitude $\pT$, the transverse momentum angle $\phi$, and the relative rapidity $\esp$. Their discretized sizes are $\sim30,000$, $15$, $48$, and $40$, respectively. The total sampling space size is the product of these numbers, which will be divided into groups. Recall that the sampling of groups is done using the efficient inverse CDF sampling method, thus the larger the groups the more efficient the process becomes (see Sec.~\ref{subsec:iSS:sampling_efficiency}). With the memory limitations and the desire for code simplicity, the largest group can be formed using the surface index and the relative rapidity, leaving out the transverse momentum degrees of freedom. Such a grouping gives groups of size $\sim 30,000\times 40$ which easily fit into memory, and their partially integrated emission function
\begin{equation} \label{eq:iSS:dN_dxtdetady}
  \frac{dN}{dy d\esp}\left(\foi, \esp, y\right) = g \tau(\foi) d^2\ts_{\mu}(\foi) \int \frac{\pT d\pT d\phi}{(2\pi)^3} p^{\mu} f(x,p)
\end{equation}
will be used in the inverse CDF sampling.

Each emitted particle is generated by first sampling the group it belongs to using the inverse CDF method, which gives a conversion surface cell index $\foi$ and a relative rapidity $\esp$. Once having them, the transverse momentum of this particle is then sampled using the direct PDF sampling method, which is efficient since it is the transverse momentum information $\vpT$ for only {\em one} particle that needs to be sampled, which falls into the category of ``generating a small number of samples'', see Sec.~\ref{subsec:iSS:sampling_efficiency}.

Another small advantage is that direct sampling with the PDF generates samples with continuous $\vpT$ distributions, which reduces the discretization error for quantities that depend sensitively on angular distributions, for example the high-order anisotropic flows.


The efficiency of the direct PDF sampling relies strongly on the estimate of the maximum (Sec.~\ref{subsec:iSS:sampling_efficiency}). These maxima for the $\pT$-weighted differential emission function can be approximated by their maxima on the discrete lattice in momentum space, and they are stored during the calculation of the total multiplicity.

Other grouping configurations have also been tried, including one with a single inverse CDF function for the whole differential emission function. Indeed the efficiency from inverse CDF sampling method is the fastest, and it is $30\%\sim50\%$ faster than the one proposed above, when clocked for pure sampling without writing to files. When writing to files is turned on, the speed difference is no more than $\sim10\%$, and $\sim60\%$ of the time is spent on writing to files. In fact, significantly more time would be spent on writing to files had there not been a manually constructed ``buffer'' for file writing, which is explained in Sec.~\ref{subsec:iSS:optimizations_and_models}.


\subsection{Semi-analytic approach}
\label{subsec:iSS:analytic_approach}

For a boost-invariant system, the flow rapidity is equal to the space-time rapidity, and it follows that the distribution function is a function of only $\esp$ instead of both $y$ and $\eta_s$. By changing variables, the invariant integral measure $d^3p/E$ can also be written as $d^3p/E = dy\,\pT d\pT d\phi = -d\esp\,\pT d\pT d\phi$. Therefore the emission function calculated in each cell of the conversion surface:
\begin{equation} \label{eq:iSS:dN_dxtdy}
  \frac{dN}{dy}\left(\foi, y\right) = g \tau(\foi) d^2\ts_{\mu}(\foi) \int d\esp \int \frac{\pT d\pT d\phi}{(2\pi)^3} p^{\mu} f(x,p)
\end{equation}
is a scalar under boosts. The semi-analytic approach takes advantage of this invariance and evaluates $dN/dy$ in the local fluid rest frame.

First it can be shown that the integral of $\delta f$ is zero. To see this\footnote{As an alternative argument, first write the integral measure as $d^3p$, then it is clear that the integral involving $\delta f$ is proportional to:
\begin{equation*} \label{eq:iSS:dN_integral_1}
  \int \frac{d^3p}{E(2\pi)^3} \chi(p^2) p^{\mu} p^{\rho}p^{\sigma}\pi_{\rho\sigma} f_0(1\mp f_0) = \int \frac{d^3p}{(2\pi)^3} \chi(p^2) p^{\rho}p^{\sigma}\pi_{\rho\sigma} g^{\mu 0} f_0(1\mp f_0)
\end{equation*}
since for $\mu\neq0$ the integrand is odd in $p^i$ and thus integrates to $0$. Similarly $\rho$ has to be the same as $\sigma$ to produce non-zero results. Since in the local rest frame $\pi^{\rho0}=\pi^{0\sigma}=0$, the integral in \equ{iSS:dN_integral_1} for $\mu=0$ can thus be written as:
\begin{equation*} \label{eq:iSS:dN_integral_2}
  \int \frac{dp_x dp_y dp_z}{(2\pi)^3} \chi(p^2)\, \left(p_x^2 \pi_{xx} + p_y^2 \pi_{yy} + p_z^2 \pi_{zz} \right) f_0(1\mp f_0)
\end{equation*}
Next notice that by symmetry, $p_x^2$, $p_y^2$, and $p_z^2$ can be replaced by $p^2/3$ and then it follows from the traceless condition of $\pit$ that the integral over $\delta f$ is zero.}, write the integral as
\begin{equation} \label{eq:iSS:dN_integral_1}
  \left(\int \frac{d^3p}{E(2\pi)^3} \chi(p^2) p^{\mu} p^{\rho}p^{\sigma} f_0(1\mp f_0) \right)\pi_{\rho\sigma}
\end{equation}
The quantity in the parentheses is a contravariant rank-3 tensor that depends only on $u^\mu$, therefore it can only be a sum of tensors composed from $u^\mu$ and $g^{\mu\nu}$. If the indices $\rho$ or $\sigma$ is carried by $u$, then by transversality of $u^{\mu}$ and $\pit$, the corresponding contraction upon $\rho$ or $\sigma$ gives zero; otherwise the term contains $g^{\rho\sigma}$ which also leads to zero when contracted with $\pi_{\rho\sigma}$ by the tracelessness of $\pit$.

Thus the full distribution function $f$ can be replaced by the equilibrium distribution in the local rest frame:
\begin{equation} \label{eq:iSS:dN_integral_3}
  f_0(x,p) = \frac{1}{e^{(p^\nu u_\nu - \mu)/T} \pm 1} = \frac{1}{e^{(E - \mu)/T} \pm 1}
\end{equation}
where the last equality used that $u^\mu=(1,0,0,0)$ in local rest frame. From the symmetry of $p^\mu$, it is clear that:
\begin{equation} \label{eq:iSS:dN_integral_4}
  \begin{aligned}
  \int \frac{d^3p}{E(2\pi)^3} p^{\mu} f_0(x,p)
  =& \int \frac{d^3p}{E(2\pi)^3} p^0 \frac{1}{e^{(p^\nu u_\nu - \mu)/T} \pm 1} g^{\mu 0} \\
  =& \int \frac{d^3p}{(2\pi)^3} \frac{1}{e^{(E - \mu)/T} \pm 1} g^{\mu 0}
  \end{aligned}  
\end{equation}
And finally,
\begin{equation} \label{eq:iSS:dN_integral_5}
  \int \frac{d^3p}{(2\pi)^3} \frac{1}{e^{(E - \mu)/T} \pm 1}
  = \frac{T^3}{2\pi} \int d(p/T) \frac{(p/T)^2}{e^{\sqrt{(p/T)^2 + (m/T)^2} - (\mu/T)} \pm 1}
\end{equation}
and the last integral can be pre-calculated and tabulated for rescaled mass $m/T$ and rescaled chemical potential $\mu/T$.

Furthermore in the local rest frame for any $4$-vector $A$, its $0$-th component can be written as $A^0=A^0 u_0=A^\mu u_\mu$ where the last expression can be evaluated in any frame; thus in the local rest frame, $d^2\ts_0$ becomes $d^2\ts_{\mu} u^\mu$ in the lab frame. Therefore \equ{iSS:dN_dxtdy} becomes:
\begin{equation} \label{eq:iSS:dN_dxtdy_semi_analytic}
  \frac{dN}{dy}\left(\foi, y\right) = g \tau(\foi) d^2\ts_{\mu}(\foi) u^\mu \frac{T^3}{2\pi} \int d(p/T) \frac{(p/T)^2}{e^{\sqrt{(p/T)^2 + (m/T)^2} - (\mu/T)} \pm 1}
\end{equation}

Using \equ{iSS:dN_dxtdy_semi_analytic}, the emission function on each conversion surface cell in \equ{iSS:dN_dxtdy} can be calculated without preforming the numerical integral, but by evaluating an algebraic expression with results read from a pre-calculated table. This is the most important step in this fast semi-analytic sampling approach.

As mentioned, the variable $dN$ depends on a tuple of $4$ quantities: conversion surface cell index $\foi$, transverse momentum magnitude $\pT$, transverse momentum angle $\phi$, and relative rapidity $\esp$. The semi-analytic sampling approach is a grouping sampling method that divides the full $(\foi,\pT,\phi,\esp)$ sampling space into groups labeled by $\foi$. The probability for a sampled particle to belong to each group is given by $dN/dy$ in \equ{iSS:dN_dxtdy_semi_analytic}. The groups are sampled by the inverse CDF method, which gives the conversion surface cell index from which a particle is emitted; once a particle is known to be emitted from a certain conversion surface cell, its $(\pT,\phi,\esp)$ dependence is sampled using the direct PDF sampling method.

Recall from Sec.~\ref{subsec:iSS:sampling_efficiency} that the efficiency of the direct PDF sampling method depends intensively on the closeness of the estimated maximum of the PDF to the real one. Compared to the purely numerical approach (Sec.~\ref{subsec:iSS:numerical_approach}), the maximum of the differential $dN$ {\em cannot} be extracted numerically since there is no more numerical integration thus no more numerical scan of the integrand over the sampling space, and it has to be estimated analytically.

In the following derivation, the maximum is closely related to the function:
\begin{equation} \label{eq:iSS:maximum_core_equ_1}
  G(E; A) = \frac{E^A}{e^{\beta(E-\mu)} \pm 1},\,A>0
\end{equation}
where $\beta=T^{-1}$. By setting its derivative to zero the extrema can be found by solving:
\begin{equation} \label{eq:iSS:maximum_core_equ_2}
  (1 \mp f_0) = \frac{A}{\beta E} \Longleftrightarrow
  \left\{
  \begin{aligned}
    & x e^x = y;\,x=\beta E-A,\,y=A e^{\beta\mu-A}, \mbox{fermions (upper)} \\
    & x e^{-x} = y;\,x=A-\beta E,\,y=A e^{\beta\mu-A}, \mbox{bosons (lower)} \\
  \end{aligned}
  \right.
\end{equation}
This equation is transcendental and it cannot be solved algebraically; however, the solution to the equations expressed in terms of $x$ and $y$ can be pre-calculated and tabulated. For fermions (upper sign), the solution always exists and it is expressed by the Lambert W-function; for bosons (lower sign) the equation only has real solutions when $y<1/e$, and it yields two solutions; the one needed is the one satisfying $x\in[0,1]$. In the following, the solution to this equation will be denoted as $\Emax^\pm$ when it exists.

The maximum of $G(E;A)$ with constraint $E\geq m$ will be denoted as $\Gmax^{(A)}$ and it depends on several conditions:
\begin{enumerate}
  \item For fermions (upper sign), $G(E)$ has a single peak at $\Emax^+$ and the constraint maximum is $G(\Emax^+)$ if $\Emax^+>m$ and otherwise is $G(m)$.
  \item For bosons (lower sign) with $A e^{\beta\mu-A} > 1/e$, \equ{iSS:maximum_core_equ_2} has no solution and the maximum is $G(m)$.
  \item For bosons (lower sign) with $A e^{\beta\mu-A} \leq 1/e$, $G(E)$ has two extrema in $(\mu,\infty)$ with the larger one being the maximum and given by $\Emax^-$. If $\Emax^-<m$ then the maximum is $G(m)$; otherwise the maximum is the larger one between $G(m)$ and $G(\Emax^-)$.
\end{enumerate}

In the following, an estimate is given of the maximum of the differential emission function:
\begin{equation} \label{eq:iSS:dN_differential}
  F(x,p) = d^2\ts_{\mu} p^{\mu} f(x,p)
\end{equation}

Write $F=F_0+\delta F$ corresponding to $f_0$ and $\delta f$, respectively. Note that $F_0$ and $\delta F$ are scalars whose evaluations, including their maximum, are frame independent. Their maxima are thus estimated in the local rest frame, but expressed using quantities in the lab frame.

For a boost-invariant system, $d^2\ts_3:=d^2\ts_{\eta_s}=0$ and the conversion surface cell volume in the local rest frame can be estimated as:
\begin{equation} \label{eq:iSS:fz_volume_estimate_1}
  \begin{aligned}
  d^2\ts_{\mu} p^{\mu} & = d^2\ts_0 E + d\vec{\ts}_\perp \cdot \vpT \\
  &\leq E \left( |d^2\ts_0^\mr{L.R.F}| + |d^2\ts_\perp^\mr{L.R.F}| \right) = E \left( |d^2\ts_0^\mr{L.R.F} u^0| + |d^2\ts_\perp^\mr{L.R.F}| \right)
  \end{aligned}
\end{equation}
As explained earlier in this section, in the lab frame $d^2\ts_0^\mr{L.R.F} u^0$ should be replaced by $d^2\ts_\mu u^\mu$. The quantity $d^2\ts_\perp^\mr{L.R.F}$ in the lab frame has the form $\sqrt{d\sigma_\mu d\sigma_\nu \Delta^{\mu\nu}}$, where $\Delta^{\mu\nu}=u^\mu u^\nu - g^{\mu\nu}$ is the spacial projection operator, and it can be explicitly calculated by an actual boost from $d\ts$ in the lab frame as:
\begin{equation} \label{eq:iSS:fz_volume_estimate_2}
\begin{aligned}
  \left( d^2\ts_\perp^\mr{L.R.F} \right)^2
  &= d^2\ts_0^2 \left( (u^1)^2+(u^2)^2 \right) + d^2\ts_1^2 \left( 1+(u^1)^2 \right) + d^2\ts_2^2 \left( 1+(u^2)^2 \right) \\
  & + 2 d^2\ts_0 d^2\ts_1 u^0 u^1 + 2 d^2\ts_0 d^2\ts_2 u^0 u^2 + 2 d^2\ts_1 d^2\ts_2 u^1 u^2
\end{aligned}
\end{equation}

For the equilibrium contribution $F_0$, it is clear from \equ{iSS:fz_volume_estimate_1} and \equ{iSS:dN_differential} that the remaining part is to calculate the maximum for the function:
\begin{equation} \label{eq:iSS:F0_estimate_1}
  E f_0 = \frac{E}{e^{(E-\mu)/T} \pm 1} = G(E; 1)
\end{equation}
and the solution is $\Gmax^{(1)}$.

Next I give an estimate to the maximum of the off-equilibrium part with choice $\chi(p^2)=p^\beta$:
\begin{equation} \label{eq:iSS:deltaF_estimate_1}
  \delta F = d^2\ts_{\mu} p^{\mu} f_0(1\mp f_0) \frac{p^\alpha p^\mu p^\nu \pi_{\mu\nu}}{2T^2(e+p)}
\end{equation}
The conversion surface cell volume is again estimated as in \equ{iSS:fz_volume_estimate_1} and \equ{iSS:fz_volume_estimate_2}. For a boost-invariant system $\pi_{3i}=0$; in the local rest frame, the rotational freedom in the transverse direction can be used to set $\pi_{xy}=0$; it then follows from the H\"older inequality and the fact that $|p_{x,y,z}| \leq p \leq E$ that
\begin{equation} \label{eq:iSS:deltaF_estimate_2}
\begin{aligned}
  p^\mu p^\nu \pi_{\mu\nu} &= p_x^2 \pi_{xx} + p_y^2 \pi_{yy} + p_z^2 \pi_{zz} \leq E \left( |p_x| |\pi_{xx}| + |p_y| |\pi_{yy}| + |p_z| |\pi_{zz}| \right) \\ &\leq E \left( E \sqrt{\pi_{xx}^2 + \pi_{yy}^2 + \pi_{zz}^2} \right) = E^2 \sqrt{\pi^{\mu\nu}\pi_{\mu\nu}}
\end{aligned}
\end{equation}
The quantity $\pi^{\mu\nu}\pi_{\mu\nu}$ is a scalar and thus can be evaluated using $\pi^{\mu\nu}$ from the lab frame. Using that $p^\beta \leq E^\beta$ and after combining all the power of $E$'s and assuming that $f_0<1$, the rest is straightforward:
\begin{equation} \label{eq:iSS:deltaF_estimate_3}
  E^{\beta+3} f_0 (1\mp f_0) \leq \alpha E^{\beta+3} f_0 = \alpha G(E;\beta+3) \leq \alpha \Gmax^{(\beta+3)}
\end{equation}
where $\alpha=1$ for fermions (upper sign) and $\alpha=2$ for bosons (lower sign).

To summarize, the maximum for the differential $dN$ that integrates to the one given in \equ{iSS:dN_dxtdy} is estimated as:
\begin{equation} \label{eq:iSS:maximum_summary}
  \left(dN/dy\right)_\mr{max} = \frac{g}{(2\pi)^3} \tau \left( |d^2\ts_\mu u^\mu| + |d^2\ts_\perp^\mr{L.R.F}| \right) \left( \Gmax^{(1)} + \frac{\sqrt{\pi^{\mu\nu}\pi_{\mu\nu}}}{2T^2(e+p)} \Gmax^{(\beta+3)} \right)
\end{equation}
where $d^2\ts_\perp^\mr{L.R.F}$ is given in \equ{iSS:fz_volume_estimate_2}.

The validity of the assumption $f_0<1$ depends on the choice of the conversion surface cell temperature, chemical potential, etc. It is practically true almost all the time, although there are instances where it is violated. If a more rigorous result is desired, the inequality \ref{eq:iSS:deltaF_estimate_3} can be replaced by the following one:
\begin{equation} \label{eq:iSS:deltaF_estimate_4}
  E^{\beta+3} f_0 (1\mp f_0) \leq |E^{\beta+3} f_0| + |E^\gamma f_0| |E^{\beta+3-\gamma} f_0| \leq \Gmax^{\beta+3} + \Gmax^\gamma \Gmax^{\beta+3-\gamma}
\end{equation}
where $0\leq \gamma \leq \beta+3$ is a parameter one can tune to sharpen the inequality. 

The efficiency of direct PDF sampling with maximum given by \equ{iSS:maximum_summary} can be tested numerically by studying the ratio between the real maximum encountered in actual sampling and  $(dN/dy)_\mr{max}$. This ratio varies: it is $80\%\sim 95\%$ for pions, which take most of the sampling time due to their large multiplicity, and it can be as low as $\sim 50\%$ for some heavy particles.

There is yet another approximate treatment for the maximum: the observed numerical maximum from the large number of samples is very close to the real maximum;  therefore when a huge number of samples are needed, the observed numerical maximum from some portion of the samples can be used to approximate the real distribution maximum for the rest of the sampling process. Such a ``dynamic maximum'' mechanism has been implemented and it can be enabled by switching parameters, as explained in detail in Sec.~\ref{sec:iSS:running_code}.

\subsection{Comparison between the approaches}
\label{subsec:iSS:comparison_between_approaches}

Both methods use the grouping sampling method, with groups sampled by the inverse CDF method and sampling within groups by the direct PDF sampling method. The grouping between these two approaches are different, and the main difference is how the PDF for the probability distribution of the group random variables is constructed: the numerical approach constructs it numerically and the semi-analytic approach construct it semi-analytically. The total amount of time spent on sampling can be split into two parts: the part for calculating the total multiplicity and constructing the PDF for the group variable and its inverse CDF, and the part for the actual sampling.

For the construction of the group variable PDF and its inverse CDF, the numerical approach takes a much longer time, due to the numerical summation over a huge array. As a comparison, the semi-analytic approach takes almost no time at this stage; in fact most of the time it takes is to read the pre-calculated tables from files.

For the actual sampling, the numerical approach takes a much shorter time per particle sample compared to the semi-analytic approach, and the reasons are explained in the following.

The first reason is that due to the different grouping scheme, the semi-analytic approach needs to sample one more degree of freedom $\esp$ compared to the numerical approach.


The second reason is that the numerically estimated maximum in the numerical approach is closer to the real maximum compared to the analytically estimated maximum in the semi-analytic approach. However, it should be warned that the numerically estimated maximum in the numerical approach is from a discrete $\vpT$ lattice and it is not the exact maximum on the continuum $\vpT$ space, although the actual difference is so tiny that it generates only $< 10^{-6}$ portion of defects.

Because of the different characteristics of the two approaches, the numerical approach is suitable for a large number of repeated sampling while the semi-analytic approach is suitable for a small number of repeated sampling. For clarity, I define $M$-repeated sampling such that with a given conversion surface, $M\times N$ number of particles will be sampled, where $N$ is the total number of particle calculated from the Cooper-Frye formula; thus a large number of repeated sampling means $M\gg 1$ and a small number of repeated sampling means $M\approx 1$.

The sampling time is tested with a realistic conversion surface containing $34,516$ conversion surface cells. The calculated thermal $\pi^0$ yield is $\sim 136$. The test is done using one CPU core with $2.2$GHz frequency.

First the two approaches are tested with $500$ repeated sampling. The numerical approach takes $16.67$s to calculate the group variable PDF and its inverse CDF and the total multiplicity for $\pi^0$; a similar amount of time is spent for each of the rest of the $320$ species of the particles\footnote{Another ``grouping optimization'' is used here to speed up the calculations so the actual time spent is $\sim 16.67\times 120$, see Sec.~\ref{subsec:iSS:optimizations_and_models}.}. The sampling for $\pi^0$ takes $0.74$s and much less for heavy particles due to the smaller number yields. The total execution time for all species of particles is $1530$s.

The semi-analytic approach takes $17.78$s to calculate the group variable PDF and its inverse CDF\footnote{A high precision table for the integral in \equ{iSS:dN_dxtdy_semi_analytic} is used here; when a low-precision ($\sim 5$ times larger table spacing when tabulating) table is used it takes $\sim 2$s.} for all $320$ species of particles. It then take $\sim 13.1$s to sample $\pi^0$, and much less time for the heavy particle species due to their rareness. The total execution time is $83$s.

Next, they are tested for $200$ repeated samplings. As expected, the running time for the numerical approach does not change much, since most of its time is spent on calculating the group variable PDF and its inverse CDF, and the run time for the semi-analytic approach becomes $17.78+33.05$s.

Finally they are tested for $50,000$ repeated samplings. The time spent on constructing the group variable PDF and its inverse CDF does not change and the numerical approach spent $67$s on sampling $\pi^0$, while the semi-analytic approach spent $1340$s on sampling $\pi^0$.\footnote{The rest of the species are not tested, but we expect the sampling time ratio between the two sampling methods to be roughly the same for other species of particles.}

The comparison is summarized in table \ref{tab:numerical_vs_analytical}.

\begin{table}[h]
 \begin{center}
  \begin{tabular}{c|c|c}
   \hline \hline
   $500$ repeated sampling & pure numerical & semi-analytic \\
   \hline
   calculate $dN/dy$ matrices & $16.67$s for $\pi^0$ & $17.78$s for all particles \\ & $\sim\times 100$ for rest of particles & ($15$s for loading tables) \\
   \hline
   actual sampling & $0.74$s for $\pi^0$ & $13.1$s for $\pi^0$ \\
   & faster for heavier particles & faster for heavier particles \\
   \hline
   total & $1530$s & $83$s \\
   \hline \hline
   $200$ repeated total & $\sim$ same & $17.78+33.0$s \\
   \hline \hline
   $50,000$ repeated $\pi^0$ & $67$s & $1,340$s \\
   \hline \hline
  \end{tabular}
 \end{center}
\label{tab:numerical_vs_analytical}
\caption{Efficiency comparison between pure numerical and semi-analytic methods. The test case has $34,516$ conversion surface cells and $dN/dy|_{\pi^0}\sim 136$. See the text for detailed explanations.}
\end{table}

As a quick summary, for a small number of repeated samplings, the semi-analytic approach is $\sim18$ times faster and should be used; for a large number of repeated samplings, the numerical approach is $\sim20$ times faster and should be used.

Another subtlety is that due to numerical inaccuracy, especially with fluctuating initial conditions, the trace of $\pi^{\mu\nu}$ is not exactly zero. Therefore the $dN/dy$ for one piece of the conversion surface calculated from the semi-analytic approach, which assumes the traceless of the $\pi^{\mu\nu}$, is not exactly the same as that calculated from the numerical approach. However, with our $\pi^{\mu\nu}$ regulation treatment in the hydrodynamical simulation (\sec{misc:pimunu_regulation}), the difference is small. It has been checked that only for small $dN/dy$ can the difference become as large as $\sim 2\%$ and only for extremely small $dN/dy$ can the difference becomes as large as $\sim 10\%$.

\subsection{Other optimizations and implemented models}
\label{subsec:iSS:optimizations_and_models}

\subsubsection{Particle grouping}
Within the Cooper-Frye formalism, two particles that are both fermions or both bosons are considered the same if they have the same mass, chemical potential, and spin degeneracy. Such particles can be grouped and the emission function needs only to be calculated once for each group. At least a particle and its anti-particle belong to the same group, which cuts the total number of particles that need to be calculated from $320$ to $\sim 150$. Furthermore, if high accuracy is not necessary, particles with mass and chemical potential difference within a certain threshold can also be grouped to speed up the calculation. These particles still need to be of the same type, meaning that they are both fermions or both bosons, and have the same spin degeneracy.

An automatic particle grouping treatment is implemented. Such a treatment first orders all particles according to their mass, then during the loop over particle species for calculating the emission function, if two successive particles with the same type have mass and chemical potentials with the given threshold, the previous result from the first particle will be reused.

\subsubsection{Determining total number yield}
The Cooper-Frye formula gives average particle yields, which usually is not an integer; however the actually particle yield in each sampling has to be an integer. The way to get the integer number yield from the average value is model dependent. So far three models have been implemented.

The first model separates the average particle yield into an integer and a fractional part, and the factional decimal is used as the probability to sample one more particle. For example, if the average pion yield is $15.2$, then during the actual sampling $20\%$ of the time $16$ pions are generated instead of $15$.

The second model also separates the average particle yield into an integer and a factional part. However, instead of being used as the probability of sampling one more particle, the fractional part is used as the mean number of particle in a negative binomial distribution. The actual yield of the particle then is the integer yield plus a number sampled from the negative binomial distribution.

The third model uses the average particle yield (not only the fractional part) as the mean for a negative binomial distribution to sample the actual integer number yield.

These three models yield the same average yields of particles theoretically, however when performing code checks, one should bare in mind the issue mentioned in Sec.~\ref{subsec:iSS:possible_issues}.

How to switch between these models and to tune their associated parameters are explained in Sec.~\ref{sec:iSS:running_code}.

There are other models that can be used to determine the number yield. For example, the total energy on the conversion surface can be used as one such constraint. Such a sampling model is only partially implemented, but can be completed relatively easily\footnote{The advantage of using total energy to constrain the total number yield is that only one numerical integral needs to be done on the conversion surface instead of for all $320$ species of particles, and it is computationally cheap. However, after the discovery of the ``semi-analytic" (Sec.~\ref{subsec:iSS:analytic_approach}) approach, the total number yield calculation does not cost heavy computation time anymore and the sampling using the total energy method became less appealing to us.}.

It should be emphasized again that the purpose of the iSS code is to create an easy-to-use framework in which additional physics models can be implemented and tested relatively easily. For this reason, the currently implemented models are more for demonstration purpose: the justification and comparison between models are left for future work.

\subsubsection{File-writing buffer}
When using the numerical approach to sample a large amount of samples, the efficiency of writing particle information into files becomes the bottleneck of the sampling subroutine. It was once tested on a scenario that the sampling itself takes only $14$s, while without properly handling, the process of writing to files takes $66$s. The frequent access to files causes a drop in efficiency, and an internal string buffer is then created to temporarily store the samples, and the buffer is only written to files when a large amount of data is accumulated. This string buffer mechanism is tested to be faster than the default buffer provided by the operating system, and its size can be tuned. With optimized buffer size, the time for writing for the example mentioned above is reduced to $23$s.

\section{Code verification}
\label{sec:iSS:code_verification}

The sampling algorithm is tested by comparing quantities calculated directly from the emission function with those calculated using actual generated samples. The results are shown in this section, but the subtlety regarding locally-negative emission functions needs to be elaborated on first. For clarity, the quantities calculated directly from the emission function in this section are referred to as ``theoretical results'' and the quantities calculated from actual generated samples are referred to as ``results from samples''.

The emission function used in the sampling needs to be non-negative, but its actual calculated value from the Cooper-Frye formula may not be necessarily so, and when it is not, it physically describes particles that flow from the hadron gas phase into the quark-gluon plasma phase. These are rare and usually are ignored during calculation, although they are required for ensuring energy conservation for the whole system. The different treatments of the negative emission function therefore create only tiny differences; but in a code checking for the sampling procedure, such systematic uncertainties had better be excluded. In the following, two different treatments for negative emission functions are given, and it will be explained that a slightly unphysical treatment is the one most suitable for a code check. (After the code has been verified, it was only used in the physically correct mode.)

The simplest treatment is to set the differential emission function given in \equ{iSS:Cooper_Frye_code_style} to zero whenever it is negative, which is referred to as the ``enforced-positivity'' treatment. Such a treatment ignores those particles flying back into the quark-gluon plasma phase, and it is considered to be slightly unphysical; but then the emission function is a perfectly legitimate PDF from which particles can be sampled. For comparison, the same treatment needs to be applied when calculating theoretical results, which means that there then has to be a theta function enforcing the positivity of the emission function in any integrals involving the emission function. The conclusion is that an accurate code check for the sampling {\em algorithm} is possible with such a positivity requirement as long as the theoretical calculations are modified accordingly.

Note that this treatment for the emission function is not commonly used in previous calculations of theoretical results, where the negative emission function is kept ``as-it-is'' to maintain energy conservation. The integrated elliptic flow calculated from these two treatments has been checked to give $\sim 1\%$ difference.

There are two approaches used for the sampling (see Sec.~\ref{sec:iSS:sampling_emission_function}). The enforced-positivity method is possible within the numerical approach, where the differential emission function is numerically calculated and thus open to modification. This treatment is not possible within the semi-analytic approach, because the emission function is enclosed in analytic integrals and it needs to stay unmodified in order for the integrals to be solved analytically.

Another treatment is to keep the sign of the emission function ``as long as possible'', and only enforcing it to be positive when necessary. This treatment will be referred as the ``semi-positivity'' treatment in the following. For the current two sampling approaches, which both use the grouping sampling method (Sec.~\ref{subsec:iSS:grouping}), this treatment means that the emission probabilities from group variables are calculated, either as analytic integrals or numerical sums, allowing negative contributions. It is only after the group variable for particle emission has been sampled that during the next stage of the sampling (within groups) the particles are sampled with the enforced-positivity emission function.

However, no theoretical calculations can accommodate the semi-positivity requirement for the emission function used in the sampling. This is because in the sampling procedure, the emission function is first used unmodified in the integration and then enforced to be positive later during the sampling of the integrand, and such a ``timing'' treatment has no associated analytic modifications to the theoretical integrals. Because of this, the theoretical integrals need to be solved at least in the same order, such that the same sequence of treatments can be applied for an honest comparison; for example, for elliptic flows where the spatial integral is performed before the momentum integral, the semi-positivity requirement cannot be implemented because the sampling procedure solves the momentum integral first.

Note that for the first approach, the emission probabilities for group variables are calculated by integrals of only the positive part of the emission function, and they are different from those calculated from the second approach, which has no such constrains. This means that the different treatment not only alters the distributions of the particle samples within each group, that is, the momentum distributions of particles at given spacial locations, but it also alters the distribution of group variables, which is the spacial distribution of the particles.

For code comparison, the enforced-positivity treatment for negative emission functions is used. As explained above, this means that only the purely numerical sampling approach (Sec.~\ref{subsec:iSS:numerical_approach}) is feasible. The comparison is performed using an averaged Monte-Carlo Glauber initial condition corresponding to RHIC collisions in the $30-40\%$ centrality bin. Samples are generated from $50,000$ repeated samplings.

\begin{figure}[h!]
  \begin{center}
  \includegraphics[width=0.75\linewidth]{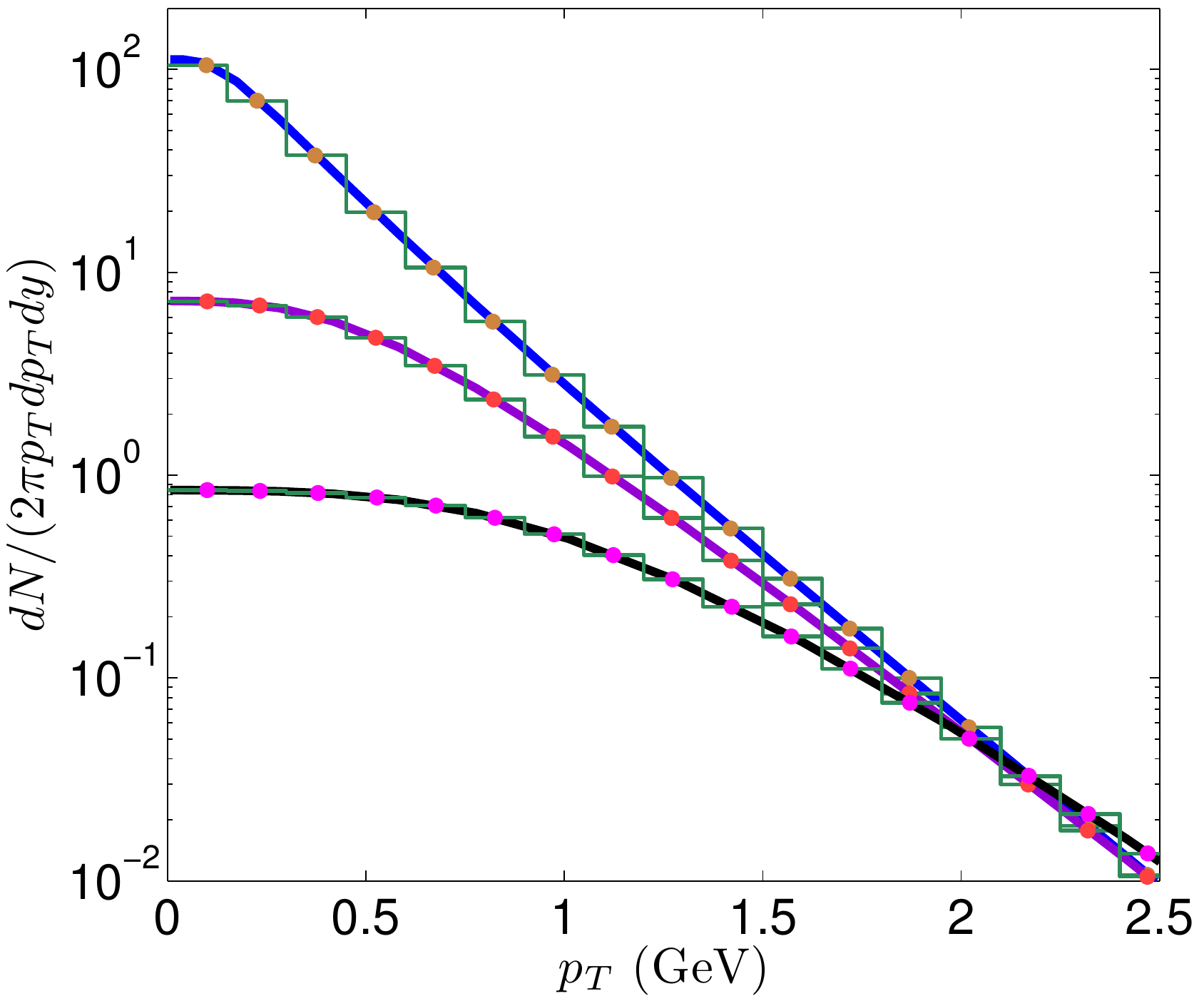}
  \end{center}
  \caption{$\pT$-spectra comparison for thermal pions, kaons, and protons, calculated by integrating emission function (solid curves) or from samples (dots). For other details, see the text.}
  \label{fig:iSS:1}
\end{figure}

The $\pT$-spectra from theoretical results and samples for thermal pions, kaons, and protons are compared in \Fig{iSS:1}.

The spectra calculated from samples agree with the theoretically calculated spectra up to $\pT\sim 2.5$ GeV, beyond which the signal becomes noisy due to low statistics.

\begin{figure}[h!]
  \begin{center}
  \includegraphics[width=0.75\linewidth]{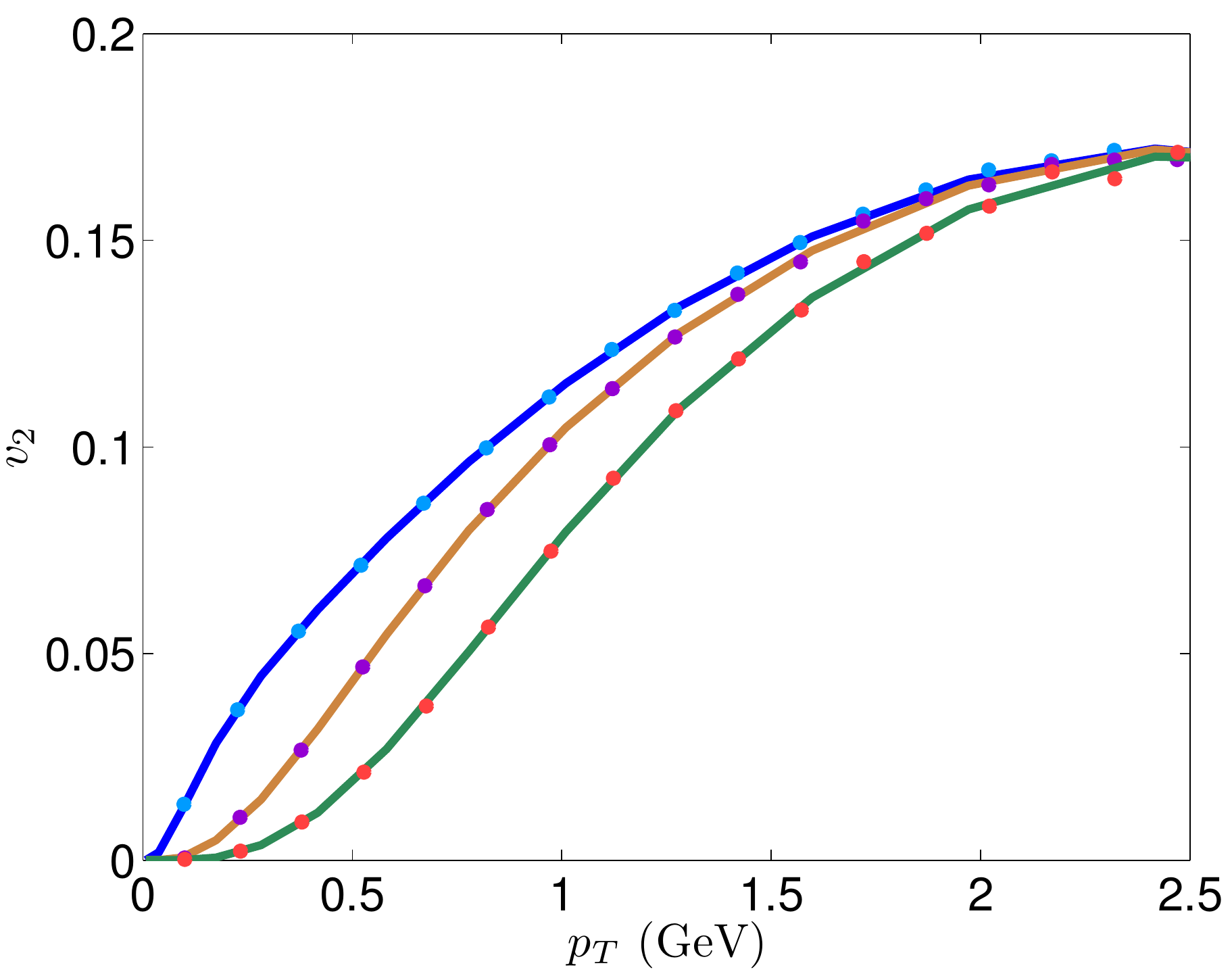}
  \end{center}
  \caption{Differential elliptic flow comparison for thermal pions, kaons, and protons, calculated theoretically (solid curves) or from samples (dots). See the text for details.}
  \label{fig:iSS:2}
\end{figure}

The differential elliptic flow from theoretical results and samples for thermal pions, kaons, and protons are compared in \Fig{iSS:2}.

The differential elliptic flow calculated from samples agrees with the theoretically calculated flows up to $\pT\sim 2.5$ GeV, where the signal starts to become noisy due to low statistics. For the integrated $v_2$, the sampled results agree with theoretical results to first $5$ significant decimals for pions, $4$ for kaons, and $3$ for protons due to the decrease in statistics; such highly accurate agreements give confidence in the verification of the codes and the sampling algorithm.

\begin{figure}[h!]
  \begin{center}
  \includegraphics[width=0.75\linewidth]{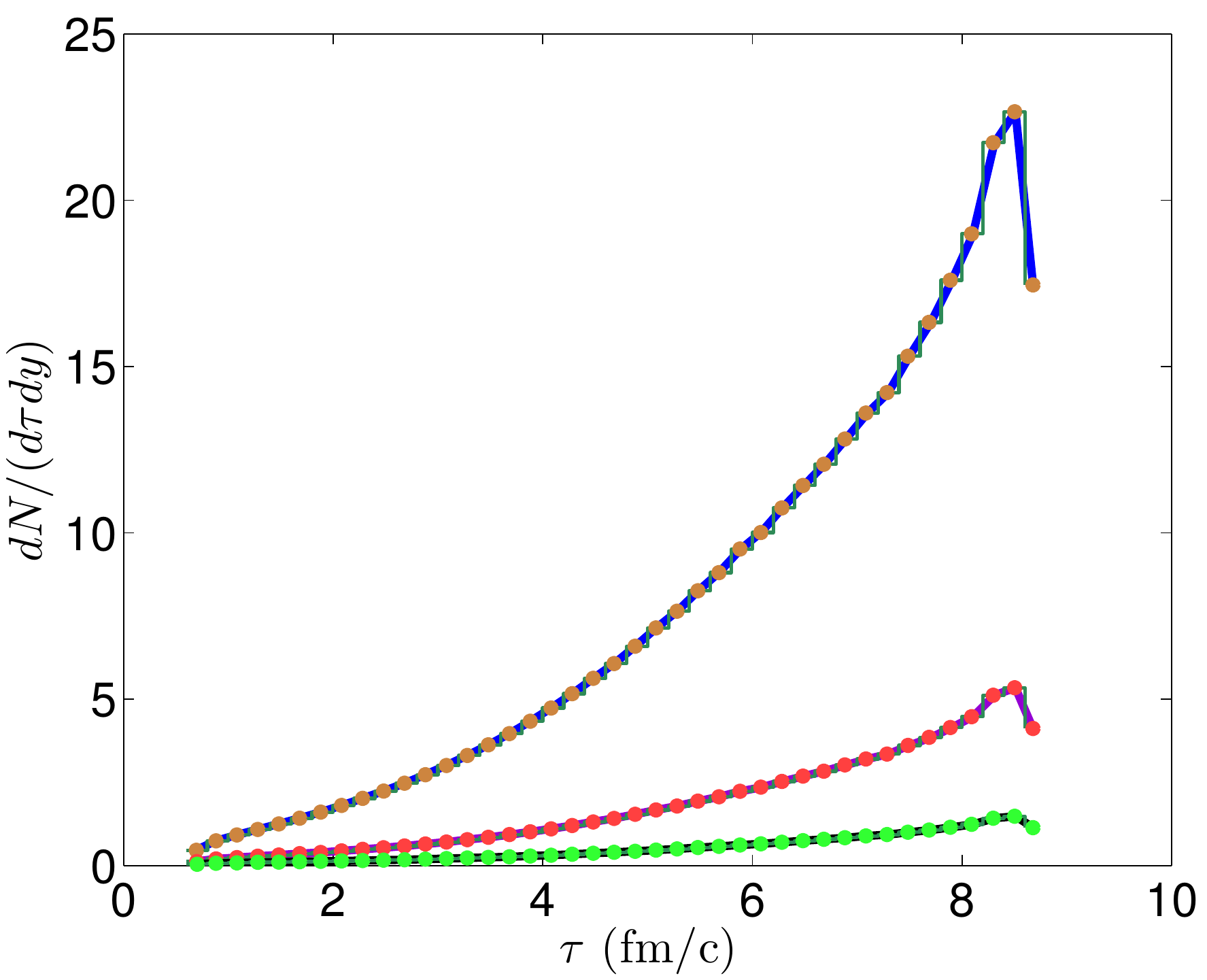}
  \end{center}
  \caption{Emission rate $dN/(d\tau dy)$ for thermal pions, kaons, and protons, calculated theoretically (solid curves) and from samples (dots). See the text for details.}
  \label{fig:iSS:3}
\end{figure}

The emission rate as a function of the proper time $dN/(d\tau dy)$ for thermal pions, kaons, and protons are compared in \Fig{iSS:3}. The agreement between the theoretical results and those from the samples are excellent.

\begin{figure}[h]
  \begin{center}
  \includegraphics[width=0.8\linewidth]{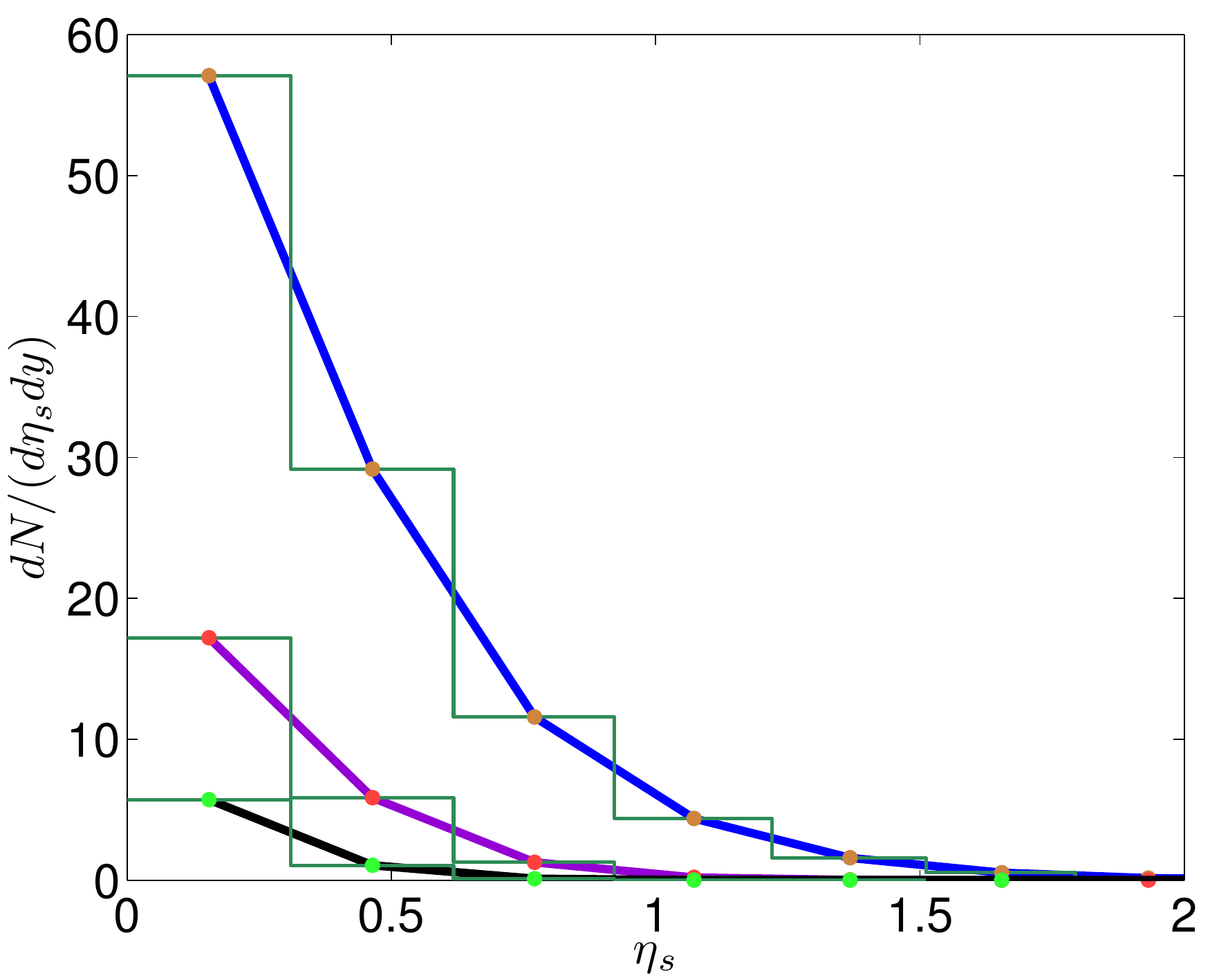}
  \end{center}
  \caption{Distribution of the emission functions in terms of $\eta_s$ at $y=0$ for thermal pions, kaons, and protons, calculated theoretically (solid curves) and from samples (dots). See the text for details.}
  \label{fig:iSS:4}
\end{figure}

The distributions of the emission functions in terms of $\eta_s$ at $y=0$ calculated theoretically and from samples are compared in \Fig{iSS:4} using thermal pions, kaons, and protons. The results from samples agree with those calculated theoretically. Note that the bins used to average the samples need to be chosen according to those used in the theoretical calculations (see Sec.~\ref{subsec:iSS:possible_issues}), where they are separated by the location of the points used by the Gauss quadrature (see also Sec.~\ref{sec:iSS:running_code}).

\begin{figure}[h]
  \begin{center}
  \includegraphics[width=0.8\linewidth]{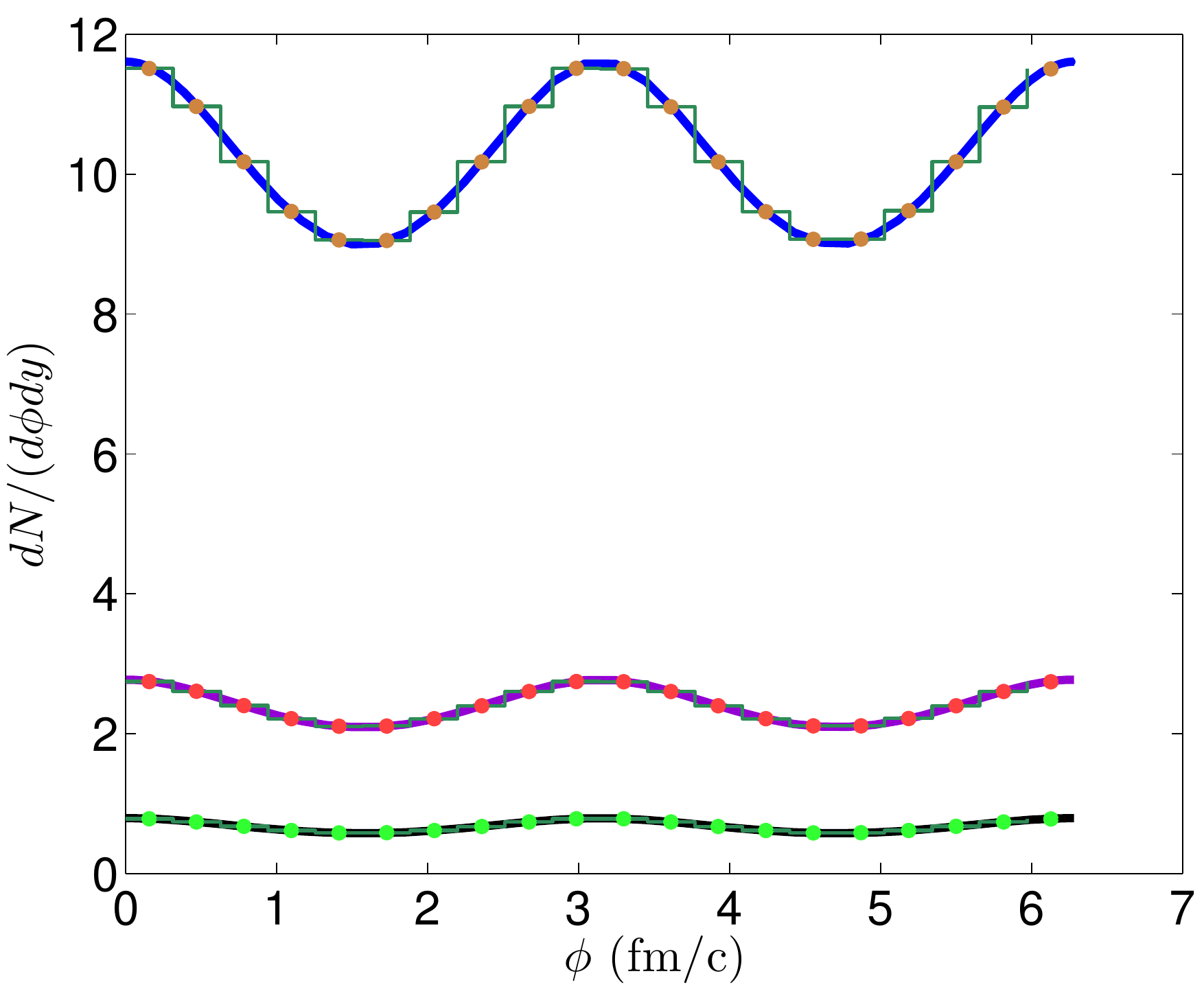}
  \end{center}
  \caption{Angular distribution of the emission function for thermal pions, kaons, and protons, calculated theoretically (solid curves) and from samples (dots). See the text for details.}
  \label{fig:iSS:5}
\end{figure}

In \Fig{iSS:5}, the angular distribution of the emission function $dN/(d\phi dy)$ calculated theoretically and from samples are compared for thermal pions, kaon, and protons. The results from samples agree well with those calculated theoretically. Here again the bins used to average samples need to be chosen according to those used in the theoretical calculations (see Sec.~\ref{subsec:iSS:possible_issues}), which are separated by the location points for Gauss quadrature (see also Sec.~\ref{sec:iSS:running_code}).

As explained earlier in this section, only the pure numerical sampling approach is appropriate for a high-accuracy code check. For the semi-analytic sampling approach, the semi-analytic results for the emission probability from each conversion surface cells have been checked to agree well with the numerical results, except when the emission function is tiny. Possible issues stem from the non-positivity of the differential emission function and from the non-tracelessness of $\pit$ produced by the hydrodynamics code (see more details in Sec.~\ref{subsec:iSS:comparison_between_approaches}).

\begin{figure}[h]
  \begin{center}
  \includegraphics[width=0.44\linewidth]{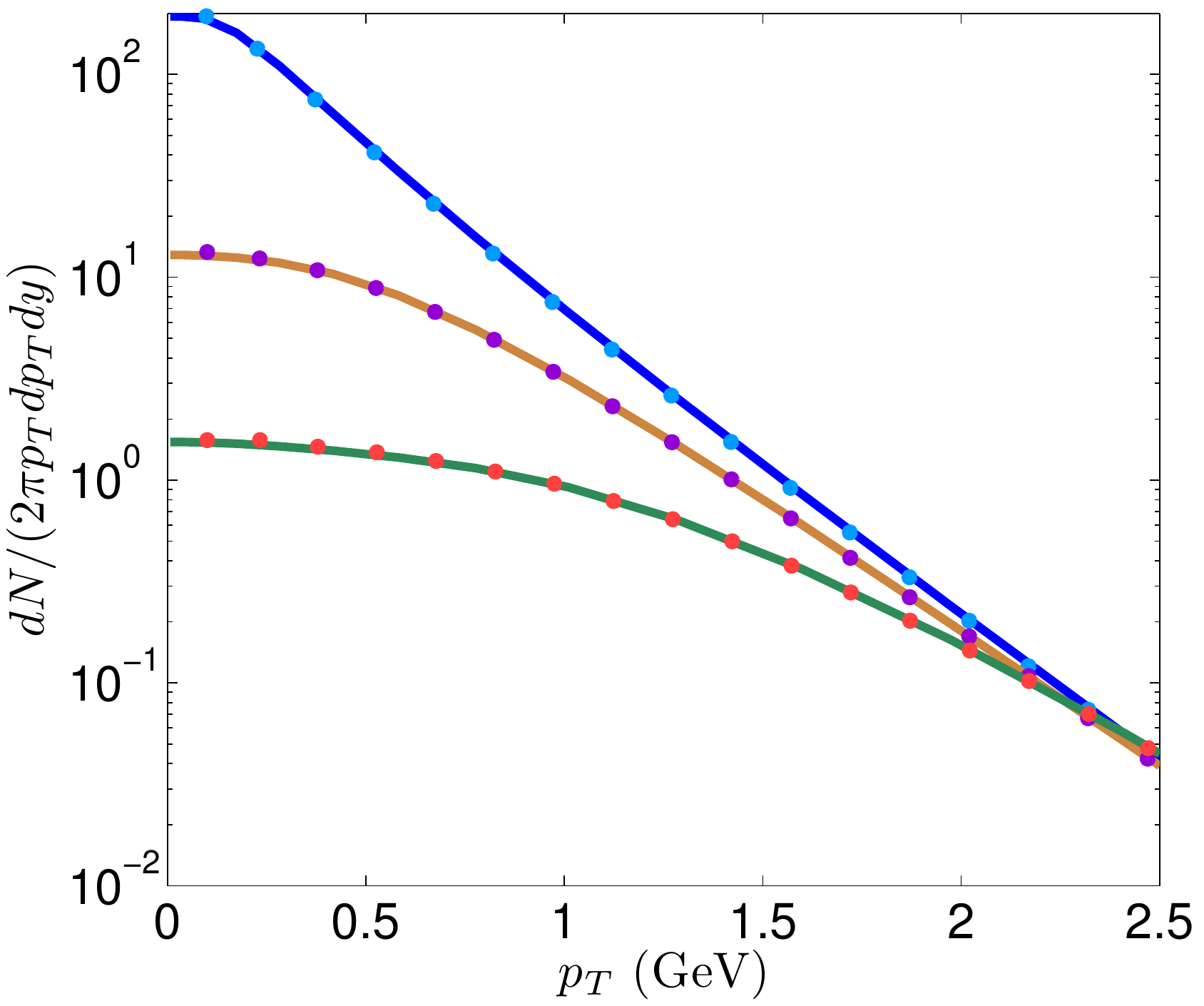}
  \includegraphics[width=0.47\linewidth]{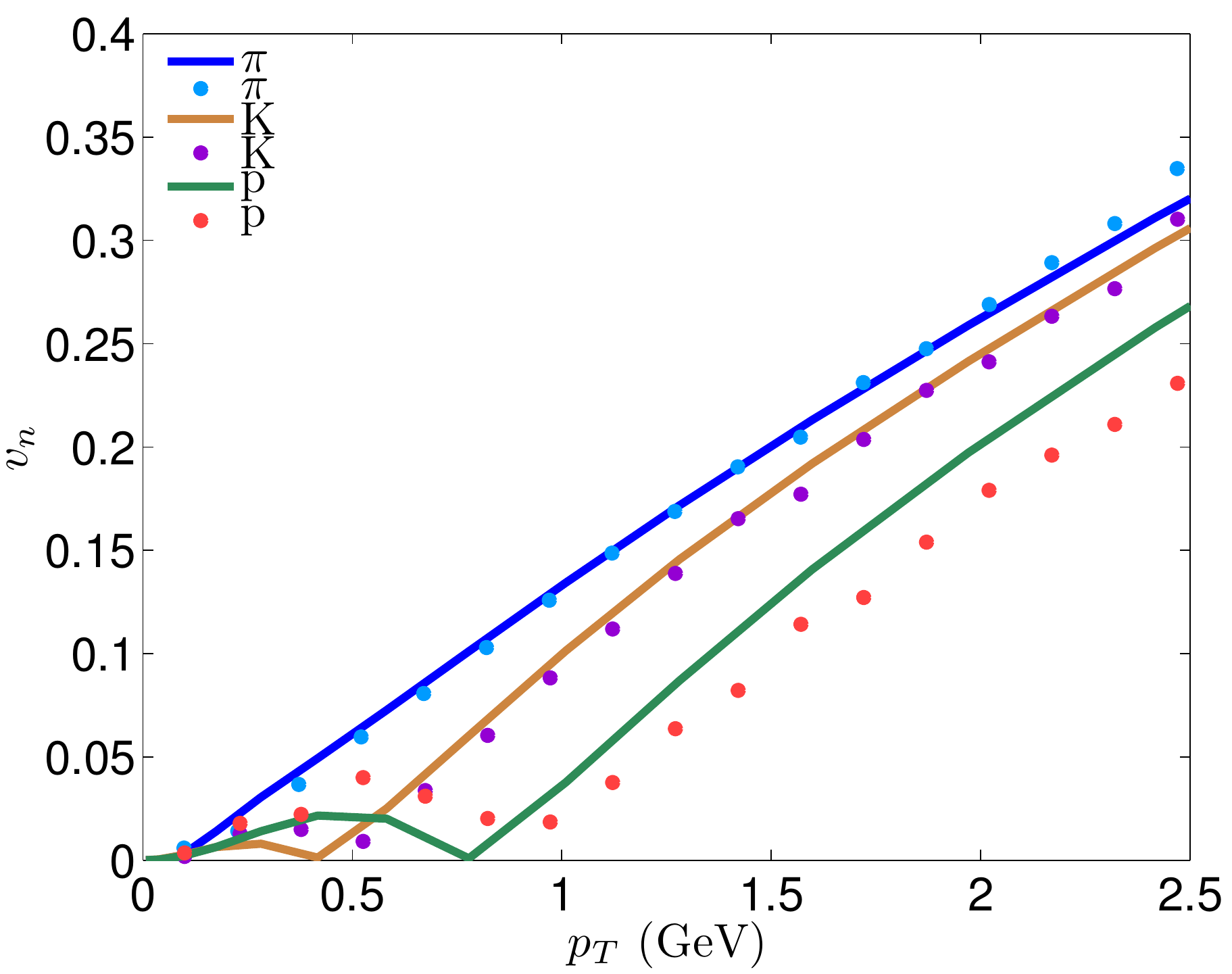}
  \end{center}
  \caption{$\pT$-spectra (left) and differential elliptic flow (right) for thermal pions, kaons, and protons, calculated by integrating the emission function (solid curves) or from samples using the semi-analytic approach (dots). For other details, see the text.}
  \label{fig:iSS:6}
\end{figure}

Still, as a physical comparison rather than a code check, the results from samples generated from the semi-analytic approach can be compared to the theoretical ones. For one bumpy initial condition and $5,000$ samples, the spectra and the differential flows are compared in \Fig{iSS:6}. Here the spectra and differential flows are calculated without any modifications to the emission function, and samples are generated by semi-analytic approach with naturally the semi-positivity modification as explained. The agreement on spectra is good, but the agreement on differential flow for kaons and protons are bad. This is expected since, for bumpy initial conditions, the emission function can become negative at random locations, which (by enforcing it to be positive) generates numeric noise in the signal for the momentum anisotropy that influences the elliptic flow more than the spectra, which rely only on the averaged flow.

\chapter{Miscellaneous topics}
\label{chap:misc}

In this chapter, we discuss some topics addressing questions that occurred during my research. All sections in this chapter are independent from each other.

\section{Comparison between eccentricities defined with $r^2$ and $r^n$ weights}
\label{sec:1104.0650:appa}

%
\begin{figure*}
 \includegraphics[width=0.32\linewidth]{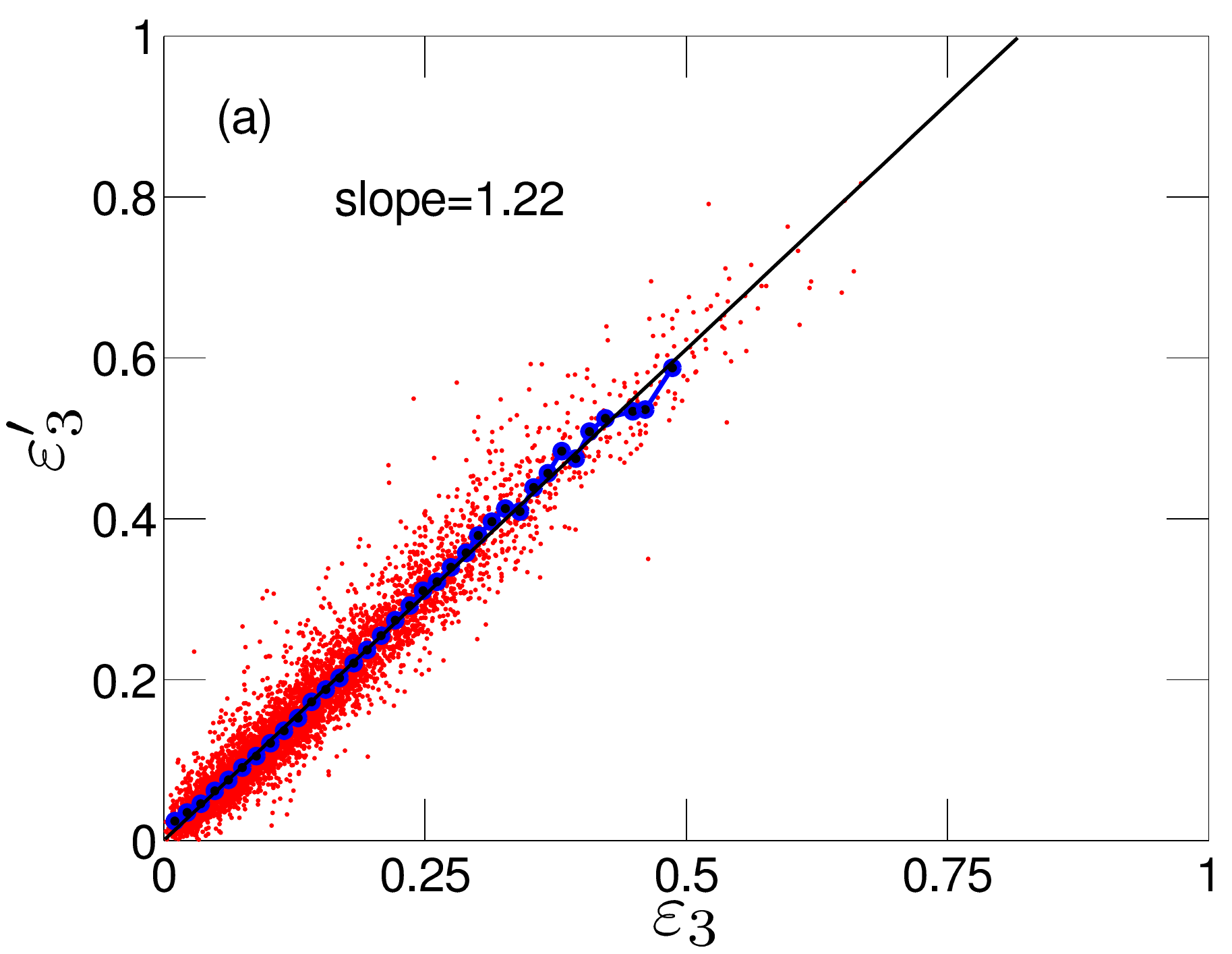}
 \includegraphics[width=0.32\linewidth]{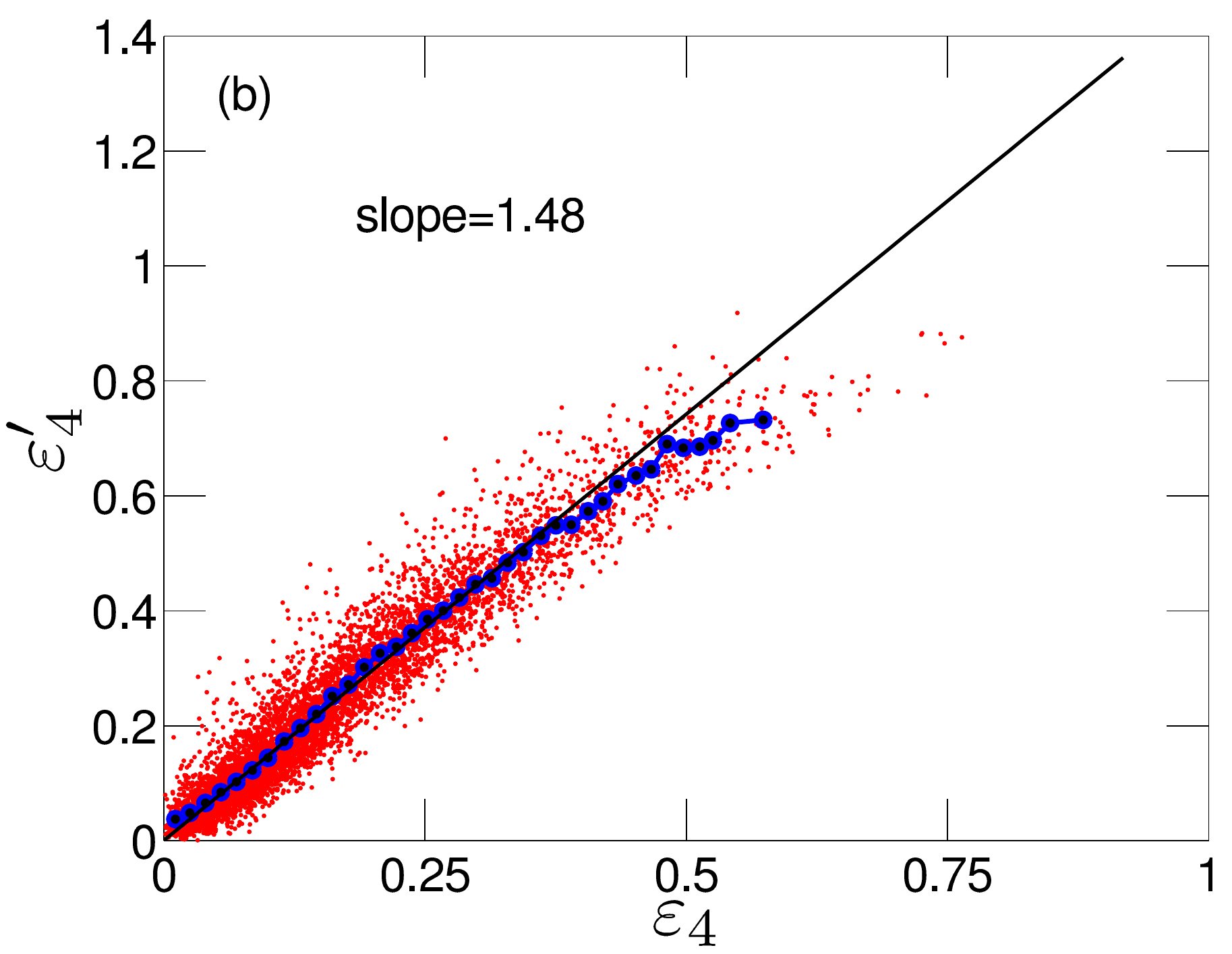}
 \includegraphics[width=0.32\linewidth]{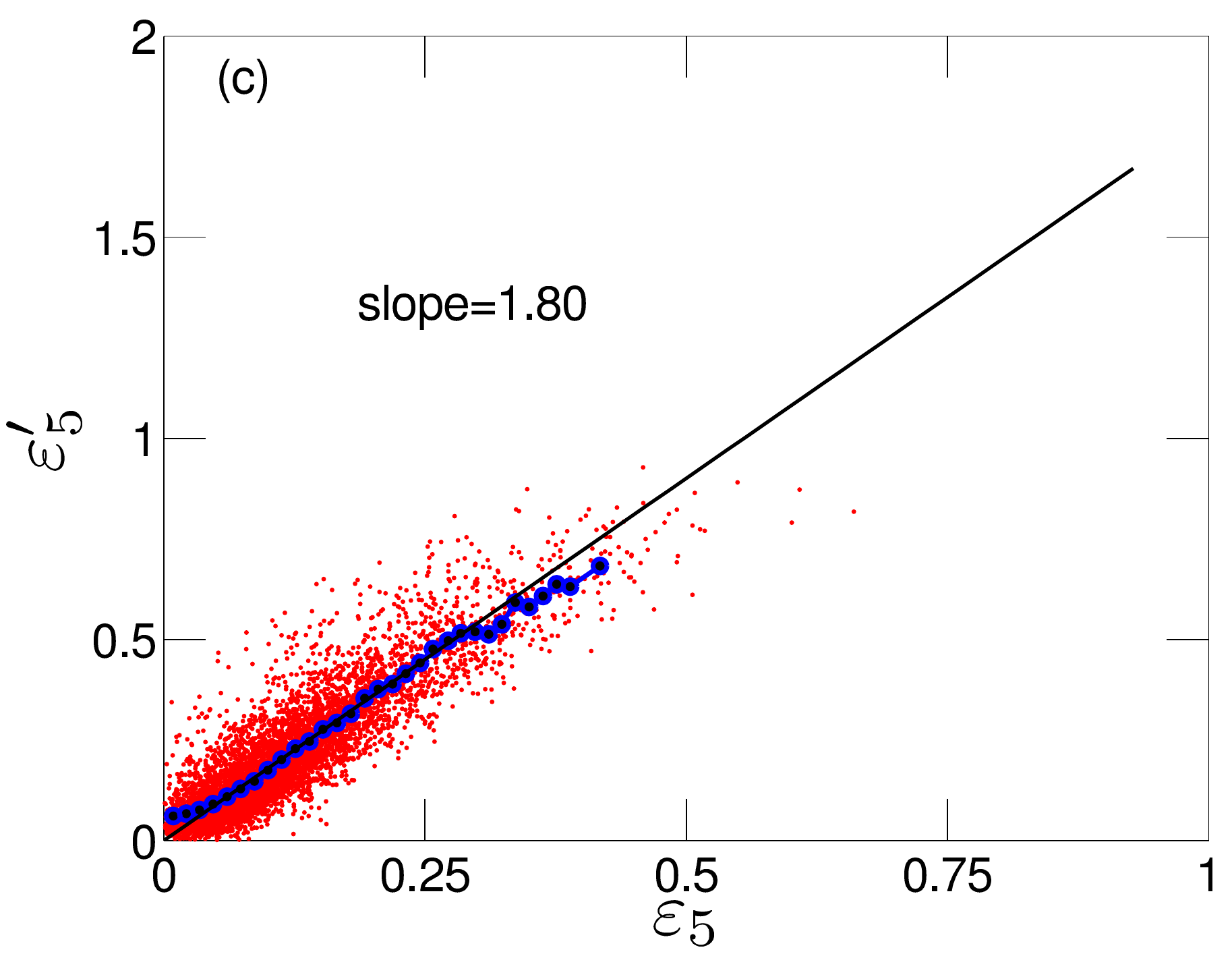}
 \caption{Correlation between $\ecc_n$ and $\ecc'_n$,
   for $n \eq 2,\,3,\,4$ (panels (a-c)). The blue dots are bin averages
   for bins that contain more than 10 events. The thick black lines are 
   linear fits.
 \label{1104.0650:F16}
 }
\end{figure*}
%

%
\begin{figure*}
 \includegraphics[width=0.32\linewidth]{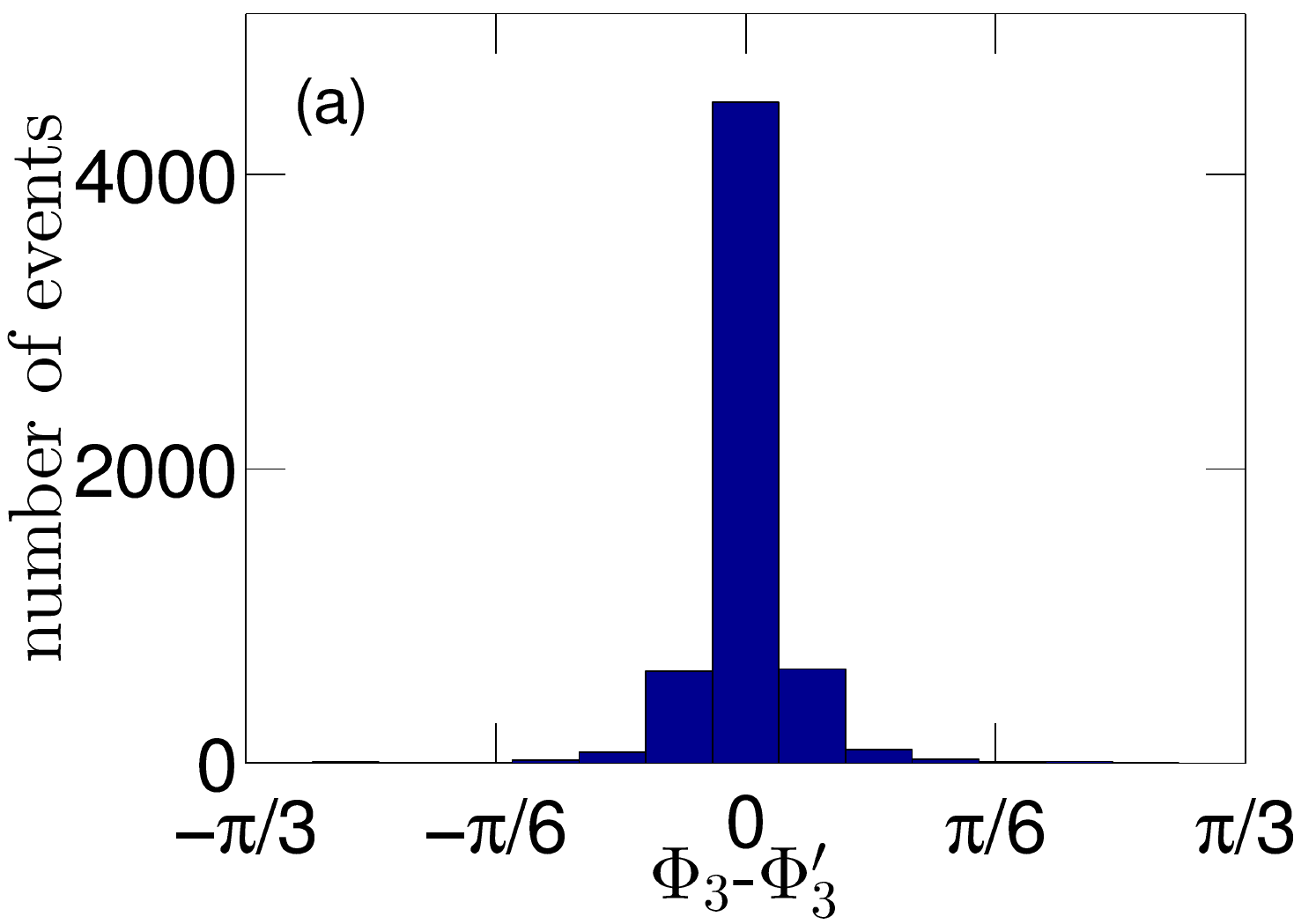}
 \includegraphics[width=0.32\linewidth]{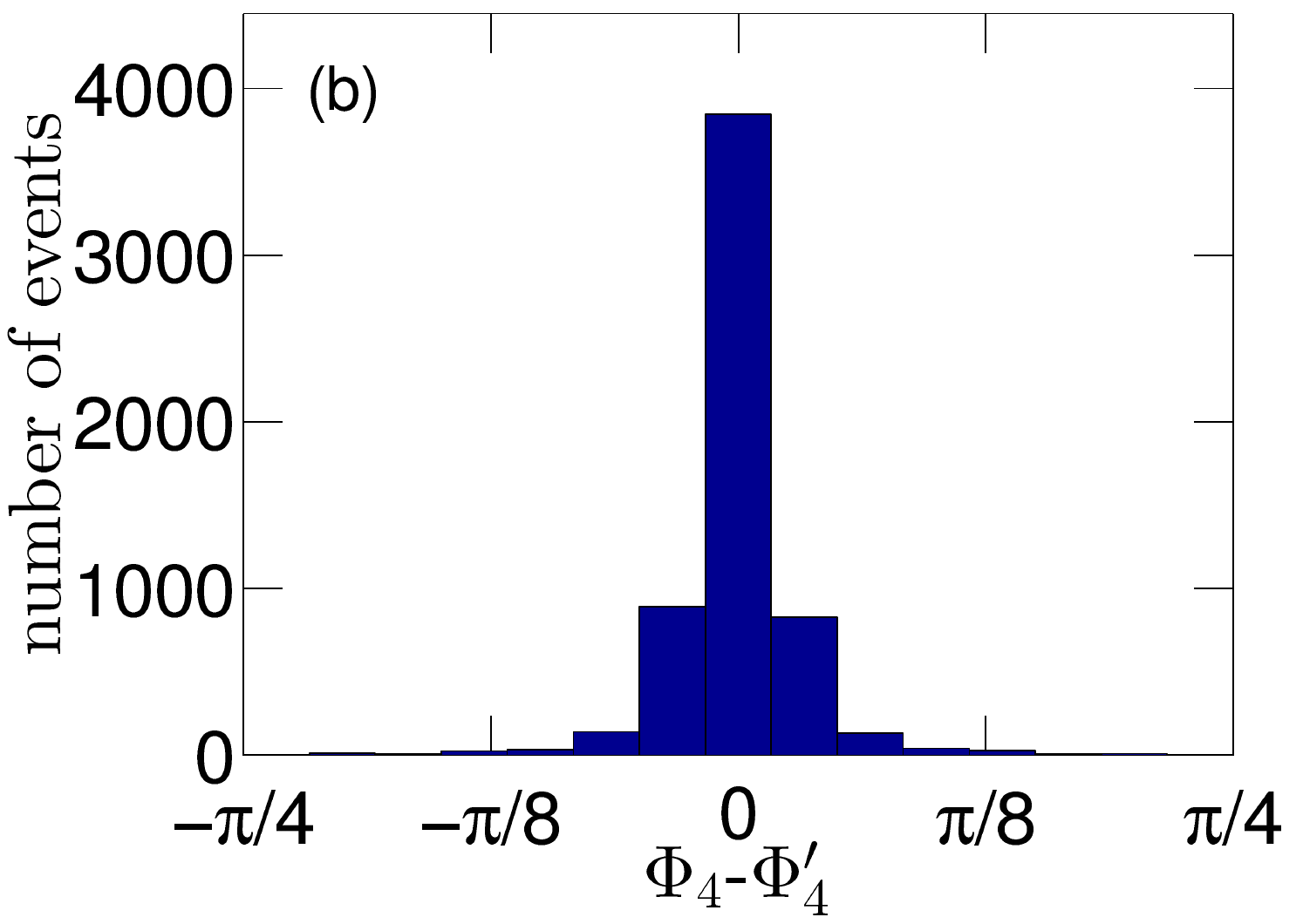}
 \includegraphics[width=0.32\linewidth]{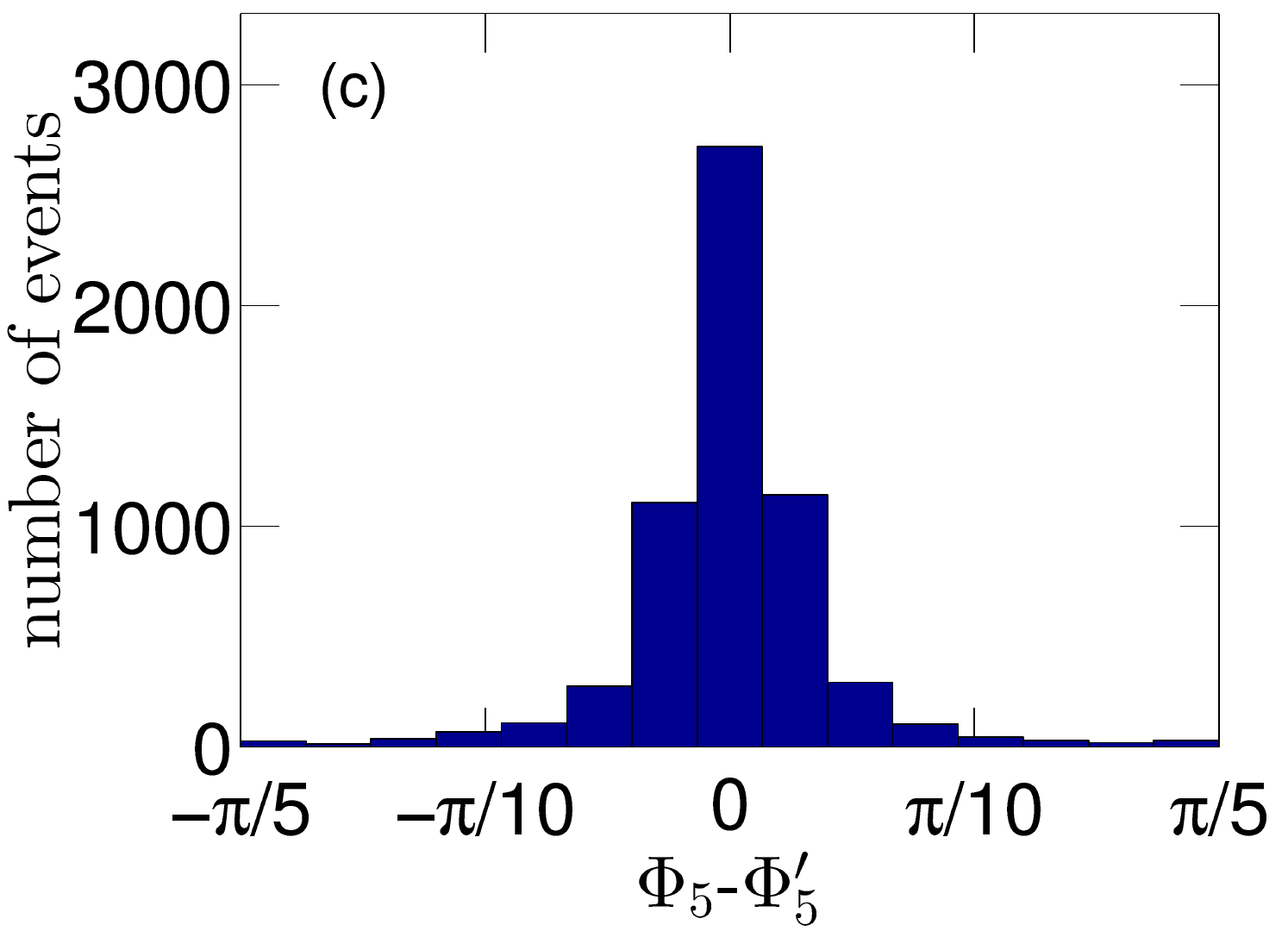} 
 \caption{Event-by-event correlation between the participant 
    plane angles associated with $r^2$- and $r^n$-weighted eccentricities
    for harmonic orders $n \eq 3$, 4, and 5 (panels (a)-(c)). From \cite{Qiu:2011iv}.
 \label{1104.0650:F17}
 }
\end{figure*}
%

In this section we present a brief comparison between the $r^2$-weighted eccentricity 
coefficients $\ecc_n$ (Eq.~(\ref{eq:1104.0650:15})) and the $r^n$-weighted $\ecc'_n$ 
(Eq.~(\ref{eq:1104.0650:15})), as well as their associated angles $\PsiPP{n}$ 
and $\PsiPPP{n}$. Fig.~\ref{1104.0650:F16}
shows a scatter plot of $\ecc'_n$ vs.\ $\ecc_n$ for $n \eq 3,4,5$. One observes
approximate proportionality ($\ecc'_3{\,\approx\,}1.22\,\ecc_3$, 
$\ecc'_4{\,\approx\,}1.48\,\ecc_4$, $\ecc'_5{\,\approx\,}1.80\,\ecc_5$)
over most of the eccentricity range, with slopes that increase with $n$.
So whereas Fig.~\ref{1104.0650:F5} shows a decrease of $\ecc_n$ with increasing $n$
at large impact parameters, the same is not necessarily true for the $\ecc'_n$
\cite{Qin:2010pf}. On the other hand, the linear relations between 
$\ecc'_n$ vs. $\ecc_n$ imply that the relations between $v_n$ and 
$\ecc'_n$ will look qualitatively the same as those between $v_n$ and
$\ecc_n$ in Fig.~\ref{1104.0650:F9}, with appropriately rescaled horizontal axes.

At the same time, the participant plane angles associated with 
$r^2$-weighted and $r^n$-weighted eccentricities are tightly correlated, 
as shown in Fig.~\ref{1104.0650:F17}. 
For given $n$, the angles $\PsiPP{n}$ and $\PsiPPP{n}$
fluctuate around each other, with a relative variance that increases with 
$n$, on account of the decreasing values of $\ecc_n$. From a practical
point of view, both definitions are equivalent,
and choosing between them is a matter of personal preference.

\section{Robust viscous hydrodynamics}
\label{sec:misc:pimunu_regulation}

Viscous hydrodynamics is a theory perturbative around the ideal-fluid theory, where the energy-momentum tensor is expanded as:
\begin{equation*}
  T^{\mu\nu} = T_{0}^{\mu\nu} + \Pi^{\mu\nu} = e u^\mu u^\nu - (p+\Pi)\Delta^{\mu\nu} + \pit + W^{\mu} u^\nu + W^\nu u^\mu.
\end{equation*}
Here $e$ and $p$ are the local energy density and pressure, $u^\mu$ is the flow 4-velocity, $T_{0}^{\mu\nu} = e u^\mu u^\nu - p\Delta^{\mu\nu}$ is the energy-momentum tensor assuming local equilibrium, and $\Delta^{\mu\nu} = g^{\mu\nu} - u^\mu u^\nu$ is the spatial projection operator in the local rest frame, $W^\mu = \frac{e+p}{n} V^\mu + q^\mu$ is the energy flow vector, with $n$ the baryon density, $V^\mu$ the baryon flow vector, and $q^\mu$ the heat flow vector. Choosing the Landau frame and working in the case $W^\mu = q^\mu = 0$ and $\Pi = 0$, the only viscous effects come from the shear pressure tensor $\pit$.

There are several criteria that $\pit$ needs to satisfy.
\begin{enumerate}
	\item Since viscous hydrodynamics is a perturbative theory, it is only reasonable as long as the $\pit$ term is smaller than the $T_{0}^{\mu\nu}$ term. A component-wise comparison between them is not appropriate since this would be frame-dependent; $\pit$ and $T^{\mu\nu}$ can only be compared after they are contracted into scalars. We choose:
	\begin{equation*}
	\mathrm{Tr}(\pi^2) = \pit\pi_{\mu\nu} \mbox{ and } T_{0}^{\mu\nu} {T_0}_{\mu\nu} = e^2+3p^2.
	\end{equation*}
The positivity of the former can be checked easily in the local rest frame. Therefore from the perturbative nature of the theory we require:
\begin{equation} \label{eq:A1}
	\pit\pi_{\mu\nu} \ll e^2+3p^2 
\end{equation}
	\item $\pit$ should be traceless:
	\begin{equation*} \pi^{\mu}_{\ \mu} = 0 \end{equation*}
	\item $\pit$ should be perpendicular to $u^\mu$:
	\begin{equation*} \pit u_\nu=0 \end{equation*}
\end{enumerate}

In actual calculations, there are limits to the numerical accuracy so we choose a small number $\xi_0\ll 1$ to be the ``numerical zero" and conditions 2 and 3 become:
\begin{equation} \label{eq:A2}
	\pi^{\mu}_{\ \mu} \leq \xi_0 \sqrt{\pit\pi_{\mu\nu}} \mbox{ and } \pit u_\nu \leq \xi_0 \sqrt{\pit\pi_{\mu\nu}}, \forall \nu
\end{equation}
The vector $\pit u_\nu$ should be component-wise zero (in any frame), therefore all its components should be compared to the ``numerical zero", multiplied by $\sqrt{\pit\pi_{\mu\nu}}$. Here we use the scalar $\pit\pi_{\mu\nu}$ as a measure for the magnitude of the $\pit$ tensor that sets the scale (via the factor $\xi_0$) for how close the numerical result is to zero.


In practice, to ensure that \equ{A1} is satisfied, we choose a number $\rho_\mathrm{max} \lesssim 1$ and require that\footnote{$\rho_\mathrm{max}\ll 1$ corresponds to the required ``$\ll$" condition in \equ{A2}; $\rho_\mathrm{max}=\infty$ corresponds to no constraint at all.}:
\begin{equation} \label{eq:A3}
	\sqrt{\pit\pi_{\mu\nu}} \leq \rho_\mathrm{max}\sqrt{e^2+3p^2}.
\end{equation}
This, together with \equ{A3}, then implies
\begin{equation} \label{eq:A3.1}
	\begin{aligned}
	\pi^{\mu}_{\ \mu} &\leq \xi_0\, \rho_\mathrm{max} \sqrt{\pit\pi_{\mu\nu}}, \\
	\pit u_\nu &\leq \xi_0\, \rho_\mathrm{max} \sqrt{\pit\pi_{\mu\nu}}. \\
	\end{aligned}
\end{equation}

These requirements may be violated during the hydrodynamic evolution for various reasons. For example, the $\pit$ could be initialized to be unphysically large (we use the Navier-Stokes condition to initialize $\pit$), so the relations would be violated at early time, or the initial profile could be not smooth enough, so the numerical derivative gives artificially large (and possibly wrong) numerical values that violate these conditions near discontinuities of the profile, etc. In principle, once the system is sufficiently close to thermal equilibrium, a properly generated $\pit$, evolved on a super-fine lattice, should avoid such problems; in practice however, a super-fine lattice is computationally infeasible and (what is worse) the Isreal-Stewart equations are not causal on a super-fine lattice \cite{Romatschke:2009im}. In practice, fluctuating initial conditions and early hydrodynamic starting times lead to patches in the numerical grid where one or several of the conditions \ref{eq:A3} are violated. For our simulations, we found, however, that the region where these conditions are violated are mostly at the early stage and/or in the dilute regions outside the freeze-out surface. The violations in these regions do not have much influence on the dynamical behavior of the QGP in the physical region inside the freeze-out surface; however if they are left untreated, these violations lead to accumulating numerical errors that cause the evolution code to break down later. For these reasons, in the following we develop a systematic treatment that suppresses large viscous terms. This stabilizes the code at a low price to pay: we effectively change the evolution equations in the fireball region where we have very large viscous corrections to the ideal fluid $T_0^{\mu\nu}$.

We enforce a continuous systematic regulation on $\pit$ in each time step on the whole lattice by replacing $\pit$ by $\hat{\pi}^{\mu\nu}$:
\begin{equation}
 \pit \rightarrow \hat{\pi}^{\mu\nu} \equiv \pit \frac{\tanh(\rho)}{\rho}
\end{equation}
where $\rho$ is the largest quantity at each lattice point among the following:
\begin{equation*}
\frac{\sqrt{\pit\pi_{\mu\nu}}}{\rho_\mathrm{max}\sqrt{e^2+3p^2}}, \frac{\pi^{\mu}_{\ \mu}}{\xi_0\,\rho_\mathrm{max}\,\sqrt{\pit\pi_{\mu\nu}}}, \mbox{ and } \frac{\pit u_\nu}{\xi_0 \rho_\mathrm{max}\,\sqrt{\pit\pi_{\mu\nu}}}, \forall \nu
\end{equation*}
It is easy to check that $\hat{\pi}^{\mu\nu}$ satisfies \equ{A3}, and that it is close to $\pit$ when no modifications are needed; that is, when the left hand sides of the inequalities in \equ{A3} are small compared to their right hand sides, the regulated quantity is close to its original value. Only those grid points where $\pit$ violates or is close to violating the inequalities (\ref{eq:A3}) will be strongly modified, in which case all components of $\pit$ are suppressed by the same factor. 

Because smoother flow velocity profiles give smaller $\pit$, the systematic suppression of $\pit$ can be understood as locally replacing sharp jumps in the flow profile by flatter pieces; the regulation process is therefore an implicit and automatic way of smoothing profiles. This treatment allows us to perform hydrodynamic calculations using very bumpy initial conditions, including those using disk-like nucleons that have density discontinuities. Without this regularization VISH2+1 breaks down for such initial conditions.

In our calculations, we take $\xi_0=0.001$, which is reasonable because the relative discretization error from solving transport equations is of the order of a few powers of the lattice spacing and our lattice spacing is chosen to be $0.1$ fm. During our simulations we found that $\rho_\mathrm{max}$ is best chosen to be a value between $1-10$ in order to enable the simulation to finish, and to keep the momentum distribution generated from the QGP almost unmodified. By choosing $\rho_\mathrm{max}$ of order unity or larger, we modify the effective evolution equations only in regions where the conditions of validity of the Isreal-Stewart equations are badly violated.

\section{Time evolution of the event plane orientations}
\label{sec:misc:time_evolution}

\begin{figure}[h]
  \center{
        \includegraphics[height=0.32\textwidth]{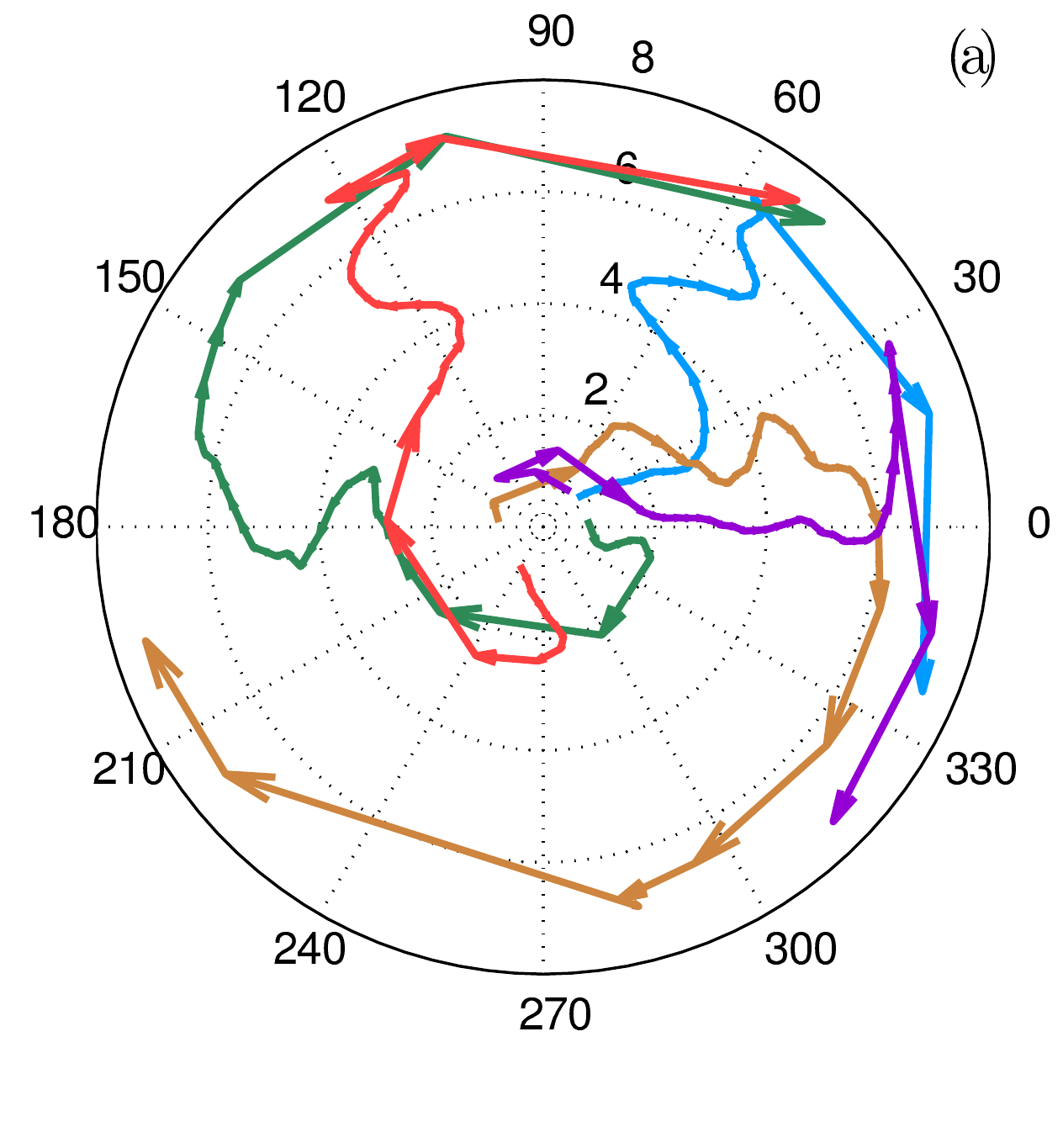}
        \includegraphics[height=0.32\textwidth]{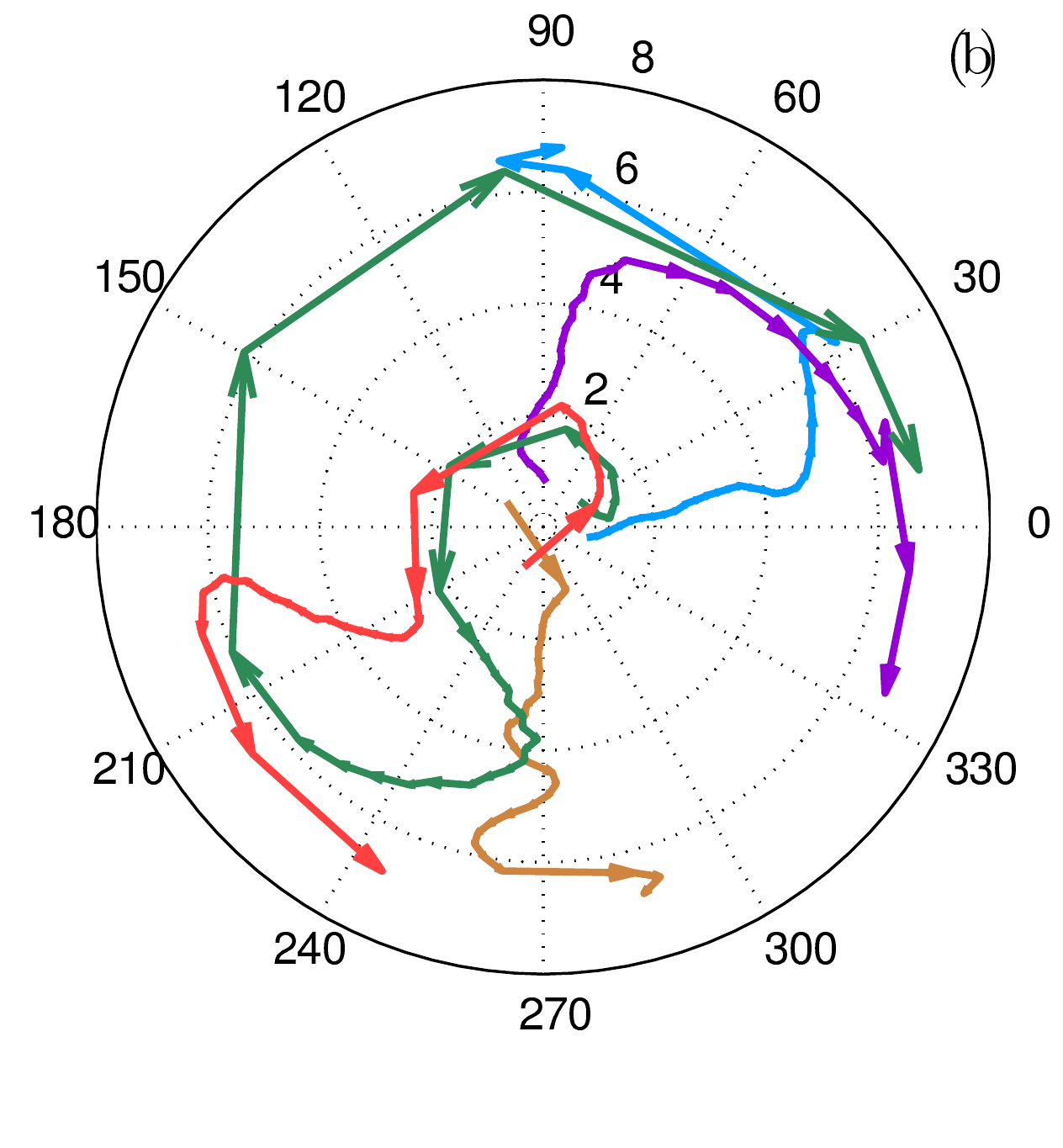}
        \includegraphics[height=0.32\textwidth]{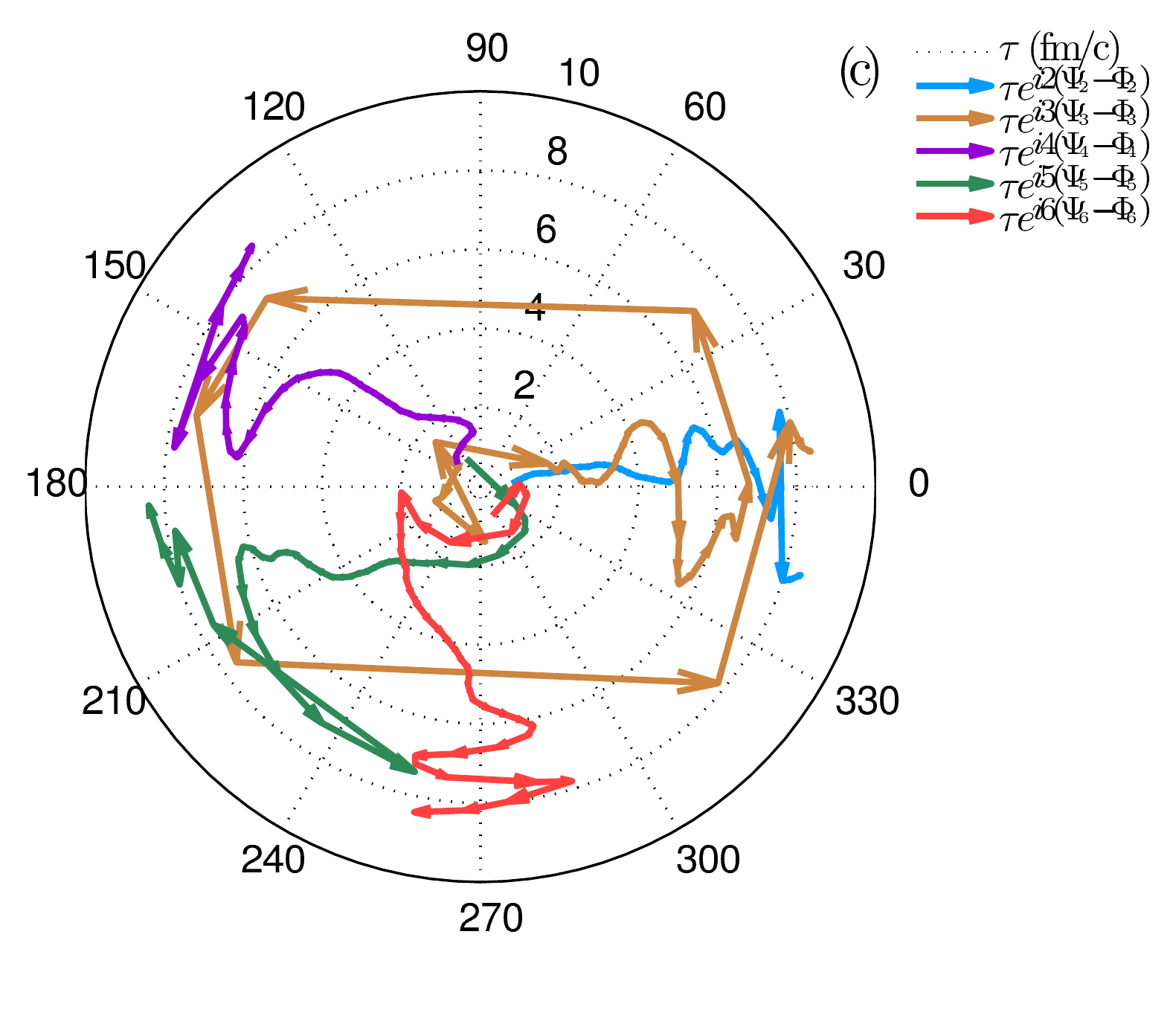}
        }
  \caption{Time evolution of the anisotropic flow angles. See text for more explanation.}
  \label{fig:ebe2:40}
\end{figure}

The orientations of the event planes are not fixed during the evolution of the fireball: one of the reasons for their changing is the mode-coupling effect, by which the orientation of an event plane of a particular harmonic order receives contributions from geometric deformations of the fireball of a different harmonic order. To demonstrate such effects visually, we take 3 randomly selected events, and divide their freeze-out surface into strips with time step $0.2$ fm/c. Each strip of the freeze-out surface is then used to calculate the anisotropy flows formed by particles emitted only from this part of the freeze-out surface using the Cooper-Frye formula. The resulting anisotropic flow angles from different pieces are then studied; their evolution is interpreted as the time evolution of the anisotropic flow angles during the lifetime of the fireball. Typical results are shown in Fig.~\ref{fig:ebe2:40}.

Here the anisotropy angles are the $n$th-order flow angles $\PsiEP{n}$ (see \equ{1104.0650:20}) calculated from the spectra emitted by the strips of the freeze-out surface, measured relative to the participant plane angles $\PsiPP{n}$. To make the time evolution manifest, we interpret $(\tau, n(\PsiEP{n}-\PsiPP{n}))$ as polar coordinates $(r, \theta)$, i.e.\ we plot the complex quantities $\tau e^{i n (\PsiEP{n}-\PsiPP{n})}$ as trajectories (as a function of $\tau$) in the complex plane. The arrows point to the positive $\tau$ direction along the trajectories.

An anisotropy angle is said to be ``in-plane" if $\Psi_n-\Phi_n$ is close to $0^\circ$ and ``out-of-plane" if $\Psi_n-\phi_n$ is close to $180^\circ$. The most dramatic rotation happens if the angle changes from one sector to the opposite sector that differs by $180^\circ$.

It is clear that the 3rd- through 6th-order anisotropy angles in the 1st event (Fig.~\ref{fig:ebe2:40} left), the 5th- and 6th- order anisotropy angles in the 2nd event (Fig.~\ref{fig:ebe2:40} middle), and the 3rd-order anisotropy angle in the 3rd event (Fig.~\ref{fig:ebe2:40} right) all have undergone dramatic rotations during the evolution of the fireball. The 2nd- and 3rd-order anisotropy angles are relatively stable during most of the early evolution, reflecting the relative smallness of the mode-coupling effects; the angles only rotate near the end of the evolution of the fireball where the remnant of the fireball becomes dilute and irregular and the orientation of all harmonics becomes somewhat random.

\section{Best estimator for flows using eccentricities}
\label{sec:misc:best_estimator}

The authors of \cite{Gardim:2012dc} use regression methods to study the flow response from the eccentricity predictors, where they use a quantity similar to the coefficient of determination to find the ``best estimator'' and to study the second-order mode-coupling effects. Their calculations are based on simulations with NeXSPheRIO which uses ideal hydrodynamics, with flow calculated from Monte-Carlo sampled particles. Here we extend their work in three ways: 1) by including viscosity, 2) by using the continuous particle emission formalism, 3) and by including higher-order mixing terms. Including higher-order terms will always improve the fit, but the degree of improvement depends on whether the expansion is made with eccentricities defined using moments or cumulants \cite{Teaney:2012ke}.

\begin{figure}[h!]
  \center{
     \includegraphics[width=0.9\textwidth]{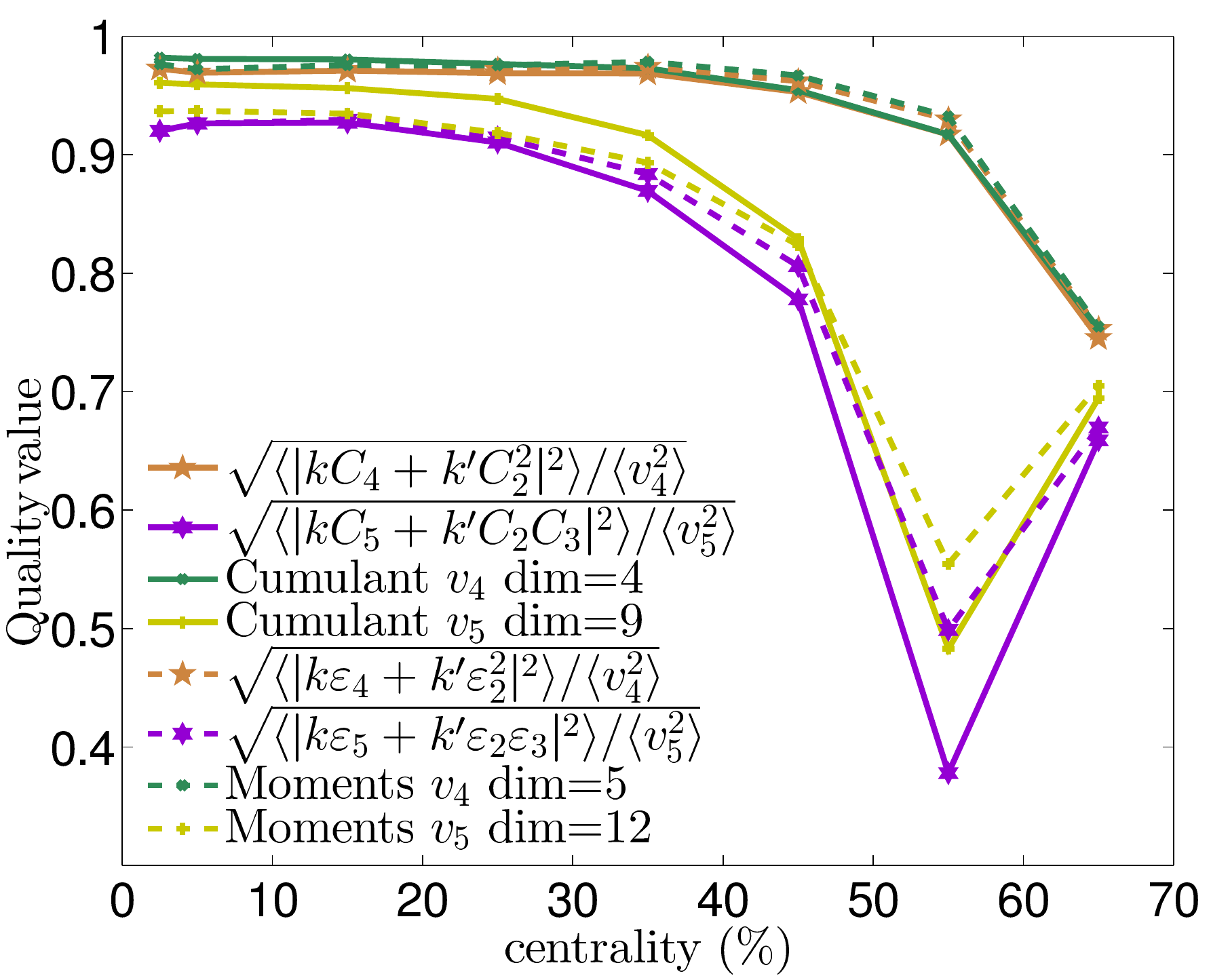}
  }
  \caption{(Color online) The ``Quality values'' (equation \ref{eq:O2}, \ref{eq:O4}; see also \cite{Gardim:2012dc}) giving the goodness of description of the flows in terms of a combination of eccentricities, as functions of centrality. The first 4 curves (solid) use the cumulant expansion and the last 4 curves (dashed) use the moment expansion. The curve labeled by ``cumulant $v_4$ dim=4'' means that the curve is the ``Quality value'' for $v_4$ using cumulant-defined eccentricities \cite{Teaney:2012ke} which includes 4 terms in the expansion, etc. See the text for details.}
  \label{fig:ebe2:42}
\end{figure}

We used the same pure event-by-event hydrodynamic simulation data tuned for Pb-Pb collision at 2.76 ATeV as reported in \cite{Qiu:2012uy} (and \chap{correlation}), and plotted the ``Quality value'' \cite{Gardim:2012dc} for various settings as functions of centrality in Fig.~\ref{fig:ebe2:42}. The ``Quality value'' is so defined that it is between 0 and 1; the closer it is to 1, the better the approximation. When approximating $v_n$ using only linear terms (equation (2) in \cite{Gardim:2012dc})
\begin{equation} \label{eq:O1}
  v_n e^{i n \PsiEP{n}} = k \ve_n e^{i n \PsiPP{n}} + \mathcal{E}
\end{equation}
where $\mathcal{E}$ is the ``error'', the ``Quality value'' is defined to be (equation (3) in \cite{Gardim:2012dc}):
\begin{equation} \label{eq:O2}
  Quality = k \frac{\sqrt{\la \ve_n^2 \ra}}{\sqrt{\la v_n^2 \ra}}
\end{equation}
using the best-fit parameter $k$ from \equ{O1}.
When approximating $v_n$ including two terms (equation (5) in \cite{Gardim:2012dc})
\begin{equation} \label{eq:O3}
  v_n e^{i n \PsiEP{n}} = k \ve_n e^{i n \PsiPP{n}} + k' \ve_{n'} e^{i n \PsiPP{n'}} + \mathcal{E}
\end{equation}
the ``Quality value'' is defined to be (equation (6) in \cite{Gardim:2012dc}):
\begin{equation} \label{eq:O4}
  Quality = \frac{\la | k \ve_n e^{i n \PsiPP{n}} + k' \ve_{n'} e^{i n \PsiPP{n'}} |^2 \ra}{ \la v_n^2 \ra}
\end{equation}
using the best-fit parameters $k$ and $k'$ from \equ{O3}. More terms can be included in the expansion and the ``Quality value'' is always defined to be the ratio between the fit and the actual value.

The differences from \cite{Gardim:2012dc} include that we use continuous Cooper-Frye particle emission to calculate flows, and that we include viscosity, although the viscosity was found to play little role, so only the results using MC-KLN initial condition with $\eta/s=0.2$ are shown here.

The elliptic and triangular flows are known to be well approximated as linear functions of the eccentricities of the same order, so we skip them here and only report on the results for the quadrangular and pentangular flows which receive large contributions from mode-coupling effects. More specifically, various studies including \cite{Gardim:2012dc} show that the quartic flow $v_4$ mainly receives a contribution from $\ve_2^2$ and the pentangular flow $v_5$ receives a contribution from $\ve_2 \ve_3$, and it is shown in \cite{Gardim:2012dc} that by including these two quadratic terms in the expansions the approximations are improved, but the approximations are still not good enough, especially in very peripheral collisions.

Here the four curves labeled as $\sqrt{.../\la v_{4,5}^2 \ra}$ are those ``Quality values'' calculated using expansions that contain only two terms: for $v_4$, they are $\ve_4$ and $\ve_2^2$, and for $v_5$, they are $\ve_5$ and $\ve_2 \ve_3$\footnote{In fact, it is the complex product of the complex eccentricity vectors that are used (see \cite{Gardim:2012dc}).}; for the cumulant expansion, $\ve_n$ is replaced by $C_n$ \cite{Teaney:2012ke}. It is seen that when including two terms in the expansion, both the moment- and the cumulant-defined eccentricities give very similar results in terms of the goodness of the fit.

Next we allow more terms in the expansion to test the convergence. The terms we included in the linear combinations are summarized in the following. Here we use the complex eccentricity notation $\Ecc_n = \ecc_n e^{i n \PsiPP{n}}$ (see \chap{1104.0650}), with $\Ecc_n$ replaced by $C_n$ in the case for cumulants, and ``*'' for complex conjugate.

\begin{itemize}
 \item $v_4$ using 5 moment-defined eccentricities: $\Ecc_4$, $\Ecc_2^2$, $\Ecc_1\Ecc_3$, $\Ecc_3^2\Ecc_2^*$, $\Ecc_6\Ecc_2^*$.
 \item $v_5$ using 12 moment-defined eccentricities: $\Ecc_5$, $\Ecc_2\Ecc_3$, $\Ecc_2^2\Ecc_1$, $\Ecc_1\Ecc_4$, $\Ecc_3^2\Ecc_1^*$, $\Ecc_2^3\Ecc_1^*$, $\Ecc_4^2\Ecc_3^*$, $\Ecc_2^4\Ecc_3^*$, $\Ecc_3^3\Ecc_4^*$, $\Ecc_7\Ecc_2^*$, $\Ecc_8\Ecc_3^*$, $\Ecc_9\Ecc_4^*$.
 \item $v_4$ using 4 cumulant-defined eccentricities: $\Ecc_4$, $\Ecc_2^2$, $\Ecc_1\Ecc_3$, $\Ecc_3^2\Ecc_2^*$.
 \item $v_5$ using 9 cumulant-defined eccentricities: $\Ecc_5$, $\Ecc_2\Ecc_3$, $\Ecc_2^2\Ecc_1$, $\Ecc_1\Ecc_4$, $\Ecc_3^2\Ecc_1^*$, $\Ecc_2^3\Ecc_1^*$, $\Ecc_4^2\Ecc_3^*$, $\Ecc_2^4\Ecc_3^*$, $\Ecc_3^3\Ecc_4^*$.
\end{itemize}

The curves labeled by ``moment/cumulant $v_{4,5}$ dim=...'' show the resulting ``Quality values'' by allowing more terms in the expansion listed above. It is seen that the cumulant-defined eccentricities give faster convergence (the ``k values" for both expansion methods converge to 1 if an infinite number of terms are included) than moment-defined eccentricities, but neither is good enough to be used as a quantitative replacement of the hydrodynamics for $v_5$ (or higher flows), even when higher order couplings are taken into account. This suggests that, in terms of a quantitative study for higher-order flows, the role of hydrodynamics cannot be replaced by simply looking at the response of the flow to various eccentricities.

\chapter{Summary}
\label{chap:summary}

In this thesis, I summarized my Ph.D. work on event-by-event hydrodynamic simulations for relativistic heavy-ion collisions and related fields.

In \chap{1104.0650}, we compared the previously dominating single-shot hydrodynamics with the current and future mainstream event-by-event hydrodynamic simulations. The event-by-event simulations are more realistic, but they are very time consuming; the single-shot simulation is economical, but then the question arises as to whether it can be used as a sufficiently precise replacement for the event-by-event simulations. The answer to this question depends on the observable studied. We showed that for the event average of the multiplicity and elliptic and triangular flow, the time consuming event-by-event hydrodynamic simulations can, to a good approximation, be replaced by the single-shot ones, when using properly constructed, participant-plane rotated and averaged initial conditions. For azimuthally averaged $p_T$ spectra the single-shot simulation can still give results not very different from those given by event-by-event simulations, especially if the shear viscosity is not very small. For higher-order flows, such as $v_{4,5}$, the single-shot simulations are shown to be incapable of reproducing those from event-by-event simulations. The main reason for this failure are nonlinear mode-coupling effects that are very sensitive to initial-state density fluctuations. In this chapter, I also showed related quantities, such as the distributions of eccentricities and anisotropic flows from event-by-event calculations. These can be used as calibrations for future studies.

In \chap{e2e3}, we showed that the elliptic and triangular flows can be studied using the much more economical single-shot hydrodynamic simulations to a good ($\sim 10\%$) approximation. We further showed that the eccentricity-normalized elliptic and triangular flows are ``definition independent": $\vt2/\et2$ is very close to $\vf2/\ef2$, $\la v_2 \ra/\la \ve_2 \ra$, and $\bar{v}_2/\bar{\ve}_2$; likewise, $\vt3/\et3$ is very close to $\la v_3 \ra/\la \ve_3 \ra$ and $\bar{v}_3/\bar{\ve}_3$. The eccentricity normalized flows from single-shot hydrodynamic simulations can thus be used to study the experimental flow data when normalized by similarly constructed eccentricities. This is particularly important when comparing experimental results (which are always affected by event-by-event fluctuations) with results from single-shot hydrodynamics (which does not account for fluctuations). The eccentricity normalized flows can be used individually to extract the specific shear viscosity for a given initial condition model or, when combined, can be used to reduce the ambiguity introduced by different initial condition models with varying initial fluctuation spectra. We followed this strategy and showed that the elliptic and triangular flow data measured by the ALICE collaboration at the LHC prefer a small specific shear viscosity close to $\eta/s=0.08$, when considering the MC-Glauber and MC-KLN  models. In order to allow for a much larger $\eta/s$ value, the initial condition model must feature triangularity values $\sim 50\%$ larger than the ones provided by the MC-Glauber and MC-KLN models. This was found to be the case in the IP-Glasma model \cite{Schenke:2012wb}.

\chap{correlation} focused on correlations between event-plane angles. We showed that the event-plane angle correlation measurements by the ATLAS  collaboration can be explained using results from the same set of hydrodynamic simulations tuned for spectra and flow observables. The same correlation patterns cannot be explained directly from the initial conditions, implying that they are generated dynamically during the evolution of the medium. Some of the strong correlation patterns, like the one between harmonic orders $2$ and $4$, and the one among orders $2$, $3$, and $5$, are direct evidence for strong mode-coupling effects. The correlations serve as another set of observables, complementing the spectra and anisotropic flow coefficients $v_n$, and the success of their hydrodynamic explanation adds to the evidence that the medium created in heavy-ion collisions is fact hydrodynamical. We also note that the strong mode-coupling effects cannot be reproduced from a linear response treatment of hydrodynamics.

In \chap{resonance}, we showed that including only $\sim 20$ out of $319$ carefully chosen resonances can already yield spectra and flow results within $1\%$ relative error. Such a treatment can be used to shorten the currently lengthy ($\sim$ 3 hours) resonance decay calculations by a factor of ten --- since these calculations are the current bottleneck for event-by-event hydrodynamic simulations --- and consequently lower the cost of performing event-by-event simulations by the same factor. We provided a complete table listing the contribution of resonances to stable particles for all major stable particles.

\chap{flow_angle} focused on various ways of calculating (theoretically) and measuring (experimentally) anisotropic flows, and their comparisons. We pointed out that the event-plane angle $\PsiEP{n}$ fluctuates from event to event, and that the differential event-plane angle $\PsiEP{n}(p_T)-\PsiEP{n}$ also fluctuates from event to event. We showed that traces of such angular fluctuations can be measured experimentally, and explained how such measurements allow the study of the fluctuation of the orientation of flows in addition to the fluctuation of the magnitude of flows. We showed that the effects are larger for heavy particles, in near-central collisions, and for small $p_T\lesssim m$. In particular, we proposed a precise measurement and comparison of $\la v_n(\pT)\rangle$, $v_n\{\EP\}(\pT)$, $v_n[2](\pT)$, and $v_n\{2\}(\pT)$ for identified pions, kaons and protons with transverse momenta $\pT{\,<\,}2$\,GeV to confirm the hydrodynamically predicted effects from flow angle fluctuations. We also pointed out the flow angle fluctuations are responsible for half of the factorization breaking effects, the other half being caused by fluctuations in the magnitudes $v_n(\pT)$ of the anisotropic flows.

In \chap{iSS}, we gave a thorough description of the general sampling methodology, and showed that it can be applied in an efficient way to sample particles whose emission is determined by the Cooper-Frye formula, using the simplification of longitudinal boost invariance. We explained several sampling algorithms and analyzed their strengths and weaknesses, focusing heavily on sampling efficiencies. This chapter serves as a study in sampling methodology as well as a technical document that explains the actual sampling implementation used in the iSpectraSampler program.

We also included discussions of a number of other miscellaneous topics in \chap{misc}. We compared the $r^2$- and $r^n$-weighted eccentricities. We described how we regulate unphysically large $\pit$ values that can arise especially during the early stage or in the dilute tail of the density distribution of the fireball hydrodynamic evolution. In addition, we showed that the event plane angles determined using particles emitted at different proper times are not necessarily the same, and that their time evolution can be dramatic. Finally, we showed quantitatively that including nonlinear combinations of eccentricities can improve the prediction of $v_4$ and $v_5$ as the corresponding responses.

\appendix


\chapter{Choice of parameters used in the simulations}
\label{sec:appa:parameters}

This section contains tables for the choice of parameters used in our simulations. Unless for academic study (\chap{1104.0650}), all our simulations are tuned to explain experimental data, which requires a choice of parameters according to the following tables. These tables should be used for pure hydrodynamic simulations only, not for hybrid simulations (see \chap{introduction}).

\begin{table}[H] 
\caption{Choice of parameters for Au+Au at $200$ AGeV}
\label{tab:para:1} 
\begin{center}
\begin{tabular}{cccccc}
  \hline \hline
  Model & initial time & decoupling temp. & $\eta/s$ & model parameter & norm. \\
  \hline
  MC-Glb & 0.6 (fm/c) & 120 MeV & 0.08 & $\alpha = 0.14$ & 28.66 \\
  MC-KLN & 0.6 (fm/c) & 120 MeV & 0.2 & $\lambda = 0.218$ & 7.42 \\
  \hline
\end{tabular}
\end{center}
\end{table}

Here the ``initial time" is the initial proper time $\tau_0$ when hydrodynamics starts. The ``decoupling temp." is the decoupling temperature under which hadrons cease to interact and fly into the detector via free streaming. The ``model parameter" for the MC-Glauber model is the soft-hard ratio $\alpha$ ($x$ in \equ{Glauber}, $\delta$ in eq.~(20) in \cite{Hirano:2009ah}), and for the MC-KLN model it is the $\lambda$ parameter used when determining the saturation scale (Eq.~(9) in \cite{Drescher:2006ca}). Finally the ``norm." is the normalization factor that, when multiplied with the profile directly generated from the initial condition models, gives the initial entropy density profile that can be used in hydrodynamic simulations.\footnote{For the MC-Glauber model, it is the $\kappa$ parameter in \equ{Glauber} or $C$ in eq.~(20) in \cite{Hirano:2009ah}. For the MC-KLN model, we take $dN/dy = (\mathrm{norm.}) dN_g/dy$.}

\begin{table}[H] 
\caption{Choice of parameters for Pb+Pb at $2760$ AGeV}
\label{tab:para:2} 
\begin{center}
\begin{tabular}{cccccc}
  \hline \hline
  Model & initial time & decoupling temp. & $\eta/s$ & model parameter & norm. \\
  \hline
  MC-Glb & 0.6 (fm/c) & 120 MeV & 0.08 & $\alpha = 0.118$ & 56.76 \\
  MC-KLN & 0.6 (fm/c) & 120 MeV & 0.2 & $\lambda = 0.138$ & 9.92 \\
  \hline
\end{tabular}
\end{center}
\end{table}

\chapter{Table of parameters for cutting centralities}
\label{sec:appa:table}

%
%
%
%

When using the optical Glauber model, the centrality class can be determined and cut in terms of ranges in impact parameter $b$; \qtable{1} lists the choice of impact parameters used in cutting centralities this way. A given centrality class includes events with impact parameters ranging from $b_\mathrm{min}$ to $b_\mathrm{max}$; the event-averaged impact parameter is $\bar{b}$ and the averaged $N_\mathrm{part}$ is $\bar{N}_\mathrm{part}$. This table was used in the study presented in \chap{1104.0650}.

In later studies, we use centrality classes determined by $N_\mathrm{part}$; such centrality class cuts are contained in tables \ref{tab:cen:1}-\ref{tab:cen:7}. For a given centrality bin, each table includes the $N_\mathrm{part, min}$ and $N_\mathrm{part, max}$ values which define the bin, the average $N_\mathrm{part,avg}$, the range in impact parameter $b_\mathrm{min}$ to $b_\mathrm{max}$, and the average $b_\mathrm{avg}$. 

\begin{table}[H] 
\caption{Centrality table for Au+Au at $200\,A$\,GeV \cite{Hirano:2009ah}.
         \label{T1}}
\label{tab:1} 
\begin{center}
\begin{tabular}{ccccccccc}
  \hline \hline
  centrality &\vline & $b_\mathrm{min}$(fm) &\vline & $b_\mathrm{max}$(fm) 
  &\vline & $\bar{b}$ (fm) &\vline & $\bar{N}_{\mathrm{part}}$ \\
  \hline
  0-5\%   &\vline & 0.0 &\vline  & 3.3  &\vline & 2.2   &\vline   & 352.2 \\
  5-10\%  &\vline & 3.3 &\vline  & 4.7  &\vline & 4.04  &\vline   & 294.7 \\
  10-15\% &\vline & 4.7 &\vline  & 5.8  &\vline & 5.27  &\vline   & 245.6 \\
  15-20\% &\vline & 5.8 &\vline  & 6.7  &\vline & 6.26  &\vline   & 204.2 \\
  20-30\% &\vline & 6.7 &\vline  & 8.2  &\vline & 7.48  &\vline   & 154.5 \\
  30-40\% &\vline & 8.2 &\vline  & 9.4  &\vline & 8.81  &\vline   & 103.8 \\
  40-50\% &\vline & 9.4 &\vline  & 10.6 &\vline & 10.01 &\vline   & 64.9 \\
  50-60\% &\vline & 10.6 &\vline & 11.6 &\vline & 11.11 &\vline   & 36.6 \\
  60-70\% &\vline & 11.6 &\vline & 12.5 &\vline & 12.06 &\vline   & 18.8\\
  70-80\% &\vline & 12.5 &\vline & 13.4 &\vline & 12.96 &\vline   & 7.5\\
  80-90\% &\vline & 13.4 &\vline & 14.3 &\vline & 13.85 &\vline   & 4.4\\
  \hline
\end{tabular}
\end{center}
\end{table}

\begin{table}[H]
\caption{Centrality table for Au+Au at $63$ AGeV (1,000,000 initial conditions)}
\label{tab:cen:1} 
\begin{center}
\begin{tabular}{c|c|c|c|c|c|c}
\hline \hline
centrality (\%) & $N_\mathrm{part, min}$ & $N_\mathrm{part, max}$ & $N_\mathrm{part,avg}$ & $b_\mathrm{min}$ (fm) & $b_\mathrm{max}$ (fm) & $b_\mathrm{avg}$ (fm) \\
\hline
 0- 5\% & 322.0 & 391.0  & 351.0 &   0.0 &   4.0 &  2.1 \\
 5-10\% & 273.0 & 322.0  & 296.7 &   2.2 &   5.3 &  3.8 \\
10-15\% & 232.0 & 273.0  & 252.0 &   3.6 &   6.2 &  5.0 \\
15-20\% & 197.0 & 232.0  & 214.2 &   4.7 &   7.1 &  5.9 \\
20-25\% & 167.0 & 197.0  & 181.5 &   5.5 &   8.0 &  6.7 \\
25-30\% & 140.0 & 167.0  & 153.1 &   6.2 &   8.6 &  7.4 \\
30-35\% & 117.0 & 140.0  & 128.2 &   6.7 &   9.3 &  8.0 \\
35-40\% &  96.0 & 117.0  & 106.3 &   7.3 &  10.0 &  8.6 \\
40-45\% &  79.0 & 96.0  &  87.3 &   8.0 &  10.6 &  9.2 \\
45-50\% &  63.0 & 79.0  &  70.8 &   8.4 &  11.3 &  9.7 \\
50-55\% &  50.0 & 63.0  &  56.4 &   8.7 &  11.8 & 10.2 \\
55-60\% &  39.0 & 50.0  &  44.1 &   9.2 &  12.4 & 10.7 \\
60-65\% &  29.0 & 39.0  &  33.7 &   9.6 &  13.0 & 11.1 \\
65-70\% &  21.0 & 29.0  &  25.0 &   9.8 &  13.5 & 11.6 \\
70-75\% &  15.0 & 21.0  &  18.0 &  10.3 &  14.2 & 12.0 \\
75-80\% &  10.0 & 15.0  &  12.5 &  10.4 &  14.7 & 12.4 \\
80-85\% &   7.0 & 10.0  &   8.4 &  10.8 &  15.4 & 12.9 \\
85-90\% &   4.0 &  7.0  &   5.3 &  10.9 &  16.2 & 13.3 \\
90-95\% &   2.0 &  4.0  &   3.1 &  11.0 &  17.5 & 13.7 \\
\hline \hline
\end{tabular}
\end{center}
\end{table}

\begin{table}[H]
\caption{Centrality table for Au+Au at $200$ AGeV (1,000,000 initial conditions)}
\label{tab:cen:2} 
\begin{center}
\begin{tabular}{c|c|c|c|c|c|c}
\hline \hline
centrality (\%) & $N_\mathrm{part, min}$ & $N_\mathrm{part, max}$ & $N_\mathrm{part,avg}$ & $b_\mathrm{min}$ (fm) & $b_\mathrm{max}$ (fm) & $b_\mathrm{avg}$ (fm) \\
\hline
 0- 5\% & 327.0 & 393.0  & 355.4 &   0.0 &   4.1 &  2.1 \\
 5-10\% & 278.0 & 327.0  & 302.0 &   2.5 &   5.3 &  3.9 \\
10-15\% & 237.0 & 278.0  & 257.1 &   3.8 &   6.4 &  5.0 \\
15-20\% & 202.0 & 237.0  & 218.8 &   4.7 &   7.2 &  5.9 \\
20-25\% & 171.0 & 202.0  & 185.7 &   5.5 &   8.0 &  6.7 \\
25-30\% & 144.0 & 171.0  & 156.8 &   6.3 &   8.7 &  7.4 \\
30-35\% & 120.0 & 144.0  & 131.3 &   6.8 &   9.4 &  8.1 \\
35-40\% &  99.0 & 120.0  & 109.1 &   7.3 &  10.3 &  8.7 \\
40-45\% &  81.0 & 99.0  &  89.7 &   7.9 &  10.7 &  9.2 \\
45-50\% &  65.0 & 81.0  &  72.7 &   8.3 &  11.2 &  9.8 \\
50-55\% &  51.0 & 65.0  &  58.0 &   8.9 &  11.9 & 10.3 \\
55-60\% &  40.0 & 51.0  &  45.5 &   9.2 &  12.4 & 10.8 \\
60-65\% &  30.0 & 40.0  &  34.8 &   9.4 &  13.2 & 11.2 \\
65-70\% &  22.0 & 30.0  &  25.9 &   9.9 &  14.2 & 11.7 \\
70-75\% &  16.0 & 22.0  &  18.7 &  10.1 &  14.1 & 12.1 \\
75-80\% &  11.0 & 16.0  &  13.0 &  10.5 &  14.8 & 12.5 \\
80-85\% &   7.0 & 11.0  &   8.6 &  10.5 &  15.4 & 12.9 \\
85-90\% &   4.0 &  7.0  &   5.4 &  11.1 &  16.2 & 13.4 \\
90-95\% &   2.0 &  4.0  &   3.2 &  11.0 &  17.8 & 13.8 \\
\hline \hline
\end{tabular}
\end{center}
\end{table}

\begin{table}[H]
\caption{Centrality table for Cu+Cu at $63$ AGeV (1,000,000 initial conditions)}
\label{tab:cen:3} 
\begin{center}
\begin{tabular}{c|c|c|c|c|c|c}
\hline \hline
centrality (\%) & $N_\mathrm{part, min}$ & $N_\mathrm{part, max}$ & $N_\mathrm{part,avg}$ & $b_\mathrm{min}$ (fm) & $b_\mathrm{max}$ (fm) & $b_\mathrm{avg}$ (fm) \\
\hline
 0- 5\% & 100.0 & 125.0  & 107.8 &   0.0 &   3.2 &  1.5 \\
 5-10\% &  86.0 & 100.0  &  92.5 &   0.8 &   4.2 &  2.6 \\
10-15\% &  74.0 & 86.0  &  79.5 &   1.7 &   4.8 &  3.4 \\
15-20\% &  63.0 & 74.0  &  68.2 &   2.9 &   5.6 &  4.1 \\
20-25\% &  54.0 & 63.0  &  58.3 &   3.4 &   6.1 &  4.6 \\
25-30\% &  46.0 & 54.0  &  49.6 &   3.8 &   6.8 &  5.1 \\
30-35\% &  38.0 & 46.0  &  41.9 &   4.1 &   7.2 &  5.6 \\
35-40\% &  32.0 & 38.0  &  35.2 &   4.5 &   7.9 &  6.0 \\
40-45\% &  27.0 & 32.0  &  29.3 &   5.0 &   8.1 &  6.4 \\
45-50\% &  22.0 & 27.0  &  24.2 &   5.0 &   8.8 &  6.7 \\
50-55\% &  18.0 & 22.0  &  19.7 &   5.3 &   9.0 &  7.1 \\
55-60\% &  14.0 & 18.0  &  15.9 &   5.6 &   9.8 &  7.4 \\
60-65\% &  11.0 & 14.0  &  12.7 &   5.8 &  10.2 &  7.8 \\
65-70\% &   9.0 & 11.0  &   9.9 &   5.9 &  10.7 &  8.1 \\
70-75\% &   7.0 &  9.0  &   7.7 &   6.3 &  11.1 &  8.4 \\
75-80\% &   5.0 &  7.0  &   5.8 &   6.3 &  11.9 &  8.7 \\
80-85\% &   4.0 &  5.0  &   4.4 &   6.2 &  12.8 &  9.0 \\
85-90\% &   3.0 &  4.0  &   3.2 &   6.1 &  13.1 &  9.3 \\
90-95\% &   2.0 &  3.0  &   2.1 &   6.8 &  14.0 &  9.7 \\
\hline \hline
\end{tabular}
\end{center}
\end{table}

\begin{table}[H]
\caption{Centrality table for Cu+Cu at $200$ AGeV (1,000,000 initial conditions)}
\label{tab:cen:4} 
\begin{center}
\begin{tabular}{c|c|c|c|c|c|c}
\hline \hline
centrality (\%) & $N_\mathrm{part, min}$ & $N_\mathrm{part, max}$ & $N_\mathrm{part,avg}$ & $b_\mathrm{min}$ (fm) & $b_\mathrm{max}$ (fm) & $b_\mathrm{avg}$ (fm) \\
\hline
 0- 5\% & 102.0 & 125.0  & 110.3 &   0.0 &   3.4 &  1.5 \\
 5-10\% &  88.0 & 102.0  &  95.1 &   0.9 &   4.2 &  2.7 \\
10-15\% &  76.0 & 88.0  &  81.9 &   2.1 &   5.0 &  3.5 \\
15-20\% &  65.0 & 76.0  &  70.4 &   2.8 &   5.5 &  4.1 \\
20-25\% &  56.0 & 65.0  &  60.3 &   3.5 &   6.0 &  4.7 \\
25-30\% &  47.0 & 56.0  &  51.3 &   3.6 &   6.9 &  5.2 \\
30-35\% &  40.0 & 47.0  &  43.5 &   4.3 &   7.3 &  5.6 \\
35-40\% &  33.0 & 40.0  &  36.6 &   4.6 &   7.8 &  6.1 \\
40-45\% &  28.0 & 33.0  &  30.5 &   4.9 &   8.2 &  6.4 \\
45-50\% &  23.0 & 28.0  &  25.1 &   5.3 &   9.0 &  6.8 \\
50-55\% &  18.0 & 23.0  &  20.5 &   5.3 &   9.4 &  7.2 \\
55-60\% &  15.0 & 18.0  &  16.5 &   5.7 &  10.0 &  7.5 \\
60-65\% &  12.0 & 15.0  &  13.2 &   5.7 &  10.4 &  7.8 \\
65-70\% &   9.0 & 12.0  &  10.3 &   5.9 &  10.8 &  8.2 \\
70-75\% &   7.0 &  9.0  &   8.0 &   6.3 &  11.5 &  8.5 \\
75-80\% &   5.0 &  7.0  &   6.0 &   6.3 &  11.6 &  8.8 \\
80-85\% &   4.0 &  5.0  &   4.5 &   6.6 &  12.3 &  9.1 \\
85-90\% &   3.0 &  4.0  &   3.3 &   6.7 &  13.0 &  9.4 \\
90-95\% &   2.0 &  3.0  &   2.2 &   6.8 &  14.0 &  9.8 \\
\hline \hline
\end{tabular}
\end{center}
\end{table}

\begin{table}[H]
\caption{Centrality table for Pb+Pb at $2760$ AGeV (10,000,000 initial conditions)}
\label{tab:cen:5} 
\begin{center}
\begin{tabular}{c|c|c|c|c|c|c}
\hline \hline
centrality (\%) & $N_\mathrm{part, min}$ & $N_\mathrm{part, max}$ & $N_\mathrm{part,avg}$ & $b_\mathrm{min}$ (fm) & $b_\mathrm{max}$ (fm) & $b_\mathrm{avg}$ (fm) \\
\hline
 0- 5\% & 357.0 & 416.0  & 384.4 &   0.0 &   4.4 &  2.2 \\
 5-10\% & 305.0 & 357.0  & 330.3 &   2.3 &   5.7 &  4.1 \\
10-15\% & 261.0 & 305.0  & 282.6 &   3.8 &   6.8 &  5.3 \\
15-20\% & 223.0 & 261.0  & 241.4 &   4.8 &   7.7 &  6.2 \\
20-25\% & 189.0 & 223.0  & 205.4 &   5.6 &   8.7 &  7.1 \\
25-30\% & 159.0 & 189.0  & 173.9 &   6.4 &   9.4 &  7.8 \\
30-35\% & 133.0 & 159.0  & 146.1 &   7.2 &  10.1 &  8.5 \\
35-40\% & 110.0 & 133.0  & 121.7 &   7.7 &  10.7 &  9.1 \\
40-45\% &  90.0 & 110.0  & 100.2 &   8.3 &  11.4 &  9.7 \\
45-50\% &  73.0 & 90.0  &  81.4 &   8.7 &  12.1 & 10.3 \\
50-55\% &  58.0 & 73.0  &  65.0 &   9.3 &  12.9 & 10.8 \\
55-60\% &  45.0 & 58.0  &  50.9 &   9.6 &  13.4 & 11.3 \\
60-65\% &  34.0 & 45.0  &  39.0 &  10.1 &  14.1 & 11.8 \\
65-70\% &  25.0 & 34.0  &  29.0 &  10.2 &  14.7 & 12.3 \\
70-75\% &  17.0 & 25.0  &  20.9 &  10.4 &  15.2 & 12.7 \\
75-80\% &  12.0 & 17.0  &  14.4 &  11.0 &  16.1 & 13.2 \\
80-85\% &   7.0 & 12.0  &   9.5 &  11.3 &  16.4 & 13.6 \\
85-90\% &   5.0 &  7.0  &   5.9 &  11.5 &  17.2 & 14.1 \\
90-95\% &   2.0 &  5.0  &   3.4 &  11.8 &  18.5 & 14.5 \\
\hline \hline
\end{tabular}
\end{center}
\end{table}

\begin{table}[H]
\caption{Centrality table for Pb+Pb at $5500$ AGeV (1,000,000 initial conditions)}
\label{tab:cen:6} 
\begin{center}
\begin{tabular}{c|c|c|c|c|c|c}
\hline \hline
centrality (\%) & $N_\mathrm{part, min}$ & $N_\mathrm{part, max}$ & $N_\mathrm{part,avg}$ & $b_\mathrm{min}$ (fm) & $b_\mathrm{max}$ (fm) & $b_\mathrm{avg}$ (fm) \\
\hline
 0- 5\% & 360.0 & 416.0  & 386.9 &   0.0 &   4.3 &  2.3 \\
 5-10\% & 309.0 & 360.0  & 333.6 &   2.5 &   6.1 &  4.1 \\
10-15\% & 265.0 & 309.0  & 286.2 &   3.9 &   6.8 &  5.3 \\
15-20\% & 226.0 & 265.0  & 244.9 &   4.9 &   7.9 &  6.3 \\
20-25\% & 192.0 & 226.0  & 208.4 &   5.9 &   8.4 &  7.1 \\
25-30\% & 162.0 & 192.0  & 176.5 &   6.6 &   9.3 &  7.9 \\
30-35\% & 135.0 & 162.0  & 148.3 &   7.2 &  10.0 &  8.6 \\
35-40\% & 112.0 & 135.0  & 123.5 &   8.0 &  10.8 &  9.2 \\
40-45\% &  92.0 & 112.0  & 101.8 &   8.3 &  11.4 &  9.8 \\
45-50\% &  74.0 & 92.0  &  82.7 &   9.0 &  11.9 & 10.3 \\
50-55\% &  59.0 & 74.0  &  66.1 &   9.4 &  12.6 & 10.9 \\
55-60\% &  45.0 & 59.0  &  51.8 &   9.8 &  13.1 & 11.4 \\
60-65\% &  34.0 & 45.0  &  39.7 &  10.2 &  13.7 & 11.9 \\
65-70\% &  25.0 & 34.0  &  29.6 &  10.6 &  14.2 & 12.4 \\
70-75\% &  18.0 & 25.0  &  21.3 &  10.8 &  15.0 & 12.8 \\
75-80\% &  12.0 & 18.0  &  14.7 &  11.4 &  15.4 & 13.3 \\
80-85\% &   8.0 & 12.0  &   9.7 &  11.7 &  16.1 & 13.7 \\
85-90\% &   5.0 &  8.0  &   6.0 &  11.8 &  17.5 & 14.1 \\
90-95\% &   2.0 &  5.0  &   3.5 &  12.2 &  18.1 & 14.6 \\
\hline \hline
\end{tabular}
\end{center}
\end{table}

\begin{center}
\begin{longtable}{c|c|c|c|c|c|c}
\caption[]{Centrality table for Pb+Pb at $2760$ AGeV} \label{tab:cen:7} \\

\hline \hline
\multicolumn{1}{c|}{\textbf{centrality (\%)}} & 
\multicolumn{1}{c|}{\textbf{$N_\mathrm{part, min}$}} & 
\multicolumn{1}{c|}{\textbf{$N_\mathrm{part, max}$}} & 
\multicolumn{1}{c|}{\textbf{$N_\mathrm{part,avg}$}} &
\multicolumn{1}{c|}{\textbf{$b_\mathrm{min}$ (fm)}} &
\multicolumn{1}{c|}{\textbf{$b_\mathrm{max}$ (fm)}} &
\multicolumn{1}{c}{\textbf{$b_\mathrm{avg}$ (fm)}} \\
\hline
\endfirsthead

\multicolumn{7}{c}%
{{\bfseries \tablename\ \thetable{} -- continued from previous page}} \\
\hline \multicolumn{1}{c|}{\textbf{centrality (\%)}} & 
\multicolumn{1}{c|}{\textbf{$N_\mathrm{part, min}$}} & 
\multicolumn{1}{c|}{\textbf{$N_\mathrm{part, max}$}} & 
\multicolumn{1}{c|}{\textbf{$N_\mathrm{part,avg}$}} &
\multicolumn{1}{c|}{\textbf{$b_\mathrm{min}$ (fm)}} &
\multicolumn{1}{c|}{\textbf{$b_\mathrm{max}$ (fm)}} & \multicolumn{1}{c}{\textbf{$b_\mathrm{avg}$ (fm)}} \\ \hline 
\endhead

\hline \multicolumn{7}{r}{{Continued on next page}} \\
\endfoot

\hline \hline
\endlastfoot
 0- 1\% & 401.0 & 416.0  & 406.6 &   0.0 &   2.8 &  1.1 \\
 1- 2\% & 390.0 & 401.0  & 395.7 &   0.0 &   3.2 &  1.8 \\
 2- 3\% & 379.0 & 390.0  & 384.6 &   0.9 &   3.7 &  2.4 \\
 3- 4\% & 368.0 & 379.0  & 373.3 &   1.4 &   4.2 &  2.8 \\
 4- 5\% & 357.0 & 368.0  & 362.1 &   1.9 &   4.4 &  3.2 \\
 5- 6\% & 346.0 & 357.0  & 351.1 &   2.3 &   4.8 &  3.5 \\
 6- 7\% & 335.0 & 346.0  & 340.3 &   2.7 &   5.3 &  3.8 \\
 7- 8\% & 325.0 & 335.0  & 330.0 &   2.9 &   5.3 &  4.1 \\
\end{longtable}
\end{center}

\chapter{Feed down contribution tables for $\pi$, $K$, $p$, $\Lambda$, $\Sigma^+$, and $\Xi^-$ \cite{Qiu:2012tm}}
\label{sec:1210.7010:6}
%

This section contains tables for the resonance decay contributions as explained in \chap{resonance}. For all the tables listed in this section, $T_\mathrm{conv}{\,=\,}120$\,MeV.

\begin{center}
\begin{longtable}{c|c|c|c}
\caption[]{Resonance contribution list for $\pi^+$} \label{tab:reso:1} \\

\hline \multicolumn{1}{c|}{\textbf{name}} & \multicolumn{1}{c|}{\textbf{mass (GeV)}} & \multicolumn{1}{c|}{\textbf{total contribution (\%)}} \\ \hline 
\endfirsthead

\multicolumn{3}{c}%
{{\bfseries \tablename\ \thetable{} -- continued from previous page}} \\
\hline \multicolumn{1}{c|}{\textbf{name}} &
\multicolumn{1}{c|}{\textbf{mass (GeV)}} &
\multicolumn{1}{c|}{\textbf{total contribution (\%)}} \\ \hline 
\endhead

\hline \multicolumn{4}{r}{{Continued on next page}} \\
\endfoot

\hline \hline
\endlastfoot

$\omega$  &     0.78259 &     15.398  &    \\
$\rho^{0}$  &      0.7758 &     11.179  &    \\
$\rho^{+}$  &      0.7758 &     11.098  &    \\
$K^{*+}(892)$  &     0.89166 &       5.54  &    \\
$\bar{K}^{*0}(892)$  &      0.8961 &      5.355  &    \\
$\eta$  &     0.54775 &      4.682  &    \\
$\bar{\Delta}^{-}  (1232)$  &       1.232 &      2.613  &    \\
$\Delta^{++} (1232) $  &       1.232 &      2.606  &    \\
\end{longtable}
\end{center}

\begin{center}
\begin{longtable}{c|c|c|c}
\caption[]{Resonance contribution list for $K^+$} \label{tab:reso:2} \\

\hline \multicolumn{1}{c|}{\textbf{name}} & \multicolumn{1}{c|}{\textbf{mass (GeV)}} & \multicolumn{1}{c|}{\textbf{total contribution (\%)}} \\ \hline 
\endfirsthead

\multicolumn{3}{c}%
{{\bfseries \tablename\ \thetable{} -- continued from previous page}} \\
\hline \multicolumn{1}{c|}{\textbf{name}} &
\multicolumn{1}{c|}{\textbf{mass (GeV)}} &
\multicolumn{1}{c|}{\textbf{total contribution (\%)}} \\ \hline 
\endhead

\hline \multicolumn{4}{r}{{Continued on next page}} \\
\endfoot

\hline \hline
\endlastfoot

$K^{*0}(892)$  &      0.8961 &     35.857  &    \\
$K^{*+}(892)$  &     0.89166 &      18.52  &    \\
$\phi(1020)$  &      1.0195 &     16.036  &   60\% \\ \hline
$K_1^{+} (1270) $  &       1.273 &      3.631  &    \\
$K_1^{0} (1270) $  &       1.273 &      3.287  &    \\
$a_0^{+}(980)$  &      0.9847 &      1.807  &    \\
$f_1 (1420)$  &      1.4263 &      1.754  &  80\%  \\ \hline
$K_2^{*0} (1430) $  &      1.4324 &       1.61  &    \\
\end{longtable}
\end{center}

\begin{center}
\begin{longtable}{c|c|c|c}
\caption[]{Resonance contribution list for $p$} \label{tab:reso:3} \\

\hline \multicolumn{1}{c|}{\textbf{name}} & \multicolumn{1}{c|}{\textbf{mass (GeV)}} & \multicolumn{1}{c|}{\textbf{total contribution (\%)}} \\ \hline 
\endfirsthead

\multicolumn{3}{c}%
{{\bfseries \tablename\ \thetable{} -- continued from previous page}} \\
\hline \multicolumn{1}{c|}{\textbf{name}} &
\multicolumn{1}{c|}{\textbf{mass (GeV)}} &
\multicolumn{1}{c|}{\textbf{total contribution (\%)}} \\ \hline 
\endhead

\hline \multicolumn{4}{r}{{Continued on next page}} \\
\endfoot

\hline \hline
\endlastfoot

$\Delta^{++} (1232) $  &       1.232 &     29.842  &    \\
$\Delta^{+} (1232) $  &       1.232 &     19.816  &    \\
$\Delta^{0} (1232) $  &       1.232 &      9.813  &    \\
$\Delta^{++} (1600) $  &         1.6 &      2.787  &  60\%  \\ \hline
$n (1520)$  &        1.52 &      2.487  &    \\
$p (1520)$  &        1.52 &      2.169  &    \\
$\Delta^{+} (1600) $  &         1.6 &      2.049  &    \\
$p (1440)$  &        1.44 &      2.034  &    \\
\end{longtable}
\end{center}

\begin{center}
\begin{longtable}{c|c|c|c}
\caption[]{Resonance contribution list for $\Lambda$} \label{tab:reso:4} \\

\hline \multicolumn{1}{c|}{\textbf{name}} & \multicolumn{1}{c|}{\textbf{mass (GeV)}} & \multicolumn{1}{c|}{\textbf{total contribution (\%)}} \\ \hline 
\endfirsthead

\multicolumn{3}{c}%
{{\bfseries \tablename\ \thetable{} -- continued from previous page}} \\
\hline \multicolumn{1}{c|}{\textbf{name}} &
\multicolumn{1}{c|}{\textbf{mass (GeV)}} &
\multicolumn{1}{c|}{\textbf{total contribution (\%)}} \\ \hline 
\endhead

\hline \multicolumn{4}{r}{{Continued on next page}} \\
\endfoot

\hline \hline
\endlastfoot
$\Sigma ^{0}$  &      1.1926 &     24.775  &    \\
$\Sigma ^{+}(1385) $  &      1.3828 &     17.893  &    \\
$\Sigma^{-} (1385) $  &      1.3872 &     17.346  &   60\% \\ \hline
$\Sigma^{0} (1385) $  &      1.3837 &     16.555  &    \\
$\Lambda (1405)$  &      1.4065 &      3.103  &    \\
$\Lambda (1520)$  &      1.5195 &       2.18  &   80\% \\ \hline
$\Sigma ^{-} (1670)$  &        1.67 &      1.304  &    \\
$\Sigma ^{+} (1670)$  &        1.67 &      1.297  &    \\
\end{longtable}
\end{center}

\begin{center}
\begin{longtable}{c|c|c|c}
\caption[]{Resonance contribution list for $\Sigma^+$} \label{tab:reso:5} \\

\hline \multicolumn{1}{c|}{\textbf{name}} & \multicolumn{1}{c|}{\textbf{mass (GeV)}} & \multicolumn{1}{c|}{\textbf{total contribution (\%)}} \\ \hline 
\endfirsthead

\multicolumn{3}{c}%
{{\bfseries \tablename\ \thetable{} -- continued from previous page}} \\
\hline \multicolumn{1}{c|}{\textbf{name}} &
\multicolumn{1}{c|}{\textbf{mass (GeV)}} &
\multicolumn{1}{c|}{\textbf{total contribution (\%)}} \\ \hline 
\endhead

\hline \multicolumn{4}{r}{{Continued on next page}} \\
\endfoot

\hline \hline
\endlastfoot
$\Lambda (1405)$  &      1.4065 &     25.159  &    \\
$\Lambda (1520)$  &      1.5195 &     10.121  &    \\
$\Sigma^{+} (1385) $  &      1.3828 &      9.288  &    \\
$\Sigma^{0} (1385) $  &      1.3837 &      9.179  &    \\
$\Sigma ^{0} (1670)$  &        1.67 &      6.648  &  60\%  \\ \hline
$\Sigma ^{+} (1670)$  &        1.67 &      6.618  &    \\
$\Xi ^{0} (1690)$  &        1.69 &      4.578  &    \\
$\Sigma ^{+} (1750)$  &        1.75 &      4.425  &    \\
\end{longtable}
\end{center}

\begin{center}
\begin{longtable}{c|c|c|c}
\caption[]{Resonance contribution list for $\Xi^-$} \label{tab:reso:6} \\

\hline \multicolumn{1}{c|}{\textbf{name}} & \multicolumn{1}{c|}{\textbf{mass (GeV)}} & \multicolumn{1}{c|}{\textbf{total contribution (\%)}} \\ \hline 
\endfirsthead

\multicolumn{3}{c}%
{{\bfseries \tablename\ \thetable{} -- continued from previous page}} \\
\hline \multicolumn{1}{c|}{\textbf{name}} &
\multicolumn{1}{c|}{\textbf{mass (GeV)}} &
\multicolumn{1}{c|}{\textbf{total contribution (\%)}} \\ \hline 
\endhead

\hline \multicolumn{4}{r}{{Continued on next page}} \\
\endfoot

\hline \hline
\endlastfoot
$\Xi^{0} (1530)$  &      1.5318 &     62.049  &   60\% \\  \hline
$\Xi ^{-} (1530)$  &       1.535 &     30.103  &    90\% \\ \hline
$\Xi ^{0} (1820)$  &       1.823 &       2.42  &    \\
$\Xi ^{-} (1820)$  &       1.823 &      2.334  &    95\% \\ \hline
$\Xi^{0} (1950)$  &        1.95 &      1.427  &     98\% \\ \hline
$\Omega(2250)$  &       2.252 &      0.957  &  99\%  \\ \hline
$\Xi ^{-} (1950)$  &        1.95 &       0.71  &    100\% \\
\end{longtable}
\end{center}


\chapter{Compiling, running, and tuning iSS}
\label{sec:iSS:running_code}

\section{Compiling}
The iS and iSS\footnote{They can be found from the iEBE package at \iEBEURL.} codes are easily compiled using the provided {\it Zmake.sh} script files. These are not real makefiles, but are actually short bash scripts that compile the code in one line. Makefiles are convenient for compiling large scale programs whenever keeping the intermediate output binary files saves compiling time. For small scale programs, however, a cleaner direct compile is preferable. The script also tries to compile with Intel compilers when available, to generate faster executables.

\section{Running}
After compilation, executables with the name ``iS.e'' or ``iSS.e'' are generated. They can be executed either without supplying any command line parameters, or with specific parameter sets using the {\it variable=value} syntax. These programs read input data files from the {\it results} directory, and write output data files to the same directory; details for this are given in the following sections. The iSS program reads {\it parameters.dat} for all the parameter assignments, and overwrites them with any additional parameter assignments read from the command line arguments. The structure of the parameter file and explanations for parameters it contains are presented in the following sections.

The number of outputs can be modified by editing the {\it AMOUNT\_OF\_OUTPUT} macro in the {\it emissionfunction.cpp} file.

\section{Input and output files}
Currently\footnote{By \today, iSS has version 2.3.0.2 and iS has version 1.2.1.12.} both programs read as input the freeze-out surface information data files {\it surface.dat} and {\it decdat2.dat}, and the chemical potential file {\it decdat\_mu.dat} from the {\it results} directory, which are all output files from the VISH2+1 hydrodynamical simulation program. For the format of these files, refer to the corresponding documents.

Both programs also need equation of state (EOS) information and particle data information, which are given by files in the {\it EOS} directory: {\it EOS\_particletable.dat}, {\it pdg.dat}, {\it resoweak.dat}. Another file {\it chosen\_particles.dat} in the same directory is a one-column list of standard particle Monte-Carlo indices, and only particles whose Monte Carlo index is included in this file will be processed by the programs; others will be skipped. The Monte-Carlo indices can be looked up in the {\it pdg.dat} file, for example, $\pi^0$ is indexed as $211$.

When the program finishes, depending on the operations that were performed, different sets of files will be generated in the {\it results} folder.

If spectra and flow calculations are enabled (optional in iSS, mandatory in iS), files with names like {\it thermal\_xxxxx\_vn.dat} and {\it thermal\_xxxxx\_integrated\_vn.dat} will be generated. Here {\it xxxxx} refers to the standard Monte-Carlo particle index given by the particle data book ({\it pdg.dat}).

The files {\it thermal\_xxxxx\_integrated\_vn.dat} record $\pT$-integrated flow results. These files contain $n$ rows of output, where $n$ is the largest harmonic flow coefficient calculated, starting from $0$. Each row has $6$ columns: the first column is the order of the flow, the second and the third columns contain the real and imaginary part of the numerator on the right hand side of \equ{iSS:anisotropic_flows}\footnote{With an additional factor $1/2\pi$; therefore the second quantity in the first row is the total multiplicity. These two columns are mainly for debugging purposes.}, and the fourth, fifth, and sixth columns contain the real and imaginary parts, and the magnitude of the complex flow vector $v_n e^{in\,\PsiEP{n}}$.

The files {\it thermal\_xxxxx\_vn.dat} store differential flow results. These are block-shaped data files that store the $dN/(\pT d\pT d\phi)$ matrix for given $\pT$ and $\phi$ indices. The row index corresponds to the $\pT$ index and the column index to the $\phi$ index. For each $\pT$ or $\phi$ index, the actual $\pT$ or $\phi$ value can be looked up from the corresponding tables in the {\it tables} directory, which by default are the {\it pT\_gauss\_table.dat} or {\it phi\_gauss\_table.dat} files. The {\it tables} directory will be explained below.

If the sampling of particles is enabled (optional with iSS, not possible with iS), files with names like {\it samples\_xxxxx.dat}, {\it samples\_control\_xxxxx.dat}, and {\it samples\_format.dat} are generated, with {\it xxxxx} the Monte-Carlo particle indices.

The files {\it samples\_control\_xxxxx.dat} contain a one-column listing of the actual number of sampled particles from each of the given number of repeated samplings. The sum of all its elements gives the total number of sampled particles over the specified rapidity range whose average is given by the $\pT$-integrated Cooper-Frye formula.

The files {\it samples\_xxxxx.dat} contain the information of the generated samples, with each line corresponding to one particle and containing all its spatial and momentum information. The meaning of elements in a particular column varies from version to version, so they are recorded in the {\it samples\_format.dat} file to ease the reading process. The {\it samples\_format.dat} file is written with ``equal-sign assignment" syntax, which can be conveniently read back using the \itt{ParameterReader} class (\ref{subsec:iSS:class_parameter_reader}). For example, the line {\it tau = 2} means that the $\tau$ information of the particles are written in the second column.

The iSS program also supports writing to OSCAR format\footnote{http://karman.physics.purdue.edu/OSCAR-old/}, in which case the file {\it OSCAR\_header.txt} is copied to as the header of the generated OSCAR file.

\section{Parameter file}
The {\it Parameters.dat} file is an ``equal-sign assignment'' file storing the major tunable parameters used by the iSS program. Such a file can be easily read using the \itt{ParameterReader} class (\S\ref{subsec:iSS:class_parameter_reader}).

All the parameters are accompanied by detailed explanations. The most important parameters are:
\begin{itemize}
  \item {\it calculate\_vn}: When set to $1$, spectra and flows will be calculated.
  \item {\it MC\_sampling}: When set to $0$, no sampling will be done; setting it to $1$ generates samples using the numerical sampling approach (\S\ref{subsec:iSS:numerical_approach}); setting it to $2$ generates samples using the semi-analytic approach (\S\ref{subsec:iSS:analytic_approach}).
  \item {\it number\_of\_repeated\_sampling}: How many repeated sampling should be performed.
  \item {\it dN\_dy\_sampling\_model}: Used to switch between different models for generating the integer $dN/dy$ from the averaged value, see \ref{subsec:iSS:optimizations_and_models}.
  \item {\it use\_dynamic\_maximum}: Whether to turn on the ``dynamic maximum'' mentioned in \S\ref{subsec:iSS:analytic_approach}.
  \item {\it grouping\_particles} and {\it grouping\_tolerance}: if {\it grouping\_particles} is set to $1$, particles with similar mass and chemical potentials will be grouped together for calculation, and the threshold for determining such similarity is controlled by the {\it grouping\_tolerance} parameter, see \S\ref{subsec:iSS:optimizations_and_models}.
\end{itemize}
Other parameters are explained by their comments in the file.

\section{Tables}

Tables (\S\ref{subsec:iSS:classes_table_and_table_function}) are used extensively throughout the iS and iSS programs. They are read from block-shaped data files located under the {\it tables} directory using \itt{Table} or \itt{TableFunction} class.

In particular, all the numerical integrals are performed by evaluating the sum:
\begin{equation} \label{eq:iSS:numerical_sum}
  \int_a^b f(x) = \sum_i f(x_i) w_i, \,x_i\in[a,b]
\end{equation}
with the location-weight pair $(x_i, w_i)$ specified by some table. For example, it is set in the {\it main.cpp} file that the $\pT$ integral should be done using the {\it pT\_gauss\_table.dat} file, which stores location and weight information for integration using Gauss quadrature. Other integration methods (e.g. Simpson's method etc.), can be conveniently substituted by simply switching the integration table file, without changing the source program. Similarly the file {\it phi\_gauss\_table.dat} and {\it eta\_gauss\_table\_20\_full.dat} are currently used to perform the $\phi$ and $\eta_s$ integrals.

Files {\it p\_integral\_table\_0.02.dat} and {\it m\_integral\_table\_0.02.dat} are the pre-calculated tables used to evaluate the integral in \equ{iSS:dN_integral_5}. The {\it lambertw\_function.dat} and {\it z\_exp\_m\_z.dat} files contain tables solving the transcendental equations (\ref{eq:iSS:maximum_core_equ_2}).

There are other files for tables used during the binning process of the samples.

\chapter{Introduction to iSS support classes}
\label{sec:iSS:support_classes}

As mentioned in the introduction of \chap{iSS}, common tasks are encapsulated into classes for re-usability. In this section, such classes are introduced briefly; details can be found in the codes.

\section{Table class and table-function classes}
\label{subsec:iSS:classes_table_and_table_function}

It is common to load a block of data of unknown size into memory to analyze. The \itt{Table} class is made for such a purpose. It can be initialized from block-form data file or double array. It provides an interface to read and write elements in the table. The size of the table is dynamically allocated and it can change automatically once a write-action exceeds the current data boundary. It can perform interpolation between any two specified columns of data using nearest-neighbor, linear, or cubic methods, and it can perform bilinear interpolation using the whole block as a matrix as well. The \itt{Table} class is used extensively throughout the iSS program to read in replaceable tables, like those used for integration or pre-tabulations.

The \itt{TableFunction} class is oriented for 2-column tables that represent functions. It can be initialized from a file or directly, and it provides interpolation and inverse search interfaces. This class is intended to be used as the underlying class for representing numerical functions using tables.

\section{Classes related to random variables}
\label{subsec:iSS:classes_random_variables}

The sampling of random variables are implemented in the base class \itt{RandomVariable}, which supports direct PDF sampling, sampling using the inverse CDF, and sampling with envelope distribution methods.

The PDF function and the inverse CDF function for the random variable being sampled and for the envelope random variables are implemented by default using a table through \itt{TableFunction} class, which can conveniently return results using several methods of interpolation, and which can be initialized  from file, double array, or generated directly. All these distribution functions are declared as virtual, meaning that, when the class is inherited, these functions can be overloaded and implemented in any desired way. For example, one implemented derived class \itt{NBD} overloads the PDF function by an analytic expression, since the PDF depends on several parameters and it is more naturally expressed analytically, rather than in tabulated form.

Once the function required for one sampling method is loaded, the sampling of the random variable can be done by calling the corresponding sampling member function. The class also has a member function that can calculate the inverse CDF function from the PDF function, and a function that can automatically generate a stair-shaped envelope function (see \S\ref{subsec:iSS:automatic_envelope}) given the mode and standard deviation.

The \itt{RandomVariable} class is meant to be the base class for the most general 1d random variable sampling. One derived class based on it is the \itt{NBD} class which samples the negative binomial distribution (NBD) using an automatically constructed stair-shaped envelope function (see \S\ref{subsec:iSS:automatic_envelope} and code). The PDF for the NBD is overloaded as an analytic function; for efficiency, the automatically generated envelope function is recalculated only when the parameters for NBD change.

There are several other simpler classes \itt{RandomVariable1DArray}, \itt{RandomVariable2DArray}, and \itt{RandomVariableNDArray} that are designed for sampling discrete PDFs given as double arrays, using the inverse CDF sampling method. They are less general compared to the \itt{RandomVariable} class, but they are optimized for index sampling and they require less memory.

\section{Parameter reader class}
\label{subsec:iSS:class_parameter_reader}

It is very common in scientific programming that results depend on many parameters. It would be convenient if they could be stored and loaded from files, which are then passed between functions via references instead of by values. The \itt{ParameterReader} class is created for such a purpose. Parameters can be assigned using ``equal-sign assignments" syntax throughout the file, from command line arguments, or by calling member functions directly. After initialization, the parameters are stored and can be extracted easily. With such a class, parameters can be passed between functions and classes using an instance of this class, instead of being passed directly, since this would be tedious to maintain, and prone to errors.

The most common usage of this class is to read parameters stored in a file, which can be done through a member function or simply through the constructor. In such cases, the file has to be written with ``equal-sign assignment'' syntax, meaning that each line of the file can contain an equal-sign assignment, a piece of comment beginning with the character ``\#'', or neither, or both. For example, the line:

{\it MC\_sampling = 2                        \# 0/1/2: whether to perform Monte-Carlo sampling}

\noindent creates a variable with string name ``MC\_sampling'' and assigns the initial value ``2'' to it, which can be conveniently extracted by the \itt{getVal} member function. The part of the line after the ``\#'' symbol is treated as a comment and is discarded during the reading process.

There is another member function that reads the command line arguments for parameter assignments. In this case, each argument needs to be an ``equal-sign assignment'', without blanks or comments; for details, see the header file.

Parameters read later will overwrite any parameters with the same name read earlier; otherwise, new variables will be created. Parameters can also be assigned directly using a member function. For other functionalities and details, refer to the header file and comments directly following the implementation of each function.

\chapter{The iEBE package}
\label{sec:iEBE}

The iEBE package is a convenient package for automating event-by-event hybrid calculations. It divide calculations into ``jobs", where each job consists of multiple ``ebe-calculations". Each ``ebe-calculation" is a complete hybrid calculation that in execution order performs: heavy-ion event generation (superMC), hydrodynamics simulation (VISHNew), particle emission sampling (iSS), hadron rescattering simulation (osc2u and urqmd), flow calculation (binUtilities), and finally, collection and storing of important results in databases (EbeCollector). Each ``job" runs the given number of ``ebe-calculations" sequentially, and ``jobs" are run in parallel. The package has utility scripts that can combine the generated SQLite database files from different jobs into one, which can be analyzed later.

The main programs are contained in the subfolder ``EBE-Node", which is used to perform one job, and which will be duplicated when multiple jobs are run. The package needs two locations to perform multi-job calculations: one folder is used to store duplications of ``EBE-Node" and intermediate results generated during the calculation (refer to as ``working folder" in the following), and another folder is used to store final results (refer to as ``result folder" in the following). By default, the working folder is named ``PlayGround" and the result folder is named ``RESULTS", both in the root directory of the package.

\section{How to use the package to perform multi-job calculations}

This section explains how to use the highest-level scripts provided by the package to perform event-by-event hybrid calculations. Any one using the package should read this section thoroughly, even those who are not interested in modifying the package.

{\em VERY IMPORTANT:} Make sure you have Python 2.7+ (or Python 3) installed before proceeding.

In the following, all paths are relative to the root directory of the package.

\textbf{Step 1} Generate jobs using the ./generateJobs.py script.

To generate jobs, use the generateJobs.py script in the root directory. Most of the runnable scripts in this package provide the feature that if you run them without additional arguments, they will print the usage echo, for example:

\$ ./generateJobs.py

\noindent And you should see the output:

Usage: generateJobs.py number\_of\_jobs number\_of\_events\_per\_job

[working\_folder=``./PlayGround''] [results\_folder=``./RESULTS'']

[walltime=``03:00:00'' (per event)] [compress\_results\_folder=``yes'']

\noindent The echo says that the 1st argument for the script should specify the number of jobs you want to generate; the 2nd argument specifies the number of ebe-calculations for each job; the 3rd argument points to the result folder; the 4th argument specifies the ``wall time" (used in torque system, explained later); the 5th argument points to the working folder; and the 6th argument is for whether to compress final results. Except for the first two, all other arguments have default values. The simplest way to generate jobs is just to accept the default values. As an example, to generate 2 jobs, each performing 5 ebe-calculations, simply do the following:

\$ ./generateJobs.py 2 5

\noindent This script will first check required libraries, compile any programs not existing already, and generate the actual folders for jobs.

After you see the echo ``Jobs generated.", you should see the working folder ``PlayGround" and the result folder ``RESULTS" in the root directory.

\textbf{Step 2} Submit jobs.

The way to submit jobs depends on the system. For a cluster that has the ``Torque" scheduling system (therefore the ``qsub" command is available), submit jobs using the submitJobs\_qsub.py script; for a local computation, use the submitJobs\_local.py script. The difference is that the local computation is only paralleled for the local CPUs and calculation on the cluster, via the torque system, will be distributed (trivailly) to multiple nodes. To submit a local calculation, simply type (the script knows how to get the location of the working folder automatically):

\$ ./submitJobs\_local.py

\noindent You should see some feedbacks listing the jobs that have been submitted.

\textbf{Step 3} Checking progress.

Progress for all jobs can be checked by the progressReport.py script in the root directory:

\$ ./progressReport.py

\noindent It will list the current progress for all jobs.

\textbf{Step 4} Combining databases.

Once all calculations are finished, the generated database files from all events will be combined automatically, and a single file ``collected.db" will be generated in the results folder.

\section{How to analyze generated data}

The ``collected.db" generated from previous steps is the SQLite database file that can be analyzed by any desired means. The recommended way is to use the uhg.py script in the utilities folder. This script not only reads the database, but also performs additional analyses, like interpolation along $p_T$, calculation of the mean, etc. It can either be run from the command line to evaluate a single expression or interactively from a shell. To evaluate a single expression, run the uhg.py script in the utilities folder:

\$ ./uhg.py database\_filename ``expression to evaluate"

A more convenient way to evaluate multiple expressions as well as perform additional analyses is to run the uhg.py script interactively. For example:

\$ python -ic ``from uhg import *"

The interactive mode will also print out a simple help showing recognizable symbols that can be included in the expression.

Another way is to use the databaseQuery.py script to evaluate a single piece of SQL query from the command line:

\# ./databaseQuery.py database\_filename ``SQL\_query"

A third way is to use unpackDatabase.py under /EBE-Node/EbeCollector/ to dump the whole database into separated space-separated text files, each for individual tables. For example, running the following under /RESULTS/:

\$ ../EBE-Node/EbeCollector/unpackDatabase.py ./collected.db .

\noindent will generate several ``.dat" files, each containing data for the corresponding type. Each file has a one-line header to indicate what data each column records, and the rest are data separated by spaces.

For more details about the structure of the database and the uhg.py script, see /EBE-Node/EbeCollector/EbeCollector\_readme.txt.

\section{How to tune parameters}

After you have familiarized yourself with how to perform multi-job hybrid calculations, you are finally ready to tune parameters for the simulations. The most commonly tuned parameters are in the ParameterDict.py file in the directory, which should be the only file used to direct the simulations. This file will be copied to the result folder for record-keeping purposes when generating jobs.

\end{document}